\documentclass[numberedappendix,twocolappendix,iop,apj]{emulateapj}
\usepackage{psfig,amsfonts,amsmath,graphicx,natbib,apjfonts,nicefrac,hyperref,breakurl,booktabs}

\hypersetup{colorlinks=true, urlcolor=blue, citecolor=blue}

\newcommand{\Swift}{\textit{Swift}}

\newcommand{\EK}{\ensuremath{E_{\rm K}}}
\newcommand{\EKiso}{\ensuremath{E_{\rm K,iso}}}
\newcommand{\Egamma}{\ensuremath{E_{\gamma}}}
\newcommand{\Egammaiso}{\ensuremath{E_{\gamma,\rm iso}}}
\newcommand{\epse}{\ensuremath{\epsilon_{\rm e}}}
\newcommand{\epsb}{\ensuremath{\epsilon_{B}}}
\newcommand{\dens}{\ensuremath{n_{0}}}
\newcommand{\Astar}{\ensuremath{A_{*}}}
\newcommand{\tdec}{\ensuremath{t_{\rm dec}}}
\newcommand{\tjet}{\ensuremath{t_{\rm jet}}}
\newcommand{\thetajet}{\ensuremath{\theta_{\rm jet}}}

\newcommand{\AV}{\ensuremath{A_{\rm V}}}

\newcommand{\pcmsq}{\ensuremath{{\rm cm}^{-2}}}
\newcommand{\pcc}{\ensuremath{{\rm cm}^{-3}}}
\newcommand{\emcee}{\textsc{emcee}}
\newcommand{\me}{GRB~120326A}

\newcommand{\nua}{\ensuremath{\nu_{\rm a}}}
\newcommand{\nusa}{\ensuremath{\nu_{\rm sa}}}
\newcommand{\nuac}{\ensuremath{\nu_{\rm ac}}}
\newcommand{\numax}{\ensuremath{\nu_{\rm m}}}
\newcommand{\nuc}{\ensuremath{\nu_{\rm c}}}

\newcommand{\nuX}{\ensuremath{\nu_{\rm X}}}
\newcommand{\nuopt}{\ensuremath{\nu_{\rm opt}}}
\newcommand{\nuNIR}{\ensuremath{\nu_{\rm NIR}}}

\def\cfa{1}
\def\caltech{2}

\def\racah{3}
\def\einsteinfellow{4}
\def\az{5}

\begin{document} 

\submitted{Submitted to ApJ}

\title{\textsc{Energy Injection in Gamma-ray Burst Afterglows}}

\author{
Tanmoy Laskar\altaffilmark{\cfa},
Edo Berger\altaffilmark{\cfa},
Raffaella Margutti\altaffilmark{\cfa},
Daniel Perley\altaffilmark{\caltech},\\
B.~Ashley Zauderer\altaffilmark{\cfa},
Re\'em Sari\altaffilmark{\racah}
and Wen-fai Fong\altaffilmark{\einsteinfellow,\az}
}

\altaffiltext{\cfa}{Harvard-Smithsonian Center for Astrophysics, 60
Garden Street, Cambridge, MA 02138}
\altaffiltext{\caltech}{Department of Astronomy, California Institute of Technology,
MC 249-17, 1200 East California Blvd, Pasadena CA 91125, USA}
\altaffiltext{\racah}{Racah Institute of Physics, Edmund J. Safra Campus, Hebrew University of 
Jerusalem, Jerusalem 91904, Israel}
\altaffiltext{\einsteinfellow}{Einstein Fellow}
\altaffiltext{\az}{University of Arizona, 933 N. Cherry Ave, Tucson, AZ 85721}

\shorttitle{Energy injection in GRBs}
\shortauthors{Laskar et al.}

\begin{abstract}
We present multi-wavelength observations and modeling of Gamma-ray Bursts (GRBs) that exhibit a 
simultaneous re-brightening in their X-ray and optical light curves, and are also detected at radio 
wavelengths. We show that the re-brightening episodes can be modeled by injection of energy into 
the 
blastwave and that in all cases the energy injection rate falls within the theoretical bounds 
expected for a distribution of energy with ejecta Lorentz factor. Our measured values of the 
circumburst density, jet opening angle, and beaming corrected kinetic energy are consistent with 
the distribution of these parameters for long-duration GRBs at both $z\sim1$ and $z\gtrsim6$, 
suggesting that the jet launching mechanism and environment of these events are similar to that of 
GRBs that do not have bumps in their light curves. However, events exhibiting re-brightening 
episodes have lower radiative efficiencies than average, suggesting that a majority of the kinetic 
energy of the outflow is carried by slow-moving ejecta, which is further supported by steep 
measured distributions of the ejecta energy as a function of Lorentz factor. We do not find 
evidence 
for reverse shocks over the energy injection period, implying that the onset of energy injection is 
a gentle process. We further show that GRBs exhibiting simultaneous X-ray and optical 
re-brightenings are likely the tail of a distribution of events with varying rates of energy 
injection, forming the most extreme events in their class. Future X-ray observations of GRB 
afterglows with \Swift\ and its successors will thus likely discover several more such events, 
while 
radio follow-up and multi-wavelength modeling of similar events will unveil the role of energy 
injection in GRB afterglows.
\end{abstract}

\keywords{gamma-ray burst: general -- gamma-ray burst: individual (GRB~100418A, 
GRB~100901A, GRB~120326A, GRB~120404A)}
 
\LongTables

\section{Introduction}
Gamma-ray bursts (GRBs) have traditionally been modeled as point explosions that inject 
$\sim10^{51}$\,erg of energy into a collimated, relativistically-expanding fireball over a period 
of a few seconds. In this model, the subsequent afterglow radiation is synchrotron emission produced 
by the interaction of the relativistic ejecta with the circumburst medium. Depending on the density 
profile of the ambient medium, usually assumed to be either uniform (`ISM-like') or falling with 
radius as $r^{-2}$ (`wind-like'), this model has several verifiable predictions: smooth light curves 
at all frequencies from the X-rays to the radio, which rise and fall as the peak of the spectral 
energy distribution evolves through the observer band; a `jet break' as the expanding ejecta 
decelerate and begin to spread sideways; and an eventual transition to the sub-relativistic regime 
where the ejecta become quasi-spherical. In this framework, the energetics of the explosion and the 
properties of the environment can be determined from fitting light curves with the synchrotron 
model, while a measurement of the jet break allows for a determination of the angle of collimation 
of the outflow and the calculation of geometric corrections to the inferred energy.

Despite its simplicity, this model was quite successful in the study of a large number of GRB 
afterglows (e.g.~\citealt{bsf+00,pk01,pk02,yhsf03}) until the launch of \Swift\ in 2004 
\citep{gcg+04}. With its rapid-response X-ray and UV/optical afterglow measurements, \Swift\ has 
revolutionized the study of GRB afterglows. Rapidly-available localizations have allowed detailed 
ground-based follow-up. X-ray observations from \Swift\ have been even more revolutionary, 
revealing light curves with multiple breaks falling into a `canonical' series, consisting of a 
steep decay, plateau, and normal decay, sometimes with evidence for jet-breaks. While the steep 
decay has been associated with the prompt emission \citep{tgc+05,owo+06} and the normal and post 
jet-break decay phases are associated with the afterglow, the plateaus cannot be explained by the 
standard model (\citealt{zfd+06}; however, see recent numerical calculations by \citealt{dm14}, 
which suggest that the plateaus may be a natural consequence of a coasting phase in the jet dynamics 
between $10^{13}$ to $10^{16}$\,cm from the progenitor). In addition, short ($\Delta t/t \ll 1$) 
flares that rise and decline rapidly and exhibit large flux variations ($\Delta F/F \gg 1$; 
\citealt{cmm+10,mgc+10}) are often seen superposed on the light curves. These flares are believed 
to be more closely associated with the GRB prompt emission than with the afterglow, and have been 
interpreted as late-time activity by the central engine 
\citep{fbl+06,rmb+06,wlm06,pmk+06,lzo+06,lp07,mgc+10,gdf+15}. Some GRBs also exhibit optical 
flares, which do not always correspond to flares in the X-ray light curves \citep{llt+12}. 

These new features of \Swift\ X-ray light curves cannot be simply explained in the traditional 
picture. New physical mechanisms such as energy injection, circumburst density enhancements, 
structured jets, viewing angle effects, varying microphysical parameters, and gravitational 
micro-lensing have been invoked to explain various features of \Swift\ X-ray light curves 
\citep{zfd+06,nkg+06,pmg+06,tiyn06,eg06,gkp06,jyfw07,sd07,kwhc10,dm14,uz14}. 
However, although a wealth of information is available from X-ray light curves in general, 
definitive statements on the physical origin of these features requires synergy with observations 
at other wavelengths. Ultraviolet (UV), optical, near infra-red (NIR), millimeter, and radio data 
probe distinct parts of the afterglow spectral energy distribution (SED), and the various physical 
mechanisms are expected to influence light curves in these bands differently. Thus, a detailed 
analysis of GRB afterglows requires multi-wavelength data and modeling.

Of the GRBs with plateaus in their X-ray light curves, there is a small class of peculiar events 
that additionally exhibit an X-ray re-brightening of a non-flaring origin ($\Delta T / T \sim 1$), 
and an even smaller class where the re-brightening appears to occur simultaneously in both 
the optical 
and X-rays \citep{mhm+07, llt+12, pvw13}. An exemplar of this latter class is GRB~120326A, which 
exhibits a peak in its well-sampled X-ray light curve at around 0.3\,d together with a simultaneous 
optical re-brightening. \cite{uht+14} reported optical and millimeter observations of the 
afterglow of GRB~120326A, and invoked synchrotron self inverse-Compton radiation from a reverse 
shock to explain the millimeter, optical, and X-ray light curves. By fitting energy-injection 
models 
in a wind-like circumburst environment to the X-ray and optical $R$-band light curves, 
\cite{hgw+14} proposed that a newborn millisecond pulsar with a strong wind was responsible for the 
re-brightening. \cite{mvg+14} present multi-band optical and NIR light curves of this event, and 
explore various physical scenarios for the re-brightening, including the onset of the afterglow, 
passage of a synchrotron break frequency through the observing band, and geometrical effects.

Here we report detailed radio observations of this event spanning 4 -- 220 GHz and 0.3 to 120\,d, 
making this the first GRB with an achromatic re-brightening and with such a rich multi-band data 
set. We perform the first broad-band modeling for this event using a physical 
GRB afterglow model \citep{gs02} using methods described in \cite{lbz+13} and \cite{lbt+14}, 
employing Markov Chain Monte Carlo procedures to characterize the blastwave shock. Our radio 
observations allow us to constrain the synchrotron self-absorption frequency, and to unambiguously 
locate the synchrotron peak frequency, together placing strong constraints on the nature of the 
X-ray/UV/optical re-brightening. We find the clear signature of a jet break at all wavelengths 
from the radio through the X-rays, allowing us to constrain the true energetics of this event.

We next consider and test various physical processes that may cause a re-brightening in the 
afterglow light curve, and argue that energy injection is the most plausible mechanism. We model 
the re-brightening as a power-law increase in blastwave energy, self-consistently accounting for 
the change in the synchrotron spectrum over the injection period, and compute the fractional 
increase in energy during the re-brightening. We interpret the energy injection process in the 
context of a distribution of ejecta Lorentz factors, and provide a measurement of the power law 
index of the Lorentz factor distribution.

To place our results in context, we search the \Swift\ X-ray light curve archive for all 
events exhibiting a similar re-brightening, and present full multi-wavelength analyses, 
complete with deduced correlations between the physical parameters, for all events with radio 
detections that exhibit simultaneous optical and X-ray re-brightenings. This selection yields three 
additional events: GRBs~100418A, 100901A, and 120404A. We collect, analyze, and report all X-ray 
and UV data from the X-ray telescope (XRT) and UV-optical telescope (UVOT) on board \Swift\ for 
these three events in addition to GRB~120326A. We also analyze and report previously un-published 
archival radio observations from the Very Large Array (VLA), Submillimeter Array (SMA), and 
Westerbork Synthesis Radio Telescope (WSRT) for events in our sample, and the first complete 
multi-wavelength model fits for GRBs~100418A and 100901A.
Finally, we compare our results to modeling efforts of a sample of GRBs ranging from $z\sim1$ to 
$z\gtrsim6$, as well as with a complete sample of plateaus in \Swift/XRT light curves, and thereby 
assess the ubiquity of the energy injection phenomenon. We infer the fractional increase in 
blastwave energy over the plateau phase using simple assumptions on the afterglow properties, and 
determine the unique characteristics of these events that result in multi-band re-brightenings. We 
conclude with a discussion of the results from this X-ray-only analysis and our full broad-band 
modeling in the context of energy injection in GRBs.

Throughout the paper, we use the following values for cosmological parameters: $\Omega_{\rm m} = 
0.27$, $\Omega_{\rm \Lambda} = 0.73$, and $H_{\rm 0} = 71\,{\rm km}\,{\rm s}^{-1}\,{\rm Mpc}^{-1}$. 
All times are in the observer frame, uncertainties are at the 68\% confidence level (1$\sigma$), 
and magnitudes are in the AB system and are not corrected for galactic extinction, unless stated 
otherwise.

\section{GRB Properties and Observations}
\label{text:GRB_Properties_and_Observations}
GRB~120326A was discovered by the \Swift\ Burst Alert Telescope \citep[BAT,][]{bbc+05} on 2012 
March 26 at 01:20:29\,UT \citep{gcn13105}. The burst duration is $T_{90} = 26.7\pm0.4$\,s, with a 
fluence of $F_{\gamma} = (1.1 \pm 0.1) \times10^{-6}$\,erg\,cm$^{-2}$ 
\citep[15--150\,keV][]{gcn13120}. A bright X-ray and UV/optical afterglow was detected by \Swift\ 
\citep{gcn13105,gcn13114} and numerous ground-based observatories. Spectroscopic observations at the 
10.4\,m Gran Telescope Canarias (GTC) provided a redshift of $z = 1.798$ \citep{gcn13118}.

The burst also triggered the Fermi Gamma-ray Burst Monitor (GBM) at 01:20:31.51\,UT 
\citep{gcn13145}. The burst duration as observed by GBM is $T_{90} = 11.8\pm1.8$\,s 
(50--300\,keV) with a fluence of $(3.26\pm0.05)\times10^{-6}$\,erg\,cm$^{-2}$ (10--1000\,keV). 
The time-averaged $\gamma$-ray spectrum is well fit by a Band function\footnote{From the Fermi GRB 
catalog for trigger 120326056 at \url{http://heasarc.gsfc.nasa.gov/db-perl/W3Browse/w3query.pl}.}, 
with break energy, $E_{\rm peak}=43.9\pm3.9$\,keV, low energy index, $\alpha=-0.67\pm0.19$, and 
high-energy index, $\beta = -2.33\pm0.09$ Using the source redshift of $z=1.798$, the inferred 
isotropic equivalent $\gamma$-ray energy in the 1--$10^4$ keV rest frame energy band is 
$\Egammaiso=(3.15\pm0.12)\times10^{52}$\,erg.

\subsection{X-ray: \Swift/XRT}
\label{text:data_analysis:XRT}
The \Swift\ X-ray Telescope \citep[XRT,][]{bhn+05} began observing the field at 69\,s after 
the BAT trigger, leading to the detection of an X-ray afterglow. The source was
localized to RA = 18h\,15m\,37.06s, Dec = +69d\,15\arcmin\,35.4\arcsec\ (J2000), with an
uncertainty radius of 1.4 arcseconds (90\% 
containment)\footnote{\url{http://www.swift.ac.uk/xrt_positions/00518626/}}. XRT continued 
observing the afterglow for 18.7\,d in photon counting mode, with the last detection at 5.2\,d.

We extracted XRT PC-mode spectra using the on-line tool on the \Swift\ website 
\citep{ebp+07,ebp+09} \footnote{\url{http://www.swift.ac.uk/xrt_spectra/00518626/}}. We analyzed the 
data after the end of the steep decay at $400$\,s using the latest version of the HEASOFT package 
(v6.14) and corresponding calibration files. We used Xspec (v12.8.1) to fit all available PC-mode 
data, assuming a photoelectrically absorbed power law model (\texttt{tbabs $\times$ ztbabs $\times$ 
pow}) and a Galactic neutral hydrogen column density of $N_{\rm H, MW} = 
6.3\times10^{20}$\,cm${}^{-2}$ (for consistency with the value used on the \Swift\ website), 
fixing the source redshift at $z=1.798$. Our best-fit model has a photon index of $\Gamma = 
1.85\pm0.04$ and excess absorption 
corresponding to a neutral hydrogen column (assuming solar metallicity) of $N_{\rm H,{\rm int}} = 
(4.1\pm0.7)\times10^{21}\,\pcmsq$ intrinsic to the host galaxy (C-stat = 497.1 for 570 degrees of 
freedom). We divided the PC-mode data into 6 roughly equal time 
bins and extracted time-resolved PC-mode spectra to test for spectral evolution. We do not find 
clear evidence for significant spectral evolution over this period. In the following analysis, we 
take the 0.3 -- 10\,keV count rate light curve from the above website together with 
$\Gamma_{\rm X}=1.85$ (corresponding to $\beta_{\rm X}\equiv1-\Gamma_{\rm X}=-0.85$) for the PC 
mode to compute the 1\,keV flux density. We combine the uncertainty in flux calibration based on 
our spectral analysis (2.7\%) in quadrature with the statistical uncertainty from the on-line light 
curve. For the WT-mode, we convert the count rate light curve to a flux-calibrated light curve 
using $\Gamma=3.57$ and a count-to-flux conversion factor of $7\times10^{-11}{\rm erg}\,\pcmsq
{\rm ct}^{-1}$ as reported on the on-line spectral analysis.

The WT-mode X-ray light curve declines rapidly as $t^{-3.3\pm0.1}$. \Swift\ switched to collecting 
data in PC-mode at 150\,s. The PC-mode data between 150 and 300\,s continue to decline rapidly 
as $t^{-3.8\pm0.4}$, followed by a plateau where the count rate evolves as $t^{-0.22\pm0.05}$. This 
part of the lightcurve is likely dominated by the high-latitude prompt emission \citep{wggo10}, and 
we therefore do not consider the rapid decline before 300\,s in our afterglow modeling.
About 0.16\,d after the trigger, the X-ray count rate begins rising and peaks at around 0.41\,d. 
This re-brightening is unusual for X-ray afterglows and we discuss this feature 
further in section \ref{text:basic_considerations:re-brightening}. The XRT count rate light curve 
after 1.4\,d can be fit by a single power law with a decline rate of $\alpha_{\rm X} = 
-2.29\pm0.16$.

\begin{deluxetable*}{cccccc}[hb]
\tabletypesize{\footnotesize}
\tablecolumns{6}
\tablewidth{0pt}
\tablecaption{\Swift/UVOT Observations of GRB~120326A\label{tab:data:UVOT}}
\tablehead{
  \colhead{$t-t_0$} &  
  \colhead{Filter} &
  \colhead{Frequency} &  
  \colhead{Flux density$^{\dag}$} &
  \colhead{Uncertainty} &
  \colhead{Detection?} \\  
  \colhead{(days)} &
  \colhead{} &
  \colhead{(Hz)} &
  \colhead{($\mu$Jy)} &
  \colhead{($\mu$Jy)} &
  \colhead{($1=$ Yes)}
  }
\startdata
0.00165 & \textit{White} & 8.64e+14 & 8.89 & 2.71 & 1 \\
0.00354 & \textit{u} & 8.56e+14 & 20.7 & 11.5 & 0 \\
0.0523 & \textit{b} & 6.92e+14 & 39 & 6.23 & 1 \\
0.0546 & \textit{White} & 8.64e+14 & 15.6 & 2.55 & 1 \\
0.0558 & \textit{uvw1} & 1.16e+15 & 10.9 & 3.29 & 1 \\
0.0582 & \textit{u} & 8.56e+14 & 18.6 & 4.24 & 1 \\
0.0594 & \textit{v} & 5.55e+14 & 42.3 & 14.2 & 0 \\
0.115 & \textit{uvw1} & 1.16e+15 & 9.16 & 1.54 & 1 \\
0.126 & \textit{u} & 8.56e+14 & 30.1 & 2.34 & 1 \\
0.134 & \textit{b} & 6.92e+14 & 68.2 & 8.04 & 1 \\
0.191 & \textit{uvw1} & 1.16e+15 & 13.8 & 2.12 & 1 \\
0.202 & \textit{u} & 8.56e+14 & 50.6 & 6.19 & 1 \\
0.258 & \textit{v} & 5.55e+14 & 134 & 11.8 & 1 \\
0.285 & \textit{uvw2} & 1.48e+15 & 1.95 & 0.545 & 1 \\
0.295 & \textit{uvm2} & 1.34e+15 & 0.607 & 0.549 & 0 \\
0.316 & \textit{v} & 5.55e+14 & 150 & 10.6 & 1 \\
0.383 & \textit{uvw1} & 1.16e+15 & 19.1 & 1.94 & 1 \\
0.393 & \textit{u} & 8.56e+14 & 59.2 & 3.07 & 1 \\
0.399 & \textit{b} & 6.92e+14 & 124 & 13.6 & 1 \\
0.46 & \textit{uvw1} & 1.16e+15 & 12.3 & 1.66 & 1 \\
0.468 & \textit{u} & 8.56e+14 & 43.1 & 4.43 & 1 \\
0.527 & \textit{v} & 5.55e+14 & 97.6 & 9.45 & 1 \\
0.715 & \textit{v} & 5.55e+14 & 83.9 & 10.4 & 1 \\
0.715 & \textit{uvw1} & 1.16e+15 & 11.7 & 1.35 & 1 \\
0.737 & \textit{u} & 8.56e+14 & 28.9 & 4.06 & 1 \\
0.782 & \textit{uvw2} & 1.48e+15 & 0.968 & 0.68 & 0 \\
0.783 & \textit{b} & 6.92e+14 & 50.1 & 4.38 & 1 \\
0.792 & \textit{White} & 8.64e+14 & 26.3 & 3.74 & 1 \\
1.65 & \textit{u} & 8.56e+14 & 11.6 & 2.31 & 1 \\
1.65 & \textit{uvw2} & 1.48e+15 & 0.339 & 0.668 & 0 \\
1.66 & \textit{uvm2} & 1.34e+15 & 2.42 & 1.71 & 0 \\
1.66 & \textit{uvw1} & 1.16e+15 & 6.29 & 2.29 & 0 \\
2.49 & \textit{uvw1} & 1.16e+15 & 3.53 & 1.67 & 0 \\
2.49 & \textit{u} & 8.56e+14 & 8.3 & 1.83 & 1 \\
2.69 & \textit{b} & 6.92e+14 & 16 & 4.01 & 1 \\
2.7 & \textit{White} & 8.64e+14 & 6.59 & 1.29 & 1 \\
2.7 & \textit{v} & 5.55e+14 & 31.9 & 8.93 & 1 \\
3.72 & \textit{u} & 8.56e+14 & 5.93 & 1.62 & 1 \\
3.73 & \textit{b} & 6.92e+14 & 7.59 & 3.43 & 0 \\
3.73 & \textit{uvw1} & 1.16e+15 & 0.779 & 0.626 & 0 \\
3.73 & \textit{White} & 8.64e+14 & 4.26 & 0.985 & 1 \\
3.74 & \textit{v} & 5.55e+14 & 5.94 & 7.16 & 0 \\
4.83 & \textit{u} & 8.56e+14 & 2.92 & 1.42 & 0 \\
4.83 & \textit{b} & 6.92e+14 & 8.72 & 3.14 & 0 \\
4.84 & \textit{White} & 8.64e+14 & 2.06 & 0.768 & 0 \\
4.84 & \textit{v} & 5.55e+14 & 5.12 & 6.37 & 0 \\
5.73 & \textit{u} & 8.56e+14 & 0.689 & 1.39 & 0 \\
5.73 & \textit{b} & 6.92e+14 & 6.41 & 3.13 & 0 \\
5.74 & \textit{White} & 8.64e+14 & 0.754 & 0.719 & 0 \\
5.74 & \textit{v} & 5.55e+14 & 0.413 & 6.26 & 0 \\
17.6 & \textit{u} & 8.56e+14 & 0.28 & 1.39 & 0 \\
19.5 & \textit{u} & 8.56e+14 & 2.34 & 1.28 & 0
\enddata
\tablecomments{$^{\dag}$In cases of non-detections, we report the formal flux density measurement 
from aperture photometry.}
\end{deluxetable*}

\subsection{UV/Optical: \Swift/UVOT}
\label{text:data_analysis:UVOT}
The \Swift\ UV/Optical Telescope (UVOT; \citealt{rkm+05}) observed \me\ using 6 filters spanning 
the central wavelength range $\lambda_{\rm c}=1928$\,\AA\ (\textit{W2}) to $\lambda_{\rm 
c}=5468$\,\AA\ (\textit{v}) beginning 67\,s after the burst. The afterglow was detected in the 
initial exposures at RA = 18h\,15m\,37.13s, Dec = +69d\,15\arcmin\,35.36\arcsec\ (J2000) 
\citep[90\% confidence,][]{gcn13114}, and exhibited a clear re-brightening concomitant 
with the peak in the X-ray light curve. We analyzed the UVOT data 
using the latest version of HEASOFT (v. 6.14) and corresponding calibration files. We 
performed photometry with a $5\arcsec$ aperture and used a $90\arcsec$ annulus with foreground 
sources masked to estimate the background. We list our derived fluxes in Table \ref{tab:data:UVOT}.

\subsection{Optical/NIR: Palomar Observations and GCN Circulars}
\label{text:data_analysis:optical}
We observed \me\ with the Wide-Field Infrared Camera \citep{weh+03} on the Palomar 200-inch 
telescope beginning on 2012 March 30.48 UT. We acquired a series of nine 75\,s $J$-band exposures, 
followed by a series of nine 75\,s $K_s$-band exposures. We reduced the data following standard IR 
imaging techniques using a modified version of the WIRCSOFT pipeline and performed aperture 
photometry of the afterglow in the combined, stacked exposures relative to 2MASS standards in the 
field.

We also carried out a series of imaging observations of \me\ with the robotic Palomar 60-inch 
telescope \citep{cfm+06} on the nights of 2012 March 31, April 02, April 03 ($r^\prime$ and 
$i^\prime$), April 04, April 08, and April 16 ($r^\prime$ only). We reduced the images using the 
P60 automated pipeline and performed aperture photometry relative to secondary standard stars 
calibrated via separate observations of \cite{lan09} standards.

Finally, we collected other optical and NIR observations of \me\ reported through the Gamma-ray 
Burst Coordinates Network (GCN) Circulars and converted all photometry to flux densities. 
We also include the optical photometry of the afterglow published by \cite{mvg+14} and 
\cite{uht+14} in our analysis. We list our complete compilation of optical observations of this 
burst, together with our P60 and P200 observations, in Table \ref{tab:data:GCN}.

\begin{deluxetable*}{ccccccccc}
\tabletypesize{\footnotesize}
\tablecolumns{9}
\tablewidth{0pt}
\tablecaption{Optical Observations of GRB~120326A\label{tab:data:GCN}}
\tablehead{
  \colhead{$t-t_0$} &  
  \colhead{Observatory} &
  \colhead{Telescope /} &
  \colhead{Filter} &
  \colhead{Frequency} &  
  \colhead{Flux density$^{\dag}$} &
  \colhead{Uncertainty} &
  \colhead{Detection?} &
  \colhead{Reference} \\  
  \colhead{(days)} &
  \colhead{} &
  \colhead{Instrument} &
  \colhead{} &
  \colhead{(Hz)} &
  \colhead{($\mu$Jy)} &
  \colhead{($\mu$Jy)} &
  \colhead{($1=$ Yes)} & 
  }
\startdata
0.000291 & TNO & ROTSE & \textit{CR} & 4.56e+14 & 4.04e+03 & 1.35e+03 & 0 & \cite{gcn13106} \\
0.001 & TNO & ROTSE & \textit{CR} & 4.56e+14 & 926 & 308 & 0 & \cite{gcn13106} \\
0.00189 & Calern & TAROT & \textit{R} & 4.56e+14 & 161 & 51.2 & 1 & \cite{gcn13108} \\
0.00389 & Liverpool & RINGO2 & \textit{R} & 4.56e+14 & 120 & 21.6 & 1 & \cite{mvg+14} \\
0.00545 & Crni\_Vrh & Cicocki & \textit{R} & 4.56e+14 & 95.1 & 13.1 & 1 & \cite{mvg+14} \\
0.00623 & Crni\_Vrh & Cicocki & \textit{R} & 4.56e+14 & 87.6 & 16.7 & 1 & \cite{mvg+14} \\
0.007 & Crni\_Vrh & Cicocki & \textit{R} & 4.56e+14 & 86.8 & 12.9 & 1 & \cite{mvg+14} \\
0.00813 & Crni\_Vrh & Cicocki & \textit{R} & 4.56e+14 & 74.9 & 14.3 & 1 & \cite{mvg+14} \\
0.00824 & Liverpool & RATCam & \textit{R} & 4.56e+14 & 97.8 & 17.6 & 1 & \cite{mvg+14} \\
0.00968 & Crni\_Vrh & Cicocki & \textit{R} & 4.56e+14 & 87.6 & 14.8 & 1 & \cite{mvg+14} \\
0.00991 & Liverpool & RATCam & \textit{r'} & 4.81e+14 & 109 & 17.2 & 1 & \cite{mvg+14} \\
0.0109 & Liverpool & RATCam & \textit{r'} & 4.81e+14 & 91.2 & 12.6 & 1 & \cite{mvg+14} \\
0.0116 & Crni\_Vrh & Cicocki & \textit{R} & 4.56e+14 & 82.9 & 8.84 & 1 & \cite{mvg+14} \\
0.0122 & Liverpool & SkyCam & \textit{R} & 4.56e+14 & 90 & 37.7 & 1 & \cite{mvg+14} \\
0.0123 & Liverpool & RATCam & \textit{r'} & 4.81e+14 & 95.5 & 3.58 & 1 & \cite{mvg+14} \\
0.0136 & Crni\_Vrh & Cicocki & \textit{R} & 4.56e+14 & 70.2 & 10.4 & 1 & \cite{mvg+14} \\
0.0142 & Liverpool & RATCam & \textit{i'} & 3.93e+14 & 86.3 & 5.75 & 1 & \cite{mvg+14} \\
0.0151 & Crni\_Vrh & Cicocki & \textit{R} & 4.56e+14 & 54.8 & 13.5 & 1 & \cite{mvg+14} \\
0.016 & Liverpool & RATCam & \textit{z'} & 3.28e+14 & 111 & 14.1 & 1 & \cite{mvg+14} \\
0.017 & Crni\_Vrh & Cicocki & \textit{R} & 4.56e+14 & 49.5 & 8.92 & 1 & \cite{mvg+14} \\
0.0178 & Liverpool & RATCam & \textit{r'} & 4.81e+14 & 72.4 & 3.41 & 1 & \cite{mvg+14} \\
0.0193 & Liverpool & RATCam & \textit{r'} & 4.81e+14 & 67.3 & 3.82 & 1 & \cite{mvg+14} \\
0.0193 & Crni\_Vrh & Cicocki & \textit{R} & 4.56e+14 & 54.3 & 12.2 & 1 & \cite{mvg+14} \\
0.0212 & Liverpool & RATCam & \textit{i'} & 3.93e+14 & 68.5 & 3.89 & 1 & \cite{mvg+14} \\
0.0228 & Calern & TAROT & \textit{R} & 4.56e+14 & 70.2 & 22.3 & 1 & \cite{gcn13108} \\
0.0254 & Liverpool & RATCam & \textit{z'} & 3.28e+14 & 103 & 9.92 & 1 & \cite{mvg+14} \\
0.0259 & DAO & Skynet & \textit{g'} & 6.29e+14 & 17.9 & 5.69 & 1 & \cite{gcn13109} \\
0.0277 & DAO & Skynet & \textit{r'} & 4.56e+14 & 41 & 13 & 1 & \cite{gcn13109} \\
0.028 & Liverpool & RATCam & \textit{r'} & 4.81e+14 & 63.1 & 3.58 & 1 & \cite{mvg+14} \\
0.0296 & Liverpool & RATCam & \textit{r'} & 4.81e+14 & 63.1 & 2.97 & 1 & \cite{mvg+14} \\
0.0311 & Liverpool & RATCam & \textit{r'} & 4.81e+14 & 64.3 & 3.65 & 1 & \cite{mvg+14} \\
0.0333 & Liverpool & SkyCam & \textit{R} & 4.56e+14 & 51.3 & 43.8 & 1 & \cite{mvg+14} \\
0.0344 & Liverpool & RATCam & \textit{i'} & 3.93e+14 & 63.1 & 4.2 & 1 & \cite{mvg+14} \\
0.0358 & DAO & Skynet & \textit{i'} & 3.93e+14 & 73.4 & 23.4 & 1 & \cite{gcn13109} \\
0.0396 & Liverpool & RATCam & \textit{z'} & 3.28e+14 & 87.1 & 8.4 & 1 & \cite{mvg+14} \\
0.0709 & T17 & CDK17 & \textit{CR} & 4.56e+14 & 77 & 15.6 & 1 & \cite{gcn13119} \\
0.32 & LOAO & 1m & \textit{R} & 4.56e+14 & 272 & 15.4 & 1 & \cite{gcn13139} \\
0.324 & LOAO & 1m & \textit{R} & 4.56e+14 & 298 & 11.2 & 1 & \cite{gcn13139} \\
0.327 & LOAO & 1m & \textit{R} & 4.56e+14 & 277 & 10.4 & 1 & \cite{gcn13139} \\
0.331 & LOAO & 1m & \textit{R} & 4.56e+14 & 224 & 6.28 & 1 & \cite{gcn13139} \\
0.335 & LOAO & 1m & \textit{R} & 4.56e+14 & 212 & 5.94 & 1 & \cite{gcn13139} \\
0.338 & LOAO & 1m & \textit{R} & 4.56e+14 & 180 & 5.03 & 1 & \cite{gcn13139} \\
0.367 & McDonald & CQUEAN & \textit{r'} & 4.81e+14 & 173 & 2.2 & 1 & \cite{uht+14}$^{\ddag}$ \\
0.371 & McDonald & CQUEAN & \textit{i'} & 3.93e+14 & 181 & 1 & 1 & \cite{uht+14}$^{\ddag}$ \\
0.375 & McDonald & CQUEAN & \textit{z'} & 3.28e+14 & 233 & 1.6 & 1 & \cite{uht+14}$^{\ddag}$ \\
0.379 & McDonald & CQUEAN & \textit{Y'} & 2.94e+14 & 264 & 4 & 1 & \cite{uht+14}$^{\ddag}$ \\
0.383 & McDonald & CQUEAN & \textit{r'} & 4.81e+14 & 164 & 0.8 & 1 & \cite{uht+14}$^{\ddag}$ \\
0.387 & McDonald & CQUEAN & \textit{i'} & 3.93e+14 & 185 & 0.5 & 1 & \cite{uht+14}$^{\ddag}$ \\
0.391 & McDonald & CQUEAN & \textit{z'} & 3.28e+14 & 231 & 2.1 & 1 & \cite{uht+14}$^{\ddag}$ \\
0.398 & McDonald & CQUEAN & \textit{r'} & 4.81e+14 & 151 & 2.7 & 1 & \cite{uht+14}$^{\ddag}$ \\
0.402 & McDonald & CQUEAN & \textit{i'} & 3.93e+14 & 190 & 2.3 & 1 & \cite{uht+14}$^{\ddag}$ \\
0.405 & McDonald & CQUEAN & \textit{z'} & 3.28e+14 & 225 & 7.3 & 1 & \cite{uht+14}$^{\ddag}$ \\
0.418 & McDonald & CQUEAN & \textit{g'} & 6.29e+14 & 115 & 3.39 & 1 & \cite{uht+14}$^{\ddag}$ \\
0.425 & McDonald & CQUEAN & \textit{r'} & 4.81e+14 & 148 & 1 & 1 & \cite{uht+14}$^{\ddag}$ \\
0.429 & McDonald & CQUEAN & \textit{i'} & 3.93e+14 & 166 & 3.9 & 1 & \cite{uht+14}$^{\ddag}$ \\
0.737 & GMG & 2.4m & \textit{R} & 4.56e+14 & 101 & 9.79 & 1 & \cite{gcn13122} \\
0.742 & GMG & 2.4m & \textit{v} & 5.55e+14 & 148 & 14.2 & 1 & \cite{gcn13122} \\
0.747 & Lulin & LOT & \textit{i'} & 3.93e+14 & 99.1 & 2.1 & 1 & \cite{uht+14}$^{\ddag}$ \\
0.747 & Lulin & LOT & \textit{g'} & 6.29e+14 & 55.3 & 0.212 & 1 & \cite{uht+14}$^{\ddag}$ \\
0.75 & GMG & 2.4m & \textit{R} & 4.56e+14 & 92.6 & 8.93 & 1 & \cite{gcn13122} \\
0.754 & GMG & 2.4m & \textit{v} & 5.55e+14 & 135 & 13 & 1 & \cite{gcn13122} \\
0.758 & Lulin & LOT & \textit{r'} & 4.81e+14 & 85.2 & 1.11 & 1 & \cite{uht+14}$^{\ddag}$ \\
0.758 & Lulin & LOT & \textit{z'} & 3.28e+14 & 132 & 8.8 & 1 & \cite{uht+14}$^{\ddag}$ \\
0.794 & Lulin & LOT & \textit{i'} & 3.93e+14 & 98.2 & 2.8 & 1 & \cite{uht+14}$^{\ddag}$ \\
0.806 & Lulin & LOT & \textit{z'} & 3.28e+14 & 109 & 8.5 & 1 & \cite{uht+14}$^{\ddag}$ \\
0.81 & Xinglong & TNT & \textit{R} & 4.56e+14 & 115 & 36.7 & 1 & \cite{gcn13150} \\
0.842 & Maisoncelles & 30cm & \textit{R} & 4.56e+14 & 111 & 10.7 & 1 & \cite{gcn13129} \\
0.92 & BBO & 320mm & \textit{R} & 4.56e+14 & 140 & 9.33 & 1 & \cite{mvg+14} \\
1.36 & McDonald & CQUEAN & \textit{r'} & 4.81e+14 & 49.5 & 0.36 & 1 & \cite{uht+14}$^{\ddag}$ \\
1.36 & McDonald & CQUEAN & \textit{i'} & 3.93e+14 & 62.9 & 0.3 & 1 & \cite{uht+14}$^{\ddag}$ \\
1.37 & McDonald & CQUEAN & \textit{z'} & 3.28e+14 & 75.6 & 0.6 & 1 & \cite{uht+14}$^{\ddag}$ \\
1.37 & McDonald & CQUEAN & \textit{Y'} & 2.94e+14 & 83.4 & 2.5 & 1 & \cite{uht+14}$^{\ddag}$ \\
1.38 & McDonald & CQUEAN & \textit{g'} & 6.29e+14 & 31 & 0.72 & 1 & \cite{uht+14}$^{\ddag}$ \\
1.39 & LOAO & 1m & \textit{R} & 4.56e+14 & 89.3 & 15.8 & 1 & \cite{uht+14}$^{\ddag}$ \\
1.43 & FLWO & PAIRITEL & \textit{K} & 1.37e+14 & 248 & 64.1 & 1 & \cite{mvg+14} \\
1.43 & FLWO & PAIRITEL & \textit{H} & 1.84e+14 & 308 & 137 & 1 & \cite{mvg+14} \\
1.43 & FLWO & PAIRITEL & \textit{J} & 2.38e+14 & 121 & 53.7 & 1 & \cite{mvg+14} \\
1.66 & Ishigakijima & MITSuME & \textit{R} & 4.56e+14 & 52.8 & 4.56 & 1 & \cite{mvg+14} \\
1.66 & Ishigakijima & MITSuME & \textit{g'} & 6.29e+14 & 21.3 & 2.93 & 1 & \cite{mvg+14} \\
1.71 & Ishigakijima & MITSuME & \textit{R} & 4.56e+14 & 47.7 & 4.12 & 1 & \cite{mvg+14} \\
1.71 & Ishigakijima & MITSuME & \textit{g'} & 6.29e+14 & 20.5 & 3.04 & 1 & \cite{mvg+14} \\
1.82 & Xinglong & TNT & \textit{R} & 4.56e+14 & 57.3 & 18.2 & 1 & \cite{gcn13150} \\
1.9 & Hanle & HCT & \textit{R} & 4.56e+14 & 41.2 & 2.74 & 1 & \cite{gcn13185} \\
2.01 & BBO & 320mm & \textit{R} & 4.56e+14 & 56.8 & 8.42 & 1 & \cite{mvg+14} \\
2.15 & Ishigakijima & MITSuME & \textit{I} & 3.93e+14 & 36.6 & 7.4 & 1 & \cite{mvg+14} \\
2.34 & McDonald & CQUEAN & \textit{r'} & 4.81e+14 & 23.8 & 3.68 & 1 & \cite{uht+14}$^{\ddag}$ \\
2.35 & McDonald & CQUEAN & \textit{i'} & 3.93e+14 & 33.3 & 5.34 & 1 & \cite{uht+14}$^{\ddag}$ \\
2.35 & McDonald & CQUEAN & \textit{z'} & 3.28e+14 & 42.7 & 15.4 & 1 & \cite{uht+14}$^{\ddag}$ \\
2.38 & McDonald & CQUEAN & \textit{Y'} & 2.94e+14 & 59.6 & 5.4 & 1 & \cite{uht+14}$^{\ddag}$ \\
2.4 & LOAO & 1m & \textit{R} & 4.56e+14 & 41.5 & 7.08 & 1 & \cite{uht+14}$^{\ddag}$ \\
2.4 & McDonald & CQUEAN & \textit{g'} & 6.29e+14 & 18.6 & 1.81 & 1 & \cite{uht+14}$^{\ddag}$ \\
2.42 & FLWO & PAIRITEL & \textit{K} & 1.37e+14 & 168 & 43.6 & 1 & \cite{mvg+14} \\
2.42 & FLWO & PAIRITEL & \textit{H} & 1.84e+14 & 123 & 71.8 & 1 & \cite{mvg+14} \\
2.42 & FLWO & PAIRITEL & \textit{J} & 2.38e+14 & 43.8 & 25.6 & 1 & \cite{mvg+14} \\
2.43 & McDonald & CQUEAN & \textit{r'} & 4.81e+14 & 26.1 & 1.3 & 1 & \cite{uht+14}$^{\ddag}$ \\
2.43 & McDonald & CQUEAN & \textit{i'} & 3.93e+14 & 30.5 & 2.71 & 1 & \cite{uht+14}$^{\ddag}$ \\
2.44 & McDonald & CQUEAN & \textit{z'} & 3.28e+14 & 42.8 & 5.47 & 1 & \cite{uht+14}$^{\ddag}$ \\
2.6 & Ishigakijima & MITSuME & \textit{R} & 4.56e+14 & 30.6 & 5.19 & 1 & \cite{mvg+14} \\
2.6 & Ishigakijima & MITSuME & \textit{g'} & 6.29e+14 & 14.3 & 4.91 & 1 & \cite{mvg+14} \\
3 & BBO & 320mm & \textit{R} & 4.56e+14 & 28 & 9.24 & 1 & \cite{mvg+14} \\
3.32 & McDonald & CQUEAN & \textit{i'} & 3.93e+14 & 21.2 & 0.864 & 1 & \cite{uht+14}$^{\ddag}$ \\
3.33 & McDonald & CQUEAN & \textit{z'} & 3.28e+14 & 28.1 & 2.03 & 1 & \cite{uht+14}$^{\ddag}$ \\
3.33 & McDonald & CQUEAN & \textit{Y'} & 2.94e+14 & 44.5 & 5.3 & 1 & \cite{uht+14}$^{\ddag}$ \\
3.34 & McDonald & CQUEAN & \textit{r'} & 4.81e+14 & 16.8 & 0.868 & 1 & \cite{uht+14}$^{\ddag}$ \\
3.38 & McDonald & CQUEAN & \textit{g'} & 6.29e+14 & 11.6 & 1.33 & 1 & \cite{uht+14}$^{\ddag}$ \\
3.43 & McDonald & CQUEAN & \textit{i'} & 3.93e+14 & 21.2 & 5.02 & 1 & \cite{uht+14}$^{\ddag}$ \\
3.49 & LOAO & 1m & \textit{R} & 4.56e+14 & 23 & 3.74 & 1 & \cite{uht+14}$^{\ddag}$ \\
3.71 & Ishigakijima & MITSuME & \textit{R} & 4.56e+14 & 16.7 & 2.47 & 1 & \cite{mvg+14} \\
3.73 & Ishigakijima & MITSuME & \textit{g'} & 6.29e+14 & 6.31 & 1.21 & 1 & \cite{mvg+14} \\
3.73 & Ishigakijima & MITSuME & \textit{I} & 3.93e+14 & 19.4 & 4.35 & 1 & \cite{mvg+14} \\
3.77 & Ishigakijima & MITSuME & \textit{R} & 4.56e+14 & 16.8 & 2.32 & 1 & \cite{mvg+14} \\
4.4 & McDonald & CQUEAN & \textit{z'} & 3.28e+14 & 13.7 & 2.3 & 1 & \cite{uht+14}$^{\ddag}$ \\
4.41 & McDonald & CQUEAN & \textit{i'} & 3.93e+14 & 10.8 & 1.01 & 1 & \cite{uht+14}$^{\ddag}$ \\
4.42 & McDonald & CQUEAN & \textit{r'} & 4.81e+14 & 7.88 & 0.742 & 1 & \cite{uht+14}$^{\ddag}$ \\
4.43 & McDonald & CQUEAN & \textit{g'} & 6.29e+14 & 5.2 & 0.32 & 1 & \cite{uht+14}$^{\ddag}$ \\
4.43 & Palomar & P200 & \textit{J} & 2.38e+14 & 31.2 & 3.65 & 1 & This work \\
4.45 & Palomar & P200 & \textit{K} & 1.37e+14 & 33.1 & 5.97 & 1 & This work \\
5.25 & P60 &  & \textit{i'} & 3.93e+14 & 6.79 & 0.519 & 1 & This work \\
5.26 & Palomar & P60 & \textit{r'} & 4.81e+14 & 6.14 & 0.409 & 1 & This work \\
5.38 & McDonald & CQUEAN & \textit{z'} & 3.28e+14 & 6.35 & 3.62 & 1 & \cite{uht+14}$^{\ddag}$ \\
5.39 & McDonald & CQUEAN & \textit{i'} & 3.93e+14 & 6.39 & 1.06 & 1 & \cite{uht+14}$^{\ddag}$ \\
5.41 & McDonald & CQUEAN & \textit{r'} & 4.81e+14 & 4.65 & 0.735 & 1 & \cite{uht+14}$^{\ddag}$ \\
5.42 & McDonald & CQUEAN & \textit{g'} & 6.29e+14 & 3.1 & 0.42 & 1 & \cite{uht+14}$^{\ddag}$ \\
5.92 & CRO & 51cm & \textit{R} & 4.56e+14 & 7.56 & 3.17 & 1 & \cite{mvg+14} \\
5.92 & CRO & 51cm & \textit{CR} & 4.56e+14 & 5.84 & 0.563 & 1 & \cite{gcn13198} \\
6.4 & McDonald & CQUEAN & \textit{r'} & 4.81e+14 & 2.96 & 0.0931 & 1 & \cite{uht+14}$^{\ddag}$ \\
6.41 & McDonald & CQUEAN & \textit{i'} & 3.93e+14 & 4.13 & 0.526 & 1 & \cite{uht+14}$^{\ddag}$ \\
6.42 & McDonald & CQUEAN & \textit{z'} & 3.28e+14 & 5.11 & 2.44 & 1 & \cite{uht+14}$^{\ddag}$ \\
6.83 & LOAO & 1m & \textit{R} & 4.56e+14 & 5.75 & 0.88 & 1 & \cite{uht+14}$^{\ddag}$ \\
7.24 & P60 &  & \textit{i'} & 3.93e+14 & 6.43 & 0.885 & 1 & This work \\
7.25 & Palomar & P60 & \textit{r'} & 4.81e+14 & 2.81 & 0.759 & 1 & This work \\
7.41 & McDonald & CQUEAN & \textit{r'} & 4.81e+14 & 1.97 & 0.3 & 1 & \cite{uht+14}$^{\ddag}$ \\
7.51 & McDonald & CQUEAN & \textit{i'} & 3.93e+14 & 2.44 & 0.29 & 1 & \cite{uht+14}$^{\ddag}$ \\
8.29 & P60 &  & \textit{i'} & 3.93e+14 & 3.37 & 0.5 & 1 & This work \\
8.3 & Palomar & P60 & \textit{r'} & 4.81e+14 & 2.47 & 0.392 & 1 & This work \\
9.29 & Palomar & P60 & \textit{r'} & 4.81e+14 & 2.44 & 0.236 & 1 & This work \\
13.3 & Palomar & P60 & \textit{r'} & 4.81e+14 & 1.98 & 0.4 & 1 & This work \\
21.4 & Palomar & P60 & \textit{r'} & 4.81e+14 & 1.64 & 0.11 & 1 & This work \\

\enddata
\tablecomments{$^{\dag}$In cases of non-detections, we report the $3\sigma$ upper limit on the 
flux density. $^{\ddag}$ We note that the photometry reported in Table 1 of 
\cite{uht+14} yields an unphysical spectral index of $\beta\approx-4.5$. We assume that these 
numbers have been scaled by the same factors as reported in Figure 4 of their paper, and divide 
the $gr^{\prime}Rizy$ light curves by a factor of 0.5, 1, 1, 2, 4, and 6, respectively. We further 
assume that the data have not been corrected for galactic extinction. 
}
\end{deluxetable*}

\subsection{Sub-millimeter: CARMA and SMA}
\label{text:data_analysis:millimeter}
We observed GRB\,120326A with the Combined Array for Research in Millimeter Astronomy (CARMA) 
beginning on 2012 March 30.56 UT (4.52 d after the burst) in continuum wideband mode with 
$\approx8$ GHz bandwidth (16 windows, 487.5 MHz each) at a mean frequency of 93 GHz. Following an 
initial detection \citep{gcn13175}, we obtained two additional epochs. Weather conditions 
were excellent for each epoch, with the first observation being taken in C configuration (maximum 
baseline $\sim$370~m), and the following observations in D configuration (maximum baseline 
$\sim$145~m). For all observations, we utilized 3C371 for bandpass, amplitude and phase gain 
calibration, and MWC349 for flux calibration. We reduced the data using standard procedures in 
MIRIAD \citep{stw95}. In the last epoch on April 10, we also observed 3C273 and 3C279 and utilized 
these observations as independent checks on the bandpass and flux calibration.
We summarize our observations in Table \ref{tab:data:120326A:mm}.

The Submillimeter Array \citep[SMA;][]{hml04} observed \me\ at a mean frequency of 222 GHz (1.3 mm; 
lower sideband centered at 216 GHz, upper sideband at 228 GHz; PI: Urata). Six epochs of 
observations were obtained between 2012 March 26.43 UT (0.37\,d after the burst) and 2012 April 
11.64 UT (16.6\,d after the burst). These observations have been reported in \cite{uht+14}. We 
carried out an independent reduction of the data using standard MIR IDL procedures for the SMA, 
followed by flagging, imaging and analysis in MIRIAD and the Astronomical Image Processing System 
\citep[AIPS;][]{gre03}. We utilized 3C279 for bandpass calibration in all but the last epoch, where 
we utilized J1924-292. We utilized J1800+784 and J1829+487 for gain calibration\footnote{The gain 
calibrators were separated by more than 20 degrees from the source; thus some decoherence, 
resulting in a systematic reduction of the observed flux is possible.}. We utilized MWC349a for 
flux calibration, determining a flux density of 1.42\,Jy for J1800+784, consistent with flux values 
in the SMA catalog at similar times. We utilized this flux throughout all observations (not every 
epoch contained useful flux calibrator scans) and scaled the gains appropriately. We note an 
uncertainty in the absolute flux density scale of $\approx15\%$. We detect a source at RA = 
18h\,15m\,37.15s $(\pm 0.02)$\,s, Dec = +69d\,15\arcmin\,35.41\arcsec\ ($\pm 0.14$; J2000). Two 
epochs obtained on March 31.40 and April 6.51 showed poor noise characteristics, and we do not 
include them in our analysis. We measure poorer noise statistics in the observations than reported 
by \cite{uht+14}, and also do not detect the source in the epoch at 3.53\,d, contrary to the 
previous analysis of this dataset. We report the results of our analysis in Table 
\ref{tab:data:120326A:mm}.

\subsection{Radio: VLA}
\label{text:data_analysis:radio}
We observed the afterglow at C (4--7\,GHz), K (18--25\,GHz), and Ka (30--38\,GHz) bands using 
the Karl G. Jansky Very Large Array (VLA) starting 5.45\,d after the burst. We detected 
and tracked the flux density of the afterglow over eight epochs until 122\,d after the burst, 
until the source had either faded below or was barely detectable at the $3\sigma$-level at all 
frequencies. Depending on the start time of the observations, we used either 3C286 or 3C48 as the 
flux and bandpass calibrator; we used J1806+6949 or J1842+6809 as gain calibrator depending on the 
array configuration and observing frequency. We carried out data reduction using the Common 
Astronomy Software Applications (CASA). We list our VLA photometry in Table 
\ref{tab:data:120326A:mm}.

\begin{deluxetable*}{lccccccc}
\tabletypesize{\footnotesize}
\tablecolumns{9}
\tablewidth{0pt}
\tablecaption{Millimeter and Radio Observations of GRB~120326A\label{tab:data:120326A:mm}}
\tablehead{
  \colhead{Date} & 
  \colhead{$t-t_0$} &  
  \colhead{Observatory$^*$} &
  \colhead{Frequency} &
  \colhead{Integration time} &
  \colhead{Flux density} &
  \colhead{Uncertainty$^{\dag}$} &
  \colhead{Detection?} \\
  \colhead{(UT)} &
  \colhead{(days)} & &
  \colhead{(GHz)} &
  \colhead{(min)} &
  \colhead{(mJy)} &
  \colhead{(mJy)} &
  \colhead{($1=$ Yes)}
  }
\startdata
2012 Mar 30.56 & 4.55  & CARMA           & 92.5 &  60.3  &  3.38   & 0.87  & 1 \\
2012 Apr 5.53  & 10.51 & CARMA           & 92.5 &  73.0  &  1.46   & 0.42  & 1 \\
2012 Apr 10.40 & 15.51 & CARMA           & 92.5 & 291.9  &  0.476  & 0.17  & 1 \\
2012 Mar 26.43 & 0.504 & SMA$^{\ddag}$   & 222  & 386    &  3.3    & 0.90  & 1 \\
2012 Mar 27.53 & 1.55  & SMA$^{\ddag}$   & 222  & 227    &  2.4    & 0.80  & 1 \\
2012 Mar 29.50 & 3.53  & SMA$^{\ddag}$   & 222  & 252    & $<1.8$  & 0.60  & 0 \\
2012 Apr 11.64 & 16.6  & SMA$^{\ddag}$   & 222  & 201    & $<2.4$  & 0.80  & 0 \\
2012 Mar 31.55 & 5.66  & VLA/C & 5.0  & 8.54 & 0.674 & 0.017 & 1 \\
2012 Mar 31.55 & 5.66  & VLA/C & 7.1  & 4.87 & 0.369 & 0.018 & 1 \\
2012 Mar 31.53 & 5.64  & VLA/C & 19.2 & 9.35 & 1.25  & 0.031 & 1 \\
2012 Mar 31.53 & 5.64  & VLA/C & 24.5 & 7.70 & 1.49  & 0.038 & 1 \\
2012 Apr 04.48 & 9.42  & VLA/C & 5.0  & 5.37 & 0.411 & 0.024 & 1 \\
2012 Apr 04.48 & 9.42  & VLA/C & 7.1  & 6.00 & 0.561 & 0.017 & 1 \\
2012 Apr 04.46 & 9.41  & VLA/C & 19.2 & 7.16 & 0.825 & 0.028 & 1 \\
2012 Apr 04.46 & 9.40  & VLA/C & 24.5 & 8.46 & 0.936 & 0.039 & 1 \\
2012 Apr 04.45 & 9.40  & VLA/C & 33.5 & 7.16 & 0.94  & 0.041 & 1 \\
2012 Apr 10.55 & 15.49 & VLA/C & 5.0  & 7.59 & 0.313 & 0.019 & 1 \\
2012 Apr 10.55 & 15.49 & VLA/C & 7.1  & 6.72 & 0.352 & 0.016 & 1 \\
2012 Apr 10.53 & 15.48 & VLA/C & 19.2 & 7.87 & 0.667 & 0.035 & 1 \\
2012 Apr 10.53 & 15.48 & VLA/C & 24.5 & 7.88 & 0.686 & 0.045 & 1 \\
2012 Apr 10.52 & 15.46 & VLA/C & 33.5 & 7.87 & 0.706 & 0.053 & 1 \\
2012 Apr 26.42 & 31.37 & VLA/C & 5.0  & 5.39 & 0.409 & 0.022 & 1 \\
2012 Apr 26.42 & 31.37 & VLA/C & 7.1  & 4.55 & 0.562 & 0.020 & 1 \\
2012 Apr 26.41 & 31.35 & VLA/C & 19.2 & 13.34& 0.458 & 0.025 & 1 \\
2012 Apr 26.41 & 31.35 & VLA/C & 24.5 & 9.43 & 0.401 & 0.032 & 1 \\
2012 Apr 26.39 & 31.34 & VLA/C & 33.5 & 7.86 & 0.277 & 0.034 & 1 \\
2012 May 27.49 & 62.44 & VLA/B & 5.0  & 6.11 & 0.217 & 0.020 & 1 \\
2012 May 27.49 & 62.44 & VLA/B & 7.1  & 5.10 & 0.283 & 0.017 & 1 \\
2012 May 27.46 & 62.40 & VLA/B & 19.2 & 20.25& 0.183 & 0.017 & 1 \\
2012 May 27.46 & 62.40 & VLA/B & 24.5 & 21.98& 0.142 & 0.019 & 1 \\
2012 Jul 26.17 & 122.1 & VLA/B & 5.0 & 9.81 & 0.057 & 0.017 & 1 \\
2012 Jul 26.17 & 122.1 & VLA/B & 7.1 & 9.77 & 0.072 & 0.014 & 1 \\
2012 Jul 26.12 & 122.1 & VLA/B & 19.2 & 31.85& 0.052 & 0.016 & 1 \\
2012 Jul 26.12 & 122.1 & VLA/B & 24.5  & 26.57& $<0.07$ & 0.023 & 0
\enddata
\tablecomments{$^*$ The letter following `VLA' indicates the array configuration. $^{\dag}$ 
$1\sigma$ statistical uncertainties from AIPS task \texttt{JMFIT} (CARMA and SMA observations) or 
CASA task \texttt{IMFIT} (VLA observations). 
$^{\ddag}$PI: Y.~Urata.}
\end{deluxetable*}

\begin{figure}
 \centering
 \includegraphics[width=\columnwidth]{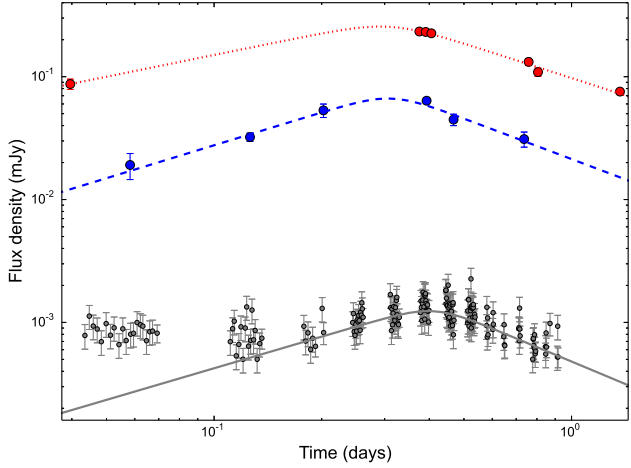}
 \caption{Broken power law fits to the X-ray (grey, solid), $U$-band (blue, dashed), and $z$-band 
(red, dotted) light curves for \me\ near the re-brightening around 0.35\,d. X-ray points before 
0.15\,d are not included in the fit. Errorbars at $z$-band are typically smaller than the size of 
the plotted symbols. The lines correspond to the independent fits at the three frequencies. The 
best 
fit parameters are listed in Table \ref{tab:120326A_bplfit}.
\label{fig:120326A_bplfit}}
\end{figure}

\section{Basic Considerations}
\label{text:basic_considerations}
We consider the X-ray to radio afterglow of \me\ in the context of the standard synchrotron 
afterglow model, where the afterglow emission is produced by synchrotron radiation from a 
non-thermal distribution of electrons. The electron energy distribution is assumed to be a power 
law 
with electron number density, $n(\gamma) \propto \gamma^{-p},$ for $\gamma > \gamma_{\rm min}$; 
here $\gamma_{\rm min}$ is the minimum Lorentz factor and the electron energy index, $p$ is 
expected to lie between 2 and 3. This electron distribution results in a synchrotron spectrum that 
can be 
described by a series of power law segments connected at `break frequencies': the self-absorption 
frequency, $\nua$, below which synchrotron self-absorption suppresses the flux, the characteristic 
synchrotron frequency, $\numax$, which corresponds to emission from electrons with 
$\gamma=\gamma_{\rm min}$, and the cooling frequency, $\nuc$, above which synchrotron cooling is 
important; the flux density at $\numax$ sets the overall flux normalization. The synchrotron model 
is described in detail in \citet{spn98}, \citet{cl00} and \citet{gs02}. 

\subsection{X-ray/UV/Optical Re-brightening at 0.4\,d}
\label{text:basic_considerations:re-brightening}
A prominent feature of the afterglow light curve of \me\ is a re-brightening at about 0.4\,d. We 
quantify the shape of the light curve at the re-brightening by fitting the X-ray data with a 
smoothly-joined broken power law of the form 
\begin{equation}
F_{\nu} = F_{\rm b} \left(
  \frac{(t/t_{\rm b})^{-y\alpha_1}+(t/t_{\rm b})^{-y\alpha_2}}{2}
\right)^{-1/y},
\end{equation}
where $t_{\rm b}$ is the break time, $F_{\rm b}$ is the flux at the break time, 
$\alpha_1$ and $\alpha_2$ are the temporal indices before and after the break, respectively, 
and $y$ is the sharpness of the break \footnote{We impose a floor of 12\% on the uncertainty of 
each data point, as explained in Section \ref{text:modeling}.}. The X-ray data before 0.15\,d 
exhibit a plateau and we therefore restrict the fit to span 0.15 to 1.25\,d. We use the Python 
function \texttt{curve\_fit} to estimate these model parameters and the associated covariance 
matrix. Our best-fit parameters are $t_{\rm b} = (0.41\pm0.02)\,\rm{d}$, $\alpha_1 = 0.85\pm0.19$, 
$\alpha_2 = -1.22\pm0.18$, and $y = 5\pm4$ (Figure \ref{fig:120326A_bplfit} and 
Table \ref{tab:120326A_bplfit}). 

The optical data are more sparsely sampled than the X-ray observations. We fit the $U$-band data 
between 0.05\,d and 1.25\,d after fixing $y=5$ as suggested by the fit to the better-sampled X-ray 
light curve. Our derived value of the peak time, $t_{\rm b} = 0.31\pm0.04$ is 
marginally earlier than, but close to the peak time of the X-ray light curve. The rise and decay 
rate in $U$-band are also statistically consistent with those derived from the X-ray light curve. 
Similarly, we fit the $z$-band light curve between 0.04\,d and 1.4\,d, fixing $y=5$. Since this 
time range includes a single point prior to the peak, the peak time and rise rate are degenerate 
in the fit. We therefore fix $t_{\rm b} = 0.31$ as derived from the $U$-band fit. The best-fit 
parameters are listed in Table \ref{tab:120326A_bplfit}, and are consistent with those derived for 
the UV and X-ray light curves.

Finally, we fit the X-ray, UV, and interpolated optical/NIR data jointly, where the three light 
curves are constrained to the same rise and decay rate, time of peak, and sharpness of the break, 
with independent normalizations in the three bands. Using a Markov Chain Monte Carlo simulation 
using \emcee\ \citep{fhlg13}, we find $t_{\rm b} = 0.40\pm0.01$, $\alpha_1 = 0.52\pm0.06$, and 
$\alpha_2 = -1.10\pm0.10$.

In summary, the X-ray/UV/optical light curves exhibit a prominent peak, nearly simultaneously 
in multiple bands (X-rays through the optical), which, therefore, cannot be related to the passage 
of a synchrotron break frequency. We explore various explanations for this behavior in Section 
\ref{text:energy_injection}.

\subsection{X-ray/radio Steep Decline: $\numax$ and $t_{\rm jet}$}
\label{text:basic_considerations:steep_decline}
The X-ray data at $\gtrsim1.4\,$d exhibit a steep decline, with $\alpha_{\rm X}=-2.29\pm0.16$. In 
the standard afterglow model a steep decline of $\alpha \approx -p \lesssim-2$ at frequencies above 
$\numax$ is expected after the `jet-break' ($t_{\rm jet}$), when the bulk Lorentz factor, $\Gamma$, 
decreases below the inverse opening angle of the jet, $1/\thetajet$, and the edges of the 
collimated outflow become visible. The $r^{\prime}$-band light curve also exhibits a 
shallow-to-steep transition at $\approx1.5$\,d, with a post-break decay rate of $\alpha_{\rm R} = 
-2.27\pm0.11$, consistent with the X-rays. These observations suggest that the jet break occurs at 
about $1.5$\,d. We now consider whether this interpretation is consistent with the radio 
observations.

\begin{figure}
 \centering
 \includegraphics[width=\columnwidth]{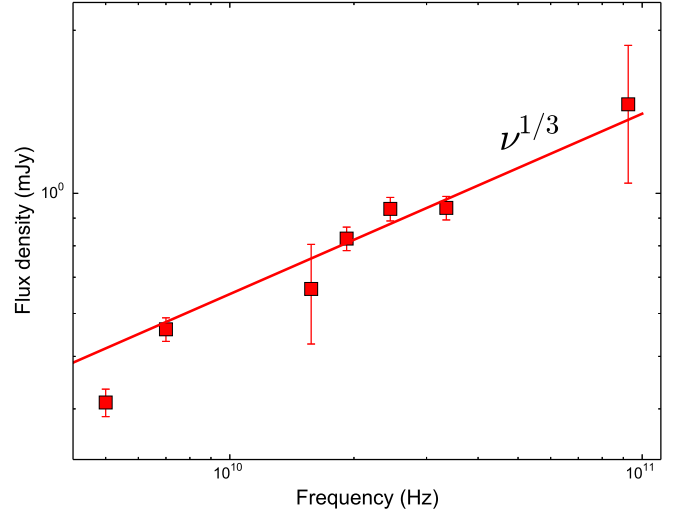}
 \caption{Afterglow SED for \me\ at 9\,d from 5\,GHz to 92.5\,GHz together with a best fit 
power law fit to the data above 7\,GHz. The spectrum is optically thin with a spectral index of 
$\nu^{0.35\pm0.03}$ from 7\,GHz to 92.5\,GHz. The 5\,GHz observation shows evidence of synchrotron 
self-absorption. \label{fig:120326A_radiosed}}
\end{figure}

The multi-wavelength radio SED at 9\,d is well-fit by a single power law with spectral index, 
$\beta = \nicefrac{1}{3}$ from 7\,GHz to 93\,GHz (Figure \ref{fig:120326A_radiosed} and Section 
\ref{text:basic_considerations:nua}), indicating that $\numax>93$\,GHz at this time. In the absence 
of a jet break, we would expect the flux density below $\numax$ to remain constant (for a wind-like 
circumburst environment) or rise with time (for a constant density circumburst environment). 
However, the radio light curves decline after 4.6\,d at all frequencies from 15\,GHz to 93\,GHz. The 
combination of $\numax>93$\,GHz at 9\,d and the declining light curve in the radio bands is only 
possible in the standard afterglow model if a jet break has occurred before 4.6\,d. This is 
consistent with the steepening observed in the X-ray and optical light curves at $\approx 1.5$\,d, 
suggesting that $t_{\rm jet} \approx 1.5$\,d.

We note that the flux at a given frequency decays steeply following a jet break only once $\numax$ 
has crossed the observing frequency; thus the steepening in the radio light curves is expected to 
be delayed past that of the steepening in the X-ray and optical light curves until $\numax$ passes 
through the radio band. We find that the 7\,GHz light curve is consistent with being flat 
to 50\,d, after which it declines rapidly with $\alpha_{\rm radio} \approx -2$. This suggests that 
$\numax$ crosses the $7$\,GHz band at around $50$\,d. Since $\numax$ is expected to decline as 
$t^{-2}$ following the jet break, we have $\nu_{\rm m}\approx8\times10^{12}$\,Hz at 1.5\,d if we 
take $t_{\rm jet}\approx1.5$\,days. Thus $\numax$ is below both the X-rays and the optical 
frequencies at $t=t_{\rm jet}$, consistent with the steepening being observed around the same time 
in the X-ray and optical bands.

To summarize, a simultaneous steepening in the X-ray and optical light curves between 1 and 2\,d 
indicates that $t_{\rm jet}\approx 1.5$\,d. Taken together with a similar steepening observed in 
the 7\,GHz radio light curve around 50\,d, we find $\numax\approx7$\,GHz at 50\,d.

\begin{deluxetable*}{lcccc}
\tabletypesize{\footnotesize}
\tablecolumns{5}
\tablewidth{0pt}
\tablecaption{Parameters for broken power law fit to X-ray/UV/optical re-brightening for  
GRB~120326A\label{tab:120326A_bplfit}}
\tablehead{
  \colhead{} &
  \colhead{XRT} &  
  \colhead{$U$-band} &
  \colhead{$z$-band} &  
  \colhead{Joint$^{\ddag}$ (MCMC)}  
  }
\startdata
Break time, $t_{\rm b}$ (d)    &  $0.41\pm0.02$ &  $0.31\pm0.04$ & $0.31^{\dag}$ & $0.40\pm0.02$ \\
Flux density at $t_{\rm b}$, 
         $F_{\rm b}$ ($\mu$Jy) &  $1.23\pm0.05$ & $66\pm7$       & $255\pm10$ &  ---  \\
Rise rate, $\alpha_1$          &  $0.85\pm0.19$ & $0.89\pm0.21$  & $0.59\pm0.03$ & $0.52\pm0.06$\\
Decay rate, $\alpha_2$         & $-1.22\pm0.18$ & $-1.10\pm0.29$ & $-0.94\pm0.05$ & $-1.10\pm0.10$\\
Smoothness, $y$                &  $5.0\pm4.0$   & $5.0^{\dag}$   & $5.0^{\dag}$&$17\pm10$
\enddata
\tablecomments{$^{\dag}$ Parameters fixed during fit (see text for details). $^{\ddag}$ We used a 
flat prior on all parameters, with the ranges $t_{\rm b}\in[0.05,1.0]$, $F_{\rm 
b,X}\in[1\times10^{-4},1\times10^{-2}]$\,mJy, $F_{\rm 
b,UV}\in[5\times10^{-3},1\times10^{-1}]$\,mJy, 
$F_{\rm b,z}\in[1\times10^{-2},2]$\,mJy,$\alpha_1 \in[0.1,5.0]$,$\alpha_2 \in[-5.0,0.0]$, and 
$y\in[0.5,30]$.}
\end{deluxetable*}

\subsection{The Circumburst Density Profile and the Location of $\nuc$}
\label{text:basic_considerations:nuc}
The density profile in the immediate (sub-pc scale) environment of the GRB progenitor impacts the 
hydrodynamic evolution of the shock powering the GRB afterglow. The evolution of the shock Lorentz 
factor is directly reflected in the afterglow light curves at all frequencies and multi-band 
modeling of GRB afterglows therefore allows us to disentangle different density profiles. Since 
the progenitors of long-duration GRBs are believed to be massive stars, the circumburst density 
structure is expected to be shaped by the stellar wind of the progenitor into a profile that falls 
of as $\rho(r)=A r^{-2}$ with radius $r$. Here $A = {\dot M_w}/4\pi V_w \equiv 
5\times10^{11}\Astar$\,g\,cm$^{-1}$ is a constant proportional to the progenitor mass-loss rate 
$\dot M_w$ (assumed constant), for a given wind speed, $V_w$ \citep{cl00}, and $\Astar$ is a 
dimensionless parametrization, corresponding to $\dot M_w = 1\times10^{-5}\,M_{\odot}\,{\rm 
yr}^{-1}$ and $V_w = 1000$\,km\,s$^{-1}$. Alternatively, the shock may directly encounter the 
uniform interstellar medium (ISM). Both wind- and ISM-like environments have been inferred for 
different events from previous observations of GRB afterglows 
(e.g., \citealt{pk00,hys+01,pk01,pk02,yfh+02,fyb+03,ccf+08,cfh+10,cfh+11,lbz+13,lbt+14}). Here, we 
explore the constraints on the progenitor environment of \me.

\begin{figure}
 \centering
 \includegraphics[width=\columnwidth]{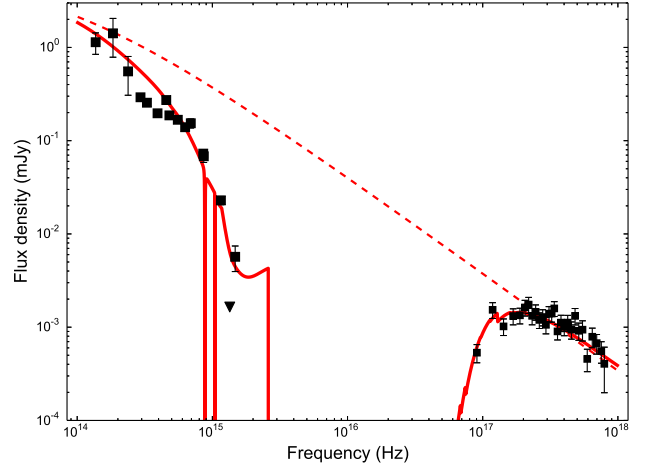}
 \caption{Afterglow SED for \me\ at 0.33\,d from the NIR to the X-rays together with the best fit 
forward shock model (red, solid). The optical and UV data exhibit a clear decrement due to 
extinction in the host galaxy. The solid line is the SED from the highest-likelihood model 
and the dashed curve indicates the SED in the absence of absorption along the line of sight to the 
GRB by the Milky Way and the GRB host (Section \ref{text:results:ISM_lowp}). The spectral break 
apparent in the model SED at $\approx10^{14}$\,Hz is $\numax$. The data above 
$\approx9\times10^{14}$\,Hz (Ly-$\alpha$ in GRB rest frame) are likely affected by absorption by 
the 
intergalactic medium (IGM) along the line of sight. \label{fig:120326A_sed}}
\end{figure}

The spectral index between the PAIRITEL $K$-band observation at 1.4\,d \citep{gcn13143} and 
the X-rays is $\beta_{\rm NIR,X}=-0.96\pm0.05$ (Figure \ref{fig:120326A_sed}), which is consistent 
with the X-ray spectral index of $\beta_{\rm X} = -0.85\pm0.04$ (Section 
\ref{text:data_analysis:XRT}) at $2\sigma$, 
suggesting that the NIR, optical, and X-ray bands are located on the same power law segment of the 
afterglow SED at 1.4\,d. At the same time, the spectral index within the NIR/optical bands is 
$\beta=-1.80\pm0.16$, indicating that extinction is present. 

Since $\beta_{\rm NIR,X}\approx\beta_{\rm X}\approx-0.90$, the cooling frequency, $\nuc$ must lie 
either below the NIR or above the X-rays at 1.4\,d. For $\nuc<\nu_{\rm IR}$, we would infer an
electron energy index, $p\approx1.8$ and a light curve decay rate of $\alpha\approx-0.85$ 
regardless of the circumburst density profile. On the other hand, $\nuc>\nuX$ requires 
$p\approx 2.8$ with $\alpha \approx -1.35$ for a constant density environment and 
$\alpha\approx-1.85$ for a wind-like environment. Actual measurements of the light curve decay 
rate at $\approx1.4$\,d are complicated by the presence of the jet break at around this time. The 
optical and X-ray light curves decline as $\alpha_{\rm R}=-2.05\pm0.13$ and $\alpha_{\rm 
X}=2.29\pm0.16$ after the jet break, with the expected decay rate being $\alpha\approx-p$ 
\citep{rho99,sph99}. This indicates $p\approx2$ (Section 
\ref{text:basic_considerations:steep_decline}) and $\nuc < \numax$ for an ISM 
model. Upon detailed investigation (Section \ref{text:modeling}), we find that a $p\approx2$ model 
with an ISM-like environment fits the data after the re-brightening well. This model additionally 
requires $\nuc<\numax$ (fast cooling) until $\approx 2$\,d. 
For completeness, we present our investigation of the wind model in 
Appendix \ref{appendix:120326A_wind}.

\subsection{Location of $\nua$}
\label{text:basic_considerations:nua}
The radio SED from 7\,GHz to 93\,GHz at $\approx9$\,days is optically thin with a spectral 
slope of $\beta = 0.33\pm0.04$ (Figure \ref{fig:120326A_radiosed}), suggesting that $\numax$ lies 
above 93\,GHz and $\nua$ lies below 7\,GHz at this time. This is consistent with the passage of 
$\numax$ through 7\,GHz at 50\,d inferred in Section 
\ref{text:basic_considerations:steep_decline}. 
The spectral index between 5\,GHz and 7\,GHz at 9\,d, $\beta = 0.92\pm0.23$, is steeper 
than $\nu^{1/3}$. This may suggest that the synchrotron self-absorption frequency, $\nua\approx 
5$\,GHz. However, this spectral index does not show a monotonic trend with time. Since this 
part of the radio spectrum is strongly affected by interstellar scintillation (ISS) in the ISM of 
the Milky Way, a unique interpretation of the observed spectral index is difficult at these
frequencies. This difficulty in constraining $\nua$ results in degeneracies in the physical 
parameters. We return to this point in Section \ref{text:results:ISM_lowp}.

\section{Multi-wavelength Modeling}
\label{text:modeling}
Although the panchromatic peak at 0.4\,d is a unique feature of \me, the X-ray, optical, 
and radio light curves of this event exhibits standard afterglow features after this time, with 
evidence for an un-broken power law spectrum extending from the optical to the X-rays (Section 
\ref{text:basic_considerations:nuc}), a $\nu^{\nicefrac{1}{3}}$ spectrum in the radio (Section 
\ref{text:basic_considerations:nua}), and evidence for a jet break 
at $t_{\rm jet} \approx 1.5$\,d. (Section \ref{text:basic_considerations:steep_decline}). We 
therefore determine the physical properties of this event by using observations after the 
X-ray/optical peak. We model the data at $\gtrsim 0.4$\,d as arising from the afterglow 
blastwave, using the smoothly-connected power law synchrotron spectra described by \cite{gs02}. We 
compute the break frequencies and normalizations using the standard 
parameters: the fractions of the blastwave energy imparted to relativistic electrons 
($\epse$) and magnetic fields ($\epsb$), the kinetic energy ($\EKiso$), and the 
circumburst density ($\dens$).  We also use the SMC extinction curve\footnote{Our previous work 
shows negligible changes in the blastwave parameters with LMC and Milky Way-like extinction curves 
\citep{lbt+14}.} \citep{pei92} to model the extinction in the host galaxy ($A_{\rm V}$). Since the 
$R$-band light curve flattens at $\approx10$ days due to contribution from the host, we fit for the 
$R$-band flux density of the host galaxy as an additional free parameter.

The various possible orderings of the spectral break frequencies (e.g., $\numax<\nuc$: `slow 
cooling' and $\nuc<\numax$: `fast cooling') give rise to five possible shapes of the afterglow SED 
\citep{gs02}. Due to the hydrodynamics of the blastwave, the break frequencies evolve with time 
and the SED transitions between spectral shapes. To preserve smooth light curves when break 
frequencies cross and the spectral shape changes, we employ the weighting schemes described in 
\cite{lbt+14} to compute the afterglow SED as a function of time. To efficiently and rapidly sample 
the available parameter space, we carry out a Markov Chain Monte Carlo (MCMC) analysis using a 
python implementation of the ensemble MCMC sampler \emcee\ \citep{fhlg13}. For a detailed 
discussion of our modeling scheme, see \cite{lbt+14}. To account for heterogeneity of 
UV/Optical/NIR data collected from different observatories, we usually impose an uncertainty floor 
of $5\%$ prior to fitting with our modeling software. For this GRB, we find that the fit is driven 
by the optical data at the expense of fits at the radio and X-ray bands, which we mitigate by 
increasing the uncertainty floor further to $12\%$. We additionally correct for the effect of 
inverse Compton cooling (Appendix \ref{appendix:IC}).

\section{Results for GRB~120326A}
\label{text:results}
\subsection{Multi-wavelength model at $\gtrsim 0.4$\,d}
\label{text:results:ISM_lowp}
\begin{figure*}
\begin{tabular}{cc}
 \centering
 \includegraphics[width=0.47\textwidth]{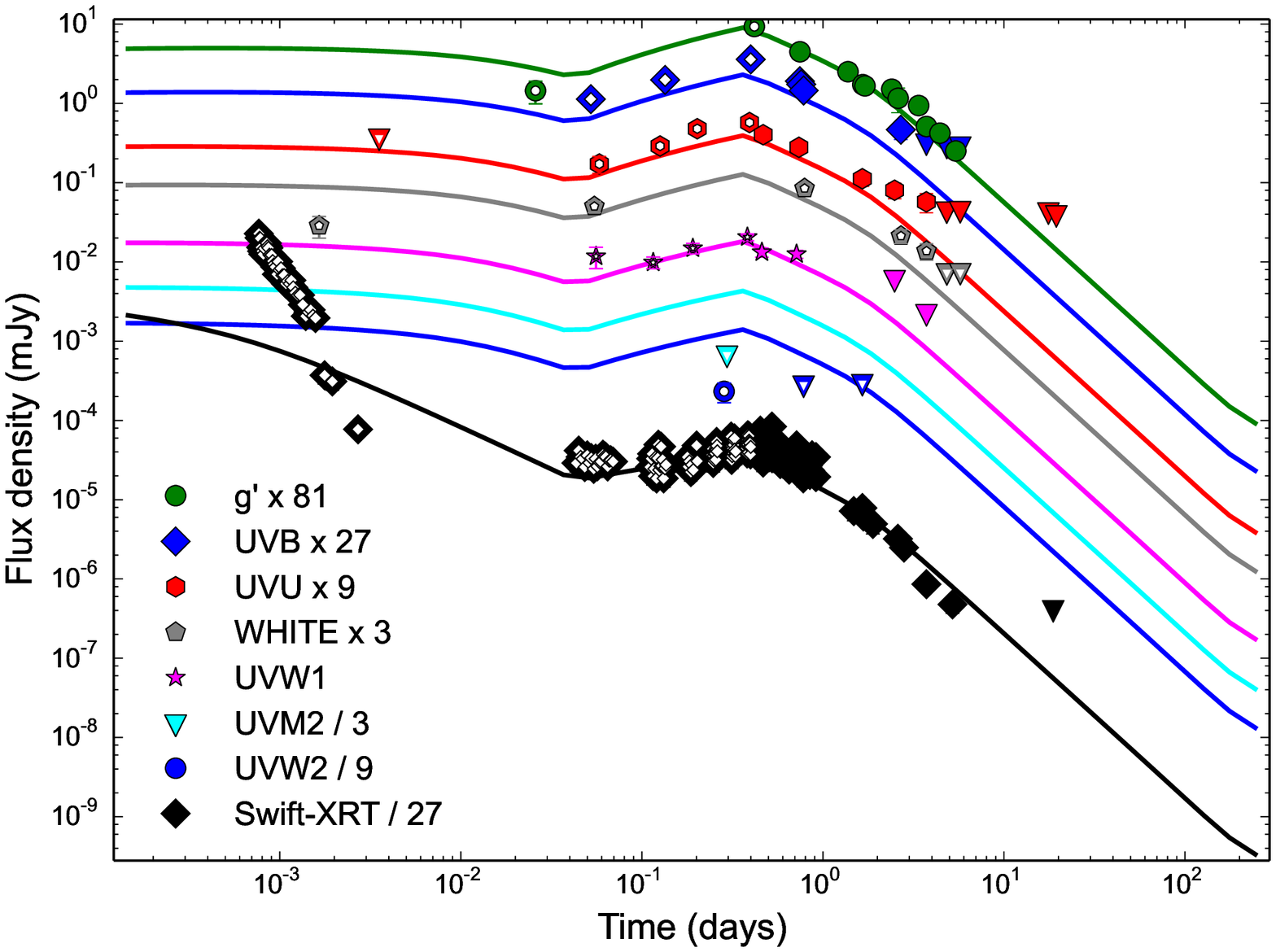} &
 \includegraphics[width=0.47\textwidth]{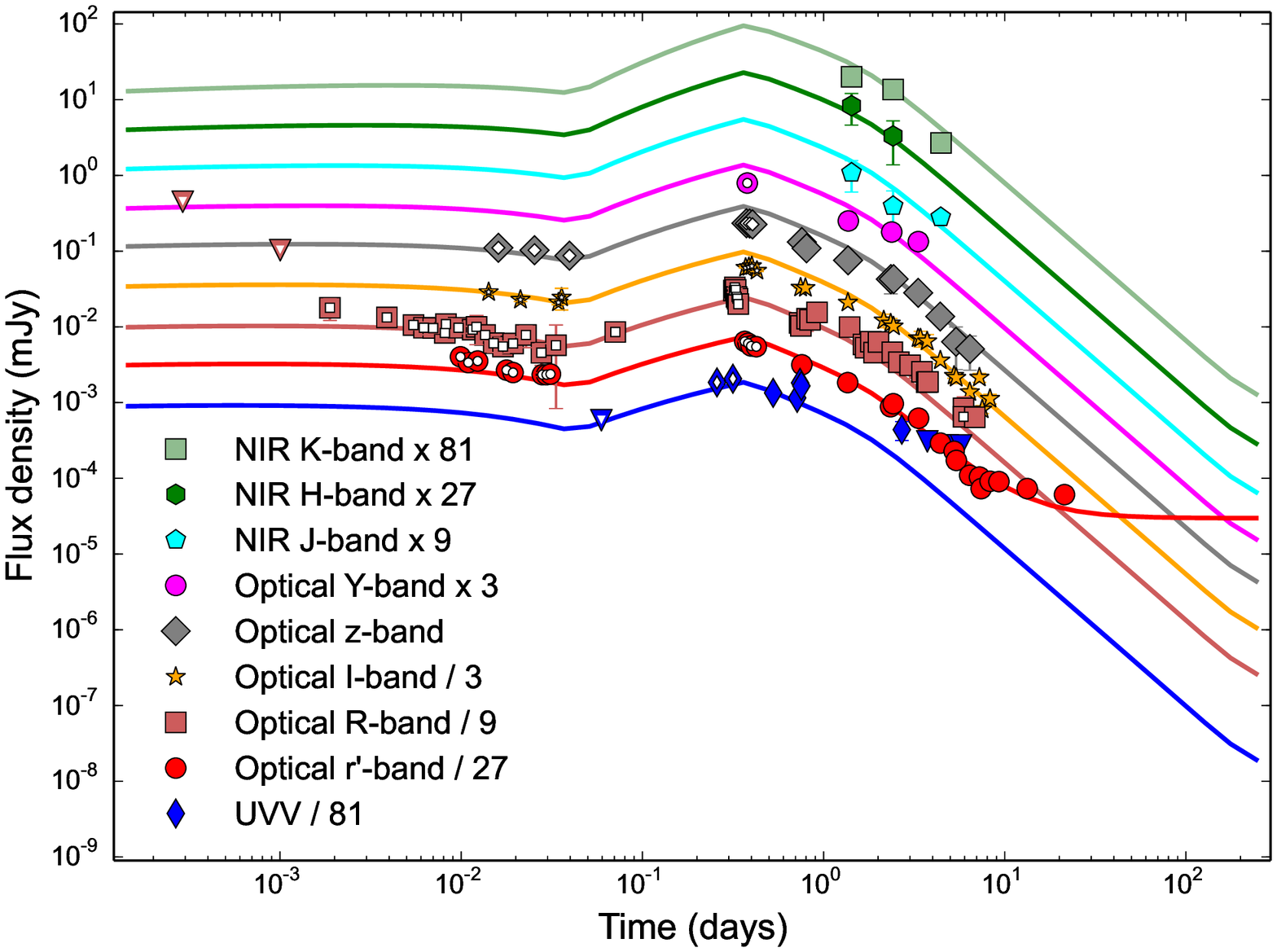} \\
 \includegraphics[width=0.47\textwidth]{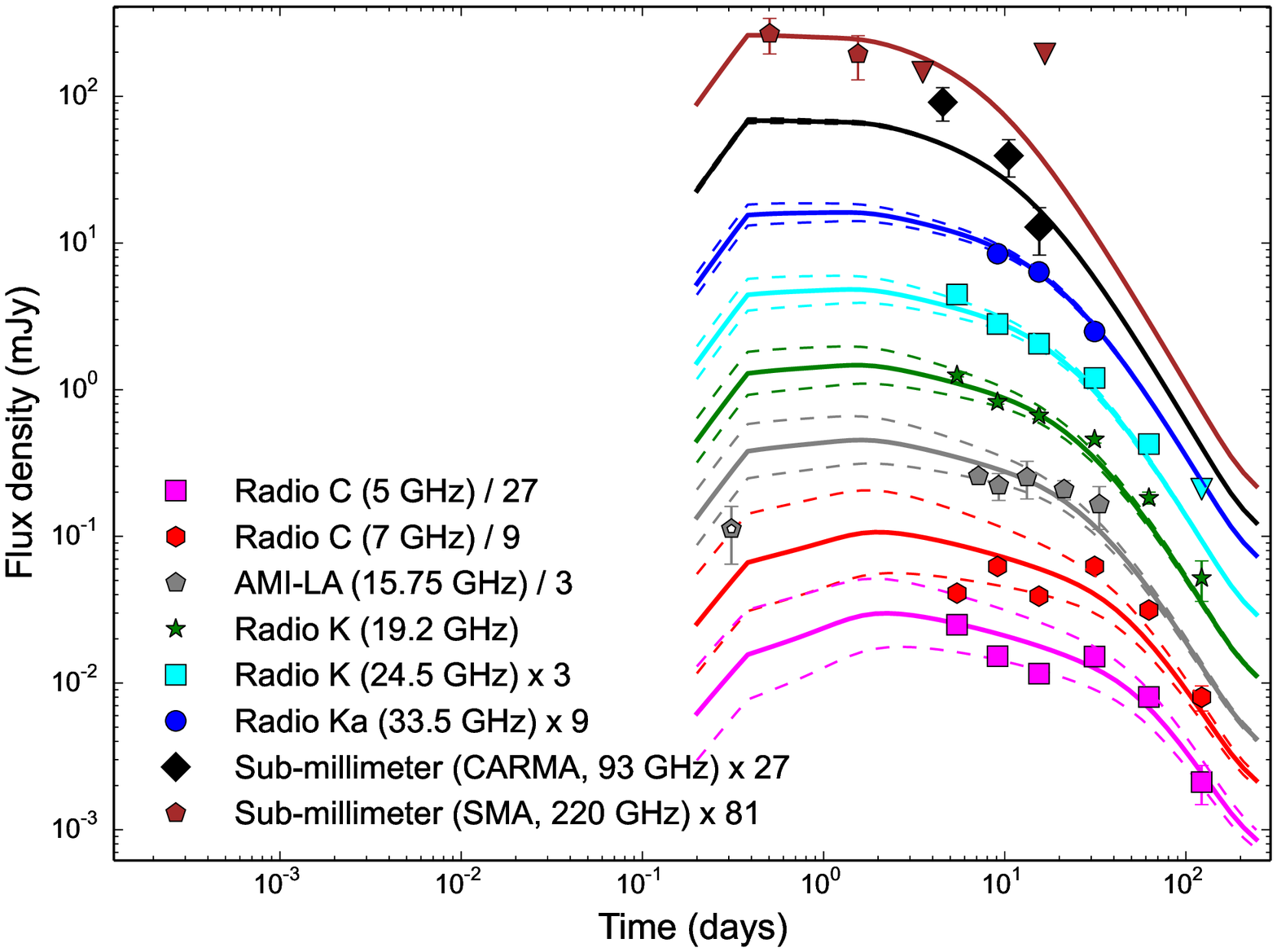} &
 \includegraphics[width=0.47\textwidth]{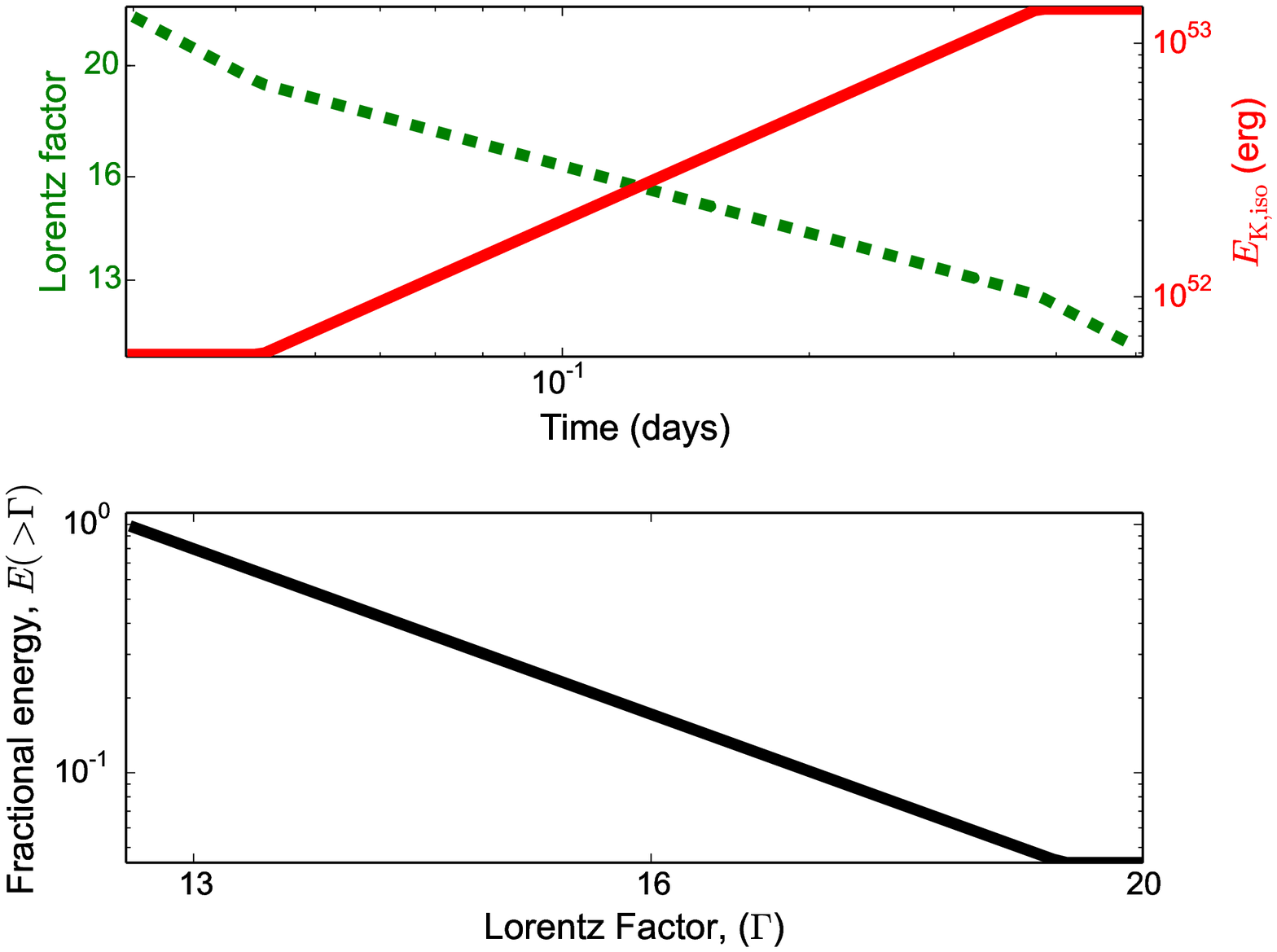} \\ 
\end{tabular}
\caption{X-ray, UV (top left), optical (top right), and radio (bottom left) light curves of 
GRB~120326A in the ISM scenario, with the full afterglow model (solid lines), including energy 
injection before 0.4\,d. The X-ray data before 0.004\,d is likely dominated by high-latitude prompt 
emission and we do not include these data in our analysis. The dashed envelopes around the radio 
light curves indicate the expected effect of scintillation at the $1\sigma$ level. The data prior 
to the end of the re-brightening at 0.4\,d (open symbols) are not used to determine the parameters 
of the forward shock in the MCMC analysis. The \Swift/UVOT data in the \textit{uvw2} and 
\textit{uvm2} bands are strongly affected by IGM absorption and are not included in the analysis. 
Bottom right: blastwave Lorentz factor (green, dashed; upper sub-panel) and isotropic equivalent 
kinetic energy (red, solid; upper sub-panel) as a function of time, together with the energy 
distribution across ejecta Lorentz factors (black, solid; lower sub-panel) as determined from 
fitting the X-ray/UV/optical re-brightening at 0.4\,d.
\label{fig:120326A_enj}}
\end{figure*}

In confirmation of the basic analysis presented in Section \ref{text:basic_considerations:nuc}, 
our highest likelihood model (Figure \ref{fig:120326A_enj}) has $p\approx2.09$, $\epse\approx0.33$, 
$\epsb\approx0.33$, $\dens\approx0.27\,\pcc$, $\EKiso\approx1.4\times10^{53}$\,erg, 
$\tjet\approx1.5$\,d, $\AV\approx0.48$\,mag, and an $r^{\prime}$-band flux density of $2.3\,\mu$Jy 
for the host galaxy. The spectrum is in fast cooling until 1.4\,d. During the fast cooling phase, 
$\nua$ splits into two distinct frequencies: $\nuac$ and $\nusa$ \citep{gs02}. The spectrum has the 
Rayleigh-Jeans shape ($\nu^2$) below $\nuac$, and a slope of $\nu^{11/8}$ between $\nuac$ and 
$\nusa$. For the highest likelihood model, the break frequencies\footnote{
$\nuc$ here is reported as $\sqrt{\nu_3\nu_{11}}$, where $\nu_3$ and $\nu_{11}$ are 
(differently-normalized) expressions for the cooling frequency in the slow cooling and fast 
cooling regimes, respectively \citep{gs02},
such that $\nu_3/\nu_{11}=10.87(p-0.46)e^{-1.16p}$. We find 
$\nu_3 = 1.2\times10^{13}$ and $\nu_{11}=7.4\times10^{12}$ at 1\,d.
\label{footnote:nuc}
}
are located at $\nuac\approx2.8\times10^{9}$\,Hz,  $\nusa\approx4.9\times10^{9}$\,Hz, 
$\nuc\approx9.3\times10^{12}$, and $\numax\approx1.5\times10^{13}$\,Hz at 1\,d and the peak flux 
density is $\approx18$\,mJy at $\nuc$. \numax\ evolves as $t^{-2}$ following the jet break at 
$\approx1.5$\,d to $\approx7.7$\,GHz at 50\,d, consistent with the basic considerations outlined in 
Section \ref{text:basic_considerations:steep_decline}. The Compton $y$-parameter is $\approx0.6$, 
indicating that cooling due to inverse-Compton scattering is moderately significant.

\begin{figure}
\begin{tabular}{ccc}
 \centering
 \includegraphics[width=0.30\columnwidth]{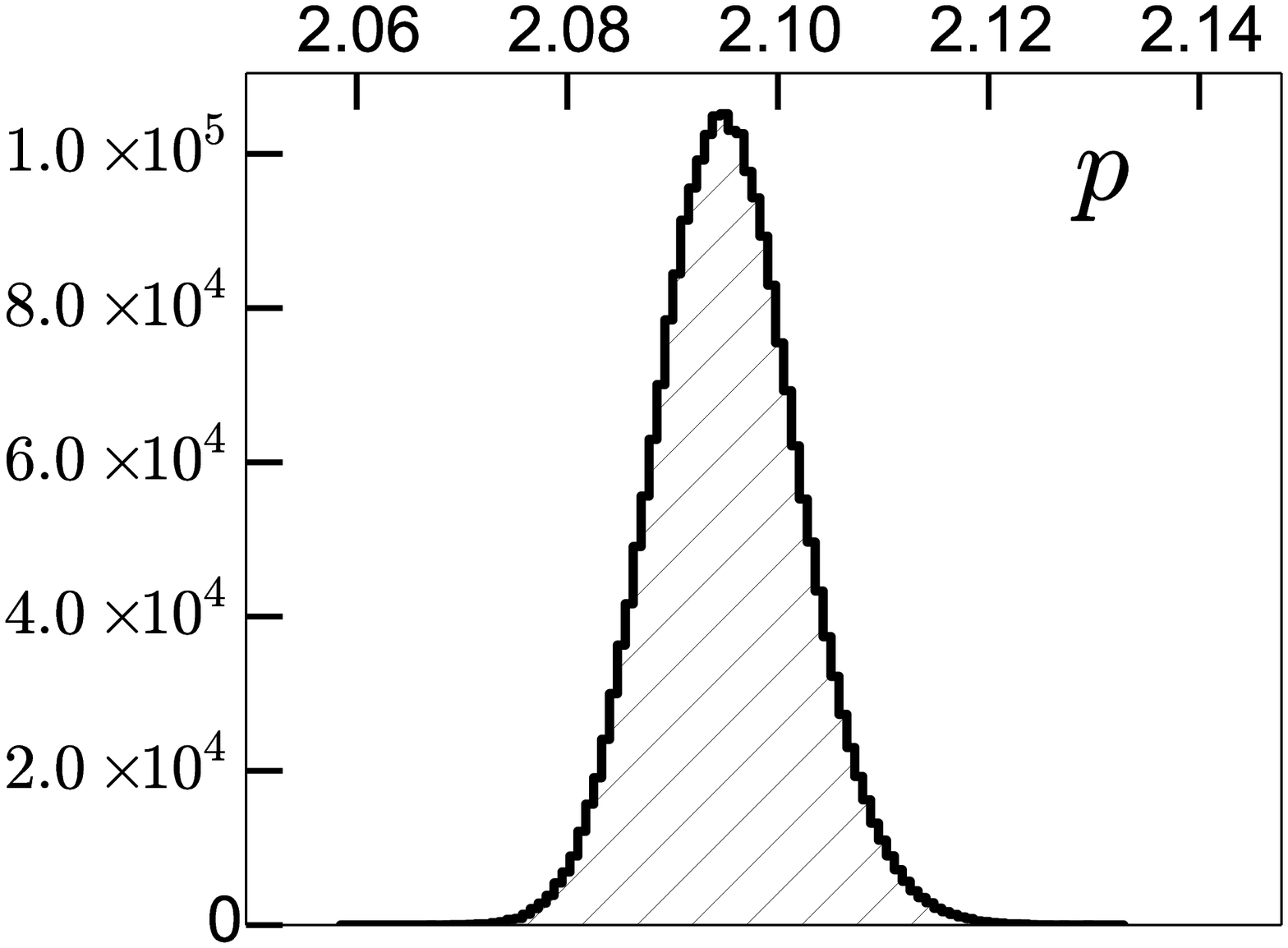} &
 \includegraphics[width=0.30\columnwidth]{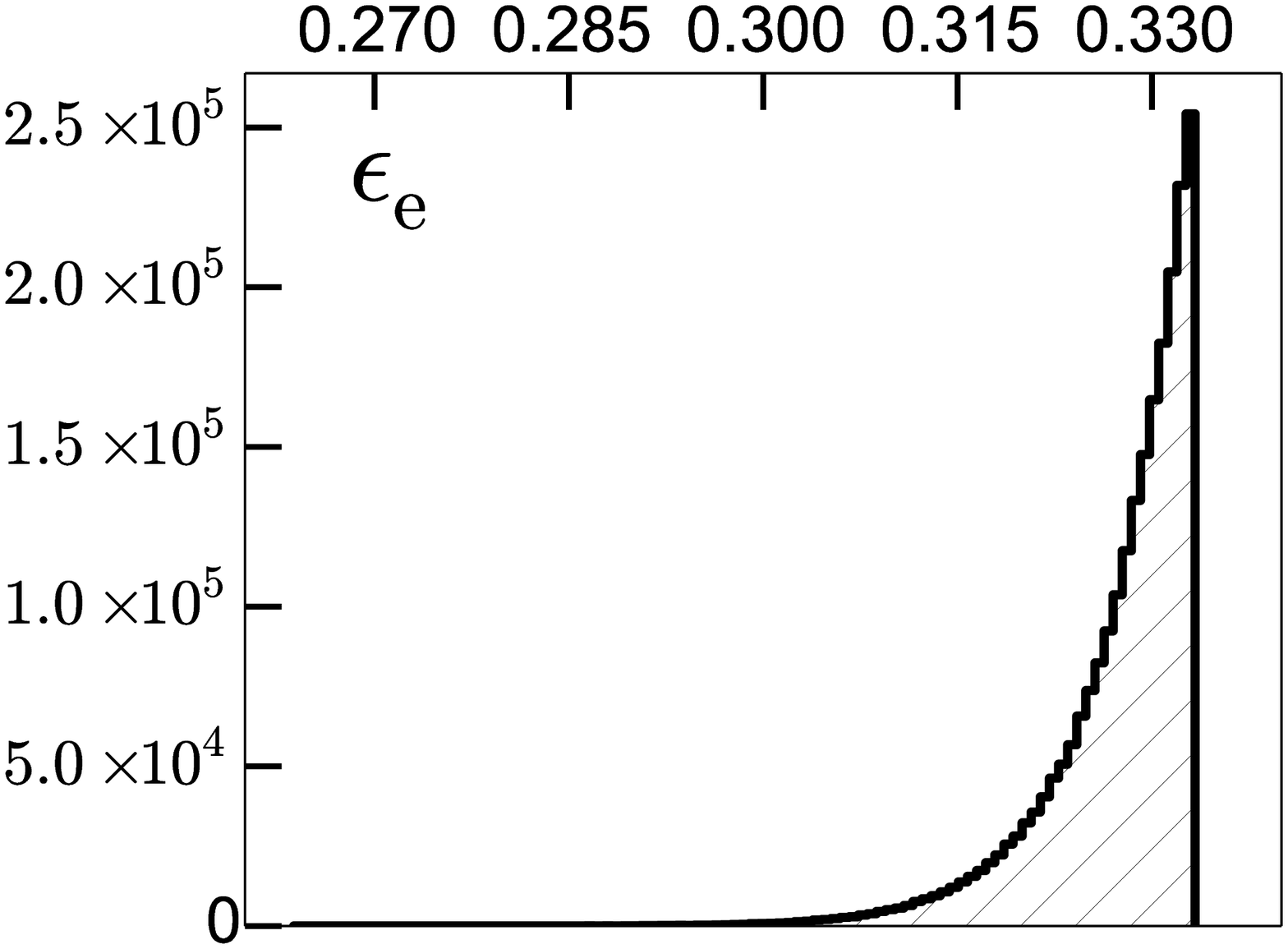} &
 \includegraphics[width=0.30\columnwidth]{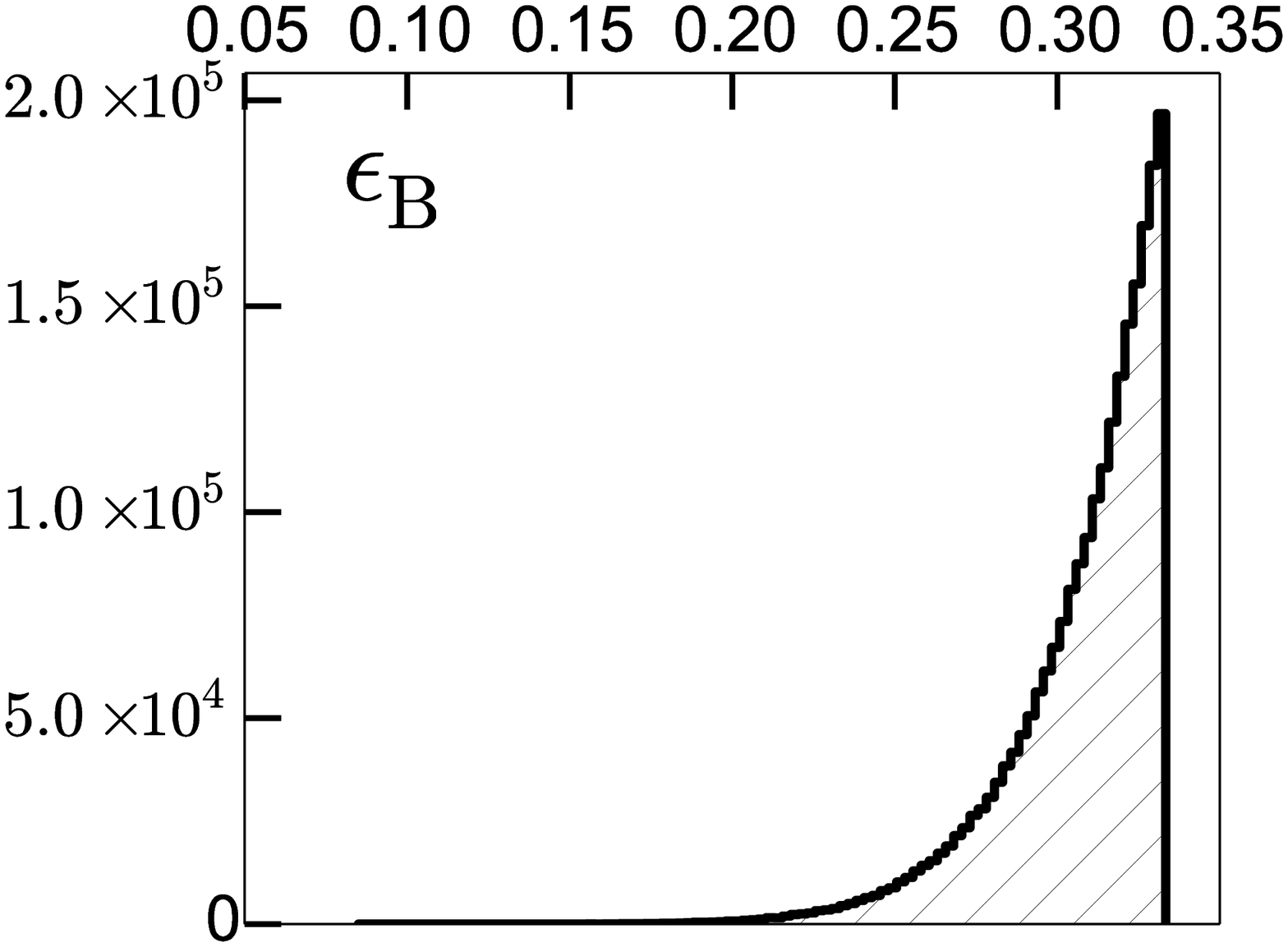} \\
 \includegraphics[width=0.30\columnwidth]{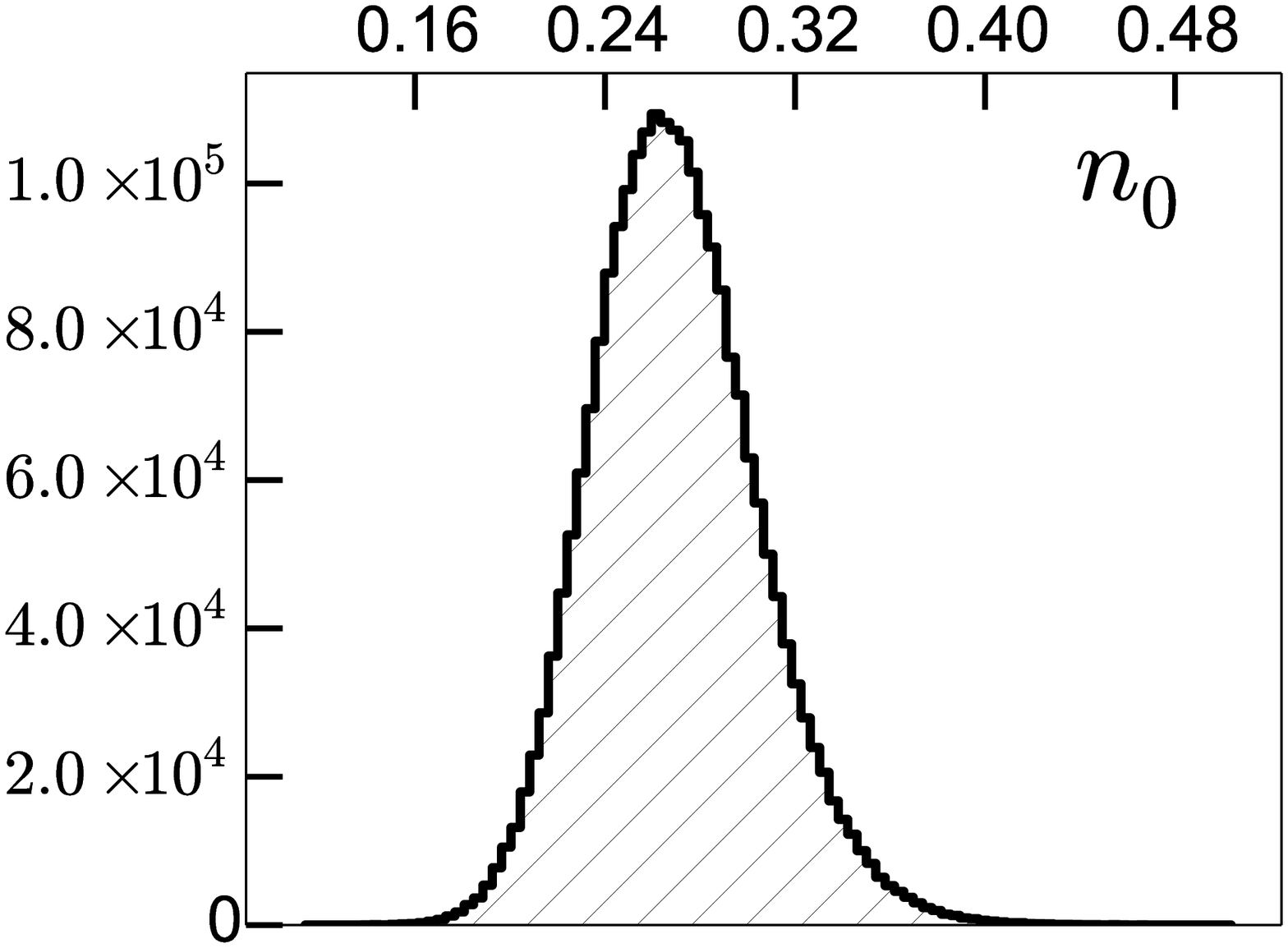} & 
 \includegraphics[width=0.30\columnwidth]{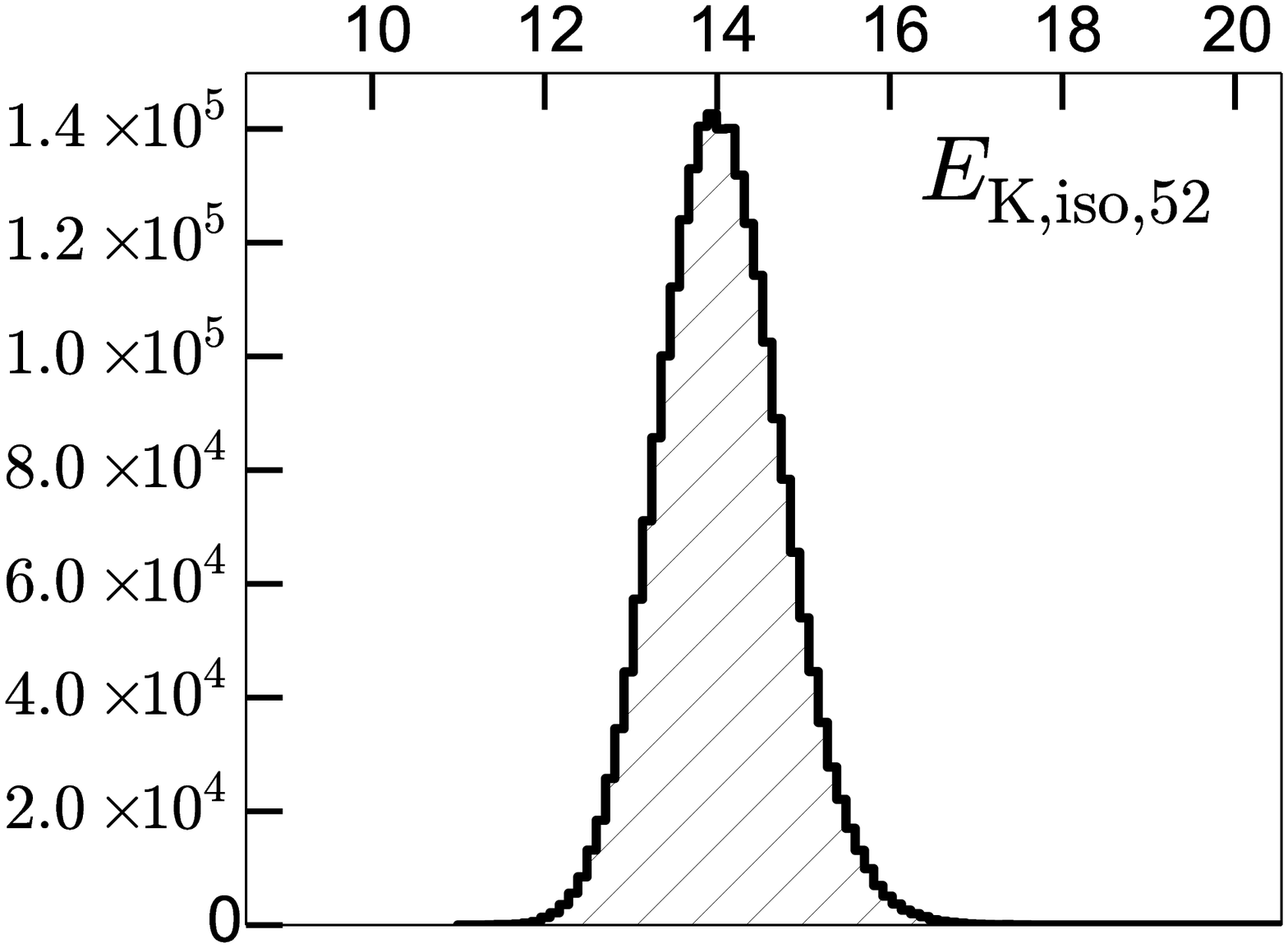} &
 \includegraphics[width=0.30\columnwidth]{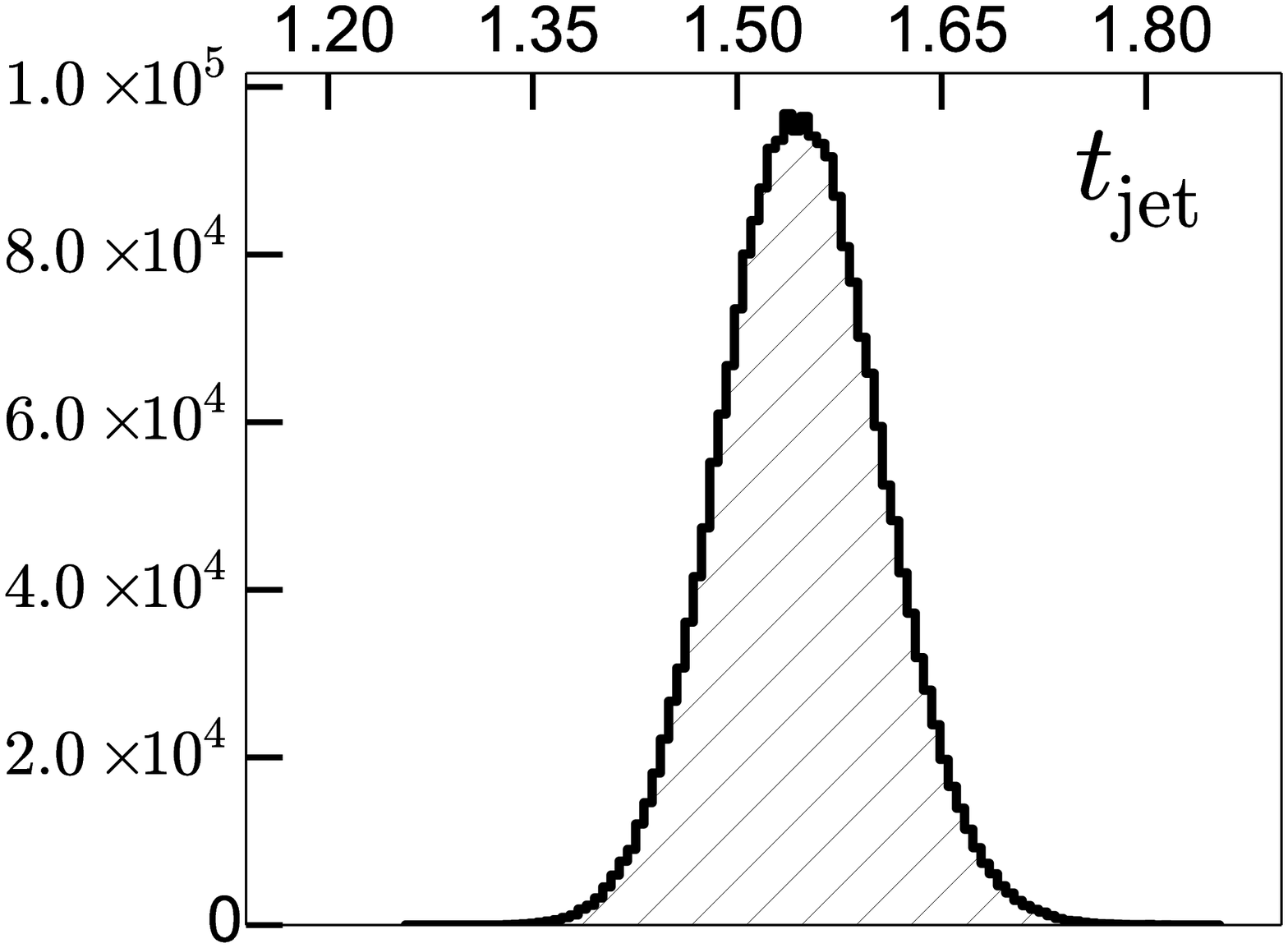} \\ 
 \includegraphics[width=0.30\columnwidth]{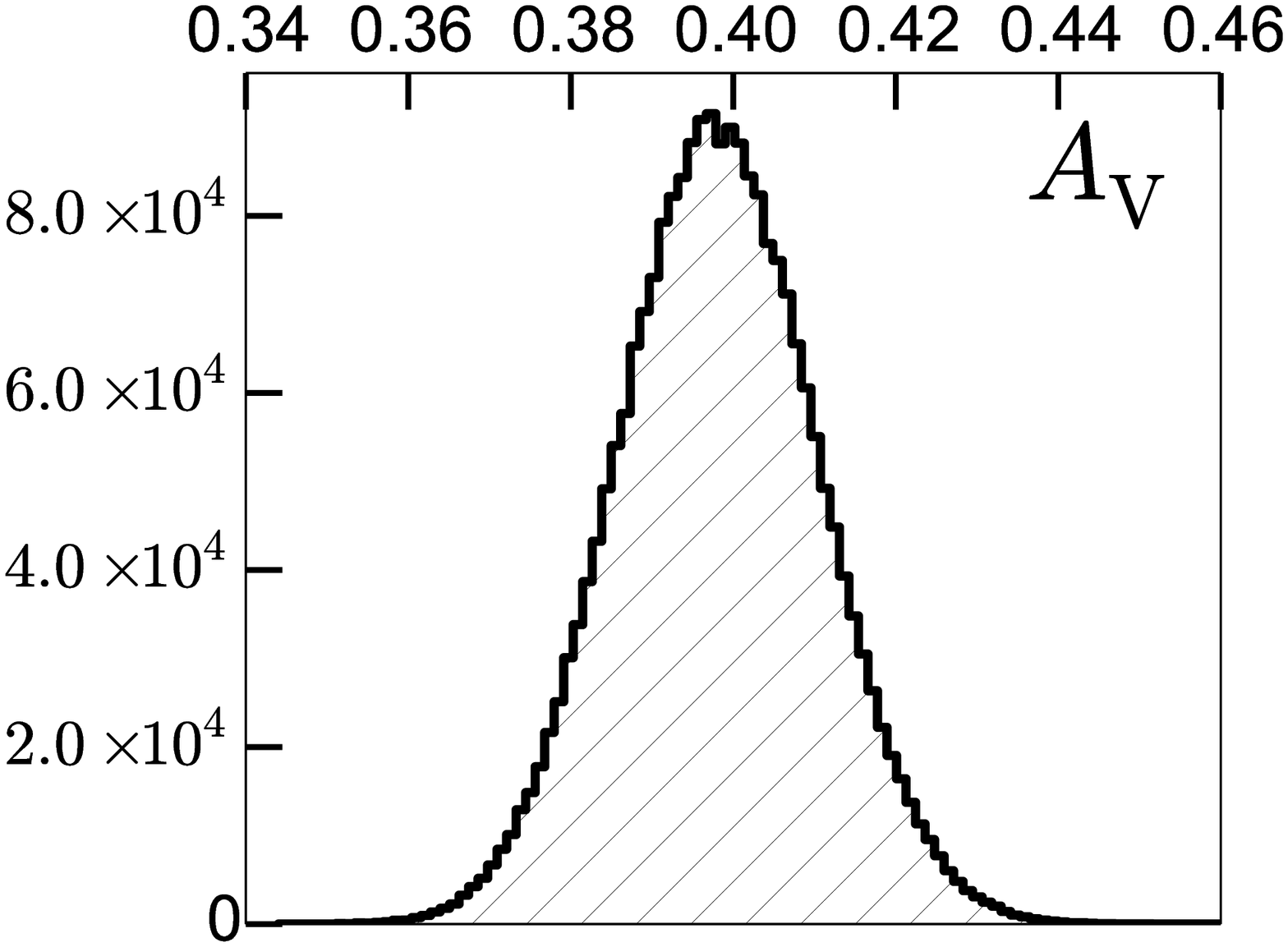} &
 \includegraphics[width=0.30\columnwidth]{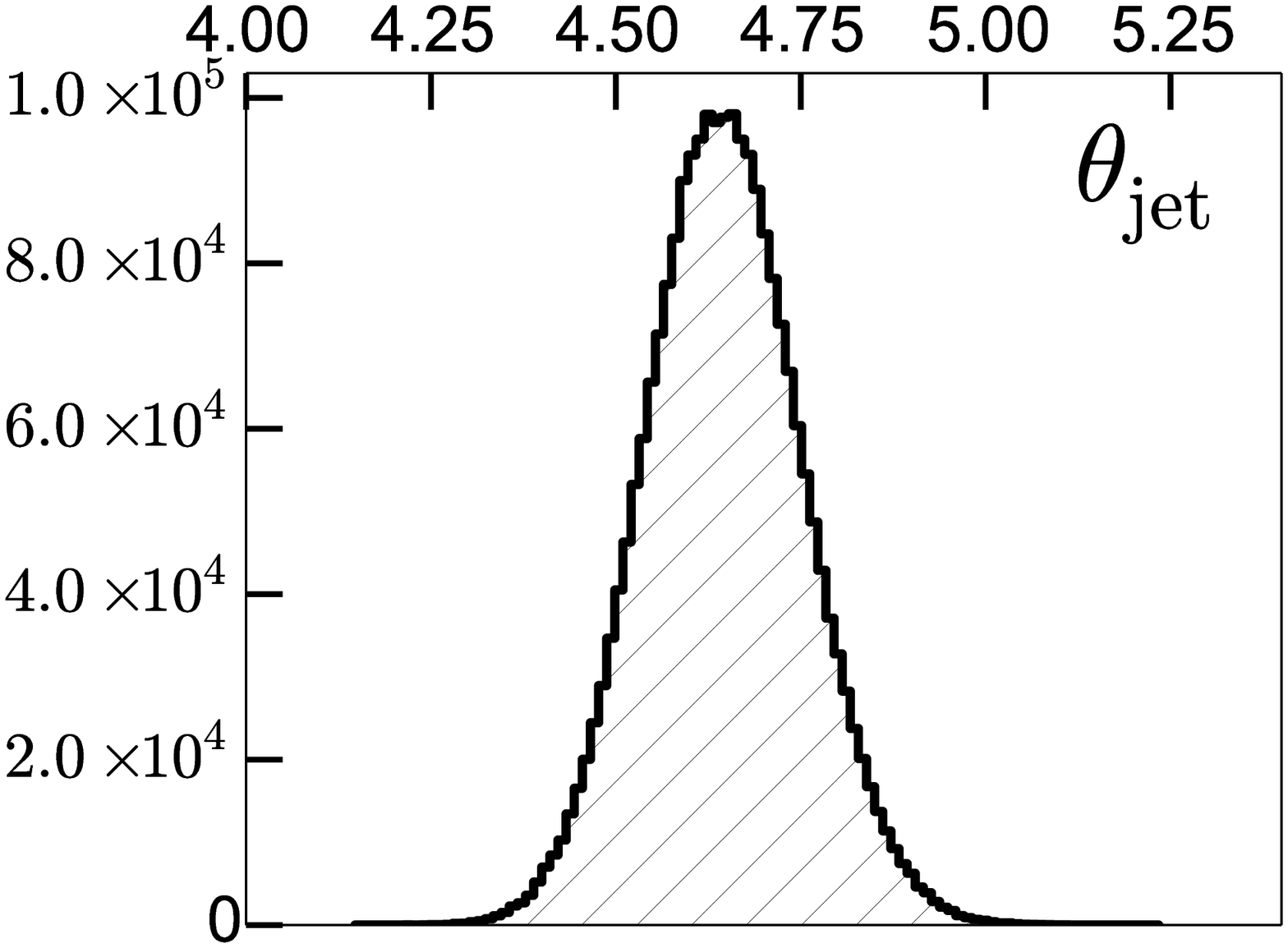}  &
 \includegraphics[width=0.30\columnwidth]{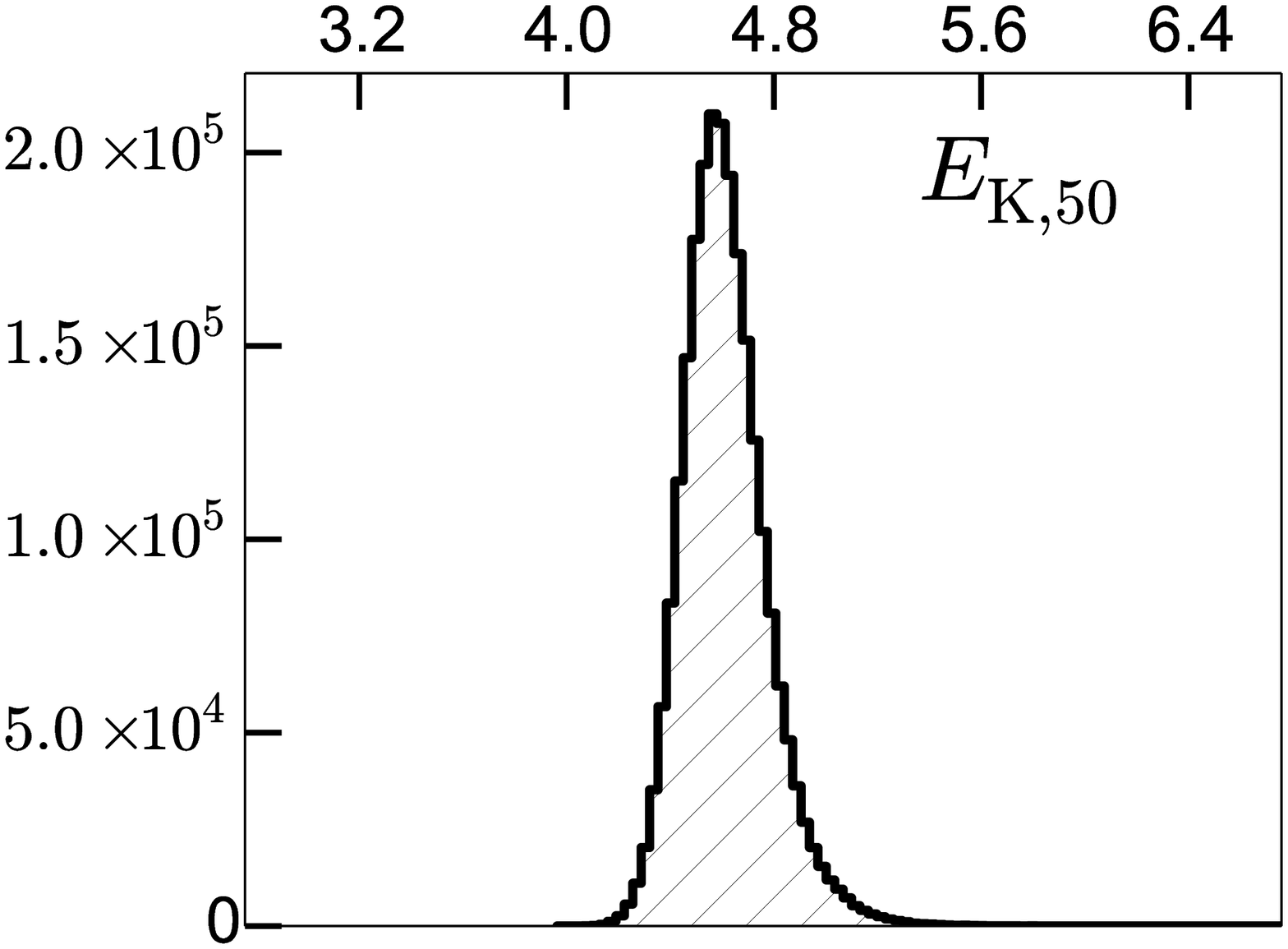} \\  
\end{tabular}
\caption{Posterior probability density functions for the physical parameters of GRB~120326A in 
the ISM model from MCMC simulations. We have restricted $\epsilon_{\rm e} < \nicefrac{1}{3}$ and
$\epsilon_{\rm B} < \nicefrac{1}{3}$.
\label{fig:120326A_ISM_hists}}
\end{figure}

\begin{figure}
\begin{tabular}{ccc}
\centering
 \includegraphics[width=0.30\columnwidth]{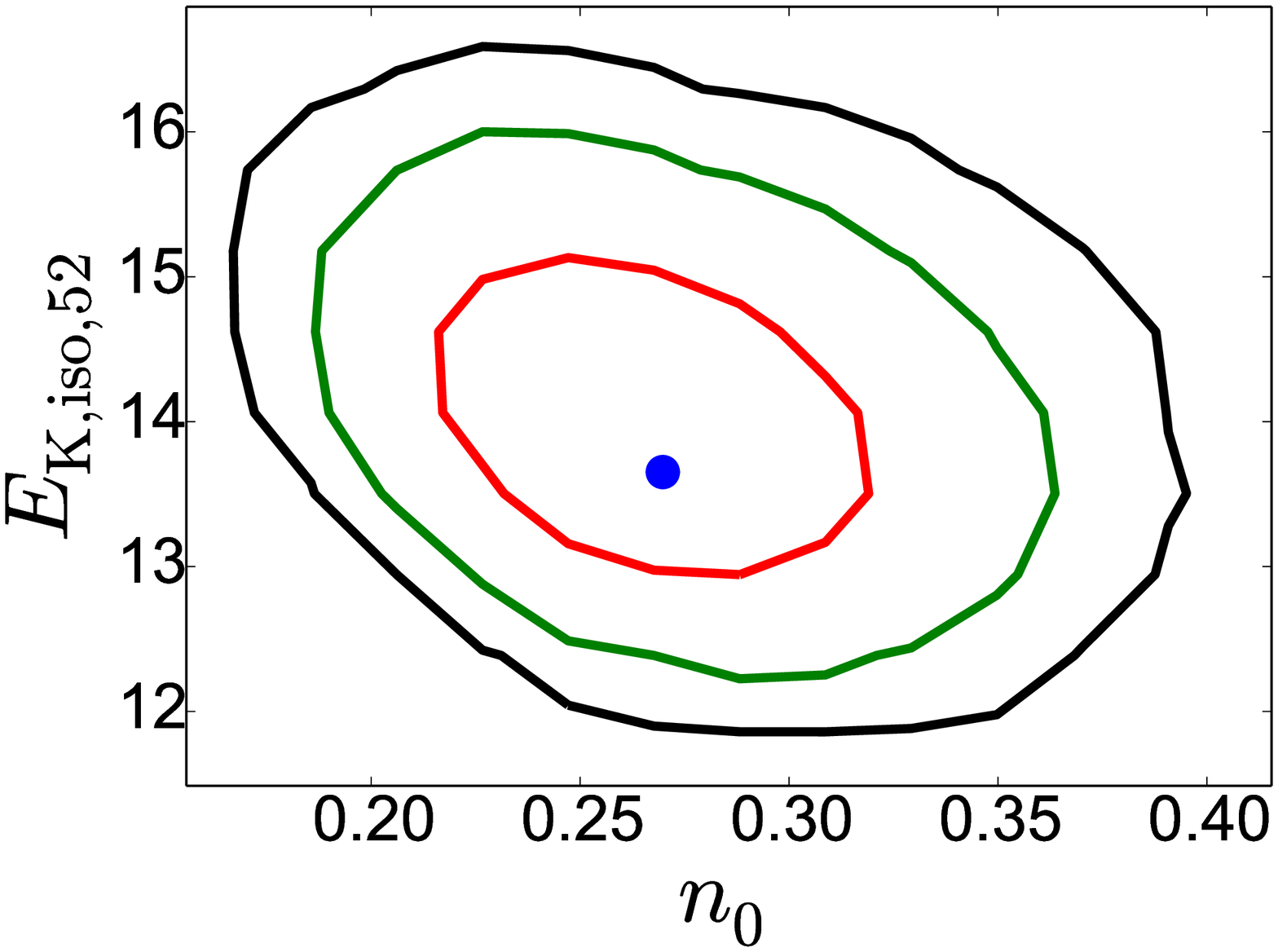} &
 \includegraphics[width=0.30\columnwidth]{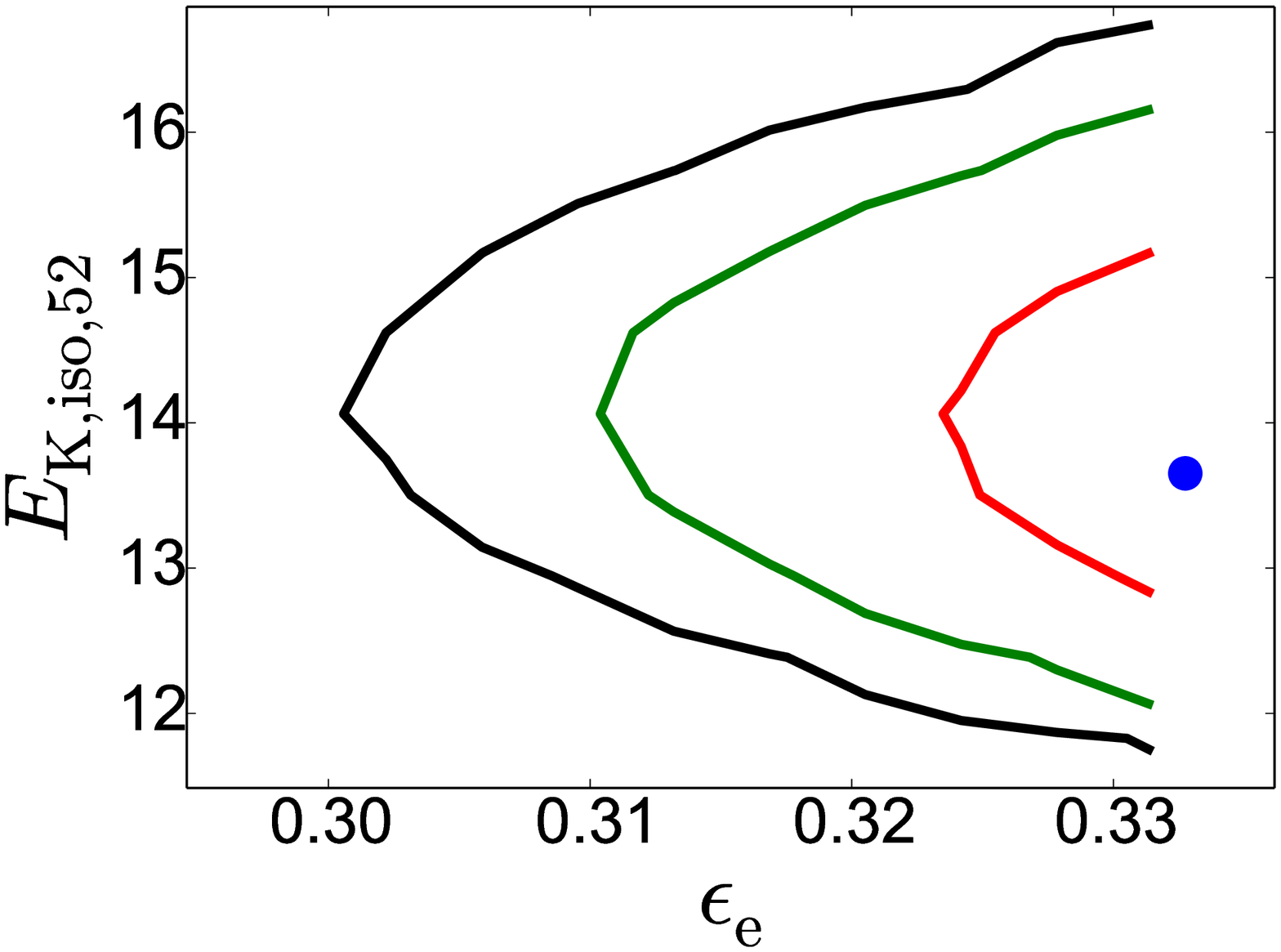} &
 \includegraphics[width=0.30\columnwidth]{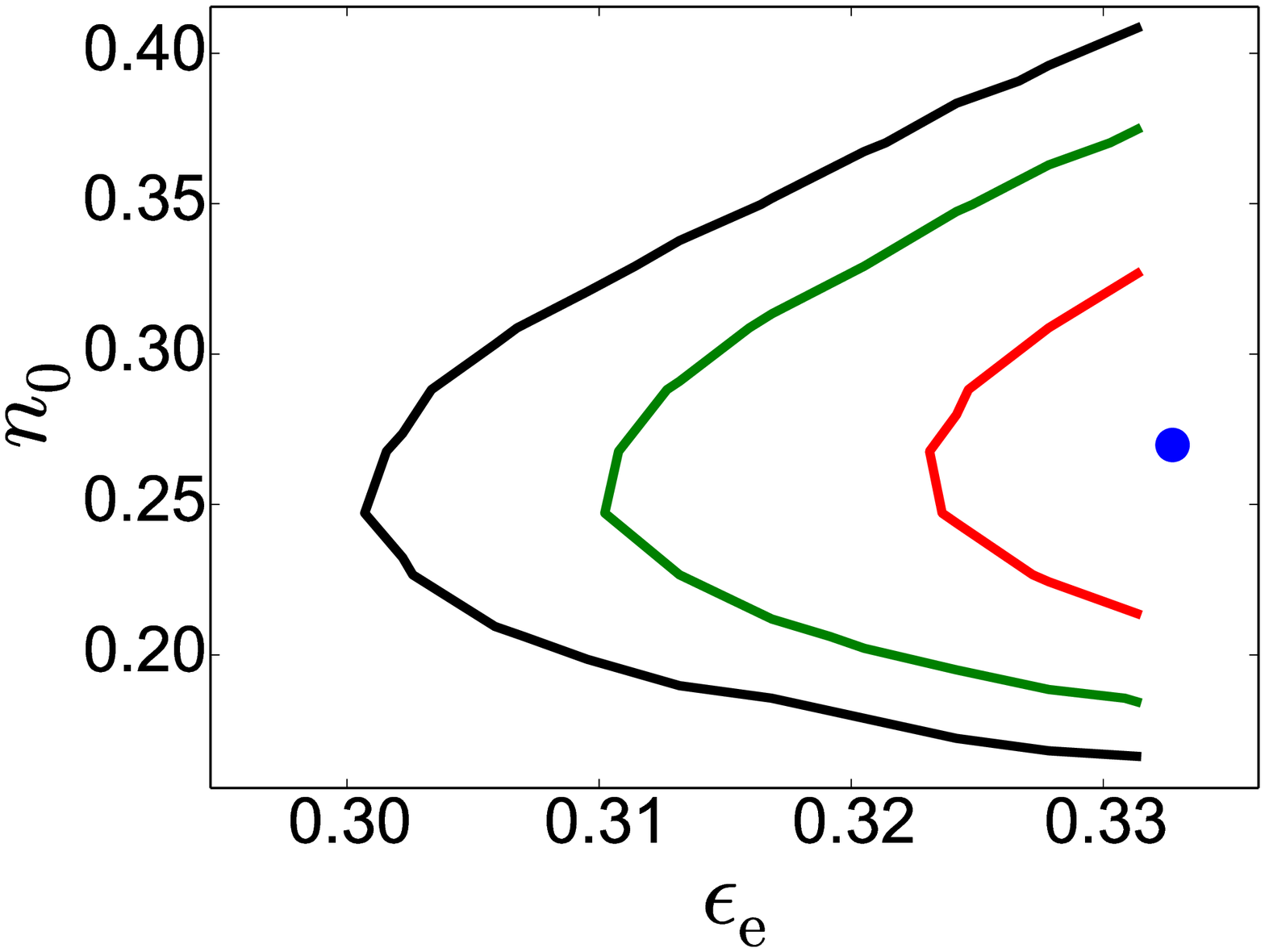} \\
 \includegraphics[width=0.30\columnwidth]{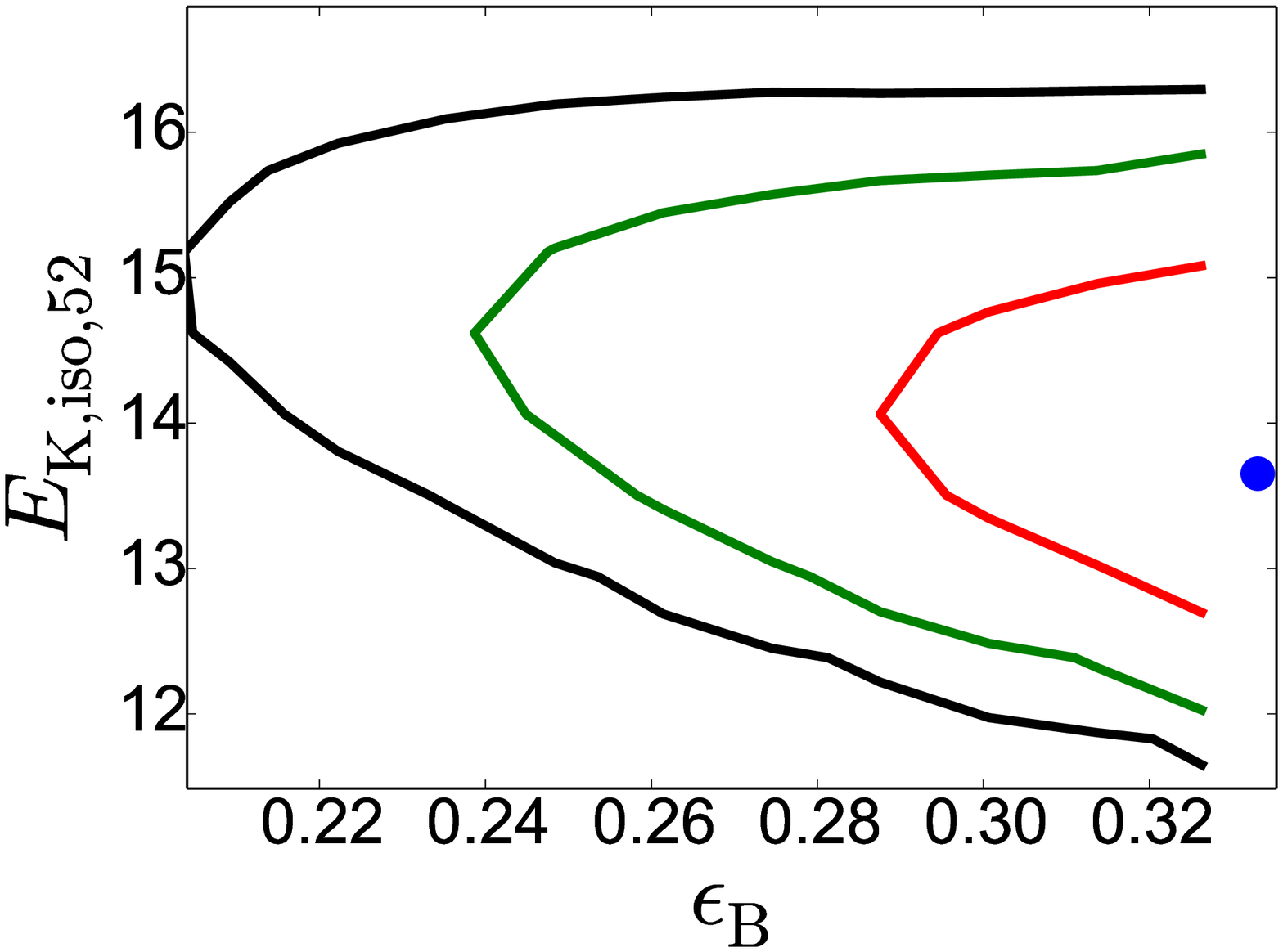} &
 \includegraphics[width=0.30\columnwidth]{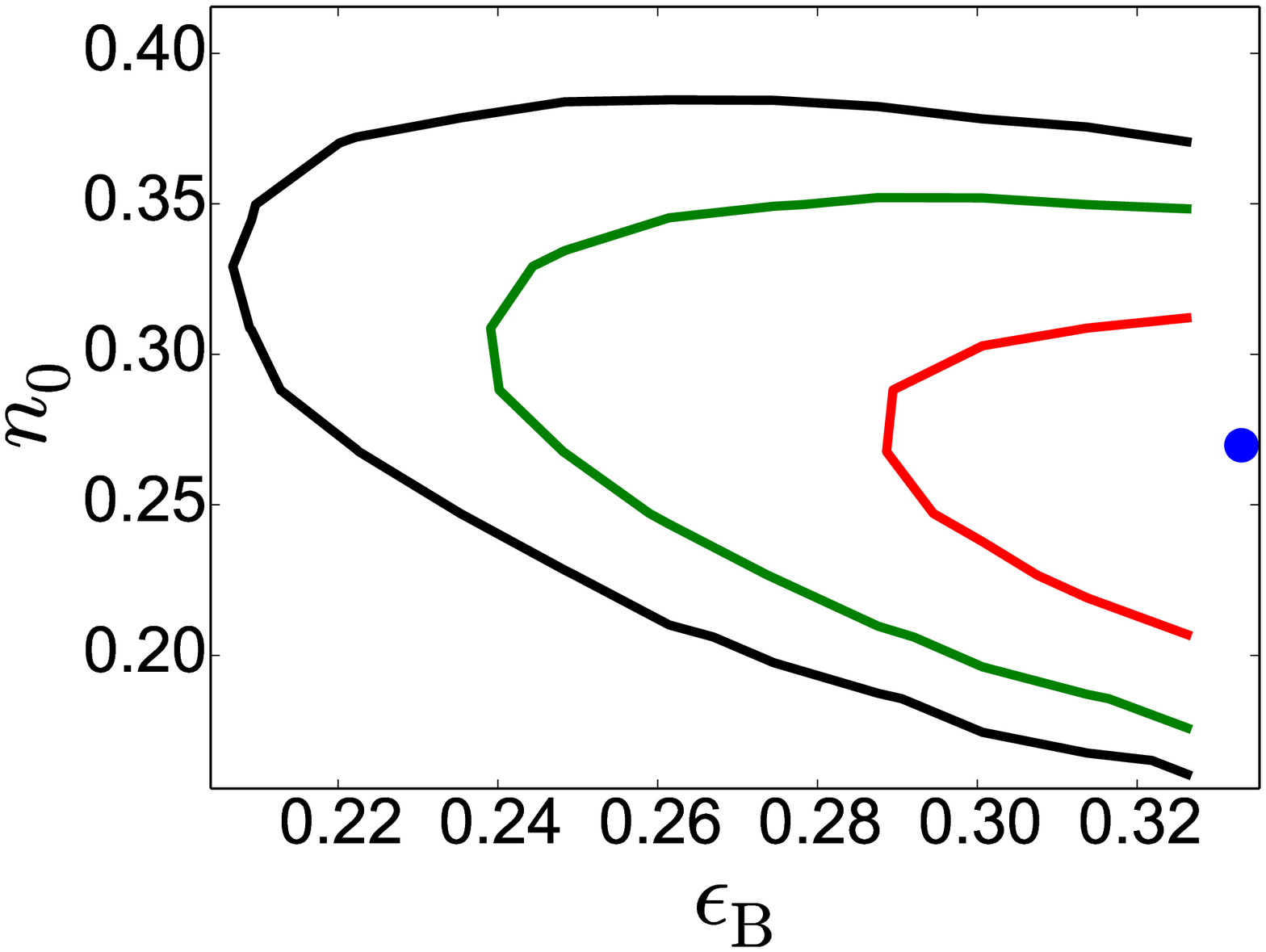} &
 \includegraphics[width=0.30\columnwidth]{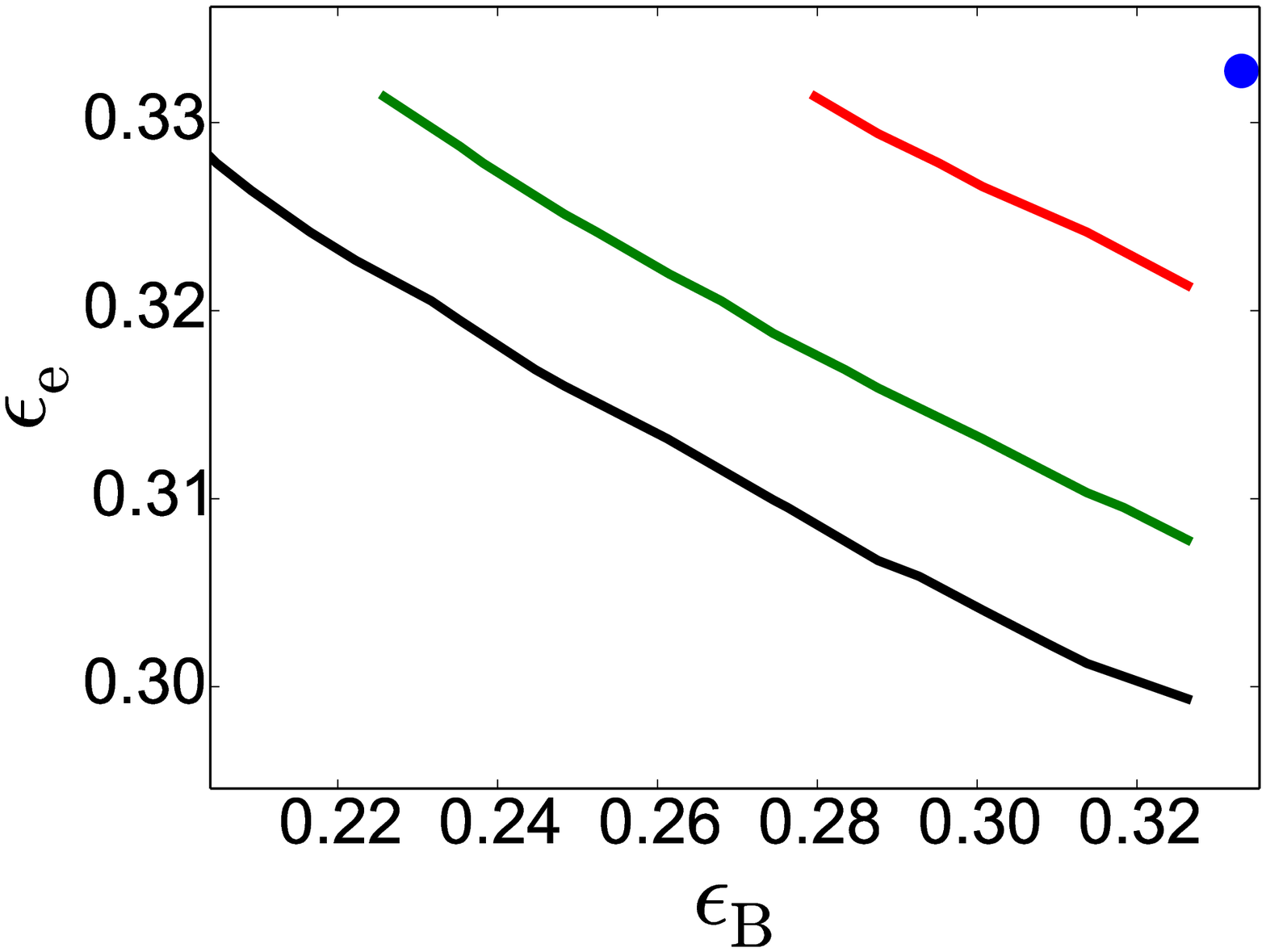} \\
\end{tabular}
\caption{1$\sigma$ (red), 2$\sigma$ (green), and 3$\sigma$ (black) contours for correlations
between the physical parameters, \EKiso, \dens, \epse, and \epsb\ for GRB~120326A, in the ISM 
model from Markov chain Monte Carlo simulations, together with the maximum-likelihood model (blue 
dot). We have restricted $\epsilon_{\rm e} < \nicefrac{1}{3}$ and $\epsilon_{\rm B} < 
\nicefrac{1}{3}$. See the on-line version of this Figure for additional plots of correlations 
between these parameters and $p$, $t_{\rm jet}$, $\thetajet$, $E_{\rm K}$, $A_{\rm V}$, and 
$F_{\nu,{\rm host,R}}$. \label{fig:120326A_ISM_corrplots}} 
\end{figure}

We present histograms of the marginalized posterior density for each parameter in Figure 
\ref{fig:120326A_ISM_hists}. Most of our radio observations are after the transition to slow 
cooling, at which time \nua\ lies below the lowest radio frequency observed, resulting in 
degeneracies between the model parameters (Figure \ref{fig:120326A_ISM_corrplots}). We summarize 
the results of our MCMC analysis in Table \ref{tab:enjsummary}.

Using the relation 
\begin{equation}
\thetajet = 0.17\left(\frac{E_{\rm K, iso,52}}{n_0}\right) ^{\frac{1}{8}} 
               \left(\frac{t_{\rm jet}/(1+z)}{1\,\rm d}\right)^{\frac{3}{8}} 
\end{equation}
for the jet opening angle \citep{sph99}, and the distributions of \EKiso, \dens, and $t_{\rm jet}$ 
from our MCMC simulations, we find $\thetajet = 4.6\pm0.2$ degrees. Applying the beaming 
correction, $E=E_{\rm iso} (1-\cos{\thetajet})$, we find $\Egamma = (1.0\pm0.1)\times10^{50}$\,erg 
(1--$10^4$\,keV; rest frame) and $\EK = \left(4.6^{+0.2}_{-0.1}\right)\times10^{50}$\,erg. 

\subsection{X-ray/UV/optical Re-brightening}
We now consider physical explanations of the unusual re-brightening between 0.1 and 1 day 
observed in the X-ray, UV, and optical bands.

Scattering by dust grains in the host galaxy of GRBs has been suggested as a potential explanation 
for the shallow decay phase of X-ray light curves \citep{sd07}. However, this mechanism is 
expected to cause a significant softening of the X-ray spectrum with time \citep{sbdk08}, and 
cannot produce light curves that rise with time, as is observed in the case of \me\ at about 
0.4\,d. 
The bump can also not be caused by the passage of a spectral break frequency, since it occurs 
almost simultaneously in the X-ray, UV, and optical bands. Four remaining potential models for the 
bump are: (1) onset of the afterglow, (2) geometric effects due to an observer located outside the 
jet (off-axis scenario), (3) a density enhancement in the circumburst environment, and (4) a 
refreshed shock (energy injection scenario). We now explore these possibilities in turn.

\subsubsection{Afterglow onset}
In this scenario, the peak of the light curve emission corresponds to emission from a reverse shock 
due to deceleration of the ejecta by the surrounding material. The ejecta are assumed to be 
composed of a conical shell segment at a single Lorentz factor. The reverse shock light curves 
before and after the deceleration time depend on the properties of the environment (ISM or wind) 
and the ejecta. Reverse shock light curves for a constant density environment depend upon whether 
the ejecta shell is thick ($\Delta_0 > l/2\Gamma_0^{\nicefrac{8}{3}}$) or thin ($\Delta_0 > 
l/2\Gamma_0^{\nicefrac{8}{3}}$), where $\Delta_0$ is the initial shell width, $\Gamma_0$ is the 
initial Lorentz factor of the ejecta and $l = (3E/4\pi n_0 m_{\rm p}c^2)^{1/3}$ is the Sedov length 
\citep{kob00}. Similarly, when the circumburst medium has a wind-like profile, the light curves 
again depend on whether $\Delta_0 > E/4\pi Ac^2\Gamma_0^4$ (thick shell) or $\Delta_0 < E/4\pi 
Ac^2\Gamma_0^4$ (thin shell; \citealt{kz03a}). 

In the thick shell case, the reverse shock crosses the ejecta in a time comparable 
to the burst duration, $t_{\gamma}$, and the light curves at all frequencies are expected to peak 
at that time scale. In the case of \me, $T_{90}$ in the \Swift/BAT band is $\approx 27$\,s, which 
is much shorter than the observed time at which the light curves peak, $t_{\rm peak} \approx 
0.4$\,d. Thus, the re-brightening cannot be explained by reverse shock emission in the thick shell 
case. 

In the thin shell case, temporal separation is expected between the GRB itself and the peak 
of the reverse shock, which occurs when the RS crosses the shell at 
$t_{\gamma}=90 E_{\rm K,iso,52}^{1/3} \dens^{-1/3} \Gamma_{0,100}^{-8/3}$\,s (ISM environment; 
\citealt{kob00}) or $t=2.9\times10^3(1+z)E_{\rm K,iso,52}\Gamma_{0,100}^{-4}A_*^{-1}$\,s (wind 
environment; \citealt{zwd05}). A reverse shock peak at $0.4$\,d would imply a rather low burst 
Lorentz factor of $\Gamma_0\approx 17$. Additionally, the light curves at all frequencies would be 
expected to rise until $t_{\gamma}$ and decline thereafter (except above $\nuc$ where the flux is 
constant until $t_{\gamma}$ and disappears after $t_{\gamma}$). The observed $R$-band light curve, 
on the other hand, declines between $100$\,s and $500$\,s, followed by a flat segment until 
$3500$\,s, followed by the rise into the bump at $0.4$\,d. This combination of a flat portion 
followed by a rise cannot be explained as being due to a reverse shock.

Thus the onset of the afterglow does not provide a viable explanation for the re-brightening 
regardless of density profile, and whether the ejecta are in the thick- or thin-shell regime. 

\subsubsection{Off-axis model}
\label{text:offaxis}
We now investigate the re-brightening in the context of viewing geometry effects. We consider a 
jet with an opening half-angle, $\thetajet$, viewed from an angle $\theta_{\rm obs}$. While small 
offsets in the observer's viewing angle relative to the jet axis do not cause significant changes 
in the light curves provided that $\theta_{\rm obs} \lesssim \thetajet$, it is possible to obtain a 
rising light curve when $\theta_{\rm obs} \gtrsim 2\thetajet$ \citep{gmp+01,gpkw02}. In this case, 
the time of the peak of the light curve, $t_{\rm p}$ is related to the jet break time for an 
on-axis 
viewer as $t_{\rm p} \approx (5+2\ln{\Theta})\Theta^2 t_{\rm jet} \gtrsim 5 t_{\rm jet}$, where 
$\Theta \equiv \theta_{\rm obs}/\thetajet - 1 \gtrsim 1$ \citep{gpkw02}. This implies that for a 
light curve that peaks at $\approx 0.4$\,d, the on-axis jet break time must have occurred earlier, 
at $\approx2$ hours.


In this case, the radio light curves are expected to rise until the observer's line of sight 
enters the beaming cone of the jet and then approximately converge with the predicted on-axis light 
curves after 0.4\,d. After the jet break, the flux density declines as $t^{-\nicefrac{1}{3}}$ for 
$\nu<\numax$ and as $t^{-p}$ for $\nu>\numax$ (for an on-axis observer; \citealt{sph99}). Thus, the 
radio light curves should be declining at all frequencies after 0.4\,d. However, the flux density 
at the 15.75\,GHz AMI-LA band rises from $(0.34\pm0.14)$\,mJy at 0.31\,d to $(0.77\pm0.08)$\,mJy at 
7.15\,d, which is inconsistent with this expectation.
This argues against the off-axis model as an explanation for the X-ray/UV/optical 
re-brightening.


\subsubsection{Density enhancement}
If the blastwave encounters an enhancement in the local density as it propagates into the 
circumburst environment, an increase in the flux density is expected. Bumps in afterglow light 
curves have been ascribed to this phenomenon in the past (e.g., \citealt{bsf+00,lrc+02,sgh+03}). 
However, the flux density at frequencies above $\nuc$ (for slow-cooling spectra) and above $\numax$ 
(for fast cooling spectra) are insensitive to variations in the circumburst density \citep{ng07}. 
In the case of \me, a re-brightening is observed both in the UV/optical and in the X-rays. The 
X-rays are located above both $\nuc$ and $\numax$ in the $p\approx2.1$ (fast cooling) model. Thus 
in our favored afterglow model, the X-ray re-brightening cannot be due to a density enhancement.
 
\subsubsection{Energy injection model}
\label{text:energy_injection}
An alternative model for a re-brightening of the afterglow is the injection of energy into the 
blastwave shock due to prolonged central engine activity, deceleration of a 
Poynting flux dominated outflow, or the presence of substantial ejecta mass (and hence kinetic 
energy) at low Lorentz factors \citep{dl98a,rm98,kp00a,sm00,zm01,zm02,gk06,zfd+06,dsg+11,uzh+12}. 
This mechanism has been invoked to explain plateaus in the observed light curves of a large 
fraction of GRB X-ray afterglows \citep{nkg+06,lzz07,yd07,mgg+10,hdm12,xpk+12}. In this section, we 
explore the energy-injection model as a possible explanation for the X-ray/UV/optical 
re-brightening at 0.4\,d.

Models involving a transfer of energy from braking radiation of millisecond magnetars into the 
forward shock predict plateaus in X-ray light curves but do not generally lead to a 
re-brightening. In particular, they require an injection luminosity, $L\propto t^{q}$ (corresponding
to an increase in the blastwave energy with time as $E\propto t^{1+q}$), with $q\le0$. In our 
$p\approx2.1$ model, the X-rays are located above the cooling frequency. In this regime, the flux 
density above $\nuc$ is $F_{\nu>\nuc}\propto\EKiso^{(2+p)/4}t^{(2-3p)/4}$ \citep{gs02}. For 
$p\approx2.1$, this reduces to $F_{\nu>\nuc}\propto\EKiso^{1.03} t^{-1.1}$. During energy 
injection, $E\propto t^{m}$, such that $F_{\nu>\nuc}\propto t^{1.03 m-1.1}$. The steep rise 
($\alpha_1 = 0.85\pm0.19$; Section \ref{text:basic_considerations:re-brightening}) requires 
$m=1.89\pm0.18$ or $q\equiv m-1=0.89\pm0.18$. Energy injection due to spin-down or gravitational 
wave radiation from a magnetar is expected to provide at best a constant luminosity ($q\le0$). Thus 
the observed re-brightening cannot be explained by energy injection from a magnetar. In the case of 
energy injection due to fall-back accretion onto a black hole, the expected accretion rate is 
$\dot{M}\propto t^{-5/3}$ \citep{phi89}, which is also insufficient to power the observed 
plateaus. Similarly, central engine activity is usually associated with flaring behavior in X-ray 
\citep{brf+05,fw05,fbl+06,pz06,paz06,zfd+06,cmr+07,cmm+10,mgc+10,bmc+11,mbbd+11} and optical 
\citep{llt+12} light curves, but it involves shorter characteristic time scales ($\Delta t / t \ll 
1$) than observed for \me\ at 0.4\,d. 

In the standard afterglow model, the ejecta are assumed to have a single Lorentz factor. We now 
relax this assumption and consider models including a distribution of Lorentz factors in the ejecta 
as a possible explanation for the late re-brightening in \me. Provided they are released over a 
time 
range small compared to the afterglow timescale, the ejecta arrange themselves in homologous 
expansion with the Lorentz factors monotonically increasing with distance from the source 
\citep{kp00a}. We follow the formalism of \cite{sm00}, where the ejecta are assumed to possess a 
continuous distribution of Lorentz factors such that a mass, $M(>\Gamma)\propto\Gamma^{-s}$ is 
moving with Lorentz factors greater than $\Gamma$ with corresponding energy, $E(> \Gamma)\propto 
\Gamma^{-s+1}$ down to some minimum Lorentz factor, $\Gamma_{\rm min}$ (also known as the 
`massive-ejecta model'). Additionally, we posit a maximum Lorentz factor for this distribution, 
$\Gamma_{\rm max}$, corresponding to the Lorentz factor of the blastwave $\Gamma_{\rm BW}$ at 
the onset of energy injection. In this model, there is a gap between the blastwave, which is 
powered 
by an initial shell of fast-moving ejecta, and the leading edge of the remaining ejecta traveling 
at $\Gamma_{\rm max}$. When these slower ejecta shells catch up with the blastwave at a time when 
$\Gamma_{\rm BW}(t)\approx\Gamma_{max}$, they begin depositing energy into the blastwave, and 
energy injection commences. This proceeds until the lowest energy ejecta located at $\Gamma_{\rm 
min}$ have transferred their energy to the blastwave, and the subsequent evolution proceeds like a 
standard afterglow powered by a blastwave with increased energy. The Lorentz factor of the 
blastwave at the end of energy injection is therefore $\Gamma_{\rm min}$, which can be determined 
from modeling the subsequent afterglow evolution and invoking the standard hydrodynamical 
framework. 

For an ambient medium with a density profile, $n\propto r^{-k}$, the energy of the blastwave 
increases as $E\propto t^m$ during the period of energy injection, where\footnote{Other authors 
have defined this in terms of luminosity, with $L\propto t^{-q}$, or $E\propto t^{1-q}$. Thus 
$m\equiv1-q$.} $m = \frac{(3-k)(s-1)}{7+s-2k}$. For a constant density medium ($k=0$) this 
translates to $m = \frac{3(s-1)}{7+s}$, which is bounded 
$\left(0 < m < 3 \textrm{ for } s\in\mathopen[1,\infty\mathclose)\right)$ while for a wind-like 
environment ($k = 2$) we have $m = \frac{s-1}{s-3}$, which is also bounded $\left(0 < m < 1\textrm{ 
for }s\in\mathopen[1,\infty\mathclose)\right)$. 
  
For \me, the optical and X-ray light curves are located above both $\numax$ and $\nuc$ and the 
spectrum is fast cooling ($\nuc < \numax$; Section \ref{text:results:ISM_lowp}). In this regime, 
the flux density depends on the blastwave energy and time as, $F_{\nu}\propto 
E^{(2+p)/4}t^{(2-3p)/4}$. Writing $E = E_0 (t/t_0)^{m}$ for the energy injection episode, this 
implies $F_{\nu}\propto t^{[(2+p)m+2-3p]/4}\propto t^{m-1}$, for $p=2$. Since the optical and X-ray 
light curves rise with temporal index, $\alpha\approx0.6$ (Section 
\ref{text:basic_considerations:re-brightening}) and $p\approx2$ (Section 
\ref{text:results:ISM_lowp}), this implies $m\approx1.6$, which corresponds to 
$s=\frac{7m+3}{3-m}\approx10$.

For a detailed analysis, we use the afterglow parameters for the maximum likelihood model listed 
in Section \ref{text:results:ISM_lowp} as our starting model (which explains all available 
multi-wavelength data after the re-brightening). We then reduce the energy at earlier times as a 
power law with time, 
\begin{equation}
\EKiso(t) = 
  \begin{cases}
      E_{\rm K,iso,f}, & t \ge t_{\rm 0}\\
      E_{\rm K,iso,f}\left(\frac{t}{t_{\rm 0}}\right)^{m},  & t_1 < t < t_{\rm 0} \\      
      E_{\rm K,iso,i}\equiv E_{\rm K,iso,f}\left(\frac{t_1}{t_{\rm 0}}\right)^{m} & t \le t_1,\\
  \end{cases}
\end{equation}
where $E_{\rm K,iso,i}$ and $E_{\rm K,iso,f}$ are the initial and final energy of the blastwave, 
respectively, with the latter fixed to the value determined by our MCMC analysis of the data at 
$\ge 0.4$\,d. Energy injection 
into the blastwave at the rate $E(t) \propto t^m$ begins at $t_1$ and ends at $t_0$, with $t_0$, 
$t_1$, and $m$ being free parameters in this model. We compute the spectrum at each time according 
to the instantaneous energy in the blastwave, adjusting the parameters $t_0$ and $t_1$ by hand 
until the resulting light curves match the observations.

We find that the X-ray and optical light curves can be modeled well by a single period of energy 
injection, with $\EKiso=1.4\times10^{53}\,{\rm erg}(t/0.38\,{\rm d})^{1.44}$ between 
$t_1 = 4.3\times10^{-2}$\,d and $t_0 = 0.38$\,d (Figure \ref{fig:120326A_enj}). In this model, 
the blastwave energy at the start of energy injection is $E_{\rm 
K,iso,i}\approx5.9\times10^{51}$\,erg. Thus $\approx96\%$ of the final energy of the 
blastwave is injected during this episode. In comparison, $\Egammaiso = 
(3.15\pm0.12)\times10^{52}$\,erg (Section \ref{text:basic_considerations}).
The blastwave Lorentz factor can be computed from the 
expression $\Gamma= 3.65{E_{\rm K,iso,52}}^{1/8}\dens^{-1/8}t^{-3/8}(1+z)^{3/8}$ \citep{gs02} and 
is $\Gamma \approx 19$ at the start of energy injection, decreasing to 
$\Gamma\approx13$ at $0.38$\,d. Interpreted in the context of the massive-ejecta model, the energy 
injection rate of $m=1.44$ would imply $s=8.4$, corresponding to an ejecta distribution with 
$E(>\Gamma)\propto \Gamma^{1-s}\propto \Gamma^{-7.4}$ for $13 \lesssim \Gamma \lesssim 19$.

\begin{deluxetable*}{lcccc}[t]
\tabletypesize{\footnotesize}
\tablecolumns{5}
\tablewidth{0pt}
\tablecaption{Energy injection in GRB afterglows\label{tab:enjsummary}}
\tablehead{
\colhead{GRB} & \colhead{100418A} & \colhead{100901A} &
                \colhead{120326A} & \colhead{120404A}
  }
\startdata
Redshift, $z$        & 0.6235     & 1.408      
                     & 1.798      & 2.876   \\ [3 pt]
$T_{\rm 90}$ (s)     & $8.0\pm2.0$ & $439\pm33$  & $11.8\pm1.8$ & $38.7\pm4.1$     \\ [3 pt]
$\Egammaiso$\,(erg; 1--$10^4$\,keV, rest frame)  & $9.9^{+6.3}_{-3.4}\times10^{50}$ & 
$(8\pm1)\times10^{52}$
                     & $(3.2\pm0.1)\times10^{52}$   & $(9\pm4)\times10^{52}$\\ [3 pt]
\cmidrule{2-5} \\ [3 pt]
& \multicolumn{4}{c}{\textbf{Best fit model$^{*}$}} \\ [3 pt]
\cmidrule{2-5} \\ [3 pt]
section in text      & \ref{text:100418A:FS} & \ref{text:100901A:FS}
                     & \ref{text:results:ISM_lowp}  & \ref{text:120404A:FS} \\ [3 pt]
circumburst environment   & ISM & ISM   & ISM                         & ISM \\ [3 pt]
$p$                  & $2.14$        & $2.03$
                     & $2.09$        & $2.06$ \\ [3 pt]

\epse                & $0.12$     & $0.33$
                     & $0.33$     & $0.27$\\ [3 pt]
		       
\epsb                & $1.1\times10^{-2}$     & $0.32$
                     & $0.33$     & $0.16$\\ [3 pt]
		       
\dens\,(\pcc)        & $1.4$     & $3.2\times10^{-3}$ 
                     & $0.27$     & $2.8\times10^{2}$\\ [3 pt] 
		       
$\EKiso$\,($10^{52}$\,erg) 
                     & $3.36$     & $29.7$
                     & $13.7$     & $12.3$\\ [3 pt]
		       
\AV\ (mag)           & $\lesssim0.1$     & $0.1$
                     & $0.40$     & $0.13$\\ [3 pt]
\tjet\ (days)        & $16.9$     & $0.96$
                     & $1.54$     & $6.6\times10^{-2}$\\ [3 pt]
\thetajet\ (deg)     & $20.4$     & $2.1$
                     & $4.7$     & $3.1$ \\ [3 pt]

\EK ($10^{50}$\,erg) & $21.1$     & $2.1$
                     & $4.5$     & $1.7$\\ [3 pt]
		       
\Egamma ($10^{50}$\,erg; 1--$10^4$\,keV, rest frame)  & $0.62$          & $0.5$
                          & $1.0$          & $1.3$ \\ [3 pt]
Peak time of re-brightening (d) 
                     & $0.75\pm0.13$     	     & $0.36\pm0.02$
                     & $0.40\pm0.01$                 & $(2.8\pm0.4)\times10^{-2}$\\ [3 pt]
$E_{\rm K,iso,initial}$\,(erg)
                     & $1.0\times10^{50}$	     & $5.4\times10^{52}$
                     & $3.1\times10^{51}$            & $4.4\times10^{51}$\\ [3 pt]
$E_{\rm K,initial}$\,(erg)
                     & $6.3\times10^{48}$	     & $3.8\times10^{49}$
                     & $8.7\times10^{48}$            & $6.4\times10^{48}$\\ [3 pt]

\cmidrule{2-5} \\ [3 pt]
& \multicolumn{4}{c}{\textbf{MCMC Results$^{*}$}} \\ [3 pt]
\cmidrule{2-5} \\ [3 pt]
$p$                  & $2.13^{+0.02}_{-0.01}$	     & $2.05\pm0.01$
                     & $2.095\pm0.007$                 & $2.07\pm0.02$ \\ [3 pt]

\epse                & $1.2^{+0.3}_{-0.2}\times10^{-1}$  & $0.30^{+0.03}_{-0.05}$
                     & $0.329^{+0.003}_{-0.007}$         & $0.27^{+0.04}_{-0.05}$\\ [3 pt]
		       
\epsb                & $9.4^{+3.5}_{-3.1}\times10^{-3}$    & $0.12^{+0.12}_{-0.08}$
                     & $0.31^{+0.01}_{-0.03}$        & $0.13^{+0.12}_{-0.08}$\\ [3 pt]
		       
\dens\,(\pcc)        & $1.6^{+0.4}_{-0.3}$           & $7.2^{+6.1}_{-3.1}\times10^{-3}$ 
                     & $0.27^{0.04}_{-0.03}$         & $3.5^{+4.9}_{-1.8}\times10^{2}$\\ [3 pt] 
		       
$\log{(n_{\rm 0})}^{\dag}$  & $0.22\pm0.09$          & $-2.1^{+0.3}_{-0.2}$
                     & $-0.57\pm0.06$                & $2.5^{+0.4}_{-0.3}$\\ [3 pt]
		       
$\EKiso$\,($10^{52}$\,erg) 
                     & $3.6^{+1.0}_{-0.7}$           & $31.1^{+8.6}_{-4.3}$
                     & $14.0\pm0.07$            & $13.3^{+3.5}_{-2.0}$\\ [3 pt]
		       
\AV\ (mag)           & $\lesssim0.1$ 		     & $0.09\pm0.01$
                     & $0.40\pm0.01$                 & $0.13\pm0.01$\\ [3 pt]
\tjet\ (days)        & $17.3\pm1.0$		     & $0.97\pm0.02$
                     & $1.55\pm0.06$                 & $(6.9\pm0.8)\times10^{-2}$\\ [3 pt]
\thetajet\ (deg)     & $20.9\pm0.5$		     & $2.4\pm0.1$
                     & $4.6\pm0.2$                   & $3.1\pm0.3$ \\ [3 pt]

\EK ($10^{50}$\,erg) & $24^{+6}_{-4}$		     & $2.6^{+1.1}_{-0.4}$
                     & $4.6^{+0.2}_{-0.1}$           & $2.0^{+0.9}_{-0.5}$\\ [3 pt]
		       
\Egamma ($10^{50}$\,erg; 1--$10^4$\,keV, rest frame)  & $0.66\pm0.33$            & $0.70\pm0.10$
                          & $1.0\pm0.10$            & $1.3\pm0.6$ 
\enddata
\tablecomments{$^{*}$ Note that the parameters for the best-fit model may differ 
slightly from the results from the MCMC analysis. This is because the former is the peak of the 
likelihood distribution (and is appropriate for generating model light curves), while the latter is 
summarized here as $68\%$ credible intervals about the median of the marginalized posterior density 
functions for each parameter. The posterior density serves as our best estimate for the value of 
each parameter, and may be asymmetric about the value corresponding to the highest-likelihood 
model.\\$^{\dag}$ In instances where the measured value of \dens\ spans more than about a factor 
of 2, $\log{n_0}$ is a more meaningful quantity. We therefore report both \dens\ and $\log{n_0}$ 
for all cases.}
\end{deluxetable*}

\section{Panchromatic Re-brightening Episodes in Other GRBs}
Whereas individual flattening or re-brightening episodes have been seen both in optical and X-ray 
observations of GRB afterglows \citep{lzz07,llt+12}, simultaneous re-brightening of the afterglow 
in multiple bands spanning both the optical and the X-rays as seen in GRB~120326A is quite 
rare. The X-ray and $R$-band light curves of GRB~970508 exhibited a bump at around 60\,ks, lasting 
for about 100\,ks \citep{paa+98}. Multiple episodes of flux variations of $\approx 30\%$ on time 
scales of $\Delta t/t\approx 1$ between 1 and 10\,d since the burst were detected superposed on a 
power law decline in multi-band X-ray and \textit{BVRI}-band data for GRB~030329 \citep{log+04}, 
while \cite{dupfj+07} detected an X-ray and optical re-brightening in GRB~050408 around 260\,ks. 
The short burst GRB~050724 had a large re-brightening at around 42\,ks, simultaneous with an 
optical flare \citep{bpc+05,mcd+07,mcg+11}. GRB~060206 exhibited a simultaneous X-ray and optical 
re-brightening at 3.8\,ks \citep{mkg+06,lwl08}. GRB~070311 exhibited a re-brightening at $\approx 
100\,$ks in the X-rays and in $R$-band, but the peak in the X-rays is not well-sampled and does not 
appear to have occurred at the same time in the two bands \citep{gvs+07}. \cite{cdk+08} report a 
simultaneous X-ray and $RJHK$-band re-brightening for GRB~071010A at $\approx52$\,ks. GRB~060614 
exhibited a prominent brightening in the \textit{UBVR}-bands simultaneous with an X-ray plateau 
\citep{mhm+07}. Similarly, the $z\approx6$ GRB~120521C exhibited an X-ray plateau at the same time 
as a bump in the $z$-band light curve \citep{lbt+14}. GRB~081028 was one of the first events with a 
well-sampled rising X-ray light curve following the steep decay phase with a concomitant optical 
re-brightening \citep{mgg+10}. An X-ray re-brightening simultaneous with a re-brightening in 
\Swift/UVOT observations was seen for GRB~100418A and interpreted as energy injection by 
\cite{mab+11}. \cite{gll+12} report a prominent multi-band re-brightening in GRB~100901A, whose 
X-ray light curve is remarkably similar to that of GRB~120326A. GRB~120404 was found to exhibit a 
strong re-brightening in the optical and NIR, although X-ray observations around the time of the 
re-brightening are sparse as this time fell within an orbital gap of \Swift\ \citep{gmh+13}.

Of these instances of X-ray or optical re-brightening events seen in GRB afterglow light 
curves, we select those objects with multi-band (at least X-ray and optical) datasets that exhibit
a simultaneous re-brightening in both X-rays (following the steep decay phase) and optical/NIR. 
Since radio observations are vital for constraining the physical parameters, we restrict our sample 
to events with radio detections: GRBs~100418A, 100901A, and 120404A. We perform a full 
multi-wavelength afterglow analysis for these events and compare our results with 
those for \me\ in Section \ref{text:discussion}.

\subsection{GRB~100418A}
\subsubsection{GRB properties and basic considerations}
\label{text:100418A:basic_considerations}
GRB~100418A was detected and localized by the \Swift\ BAT on 2010 April 18 at 21:10:08\,UT 
\citep{gcn10612}. The burst duration is $T_{90} = 8.0\pm2.0$\,s, with a fluence of $F_{\gamma} = 
(3.4 \pm 0.5)\times10^{-7}$\,erg\,cm$^{-2}$ \citep[15--150\,keV observer frame;][]{mab+11}. 
The afterglow was detected in the X-ray and optical bands by XRT, UVOT, and various
ground-based observatories \citep{gcn10617,gcn10619,gcn10625}, as well as in the radio by WSRT, 
VLA, the Australia Telescope Compact Array (ATCA), and the Very Long Baseline Array (VLBA) 
\citep{gcn10647,gcn10650,gcn10832}. Spectroscopic observations with the ESO Very Large Telescope 
(VLT) yielded a redshift of $z=0.6235$ \citep{gcn10620}. By fitting the $\gamma$-ray spectrum with 
a Band function, \cite{mab+11} determined the isotropic equivalent $\gamma$-ray energy of this 
burst to be $\Egammaiso=\left(9.9^{+6.3}_{-3.4}\right)\times10^{50}$\,erg (1--$10^4$\,keV; rest 
frame).

The X-ray light curve for this burst gently rises to 0.7\,d, while the optical light curves 
exhibit a slow brightening at the same time. Both the X-ray and optical light curves break into a 
power law decline following the peak. Using XRT and UVOT \textit{White}-band observations, 
\cite{mab+11} find that the X-ray and optical bands are located on the same part of the synchrotron 
spectrum after the peak at 0.7\,d. By fitting the X-ray and optical decline rates and 
X-ray-to-optical SED, they find that both $\numax$ and $\nuc$ are located below the optical, 
requiring $p\approx2.3$. In this regime, the light curves are insensitive to the density profile of 
the circumburst environment. In the absence of radio data, \cite{mab+11} assume an ISM model and 
determine that the non-detection of a jet break out to $23$\,d requires a high beaming corrected 
energy, $\EK\ge3\times10^{52}E_{\rm K, iso, 54}^{3/4}\dens^{1/4}$. 
They investigate the off-axis model, and compute an intrinsic burst duration for an on-axis 
observer of $T_{90}^{\rm on}= \mathcal{D}(\theta=\theta_{\rm obs}-\thetajet) / 
\mathcal{D}(\theta = 0)T_{\rm 90}\approx$10\,ms, where 
$\mathcal{D}\equiv[\gamma_0(1-\beta_0\cos{\theta})]^{-1}$ is the Doppler factor, $\theta_{\rm 
obs}$ is the off-axis viewing angle, $\thetajet$ is the jet opening half-angle, and $T_{90}$ is the 
observed burst duration. Based on this short on-axis $T_{90}^{\rm on}$ and large intrinsic energy 
output, they argue against the off-axis scenario.

\cite{mab+11} also consider the energy injection model. For the magnetar model involving injection 
of energy into the blastwave via Poynting flux, they find $q=-0.23^{+0.13}_{-0.14}$, which is 
marginally consistent with the $q=0$ case. For models with a Lorentz factor distribution, they use 
their measured value of $q$ to calculate $s=6.4^{+1.3}_{-1.0}$ (ISM model), or $s<0$ (wind model, 
which is unphysical). They argue that since the ISM value is too different from the value 
$s\approx2.5$ indicated by X-ray plateaus in \Swift\ observations \citep{nkg+06}, this mode of 
energy injection is unlikely.

\cite{mcm+13} reported multi-wavelength radio observations of GRB~100418A from 5.5 to 9.0\,GHz from 
ATCA, VLA and VLBA. Using upper limits on the expansion rate of the ejecta derived from the VLBA 
observations, they report an upper limit on the average ejecta Lorentz factor of $\Gamma \lesssim 
7$ and an upper limit on the ejecta mass of $M_{\rm ej} \lesssim 0.5\times10^{-3}\,\rm M_{\odot}$. 
They use this limit on $\Gamma$ to suggest that the contribution of low-Lorentz factor ejecta to 
the blastwave energy must be negligible during the period of energy injection, and that a separate 
injection mechanism is required. We include their radio observations in our analysis
(Section \ref{text:100418A:FS}) 
and address the constraints derived from the VLBA data in Section \ref{text:100418A:enj}.


\cite{mab+11} fit multi-band UVOT and XRT spectra at three different epochs ($7\times10^{-2}$, 
$0.8$, and 7.3\,d) and found that a single power law with spectral index $\beta\approx-1.0$ fit all 
three epochs well. We follow the procedures outlined by \cite{ebp+07}, \cite{ebp+09}, and 
\cite{mzb+13} to obtain time-resolved XRT spectra for this burst. The spectra are well fit by an 
absorbed single power law model with $N_{\rm H,Gal}=4.78\times 10^{20}\,\pcmsq$ \citep{kbh+05}, and 
intrinsic hydrogen column $N_{\rm H,int} = (0.7^{+1.5}_{-0.7})\times 10^{21}\,\pcmsq$, with 
$\Gamma=3.35\pm0.09$ and $\Gamma = 3.9\pm0.3$ during the early (85--130\,s; $\chi^2=1.74$ for 123 
degrees of freedom) and late (130\,s--180\,s; $\chi^2 = 0.97$ for 55 degrees of freedom) WT-mode 
steep decay, respectively, followed by $\Gamma_{\rm X} = 2.08\pm0.18$ during the plateau phase 
($2.3\times10^{-3}$\,d--$1.8\,$d; $\chi^2=0.75$ for 79 degrees of freedom), and $\Gamma_{\rm X} = 
1.81\pm0.20$ for the remainder of the observations (1.8--34.5\,d; $\chi^2=0.70$ for 89 degrees of 
freedom). We obtain a flux light curve in the 0.3--10\,keV XRT band following the methods reported 
by \cite{mzb+13}, which we convert to a flux density light curve at 1\,keV using the above 
time-resolved spectra. We also analyze all UVOT photometry for this burst and report our results in 
Table \ref{tab:data:100418A}.

\begin{deluxetable}{cccccc}
\tabletypesize{\footnotesize}
\tablecolumns{9}
\tablewidth{0pt}
\tablecaption{Swift/UVOT Observations of GRB~100418A\label{tab:data:100418A}}
\tablehead{
  \colhead{$t-t_0$} &    
  \colhead{Filter} &
  \colhead{Frequency} &  
  \colhead{Flux density} &
  \colhead{Uncertainty} &
  \colhead{Detection?} \\  
  \colhead{(days)} &  
  \colhead{} &
  \colhead{(Hz)} &
  \colhead{($\mu$Jy)} &
  \colhead{($\mu$Jy)} &
  \colhead{($1=$ Yes)}  
  }
\startdata
0.00188  & \textit{White} & 8.64e+14 & 11.5 & 2.47 & 1\\
0.00499  & \textit{White} & 8.64e+14 & 10.6 & 2.81 & 1\\
0.00591  & \textit{u} & 8.56e+14 & 8.41 & 2.74 & 0\\
0.00753  & \textit{b} & 6.92e+14 & 74.7 & 22.3 & 0\\
0.011  & \textit{White} & 8.64e+14 & 10.8 & 3.25 & 1\\
0.0119  & \textit{v} & 5.55e+14 & 123 & 35.3 & 0\\
0.0125  & \textit{uvw1} & 1.16e+15 & 19.9 & 5.85 & 0\\
0.0126  & \textit{uvw2} & 1.48e+15 & 12.9 & 3.62 & 0\\
0.0152  & \textit{u} & 8.56e+14 & 50.9 & 14.8 & 0\\
0.0155  & \textit{b} & 6.92e+14 & 101 & 31.5 & 0\\
0.0158  & \textit{White} & 8.64e+14 & 22.8 & 7.49 & 0\\
0.0441  & \textit{uvm2} & 1.34e+15 & 10.1 & 3.15 & 0\\
0.0465  & \textit{uvm2} & 1.34e+15 & 9.69 & 2.95 & 0\\
0.0604  & \textit{v} & 5.55e+14 & 67.3 & 19.7 & 0\\
0.0699  & \textit{b} & 6.92e+14 & 34.5 & 10.2 & 1\\
0.0711  & \textit{uvm2} & 1.34e+15 & 7.29 & 2.32 & 1\\
0.0723  & \textit{White} & 8.64e+14 & 15.8 & 3.22 & 1\\
0.0735  & \textit{uvw1} & 1.16e+15 & 8.37 & 2.71 & 0\\
0.0746  & \textit{uvw2} & 1.48e+15 & 7.68 & 2.23 & 0\\
0.0757  & \textit{u} & 8.56e+14 & 18.3 & 5.7 & 1\\
0.077  & \textit{v} & 5.55e+14 & 67 & 21.9 & 0\\
0.602  & \textit{White} & 8.64e+14 & 33.5 & 4.57 & 1\\
1.03  & \textit{White} & 8.64e+14 & 22.6 & 3.15 & 1\\
1.77  & \textit{White} & 8.64e+14 & 11.7 & 1.68 & 1\\
2.8  & \textit{White} & 8.64e+14 & 8.74 & 1.3 & 1\\
3.45  & \textit{White} & 8.64e+14 & 5.67 & 0.829 & 1\\
4.42  & \textit{White} & 8.64e+14 & 6.22 & 0.951 & 1\\
5.22  & \textit{uvw1} & 1.16e+15 & 6.28 & 1.76 & 0\\
5.22  & \textit{u} & 8.56e+14 & 13 & 3.96 & 0\\
5.22  & \textit{b} & 6.92e+14 & 27.5 & 8.56 & 0\\
5.23  & \textit{uvw2} & 1.48e+15 & 2.87 & 0.819 & 0\\
5.23  & \textit{v} & 5.55e+14 & 60.2 & 18.9 & 0\\
5.23  & \textit{uvm2} & 1.34e+15 & 2.92 & 0.967 & 0\\
5.43  & \textit{White} & 8.64e+14 & 4.05 & 0.735 & 1\\
7.64  & \textit{uvw1} & 1.16e+15 & 3.93 & 1.28 & 0\\
7.64  & \textit{u} & 8.56e+14 & 5.55 & 1.84 & 0\\
7.64  & \textit{b} & 6.92e+14 & 12.1 & 4.02 & 0\\
7.64  & \textit{White} & 8.64e+14 & 3.74 & 1.13 & 1\\
7.64  & \textit{uvw2} & 1.48e+15 & 2.98 & 0.995 & 0\\
7.64  & \textit{v} & 5.55e+14 & 38.2 & 12.5 & 0\\
7.65  & \textit{uvm2} & 1.34e+15 & 4.05 & 1.27 & 0\\
8.64  & \textit{uvw1} & 1.16e+15 & 3.29 & 1.06 & 0\\
8.64  & \textit{u} & 8.56e+14 & 5.05 & 1.64 & 0\\
8.64  & \textit{b} & 6.92e+14 & 11.3 & 3.68 & 0\\
8.65  & \textit{White} & 8.64e+14 & 2.89 & 1.05 & 0\\
8.65  & \textit{uvw2} & 1.48e+15 & 2.61 & 0.79 & 0\\
8.65  & \textit{v} & 5.55e+14 & 31.4 & 10.3 & 0\\
8.65  & \textit{uvm2} & 1.34e+15 & 3.83 & 1.14 & 0\\
9.58  & \textit{White} & 8.64e+14 & 2.89 & 0.557 & 1\\
10.4  & \textit{White} & 8.64e+14 & 2.73 & 0.482 & 1\\
11.4  & \textit{White} & 8.64e+14 & 2.61 & 0.51 & 1\\
12.4  & \textit{White} & 8.64e+14 & 4.04 & 0.635 & 1\\
13.4  & \textit{White} & 8.64e+14 & 2.74 & 0.468 & 1\\
14.4  & \textit{White} & 8.64e+14 & 3.2 & 0.545 & 1\\
15.4  & \textit{White} & 8.64e+14 & 3.11 & 0.525 & 1\\
16.9  & \textit{White} & 8.64e+14 & 2.28 & 0.425 & 1\\
17.6  & \textit{White} & 8.64e+14 & 3.63 & 0.587 & 1\\
19.7  & \textit{White} & 8.64e+14 & 2.55 & 0.464 & 1\\
22.4  & \textit{White} & 8.64e+14 & 2.07 & 0.502 & 1\\
23.5  & \textit{White} & 8.64e+14 & 3.5 & 0.62 & 1\\
24.8  & \textit{White} & 8.64e+14 & 2.79 & 0.587 & 1\\
25.6  & \textit{White} & 8.64e+14 & 3.01 & 0.591 & 1\\
26.7  & \textit{White} & 8.64e+14 & 1.82 & 0.594 & 1\\
34.4  & \textit{White} & 8.64e+14 & 1.97 & 0.401 & 1
\enddata
\end{deluxetable}

\begin{figure}
 \centering
 \includegraphics[width=\columnwidth]{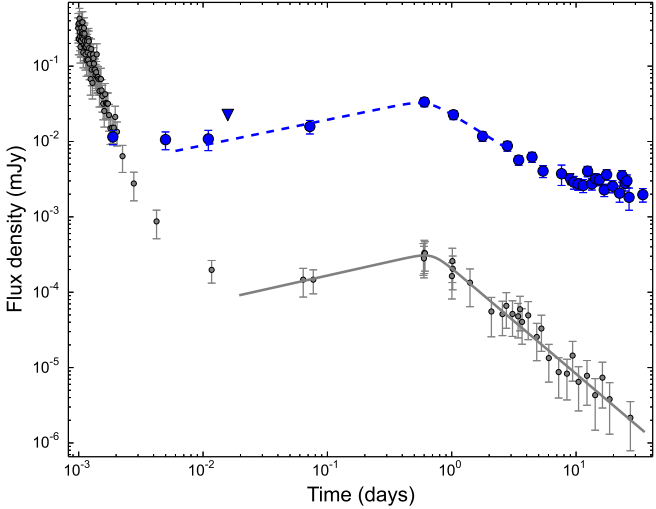}
 \caption{Broken power law fit to the X-ray (grey, solid) and \Swift/UVOT \textit{White}-band light 
curves for GRB~100418A near the re-brightening around 0.6\,d. The X-ray fit includes points between 
0.02 and 50\,d, while the White-band fit includes points between $6\times10^{-3}$ and 3.1\,d. 
A joint fit yields the best fit parameters $t_{\rm b} = 0.58^{+0.37}_{-0.34}$\,d, 
$\alpha_1 = 0.42^{+0.31}_{-0.13}$, and $\alpha_2 = -1.10^{+0.15}_{-0.21}$, with a fixed value of 
$y=5.0$ (Section \ref{text:100418A:basic_considerations}).
\label{fig:100418A_bplfit}}
\end{figure}

The X-ray light curve up to $8\times10^{-3}$\,d declines steeply as $\alpha_{\rm X}=-4.52\pm0.14$, 
transitioning to a plateau, at about 0.01\,d, followed by a re-brightening. The X-ray observations 
between 0.02\,d and 5.5\,d can be well fit with a broken power law model, with $t_{\rm 
b}=0.54\pm0.18$\,d, $F_{\nu, \rm X}(t_{\rm b}) = 0.32\pm0.08\,\mu$Jy, $\alpha_1=0.46\pm0.20$, and 
$\alpha_2 = -1.15\pm0.10$, with the smoothness parameter fixed at $y=5.0$. A broken power law fit 
to the UVOT \textit{White}-band data between $6\times10^{-3}$\,d and 3.1\,d yields $t_{\rm 
b}=0.64\pm0.17$\,d, $F_{\nu, \rm White}(t_{\rm b}) = 32\pm4\,\mu$Jy, $\alpha_1=0.34\pm0.09$, and 
$\alpha_2 = -1.02\pm0.18$, with $y=5.0$ (Figure \ref{fig:100418A_bplfit}). The decline rate in the 
\textit{White} band following the re-brightening is consistent with the decline rate in the 
$R$-band \footnote{The BYU/WMO $R$-band detection at 35.25\,d reported by \cite{gcn10665} is about 
0.5\,mag brighter than observations from KPNO/SARA \citep{gcn10637} and VLT/X-shooter 
\cite{gcn10631} at a similar time, suggesting that the BYU data point may be plagued by a typo. We 
do not include the BYU/WMO $R$-band data point at 35.25\,d in our analysis.} between 0.4\,d and 
10\,d ($\alpha_{\rm R} = -0.97\pm0.08$) and in the $H$-band between the PAIRITEL observations at 
0.46\,d and 1.49\,d ($\alpha_{\rm H} = -1.0\pm0.1$). Thus, the break time and rise and decay rates 
of the re-brightening in the X-ray and optical bands are consistent. Finally, we fit the X-ray and 
UVOT \textit{White}-band data jointly, where we constrain the model light curves in the two bands to 
have the same rise and decay rate and time of peak with independent normalizations. Using a Markov 
Chain Monte Carlo simulation using \emcee\ \citep{fhlg13}, we find $t_{\rm b} = 
0.58^{+0.37}_{-0.34}$\,d, $\alpha_1 = 0.42^{+0.31}_{-0.13}$, and $\alpha_2 = -1.10^{+0.15}_{-0.21}$, 
with a fixed value of $y=5.0$.

\begin{figure}
 \centering
 \includegraphics[width=\columnwidth]{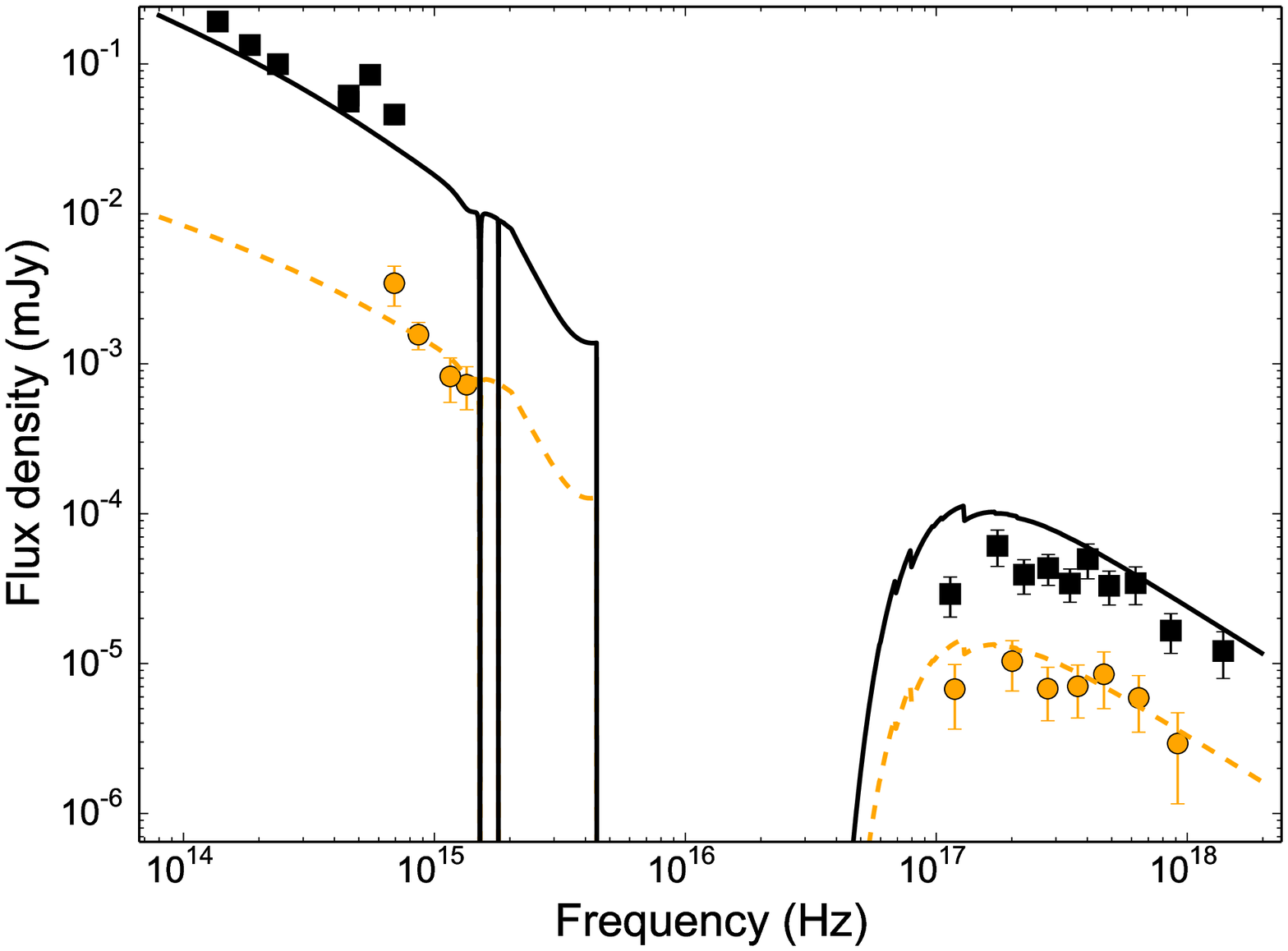}
\caption{Observed NIR to X-ray SED of the afterglow of GRB~100418A at 0.07\,d (orange circles) and 
after the re-brightening (1.5\,d; black squares), together with the best-fit ISM model (Section 
\ref{text:100418A:FS}) including energy injection (Section \ref{text:100418A:enj}) at 0.07\,d 
(orange; dashed) and 1.5\,d (black; solid). The data points and model at 
0.07\,d have been multiplied by a factor of 0.1 for clarity. 
The XRT SED at 1.5\,d has been extrapolated from 0.88\,d using the best-fit broken power law 
model to the XRT light curve (Figure \ref{fig:100418A_bplfit}; the correction factor is 
$\approx0.5$). The optical data have been extrapolated using the best fit to the \Swift/UVOT light 
curve (the corrections are small at $<5\%$). The cooling break (visible in the orange, dashed 
curve at $\approx3\times10^{14}$\,Hz) is already below the optical bands at 0.07; the NIR to 
X-ray frequencies are therefore on the same part of the afterglow SED at 1.5\,d. 
\label{fig:100418A_sed}}
\end{figure}

We extract XRT spectra at 0.06--0.08\,d (mean photon arrival 0.07\,d), and at 0.58--1.81\,d 
(mean photon arrival 0.88\,d). We extrapolate the second SED to 1.5\,d and show the complete 
optical to X-ray SED at 0.07\,d and 1.5\,d in Figure \ref{fig:100418A_sed}. The spectral index 
between the UVOT $B$-band observation at 0.07\,d and the X-rays is $\beta_{\rm opt-X} = 
-0.93\pm0.09$, which is consistent with the X-ray spectral index of $\beta_{\rm X} = -1.08\pm0.18$ 
over this period, suggesting that the optical and X-ray are located on the same part of the 
afterglow SED at 0.07\,d. The spectral index between the NIR and X-rays at 1.5\,d is $\beta_{\rm 
NIR-X} = -0.94\pm0.04$, which is consistent with the spectral index in the NIR and optical alone, 
$\beta_{\rm NIR-opt} = -0.8\pm0.1$, once again suggesting that the NIR/optical and X-rays are 
located on the same part of the afterglow SED at 1.5\,d and that intrinsic extinction is 
negligible. This implies that either $\numax < \nuNIR < \nuX < \nuc$ with $p\approx 3$ or 
$\numax,\nuc < \nuNIR < \nuX$ with $p\approx2$ at both 0.07\,d and 1.5\,d. If $\numax < \nuNIR < 
\nuX < \nuc$ at 1.5\,d, the decay rate would be $\alpha\approx-1.5$ in the ISM model or 
$\alpha\approx-2.0$ in the wind model, whereas in the latter we would expect $\alpha\approx-1$ in 
both ISM and wind scenarios. The observed UV/NIR/X-ray common decay rate after the re-brightening 
of $\alpha\approx-1$ indicates that $\numax,\nuc < \nuNIR < \nuX$ at 1.5\,d and 
$p\approx2$. 

The 8.46\,GHz light curve drops as $t^{-2.5}$ between the ATCA observation at 2\,d and the VLA 
observation at 3\,d, while the SMA 345\,GHz light curve is flat over this period (Figure 
\ref{fig:100418A_enj}). At $\approx$2\,d, the radio spectral index is $\approx0.7$ from 4.9\,GHz to 
8.46\,GHz, and $\approx0.3$ from 8.46\,GHz to 345 GHz. The rapid decline in the light curve at a 
single waveband can not be explained in the standard synchrotron model by forward shock emission. 
We consider an alternative scenario, in which the radio to millimeter emission at 2\,d is 
dominated by a reverse shock. In that case, the peak frequency of the reverse shock must pass 
through 8.46\,GHz before 2\,d with a peak flux density of at least $\approx1.3$\,mJy. 
Propagating the RS spectrum back in time assuming the peak frequency and peak flux evolve as 
$t^{-1.5}$ and $t^{-1}$ respectively \citep{ks00}, we would expect a minimum $R$-band flux density 
of $\approx2$\,Jy at 120\,s. This is more than five orders of magnitude brighter than \Swift\ 
\textit{White}-band observations at this time. Thus the rapid decline and re-brightening of the 
8.46\,GHz light curve between 2 and 20\,d cannot be explained in the standard synchrotron 
framework. It is possible that these data suffer from strong ISS, and we therefore exclude these 
data from our analysis.


\subsubsection{Forward shock model at $t\gtrsim0.4$\,d}
\label{text:100418A:FS}
\begin{figure*}
\begin{tabular}{cc}
 \centering
 \includegraphics[width=0.47\textwidth]{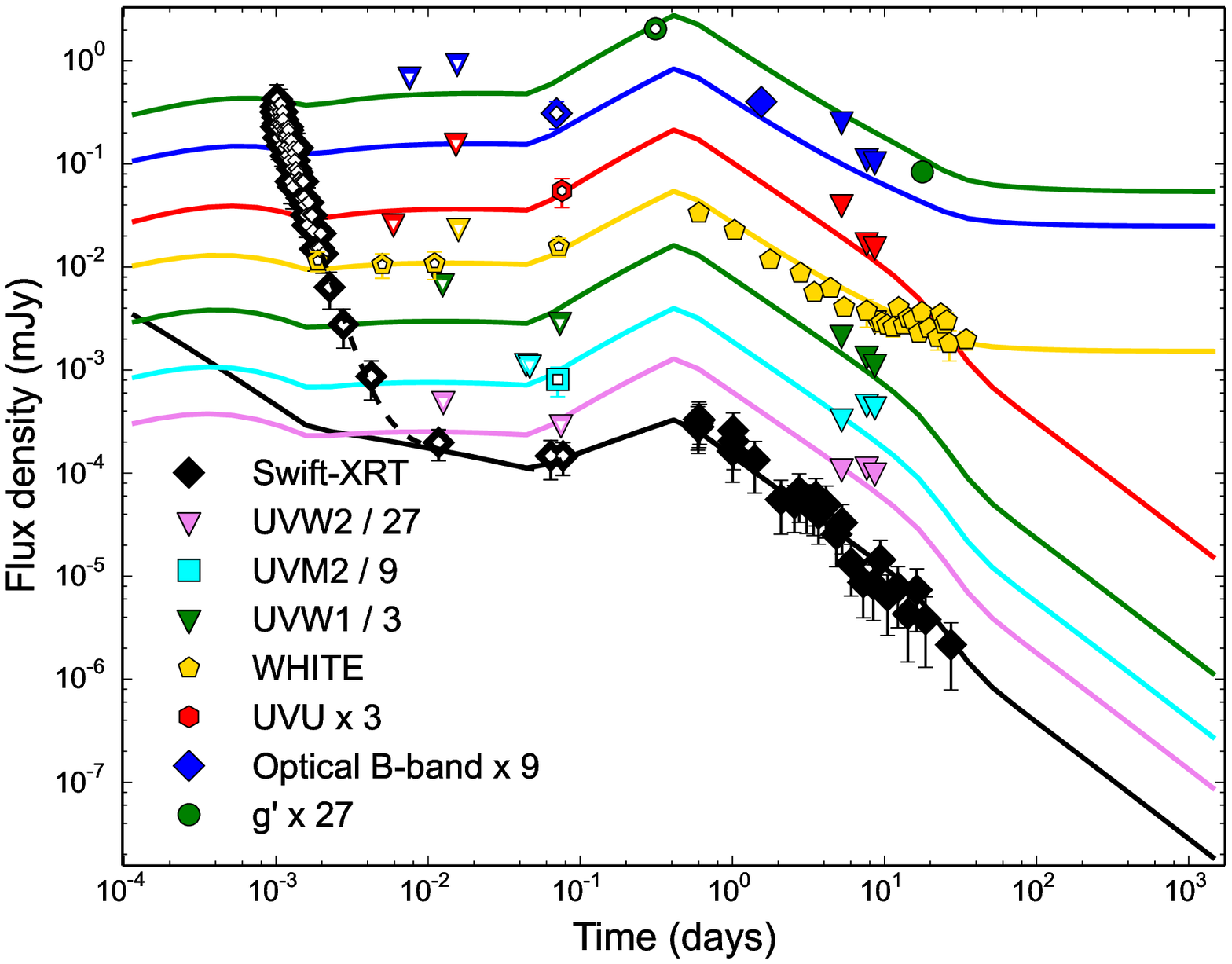} &
 \includegraphics[width=0.47\textwidth]{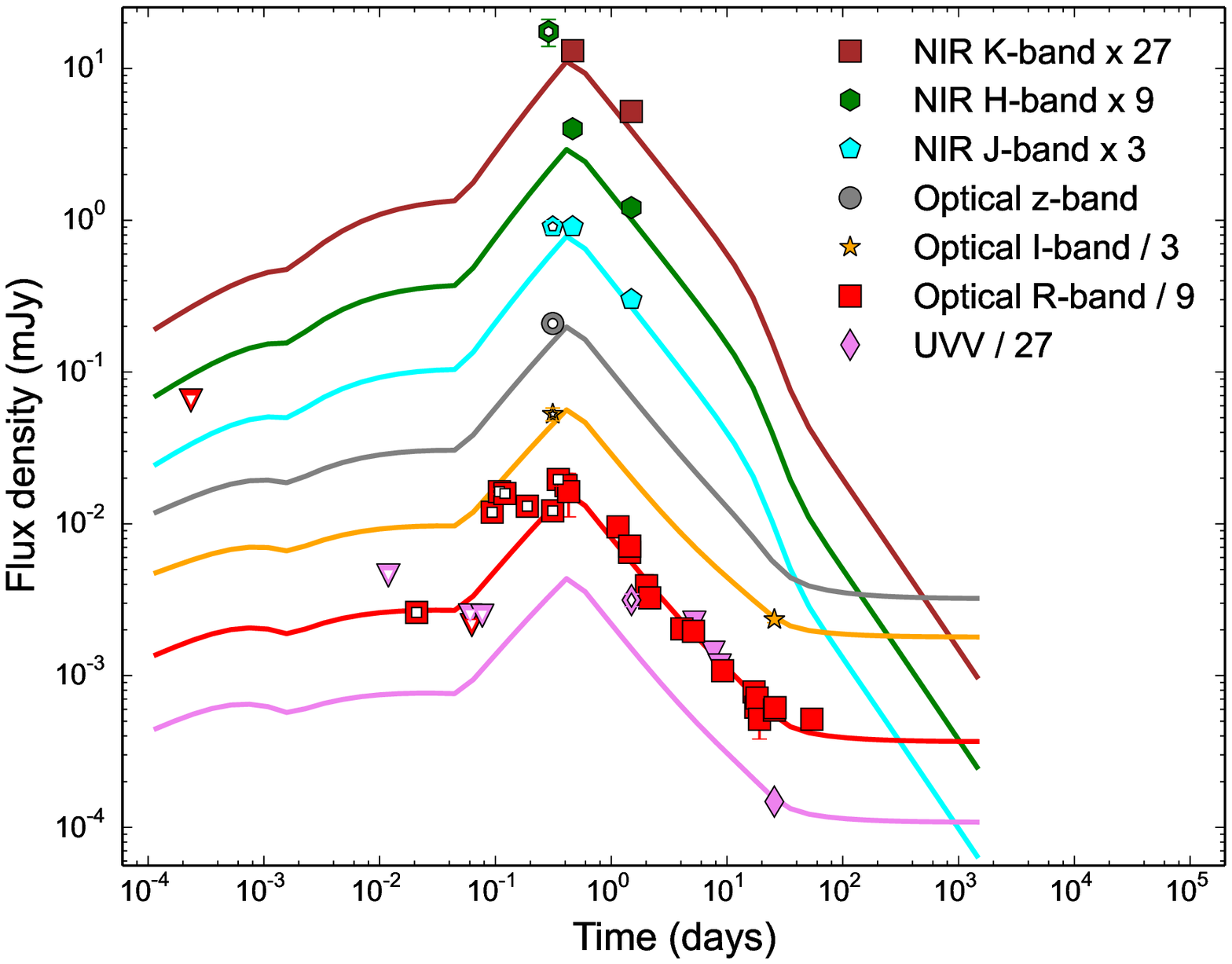} \\
 \includegraphics[width=0.47\textwidth]{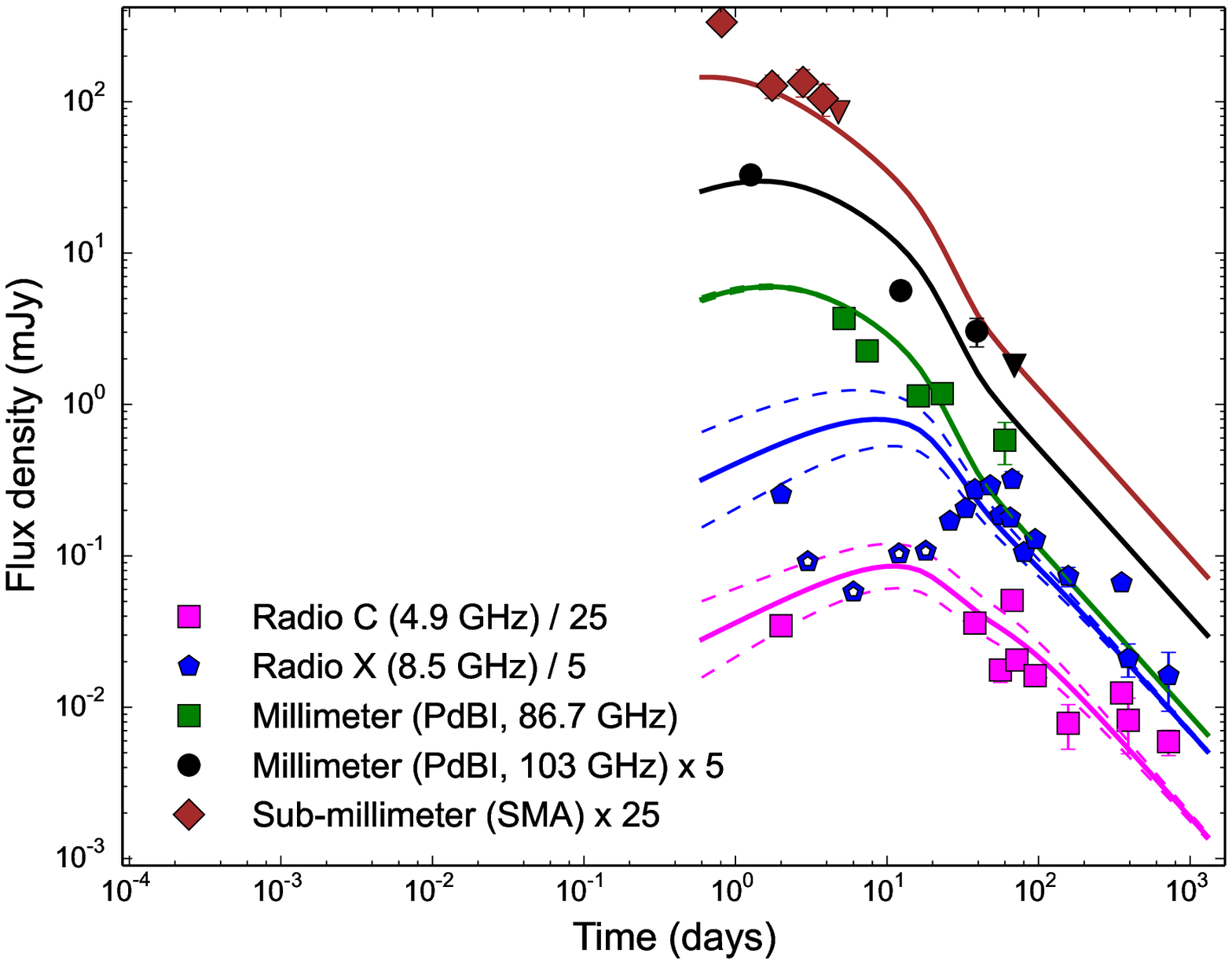} &
 \includegraphics[width=0.47\textwidth]{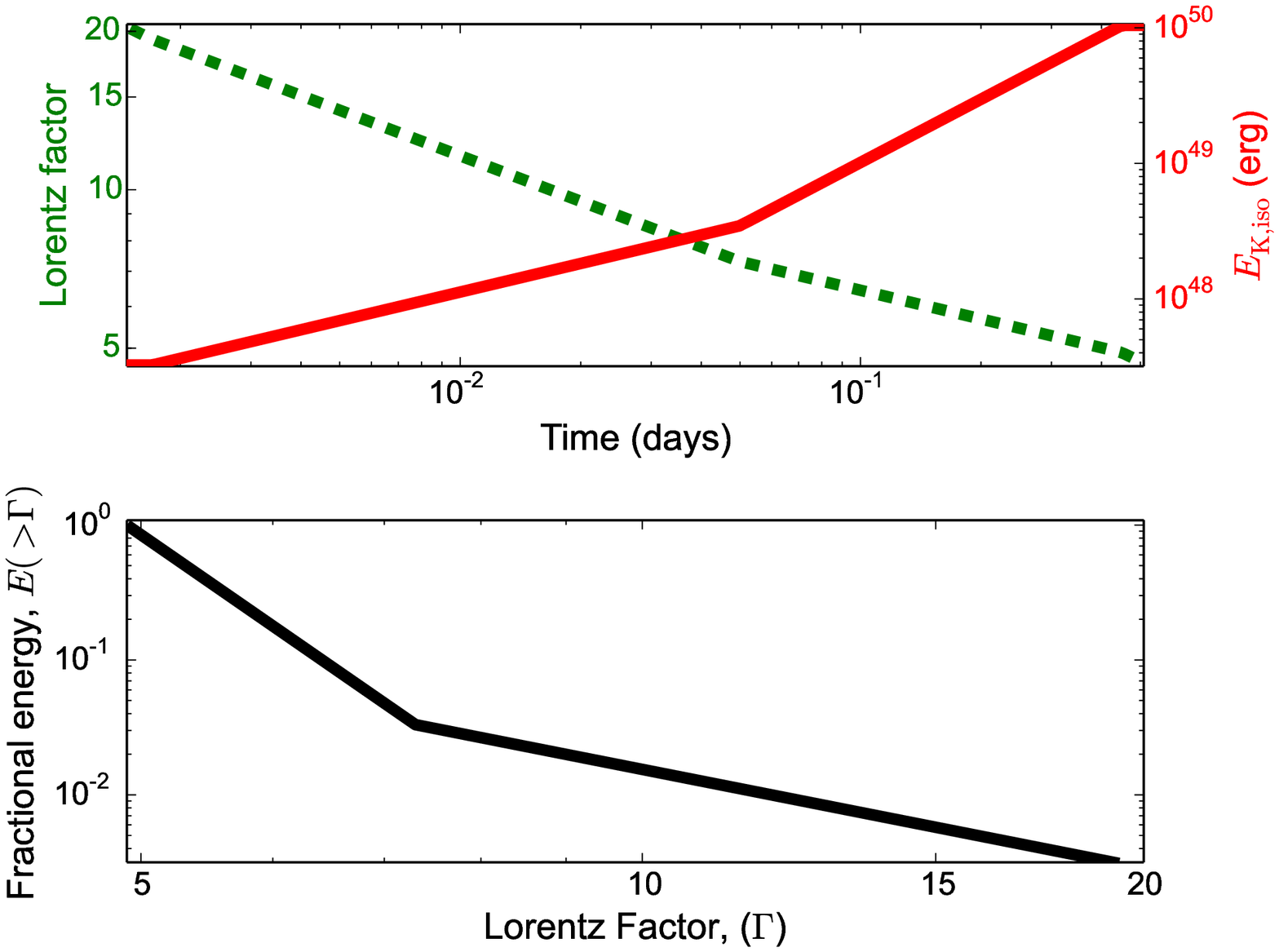} \\ 
\end{tabular}
\caption{X-ray, UV (top left), optical (top right), and radio (bottom left) light curves of 
GRB~100418A, with the full afterglow model (solid lines), including energy injection before 
0.04\,d. The X-ray data before 0.008\,d is likely dominated by high-latitude prompt emission and we 
do not include these data in our analysis; the best fit power law to the X-ray data before 0.008\,d 
added to the blastwave model is shown in the upper left panel (black, dashed). The dashed 
envelopes around the radio light curves indicate the expected effect of scintillation at the 
$1\sigma$ level. The data indicated by open symbols are not used to determine the parameters of the 
forward shock (the MCMC analysis). Bottom right: blastwave Lorentz factor (green, dashed; upper 
sub-panel) and isotropic equivalent kinetic energy (red, solid; upper sub-panel) as a function of 
time, together with the energy distribution across ejecta Lorentz factors (black, solid; lower 
sub-panel) as determined from fitting the X-ray/UV/optical re-brightening at 0.7\,d.
\label{fig:100418A_enj}}
\end{figure*}

The optical and X-ray light curves after 0.4\,d are insensitive to the circumburst density profile 
since $\numax,\nuc < \nuNIR < \nuX$ after the re-brightening. The best-sampled radio light 
curve is the composite 8.46\,GHz formed from VLA, VLBA, and ATCA observations. However, this light 
curve exhibits significant scatter about a smooth power law evolution, possibly due to either ISS 
or inter-calibration issues across the three telescopes\footnote{To avoid the radio data from 
driving the fit, we apply an uncertainty floor of 15\% to the cm- and mm-band observations.}. Thus 
we are unable to constrain the circumburst density profile from the afterglow data. For the 
remainder of this section we focus on the ISM model, discussing the possibility of a wind 
environment in Appendix \ref{appendix:100418A_wind}.

Since this event occurred at a relatively low redshift, $z=0.6235$ \citep{gcn10620}, the host 
galaxy is detected in the Sloane Digital Sky Survey (SDSS) as SDSS~J170527.10+112742.5. We 
use the SDSS photometry ($u = 24.55 \pm 1.14$, $g = 22.92 \pm 0.17$, $r = 22.45 \pm 0.17$, 
$i = 21.95 \pm 0.17$, $z = 22.54 \pm 0.83$) as fixed a-priori measurements of the host galaxy flux 
in the $griz$ bands. We compute the host flux density at $B$- and $V$-band using $B = g + 0.33(g - 
r) + 0.2$ and $V = g - 0.58(g - r) - 0.01$ \citep{jsr+05} and hold these values fixed for our MCMC 
analysis. Since there is no a-priori measurement of the host flux density in the 
UVOT/\textit{White}-band and much of the UVOT data was taken in \textit{White}-band, we keep the 
host flux density in the \textit{White}-band as a free parameter, and integrate over the GRB 
spectral energy distribution using the \textit{White}-band filter 
function\footnote{\url{http://www.swift.ac.uk/analysis/uvot/filters.php}}.

\begin{figure}
\begin{tabular}{ccc}
 \centering
 \includegraphics[width=0.30\columnwidth]{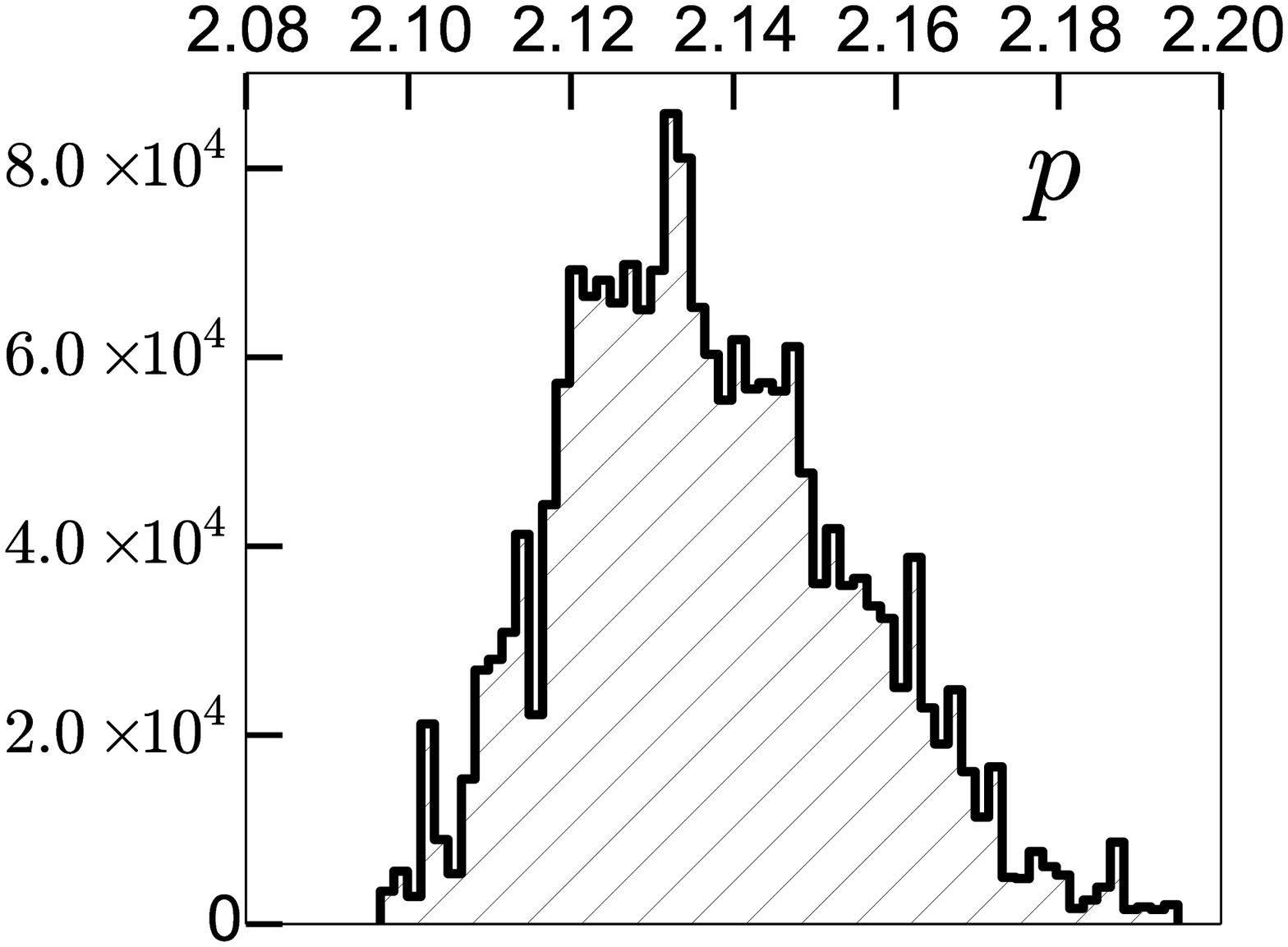} &
 \includegraphics[width=0.30\columnwidth]{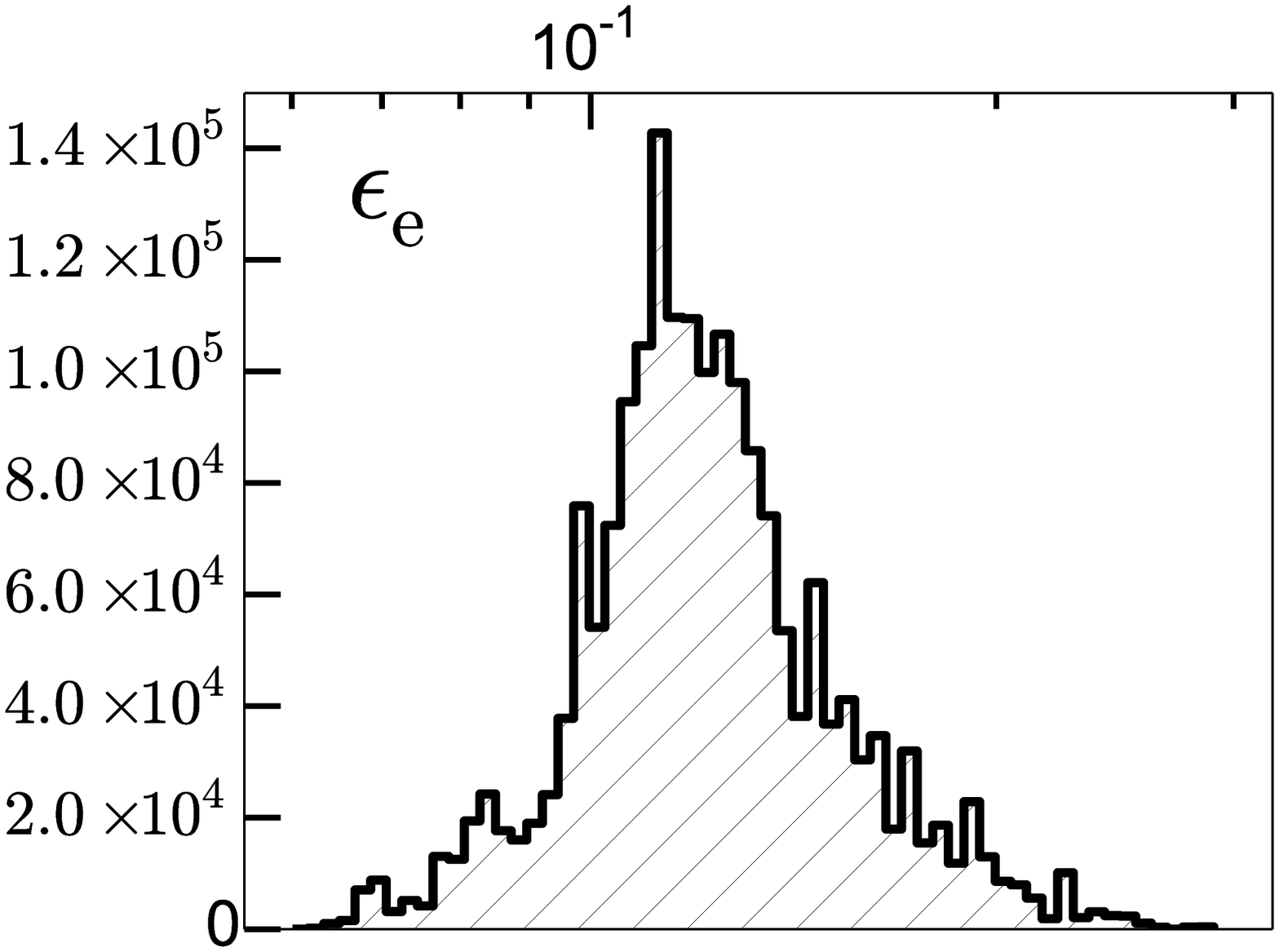} &
 \includegraphics[width=0.30\columnwidth]{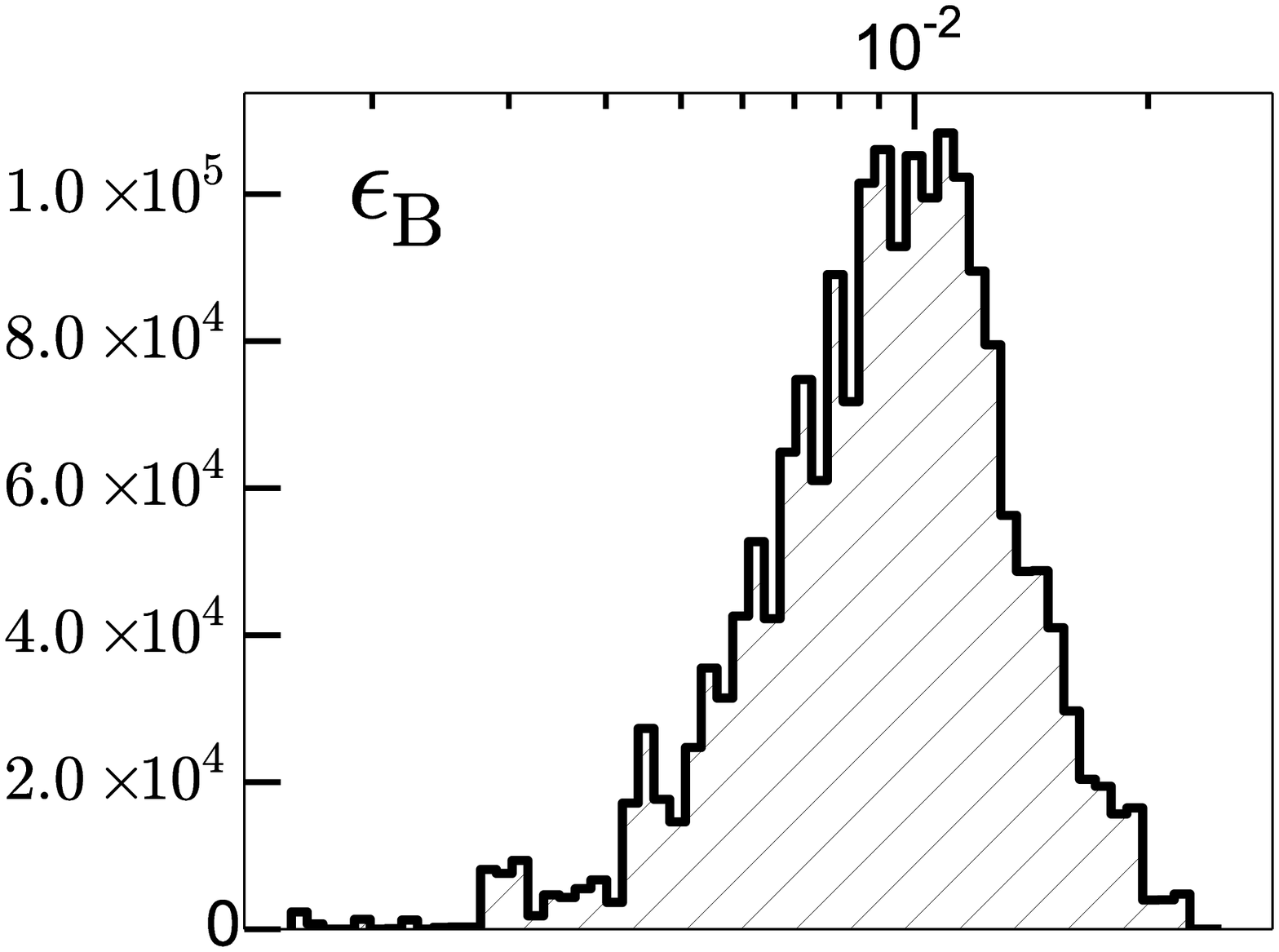} \\
 \includegraphics[width=0.30\columnwidth]{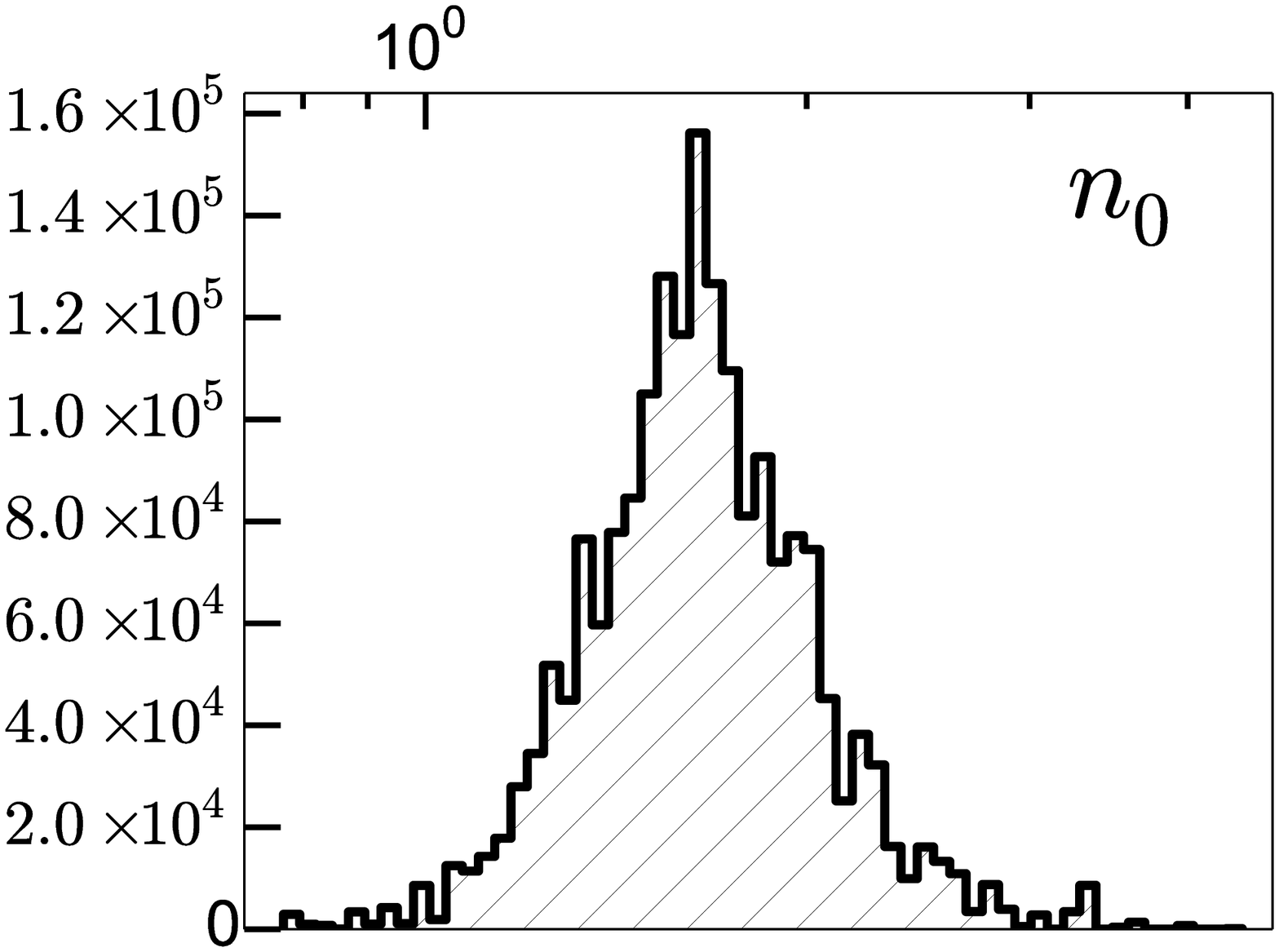} & 
 \includegraphics[width=0.30\columnwidth]{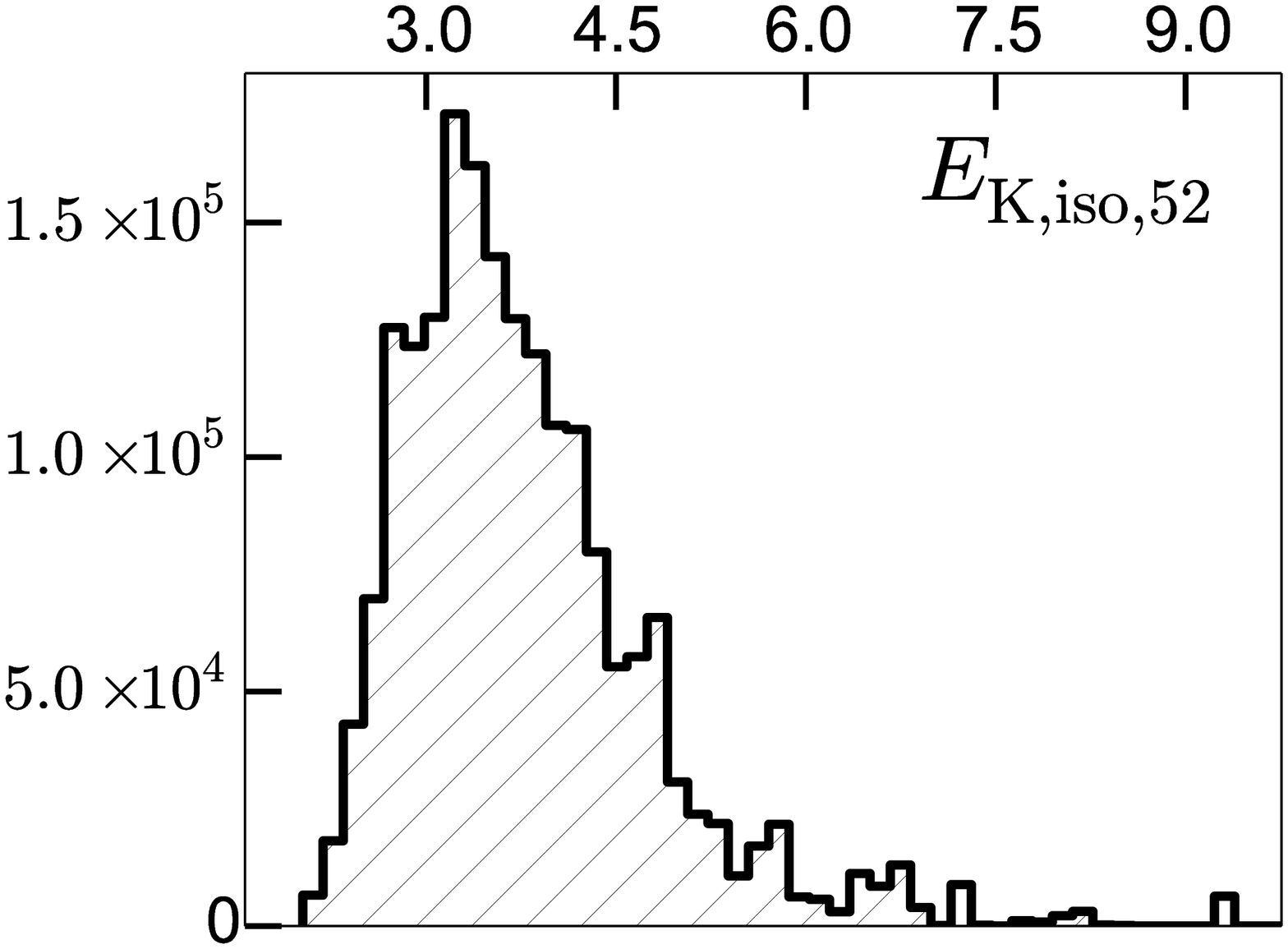} &
 \includegraphics[width=0.30\columnwidth]{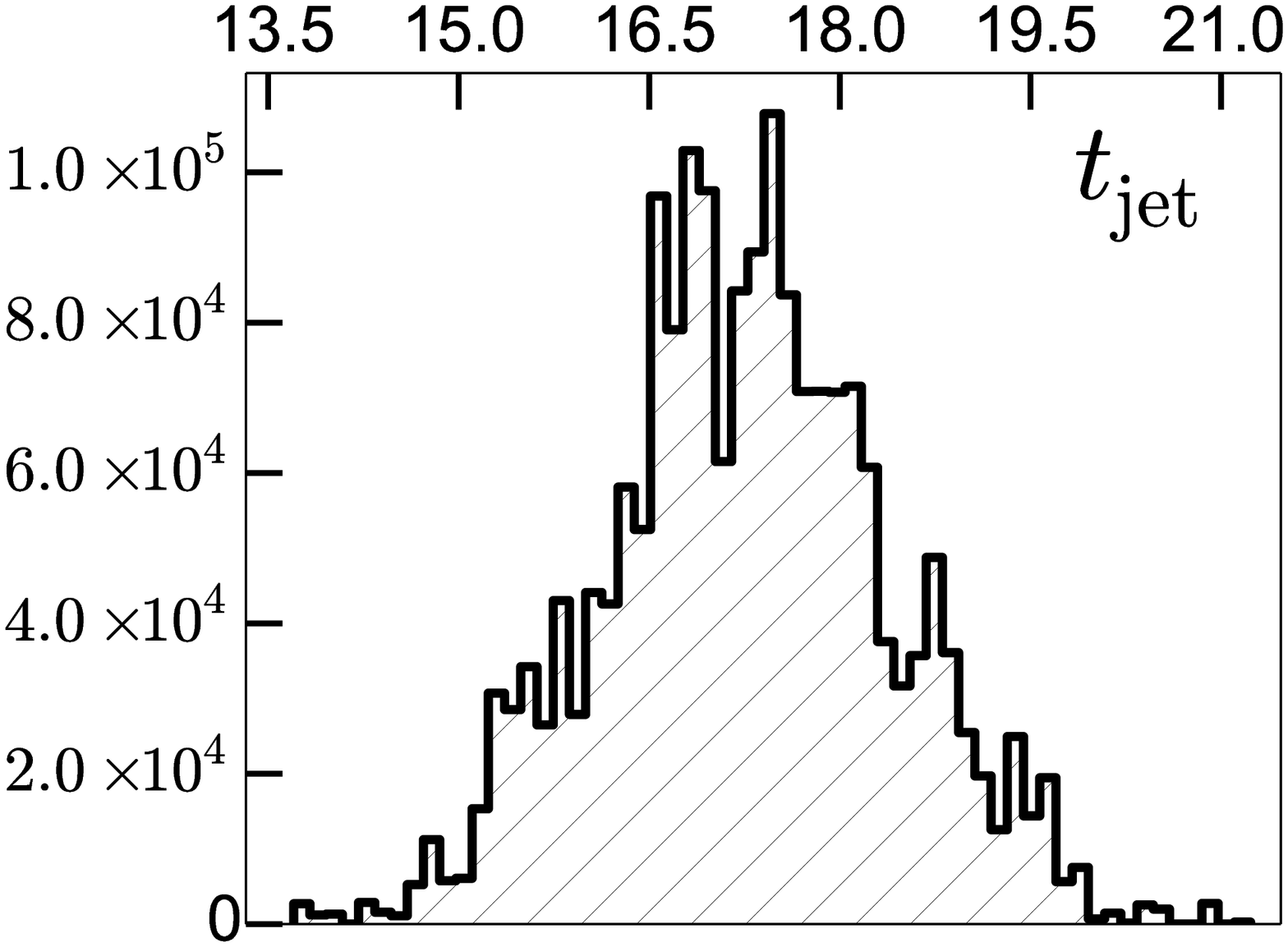} \\  
 \includegraphics[width=0.30\columnwidth]{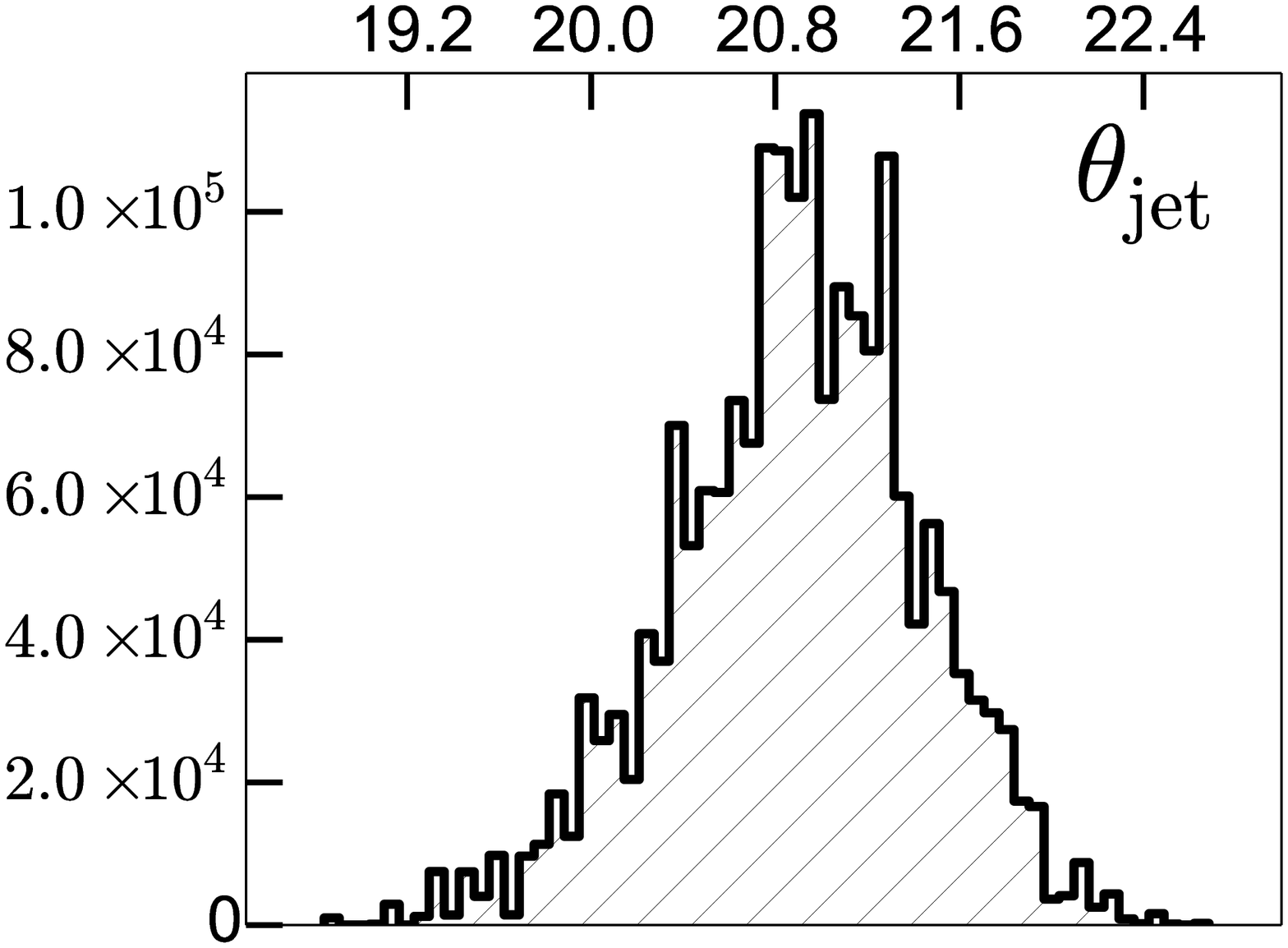}  &
 \includegraphics[width=0.30\columnwidth]{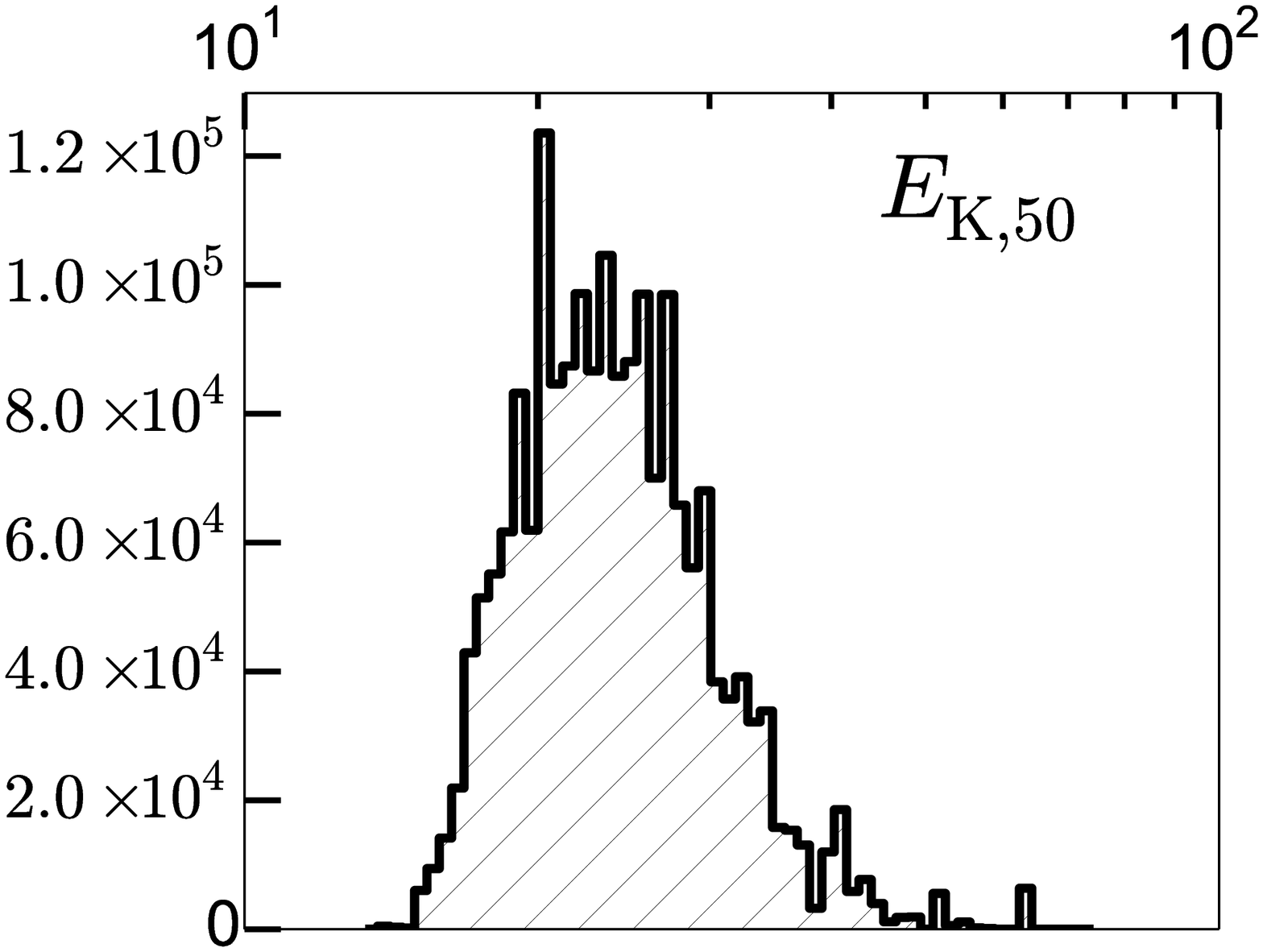} & 
 \includegraphics[width=0.30\columnwidth]{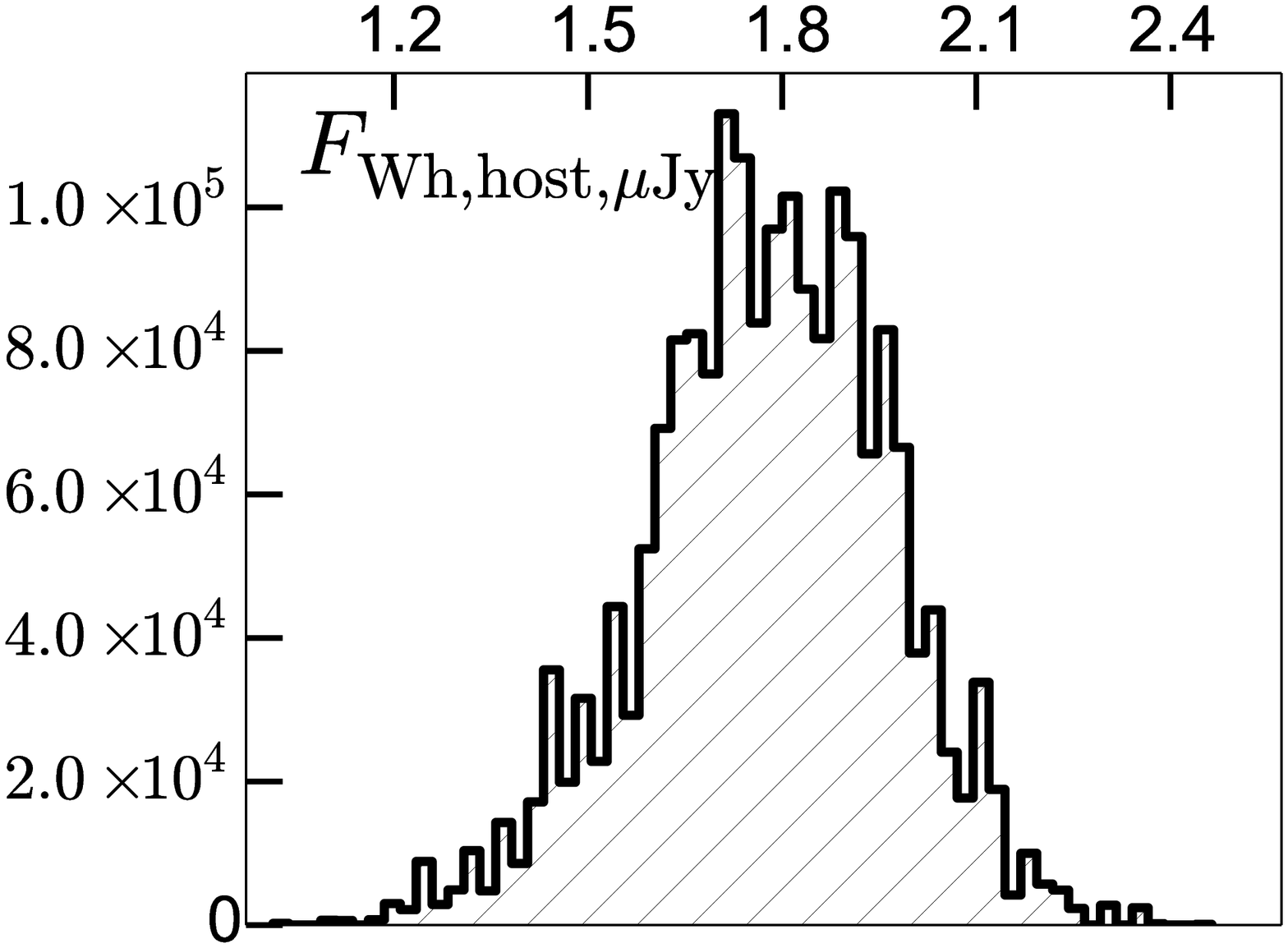}\\  
\end{tabular}
\caption{Posterior probability density functions for the physical parameters for GRB~100418A in 
a constant density environment from MCMC simulations.\label{fig:100418A_ISM_hists}}
\end{figure}

\begin{figure}
\begin{tabular}{ccc}
\centering
 \includegraphics[width=0.30\columnwidth]{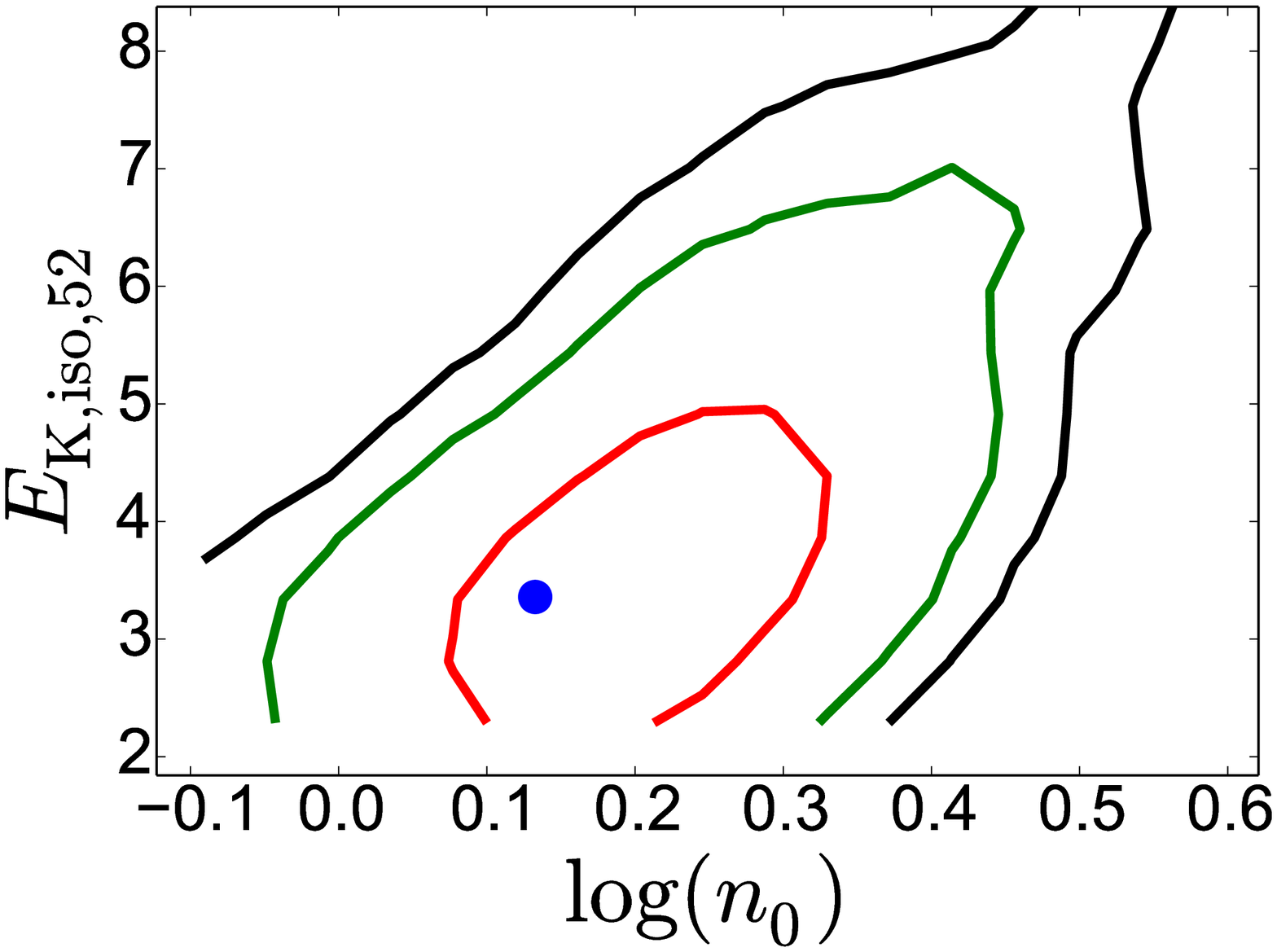} &
 \includegraphics[width=0.30\columnwidth]{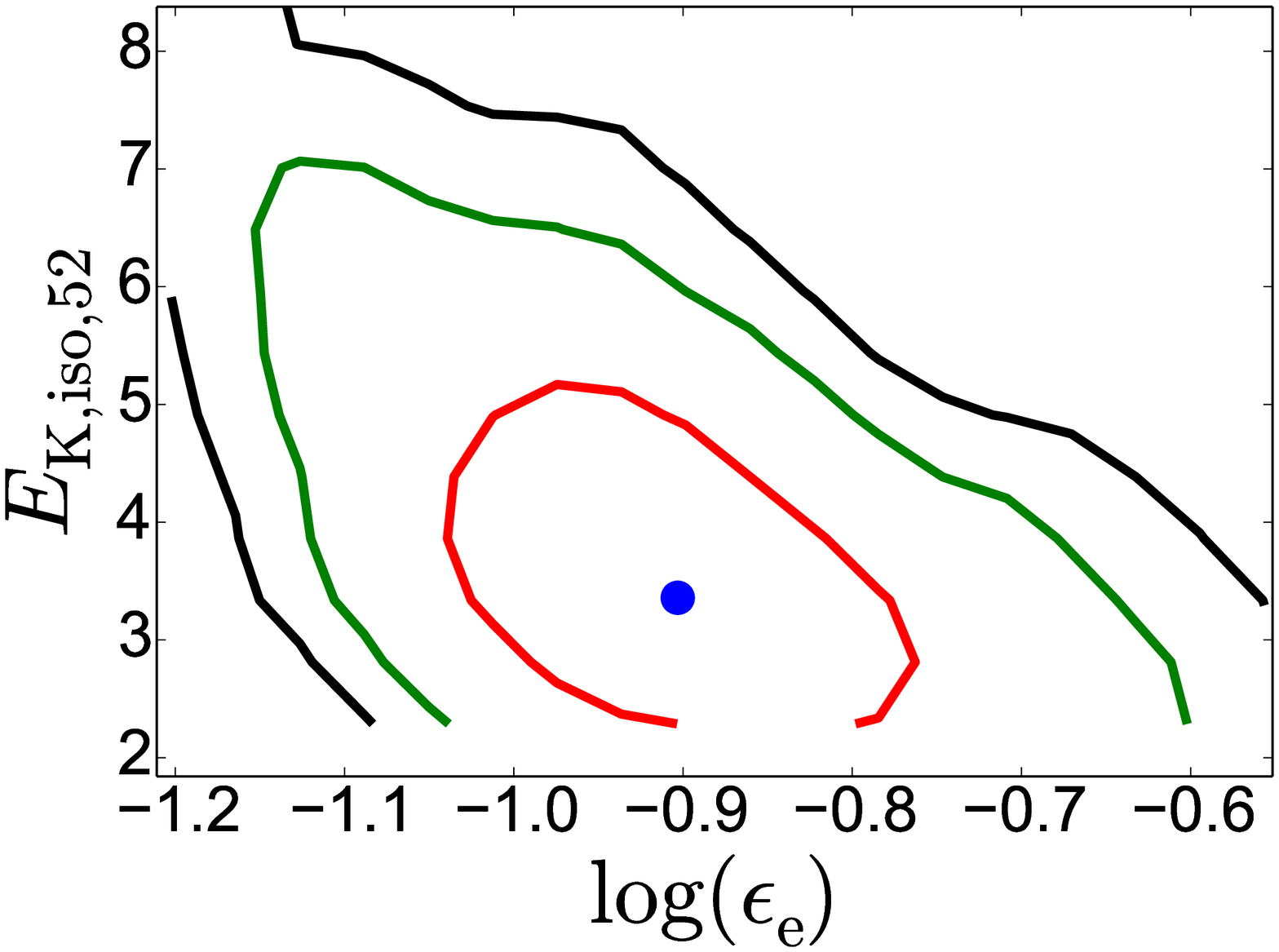} &
 \includegraphics[width=0.30\columnwidth]{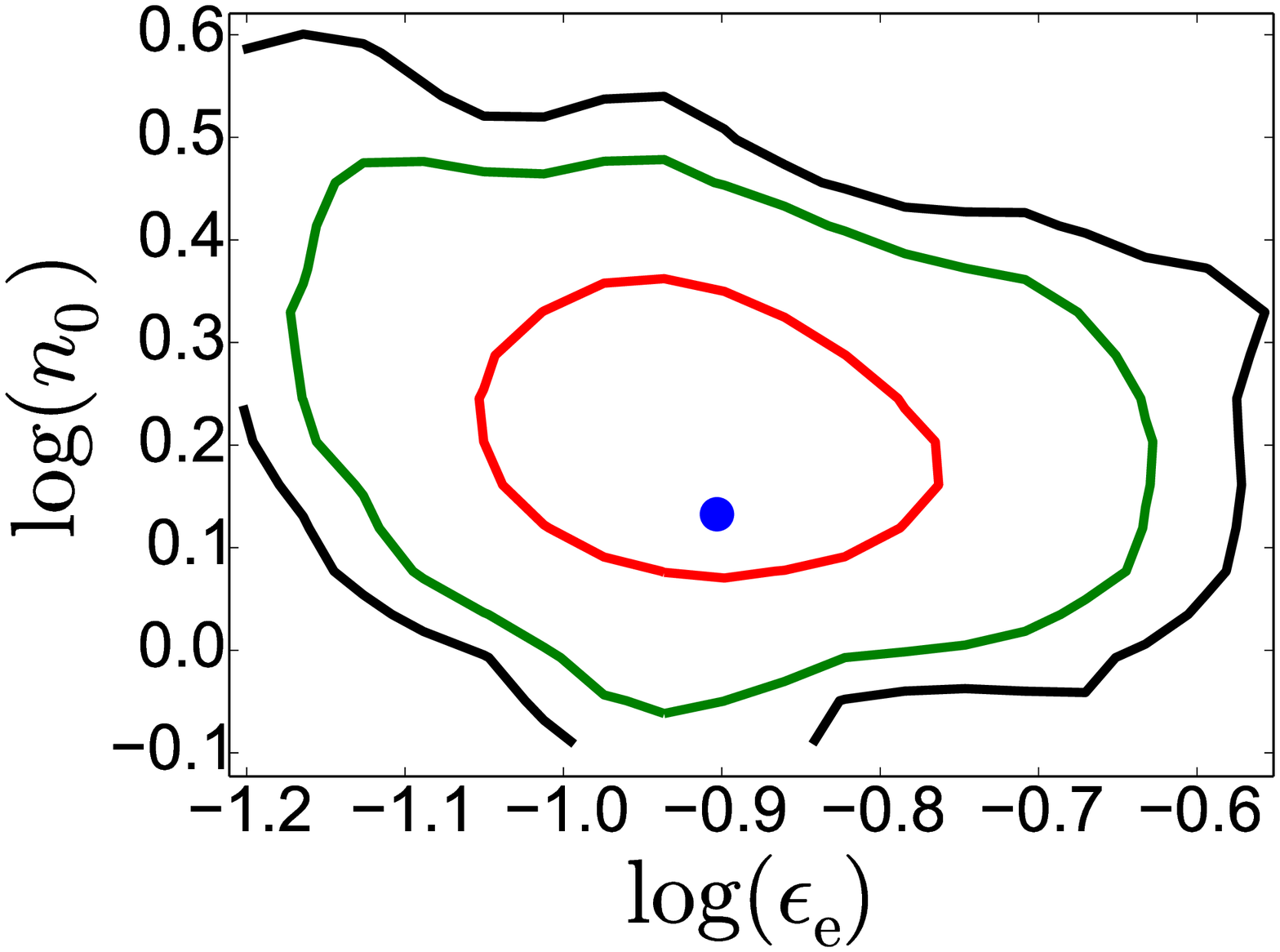} \\
 \includegraphics[width=0.30\columnwidth]{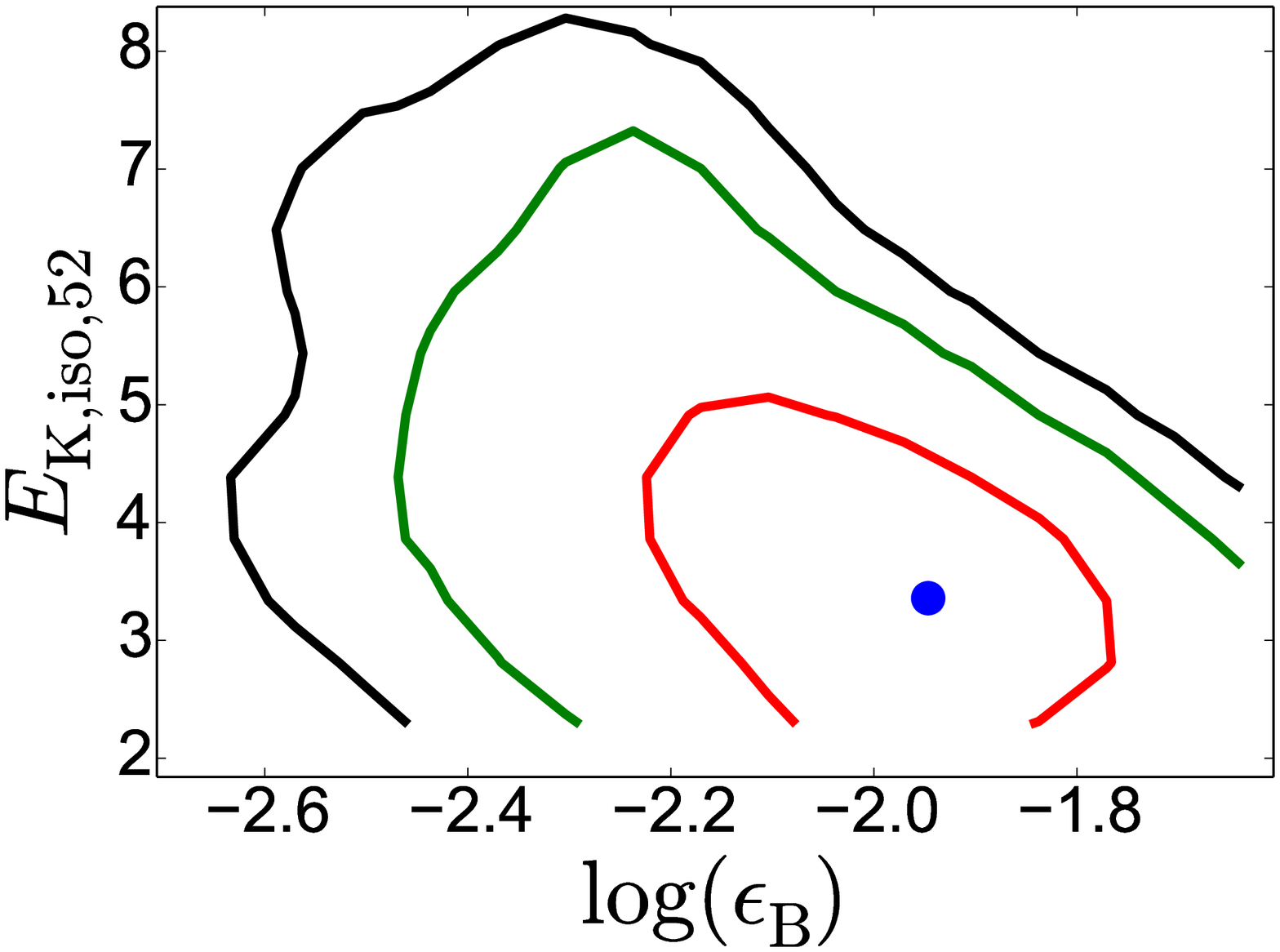} &
 \includegraphics[width=0.30\columnwidth]{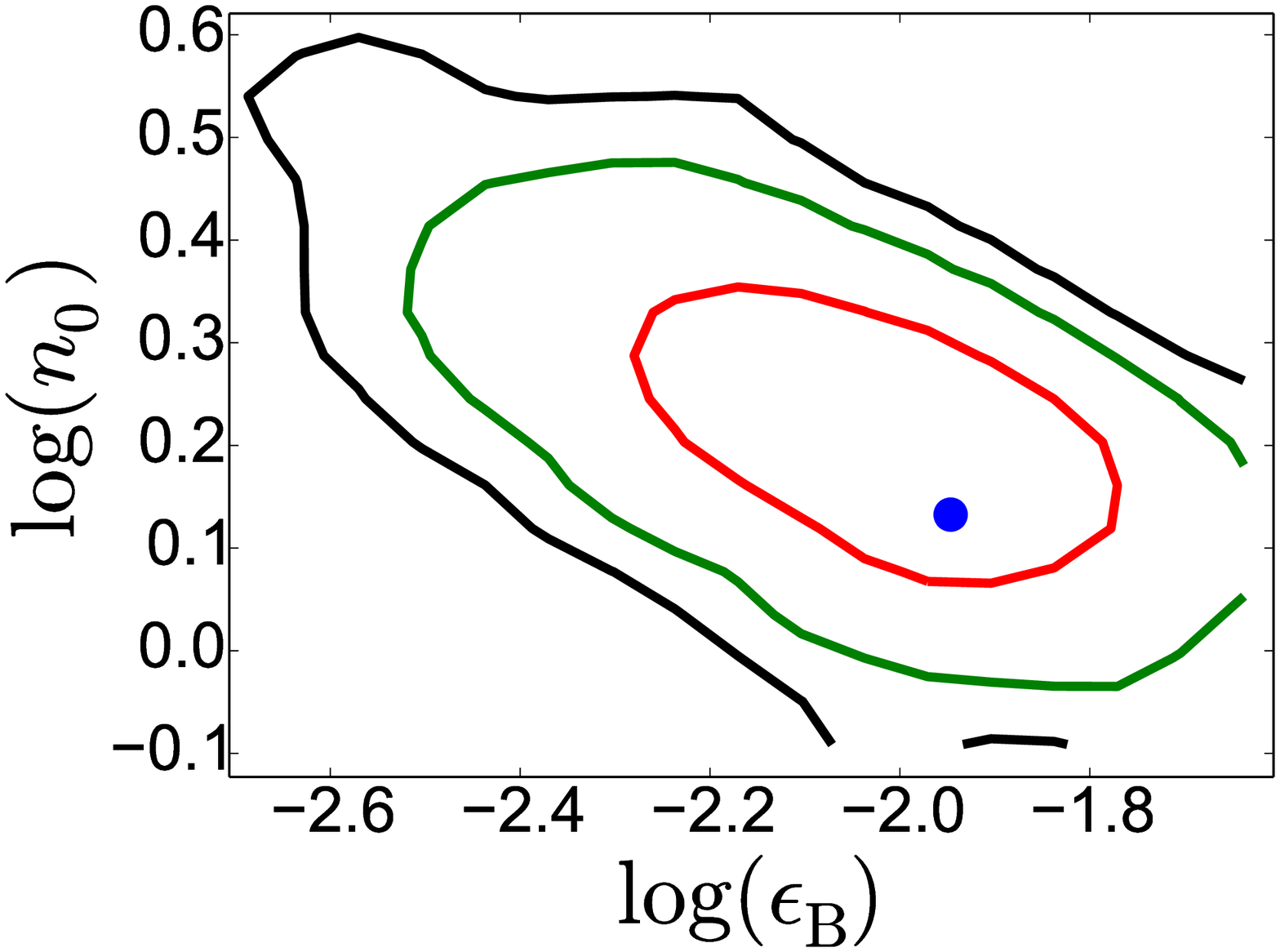} &
 \includegraphics[width=0.30\columnwidth]{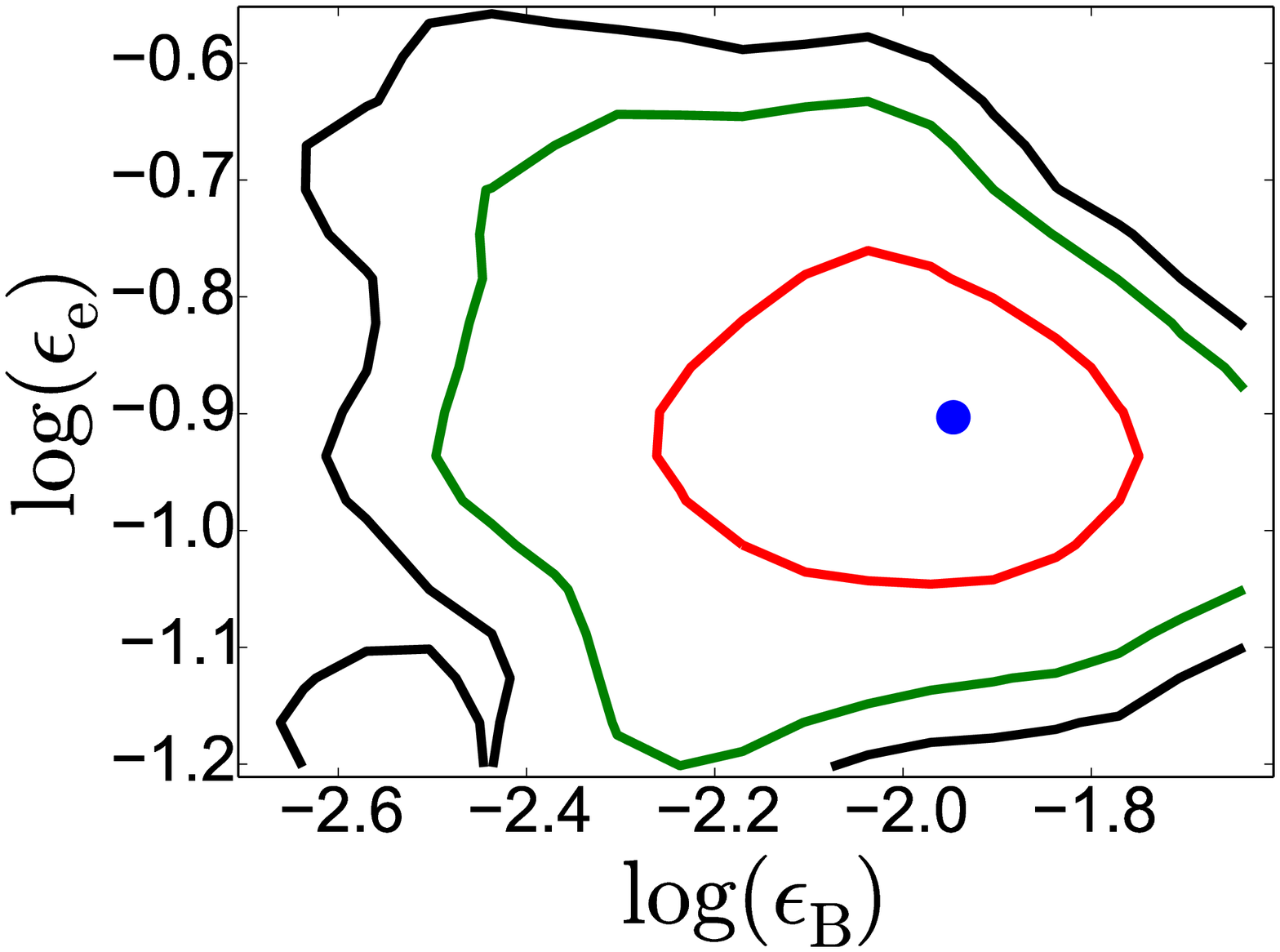} \\
\end{tabular}
\caption{1$\sigma$ (red), 2$\sigma$ (green), and 3$\sigma$ (black) contours for correlations
between the physical parameters, \EKiso, \dens, \epse, and \epsb\ for GRB~100418A, in the ISM model 
from Monte Carlo simulations. We have restricted $\epsilon_{\rm e} < \nicefrac{1}{3}$ and 
$\epsilon_{\rm B} < \nicefrac{1}{3}$. See the on line version of this Figure for additional plots 
of correlations between these parameters and $p$, $t_{\rm jet}$, $\thetajet$, $E_{\rm K}$, and 
$F_{\nu,\rm host, White}$. \label{fig:100418A_ISM_corrplots}}
\end{figure}

Using our MCMC analysis as described in Section \ref{text:modeling}, we fit the data after 
0.4\,d, and our highest-likelihood model (Figure \ref{fig:100418A_enj}) has the parameters 
$p\approx2.1$, $\epse\approx0.12$, $\epsb\approx1.1\times10^{-2}$, $\dens\approx1.4$\,\pcc, 
$\EKiso\approx3.4\times10^{52}$\,erg, $\tjet\approx17$\,d, $\AV\lesssim 0.1$\,mag, and $F_{\nu,\rm 
host, White}\approx1.6\,\mu$Jy, with a Compton $y$-parameter of $\approx2.9$, indicating IC 
cooling is important. The blastwave Lorentz factor is $\Gamma=4.9(t/1\,{\rm d})^{-3/8}$ and the jet 
opening angle is $\thetajet\approx20\degr$. The beaming-corrected kinetic energy is 
$\EK\approx2.1\times10^{51}\,{\rm erg}$, while the beaming corrected $\gamma$-ray energy is 
$\Egamma\approx6\times10^{49}\,{\rm erg}$ (1--$10^4$\,keV; rest frame), corresponding to an 
extremely low radiative efficiency of $\eta_{\rm rad}\equiv\Egamma/(\EK+\Egamma)\approx3\%$. These 
results are summarized in Table \ref{tab:enjsummary}. We present histograms of the marginalized 
posterior density for all parameters in Figure \ref{fig:100418A_ISM_hists} and contour plots of 
correlations between the physical parameters in Figure \ref{fig:100418A_ISM_corrplots}.

In concordance with the basic analysis outlined above, we find that the break 
frequencies at 1\,d for this model are located at $\nua\approx9.0$\,GHz, 
$\numax\approx3.4\times10^{11}$\,Hz, and $\nuc\approx2.3\times10^{14}$\,Hz, while the peak flux 
density is about 12\,mJy at \numax. The 
spectrum transitions from fast to slow cooling at about 230\,s after the burst, and \numax\ drops 
below \nua\ at about 25\,d. The high value of \nua\ is required to suppress the flux density at 
$8.46$\,GHz and $5$\,GHz (VLA), relative to the mm-band (PdBI and SMA). The ordering of the break 
frequencies relative to the observing bands, $\numax,\nuc<\nuNIR$ at 1\,d, ensures that the 
optical and X-ray are on the same part of the synchrotron spectrum at this time. The jet break at 
$\approx17$\,d is largely driven by the millimeter and radio data. The blastwave becomes 
non-relativistic at $t_{\rm NR}\approx40$\,d, resulting in a subsequent slow decline ($F_{\nu}\sim 
t^{0.6+3\beta}\sim t^{-1}$, with $\beta=(1-p)/2\approx-0.5$; \citealt{fwk00}) in the radio light 
curves after the peak.

\cite{mcm+13} use a VLBA limit on the size of the afterglow at 65\,d to estimate the average 
ejecta Lorentz factor. However, we note that the apparent physical size of the afterglow 
is given by $R_{\perp} = \left[2^{12-3k}(4-k)^{5-k}/(5-k)^{5-k}\right]^{1/(8-2k)}\Gamma ct_{\rm 
z}\approx2.3\Gamma ct_{\rm z}$, where $\gamma_{l}$ is the ejecta Lorentz factor of the ejecta at 
time, $t_{\rm z}=t_{\rm obs}/(1+z)$, and $k=0$ for the ISM model \citep{gs02}; therefore the VLBI 
measurement actually represents \textit{an upper limit to the ejecta Lorentz factor at the time of 
the measurement}, such that $\Gamma (t_{\rm obs}=65\,\rm d)<5.8$.  Since the blastwave becomes 
non-relativistic at around 40 days, our model does not violate the VLBA limit on the angular size of 
the afterglow.

\subsubsection{Energy injection model}
\label{text:100418A:enj}
Taking the forward shock model described in Section \ref{text:100418A:FS} as a starting point, we 
find that the X-ray and UV/optical data before the re-brightening can be explained by two 
successive periods of energy injection,
\begin{equation}
\EKiso(t) = 
  \begin{cases}
      E_{\rm K,iso,f}, & t > t_{\rm 0} = 0.45\,{\rm d} \\
      E_{\rm K,iso,f}\left(\frac{t}{t_{\rm 0}}\right)^{1.55}, &
      t_1 = 0.05\,{\rm d} < t < t_{\rm 0} \\
      E_{\rm K,iso,f}\left(\frac{t_1}{t_{\rm 0}}\right)^{1.55}\left(\frac{t}{t_1}\right)^{0.7},&
      t_2  = 1.7\times10^{-3}\,{\rm d} < t < t_1\\
      E_{\rm K,iso,f}\left(\frac{t_1}{t_{\rm 0}}\right)^{1.55}\left(\frac{t_2}{t_1}\right)^{0.7},&
      t < t_2\\
  \end{cases}
\end{equation}
In this model, the energy increases by a factor of $\approx11$ from $E_{\rm 
K,iso,i}\approx1.0\times10^{50}$\,erg at $1.7\times10^{-3}$\,d to $E_{\rm 
K,iso}\approx1.1\times10^{51}$\,erg at 0.05\,d followed by another increase by a factor of 
$\approx30$ to $E_{\rm K,iso,f}\approx3.4\times10^{52}$\,erg at 0.45\,d. The overall increase in 
energy is a factor of $\approx340$, corresponding to an injected energy fraction of $\approx 
99.7\%$ over this period. In comparison, $\Egammaiso=9.9^{+6.3}_{-3.4}\times10^{50}$ \citep{mab+11}. 
The blastwave Lorentz factor decreases from $\Gamma\approx38$ at the start of energy injection at 
$1.7\times10^{-3}$ to $\Gamma\approx11$ at $0.05$\,d, and then to $\Gamma\approx7$ at the end of 
energy injection at 0.45\,d. The values of $m$ derived above correspond to an ejecta distribution 
with $E(>\Gamma)\propto\Gamma^{1-s}$ with $s\approx9.6$ for $7\lesssim\Gamma\lesssim11$ and 
$s\approx3$ for $11\lesssim\Gamma\lesssim38$.

\subsection{GRB~100901A}
\subsubsection{GRB properties and basic considerations}
\label{text:100901A:basic_considerations}
GRB~100901A was detected and localized by the \Swift\ BAT on 2010 September 01 at 13:34:10\,UT 
\citep{gcn11159}. The burst duration is $T_{90} = (439\pm33)$\,s, with a fluence of $F_{\gamma} = 
(2.1 \pm 0.3)\times10^{-6}$\,erg\,cm$^{-2}$ \citep[15--150\,keV observer frame;][]{gcn11159}. 
The afterglow was detected in the X-rays and optical bands by XRT \citep{gcn11171}, UVOT 
\citep{gcn11176}, and by multiple ground-based observatories 
\citep[e.g.][]{gcn11170,gcn11172,gcn11173,gcn11174,gcn11175,gcn11177,gcn11178,gcn11179}, as well 
as in the radio by the WSRT and the VLA \citep{gcn11221,gcn11257}. Spectroscopic observations with
Magellan yielded a redshift of $z=1.408$ \citep{gcn11164}. Using the BAT fluence, \cite{gll+12} 
determined the isotropic equivalent $\gamma$-ray energy of this burst to be 
$\Egammaiso=6.3\times10^{52}$\,erg (1--$10^4$\,keV; rest frame). However, the 
$\gamma$-ray spectrum does not exhibit a turnover within the BAT energy range, indicating that the 
peak of the $\gamma$-ray spectrum, $E_{\rm peak}$, is outside the BAT energy range.  
\cite{mzb+13} obtain a correlation between $E_{\rm BAT,obs}$ (15--150\,keV) and $E_{\rm 
\gamma,iso}$: $\frac{\Egammaiso}{\rm erg} = 
10^{-3.7}\left[\frac{E_{\rm BAT}^{15--150}}{\rm erg}\right]^{1.08\pm0.01} $ with a 
scatter of $\sigma = 0.24$ (68\%), which we use to perform a K-correction and obtain
$\Egammaiso = (8\pm1)\times10^{52}$\,erg. Note that the error bars do not include the uncertainty
in the correlation itself.

\begin{deluxetable}{cccccc}
\tabletypesize{\footnotesize}
\tablecolumns{9}
\tablewidth{0pt}
\tablecaption{Swift/UVOT Observations of GRB~100901A\label{tab:data:100901A:UVOT}}
\tablehead{
  \colhead{$t-t_0$} &    
  \colhead{Filter} &
  \colhead{Frequency} &  
  \colhead{Flux density$^{\dag}$} &
  \colhead{Uncertainty} &
  \colhead{Detection?} \\  
  \colhead{(days)} &  
  \colhead{} &
  \colhead{(Hz)} &
  \colhead{($\mu$Jy)} &
  \colhead{($\mu$Jy)} &
  \colhead{($1=$ Yes)}  
  }
\startdata
0.00258  & \textit{White} & 8.64e+14 & 27.4 & 5.65 & 1\\
0.00498  & \textit{u} & 8.56e+14 & 146 & 11.4 & 1\\
0.00661  & \textit{b} & 6.92e+14 & 230 & 75.6 & 0\\
0.00689  & \textit{White} & 8.64e+14 & 58.6 & 18.9 & 1\\
0.00717  & \textit{uvw2} & 1.48e+15 & 67.9 & 12.1 & 0\\
0.0454  & \textit{v} & 5.55e+14 & 155 & 27 & 1\\
0.0478  & \textit{uvm2} & 1.34e+15 & 18.3 & 4.42 & 1\\
0.0502  & \textit{uvw1} & 1.16e+15 & 37.1 & 5.4 & 1\\
0.0597  & \textit{uvw2} & 1.48e+15 & 7.4 & 2.36 & 0\\
0.0609  & \textit{u} & 8.56e+14 & 83.4 & 6.21 & 1\\
0.062  & \textit{v} & 5.55e+14 & 161 & 27.1 & 1\\
0.0632  & \textit{b} & 6.92e+14 & 134 & 12.9 & 1\\
0.0644  & \textit{uvm2} & 1.34e+15 & 20.7 & 4.63 & 1\\
0.0652  & \textit{White} & 8.64e+14 & 62.2 & 9.12 & 1\\
0.0668  & \textit{uvw1} & 1.16e+15 & 17.2 & 4.39 & 1\\
0.117  & \textit{uvw2} & 1.48e+15 & 3.45 & 1.02 & 1\\
0.127  & \textit{v} & 5.55e+14 & 97.7 & 12 & 1\\
0.137  & \textit{uvm2} & 1.34e+15 & 10.1 & 2 & 1\\
0.183  & \textit{uvw1} & 1.16e+15 & 42.7 & 3.03 & 1\\
0.194  & \textit{u} & 8.56e+14 & 129 & 5.18 & 1\\
0.204  & \textit{b} & 6.92e+14 & 192 & 10.2 & 1\\
0.25  & \textit{uvm2} & 1.34e+15 & 21.3 & 2.22 & 1\\
0.261  & \textit{uvw1} & 1.16e+15 & 58.3 & 3.6 & 1\\
0.271  & \textit{u} & 8.56e+14 & 160 & 6.96 & 1\\
0.318  & \textit{uvw2} & 1.48e+15 & 11.4 & 1.37 & 1\\
0.329  & \textit{v} & 5.55e+14 & 291 & 15.5 & 1\\
0.338  & \textit{uvm2} & 1.34e+15 & 26.6 & 2.92 & 1\\
0.384  & \textit{uvw1} & 1.16e+15 & 40.4 & 8.67 & 1\\
0.534  & \textit{uvw1} & 1.16e+15 & 36.8 & 2.88 & 1\\
0.541  & \textit{u} & 8.56e+14 & 107 & 10.1 & 1\\
0.599  & \textit{uvw2} & 1.48e+15 & 6.15 & 1.19 & 1\\
0.607  & \textit{v} & 5.55e+14 & 178 & 28.3 & 1\\
0.647  & \textit{b} & 6.92e+14 & 130 & 43.4 & 0\\
0.717  & \textit{uvm2} & 1.34e+15 & 8.27 & 2.26 & 1\\
0.736  & \textit{uvw1} & 1.16e+15 & 24.4 & 2.5 & 1\\
0.743  & \textit{u} & 8.56e+14 & 63.4 & 13.4 & 1\\
0.786  & \textit{uvw2} & 1.48e+15 & 3.82 & 1.01 & 1\\
0.796  & \textit{v} & 5.55e+14 & 117 & 12.3 & 1\\
0.806  & \textit{uvm2} & 1.34e+15 & 12.7 & 2.14 & 1\\
0.853  & \textit{uvw1} & 1.16e+15 & 16.8 & 2.01 & 1\\
0.863  & \textit{u} & 8.56e+14 & 54.4 & 3.36 & 1\\
0.873  & \textit{b} & 6.92e+14 & 64.8 & 8.72 & 1\\
0.92  & \textit{uvm2} & 1.34e+15 & 6.86 & 1.49 & 1\\
0.93  & \textit{uvw1} & 1.16e+15 & 19.1 & 2.14 & 1\\
0.94  & \textit{u} & 8.56e+14 & 41.4 & 4.63 & 1\\
0.983  & \textit{uvw2} & 1.48e+15 & 7.57 & 2.11 & 0\\
1.06  & \textit{v} & 5.55e+14 & 74.5 & 11.6 & 1\\
1.13  & \textit{uvm2} & 1.34e+15 & 6.64 & 1.18 & 1\\
1.14  & \textit{b} & 6.92e+14 & 51.5 & 8.67 & 1\\
1.15  & \textit{uvw2} & 1.48e+15 & 2.53 & 0.828 & 0\\
1.16  & \textit{uvw1} & 1.16e+15 & 12.2 & 1.35 & 1\\
1.17  & \textit{u} & 8.56e+14 & 31 & 2.53 & 1\\
1.87  & \textit{uvw2} & 1.48e+15 & 1.57 & 0.51 & 0\\
1.87  & \textit{uvm2} & 1.34e+15 & 2.87 & 0.771 & 1\\
1.88  & \textit{uvw1} & 1.16e+15 & 5.66 & 1.13 & 1\\
1.93  & \textit{u} & 8.56e+14 & 15.5 & 1.74 & 1\\
1.94  & \textit{b} & 6.92e+14 & 26.9 & 3.6 & 1\\
1.94  & \textit{White} & 8.64e+14 & 10.7 & 1.69 & 1\\
1.94  & \textit{v} & 5.55e+14 & 54.1 & 8.97 & 1\\
2.8  & \textit{b} & 6.92e+14 & 17.4 & 5.28 & 1\\
2.8  & \textit{White} & 8.64e+14 & 5.82 & 1.45 & 1\\
2.81  & \textit{v} & 5.55e+14 & 35.6 & 11.7 & 0\\
2.86  & \textit{u} & 8.56e+14 & 6.97 & 2.32 & 0\\
2.93  & \textit{uvw2} & 1.48e+15 & 2.38 & 0.694 & 0\\
2.93  & \textit{uvm2} & 1.34e+15 & 3.5 & 0.917 & 0\\
2.94  & \textit{uvw1} & 1.16e+15 & 4.22 & 1.29 & 0\\
4.2  & \textit{u} & 8.56e+14 & 8.36 & 2.64 & 0\\
4.2  & \textit{uvw2} & 1.48e+15 & 3.99 & 0.955 & 0\\
4.27  & \textit{b} & 6.92e+14 & 24.6 & 7.94 & 0\\
4.27  & \textit{White} & 8.64e+14 & 5.56 & 1.89 & 0\\
4.28  & \textit{v} & 5.55e+14 & 53.8 & 17.3 & 0\\
4.34  & \textit{uvm2} & 1.34e+15 & 7.14 & 1.64 & 0\\
4.35  & \textit{uvw1} & 1.16e+15 & 8.36 & 2.24 & 0\\
4.8  & \textit{u} & 8.56e+14 & 8.03 & 2.61 & 0\\
4.8  & \textit{uvw2} & 1.48e+15 & 4.13 & 1.1 & 0\\
4.8  & \textit{b} & 6.92e+14 & 17.4 & 5.63 & 0\\
4.8  & \textit{uvm2} & 1.34e+15 & 6.04 & 1.48 & 0\\
4.81  & \textit{White} & 8.64e+14 & 3.98 & 1.36 & 0\\
4.81  & \textit{uvw1} & 1.16e+15 & 7.16 & 1.97 & 0\\
4.81  & \textit{v} & 5.55e+14 & 53.2 & 16.9 & 0\\
5.67  & \textit{u} & 8.56e+14 & 33.1 & 9.34 & 0\\
5.8  & \textit{uvw2} & 1.48e+15 & 10.1 & 2.13 & 0\\
7.24  & \textit{uvw2} & 1.48e+15 & 1.39 & 0.394 & 0\\
7.25  & \textit{uvm2} & 1.34e+15 & 1.83 & 0.542 & 0\\
7.25  & \textit{uvw1} & 1.16e+15 & 2.41 & 0.758 & 0\\
9.46  & \textit{v} & 5.55e+14 & 19.8 & 6.4 & 0\\
10.2  & \textit{u} & 8.56e+14 & 3.96 & 1.28 & 0\\
10.2  & \textit{b} & 6.92e+14 & 8.75 & 2.87 & 0\\
10.2  & \textit{White} & 8.64e+14 & 2.05 & 0.678 & 0\\
11  & \textit{u} & 8.56e+14 & 4.21 & 1.36 & 0\\
11  & \textit{b} & 6.92e+14 & 9.31 & 3.05 & 0\\
11  & \textit{White} & 8.64e+14 & 2.26 & 0.743 & 0\\
11  & \textit{v} & 5.55e+14 & 23 & 7.48 & 0\\
11.7  & \textit{u} & 8.56e+14 & 4.41 & 1.4 & 0\\
11.7  & \textit{b} & 6.92e+14 & 9.97 & 3.23 & 0\\
11.7  & \textit{White} & 8.64e+14 & 2.55 & 0.859 & 0\\
11.8  & \textit{v} & 5.55e+14 & 30.4 & 9.91 & 0\\
12.6  & \textit{u} & 8.56e+14 & 4.11 & 1.33 & 0\\
12.6  & \textit{b} & 6.92e+14 & 8.96 & 2.94 & 0\\
12.6  & \textit{White} & 8.64e+14 & 2.12 & 0.703 & 0\\
12.6  & \textit{v} & 5.55e+14 & 23.1 & 7.59 & 0\\
13.5  & \textit{b} & 6.92e+14 & 20.4 & 6.56 & 0\\
13.5  & \textit{White} & 8.64e+14 & 5.14 & 1.68 & 0\\
13.5  & \textit{v} & 5.55e+14 & 62 & 19.5 & 0\\
13.6  & \textit{u} & 8.56e+14 & 8.12 & 2.58 & 0\\
14.6  & \textit{u} & 8.56e+14 & 4.3 & 1.42 & 0\\
14.6  & \textit{b} & 6.92e+14 & 9.41 & 3.08 & 0\\
14.7  & \textit{White} & 8.64e+14 & 2.19 & 0.721 & 0\\
14.7  & \textit{v} & 5.55e+14 & 23.5 & 7.62 & 0\\
15.7  & \textit{u} & 8.56e+14 & 4.1 & 1.31 & 0\\
15.7  & \textit{b} & 6.92e+14 & 8.93 & 2.95 & 0\\
15.7  & \textit{White} & 8.64e+14 & 2.08 & 0.693 & 0\\
15.7  & \textit{v} & 5.55e+14 & 22.7 & 7.38 & 0\\
16.7  & \textit{u} & 8.56e+14 & 4.13 & 1.32 & 0\\
16.8  & \textit{b} & 6.92e+14 & 9.21 & 2.99 & 0\\
16.8  & \textit{White} & 8.64e+14 & 2.41 & 0.794 & 0\\
16.8  & \textit{v} & 5.55e+14 & 24.8 & 8.08 & 0\\
17.7  & \textit{u} & 8.56e+14 & 4.29 & 1.38 & 0\\
17.7  & \textit{b} & 6.92e+14 & 9.39 & 3.09 & 0\\
17.7  & \textit{White} & 8.64e+14 & 2.15 & 0.707 & 0\\
17.7  & \textit{v} & 5.55e+14 & 21.8 & 7.11 & 0\\
18.6  & \textit{u} & 8.56e+14 & 4.27 & 1.37 & 0\\
18.6  & \textit{b} & 6.92e+14 & 9.32 & 3.04 & 0\\
18.6  & \textit{White} & 8.64e+14 & 2.13 & 0.701 & 0\\
18.6  & \textit{v} & 5.55e+14 & 21.3 & 6.85 & 0
\enddata
\tablecomments{$^{\dag}$In cases of non-detections, we report the formal flux measurement at the 
position of the afterglow.}
\end{deluxetable}

XRT began observing the GRB during the $\gamma$-ray emission interval. The X-ray spectra from 1\,ks 
to the end of XRT observations are well fit by an absorbed single power law model with $N_{\rm 
H,Gal}=7.1\times 10^{20}\,\pcmsq$ \citep{kbh+05}, intrinsic hydrogen column $N_{\rm H,int} = 
(3.1\pm0.7)\times 10^{21}\,\pcmsq$, and photon index, $\Gamma_{\rm X} = 2.15\pm0.06$, with no 
evidence for spectral evolution during this period. We convert the count rate light curve published 
by \cite{mzb+13} to a flux density light curve at 1\,keV using their time-resolved spectra. We also 
analyze all UVOT photometry for this burst and report our results in Table 
\ref{tab:data:100901A:UVOT}, and a compilation of all photometry listed in GCN circulars 
\footnote{The KPNO/SARA $V$-band detection \citep{gcn11174} is significantly brighter than expected 
from interpolating the UVOT $V$-band light curve, implying a difference either in calibration or 
in the filter response between the two instruments. We do not include the KPNO $V$-band data point 
at 0.78\,d in our analysis.} or published elsewhere for this event in Table 
\ref{tab:data:100901A:GCN}.

\begin{deluxetable*}{ccccccccc}
\tabletypesize{\footnotesize}
\tablecolumns{9}
\tablewidth{0pt}
\tablecaption{Optical Observations of GRB~100901A\label{tab:data:100901A:GCN}}
\tablehead{
  \colhead{$t-t_0$} &  
  \colhead{Observatory} &
  \colhead{Telescope /} &
  \colhead{Filter} &
  \colhead{Frequency} &  
  \colhead{Flux density$^{\dag}$} &
  \colhead{Uncertainty} &
  \colhead{Detection?} &
  \colhead{Reference} \\  
  \colhead{(days)} &
  \colhead{} &
  \colhead{Instrument} &
  \colhead{} &
  \colhead{(Hz)} &
  \colhead{($\mu$Jy)} &
  \colhead{($\mu$Jy)} &
  \colhead{($1=$ Yes)} & 
  }
\startdata
0.00131 & MASTER &  & \textit{CR} & 4.4e+14 & 82.1 & 436 & 0 & \cite{gll+12} \\
0.00178 & MASTER &  & \textit{CR} & 4.4e+14 & 156 & 170 & 0 & \cite{gll+12} \\
0.00246 & MASTER &  & \textit{CR} & 4.4e+14 & 131 & 143 & 0 & \cite{gll+12} \\
0.00396 & MASTER &  & \textit{CR} & 4.4e+14 & 119 & 107 & 0 & \cite{gll+12} \\
0.00494 & MASTER &  & \textit{CR} & 4.4e+14 & 304 & 78.6 & 1 & \cite{gll+12} \\
0.00615 & MASTER &  & \textit{CR} & 4.4e+14 & 125 & 73.4 & 0 & \cite{gll+12} \\
0.00758 & MASTER &  & \textit{CR} & 4.4e+14 & 88.4 & 80 & 0 & \cite{gll+12} \\
0.0093 & MASTER &  & \textit{CR} & 4.4e+14 & 109 & 80.6 & 0 & \cite{gll+12} \\
0.0114 & MASTER &  & \textit{CR} & 4.4e+14 & 204 & 65 & 1 & \cite{gll+12} \\
0.012 & Okayama & MITSuME & \textit{Ic} & 3.93e+14 & 460 & 93.1 & 1 & \cite{gcn11172} \\
0.012 & Okayama & MITSuME & \textit{Rc} & 4.56e+14 & 336 & 32.4 & 1 & \cite{gcn11172} \\
0.012 & Okayama & MITSuME & \textit{g'} & 6.29e+14 & 132 & 77.1 & 1 & \cite{gcn11172} \\
0.0136 & MASTER &  & \textit{CR} & 4.4e+14 & 195 & 62.1 & 1 & \cite{gll+12} \\
0.0159 & MASTER &  & \textit{v} & 5.55e+14 & 292 & 55.8 & 1 & \cite{gll+12} \\
0.0182 & MASTER &  & \textit{CR} & 4.4e+14 & 232 & 74 & 1 & \cite{gll+12} \\
0.0227 & MASTER &  & \textit{v} & 5.55e+14 & 292 & 55.8 & 1 & \cite{gll+12} \\
0.0251 & MASTER &  & \textit{v} & 5.55e+14 & 351 & 59.4 & 1 & \cite{gll+12} \\
0.0275 & MASTER &  & \textit{CR} & 4.4e+14 & 228 & 22 & 1 & \cite{gll+12} \\
0.0299 & MASTER &  & \textit{CR} & 4.4e+14 & 224 & 12.7 & 1 & \cite{gll+12} \\
0.0323 & MASTER &  & \textit{CR} & 4.4e+14 & 202 & 11.5 & 1 & \cite{gll+12} \\
0.0334 & Okayama & MITSuME & \textit{Ic} & 3.93e+14 & 265 & 25.6 & 1 & \cite{gcn11172} \\
0.0334 & Okayama & MITSuME & \textit{Rc} & 4.56e+14 & 212 & 20.5 & 1 & \cite{gcn11172} \\
0.0334 & Okayama & MITSuME & \textit{g'} & 6.29e+14 & 209 & 20.2 & 1 & \cite{gcn11172} \\
0.0382 & MASTER &  & \textit{CR} & 4.4e+14 & 186 & 12.4 & 1 & \cite{gll+12} \\
0.0406 & MASTER &  & \textit{CR} & 4.4e+14 & 195 & 13 & 1 & \cite{gll+12} \\
0.043 & MASTER &  & \textit{CR} & 4.4e+14 & 192 & 12.8 & 1 & \cite{gll+12} \\
0.0454 & MASTER &  & \textit{CR} & 4.4e+14 & 214 & 12.2 & 1 & \cite{gll+12} \\
0.0478 & MASTER &  & \textit{CR} & 4.4e+14 & 185 & 12.3 & 1 & \cite{gll+12} \\
0.0502 & MASTER &  & \textit{CR} & 4.4e+14 & 212 & 14.1 & 1 & \cite{gll+12} \\
0.0526 & MASTER &  & \textit{CR} & 4.4e+14 & 155 & 13.4 & 1 & \cite{gll+12} \\
0.0548 & Okayama & MITSuME & \textit{Ic} & 3.93e+14 & 220 & 21.3 & 1 & \cite{gcn11172} \\
0.0548 & Okayama & MITSuME & \textit{Rc} & 4.56e+14 & 122 & 11.8 & 1 & \cite{gcn11172} \\
0.0548 & Okayama & MITSuME & \textit{g'} & 6.29e+14 & 120 & 11.6 & 1 & \cite{gcn11172} \\
0.055 & MASTER &  & \textit{CR} & 4.4e+14 & 117 & 13.6 & 1 & \cite{gll+12} \\
0.0574 & MASTER &  & \textit{CR} & 4.4e+14 & 109 & 15 & 1 & \cite{gll+12} \\
0.0599 & MASTER &  & \textit{CR} & 4.4e+14 & 183 & 14 & 1 & \cite{gll+12} \\
0.0623 & MASTER &  & \textit{CR} & 4.4e+14 & 185 & 14.1 & 1 & \cite{gll+12} \\
0.0647 & MASTER &  & \textit{CR} & 4.4e+14 & 186 & 14.3 & 1 & \cite{gll+12} \\
0.067 & MASTER &  & \textit{CR} & 4.4e+14 & 193 & 14.8 & 1 & \cite{gll+12} \\
0.0695 & MASTER &  & \textit{CR} & 4.4e+14 & 164 & 14.2 & 1 & \cite{gll+12} \\
0.0721 & MASTER &  & \textit{CR} & 4.4e+14 & 206 & 15.8 & 1 & \cite{gll+12} \\
0.0745 & MASTER &  & \textit{CR} & 4.4e+14 & 167 & 16.1 & 1 & \cite{gll+12} \\
0.0759 & Okayama & MITSuME & \textit{Ic} & 3.93e+14 & 152 & 30.8 & 1 & \cite{gcn11172} \\
0.0759 & Okayama & MITSuME & \textit{Rc} & 4.56e+14 & 134 & 12.9 & 1 & \cite{gcn11172} \\
0.0759 & Okayama & MITSuME & \textit{g'} & 6.29e+14 & 120 & 11.6 & 1 & \cite{gcn11172} \\
0.0804 & MASTER &  & \textit{CR} & 4.4e+14 & 131 & 14 & 1 & \cite{gll+12} \\
0.0828 & MASTER &  & \textit{CR} & 4.4e+14 & 110 & 15.2 & 1 & \cite{gll+12} \\
0.0852 & MASTER &  & \textit{CR} & 4.4e+14 & 101 & 16 & 1 & \cite{gll+12} \\
0.0877 & MASTER &  & \textit{CR} & 4.4e+14 & 61.2 & 18.7 & 1 & \cite{gll+12} \\
0.0901 & MASTER &  & \textit{CR} & 4.4e+14 & 106 & 16.9 & 1 & \cite{gll+12} \\
0.0926 & MASTER &  & \textit{CR} & 4.4e+14 & 141 & 16.5 & 1 & \cite{gll+12} \\
0.095 & MASTER &  & \textit{CR} & 4.4e+14 & 156 & 18.3 & 1 & \cite{gll+12} \\
0.0969 & Okayama & MITSuME & \textit{Ic} & 3.93e+14 & 152 & 30.8 & 1 & \cite{gcn11172} \\
0.0969 & Okayama & MITSuME & \textit{Rc} & 4.56e+14 & 122 & 11.8 & 1 & \cite{gcn11172} \\
0.0969 & Okayama & MITSuME & \textit{g'} & 6.29e+14 & 100 & 9.65 & 1 & \cite{gcn11172} \\
0.118 & Okayama & MITSuME & \textit{Ic} & 3.93e+14 & 167 & 33.8 & 1 & \cite{gcn11172} \\
0.118 & Okayama & MITSuME & \textit{Rc} & 4.56e+14 & 101 & 20.5 & 1 & \cite{gcn11172} \\
0.118 & Okayama & MITSuME & \textit{g'} & 6.29e+14 & 83.2 & 16.8 & 1 & \cite{gcn11172} \\
0.131 & MASTER &  & \textit{CR} & 4.4e+14 & 22 & 25.7 & 0 & \cite{gll+12} \\
0.134 & MASTER &  & \textit{CR} & 4.4e+14 & 74.9 & 17.7 & 1 & \cite{gll+12} \\
0.136 & MASTER &  & \textit{CR} & 4.4e+14 & 58.9 & 18.8 & 1 & \cite{gll+12} \\
0.139 & MASTER &  & \textit{CR} & 4.4e+14 & 54.3 & 35 & 0 & \cite{gll+12} \\
0.139 & Okayama & MITSuME & \textit{Ic} & 3.93e+14 & 183 & 17.7 & 1 & \cite{gcn11172} \\
0.139 & Okayama & MITSuME & \textit{Rc} & 4.56e+14 & 134 & 12.9 & 1 & \cite{gcn11172} \\
0.139 & Okayama & MITSuME & \textit{g'} & 6.29e+14 & 110 & 10.6 & 1 & \cite{gcn11172} \\
0.141 & MASTER &  & \textit{CR} & 4.4e+14 & 108 & 31.8 & 1 & \cite{gll+12} \\
0.144 & MASTER &  & \textit{CR} & 4.4e+14 & 152 & 30.8 & 1 & \cite{gll+12} \\
0.146 & MASTER &  & \textit{CR} & 4.4e+14 & 106 & 32.5 & 1 & \cite{gll+12} \\
0.149 & MASTER &  & \textit{CR} & 4.4e+14 & 185 & 29.3 & 1 & \cite{gll+12} \\
0.151 & MASTER &  & \textit{CR} & 4.4e+14 & 135 & 22.9 & 1 & \cite{gll+12} \\
0.154 & MASTER &  & \textit{CR} & 4.4e+14 & 193 & 18.7 & 1 & \cite{gll+12} \\
0.16 & Okayama & MITSuME & \textit{Ic} & 3.93e+14 & 319 & 30.7 & 1 & \cite{gcn11172} \\
0.16 & Okayama & MITSuME & \textit{Rc} & 4.56e+14 & 193 & 18.7 & 1 & \cite{gcn11172} \\
0.16 & Okayama & MITSuME & \textit{g'} & 6.29e+14 & 120 & 11.6 & 1 & \cite{gcn11172} \\
0.172 & Hanle & HCT & \textit{R} & 4.56e+14 & 267 & 12.6 & 1 & \cite{gcn11175} \\
0.172 & MASTER &  & \textit{R} & 4.56e+14 & 180 & 92.3 & 0 & \cite{gll+12} \\
0.178 & Hanle & HCT & \textit{I} & 3.93e+14 & 356 & 16.8 & 1 & \cite{gcn11175} \\
0.181 & Okayama & MITSuME & \textit{Ic} & 3.93e+14 & 349 & 33.7 & 1 & \cite{gcn11172} \\
0.181 & Okayama & MITSuME & \textit{Rc} & 4.56e+14 & 280 & 27 & 1 & \cite{gcn11172} \\
0.181 & Okayama & MITSuME & \textit{g'} & 6.29e+14 & 191 & 18.4 & 1 & \cite{gcn11172} \\
0.182 & Terksol & Z-600 & \textit{R} & 4.56e+14 & 282 & 7.9 & 1 & \cite{gcn11166} \\
0.183 & MASTER &  & \textit{CR} & 4.4e+14 & 202 & 36.5 & 1 & \cite{gll+12} \\
0.183 & MASTER &  & \textit{R} & 4.56e+14 & 230 & 39 & 1 & \cite{gll+12} \\
0.186 & MASTER &  & \textit{CR} & 4.4e+14 & 220 & 14.7 & 1 & \cite{gll+12} \\
0.186 & MASTER &  & \textit{R} & 4.56e+14 & 290 & 30.9 & 1 & \cite{gll+12} \\
0.188 & MASTER &  & \textit{CR} & 4.4e+14 & 239 & 13.6 & 1 & \cite{gll+12} \\
0.188 & MASTER &  & \textit{R} & 4.56e+14 & 277 & 26.7 & 1 & \cite{gll+12} \\
0.19 & MASTER &  & \textit{R} & 4.56e+14 & 321 & 24.5 & 1 & \cite{gll+12} \\
0.193 & MASTER &  & \textit{R} & 4.56e+14 & 239 & 23.1 & 1 & \cite{gll+12} \\
0.195 & MASTER &  & \textit{R} & 4.56e+14 & 358 & 23.9 & 1 & \cite{gll+12} \\
0.198 & MASTER &  & \textit{R} & 4.56e+14 & 290 & 25.1 & 1 & \cite{gll+12} \\
0.2 & MASTER &  & \textit{R} & 4.56e+14 & 321 & 21.4 & 1 & \cite{gll+12} \\
0.202 & Okayama & MITSuME & \textit{Ic} & 3.93e+14 & 383 & 37 & 1 & \cite{gcn11172} \\
0.202 & Okayama & MITSuME & \textit{Rc} & 4.56e+14 & 255 & 24.6 & 1 & \cite{gcn11172} \\
0.202 & Okayama & MITSuME & \textit{g'} & 6.29e+14 & 229 & 22.1 & 1 & \cite{gcn11172} \\
0.202 & MASTER &  & \textit{R} & 4.56e+14 & 339 & 22.6 & 1 & \cite{gll+12} \\
0.205 & MASTER &  & \textit{R} & 4.56e+14 & 269 & 20.6 & 1 & \cite{gll+12} \\
0.207 & MASTER &  & \textit{R} & 4.56e+14 & 287 & 19.1 & 1 & \cite{gll+12} \\
0.21 & MASTER &  & \textit{R} & 4.56e+14 & 324 & 21.6 & 1 & \cite{gll+12} \\
0.212 & MASTER &  & \textit{R} & 4.56e+14 & 304 & 20.2 & 1 & \cite{gll+12} \\
0.214 & MASTER &  & \textit{R} & 4.56e+14 & 318 & 18.1 & 1 & \cite{gll+12} \\
0.217 & MASTER &  & \textit{R} & 4.56e+14 & 330 & 18.7 & 1 & \cite{gll+12} \\
0.219 & MASTER &  & \textit{R} & 4.56e+14 & 301 & 17.1 & 1 & \cite{gll+12} \\
0.222 & MASTER &  & \textit{R} & 4.56e+14 & 309 & 17.6 & 1 & \cite{gll+12} \\
0.223 & MASTER &  & \textit{R} & 4.56e+14 & 304 & 38.6 & 1 & \cite{gll+12} \\
0.224 & Okayama & MITSuME & \textit{Ic} & 3.93e+14 & 383 & 37 & 1 & \cite{gcn11172} \\
0.224 & Okayama & MITSuME & \textit{Rc} & 4.56e+14 & 280 & 27 & 1 & \cite{gcn11172} \\
0.224 & Okayama & MITSuME & \textit{g'} & 6.29e+14 & 191 & 18.4 & 1 & \cite{gcn11172} \\
0.224 & MASTER &  & \textit{R} & 4.56e+14 & 321 & 18.2 & 1 & \cite{gll+12} \\
0.226 & MASTER &  & \textit{I} & 3.93e+14 & 383 & 33.1 & 1 & \cite{gll+12} \\
0.226 & MASTER &  & \textit{R} & 4.56e+14 & 355 & 20.2 & 1 & \cite{gll+12} \\
0.227 & MASTER &  & \textit{R} & 4.56e+14 & 324 & 18.4 & 1 & \cite{gll+12} \\
0.228 & MASTER &  & \textit{I} & 3.93e+14 & 397 & 34.3 & 1 & \cite{gll+12} \\
0.228 & MASTER &  & \textit{R} & 4.56e+14 & 293 & 19.5 & 1 & \cite{gll+12} \\
0.229 & MASTER &  & \textit{R} & 4.56e+14 & 400 & 18.9 & 1 & \cite{gll+12} \\
0.23 & MASTER &  & \textit{I} & 3.93e+14 & 359 & 34.6 & 1 & \cite{gll+12} \\
0.23 & MASTER &  & \textit{R} & 4.56e+14 & 333 & 18.9 & 1 & \cite{gll+12} \\
0.231 & MASTER &  & \textit{R} & 4.56e+14 & 327 & 18.6 & 1 & \cite{gll+12} \\
0.233 & MASTER &  & \textit{I} & 3.93e+14 & 432 & 33 & 1 & \cite{gll+12} \\
0.233 & MASTER &  & \textit{R} & 4.56e+14 & 352 & 20 & 1 & \cite{gll+12} \\
0.233 & MASTER &  & \textit{v} & 5.55e+14 & 410 & 43.7 & 1 & \cite{gll+12} \\
0.234 & MASTER &  & \textit{R} & 4.56e+14 & 321 & 18.2 & 1 & \cite{gll+12} \\
0.235 & MASTER &  & \textit{I} & 3.93e+14 & 369 & 31.9 & 1 & \cite{gll+12} \\
0.235 & MASTER &  & \textit{R} & 4.56e+14 & 324 & 18.4 & 1 & \cite{gll+12} \\
0.235 & MASTER &  & \textit{v} & 5.55e+14 & 371 & 43.3 & 1 & \cite{gll+12} \\
0.236 & MASTER &  & \textit{R} & 4.56e+14 & 365 & 17.2 & 1 & \cite{gll+12} \\
0.237 & MASTER &  & \textit{I} & 3.93e+14 & 369 & 31.9 & 1 & \cite{gll+12} \\
0.237 & MASTER &  & \textit{R} & 4.56e+14 & 372 & 17.5 & 1 & \cite{gll+12} \\
0.239 & MASTER &  & \textit{R} & 4.56e+14 & 382 & 18 & 1 & \cite{gll+12} \\
0.24 & MASTER &  & \textit{I} & 3.93e+14 & 366 & 31.6 & 1 & \cite{gll+12} \\
0.24 & MASTER &  & \textit{R} & 4.56e+14 & 330 & 18.7 & 1 & \cite{gll+12} \\
0.241 & MASTER &  & \textit{R} & 4.56e+14 & 330 & 18.7 & 1 & \cite{gll+12} \\
0.242 & MASTER &  & \textit{I} & 3.93e+14 & 448 & 29.8 & 1 & \cite{gll+12} \\
0.242 & MASTER &  & \textit{R} & 4.56e+14 & 330 & 18.7 & 1 & \cite{gll+12} \\
0.243 & MASTER &  & \textit{R} & 4.56e+14 & 312 & 20.8 & 1 & \cite{gll+12} \\
0.245 & MASTER &  & \textit{I} & 3.93e+14 & 456 & 30.4 & 1 & \cite{gll+12} \\
0.245 & MASTER &  & \textit{R} & 4.56e+14 & 352 & 20 & 1 & \cite{gll+12} \\
0.246 & MASTER &  & \textit{R} & 4.56e+14 & 342 & 19.4 & 1 & \cite{gll+12} \\
0.247 & MASTER &  & \textit{I} & 3.93e+14 & 412 & 27.5 & 1 & \cite{gll+12} \\
0.247 & MASTER &  & \textit{R} & 4.56e+14 & 355 & 16.7 & 1 & \cite{gll+12} \\
0.248 & MASTER &  & \textit{R} & 4.56e+14 & 336 & 19.1 & 1 & \cite{gll+12} \\
0.249 & MASTER &  & \textit{I} & 3.93e+14 & 460 & 30.7 & 1 & \cite{gll+12} \\
0.249 & MASTER &  & \textit{R} & 4.56e+14 & 342 & 16.1 & 1 & \cite{gll+12} \\
0.251 & MASTER &  & \textit{R} & 4.56e+14 & 318 & 21.2 & 1 & \cite{gll+12} \\
0.252 & MASTER &  & \textit{I} & 3.93e+14 & 412 & 27.5 & 1 & \cite{gll+12} \\
0.252 & MASTER &  & \textit{R} & 4.56e+14 & 352 & 16.6 & 1 & \cite{gll+12} \\
0.253 & MASTER &  & \textit{R} & 4.56e+14 & 309 & 20.6 & 1 & \cite{gll+12} \\
0.254 & MASTER &  & \textit{I} & 3.93e+14 & 529 & 30 & 1 & \cite{gll+12} \\
0.254 & MASTER &  & \textit{R} & 4.56e+14 & 330 & 15.5 & 1 & \cite{gll+12} \\
0.256 & MASTER &  & \textit{R} & 4.56e+14 & 358 & 20.4 & 1 & \cite{gll+12} \\
0.257 & MASTER &  & \textit{I} & 3.93e+14 & 482 & 27.4 & 1 & \cite{gll+12} \\
0.257 & MASTER &  & \textit{R} & 4.56e+14 & 372 & 17.5 & 1 & \cite{gll+12} \\
0.257 & Hanle & HCT & \textit{R} & 4.56e+14 & 355 & 16.7 & 1 & \cite{gcn11175} \\
0.258 & MASTER &  & \textit{R} & 4.56e+14 & 372 & 21.1 & 1 & \cite{gll+12} \\
0.259 & MASTER &  & \textit{I} & 3.93e+14 & 444 & 25.2 & 1 & \cite{gll+12} \\
0.259 & MASTER &  & \textit{R} & 4.56e+14 & 375 & 14.1 & 1 & \cite{gll+12} \\
0.261 & MASTER &  & \textit{R} & 4.56e+14 & 431 & 20.3 & 1 & \cite{gll+12} \\
0.261 & MASTER &  & \textit{I} & 3.93e+14 & 505 & 23.8 & 1 & \cite{gll+12} \\
0.262 & Hanle & HCT & \textit{I} & 3.93e+14 & 465 & 17.4 & 1 & \cite{gcn11175} \\
0.264 & MASTER &  & \textit{I} & 3.93e+14 & 519 & 24.5 & 1 & \cite{gll+12} \\
0.266 & MASTER &  & \textit{I} & 3.93e+14 & 440 & 25 & 1 & \cite{gll+12} \\
0.269 & MASTER &  & \textit{I} & 3.93e+14 & 496 & 28.2 & 1 & \cite{gll+12} \\
0.271 & MASTER &  & \textit{I} & 3.93e+14 & 491 & 27.9 & 1 & \cite{gll+12} \\
0.273 & MASTER &  & \textit{I} & 3.93e+14 & 482 & 27.4 & 1 & \cite{gll+12} \\
0.274 & Terksol & Z-600 & \textit{R} & 4.56e+14 & 349 & 6.48 & 1 & \cite{gcn11168} \\
0.276 & MASTER &  & \textit{I} & 3.93e+14 & 452 & 25.7 & 1 & \cite{gll+12} \\
0.278 & MASTER &  & \textit{I} & 3.93e+14 & 469 & 26.6 & 1 & \cite{gll+12} \\
0.281 & MASTER &  & \textit{I} & 3.93e+14 & 405 & 27 & 1 & \cite{gll+12} \\
0.283 & MASTER &  & \textit{I} & 3.93e+14 & 448 & 25.4 & 1 & \cite{gll+12} \\
0.286 & MASTER &  & \textit{I} & 3.93e+14 & 534 & 30.3 & 1 & \cite{gll+12} \\
0.288 & MASTER &  & \textit{I} & 3.93e+14 & 448 & 29.8 & 1 & \cite{gll+12} \\
0.29 & MASTER &  & \textit{I} & 3.93e+14 & 482 & 27.4 & 1 & \cite{gll+12} \\
0.293 & MASTER &  & \textit{I} & 3.93e+14 & 514 & 29.2 & 1 & \cite{gll+12} \\
0.295 & MASTER &  & \textit{I} & 3.93e+14 & 376 & 28.8 & 1 & \cite{gll+12} \\
0.296 & MASTER &  & \textit{v} & 5.55e+14 & 303 & 54.6 & 1 & \cite{gll+12} \\
0.297 & MASTER &  & \textit{I} & 3.93e+14 & 432 & 28.8 & 1 & \cite{gll+12} \\
0.299 & MASTER &  & \textit{v} & 5.55e+14 & 341 & 50.5 & 1 & \cite{gll+12} \\
0.3 & MASTER &  & \textit{I} & 3.93e+14 & 465 & 26.4 & 1 & \cite{gll+12} \\
0.302 & Golovec & Vega & \textit{R} & 4.56e+14 & 306 & 29.6 & 1 & \cite{gcn11177} \\
0.302 & MASTER &  & \textit{I} & 3.93e+14 & 416 & 31.8 & 1 & \cite{gll+12} \\
0.305 & MASTER &  & \textit{I} & 3.93e+14 & 452 & 34.6 & 1 & \cite{gll+12} \\
0.307 & MASTER &  & \textit{I} & 3.93e+14 & 369 & 35.6 & 1 & \cite{gll+12} \\
0.317 & THO & C-14 & \textit{CR} & 4.4e+14 & 333 & 32.1 & 1 & \cite{gcn11173} \\
0.32 & THO & C-14 & \textit{CR} & 4.4e+14 & 304 & 61.4 & 1 & \cite{gcn11173} \\
0.353 & Terksol & Meade-35 & \textit{v} & 5.55e+14 & 338 & 15.9 & 1 & \cite{gcn11200} \\
0.355 & Hanle & HCT & \textit{R} & 4.56e+14 & 339 & 12.7 & 1 & \cite{gcn11175} \\
0.37 & Terksol & Zeiss-2000 & \textit{b} & 6.92e+14 & 232 & 4.31 & 1 & \cite{gcn11200} \\
0.416 & Gnosca & 0.4m & \textit{CR} & 4.4e+14 & 304 & 96.6 & 1 & \cite{gcn11213} \\
0.427 & Gnosca & 0.4m & \textit{CR} & 4.4e+14 & 285 & 90.6 & 1 & \cite{gcn11213} \\
0.439 & Gnosca & 0.4m & \textit{CR} & 4.4e+14 & 287 & 91.4 & 1 & \cite{gcn11213} \\
0.442 & THO & C-14 & \textit{CR} & 4.4e+14 & 277 & 26.7 & 1 & \cite{gcn11173} \\
0.449 & Terksol & Zeiss-2000 & \textit{b} & 6.92e+14 & 169 & 3.15 & 1 & \cite{gcn11200} \\
0.45 & Gnosca & 0.4m & \textit{CR} & 4.4e+14 & 277 & 88.1 & 1 & \cite{gcn11213} \\
0.458 & Terksol & Zeiss-2000 & \textit{v} & 5.55e+14 & 266 & 4.94 & 1 & \cite{gcn11200} \\
0.461 & Gnosca & 0.4m & \textit{CR} & 4.4e+14 & 246 & 78.2 & 1 & \cite{gcn11213} \\
0.464 & Terksol & Zeiss-2000 & \textit{r'} & 4.56e+14 & 263 & 4.89 & 1 & \cite{gcn11200} \\
0.464 & THO & C-14 & \textit{Rc} & 4.56e+14 & 253 & 24.4 & 1 & \cite{gcn11173} \\
0.468 & THO & C-14 & \textit{b} & 6.92e+14 & 234 & 47.3 & 1 & \cite{gcn11173} \\
0.471 & Terksol & Zeiss-2000 & \textit{g'} & 6.29e+14 & 163 & 3.03 & 1 & \cite{gcn11200} \\
0.472 & Gnosca & 0.4m & \textit{CR} & 4.4e+14 & 243 & 77.5 & 1 & \cite{gcn11213} \\
0.483 & Gnosca & 0.4m & \textit{CR} & 4.4e+14 & 267 & 85 & 1 & \cite{gcn11213} \\
0.494 & Gnosca & 0.4m & \textit{CR} & 4.4e+14 & 253 & 80.4 & 1 & \cite{gcn11213} \\
0.505 & Gnosca & 0.4m & \textit{CR} & 4.4e+14 & 246 & 78.2 & 1 & \cite{gcn11213} \\
0.505 & NOT & ALFOSC & \textit{R} & 4.56e+14 & 277 & 88.1 & 1 & \cite{gcn11170} \\
0.516 & Gnosca & 0.4m & \textit{CR} & 4.4e+14 & 232 & 74 & 1 & \cite{gcn11213} \\
0.527 & Gnosca & 0.4m & \textit{CR} & 4.4e+14 & 210 & 66.9 & 1 & \cite{gcn11213} \\
0.539 & Gnosca & 0.4m & \textit{CR} & 4.4e+14 & 218 & 69.4 & 1 & \cite{gcn11213} \\
0.551 & Gnosca & 0.4m & \textit{CR} & 4.4e+14 & 216 & 68.7 & 1 & \cite{gcn11213} \\
0.562 & Gnosca & 0.4m & \textit{CR} & 4.4e+14 & 218 & 69.4 & 1 & \cite{gcn11213} \\
0.574 & Gnosca & 0.4m & \textit{CR} & 4.4e+14 & 206 & 65.6 & 1 & \cite{gcn11213} \\
0.585 & Gnosca & 0.4m & \textit{CR} & 4.4e+14 & 195 & 62.1 & 1 & \cite{gcn11213} \\
0.782 & KPNO & SARA-N & \textit{b} & 6.92e+14 & 135 & 13 & 1 & \cite{gcn11174} \\
0.785 & KPNO & SARA-N & \textit{v} & 5.55e+14 & 234 & 22.6 & 1 & \cite{gcn11174} \\
0.787 & KPNO & SARA-N & \textit{R} & 4.56e+14 & 145 & 14 & 1 & \cite{gcn11174} \\
0.79 & KPNO & SARA-N & \textit{I} & 3.93e+14 & 148 & 14.3 & 1 & \cite{gcn11174} \\
0.856 & ISON-NM & 0.45m & \textit{CR} & 4.4e+14 & 147 & 4.11 & 1 & \cite{gcn11184} \\
1.05 & Ishigakijima & Murikabushi & \textit{Ic} & 3.93e+14 & 121 & 5.71 & 1 & \cite{gcn11205} \\
1.05 & Ishigakijima & Murikabushi & \textit{Rc} & 4.56e+14 & 87.6 & 3.29 & 1 & \cite{gcn11205} \\
1.05 & Ishigakijima & Murikabushi & \textit{g'} & 6.29e+14 & 58.1 & 2.18 & 1 & \cite{gcn11205} \\
1.07 & Hiroshima & Kanata & \textit{Rc} & 4.56e+14 & 82.1 & 7.92 & 1 & \cite{gcn11190} \\
1.08 & Hiroshima & Kanata & \textit{Rc} & 4.56e+14 & 83.6 & 5.57 & 1 & \cite{gcn11190} \\
1.09 & Ishigakijima & Murikabushi & \textit{Ic} & 3.93e+14 & 105 & 5.94 & 1 & \cite{gcn11205} \\
1.09 & Ishigakijima & Murikabushi & \textit{Rc} & 4.56e+14 & 82.1 & 3.08 & 1 & \cite{gcn11205} \\
1.09 & Ishigakijima & Murikabushi & \textit{g'} & 6.29e+14 & 53 & 1.99 & 1 & \cite{gcn11205} \\
1.1 & Hiroshima & Kanata & \textit{Rc} & 4.56e+14 & 88.4 & 3.32 & 1 & \cite{gcn11190} \\
1.14 & Ishigakijima & Murikabushi & \textit{Ic} & 3.93e+14 & 96.2 & 6.41 & 1 & \cite{gcn11205} \\
1.14 & Ishigakijima & Murikabushi & \textit{Rc} & 4.56e+14 & 69.6 & 2.61 & 1 & \cite{gcn11205} \\
1.14 & Ishigakijima & Murikabushi & \textit{g'} & 6.29e+14 & 50.6 & 2.38 & 1 & \cite{gcn11205} \\
1.18 & Okayama & MITSuME & \textit{Ic} & 3.93e+14 & 96.2 & 9.28 & 1 & \cite{gcn11189} \\
1.18 & Okayama & MITSuME & \textit{Rc} & 4.56e+14 & 77 & 7.43 & 1 & \cite{gcn11189} \\
1.18 & Okayama & MITSuME & \textit{g'} & 6.29e+14 & 63.1 & 6.09 & 1 & \cite{gcn11189} \\
1.22 & Hanle & HCT & \textit{R} & 4.56e+14 & 75.6 & 7.29 & 1 & \cite{gcn11197} \\
1.24 & Hanle & HCT & \textit{I} & 3.93e+14 & 92.7 & 13.7 & 1 & \cite{gcn11197} \\
1.3 & MASTER &  & \textit{I} & 3.93e+14 & 90.2 & 3.39 & 1 & \cite{gll+12} \\
1.38 & Tautenburg & TLS1.34m & \textit{Rc} & 4.56e+14 & 66.4 & 1.86 & 1 & \cite{gcn11236} \\
1.39 & Tautenburg & TLS1.34m & \textit{Rc} & 4.56e+14 & 61.7 & 2.32 & 1 & \cite{gcn11236} \\
1.42 & Tautenburg & TLS1.34m & \textit{Rc} & 4.56e+14 & 59.5 & 1.67 & 1 & \cite{gcn11236} \\
1.43 & Tautenburg & TLS1.34m & \textit{Rc} & 4.56e+14 & 55.3 & 1.55 & 1 & \cite{gcn11236} \\
1.43 & Tautenburg & TLS1.34m & \textit{Rc} & 4.56e+14 & 61.2 & 2.29 & 1 & \cite{gcn11236} \\
1.44 & Tautenburg & TLS1.34m & \textit{Rc} & 4.56e+14 & 65.2 & 2.45 & 1 & \cite{gcn11236} \\
1.46 & Terksol & Z-600 & \textit{R} & 4.56e+14 & 62.9 & 2.96 & 1 & \cite{gcn11191} \\
1.7 & Lightbuckets & 0.61m & \textit{R} & 4.56e+14 & 46.8 & 14.9 & 0 & \cite{gcn11198} \\
2.13 & Ishigakijima & Murikabushi & \textit{Ic} & 3.93e+14 & 23.3 & 3.95 & 1 & \cite{gcn11205} \\
2.13 & Ishigakijima & Murikabushi & \textit{Rc} & 4.56e+14 & 31.8 & 1.81 & 1 & \cite{gcn11205} \\
2.13 & Ishigakijima & Murikabushi & \textit{g'} & 6.29e+14 & 18.9 & 1.44 & 1 & \cite{gcn11205} \\
2.3 & CrAO & AZT-11 & \textit{R} & 4.56e+14 & 28.7 & 2.2 & 1 & \cite{gcn11255} \\
2.41 & Terksol & Z-600 & \textit{R} & 4.56e+14 & 27.2 & 3.18 & 1 & \cite{gcn11201} \\
2.43 & Terksol & Z-600 & \textit{R} & 4.56e+14 & 23.7 & 3.76 & 1 & \cite{gcn11201} \\
2.83 & ISON-NM & 0.45m & \textit{CR} & 4.4e+14 & 24.1 & 6.25 & 1 & \cite{gcn11234} \\
2.88 & UKIRT &  & \textit{K} & 1.37e+14 & 47.6 & 9.64 & 1 & \cite{gcn11208} \\
3.31 & Tautenburg & TLS1.34m & \textit{Rc} & 4.56e+14 & 18.8 & 1.25 & 1 & \cite{gcn11236} \\
3.55 & Tautenburg & TLS1.34m & \textit{Rc} & 4.56e+14 & 16.1 & 0.604 & 1 & \cite{gcn11236} \\
3.82 & ISON-NM & 0.45m & \textit{CR} & 4.4e+14 & 13.9 & 2.81 & 1 & \cite{gcn11234} \\
4.26 & Ishigakijima & Murikabushi & \textit{Ic} & 3.93e+14 & 29.1 & 9.25 & 0 & \cite{gcn11258} \\
4.26 & Ishigakijima & Murikabushi & \textit{Rc} & 4.56e+14 & 9.26 & 2.95 & 0 & \cite{gcn11258} \\
4.26 & Ishigakijima & Murikabushi & \textit{g'} & 6.29e+14 & 8.32 & 2.65 & 0 & \cite{gcn11258} \\
4.4 & Tautenburg & TLS1.34m & \textit{Rc} & 4.56e+14 & 9.26 & 0.436 & 1 & \cite{gcn11236} \\
4.4 & CrAO & AZT-11 & \textit{R} & 4.56e+14 & 11.1 & 2.25 & 1 & \cite{gcn11255} \\
4.42 & Tautenburg & TLS1.34m & \textit{Rc} & 4.56e+14 & 9.17 & 2.48 & 1 & \cite{gcn11236} \\
4.43 & Tautenburg & TLS1.34m & \textit{Rc} & 4.56e+14 & 8.68 & 1.1 & 1 & \cite{gcn11236} \\
4.56 & Tautenburg & TLS1.34m & \textit{Rc} & 4.56e+14 & 9.51 & 0.918 & 1 & \cite{gcn11236} \\
4.87 & ISON-NM & 0.45m & \textit{CR} & 4.4e+14 & 8.06 & 1.81 & 1 & \cite{gcn11234} \\
5.55 & Tautenburg & TLS1.34m & \textit{Rc} & 4.56e+14 & 5.28 & 0.563 & 1 & \cite{gcn11246} \\
6.3 & Maidanak & AZT-22 & \textit{R} & 4.56e+14 & 4.64 & 0.401 & 1 & \cite{gcn11266}
\enddata
\tablecomments{$^{\dag}$In cases of non-detections, we report the formal flux measurement at the 
position of the afterglow.}
\end{deluxetable*}

\begin{figure}
 \centering
 \includegraphics[width=\columnwidth]{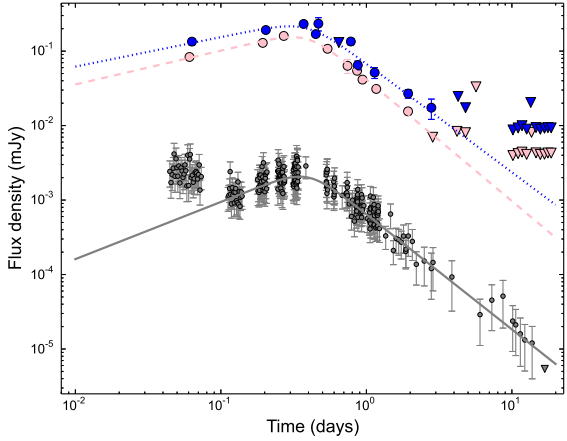}
 \caption{Broken power law fit to the X-ray (grey, solid), \Swift/UVOT $U$-band (pink, dashed), and 
\Swift/UVOT $B$-band (blue, dotted) light curves for GRB~100901A near the re-brightening around 
0.6\,d. The X-ray fit includes points between 0.08 and 20\,d, while the $U$- and $B$-band fits
include points between $0.02$\,d and 20\,d. A joint fit yields the best fit parameters $t_{\rm b} = 
0.36\pm0.02$, $\alpha_1 = 0.57\pm0.09$, and $\alpha_2 = -1.47\pm0.07$ (Section 
\ref{text:100901A:basic_considerations}).
\label{fig:100901A_bplfit}}
\end{figure}

The X-ray data before the first orbital gap at around $10^{-2}$\,d exhibits rapid flaring, ending 
in a steep decay. The earliest UVOT observations during this period detect a counterpart at 
multiple wavelengths. Owing to the contemporaneous $\gamma$-ray emission, indicating on-going 
central engine activity during this period, we do not consider data before the first orbital gap 
for 
our afterglow analysis. The X-ray re-brightening and subsequent fading between 0.08\,d and 20\,d 
can 
be well fit with a broken power law model (Equation \ref{eqn:bpl}), with $t_{\rm 
b}=0.40\pm0.02$\,d, 
$F_{\nu, \rm X}(t_{\rm b}) = 2.0\pm0.1\,\mu$Jy, $\alpha_1=0.77\pm0.15$, $\alpha_2 = -1.55\pm0.06$, 
and $y=2.3\pm1.1$  (Figure \ref{fig:100901A_bplfit}). 

A broken power law fit to the UVOT $U$-band data between $0.02$\,d and 20\,d yields $t_{\rm 
b}=0.40\pm0.03$\,d, $F_{\nu, \rm U}(t_{\rm b}) = 144\pm7\,\mu$Jy, $\alpha_1=0.45\pm0.06$, and 
$\alpha_2 = -1.63\pm0.12$, with $y=2.5$, similar to the results from fitting the X-ray 
re-brightening. Over the same time range, a broken power law fit to the UVOT $B$-band 
data yields $t_{\rm b}=0.39\pm0.08$\,d, $F_{\nu, \rm B}(t_{\rm b}) = 201\pm24\,\mu$Jy, 
$\alpha_1=0.40\pm0.06$, and $\alpha_2 = -1.46\pm0.21$, similar to the X-ray and $U$-band fits. 
Finally, we fit the X-ray, UVOT $U$-band and UVOT $B$-band data jointly, where we 
constrain the model light curves at the two frequencies to have the same rise and decay rate and 
time of peak, and use independent normalizations in the three bands. Using a Markov Chain Monte 
Carlo simulation using \emcee\ \citep{fhlg13}, we find $t_{\rm b} = 0.36\pm0.02$, $\alpha_1 = 
0.57\pm0.09$, and $\alpha_2 = -1.47\pm0.07$, with a fixed value of $y=2.5$.

The spectral index between the UKIRT $K$-band observation at 2.88\,d \citep{gcn11208} and the
X-rays is $\beta_{\rm NIR, X} = -0.78\pm0.08$. The spectral index between the NIR $K$-band and the
\Swift/$U$-band is slightly steeper, $\beta_{\rm NIR, UV} = -1.0\pm0.2$, indicating that some 
extinction may be present. The spectral slope in the X-rays following the re-brightening is 
$\beta_{\rm X}=-1.15\pm0.06$, leading to $\beta_{\rm NIR, X}-\beta_{\rm X}=0.4\pm0.1$ 
suggesting that $\nuopt < \nuc < \nuX$ at 2.88\,d.

If the X-rays indeed lie above \nuc, we would have $p=-2\beta_{\rm X} = 2.3\pm0.12$. The expected
light curve decay rate is $\alpha = (2-3p)/4=-1.2\pm0.1$ above \nuc\ and
$\alpha = 3(1-p)/4 = -1.0\pm0.1$ (ISM) or $\alpha = -1.5\pm0.1$ (wind) below \nuc.
The $U$-band light curve after the re-brightening can be fit either with a single power law with 
$\alpha = -1.5$ or with a break from $\alpha \approx -0.9$ to $\alpha \approx -1.8$ at 
$\approx1$\,d and $\alpha = -1.0\pm0.1$ between 0.2\,d and 1.0\,d. The former is consistent with 
a wind-like environment, while the latter suggests an ISM model and a jet break at 
$\approx1$\,d. We return to this point later.

Extrapolating the \textit{Lightbuckets} $R$-band data at 1.7\,d to the time of the SMA 345\,GHz 
upper limit at 1.8 days using the fit to the optical $U$-band light curve described above, we find 
a spectral index of $\beta_{\rm mm-R} \gtrsim -0.6$ from the millimeter to the optical at this 
time. 
Combined with the steeper spectrum of $\beta_{\rm NIR,X}=-0.78\pm0.08$ between the optical and 
X-rays, this suggests that the spectrum turns over above the millimeter band, indicating that 
$\numax \gtrsim 345$\,GHz at 1.8\,d.

The spectral index between the VLA 4.5\,GHz and 7.9\,GHz bands at 4.92\,d is $\beta=0.9\pm0.3$.
However, the light curve at 4.5\,GHz declines
as $\alpha = -0.17\pm0.12$ between 3\,d and 12\,d, and as $\alpha = -1.1\pm0.5$ between 12\,d and 
20\,d. The rising radio spectrum coupled with the declining light curve implies that the jet break 
has occurred before 5\,d. The steepening of the 4.5\,GHz light curve at $\approx12$\,d implies 
$\numax\ \approx4.5$\,GHz at this time. Given that $\numax>345$\,GHz at 1.8\,d, this would imply an 
evolution of $\sim \numax \propto t^{-2.3}$, consistent with the expected evolution 
of $\numax\propto t^{-2}$ following the jet break. Together, this implies that the jet break must 
have occurred before 1.8\,d, which agrees with the results from the X-ray and optical analysis 
above 
and indicates an ISM-like environment. We therefore focus on the ISM model for the remainder of 
this 
section, and discuss the wind model for completeness in Appendix \ref{appendix:100901A_wind}.

To summarize, the NIR to X-ray SED exhibits mild evidence for extinction and suggests that 
$\nuNIR < \nuc < \nuX$ at $\approx 3\,$d, with $p=2.3\pm0.1$. The radio observations indicate 
$\numax\approx5\,$GHz at $\approx12\,$d. Together, the X-ray, NIR and radio data suggest 
$\tjet\approx1$\,d and an ISM-like environment. All radio observations took place after this 
time, and are therefore insensitive to the density profile of the circumburst environment.
 
\begin{figure}
 \centering
 \includegraphics[width=\columnwidth]{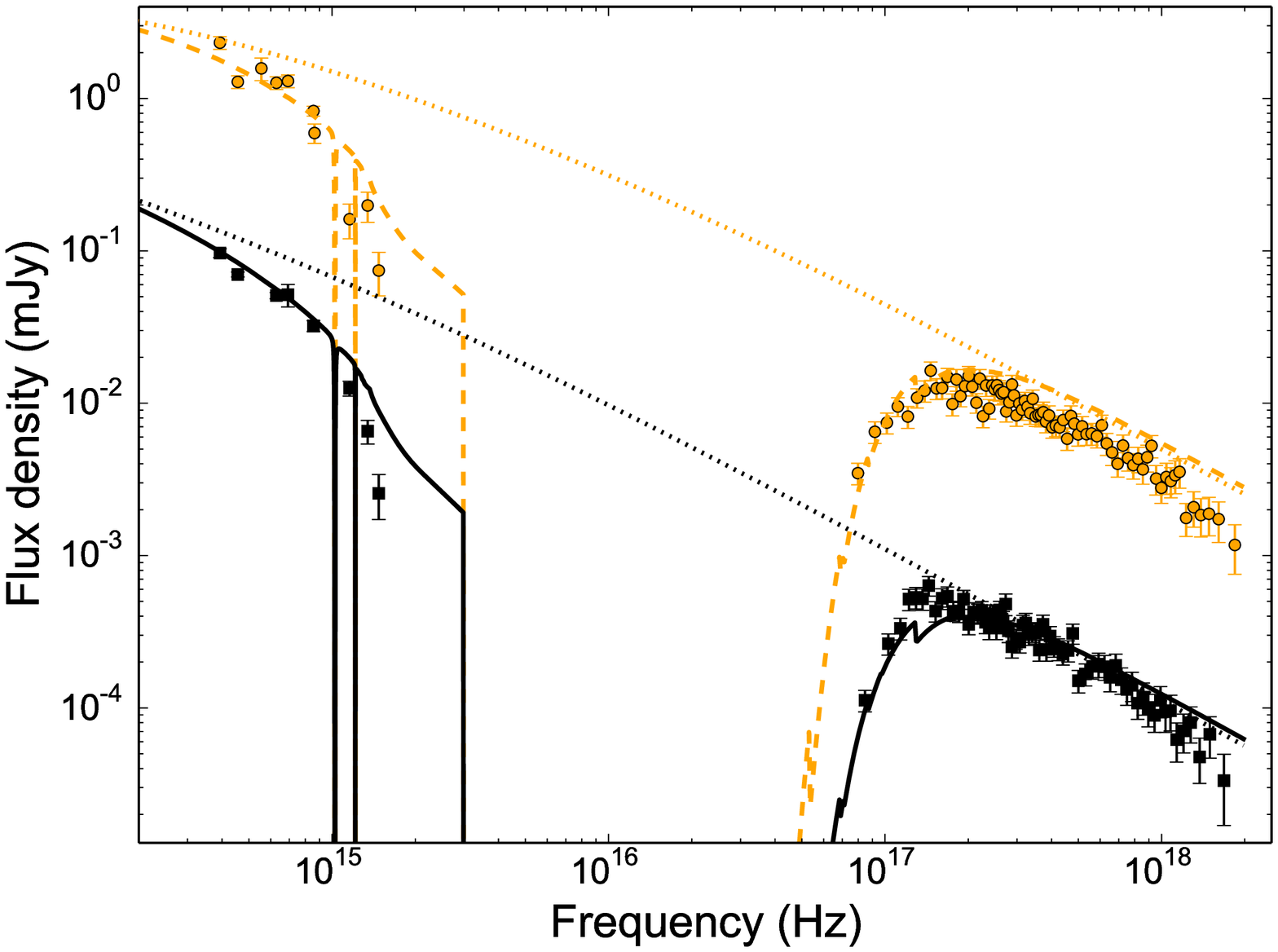}
\caption{Observed NIR to X-ray SED of the afterglow of GRB~100901A at 0.06\,d (orange circles) and 
after the re-brightening (1.14\,d; black squares), together with the best-fit ISM model (Section 
\ref{text:100901A:FS}) including energy injection (Section \ref{text:100901A:enj}) at 0.06\,d 
(orange; dashed) and 1.14\,d (black; solid). The data points and model at 
0.06\,d have been multiplied by a factor of 10. for clarity. 
The XRT SEDs at 1.14\,d has been extrapolated from 3\,d using the best-fit broken power law 
model to the XRT light curve (Figure \ref{fig:100901A_bplfit}; the correction factor is 
$\approx4.5$). The optical data have been extrapolated using the joint best fit to the 
\Swift/XRT and UVOT light curves (the corrections are small at $\lesssim5\%$).
The dotted curves indicate afterglow models with no extinction or IGM absorption. Note that 
the \Swift/UVOT $uvw1$-, $uvw2$-, and $uvm2$-band data lie blueward of Ly-$\alpha$ in the 
rest-frame of the host galaxy ($1.0\times10^{15}$\,Hz in the observer frame), and are likely 
subject to significant IGM absorption. For this reason, we do not include these bands (or the 
\Swift/UVOT \textit{White}-band, not shown here) to constrain our afterglow model.
\label{fig:100901A_sed}}
\end{figure}
 
\subsubsection{Forward shock model at $t \gtrsim 0.25$\,d}
\label{text:100901A:FS}
\begin{figure*}
\begin{tabular}{cc}
 \centering
 \includegraphics[width=0.47\textwidth]{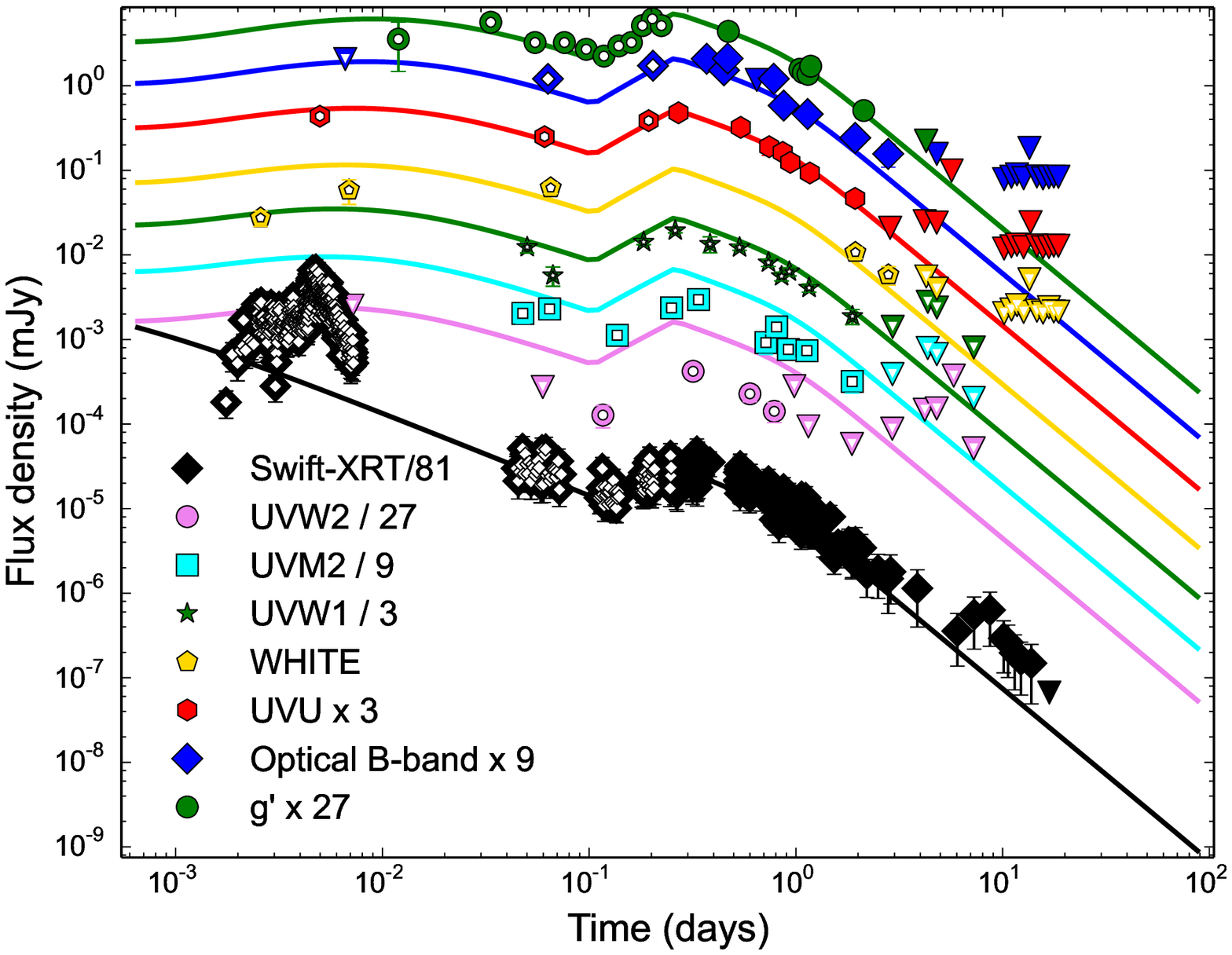} &
 \includegraphics[width=0.47\textwidth]{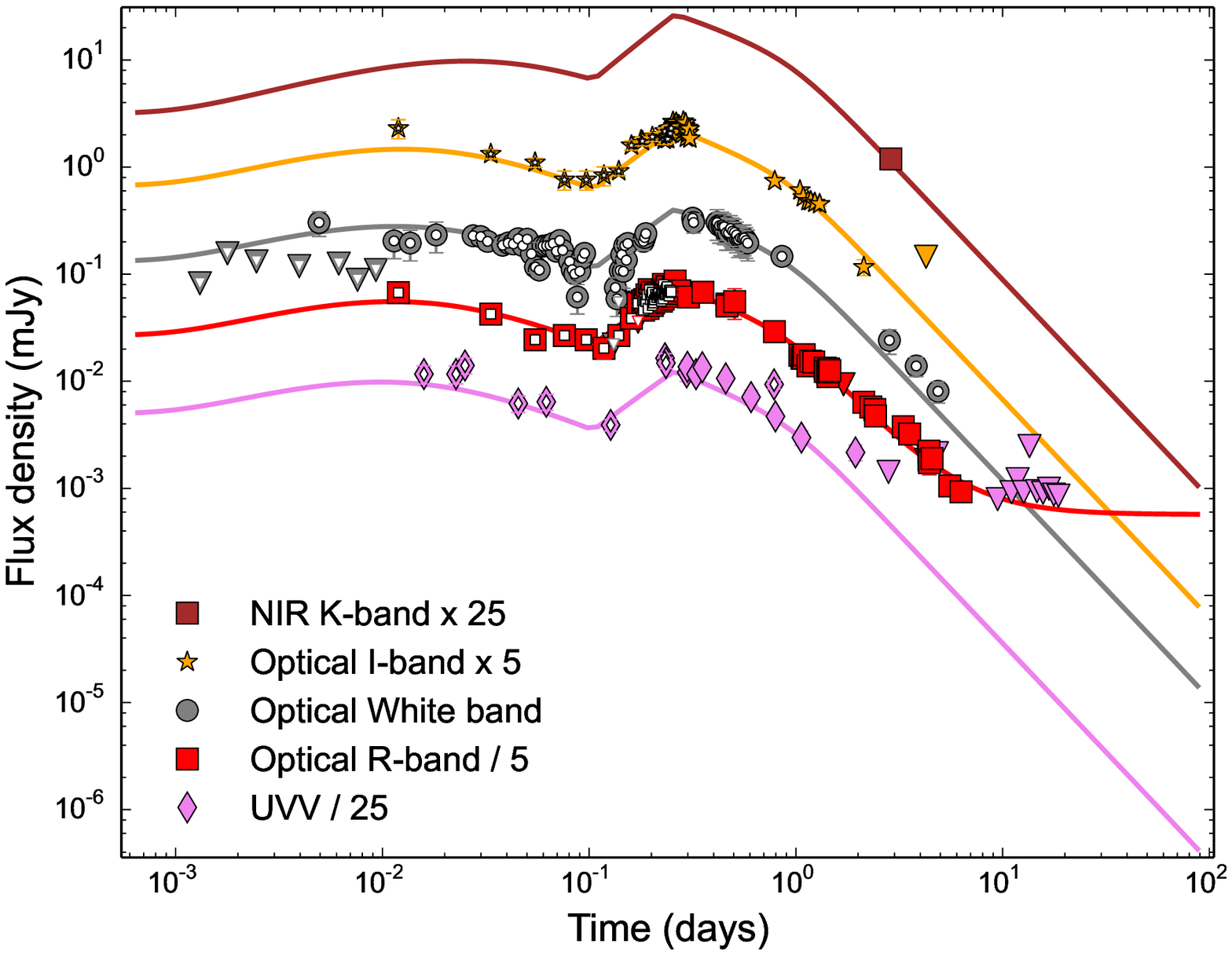} \\
 \includegraphics[width=0.47\textwidth]{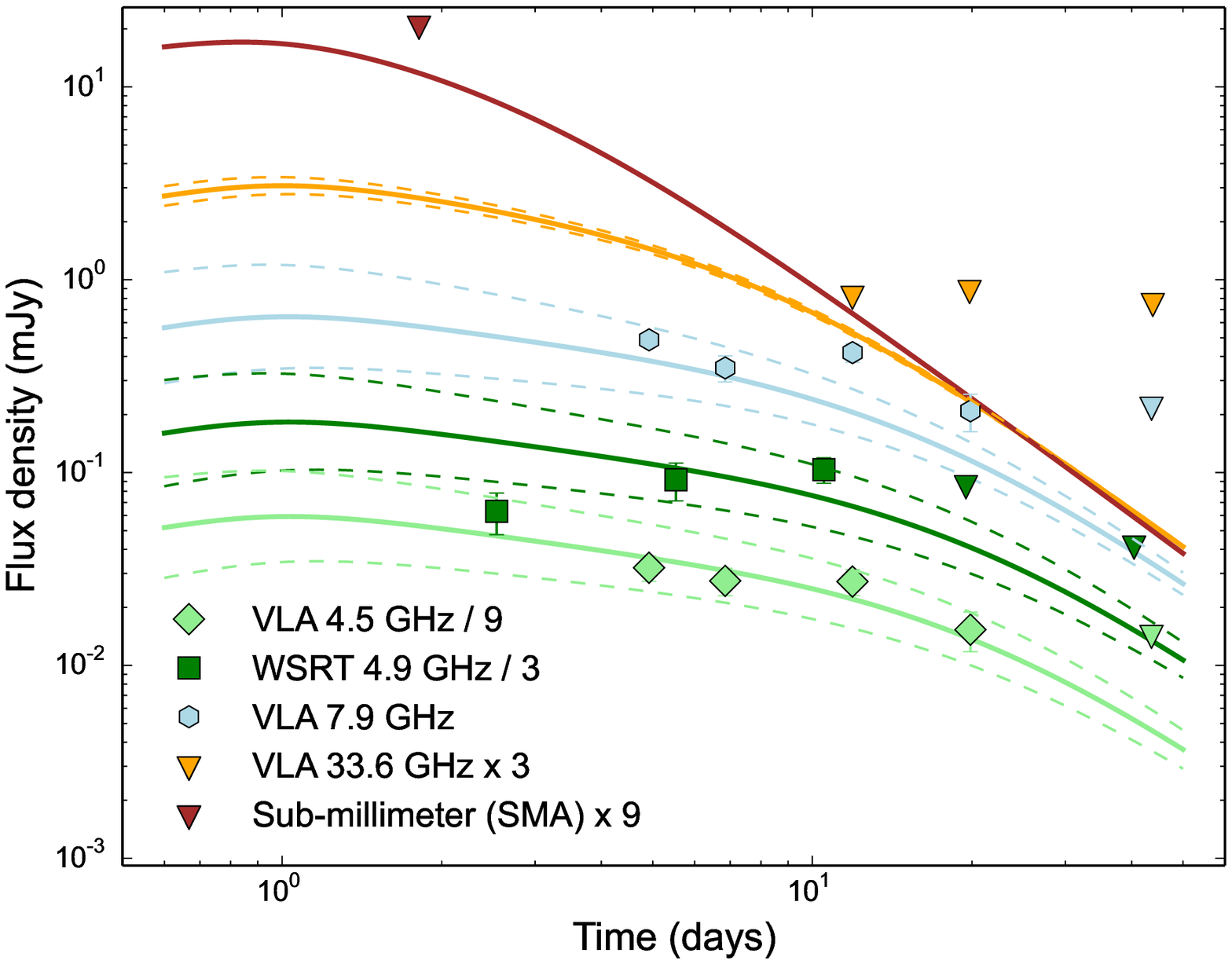} & 
 \includegraphics[width=0.47\textwidth]{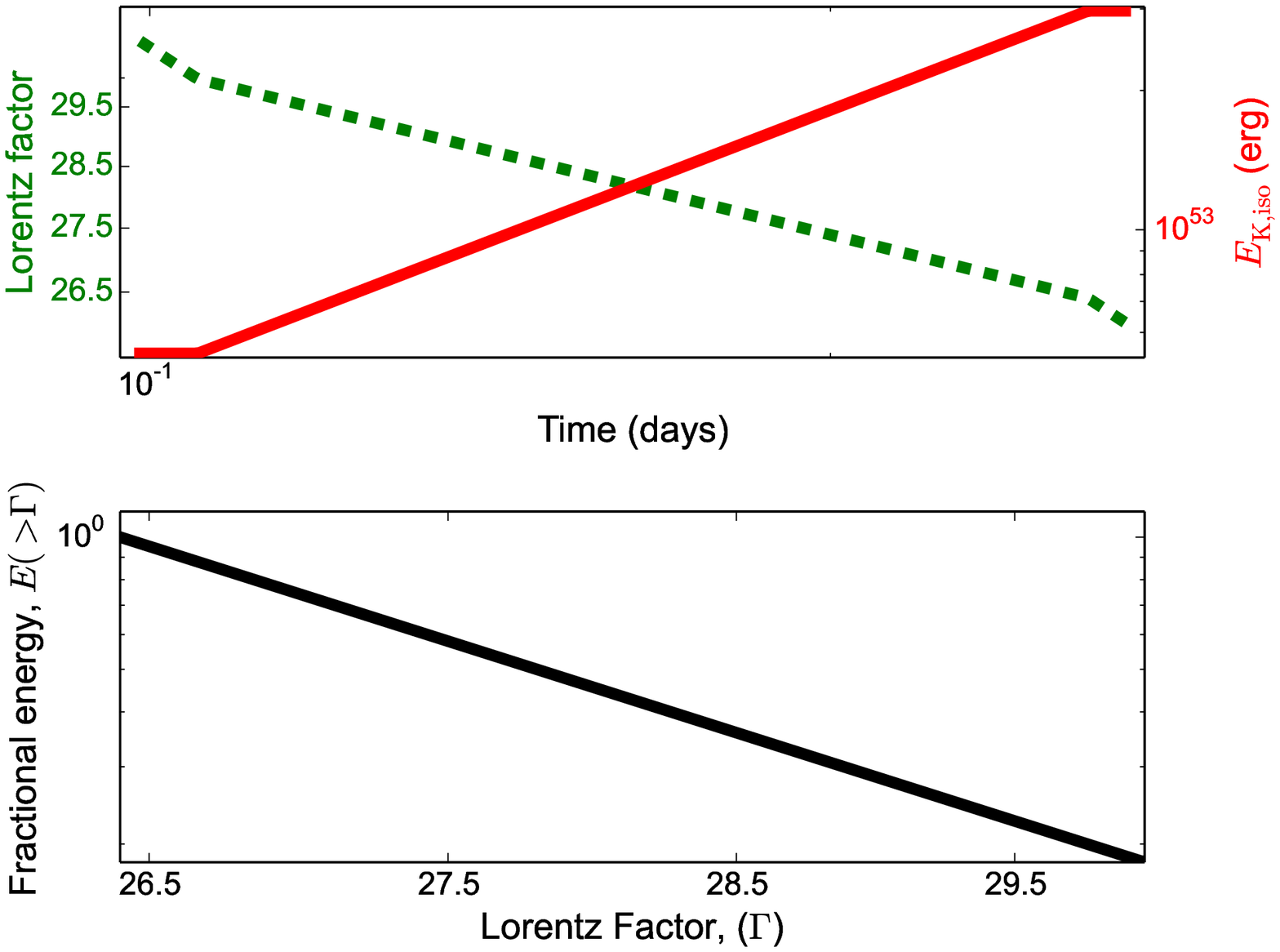} \\ 
\end{tabular}
\caption{X-ray, UV (top left), optical (top right), and radio (bottom left) light curves of 
GRB~100901A, with the full afterglow model (solid lines), including energy injection before 
0.25\,d. The X-ray data before 0.008\,d exhibits strong flaring activity and we do not include 
these data in our analysis. The dashed envelopes around the radio light curves indicate the 
expected effect of scintillation at the $1\sigma$ level. The data indicated by open symbols are not 
used to determine the parameters of the forward shock (the MCMC analysis). Bottom right: blastwave 
Lorentz factor (green, dashed; upper sub-panel) and isotropic equivalent kinetic energy (red, 
solid; 
upper sub-panel) as a function of time, together with the energy distribution across ejecta Lorentz 
factors (black, solid; lower sub-panel) as determined from fitting the X-ray/UV/optical 
re-brightening at 0.36\,d. \label{fig:100901A_enj}}
\end{figure*}

\begin{figure}
\begin{tabular}{ccc}
 \centering
 \includegraphics[width=0.30\columnwidth]{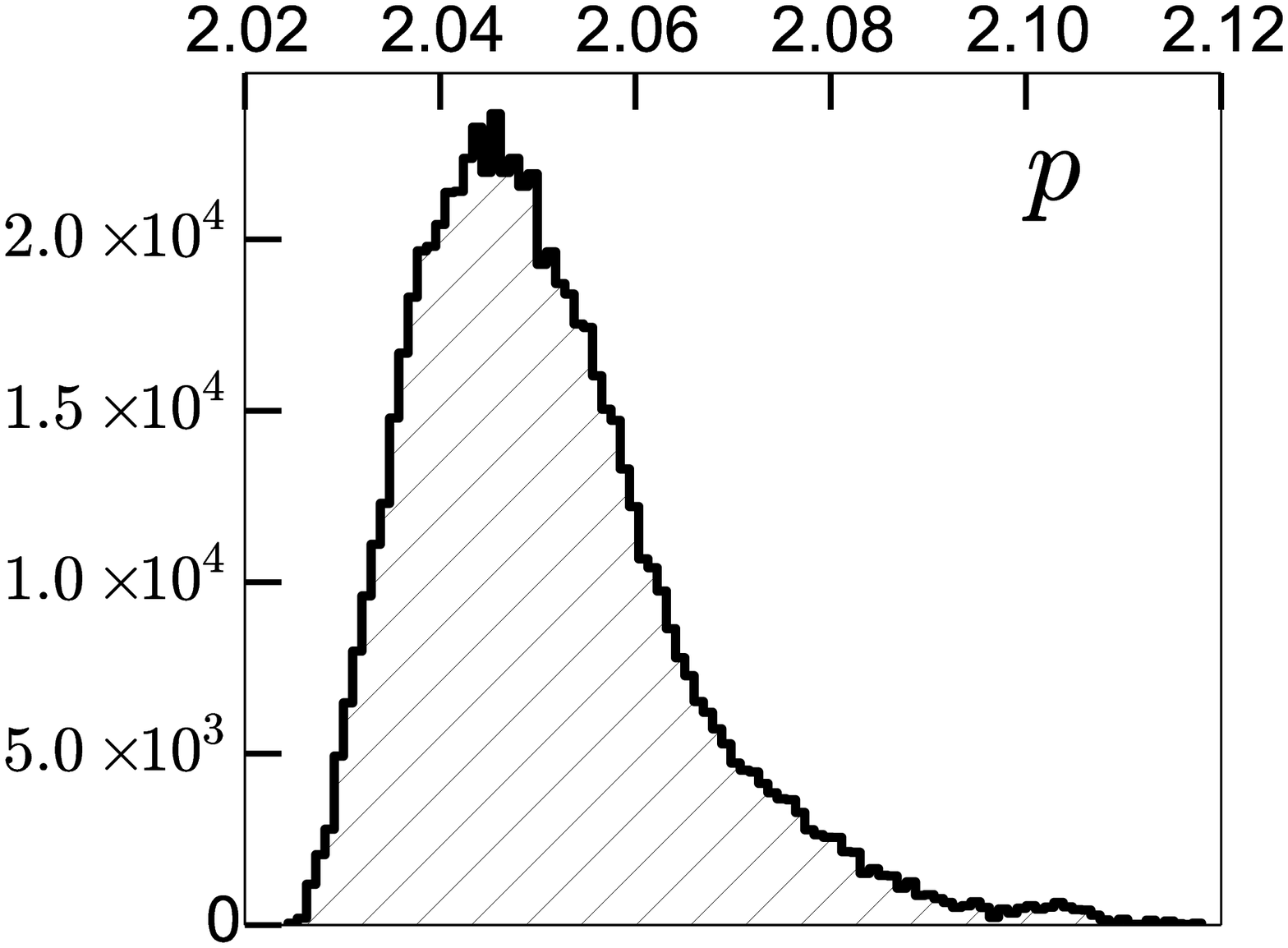} &
 \includegraphics[width=0.30\columnwidth]{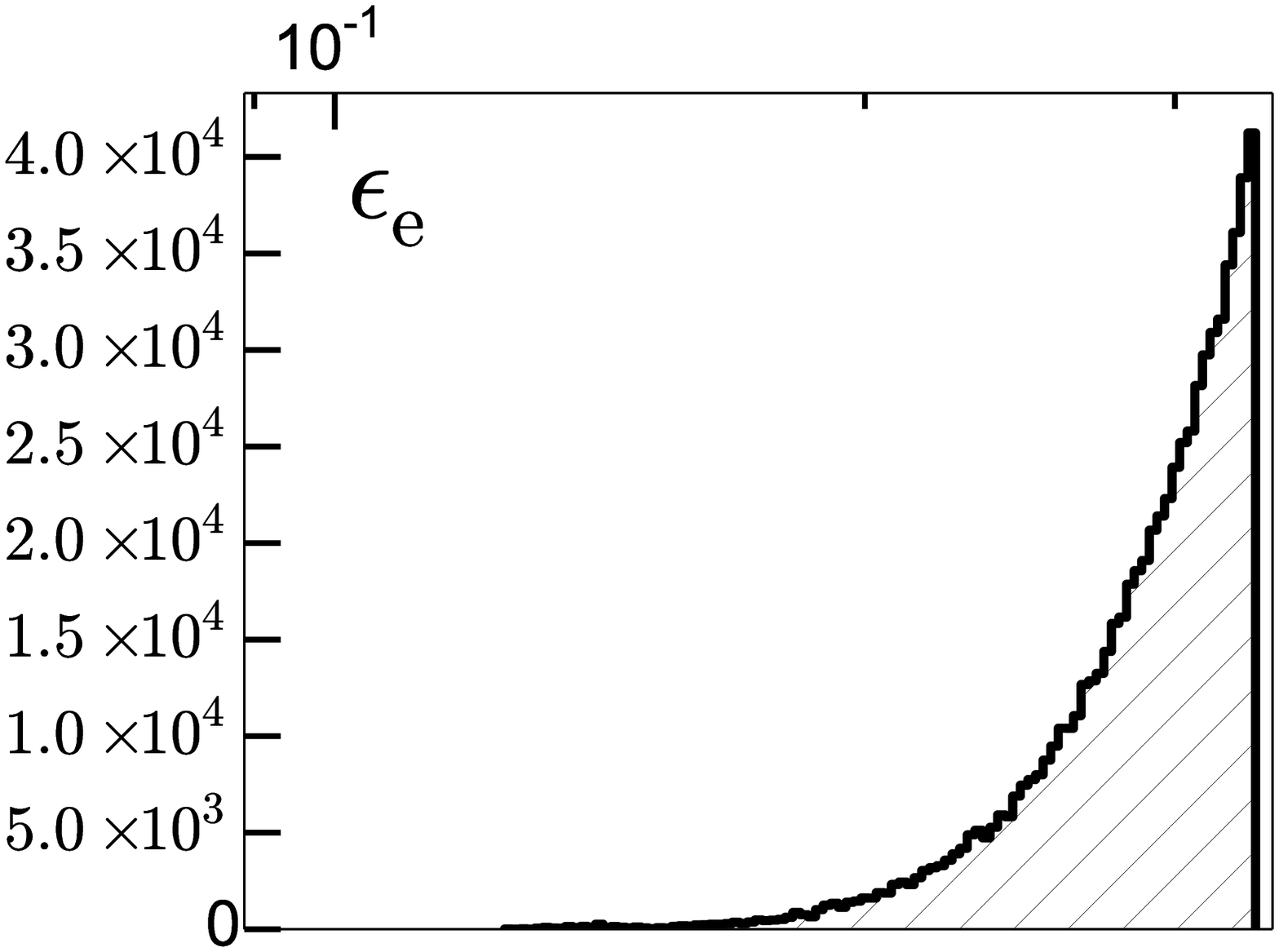} &
 \includegraphics[width=0.30\columnwidth]{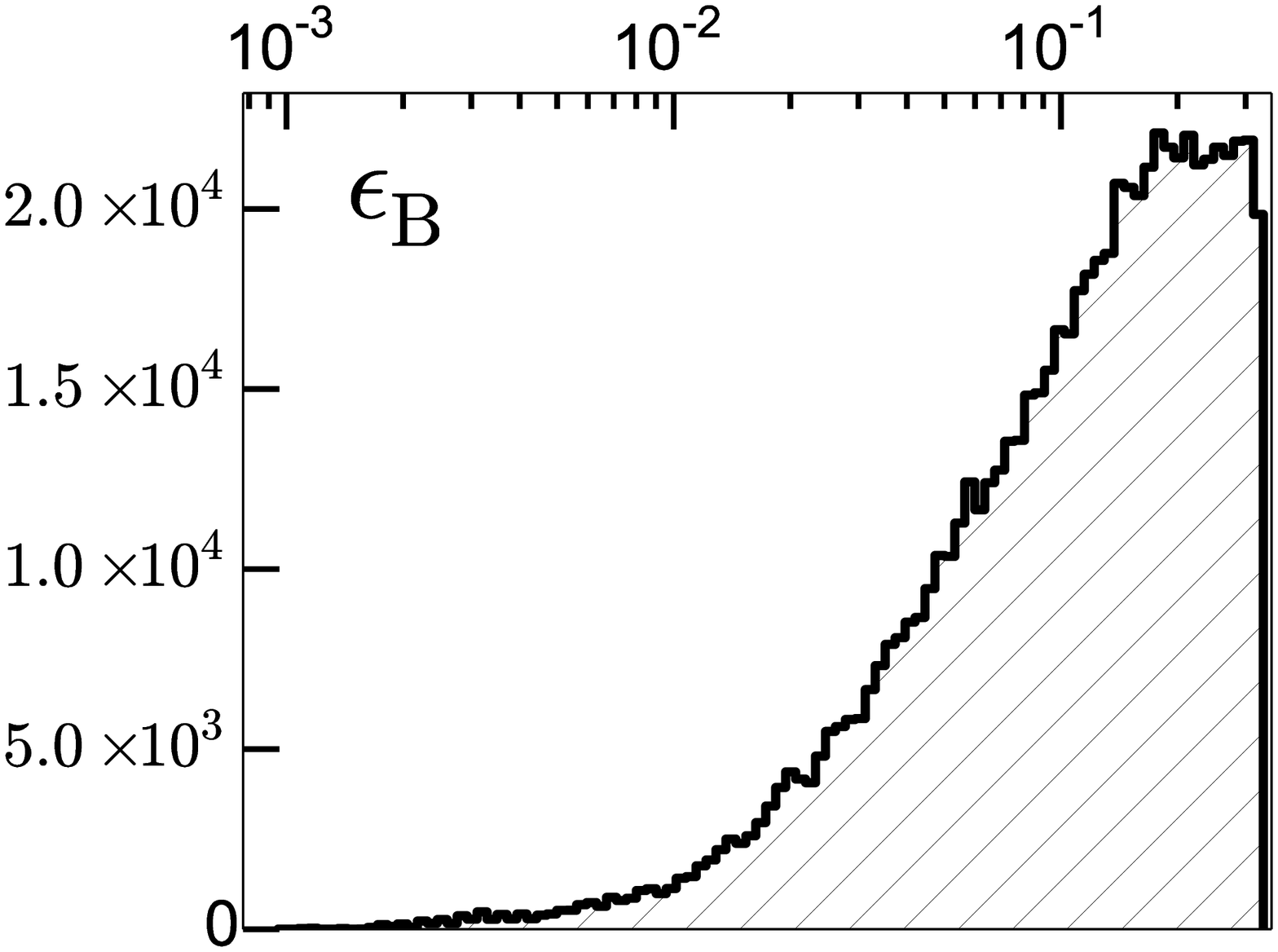} \\
 \includegraphics[width=0.30\columnwidth]{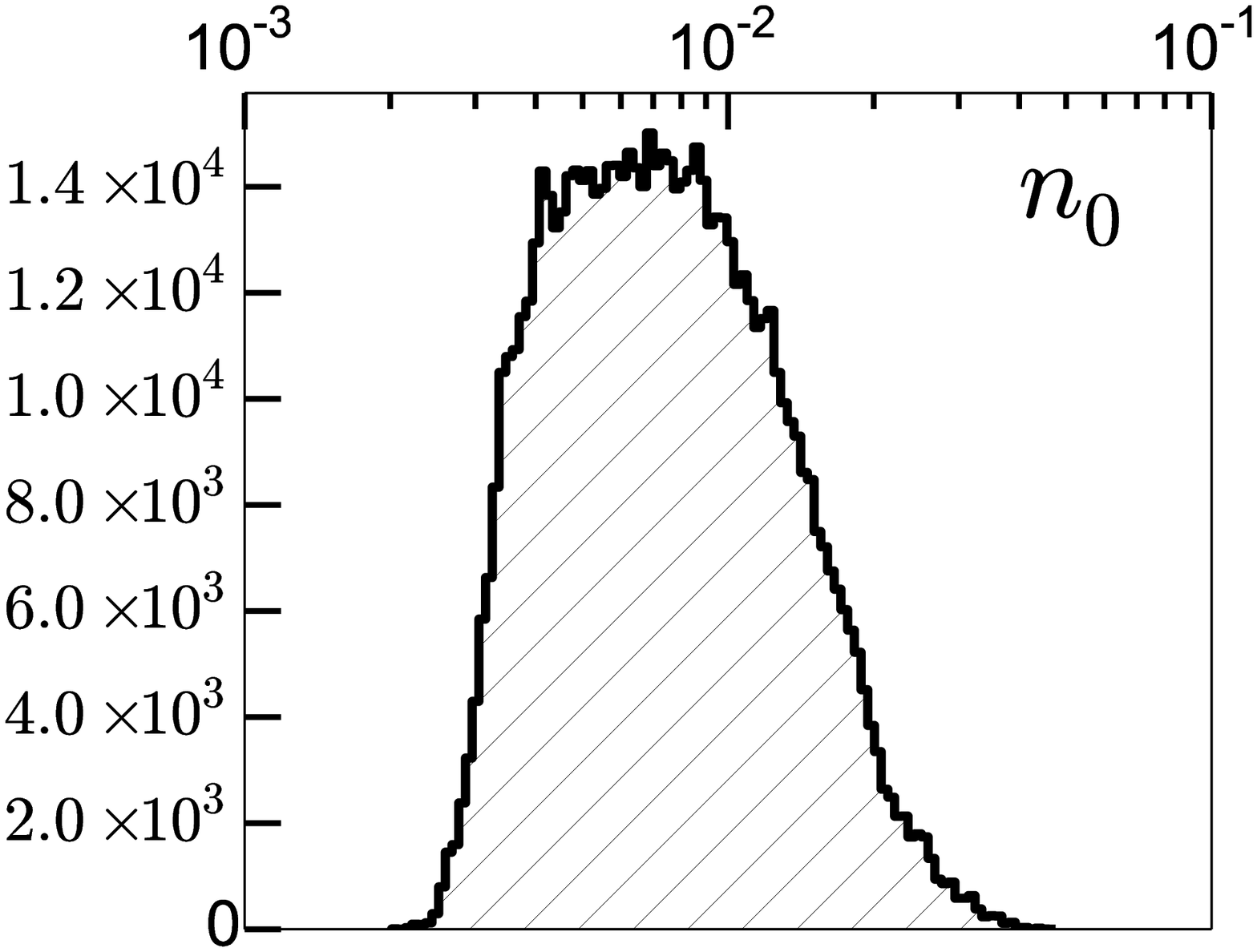} & 
 \includegraphics[width=0.30\columnwidth]{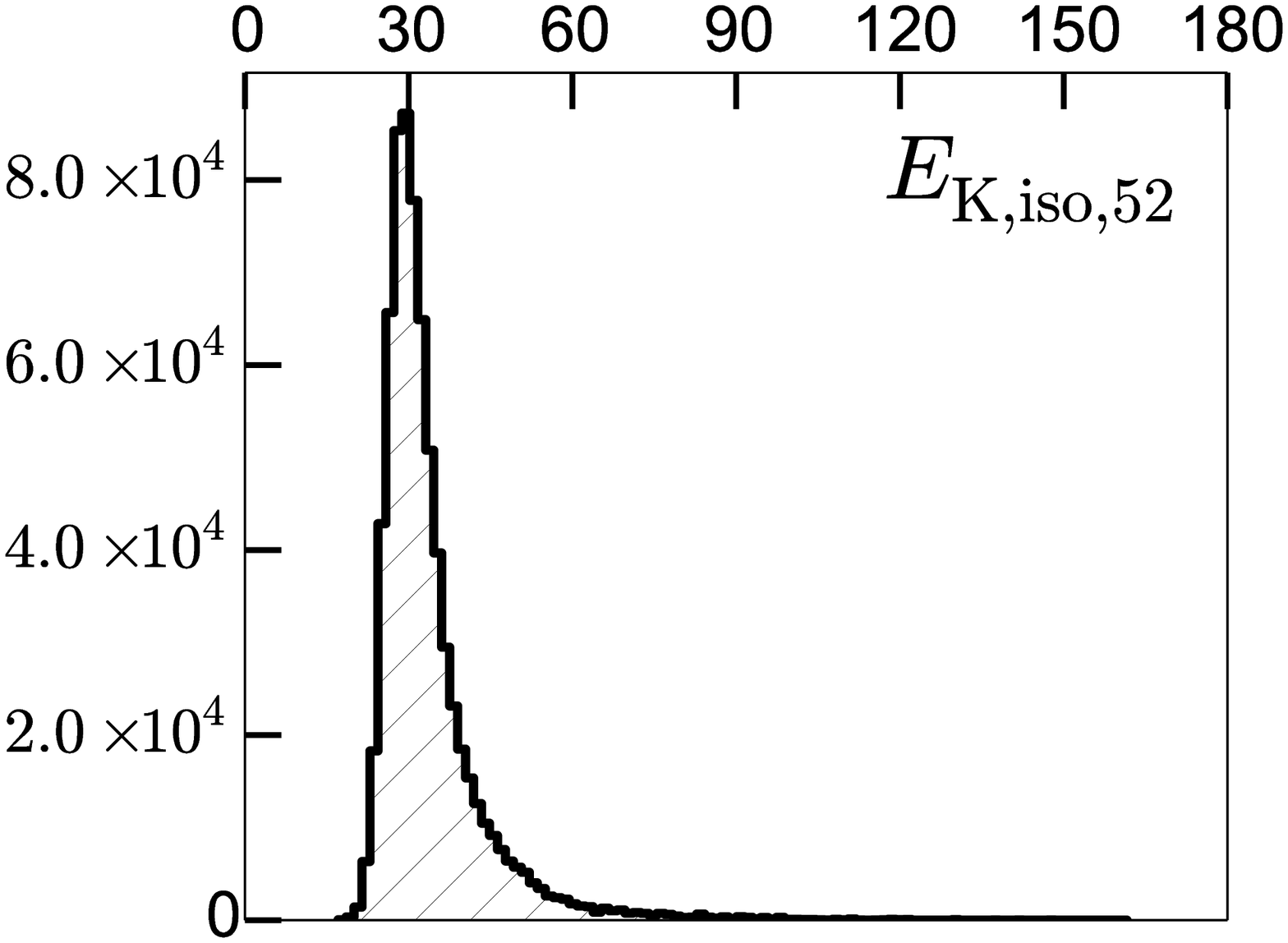} &
 \includegraphics[width=0.30\columnwidth]{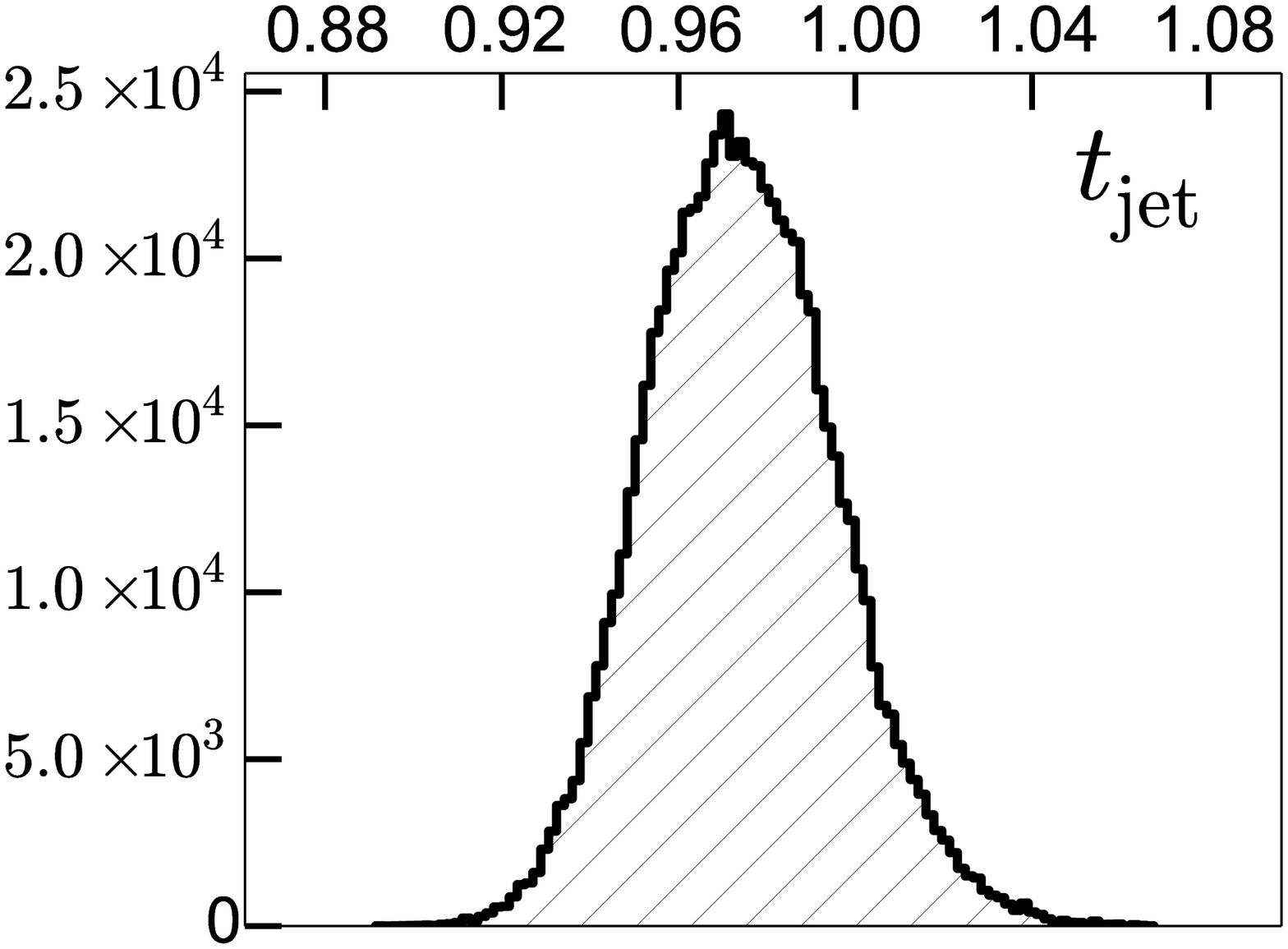} \\ 
 \includegraphics[width=0.30\columnwidth]{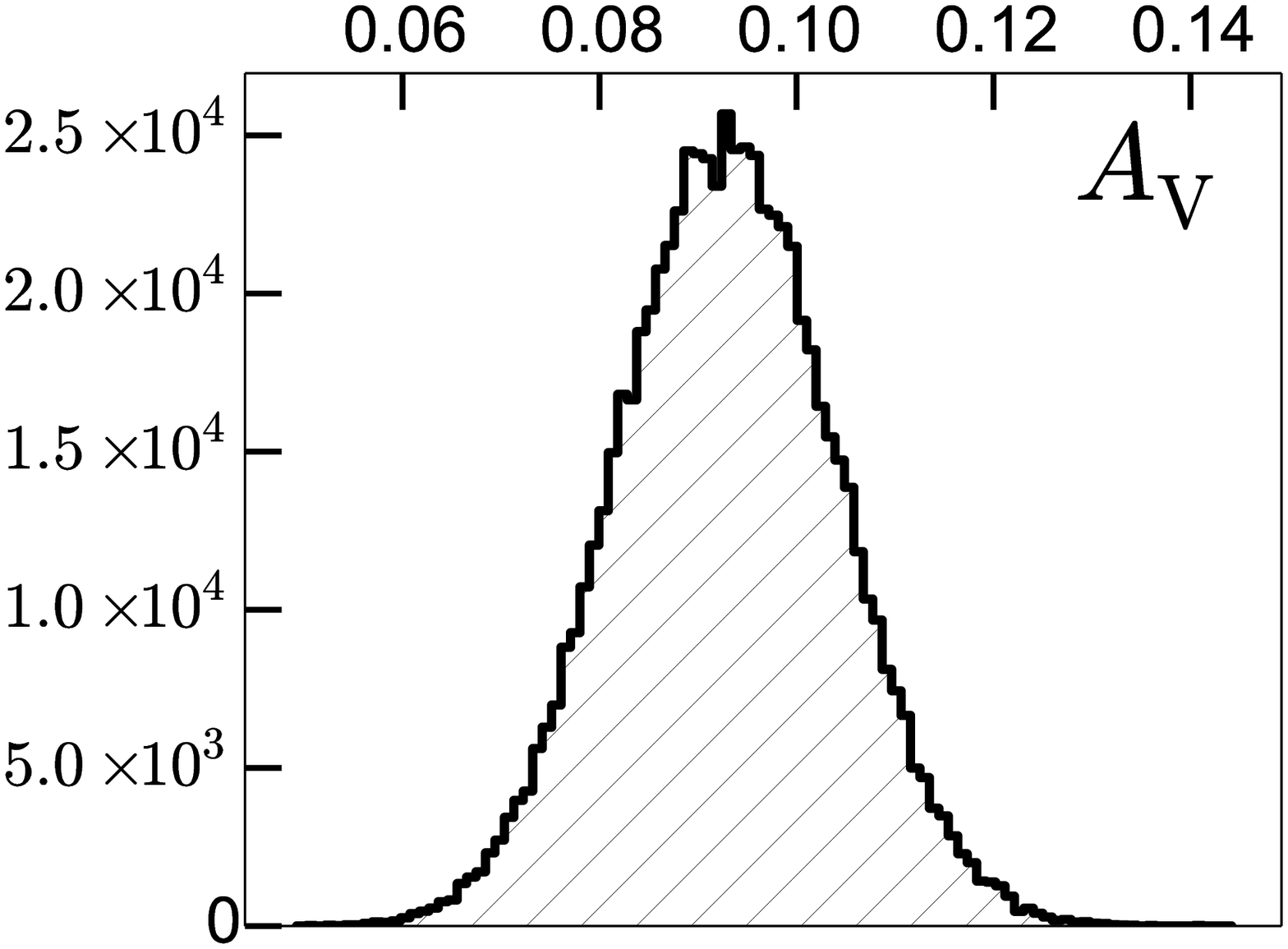} &
 \includegraphics[width=0.30\columnwidth]{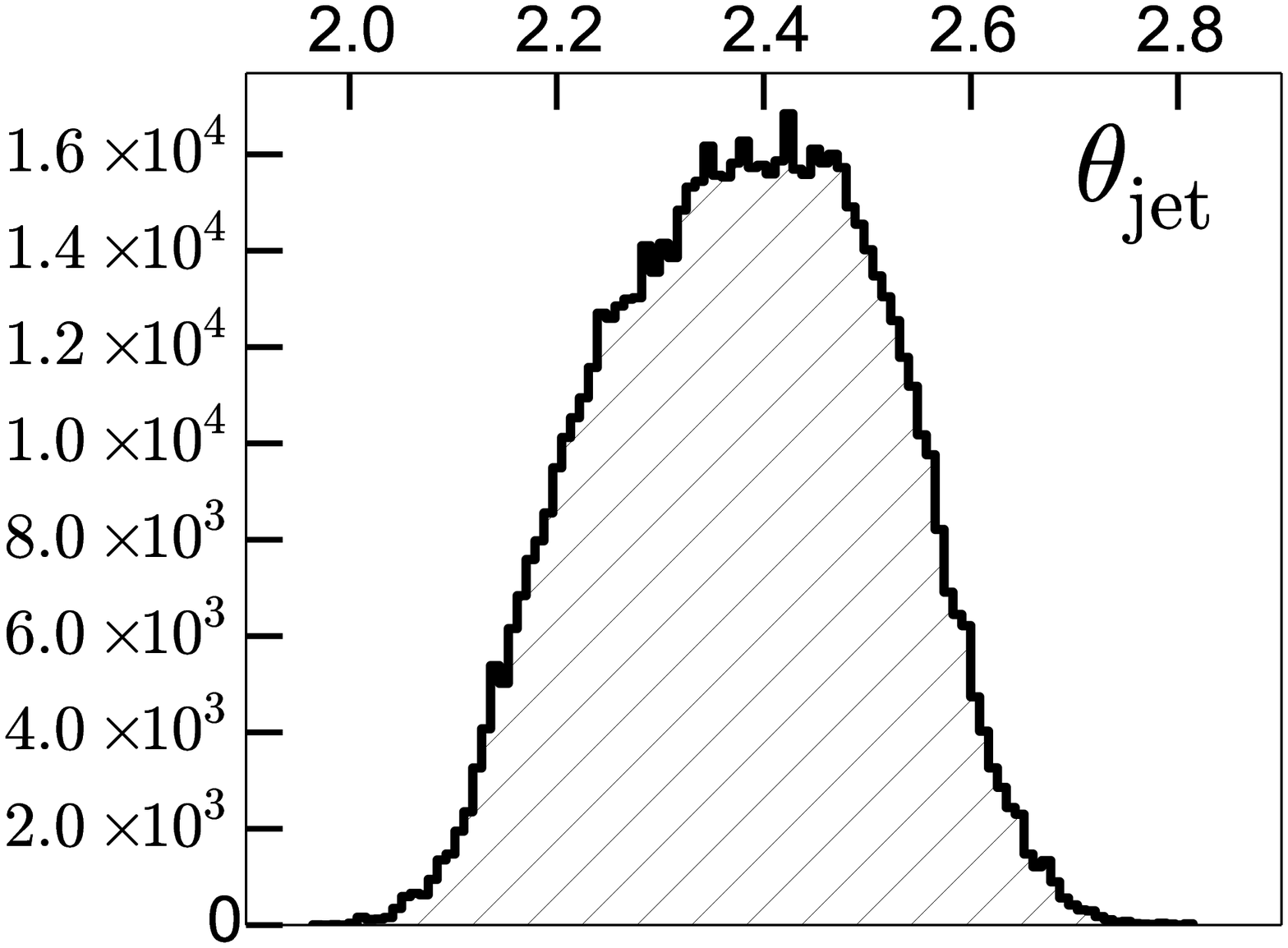}  &
 \includegraphics[width=0.30\columnwidth]{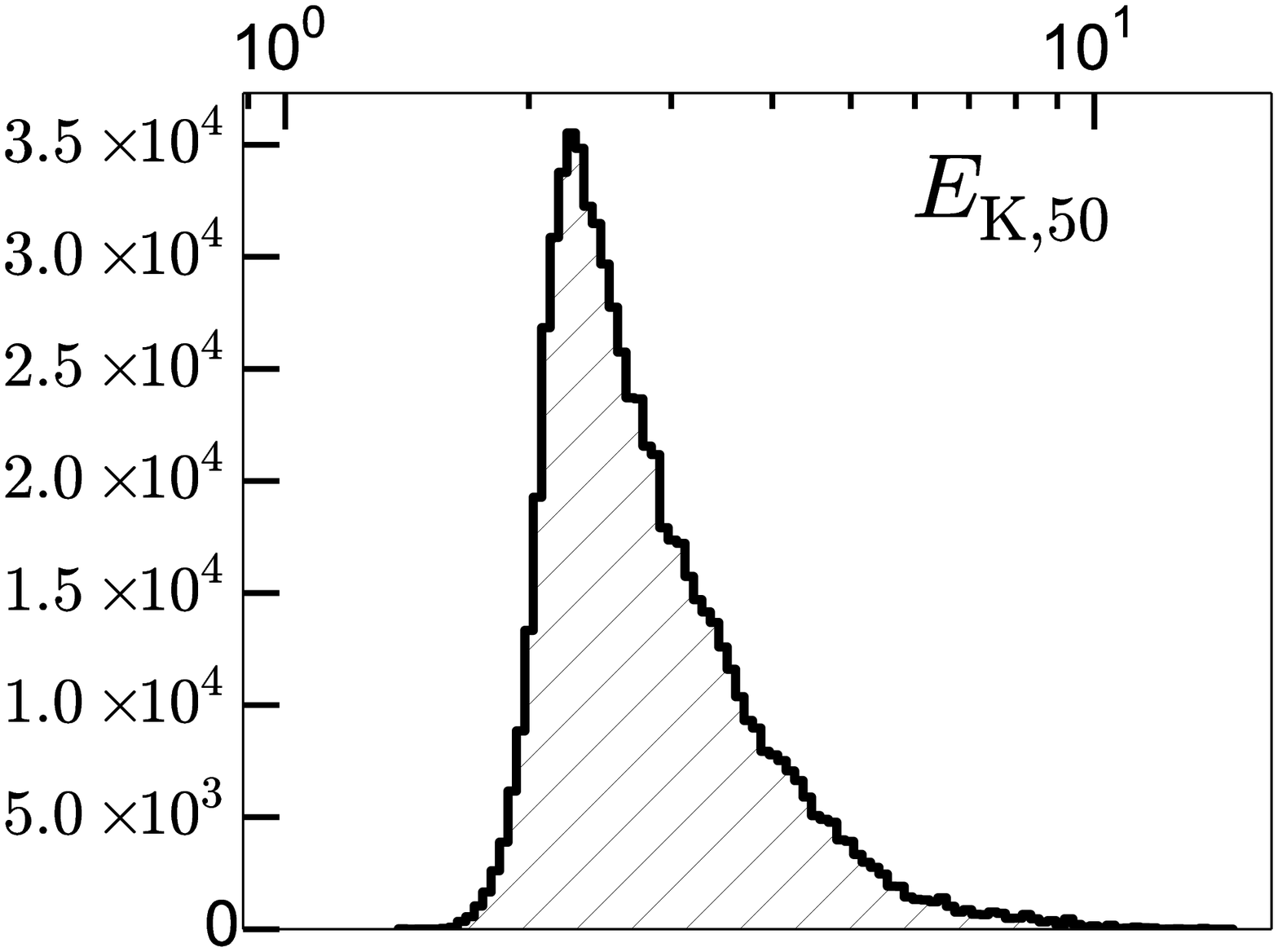} \\  
\end{tabular}
\caption{Posterior probability density functions for the physical parameters for GRB~100901A in 
a constant density environment from MCMC simulations. We have restricted $E_{\rm K, iso, 52} 
< 500$, $\epsilon_{\rm e} < \nicefrac{1}{3}$, and $\epsilon_{\rm B} < \nicefrac{1}{3}$. 
\label{fig:100901A_ISM_hists}}
\end{figure}
\begin{figure}
\begin{tabular}{ccc}
\centering
 \includegraphics[width=0.30\columnwidth]{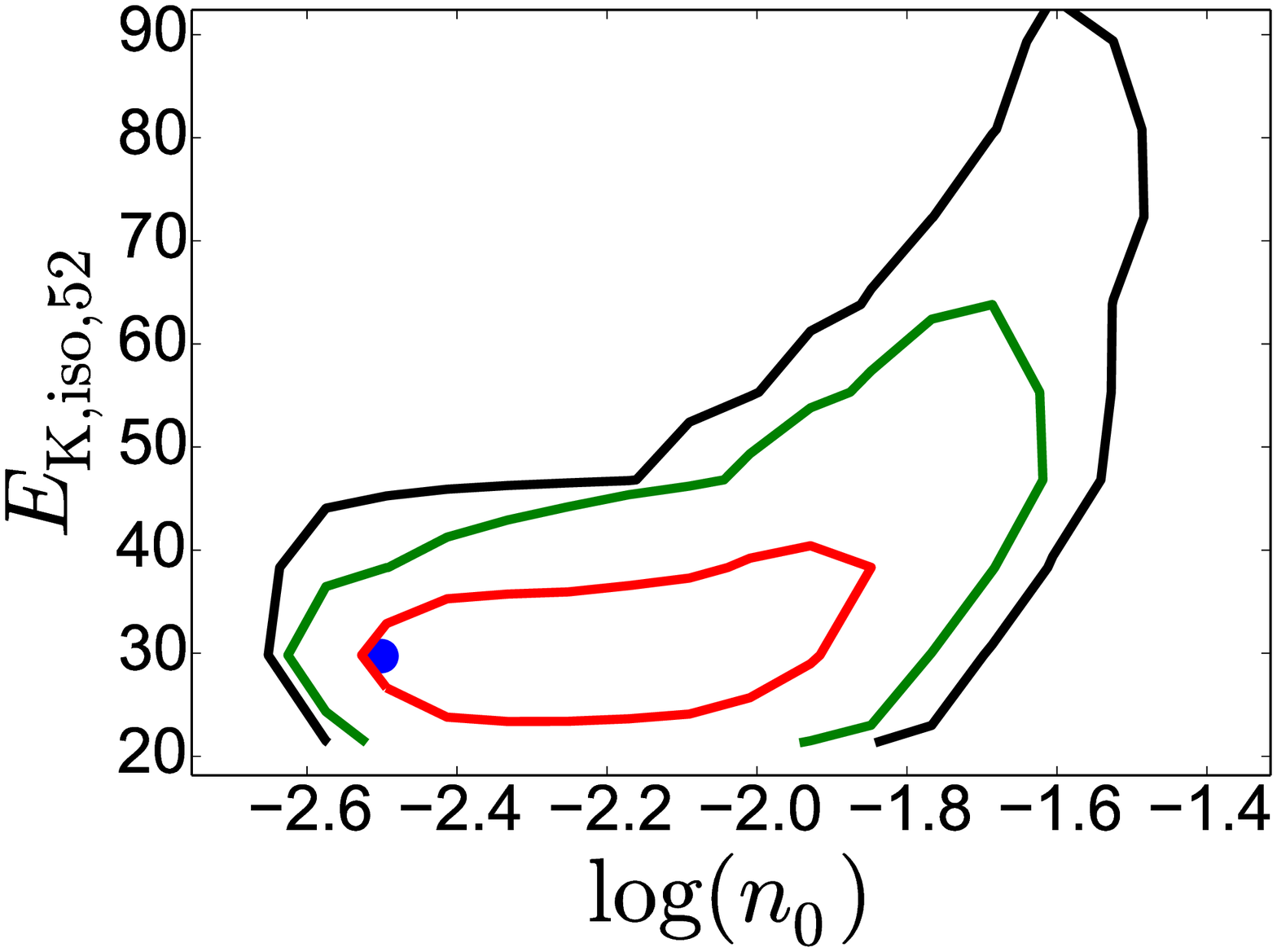} &
 \includegraphics[width=0.30\columnwidth]{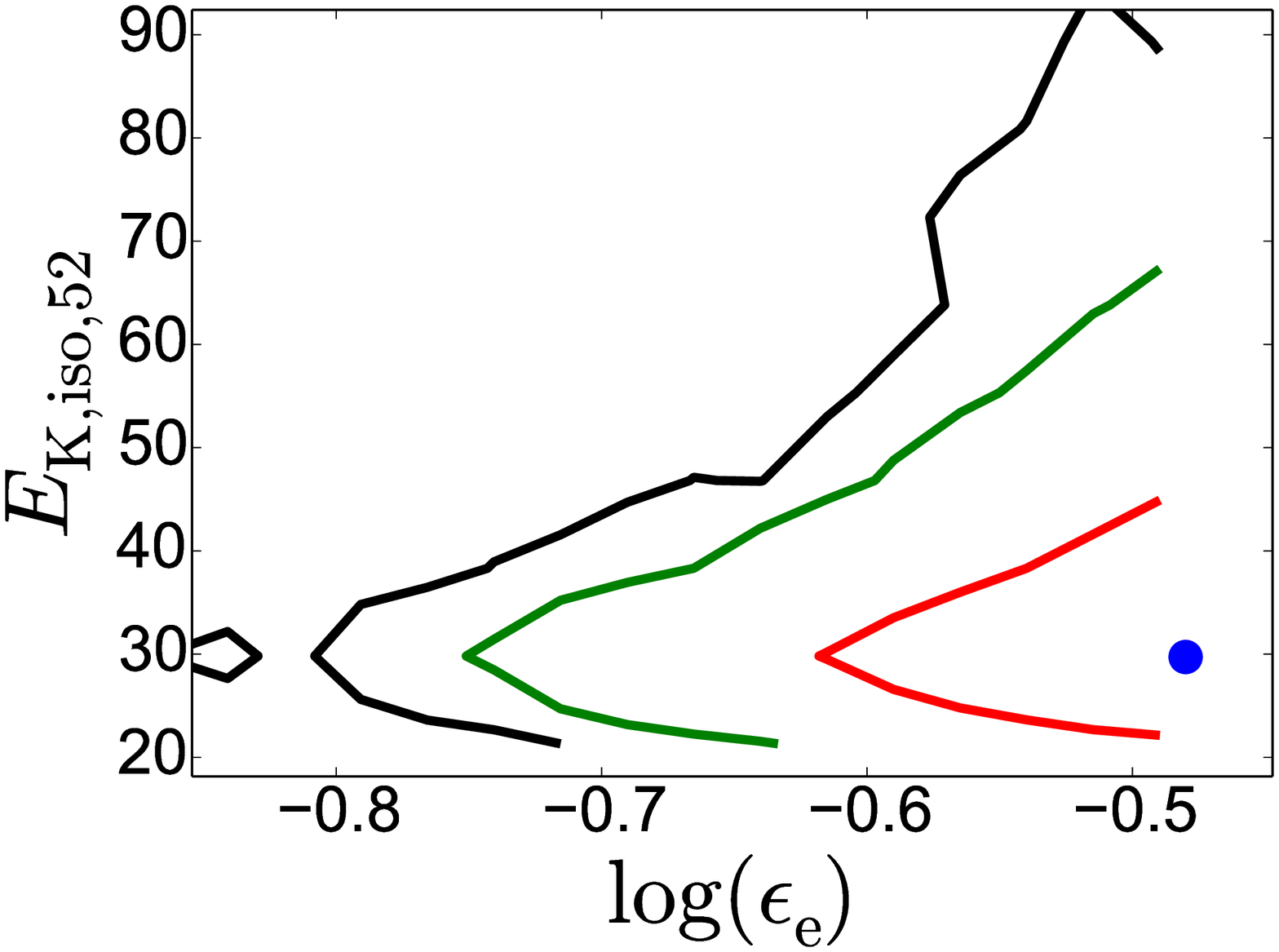} &
 \includegraphics[width=0.30\columnwidth]{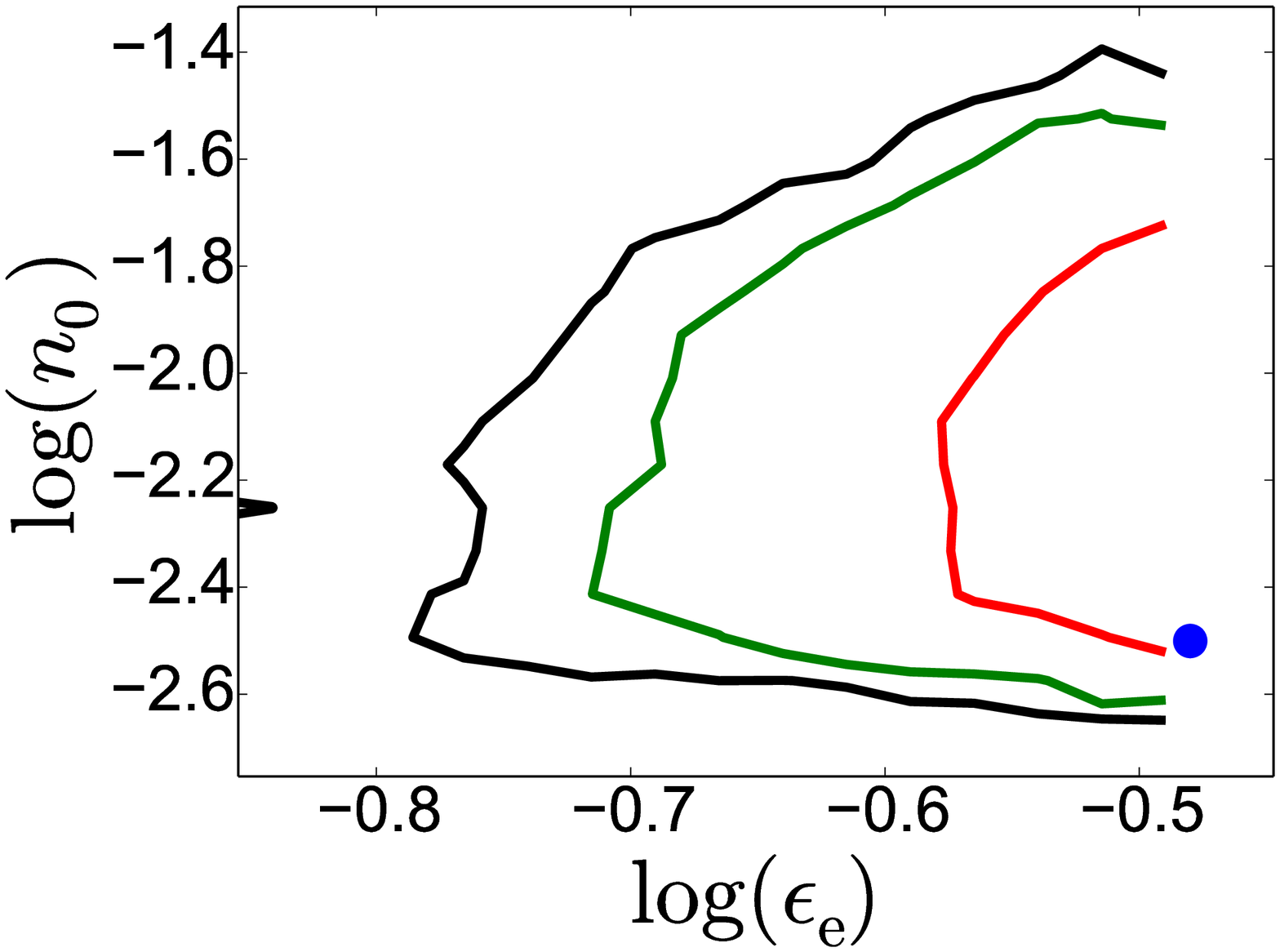} \\
 \includegraphics[width=0.30\columnwidth]{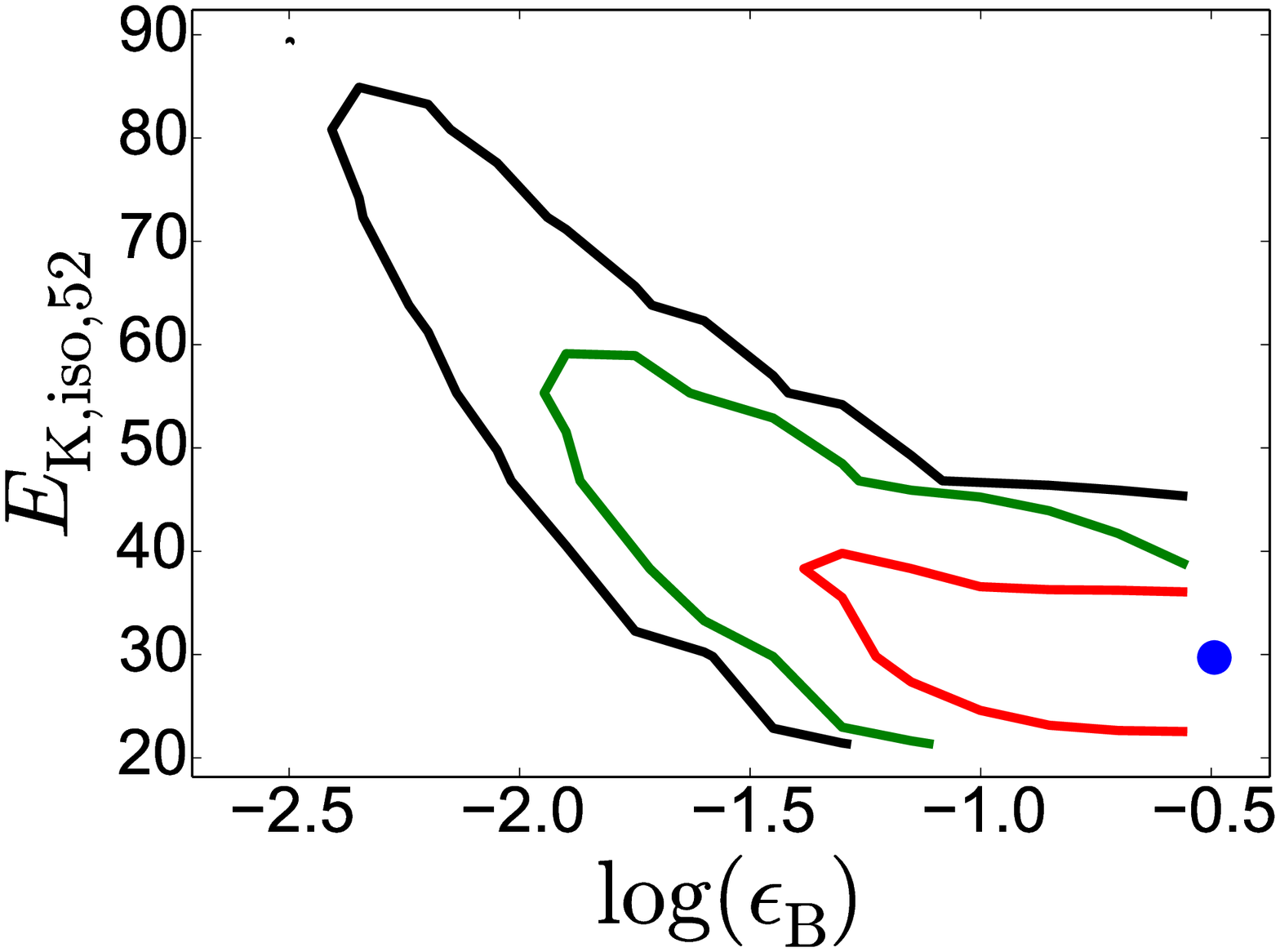} &
 \includegraphics[width=0.30\columnwidth]{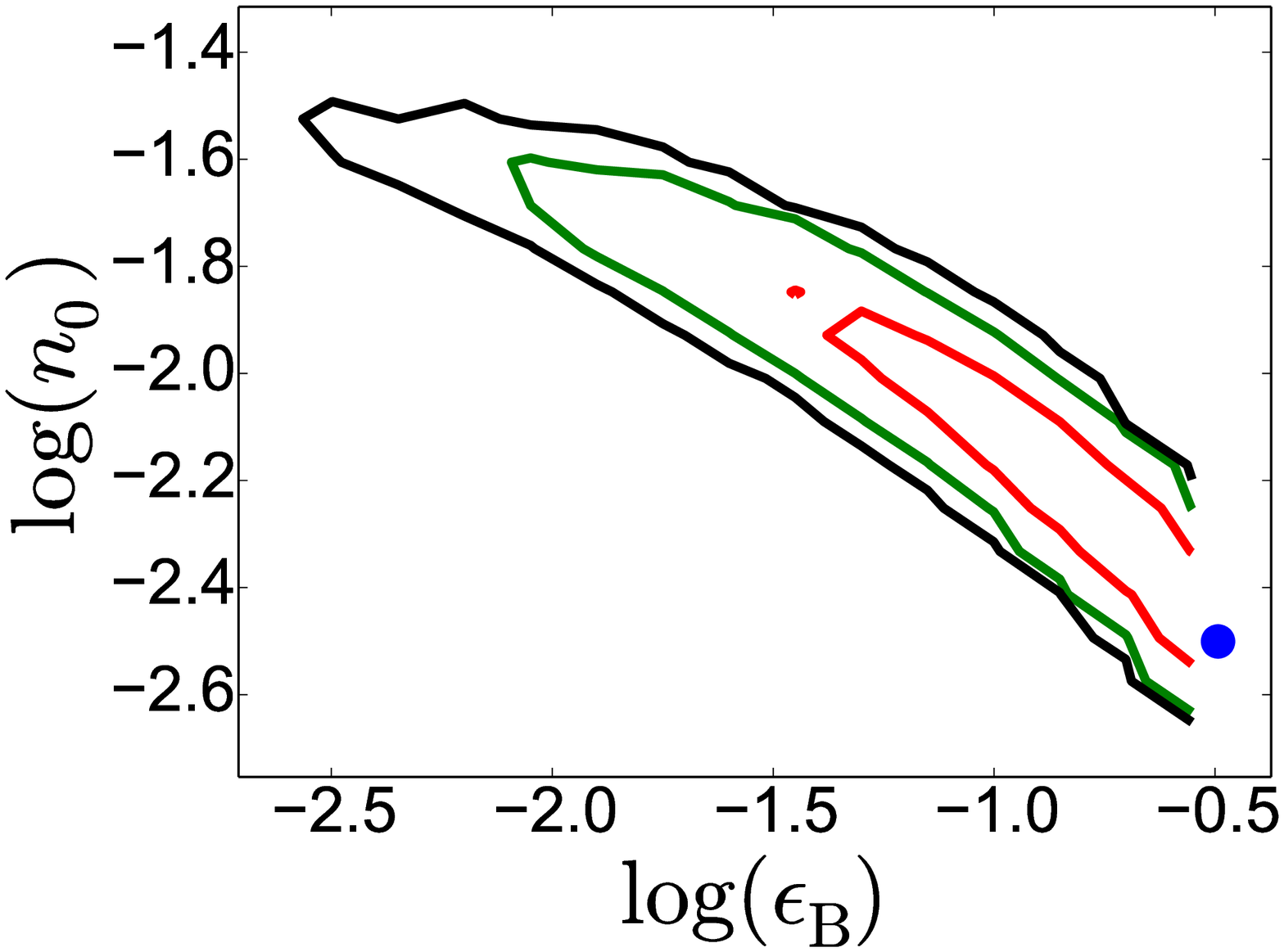} &
 \includegraphics[width=0.30\columnwidth]{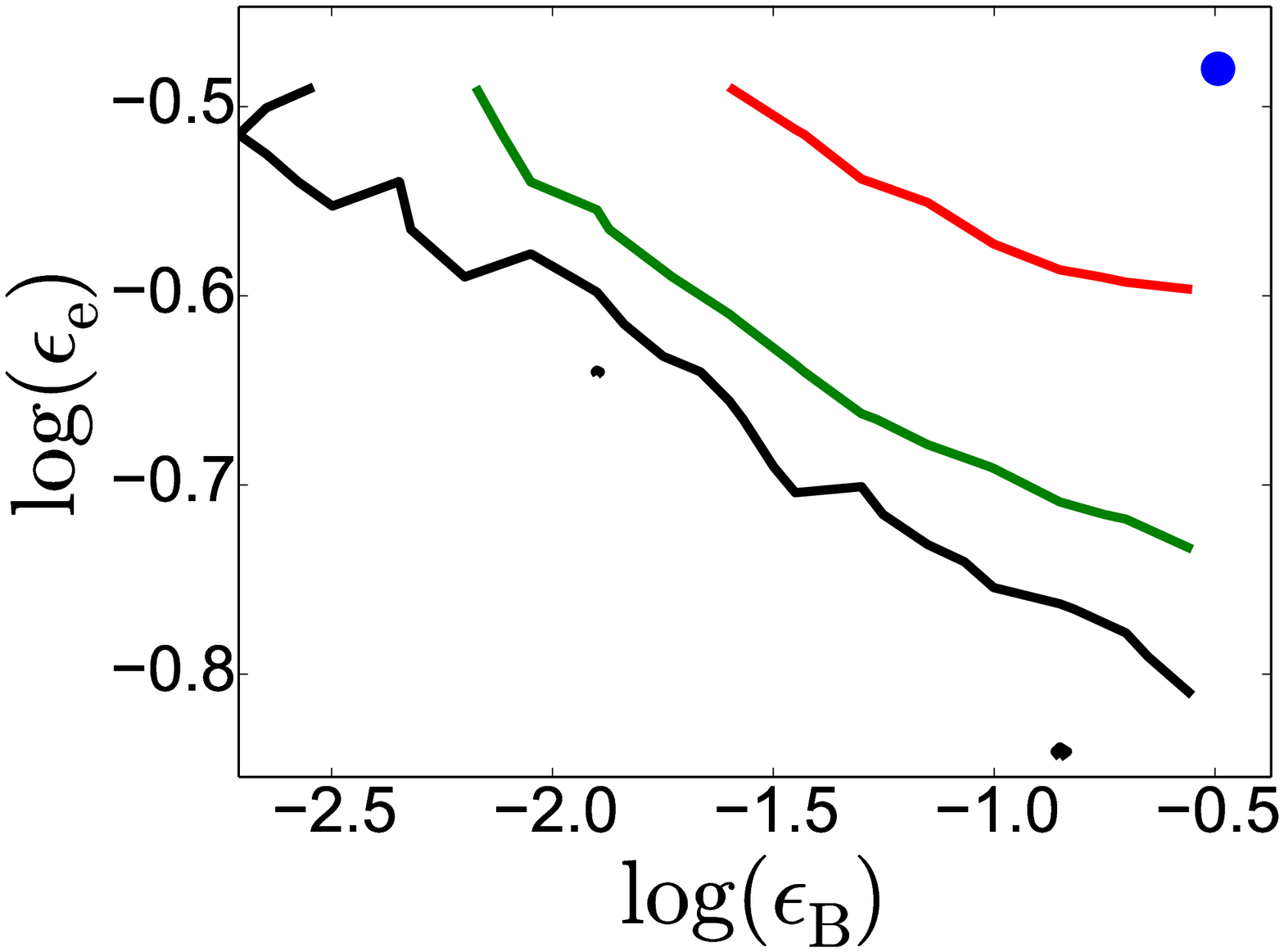} \\
\end{tabular}
\caption{1$\sigma$ (red), 2$\sigma$ (green), and 3$\sigma$ (black) contours for correlations
between the physical parameters, \EKiso, \dens, \epse, and \epsb\ for GRB~100901A, in the ISM model 
from Monte Carlo simulations. We have restricted $\epsilon_{\rm e} < \nicefrac{1}{3}$ and 
$\epsilon_{\rm B} < \nicefrac{1}{3}$. See the on line version of this Figure for additional plots 
of 
correlations between these parameters and $p$, $t_{\rm jet}$, $\thetajet$, $E_{\rm K}$, $A_{\rm 
V}$, 
and $F_{\nu,\rm host, r^{\prime}}$. \label{fig:100901A_ISM_corrplots}}
\end{figure}

We employ our MCMC analysis to fit the multi-band data for GRB~100901A after $t\approx0.25$\,d. At 
the redshift of $z=1.408$, the UVOT \textit{White}-, \textit{uvw1}-, \textit{uvw2}-, and 
\textit{uvm2}-band data are affected by IGM absorption, and we do not include these bands in our 
analysis. The highest-likelihood ISM model (Figure \ref{fig:100901A_enj}) the parameters 
$p\approx2.03$, $\epse\approx0.33$, $\epsb\approx0.32$, $\dens\approx3.2\times10^{-3}$\,\pcc, 
$\EKiso\approx3.0\times10^{53}$\,erg, $\tjet\approx0.96$\,d, $\AV\approx0.09$\,mag, and $F_{\nu,\rm 
host, r^{\prime}}\approx4.1\,\mu$Jy, with a Compton $y$-parameter of $\approx0.6$. The blastwave 
Lorentz factor is $\Gamma=15.9(t/1\,{\rm d})^{-3/8}$ and the jet opening angle is 
$\thetajet\approx2.1\degr$. The beaming-corrected kinetic energy is 
$\EK\approx2.1\times10^{50}\,{\rm erg}$, while the beaming corrected $\gamma$-ray energy 
is $\Egamma \approx 5\times10^{49}$\,erg (1--$10^4$\,keV; rest frame). We plot histograms of the 
measured parameters in Figure 
\ref{fig:100901A_ISM_hists} and correlation contours between the physical parameters in Figure 
\ref{fig:100901A_ISM_corrplots}, providing summary statistics from our MCMC analysis 
in Table \ref{tab:enjsummary}.
 
In this model, the break frequencies at 1\,d are located at $\nua\approx0.7$\,GHz, 
$\numax\approx2\times10^{12}$\,Hz, and $\nuc\approx9\times10^{14}$\,Hz, while the peak flux 
density is about 4.2\,mJy at \numax. The spectrum transitions from fast to slow cooling at about 
350\,s after the burst. The cooling frequency is located between the optical and X-rays and the 
jet break is before 1\,d, both of which are expected from the basic analysis outlined above. The 
low value of $p$ is driven by the shallow measured decline in the X-rays following the jet break, 
and the resulting spectrum remains consistent with the optical-to-X-ray SED (Figure 
\ref{fig:100901A_sed}).

\subsubsection{Energy injection model}
\label{text:100901A:enj}
Taking the forward shock model described in Section \ref{text:100901A:FS} as a starting point, we 
find that the X-ray and UV/optical data before the re-brightening can be explained by a single 
period of energy injection between $0.105$\,d and $\approx0.26$\,d:
\begin{equation}
\label{eqn:bpl}
\EKiso(t) = 
  \begin{cases}
      E_{\rm K,iso,f}, & t > t_{\rm 0} = 0.26\,{\rm d} \\
      E_{\rm K,iso,f}\left(\frac{t}{t_{\rm 0}}\right)^{1.88}, &
      t_1 = 0.105\,{\rm d} < t < t_{\rm 0} \\      
      E_{\rm K,iso,f}\left(\frac{t_1}{t_{\rm 0}}\right)^{1.88},&
      t < t_1\\
  \end{cases}
\end{equation}
In this model, the energy increases by a factor of $\approx5.5$ from $E_{\rm K,iso,i} \approx 
5.4\times10^{52}$\,erg at 0.105\,d to $E_{\rm K,iso,f} \approx3.0\times10^{53}$\,erg at 0.26\,d, 
corresponding to an injected energy fraction of $\approx 85\%$, while the Lorentz factor decreases 
from $\Gamma\approx30$ to $\Gamma\approx26$ over this period. In comparison, 
$\Egammaiso\approx8\times10^{52}$\,erg (Section \ref{text:100901A:basic_considerations}). The value 
of $m\approx1.88$ corresponds to $s\approx14.4$ for $26\lesssim\Gamma\lesssim30$.

\begin{figure*}
 \centering
 \begin{tabular}{cc}
 \includegraphics[width=0.43\linewidth]{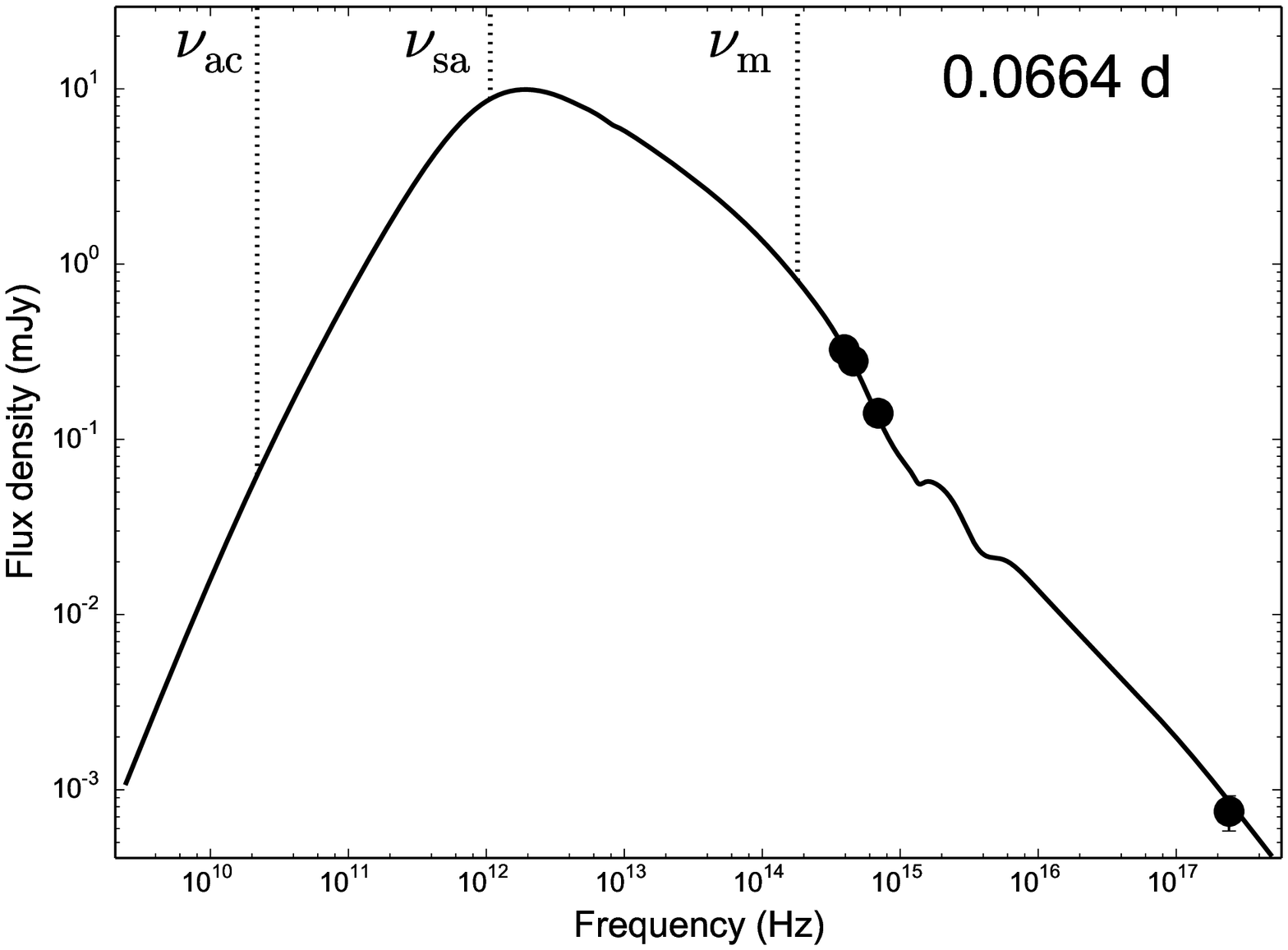} &
 \includegraphics[width=0.43\linewidth]{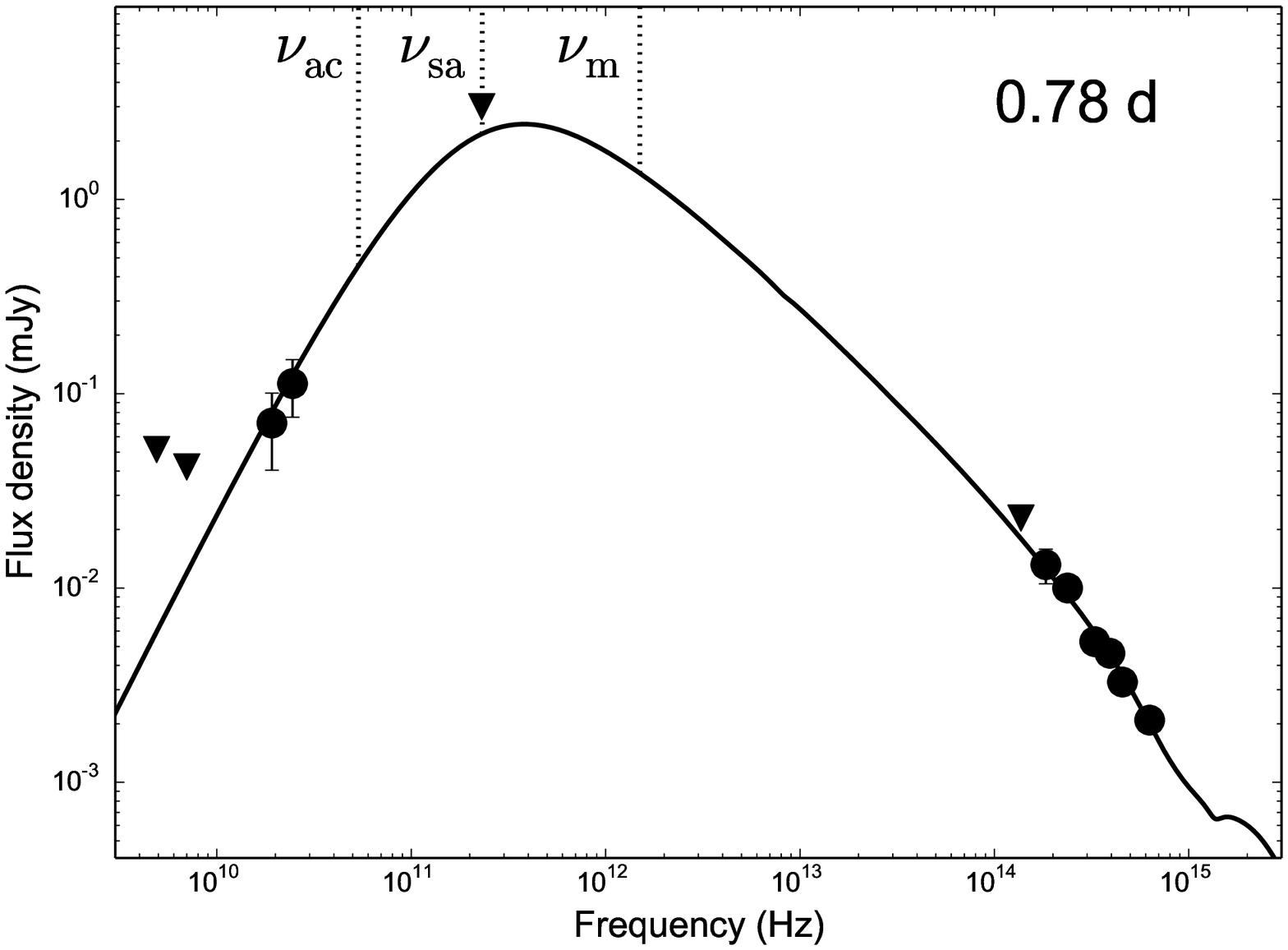}
 \end{tabular}  
 \caption{Left: Observer-frame SED of the afterglow of GRB~120404A at 0.0664\,d, together with the 
best fit model (spectrum 4 from GS02: black, solid). Right: The same as the left panel, but at 
0.78\,d. The radio detections with a steeply rising spectrum suggest that the self-absorption 
frequency is near or above the radio band, requiring a high circumburst density. Both SEDs are 
corrected for the effect of inverse Compton cooling, which is significant, with the Compton 
$y$-parameter $\approx0.9$ during the fast cooling phase. We do not show the inverse Compton 
radiation, since even at its peak (near the \Swift/XRT band), it is four orders of magnitude 
weaker than the synchrotron component. See Appendix \ref{appendix:IC} for a discussion of the 
inverse Compton effect in the context of modeling GRB afterglows.}
\label{fig:120404A_seds}
\end{figure*}

\subsection{GRB~120404A}
\subsubsection{GRB properties}
\label{text:120404A:basic_considerations}
GRB~120404A was detected and localized by the \Swift\ BAT on 2012 April 04 at 12:51:02\,UT 
\citep{gcn13208}. The burst duration was $T_{90} = 38.7\pm4.1$\,s, with a fluence of $F_{\gamma} = 
(1.6 \pm 0.1)\times10^{-6}$\,erg\,cm$^{-2}$ \citep[15--150\,keV observer frame,][]{gcn13220}. 
\Swift\ and ground-based observatories detected an afterglow in the X-rays and UV/optical 
\citep{gcn13209,gcn13221,gcn13226,gcn13230}, as well as in the radio 
\citep{gcn13231}. Spectroscopic observations with Gemini-North yielded a redshift of $z=2.876$ 
\citep{gcn13217}. The isotropic-equivalent $\gamma$-ray energy for this event is 
$\Egammaiso=(9\pm4)\times10^{52}$\,erg ($1$--$10^4$\,keV, rest frame; \citealt{gmh+13}).

This burst has been previously studied in detail by \cite{gmh+13}, who interpret the optical 
re-brightening starting around 800\,s in their well-sampled, multi-band optical light curves as due 
to the passage of the characteristic synchrotron frequency, $\numax$. They additionally invoke
reverse shock emission to explain the flat ($t^{0.0\pm0.1}$) portion of the optical light curve 
before the onset of the re-brightening. In the following, we propose an alternate model for the 
multi-band radio through X-ray light curves in the context of energy injection.


The X-ray light curve before 700\,s can be modeled as a power law decay
with $\alpha_{\rm X} \approx -2$.
This light curve phase is likely part of the high latitude emission, and we 
ignore the data before 0.008\,d in our analysis. The X-ray photon index at 0.12--0.24\,d,
$\Gamma_{X}=2.3\pm0.3$ \citep{gmh+13}, implying a spectrum, $F_{\nu}\propto\nu^{-1.3\pm0.3}$, is 
consistent with the spectral slope between the optical $i^{\prime}$-band and the X-rays at 0.07\,d, 
$\beta_{\rm opt-X} = -0.91\pm0.04$, suggesting that the optical and X-ray bands are on the same 
segment of the afterglow SED, although the large uncertainty in $\beta_{\rm X}$ leaves open the 
possibility that $\nuc$ lies between the optical and X-rays. Additionally, the spectral slope 
within 
the optical ($B$- to $i^{\prime}$-band) is $\beta_{\rm opt}=-1.3\pm0.2$ at 0.07\,d, 
indicating that extinction is present. The spectral index between the 19.2\,GHz and 24.5\,GHz 
observations at 0.75\,d is $\beta_{\rm radio}\approx2$, which indicates that $\nua$ is located 
above $24.5$\,GHz at this time (Figure \ref{fig:120404A_seds}).

The optical $R$-band light curve declines as $t^{-1.9\pm0.02}$ after 0.13\,d, consistent with the 
X-ray decline rate of $t^{-1.8\pm0.1}$ after 0.05\,d. The steep decline of $\alpha \approx -2$ is 
indicative of a jet break before $\approx0.1$\,d. A broken power law fit to the $B$-band light 
curve results in the parameters $t_{\rm b}=(2.8\pm0.4)\times10^{-2}$\,d, $F_{\nu, \rm B}(t_{\rm b}) 
= 252\pm10\,\mu$Jy, $\alpha_1 = 1.74\pm0.58$, $\alpha_2 = -1.71\pm0.17$, and $y=0.78\pm0.43$, 
making 
this the earliest re-brightening episode of the four events studied in this paper. 

\begin{figure}
\begin{tabular}{ccc}
 \centering
 \includegraphics[width=0.30\columnwidth]{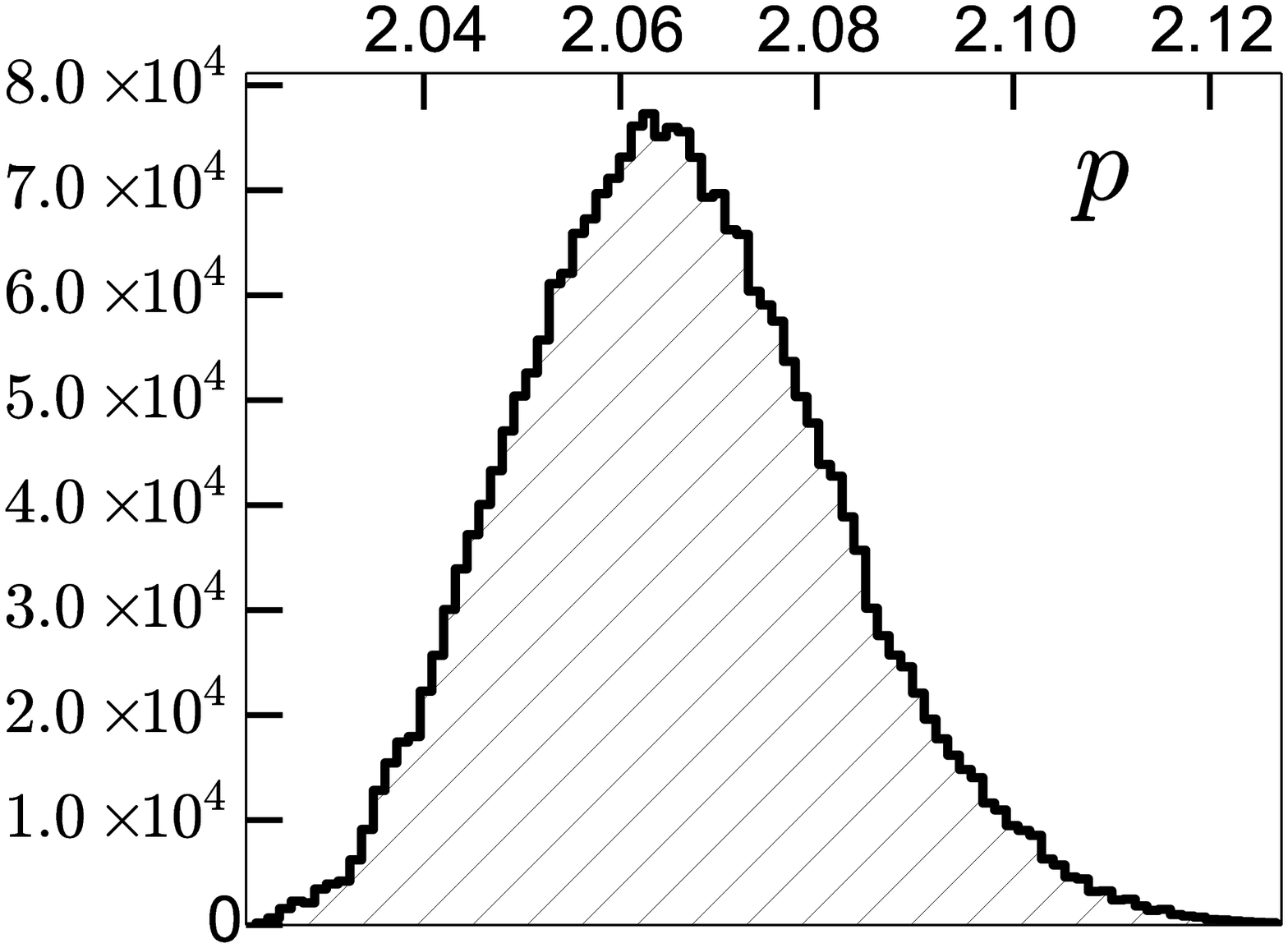} &
 \includegraphics[width=0.30\columnwidth]{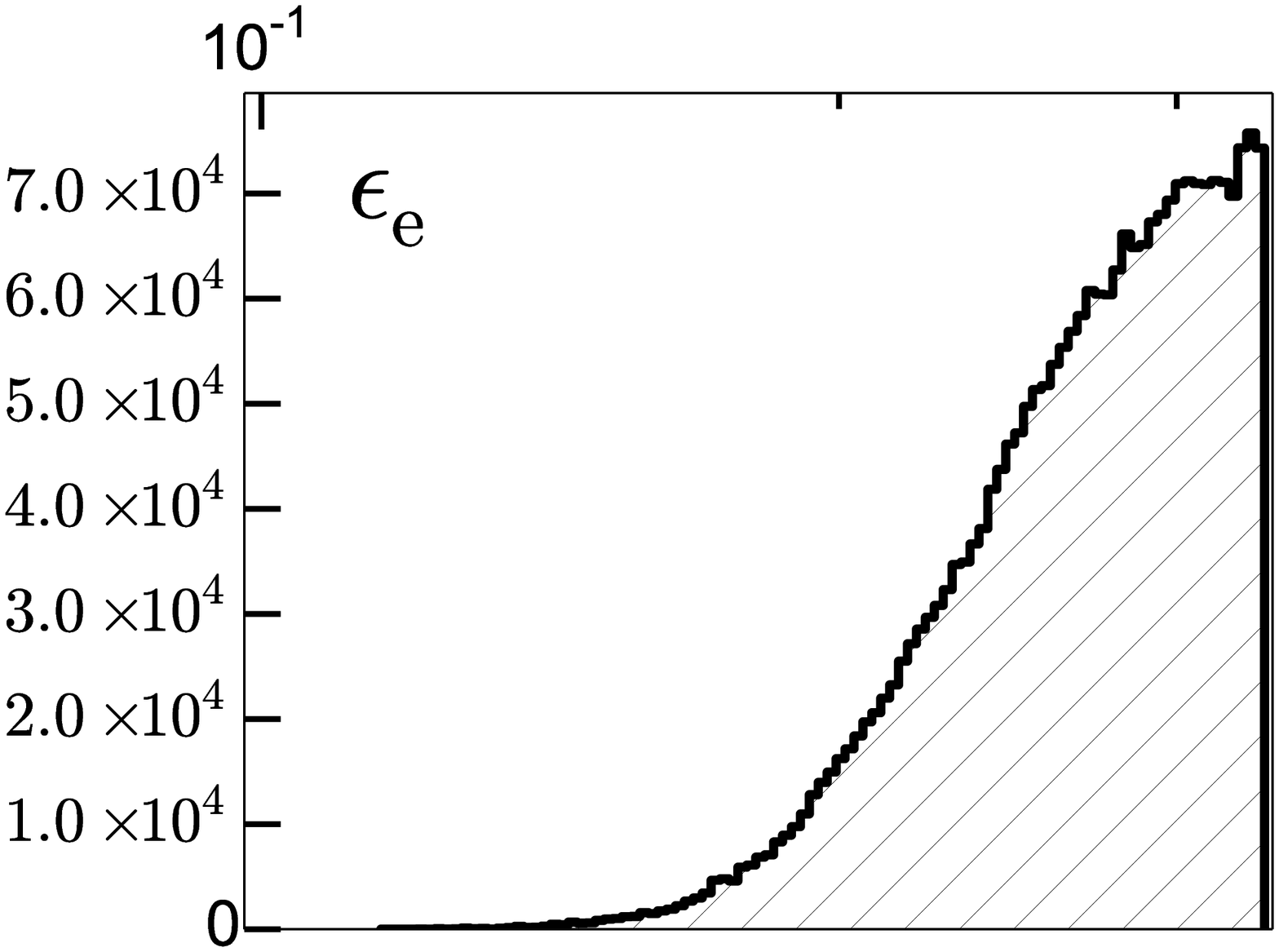} &
 \includegraphics[width=0.30\columnwidth]{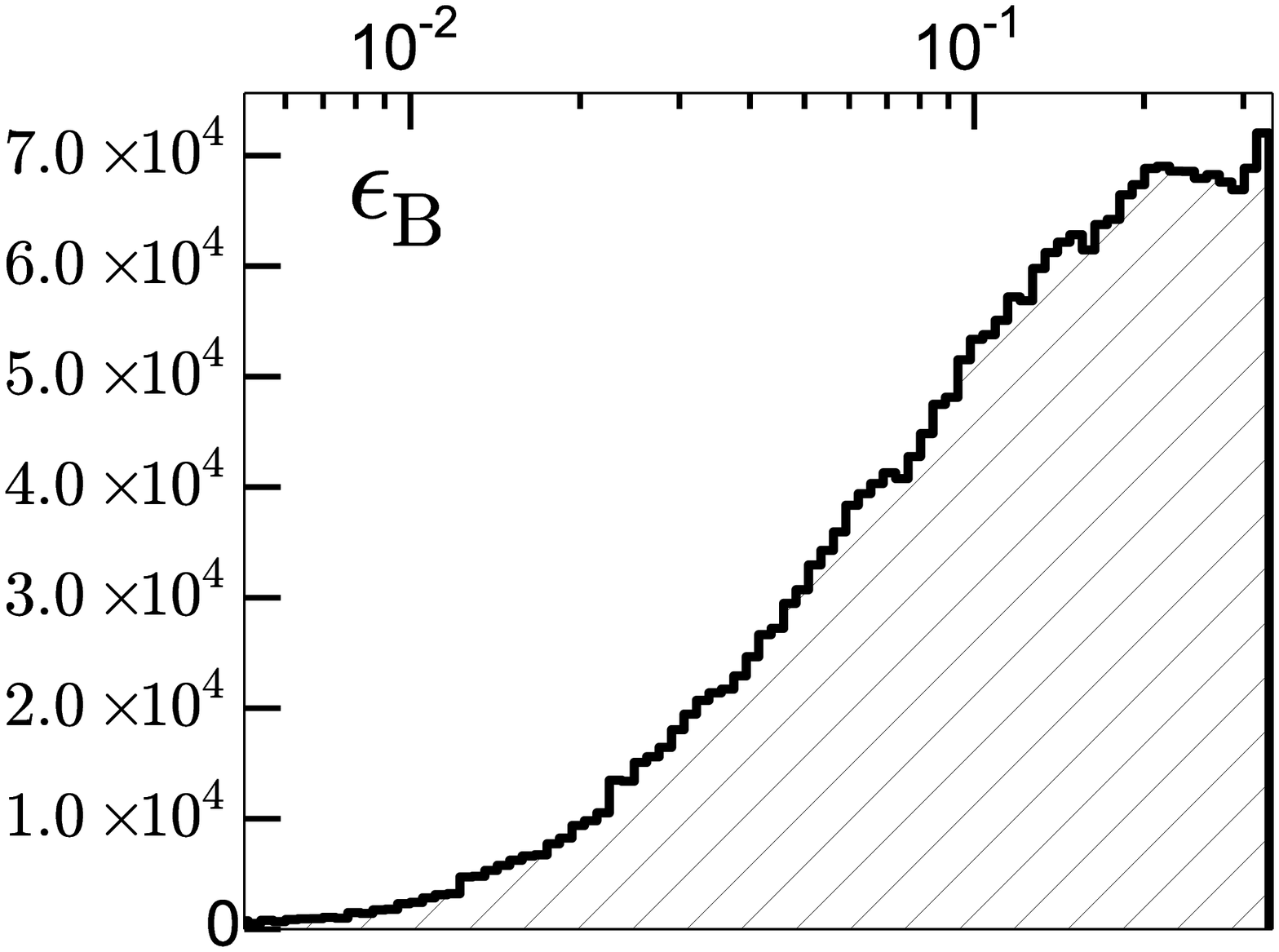} \\
 \includegraphics[width=0.30\columnwidth]{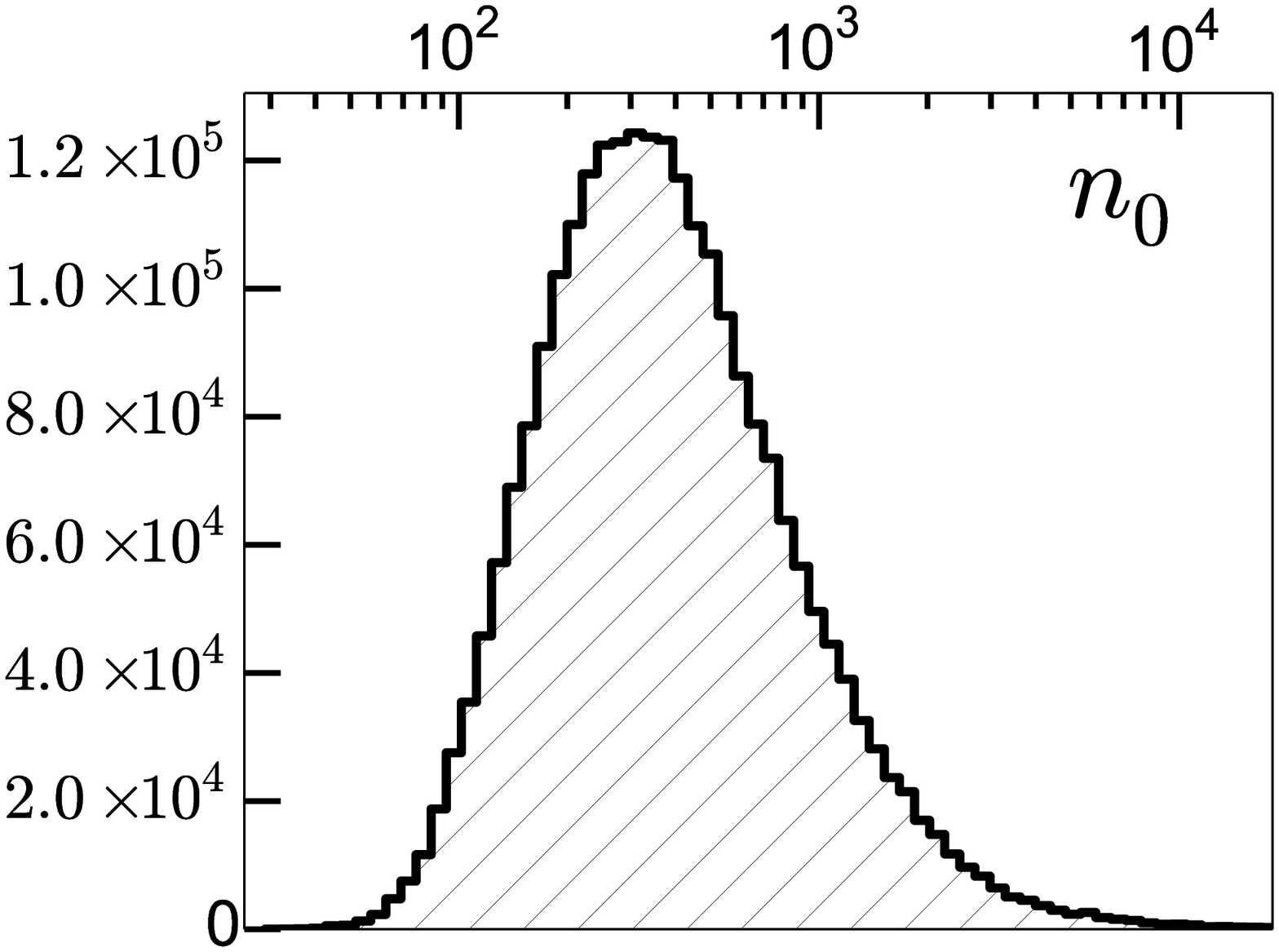} & 
 \includegraphics[width=0.30\columnwidth]{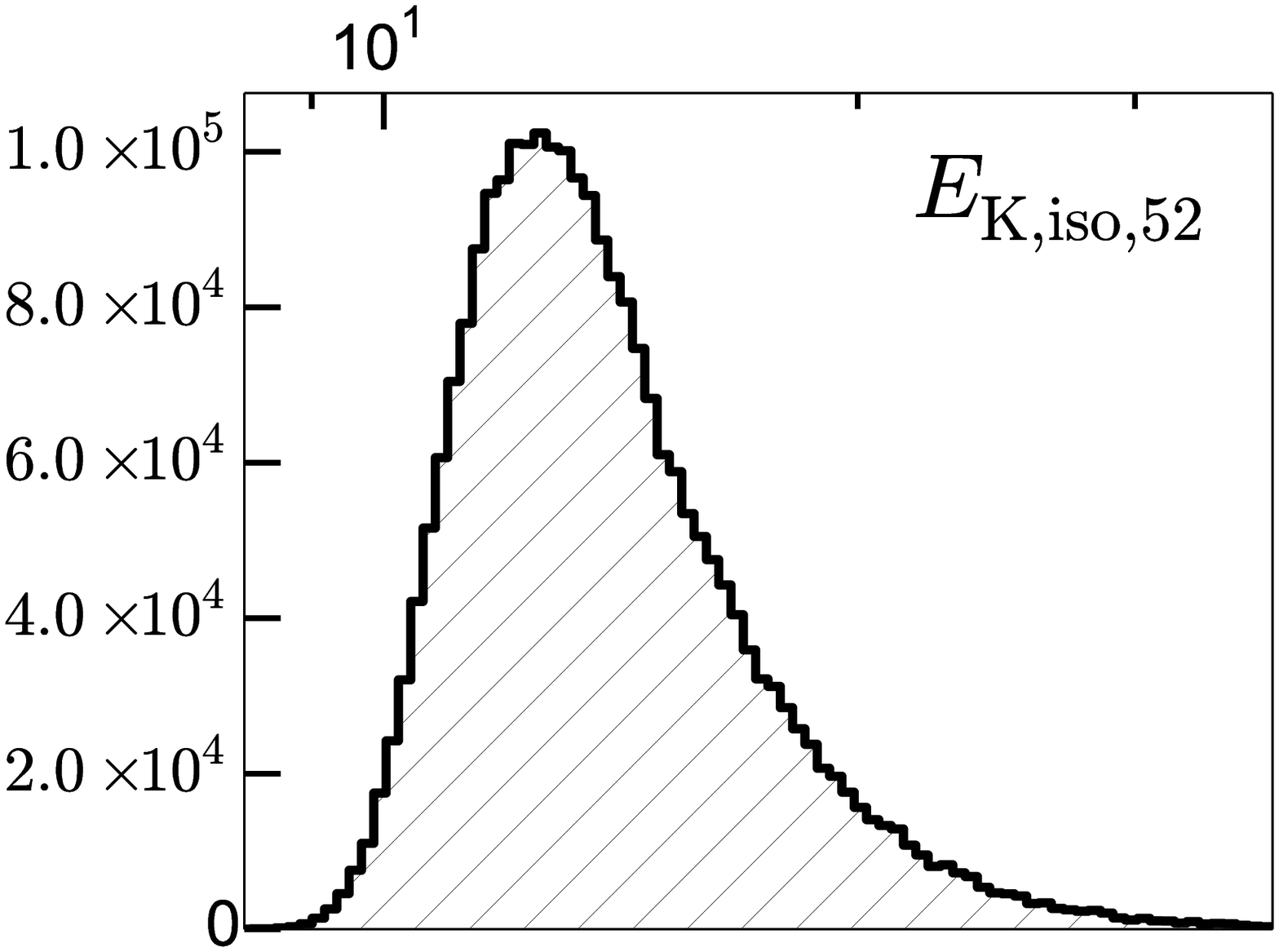} &
 \includegraphics[width=0.30\columnwidth]{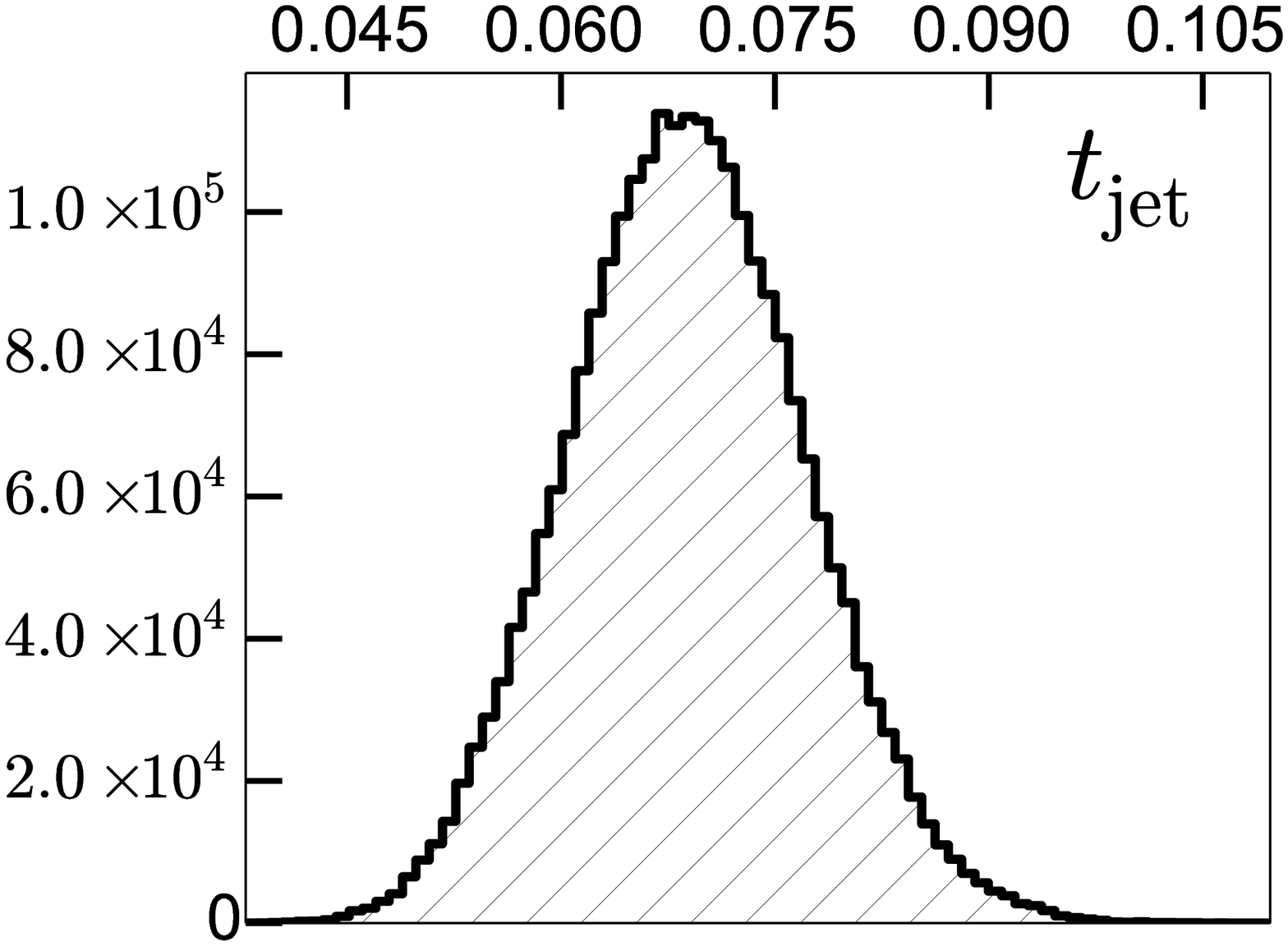} \\ 
 \includegraphics[width=0.30\columnwidth]{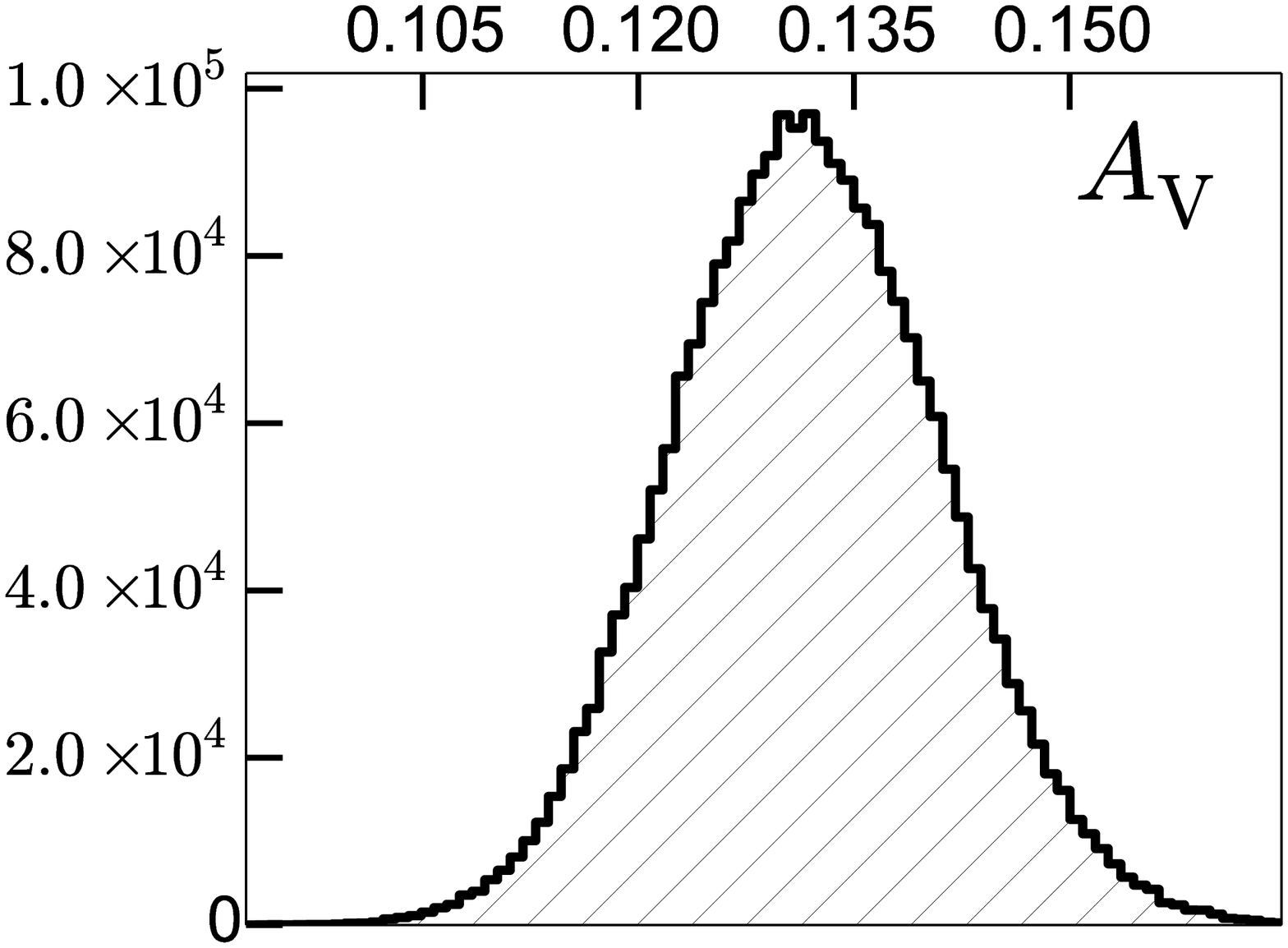} &
 \includegraphics[width=0.30\columnwidth]{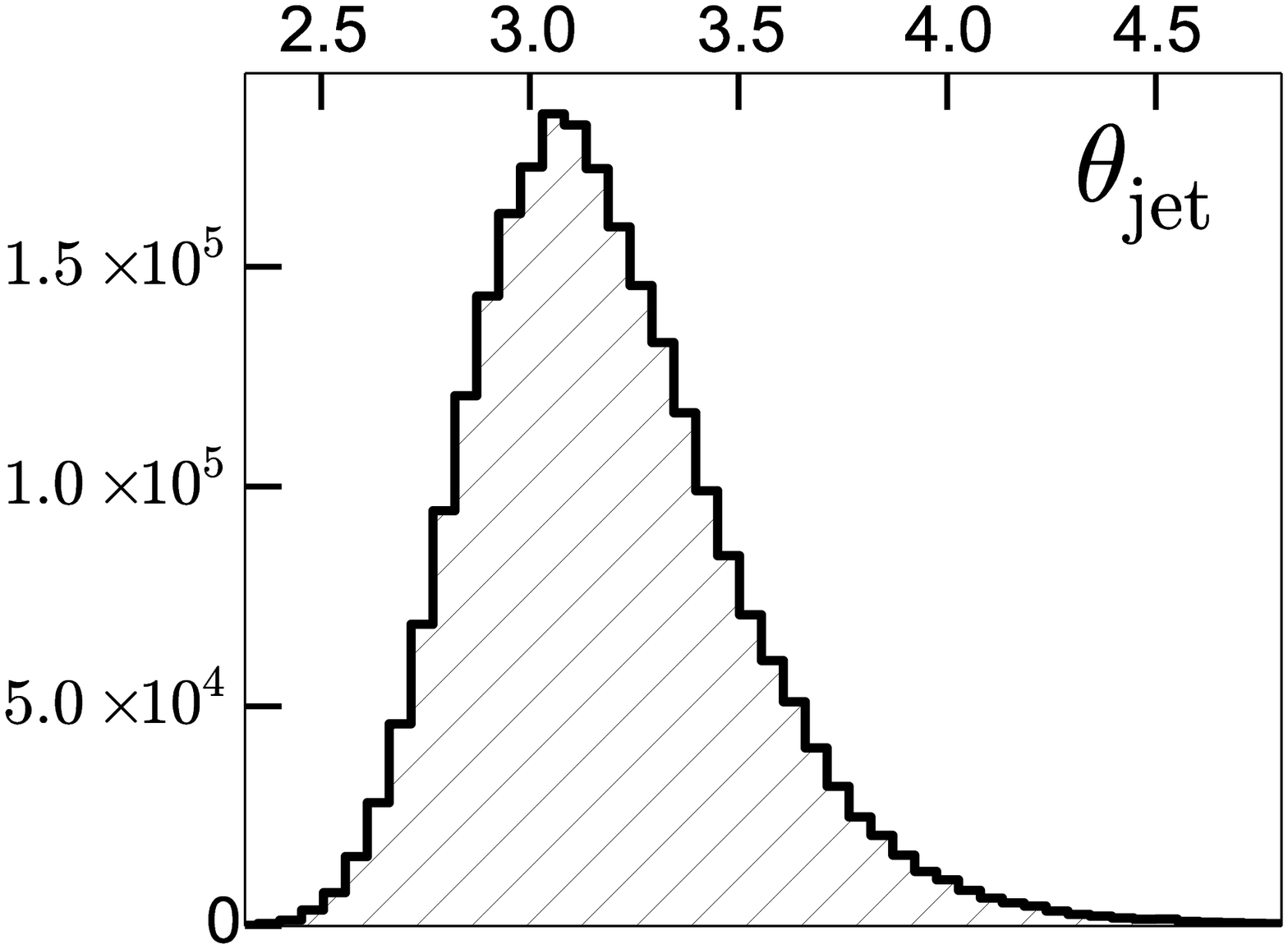}  &
 \includegraphics[width=0.30\columnwidth]{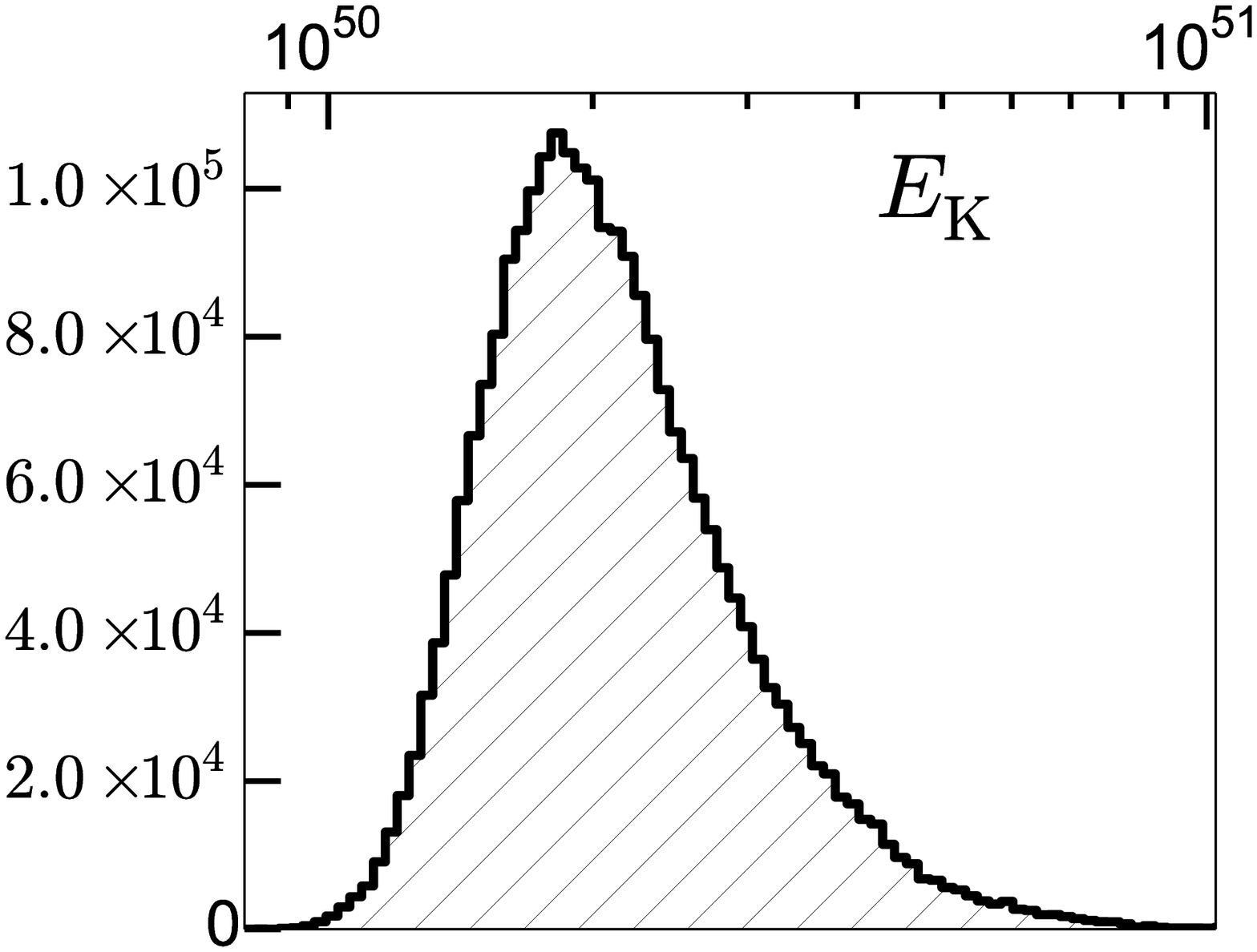} \\  
\end{tabular}
\caption{Posterior probability density functions for the physical parameters for GRB~120404A in 
a constant density environment from MCMC simulations. We have restricted $E_{\rm K, iso, 52} 
< 500$, $\epsilon_{\rm e} < \nicefrac{1}{3}$, and $\epsilon_{\rm B} < \nicefrac{1}{3}$. 
\label{fig:120404A_ISM_hists}}
\end{figure}

\begin{figure}
\begin{tabular}{ccc}
\centering
 \includegraphics[width=0.30\columnwidth]{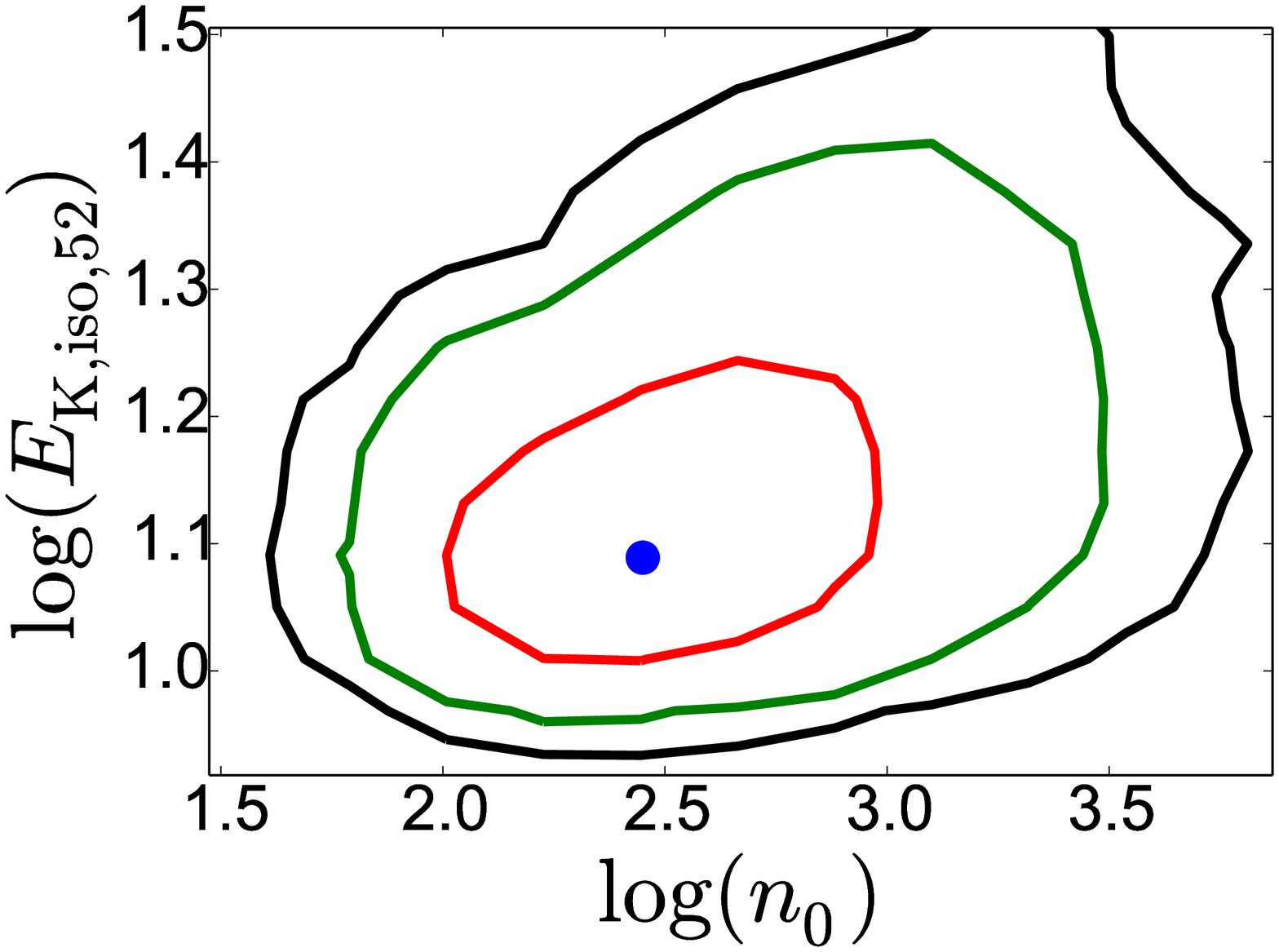} &
 \includegraphics[width=0.30\columnwidth]{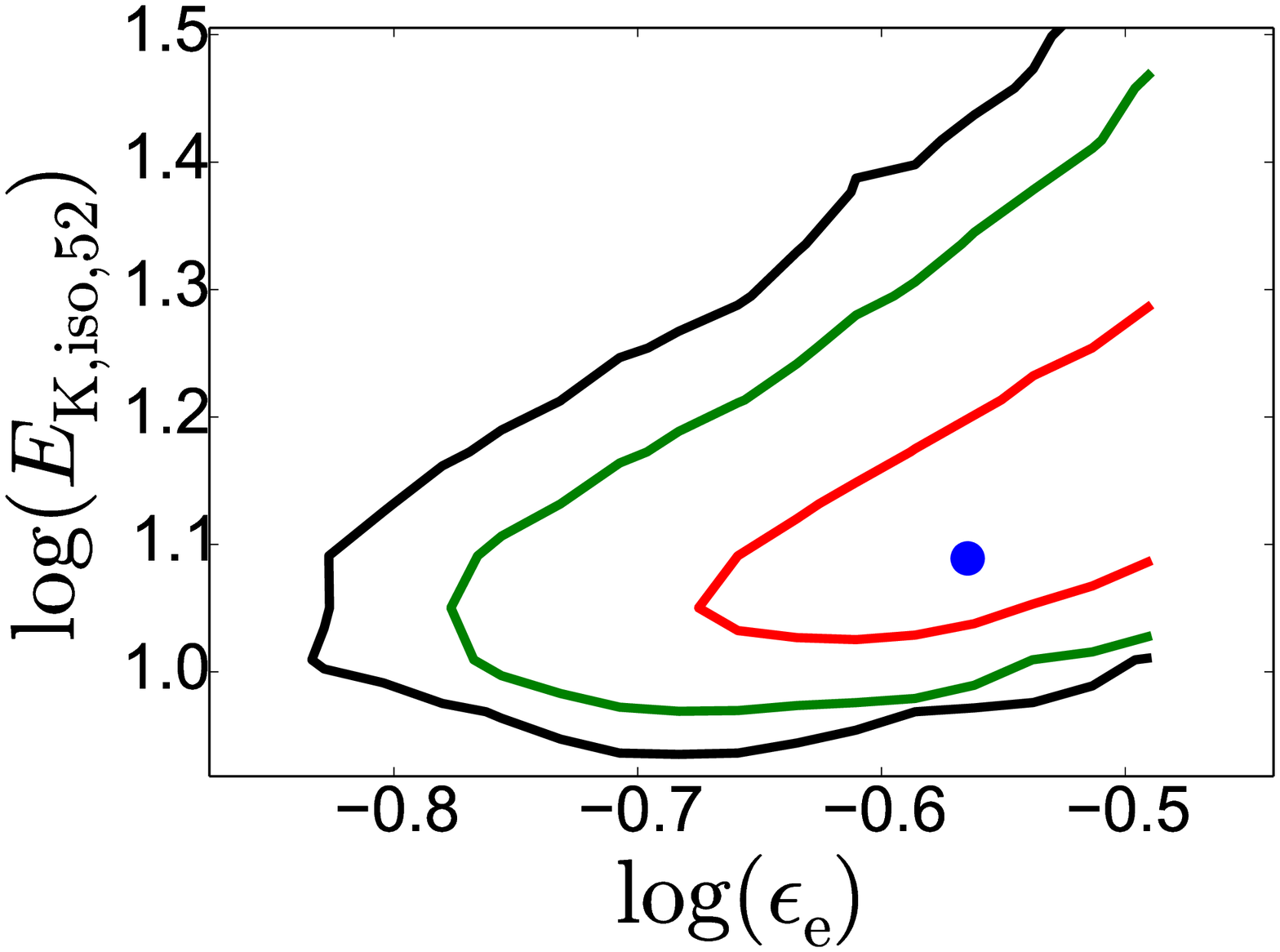} &
 \includegraphics[width=0.30\columnwidth]{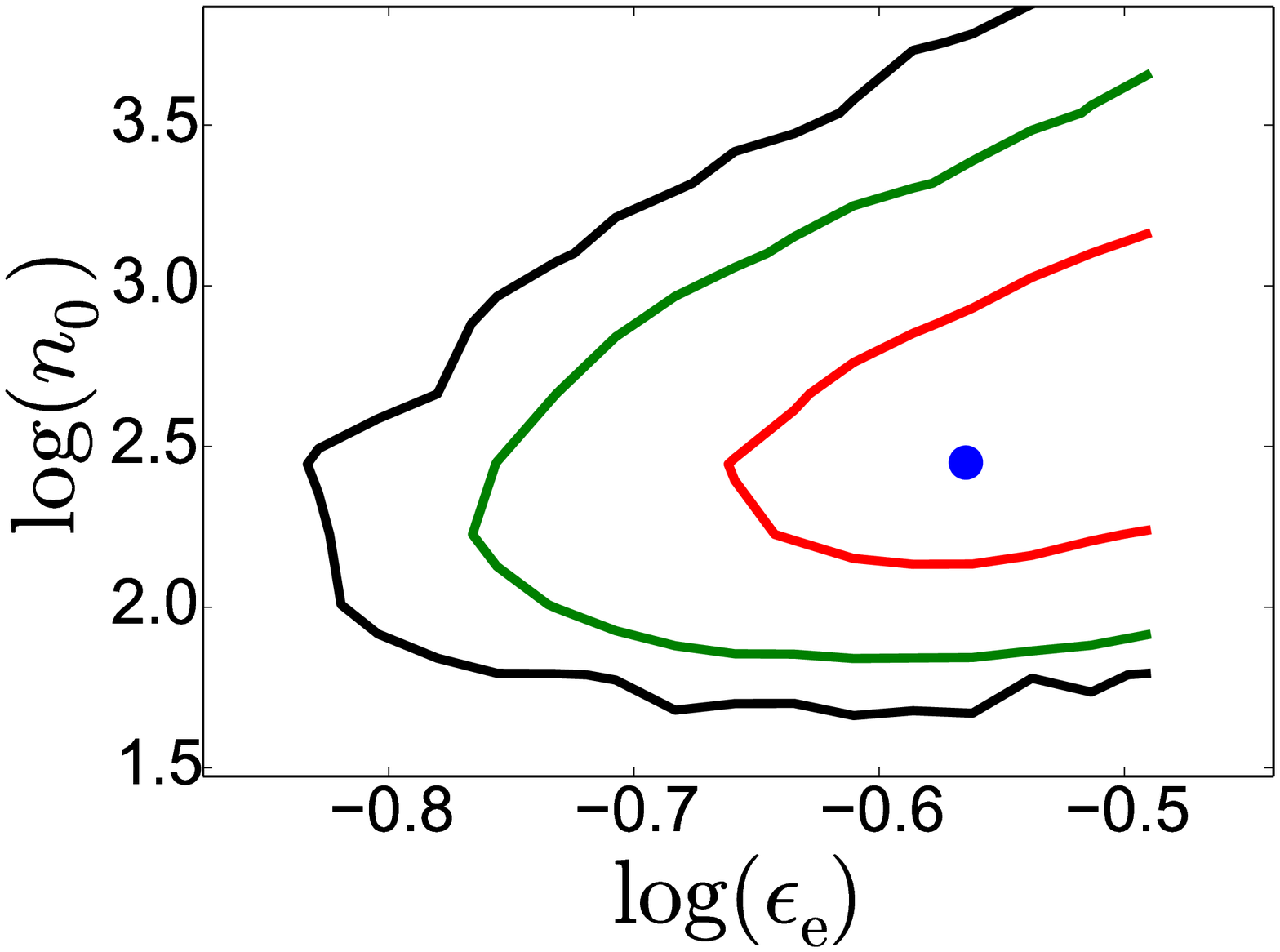} \\
 \includegraphics[width=0.30\columnwidth]{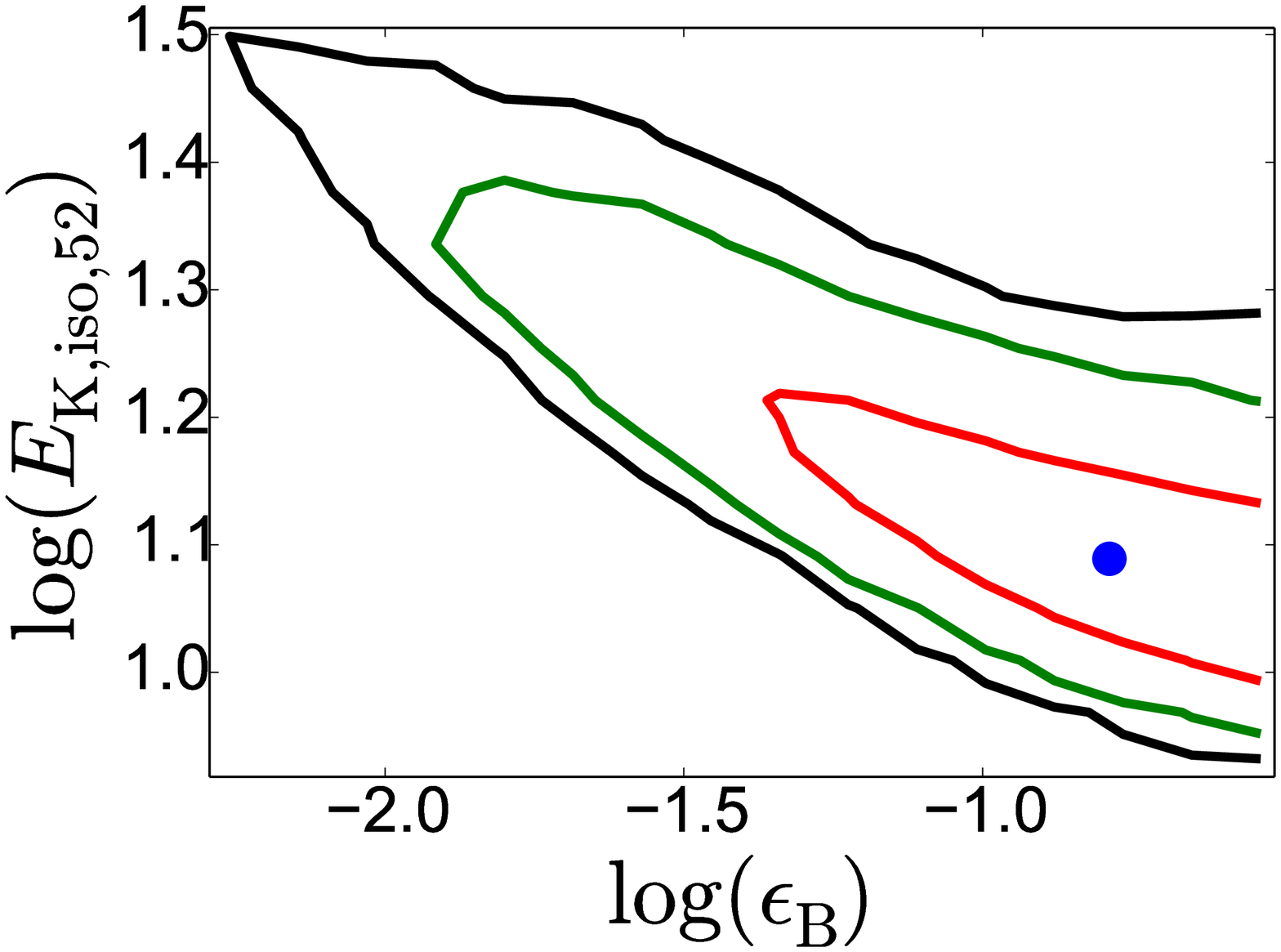} &
 \includegraphics[width=0.30\columnwidth]{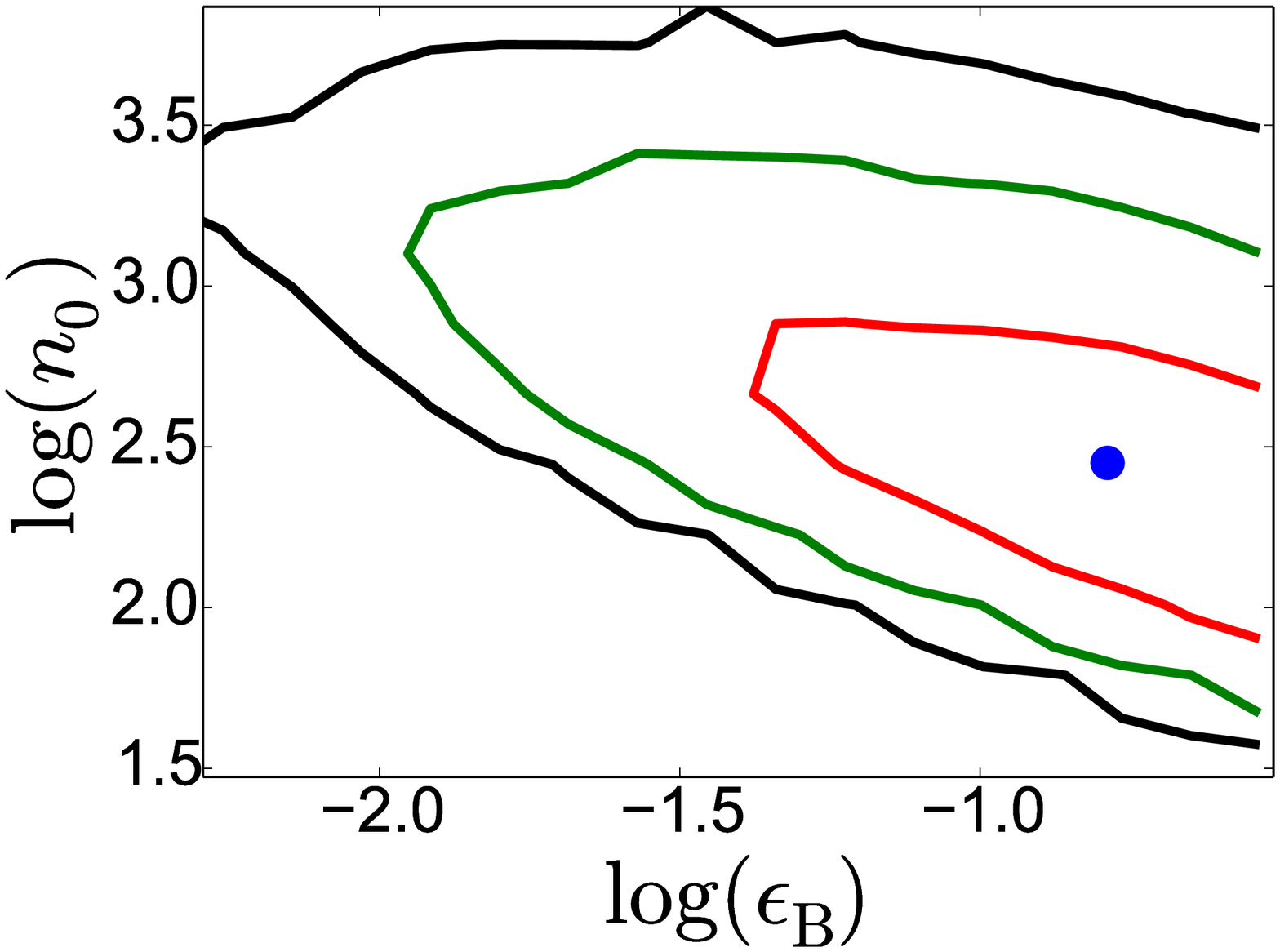} &
 \includegraphics[width=0.30\columnwidth]{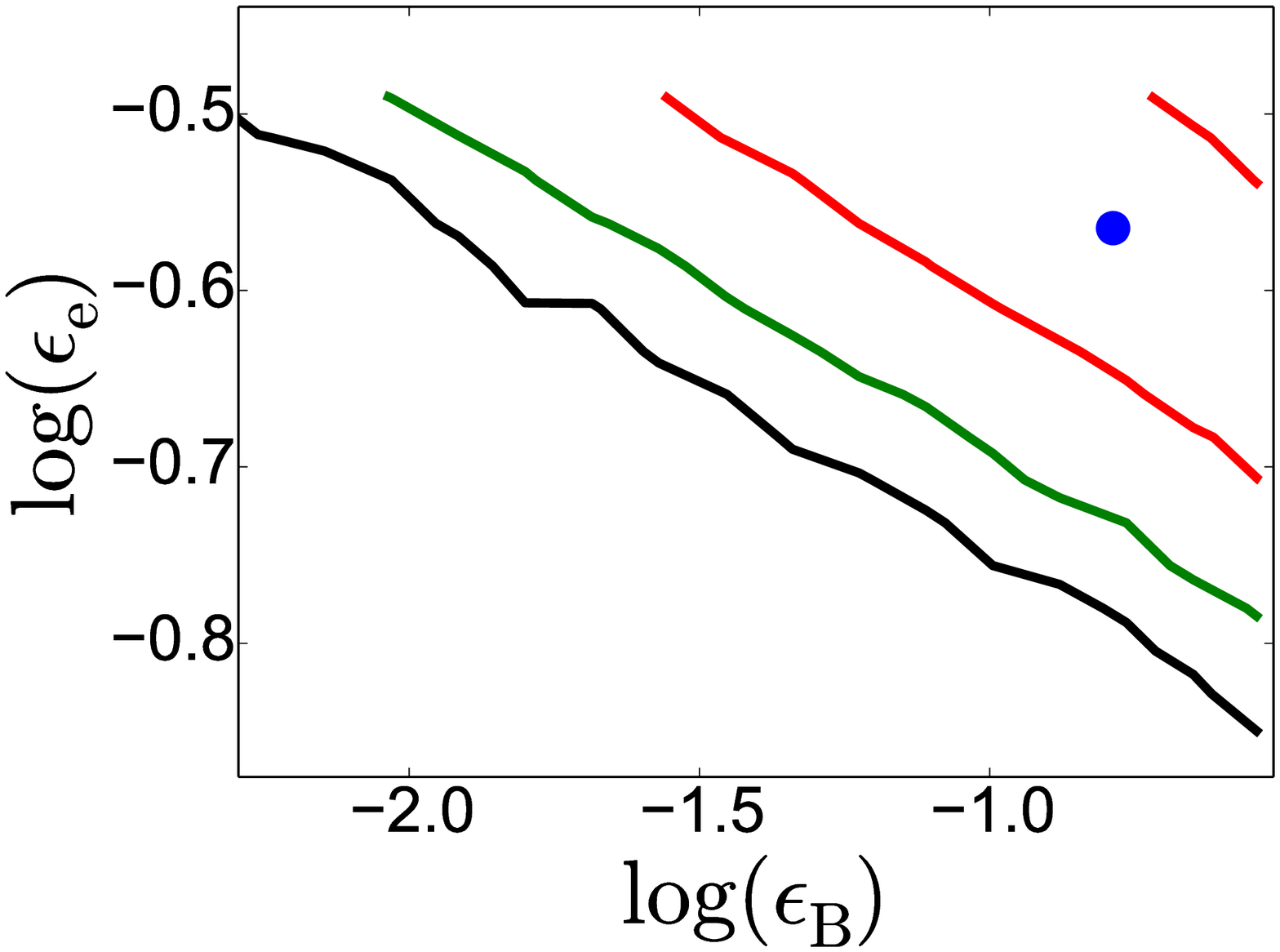} \\
\end{tabular}
\caption{1$\sigma$ (red), 2$\sigma$ (green), and 3$\sigma$ (black) contours for correlations
between the physical parameters, \EKiso, \dens, \epse, and \epsb\ for GRB~120404A, in the ISM model 
from Monte Carlo simulations. We have restricted $E_{\rm K, iso, 52} < 500$, $\epsilon_{\rm e} < 
\nicefrac{1}{3}$, and $\epsilon_{\rm B} < \nicefrac{1}{3}$. See the on line version of this Figure 
for additional plots of correlations between these parameters and $p$, $t_{\rm jet}$, $\thetajet$, 
$E_{\rm K}$, and $A_{\rm V}$. \label{fig:120404A_ISM_corrplots}}
\end{figure}

\begin{figure*}
\begin{tabular}{cc}
 \centering
 \includegraphics[width=0.47\textwidth]{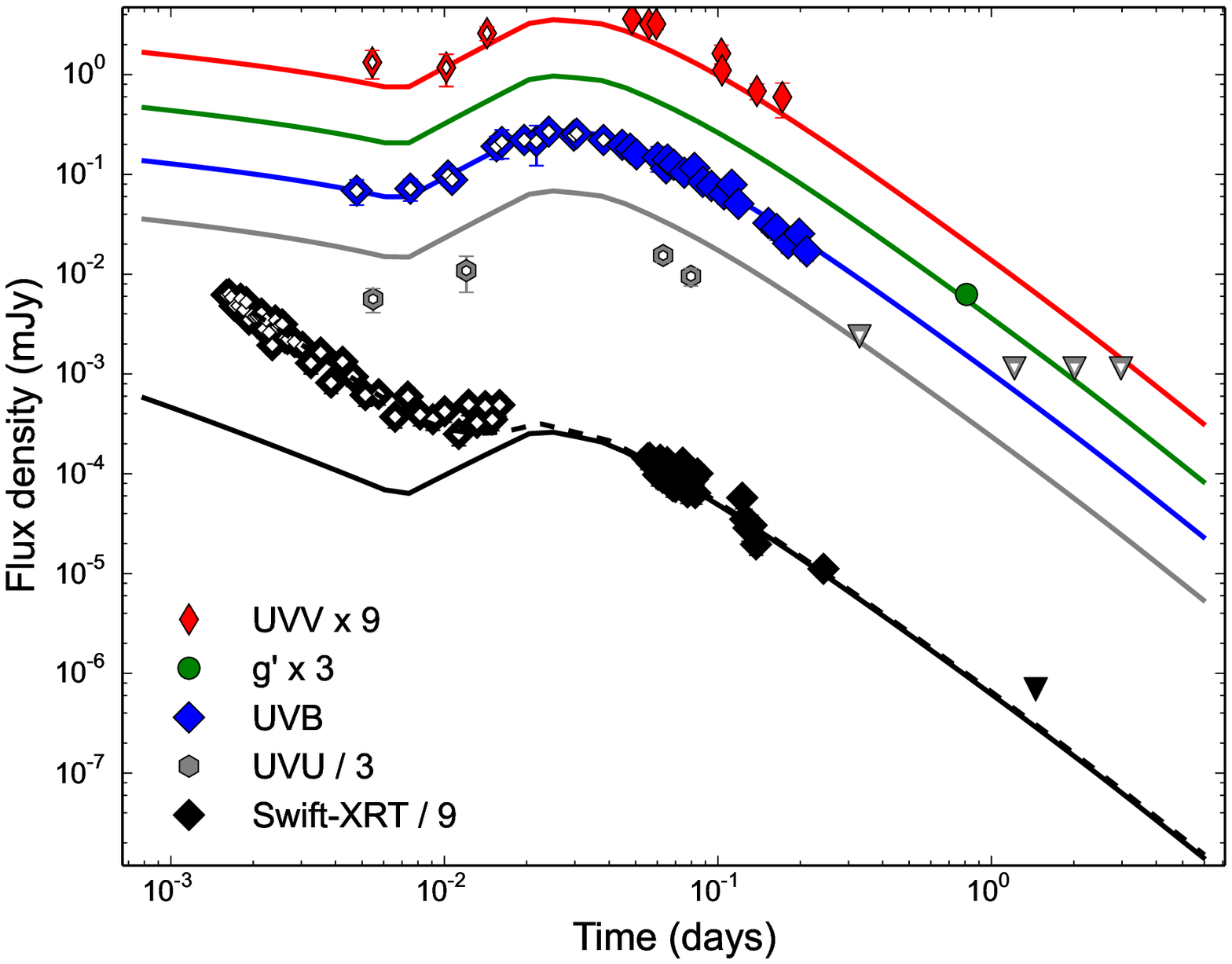} &
 \includegraphics[width=0.47\textwidth]{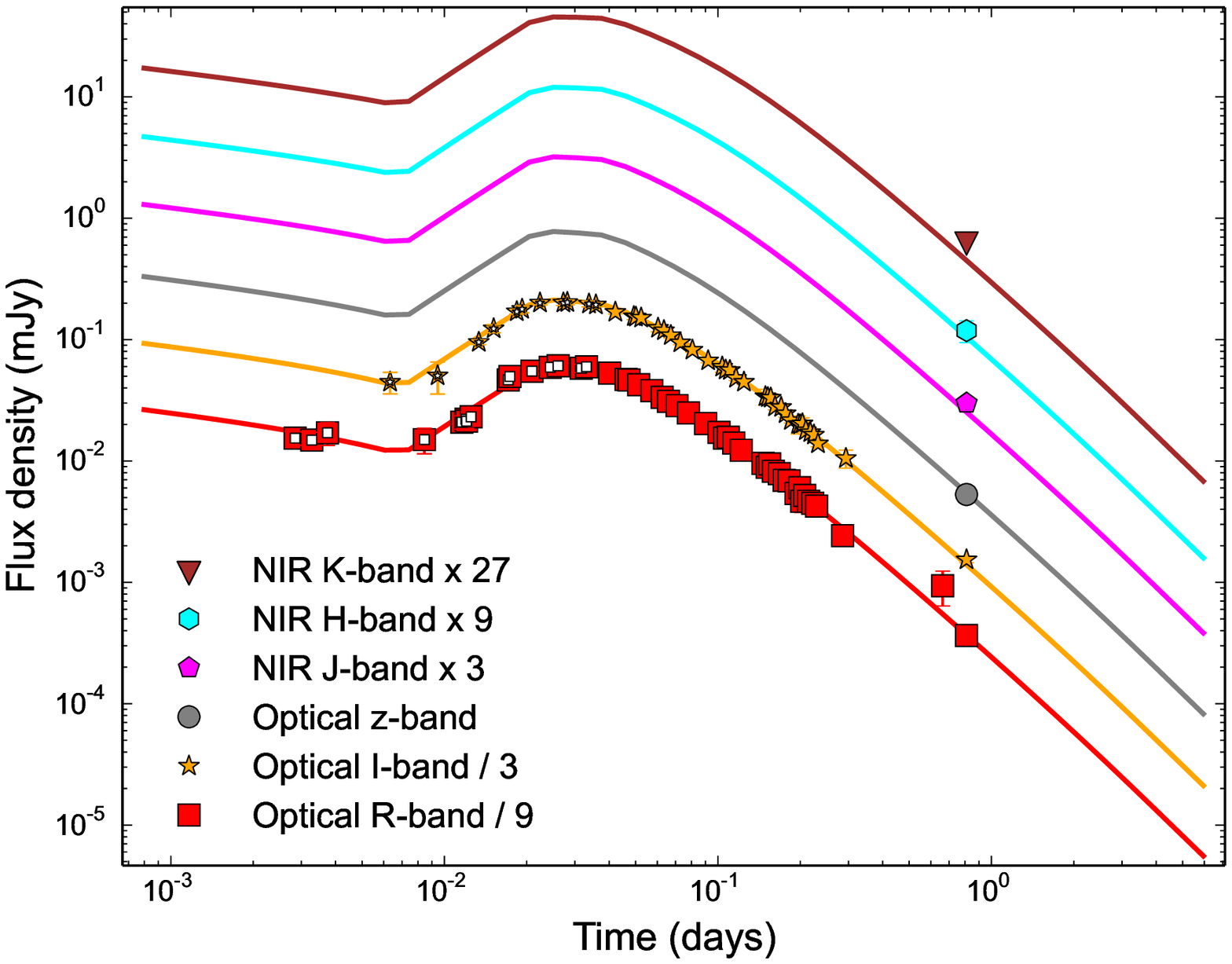} \\
 \includegraphics[width=0.47\textwidth]{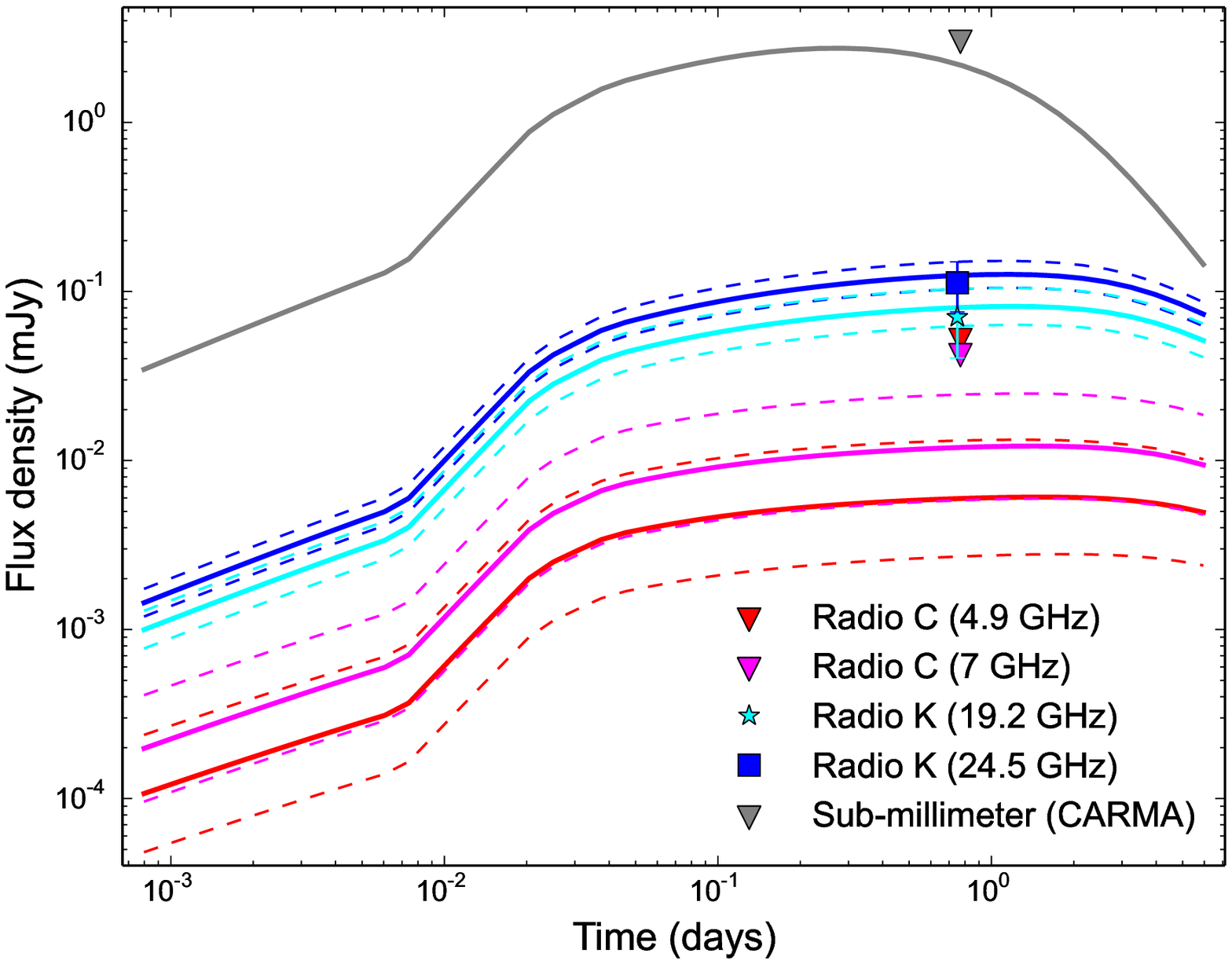} &
 \includegraphics[width=0.47\textwidth]{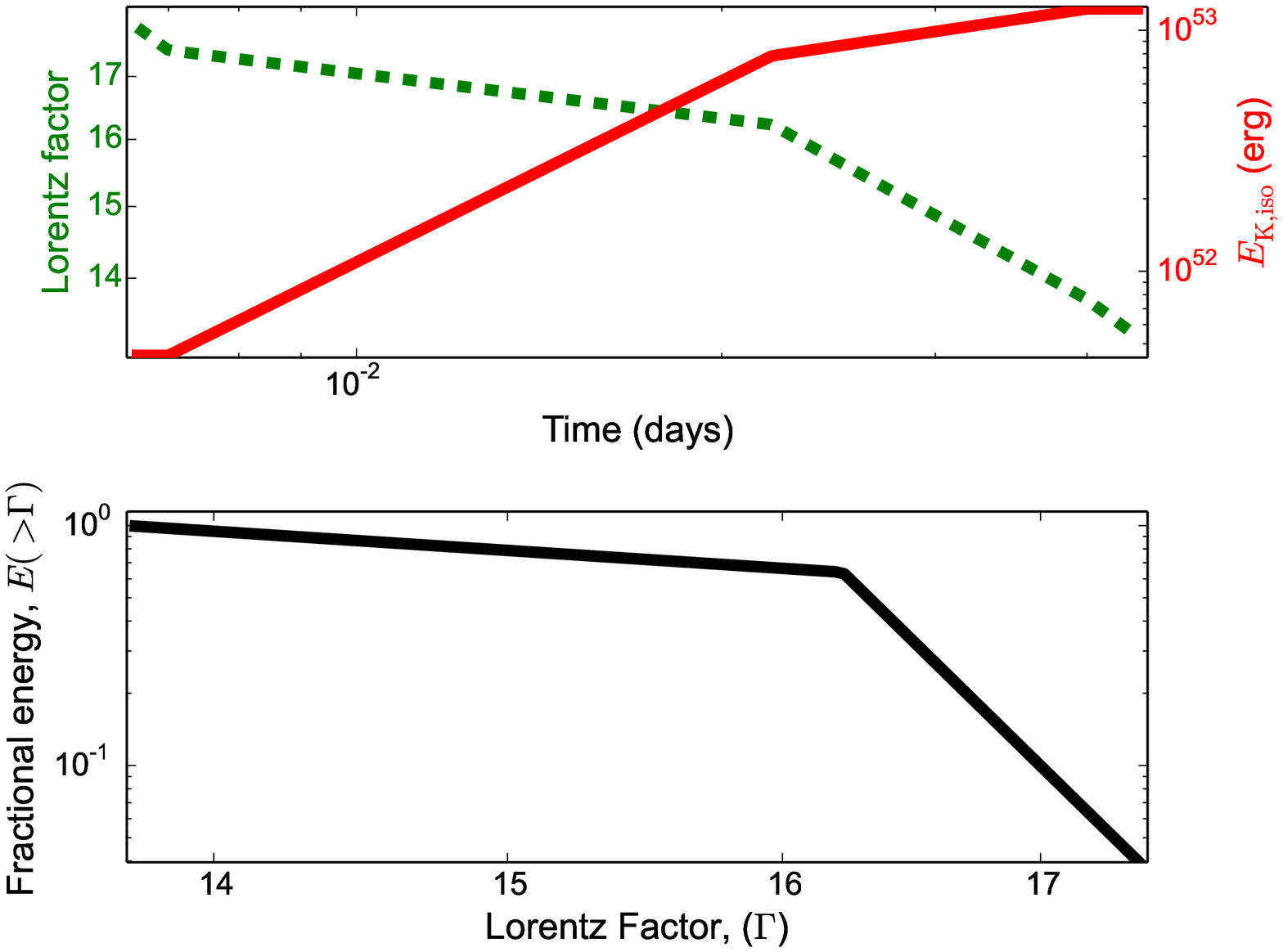} \\ 
\end{tabular}

\caption{X-ray, UV (top left), optical (top right), and radio (bottom left) light curves of 
GRB~120404A, with the full afterglow model (solid lines), including energy injection before 
0.04\,d. The X-ray data before 0.008\,d is likely dominated by high-latitude prompt emission and we 
do not include these data in our analysis; the best fit power law to the X-ray data before 0.008\,d 
added to the blastwave model is shown in the upper left panel (black, dashed). The dashed 
envelopes around the radio light curves indicate the expected effect of scintillation at the 
$1\sigma$ level. The data indicated by open symbols are not used to determine the parameters of the 
forward shock (the MCMC analysis). The $U$-band data are strongly affected by IGM absorption and 
are not included in the fit. Bottom right: blastwave Lorentz factor (green, dashed; upper 
sub-panel) 
and isotropic equivalent kinetic energy (red, solid; upper sub-panel) as a function of time, 
together with the energy distribution across ejecta Lorentz factors (black, solid; lower sub-panel) 
as determined from fitting the X-ray/UV/optical re-brightening at 0.04\,d.
\label{fig:120404A_enj}}
\end{figure*}

\subsubsection{Forward shock model at $t \gtrsim 0.04$\,d}
\label{text:120404A:FS}
We interpret the optical light curve peaks at around 2500\,s as the end of a period of energy 
injection, after which the afterglow evolves according to the standard framework with a fixed 
energy, $E_{\rm K, iso, f}$. We use the data after 0.04\,d for estimating the parameters of the 
blastwave shock and employ our MCMC tools described in Section \ref{text:modeling} to model 
the afterglow after this time.

The parameters of our highest likelihood ISM model are $p\approx2.06$, $\epse\approx0.27$, 
$\epsb\approx0.16$, $\dens\approx2.8\times10^2$\,\pcc, $\EKiso\approx1.2\times10^{53}$\,erg, 
$\tjet\approx6.6\times10^{-2}$\,d, and $\AV\approx0.13$\,mag. The Compton $y$-parameter for this 
model is $\approx0.9$. The blastwave Lorentz factor is $\Gamma=4.1(t/1\,{\rm d})^{-3/8}$ and the jet 
opening angle is $\thetajet\approx3.1\degr$. The beaming-corrected kinetic energy is 
$\EK\approx1.7\times10^{50}\,{\rm erg}$, while the beaming corrected $\gamma$-ray energy is $\Egamma 
\approx 1.2\times10^{50}$\,erg (1--$10^4$\,keV; rest frame). The MCMC analysis yields an isotropic 
equivalent kinetic energy of \EKiso$=1.3^{+0.4}_{-0.2}\times10^{53}$\,erg, which is similar to the 
value of $1.9^{+0.7}_{-0.1}\times10^{53}$\,erg derived by \cite{gmh+13}. The high 
circumburst density of $\log{(\dens)}=2.5^{0.4}_{-0.3}$ is also consistent with the value of 
$\log{(\dens)}=2.4^{+0.02}_{-0.2}$ determined by \citealt{gmh+13}), and is driven by the low flux 
density and steep spectrum in the radio. However, unlike \cite{gmh+13}, our low value of 
$p\approx2.1$ allows us to match the NIR to optical SED and the normalization of the X-ray 
light curve. We note that the high circumburst density also results in a low cooling\footnote{Once 
again, $\nuc = \sqrt{\nu_3\nu_{11}}$, with $\nu_3\approx8.0\times10^{10}$\,Hz, and 
$\nu_{11}\approx5.0\times10^{10}$\,Hz at 1\,d (see Footnote \ref{footnote:nuc}).} frequency: 
in our highest likelihood model $\nuc\ < \nua$ and the spectrum remains in the fast 
cooling regime through the duration of the observations (spectrum 4 of \citealt{gs02}), with 
$\nuac\approx5.9\times10^{10}$\,Hz, $\nusa\approx2.0\times10^{11}$\,Hz, 
$\numax\approx9.1\times10^{11}$\,Hz, and $\nuc\approx6.3\times10^{10}$\,Hz at 1\,d.

We plot the posterior density functions for the all parameters in Figure 
\ref{fig:120404A_ISM_hists} and correlation contours in Figure \ref{fig:120404A_ISM_corrplots}; we 
list our measured values for the physical parameters in Table \ref{tab:enjsummary}. In the wind and 
ISM models, light curves at all frequencies become indistinguishable following a jet break. Since 
the jet break occurs early, soon after the start of the data used for deriving the parameters of 
the 
blastwave, we expect a viable wind model to exist as well. We discuss this model in Appendix 
\ref{appendix:120404A_wind}.


\subsubsection{Energy injection model}
\label{text:120404A:enj}
We model the light curves before $t_0=0.04$\,d by injection of energy into the blastwave shock. We 
use the afterglow parameters for the highest-likelihood model (Section \ref{text:120404A:FS}) as 
the 
final parameters following the end of energy injection, with $E_{\rm 
K,iso,f}=1.2\times10^{53}$\,erg. We find that the optical and X-ray light curves can be modeled 
well 
by two subsequent periods of energy injection, beginning at $t_2 = 7\times10^{-3}$\,d (Figure 
\ref{fig:120404A_enj}):
\begin{equation}
\EKiso(t) = 
  \begin{cases}
      E_{\rm K,iso,f}, & t > t_{\rm 0} = 0.04\,{\rm d} \\
      E_{\rm K,iso,f}\left(\frac{t}{t_{\rm 0}}\right)^{0.75}, &
      t_1 = 2.2\times10^{-2}\,{\rm d} < t < t_{\rm 0} \\
      E_{\rm K,iso,f}\left(\frac{t_1}{t_{\rm 0}}\right)^{0.75}\left(\frac{t}{t_1}\right)^{2.5},&
      t_2  = 7.0\times10^{-3}\,{\rm d} < t < t_1\\
      E_{\rm K,iso,f}\left(\frac{t_1}{t_{\rm 0}}\right)^{0.75}\left(\frac{t_2}{t_1}\right)^{2.5},&
      t < t_2\\
  \end{cases}
\end{equation}
In this model, \EKiso\ increases by a factor of $\approx18$ from 
$4.4\times10^{51}$\,erg to $7.8\times10^{52}$\,erg between $7.0\times10^{-3}$\,d and 
$2.2\times10^{-2}$\,d, and then by $\approx50\%$ to $E_{\rm K,iso,f}\approx1.2\times10^{53}$\,erg 
between $2.2\times10^{-2}$\,d and $0.04$\,d. Thus $\approx95\%$ of the final kinetic energy is 
injected into the blastwave between $7.0\times10^{-3}$\,d and 0.04\,d. In comparison, $\Egammaiso 
\approx 10^{53}$\,erg is similar to the final kinetic energy. The blastwave Lorentz factor 
decreases 
from $\Gamma\approx17.4$ at $7.3\times10^{-3}$ to $\Gamma\approx16.2$ at $2.2\times10^{-2}$\,d, and 
then to $\Gamma\approx13.7$ at the end of energy injection at $0.04$\,d. The value of $m$ derived 
above corresponds to $s\approx3.7$ for $13.7\lesssim\Gamma\lesssim16.2$ and $s\approx40$ for 
$16.2\lesssim\Gamma\lesssim17.4$.

\section{Discussion}
\label{text:discussion}
\subsection{Parameter distributions}
We now turn to the question of how GRBs that exhibit simultaneous optical and X-ray re-brightening 
episodes compare with each other, and with events that do not exhibit such a feature. For this 
purpose, we use the compilation of measurements of $E_{\gamma}$, $\theta_{\rm jet}$, $E_{\rm K}$, 
and $n_{0}$ (or $A_*$) from the literature \citep{pk02,yhsf03,fb05,gngf07,cfh+10,cfh+11} reported 
in \cite{lbt+14}. This sample includes GRBs from the pre-\Swift\ era, as well as \Swift\ and 
\textit{Fermi} events.
 
\begin{figure*}[ht]
  \centering
  \begin{tabular}{cc}
   \centering
   \includegraphics[width=\columnwidth]{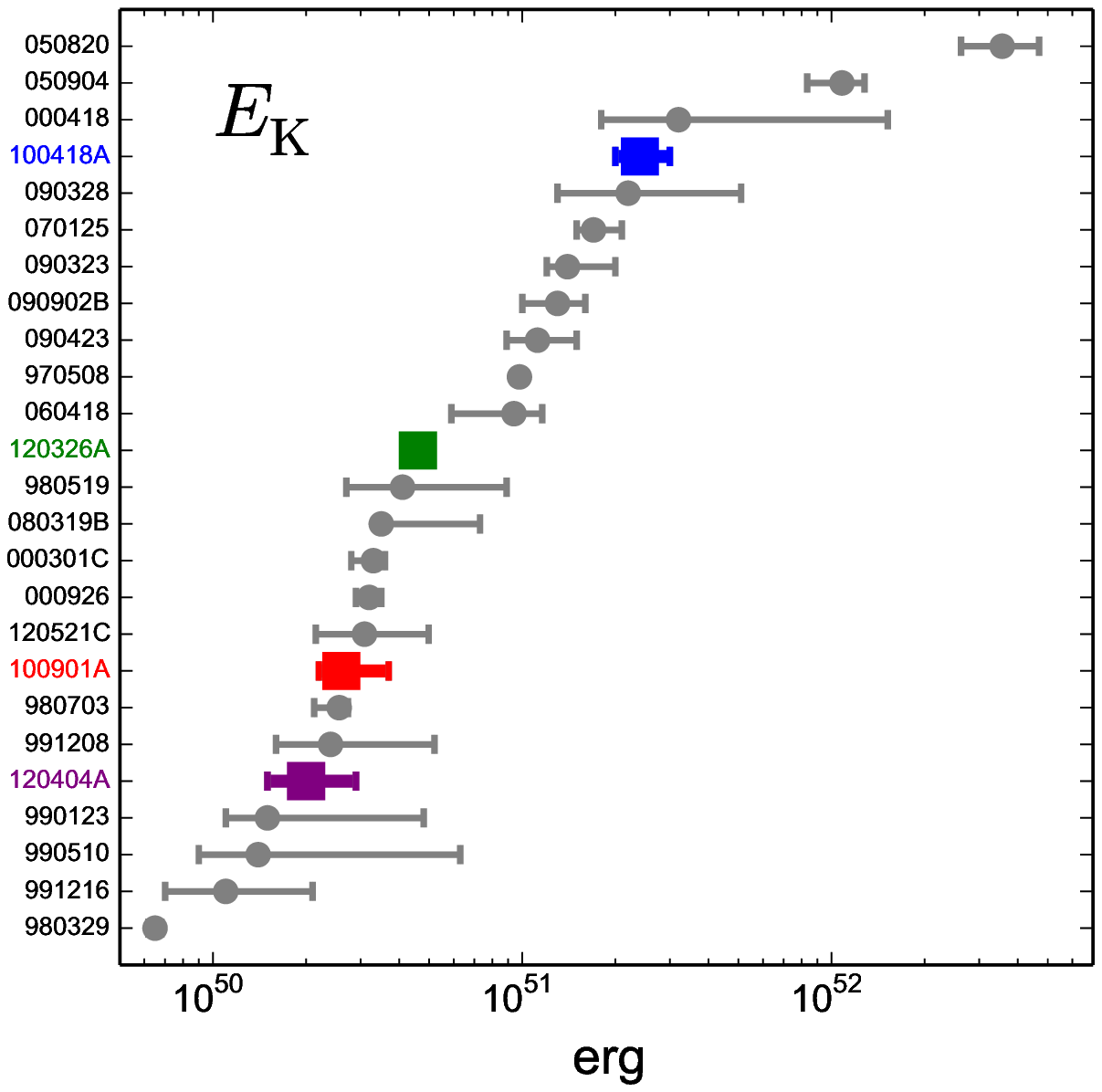} &
   \includegraphics[width=\columnwidth]{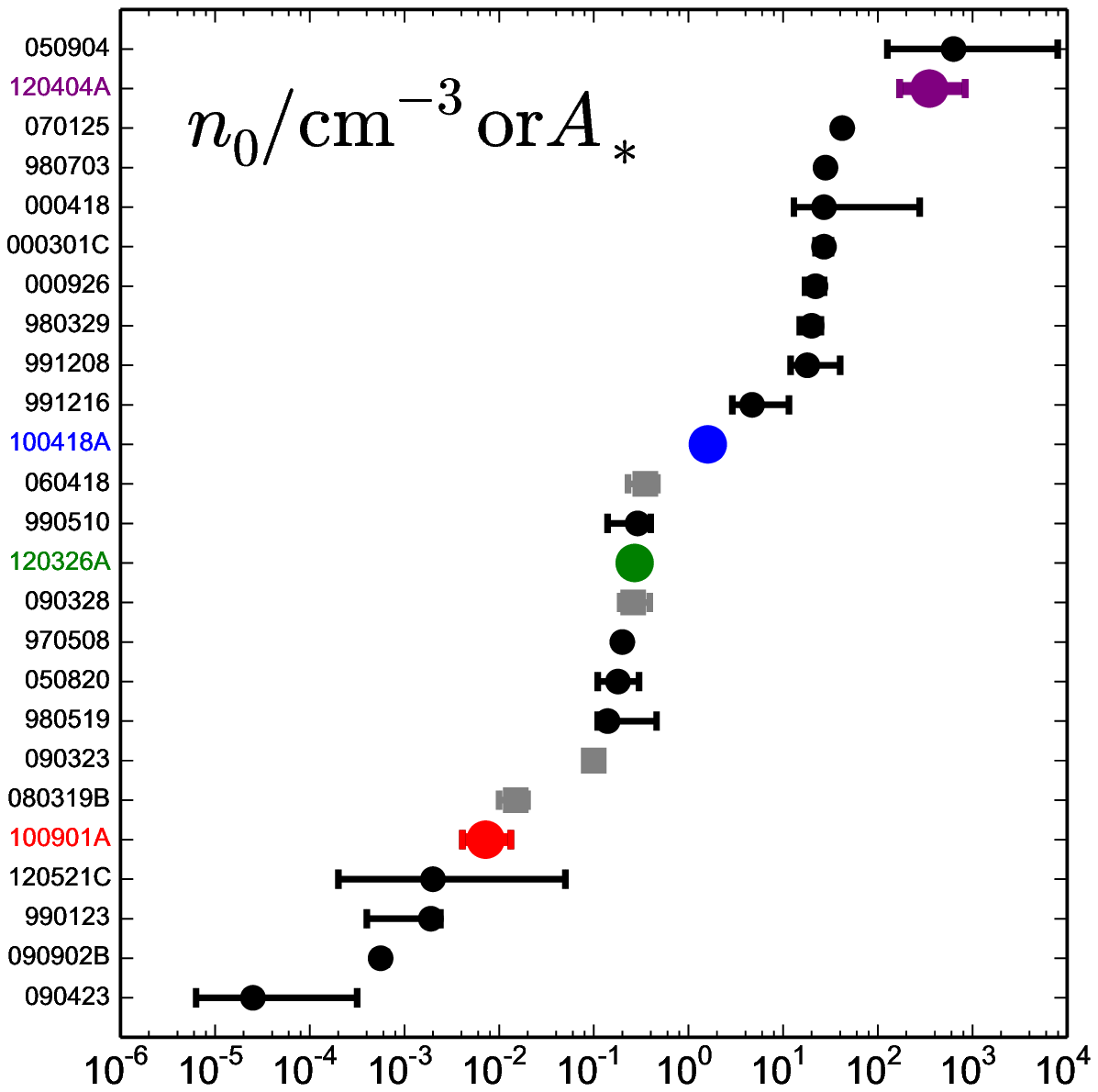}\\
  \end{tabular}
 \caption{Beaming-corrected kinetic energy (left) and circumburst density (right) for both ISM 
(black circles) and wind-like environments (grey squares). The four GRBs in our analysis 
100418A (blue), 100901A (red), 120326AA (green), and 120404A (purple), do not appear distinct from 
the comparison sample \citep[grey and black;][]{pk02, yhsf03, ccf+08, cfh+10, cfh+11,lbt+14}.}
 \label{fig:comp_EK}
\end{figure*}

The radio to X-ray observations of all four GRBs presented here can be fit by constant density 
ISM models. The best fit models yield densities from $\approx3\times10^{-3}$ to 
$\approx3\times10^{2}$\,\pcc and final beaming-corrected kinetic energies from 
$\approx2\times10^{50}$\,erg to $\approx2\times10^{51}$\,erg, spanning the full range of values 
inferred from GRB afterglow modeling (Figure \ref{fig:comp_EK}). We constrain the jet break time and 
hence the opening angle of the GRB jet in each case, and find 
$\thetajet\approx2\,\degr$--$21\,\degr$, spanning the range inferred from the comparison sample. 

\begin{figure*}[ht]
 \begin{tabular}{cc}
  \centering
  \includegraphics[width=\columnwidth]{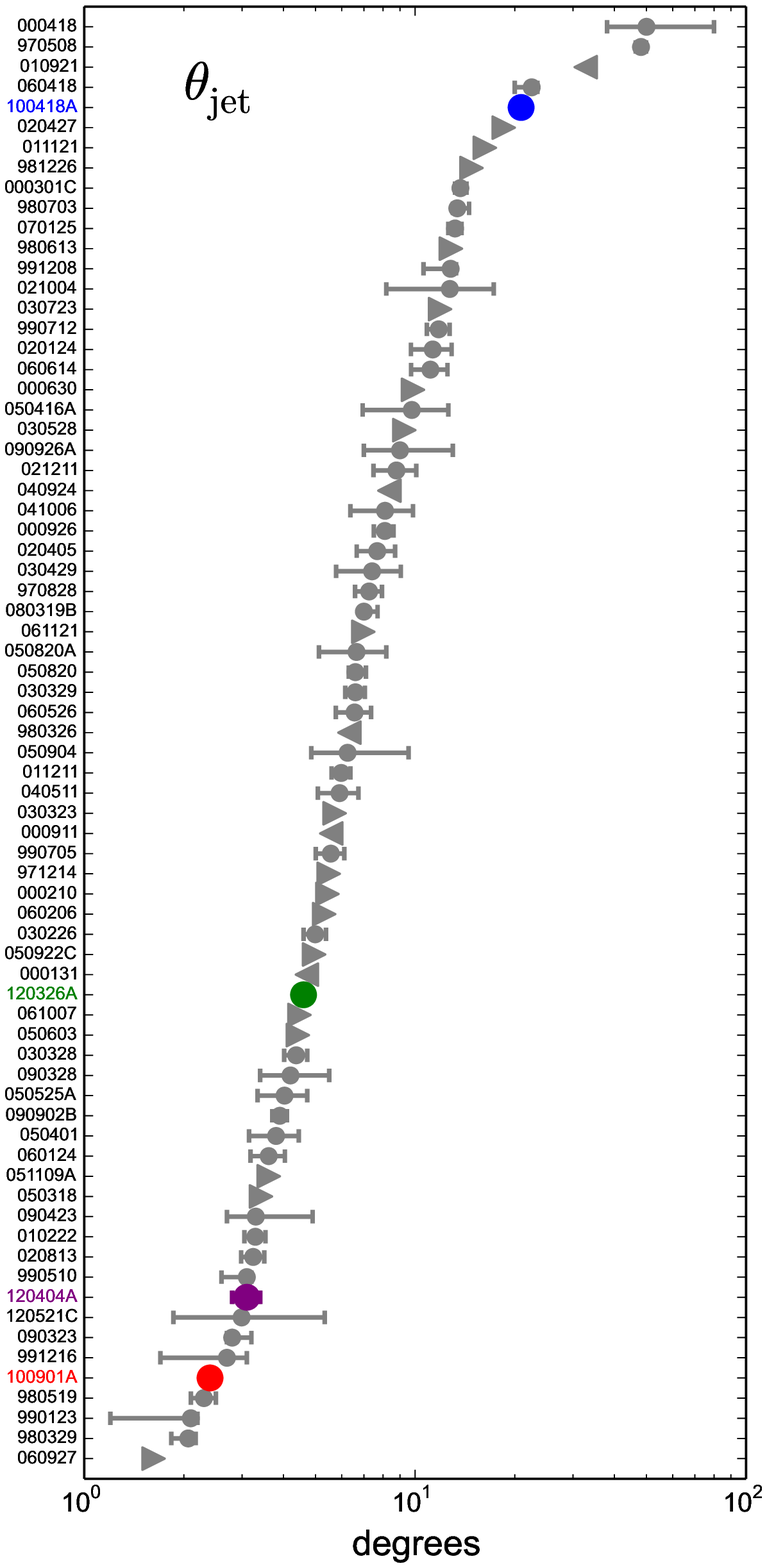} &
  \includegraphics[width=\columnwidth]{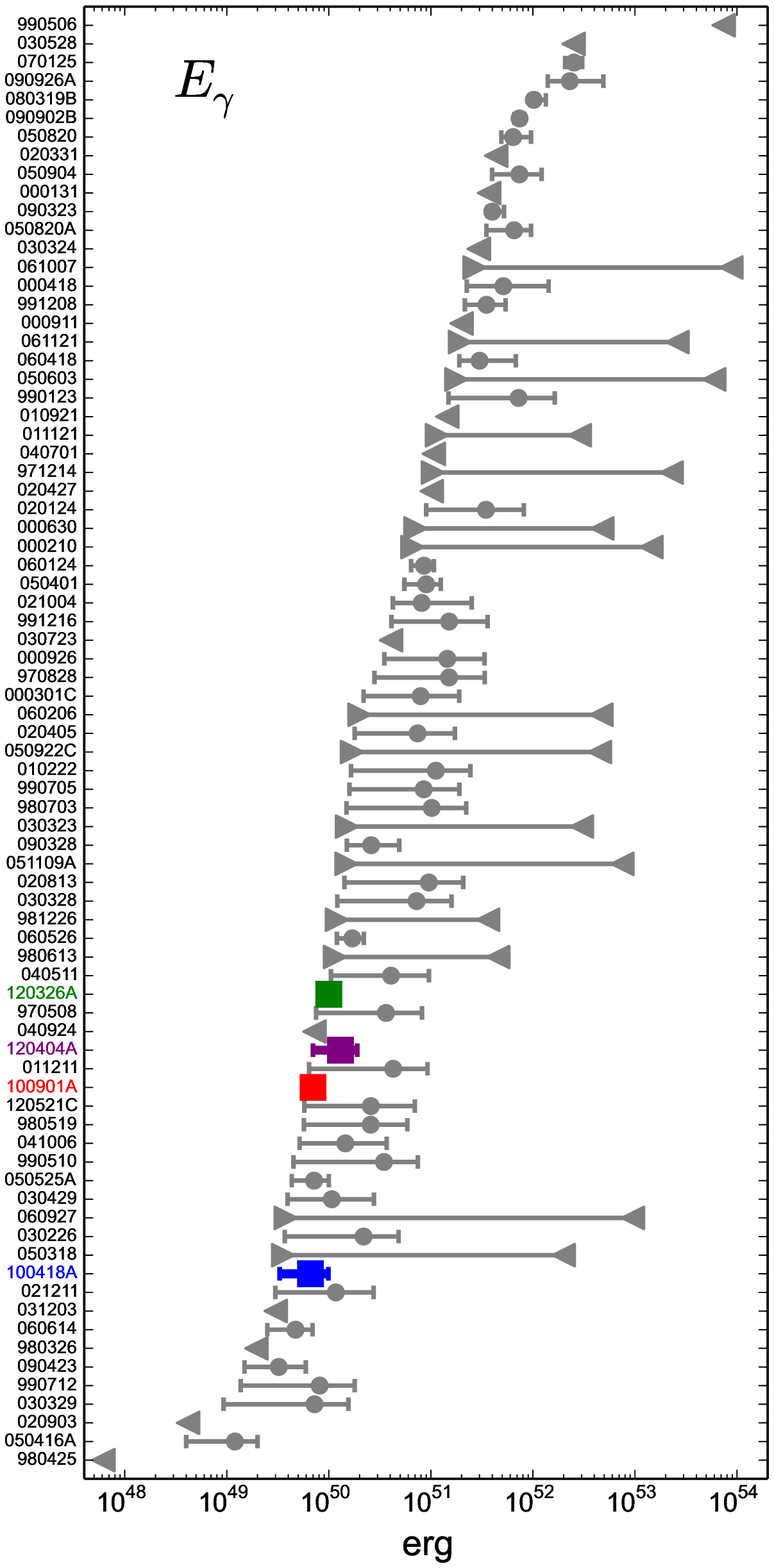}\\
 \end{tabular}
 \caption{Beaming-corrected $\gamma$-ray energy (left) and jet opening angle (right)
for the events in our analysis, GRB 100418A (blue), 100901A (red), 120326AA (green), and 120404A 
(purple), together with a comparison sample of long GRBs \citep[grey;][]{fb05, gngf07, cfh+10, 
cfh+11,lbt+14}. The isotropic-equivalent $\gamma$-ray energy for GRB~050904 is taken from 
\citet{agf+08}, and for GRB~090423 from \citet{sdvc+09}. The four GRBs exhibiting X-ray and optical 
re-brightening episodes do not appear distinct from the comparison sample in $\theta_{\rm jet}$, 
but appear to all reside at lower values of $E_{\gamma}$ than the median.} \label{fig:comp_Egamma}
\end{figure*}

The beaming-corrected $\gamma$-ray energies in the 1--$10^4$\,keV rest frame energy band of the 
events in our sample range from $5\times10^{49}$\,erg to $1.3\times10^{50}$\,erg, while the 
median\footnote{The uncertainty on the median is computed 
using Greenwood's formula for the variance of the Kaplan-Meier estimate of the cumulative 
distribution function. This method accounts for both upper and lower limits, which exist in the 
data.} beaming corrected $\gamma$-ray energy of the comparison sample is 
$\Egamma=\left(8.1^{+3.1}_{-4.1}\right)\times10^{50}$\,erg (95\% confidence interval, Figure 
\ref{fig:comp_Egamma}). Therefore, the observed values of $\Egamma$ for the events in our sample 
are all smaller than the best estimate for the median of the comparison sample.

\begin{figure}[h] 
  \centering
  \includegraphics[width=\columnwidth]{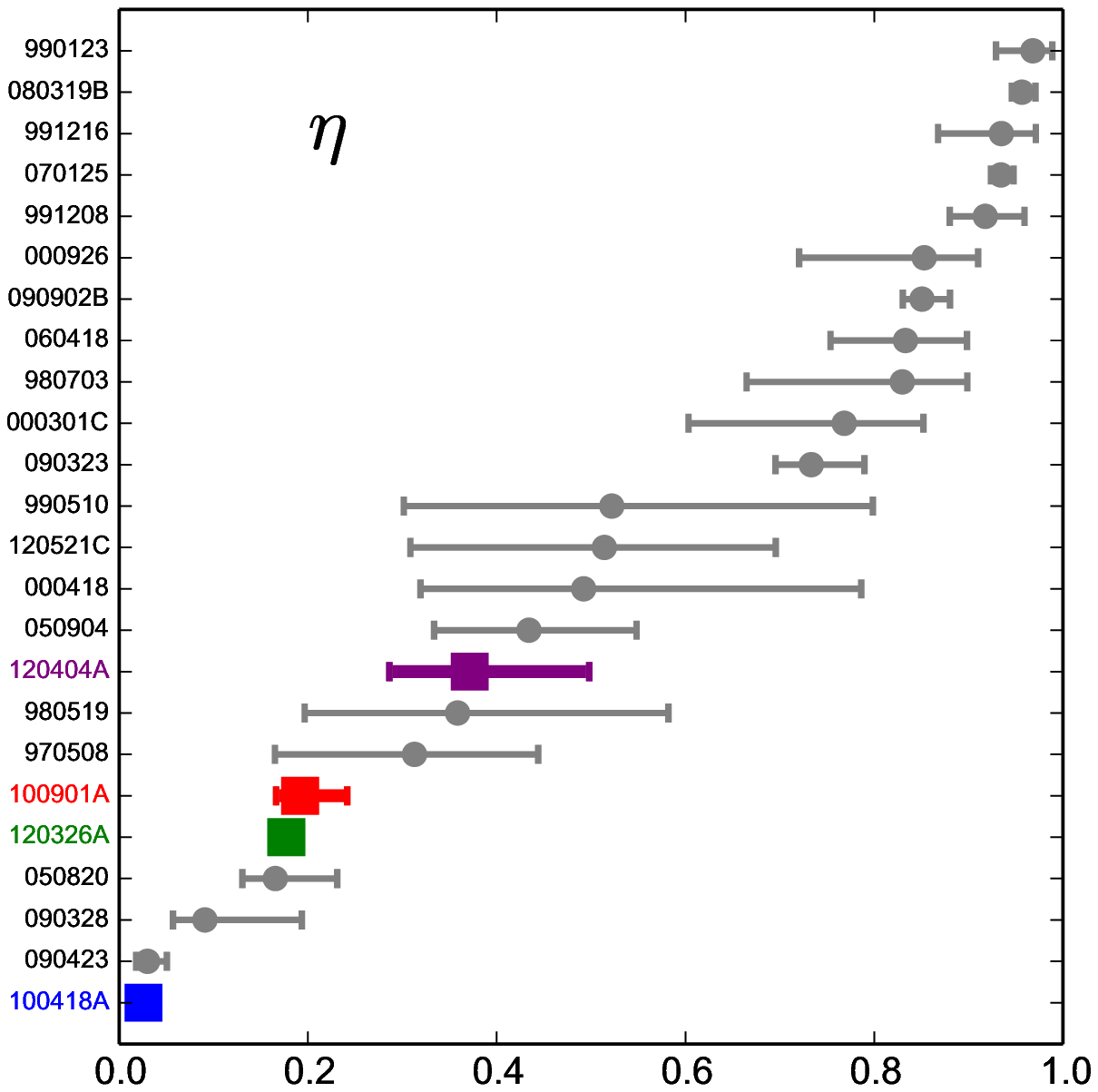}
 \caption{Radiative efficiency, $\eta$ for the events in our analysis, GRB 100418A (blue), 100901A 
(red), 120326AA (green), and 120404A (purple), together with a comparison sample of long GRBs 
\citep[grey;][]{fb05, gngf07, cfh+10, cfh+11,lbt+14}, with 68\% credible regions (error bars) about 
the median (points) computed from \EK\ and \Egamma\ using a Monte Carlo procedure. The four GRBs 
exhibiting X-ray and optical re-brightening episodes have lower radiative efficiencies than the 
median.} \label{fig:comp_eta}
\end{figure}

To further quantify this effect, we compute the radiative efficiency, $\eta \equiv 
\EK/(\Egamma+\EK)$ for each GRB and in the comparison sample. Since both \EK\, and \Egamma\ have 
associated uncertainties, while the expression for $\eta$ is non-linear in these two quantities, 
a proper accounting of the final uncertainty requires a Monte Carlo analysis. We generate 
$10^5$ Monte Carlo realizations of \EK\ and \Egamma\ for each burst, assuming a uniform 
distribution\footnote{We choose a uniform distribution instead of, say, a Gaussian 
distribution because the uncertainties on these parameters are frequently large compared to the 
mean, and a Gaussian distribution in linear space results in a tail of unphysical, negative 
values. A more detailed analysis requires the full posterior density functions for both \Egamma\ 
and \EK\ for every GRB, which are not available.} between the 1$\sigma$ error bars. We then compute 
and summarize the resulting distribution of $\eta$ using the median and $68\%$ credible intervals 
(Figure \ref{fig:comp_eta}). We find that the GRBs in our sample have systematically lower 
radiative efficiencies. This is consistent with the energy injection scenario, if the prompt 
$\gamma$-ray radiation is dominated by emission from the fast-moving ejecta, while a significant 
amount of kinetic energy is carried by slow-moving ejecta.

From this comparison, we conclude that events exhibiting simultaneous, multi-wavelength 
re-brightening episodes also span the same wide range of circumburst densities, jet opening angles, 
and beaming-corrected kinetic energies inferred from previous studies of events at $z\sim1$ and 
$z\gtrsim6$. However, the events reported here have smaller beaming-corrected $\gamma$-ray 
energies, and as a result lower radiative efficiencies than the comparison sample, suggesting
that the prompt radiation is dominated by ejecta at high Lorentz factors, while the bulk kinetic 
energy is dominated by slow-moving ejecta at least in these cases.

\subsection{Absence of reverse shock}
When energy injection into the blastwave is caused by a distribution of ejecta Lorentz factors, 
the reverse shock (RS) from the initial interaction of the leading edge of the ejecta with the 
circumburst medium is expected to continue to propagate through the ejecta until the end of the 
period of energy injection \citep{sm00}. During the period of energy injection, the afterglow SED 
is expected to be composed of contributions from both forward (FS) and reverse (RS) shocks, each 
with its three characteristic frequencies and flux normalizations. The spectral characteristics of 
the two shocks are expected to be related by $F_{\nu, \rm max, RS} (t) = F_{\nu, \rm max, 
FS} (t) \times\Gamma(t)$, $\nu_{\rm m, RS} (t) = \nu_{\rm m, FS}(t)/\Gamma^2(t)$, and $\nu_{\rm c, 
RS} (t) = \nu_{\rm c, FS} (t)$, while the two self-absorption frequencies should be related by 
$\nu_{\rm a, RS} (t) = \nu_{\rm a, FS}(t)\times\Gamma^{\lambda}(t)$, where 
$\lambda=\nicefrac{3}{5}$ when both shocks are fast cooling and $\lambda = \nicefrac{8}{5}$ when 
both shocks are slow cooling. The reverse shock is expected to last through the period of energy 
injection, whereupon all ejecta have been decelerated to a common Lorentz factor, and the two 
shocks decouple in their subsequent evolution.

A calculation of the reverse shock SED based on the above considerations, together with the Lorentz 
factor of the blastwave at the end of the period of energy injection (\tdec) and the SED of the 
forward shock at that time indicates that the reverse shock generically peaks around the millimeter 
band at \tdec\ \citep{sm00}. For GRB~100901A and 120404A, there are no data in the millimeter bands 
at the relevant time. For GRBs~120326A and 100418A where we have millimeter observations at 
$\approx\tdec$, we find that the RS light curves over-predict the observations by factors of 
$\approx2$--10. Additionally, the reverse shock would also contribute a flux density comparable to 
the forward shock in the X-rays and optical for all four events during energy injection, which would 
require suppressing the forward shock before the re-brightening by invoking even lower blastwave 
energies (and consequently requiring an even greater rate of energy injection) before the 
optical/X-ray peak. Thus the data are inconsistent with the presence of a strong reverse shock for 
these events.

We note that in our model there is a gap before the beginning of energy injection. If the energy 
injection is caused by a shell of material with a distribution of Lorentz factors (the `injective 
shell') catching up with the initial shell (`the impulsive shell'), then a long-lasting reverse 
shock in the injective shell is only expected when the two shells collide violently, with a relative 
Lorentz factor greater that the sound speed in the injective shell \citep{zm02}. If the injective 
shell is released from the central engine at roughly the same time as the impulsive shell, the 
collision between the two will be mild, and a reverse shock will not occur. This appears to be the 
case for the GRBs studied here, suggesting that the shell collisions in these events exhibiting 
X-ray and UV/optical re-brightening episodes are gentle, resulting simply in a transfer of energy to 
the blastwave. Even more fundamentally, this implies that the shells are emitted at the same time 
from near the central engine and thus the engine need not be on during the re-brightening episode.

\subsection{Energy injection: a ubiquitous phenomenon?}
\label{text:statistics}
The phenomenon of short-lived plateaus at an early time is ubiquitous in \Swift/XRT afterglow light 
curves \citep{lzz07}. Using a complete sample of \Swift/XRT light curves through 2010, 
\cite{mzb+13} find about $37\%$ of long-duration GRBs exhibit a shallow decay phase, with 
$-1\lesssim\alpha\lesssim1$. If these plateaus are caused by injection of energy into the 
forward shock, the events with X-ray/optical re-brightenings discussed in this paper might be the 
extreme tail of a distribution in energy injection factor, duration or rate. A rigorous 
exploration of these possibilities requires multi-wavelength fits to the data, but such data are 
generally not available. We therefore compute energy injection fractions based on the X-ray light 
curves alone, and compare the results for the objects where we \textit{do} have multi-wavelength 
observations as reported in this paper.

\cite{mzb+13} measured the timing of X-ray plateaus in \Swift/XRT light curves and quantified 
them by their start time, $t_1$, end time, $t_2$, and rise rate\footnote{\cite{mzb+13} use the 
convention $F(t)\propto t^{-\alpha}$, which is opposite to the convention in 
this paper.} during the plateau phase, $\alpha$. For our X-ray plateau analysis, we select the 96 
events in their study where the reported uncertainty in $\alpha$ is $<0.3$. 
We repeat their light curve decomposition analysis on the 
X-ray light curves of GRB~120326A and GRB~120404A (which are not included in their sample), and 
find $t_1=5.7\times10^{-3}$\,d, $t_2=0.62$\,d, and $\alpha=0.27\pm0.03$, for GRB~120326A and 
$t_1=9.8\times10^{-3}$\,d, $t_2=2.9\times10^{-2}$\,d, and $\alpha=0.76\pm0.23$ for 
GRB~120404A\footnote{The X-ray light curve of GRB~120404A is not well-sampled near the peak of the 
re-brightening and we therefore fix the peak time in the fit to the value inferred from the optical 
light curve ($\approx0.3$\,d; Section \ref{text:120404A:enj}).}. 

We define the plateau duration, $\Delta T \equiv t_2 - t_1$ and the fractional duration, $\xi 
\equiv (t_2 - t_1)/(t_2 + t_1)\in[0,1)$. Note that $\xi\to1$ when $t_2 \gg t_1$, which corresponds 
to the case that the plateau lasts much longer than its onset time. \cite{mzb+13} computed the start 
time of the plateau using the intersection of the steep decay phase with the best fit plateau 
model; hence they do not report an error on this quantity. We can estimate the uncertainty on $\xi$ 
using $\sigma_{\xi} = \sigma_{t_2}\partial\xi/\partial t_2 = 2t_1\sigma_{t_2}/(t_1+t_2)^2$, which 
is dimensionless as desired.

Assuming the X-ray band is located above the cooling frequency for all cases (which is indeed the 
case for the five events considered in detail here), and that $p=2$, we can compute an 
effective energy injection rate, $E\propto t^{m}$, where $m = \alpha+1$ is the rate required to 
bring the measured light curve slope to $\alpha$ from the theoretically-expected value\footnote{In 
the general case, $m = \frac{4\alpha + (3p-2)}{p+2}$ for $\nuc<\nuX$, while $m = \frac{4\alpha + 
(3p-3)}{p+3}$ (ISM) and $m = \frac{4\alpha + (3p-1)}{p+1}$ (wind) when $\nuc>\nuX$.} of 
$t^{(2-3p)/4} \propto t^{-1}$ for no energy injection. For the sake of simplicity and uniformity, 
we use $p=2$ even where we have other measurements of $p$, such as for the events reported in this 
paper. The ratio of the energy at the end of the plateau phase to the energy at the beginning of 
the 
plateau phase is then given by $\Upsilon \equiv E_2/E_1 = (t_2/t_1)^m$. 
\footnote{The uncertainty in this quantity can be estimated using
\begin{equation}
\begin{split}
\sigma^2_{\Upsilon} &= (\partial{\Upsilon}/\partial t_2)^2 \sigma^2_{t_2} + 
(\partial{\Upsilon}/\partial m)^2 \sigma^2_m \nonumber \\
&= (m\Upsilon\sigma_{t_2}/t_2)^2 + (\Upsilon\ln{(t_2/t_1)}\sigma_{\alpha})^2, \nonumber
\end{split}
\end{equation}
where we have taken $\sigma_{m} = \sigma_{\alpha}$. This assumes that $t_2$ and $\alpha$ are 
independent. However, these quantities are expected to be correlated, and a more complete analysis 
of the uncertainty in $\Upsilon$ requires the full covariance matrix between $\alpha$ and $t_2$.}

\begin{figure}
 \centering
 \includegraphics[width=\columnwidth]{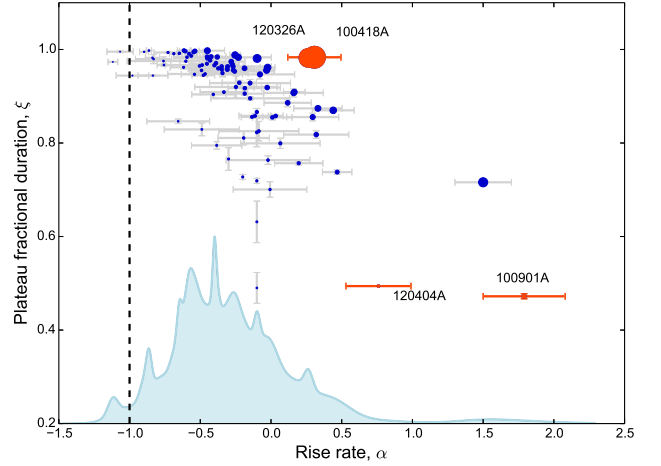}
\caption{The fractional duration of X-ray plateaus from \cite{mzb+13} as a function of the light 
curve rise rate, $\alpha$ (with the convention $F\propto t^{\alpha}$). Errors in both 
directions are computed according to the formulas described in section \ref{text:statistics}. The 
vertical dashed black line indicates the canonical light curve decay of $\alpha\approx-1$ expected 
for $p\approx2$. 21\% of events with plateaus exhibit re-brightenings ($\alpha > 0$). Events 
analyzed in this paper are shown in red. The area of each symbol is linearly proportional to the 
ratio of the energy at the end to that at the beginning of the plateau phase, $\Upsilon$. The blue 
point at $\alpha \approx-1.5$ is GRB~081028, which also exhibits an X-ray/optical re-brightening; 
however, this event does not have radio data and we therefore exclude it from our multi-wavelength 
analysis. A probability density function of the distribution of $\alpha$ (computed from a kernel 
density estimate using $\sigma_{\alpha}$ as a varying kernel bandwidth) is shown in cyan. 
\label{fig:xrt_plateau_analysis_xi_alpha}}
\end{figure}

We plot the plateau fractional duration, $\xi$ against the plateau slope, $\alpha$ in Figure 
\ref{fig:xrt_plateau_analysis_xi_alpha}, scaling the area of the symbols by $\Upsilon$. The slopes 
during the plateau phase range from $\alpha\approx-1$ to 
$\approx 1.8$. As evident from the kernel density estimate of $\alpha$, most light curves exhibit a 
gentle decay during the plateau phase (Figure \ref{fig:xrt_plateau_analysis_xi_alpha}). However, all 
four events in our sample are exceptions to this rule\footnote{Since we selected GRBs that 
exhibit an X-ray re-brightening, this is partly by sample construction.}. These four events 
also exhibit the greatest rise in their plateau phase light curves for a given normalized duration, 
$\xi$. Two out of these four (GRBs~120326A and 100418A) have the largest fractional changes in the 
kinetic energy of the entire sample of 98 events. These are also the events with the largest 
value of $\xi$ in our sample. For GRB~120404A, we note that the X-ray light curve around the peak 
of the re-brightening is missing due to a \Swift\ orbital gap. Therefore any results that rely 
solely on the X-ray data of this burst underestimate the value of $\alpha$ and the fraction of 
energy injection relative to the multi-wavelength analysis we carried out in Section 
\ref{text:120404A:enj}. 

\begin{figure}
 \centering
 \includegraphics[width=\columnwidth]{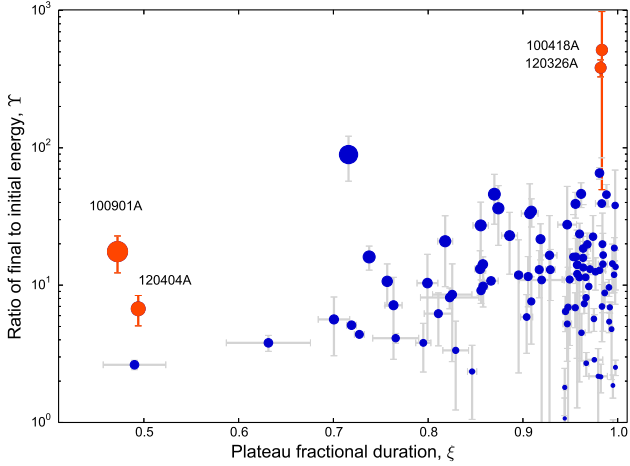}
\caption{The ratio of energy at the end to that at the beginning of the plateau phase in X-ray 
light curves of 96 events from \cite{mzb+13}, as well as GRBs~120326A and 120404A from our 
analysis, as a function of the plateau fractional duration, $\xi$, with 
error bars computed according to the prescription in section \ref{text:statistics}. The size of the 
symbols is scaled to a measure of the light curve rise rate during the plateau decay phase, 
$\alpha$, with the radius of the symbols proportional to $\alpha + 1.5$. Larger symbols denote a 
greater departure from a canonical light curve decay of $\alpha\approx-1$, indicating greater 
rates of energy injection during the plateau phase.
\label{fig:xrt_plateau_analysis_Upsilon_xi}}
\end{figure}

In Figure \ref{fig:xrt_plateau_analysis_Upsilon_xi} we plot the ratio of final to initial energy, 
$\Upsilon$ against the normalized plateau duration, $\xi$, scaling the radius\footnote{Since 
$\alpha$ can be negative while the area of the symbols in the plot is a positive definite quantity, 
we scale the radius of the symbols as $\alpha+1.5$, where the additive term accounts for the 
minimum value of $\alpha$.} of the symbols by $\alpha$. We find that for a given (normalized) 
plateau duration, the events in our sample have the largest fractional change in blastwave kinetic 
energy during the plateau phase. At the same time, they also possess the steepest rise rates in the 
sample, which is simply indicative of our selection criteria for inclusion in the present analysis. 
Finally, these events have the shortest normalized plateau durations for a given fractional change 
in energy. Together, these observations suggest that a large amount of injected 
energy is not sufficient to cause simultaneous -ray/optical re-brightenings, but that it 
\textit{must be done in a relatively small amount of time}, and thus that the defining 
characteristic of these events is a high \textit{rate} of energy injection. Physically, this 
translates to a steep distribution of ejecta Lorentz factors over a small range of $\Gamma$. 
\cite{hbdm15} recently suggested that the interaction of low-Lorentz factor ejecta with the reverse 
shock can explain flares in the X-ray light curves. Our observations provide supporting 
evidence for this hypothesis in the form of significant ejecta energy down to the requisite low 
Lorentz factors, $\Gamma \sim 10$.


\subsection{Energetics}
\label{text:energetics}
\begin{figure}
 \centering
 \includegraphics[width=\columnwidth]{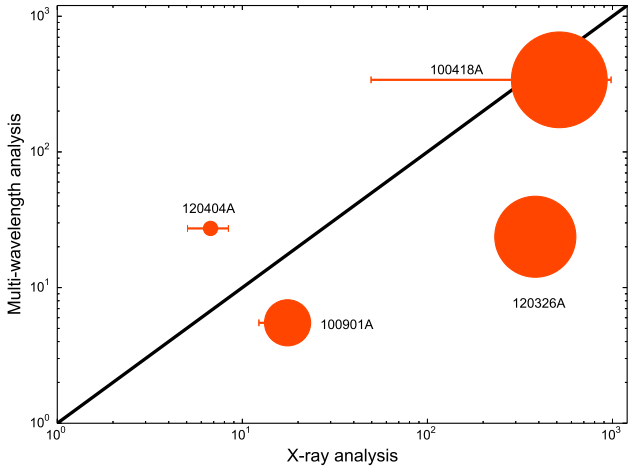}
\caption{Ratio of final energy to the energy prior to the first injection episode from the 
multi-wavelength analysis (y-axis) compared to the ratio of the energy at the end to that at the 
start of the `plateau phase', $\Upsilon$, computed in section \ref{text:statistics} for the five 
events in our sample, together with the 1:1 line (solid, black). The area of the symbols is 
linearly proportional to the physical duration of the plateaus, $t_2 - t_1$.
\label{fig:xrt_plateau_analysis_margutti_laskar}}
\end{figure}

Having compared the plateau duration, light curve rise rate, and injected energy fraction for our 
sample with a complete sample of \Swift\ events using the X-ray data alone, we now 
turn to an analysis of the results from our multi-wavelength energy injection modeling in this 
context. In Figure \ref{fig:xrt_plateau_analysis_margutti_laskar} we plot the fractional change in 
energy determined from multi-wavelength modeling against the fractional change in energy, 
$\Upsilon$, inferred from the X-ray-only analysis in section \ref{text:statistics} for the four 
events in our sample. We scale the area of the symbols with the plateau physical duration, $t_2 - 
t_1$. As expected, events with longer plateau durations have greater estimated energy injection 
fractions both from the X-ray analysis and from the full multi-wavelength study. The fractional 
change in energy from the X-ray analysis is higher than inferred from the full model for 
three out of four cases. This is due to the typically lower value of $m$ from the X-ray-only 
analysis compared to the full model. Ultimately, this can be traced to the fitting procedure: 
\cite{mzb+13} fit a sum of a steep decay and a rising light curve, the sum of which results in the 
X-ray plateau, whereas we do not subtract fits to the steep decay phase from the X-ray light curve 
prior to our multi-band modeling. In the case of GRB~120326A, where the X-ray analysis 
over-estimates $\Upsilon$ by an order of magnitude, the variance between the two techniques is due 
to differences in the plateau start time ($4.3\times10^{-2}$\,d for the multi-wavelength analysis, 
compared to $5.7\times10^{-3}$\,d in the X-ray analysis). Based on this comparison, we conclude that 
a determination of $\Upsilon$ using the X-ray data alone yields a reasonable estimate of the energy 
injection factor on average, but that multi-wavelength modeling is essential to obtain the full 
picture.

\section{Summary and Conclusions}
\label{text:conclusions}
Plateaus and re-brightening episodes are frequent in GRB afterglow X-ray light curves. However, 
X-ray data by themselves provide only a limited understanding of the physical processes underlying 
these unexpected phenomena. We perform a thorough multi-wavelength study of all long-duration GRBs 
through 2012 featuring simultaneous X-ray, and UV/Optical re-brightenings and that have radio 
detections, using a Markov Chain Monte Carlo analysis and a physical afterglow model. Our analysis 
yields the first set of models that explain the multi-wavelength afterglows for all of these events 
with the same unifying principle. In all cases, the afterglow light curves can be modeled with a 
standard forward shock model following the re-brightening episode, an ISM density profile, and a 
jet break. From our multi-wavelength analysis, we find that the circumburst densities, 
jet opening angles, and beaming-corrected kinetic energies for these events span the full range 
described by typical GRBs at $z\sim1$ and $z\gtrsim6$. 

We explore a range of possible models to understand the re-brightenings, including the onset of the 
afterglow, off-axis viewing geometry, and continuous energy injection. We are able to rule out the 
afterglow onset and off-axis jets, and find instead that injection of energy into the blastwave 
(the so-called `refreshed-shock' scenario) provides a good explanation for all events. We interpret 
energy injection in the framework of the stratified Lorentz factor model, and find that our measured 
energy injection rates \textit{always} obey the theoretical constraints relating the rate of 
injection and the distribution of ejecta Lorentz factors in the ISM model. 

We perform the first measurement of the ejecta Lorentz factor distribution index, $s$, and find 
$s\approx 3$--$40$, suggesting that a large amount of kinetic energy resides in the slowest-moving 
ejecta. This is supported by low radiative efficiencies for the events in our 
sample, indicating that the $\gamma$-ray radiation is dominated by ejecta at high 
Lorentz factors, while the kinetic energy is dominated by slower-moving ejecta. We note that 
keeping the injection rates simple power laws allows us to directly convert the injection rate to 
a Lorentz factor distribution index, but the true injection rate and also the true Lorentz factor 
distribution is likely to be a smoother function of time and Lorentz factor, respectively.

Finally, using a compilation of X-ray plateaus in GRB afterglows, we present a comparative 
discussion of this interesting sub-population of GRBs. We find that the phenomenon of energy 
injection is ubiquitous is long-duration GRBs, with re-brightening episodes likely simply extreme 
injection events. In future work, we aim to fit the light curves before the re-brightening 
episodes in a statistical sense, allowing us to estimate uncertainties on the rate and duration of 
the energy injection episodes. At the same time, radio monitoring of \Swift\ events exhibiting 
re-brightening episodes will be crucial in multi-wavelength modeling of this interesting class 
of GRBs, while ALMA observations will irrefutably establish the presence or absence of reverse 
shocks, laying to rest the question of whether the energy injection process is violent or gentle.

\acknowledgements 
We thank M. Viero and J. Viera for acquiring the P200 WIRC observations are part of our 
target-of-opportunity program. Support for CARMA construction was derived from the states of 
California, Illinois, and Maryland, the James S. McDonnell Foundation, the Gordon and Betty Moore 
Foundation, the Kenneth T. and Eileen L. Norris Foundation, the University of Chicago, the 
Associates of the California Institute of Technology, and the National Science Foundation. Ongoing 
CARMA development and operations are supported by the National Science Foundation under a 
cooperative agreement, and by the CARMA partner universities. The Submillimeter Array is a joint 
project between the Smithsonian Astrophysical Observatory and the Academia Sinica Institute of 
Astronomy and Astrophysics and is funded by the Smithsonian Institution and the Academia Sinica. 
B.A.Z. acknowledges support from NSF AST-1302954. DAP is supported by Hubble Fellowship grant 
HST-HF-51296.01-A, and by NASA through an award issued by JPL/Caltech as part of Spitzer proposal 
GO-90082. This research has made use of data obtained through the High Energy Astrophysics Science 
Archive Research Center On-line Service, provided by the NASA/Goddard Space Flight Center.

\appendix
\section{Inverse Compton Correction}
\label{appendix:IC}
The synchrotron photons produced in the GRB blastwave can Compton-scatter off the 
shock-accelerated relativistic electrons in the blastwave, producing a Comptonized spectrum 
at high energies \citep{pm98,wl98,tot98,cd99,dbc00,dcm00,pk00,bm77,se01}. We compute the IC 
spectrum by directly integrating the synchrotron spectrum over the electron Lorentz factor 
distribution (back-calculated from the synchrotron spectrum) using equation A2 in \cite{se01}. We 
find that IC emission contributes negligible flux compared to synchrotron radiation and we ignore 
this component in our analysis. However, the IC mechanism can provide a significant source of 
cooling for the shock-accelerated electrons and thereby dominate the total cooling rate, even when 
IC emission itself is not directly observable \citep{snp96,se01,zlp+07}. This effects the 
synchrotron cooling frequency, as well as the self-absorption frequency (when $\nuc<\numax$). Thus 
IC cooling should be taken into account when computing SEDs and light curves for the synchrotron 
component.

The importance of IC cooling is determined by the Compton $y$-parameter\footnote{Note that this 
formula for $Y$ does not take the Klein-Nishina correction into account. This frequency-dependent 
correction is expected to be important only at high frequencies, $\nu\gtrsim10^{18}$\,Hz at 
$t\gtrsim1$\,d \citep{fp06,zlp+07}. Upon detailed investigation, we find that a consequence of 
this effect is to reduce the overall energy required to match the light curves by up to $25\%$. The 
uncertainties arising from model selection as well as due to correlations between parameters are 
usually also of this order and we therefore do not consider this effect further in this work.}, 
\begin{equation}
 Y = \frac{-1+\sqrt{1+4\eta\epse/\epsb}}{2},
\end{equation}
where $\eta$ is the fraction of energy that has been radiated away due to synchrotron and IC 
radiation, such that $\eta = 1$ during fast cooling and $\eta = 
\left(\nuc/\numax\right)^{-(p-2)/2}$ during slow cooling \citep{se01}. Writing $\nuc = 
\nuc^{\prime}(1+Y)^{-2}$, where $\nuc^{\prime}$ is the cooling frequency of the synchrotron SED not 
corrected for IC cooling, we have
\begin{eqnarray}
 \eta & = & \left[\frac{\nuc^{\prime}(1+Y)^{-2}}{\numax}\right]^{-(p-2)/2} 
      = \left(\frac{\nuc^{\prime}}{\numax}\right)^{-(p-2)/2}(1+Y)^{p-2}\nonumber\\
      & = & H \left[\frac{1+\sqrt{1+4\eta\epse/\epsb}}{2}\right]^{p-2},
\end{eqnarray}
where $H = (\nuc^{\prime}/\numax)^{-(p-2)/2}$ is independent of $\eta$. 
We therefore obtain the following implicit equation for $\eta$,
\begin{equation}
 f(\eta) \equiv \eta - H\left[\frac{1+\sqrt{1+4\eta\epse/\epsb}}{2}\right]^{p-2} = 0,
\end{equation}
which can be solved numerically using (for instance) the Newton-Raphson method. Finally, the effect 
of IC cooling can be accounted for by scaling the spectral break frequencies and flux densities of 
the synchrotron spectrum by the appropriate powers of $1+Y$ \citep{gs02}. For convenience, we 
summarize these scaling relations in Table \ref{tab:icscalings}.

\begin{deluxetable*}{lcccc}
 \tabletypesize{\footnotesize}
 \tablecolumns{5}
 \tablecaption{Inverse Compton Corrections to Spectral Break Frequencies\label{tab:icscalings}}
 \tablehead{
   \colhead{Spectral Break} &
   \colhead{Break Frequency} &
   \colhead{Break type} &
   \colhead{Frequency scaling} &
   \colhead{Flux density scaling} \\   
   }
 \startdata
 1  & $\nusa$  & Self-absorption & $(1+Y)^0$    & $(1+Y)^0$ \\
 2  & $\numax$ & Characteristic  & $(1+Y)^0$    & $(1+Y)^0$ \\
 3  & $\nuc$   & Cooling         & $(1+Y)^{-2}$ & $(1+Y)^{p-1}$ \\
 4  & $\numax$ & Characteristic  & $(1+Y)^0$    & $(1+Y)^0$ \\
 5  & $\nusa$  & Self-absorption & $(1+Y)^0$    & $(1+Y)^0$ \\
 6  & $\nusa$  & Self-absorption & $(1+Y)^{\frac{-2}{p+5}}$ & $(1+Y)^{\frac{-5}{p+5}}$ \\
 7  & $\nuac$  & Self-absorption & $(1+Y)^{-3/5}$ & $(1+Y)^{-6/5}$ \\
 8  & $\nusa$  & Self-absorption & $(1+Y)^{-1/3}$ & $(1+Y)^{-5/6}$ \\
 9  & $\numax$ & Characteristic  & $(1+Y)^0$    & $(1+Y)^{-1}$ \\
 10 & $\nusa$  & Self-absorption & $(1+Y)^1$    & $(1+Y)^1$ \\
 11 & $\nuc$   & Cooling         & $(1+Y)^{-2}$ & $(1+Y)^0$
 \enddata
\end{deluxetable*}

\section{A Wind Model for GRB~120326A}
\label{appendix:120326A_wind}
In Sections \ref{text:results:ISM_lowp} we discussed the ISM model for \me. We now consider the 
possibility of a wind-like circumburst environment. 




The best fit model in a wind environment requires $p\approx2.52$, $\epse\approx4.9\times10^{-2}$, 
$\epsb\approx1.0\times10^{-2}$, $\Astar\approx7.0\times10^{-2}$, 
$\EKiso\approx3.6\times10^{54}$\,erg, 
$\tjet\approx19$\,d, $\AV\approx0.46$\,mag, and $F_{\nu,\rm host, r^{\prime}}\approx1.6\,\mu$Jy. 
This model transitions from fast cooling to slow cooling at $3.5\times10^{-2}$\,d. The spectral 
break frequencies at 1\,d are located at $\nua\approx6.8\times10^{8}$\,Hz, 
$\numax\approx2.2\times10^{14}$\,Hz, and $\nuc\approx4.2\times10^{15}$\,Hz. The peak of the 
spectrum 
($f_{\nu}$) is at $\numax$, with a flux density of $\approx23$\,mJy.

\begin{figure}
\begin{tabular}{ccc}
 \centering
 \includegraphics[width=0.30\columnwidth]{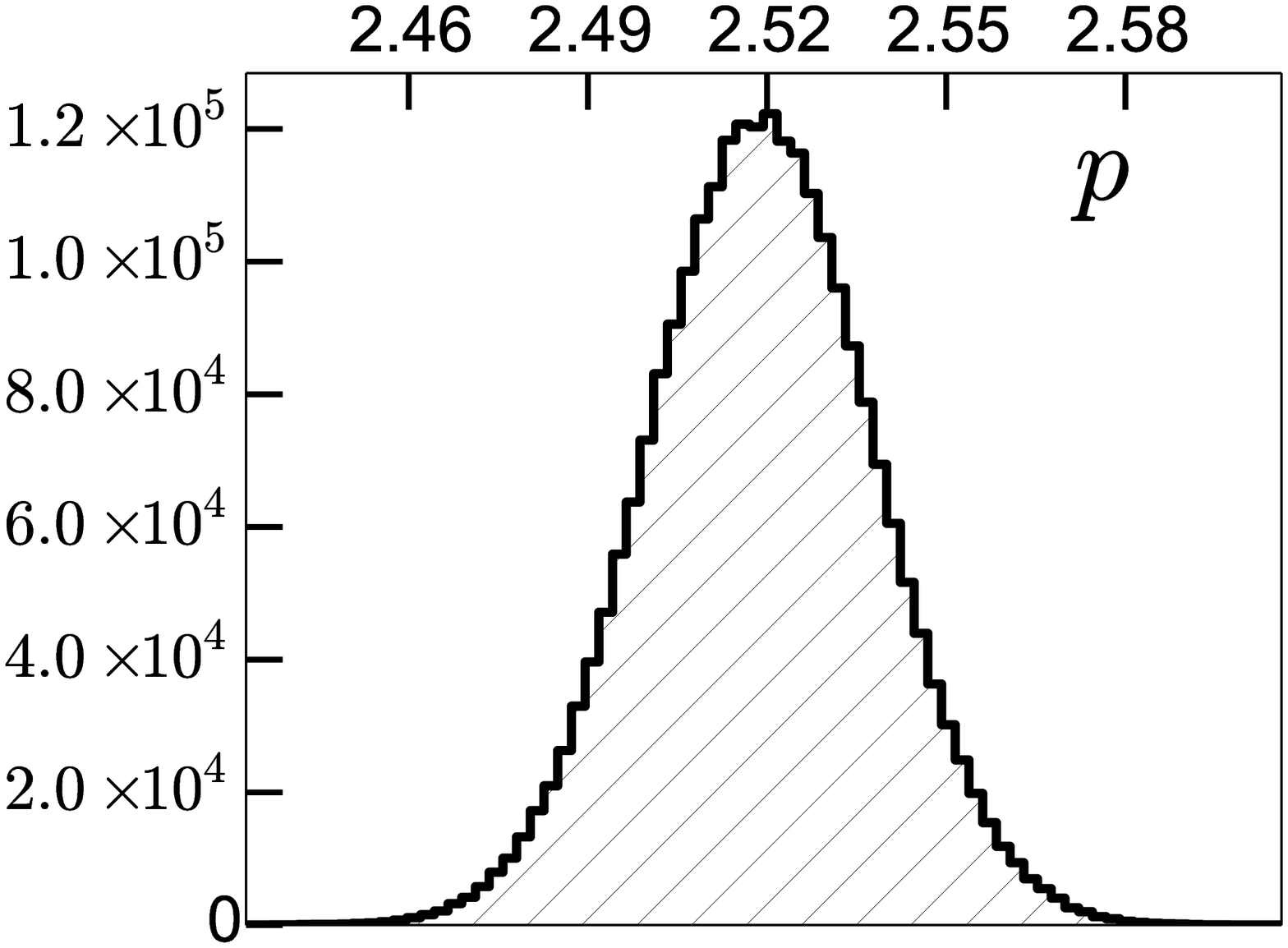} &
 \includegraphics[width=0.30\columnwidth]{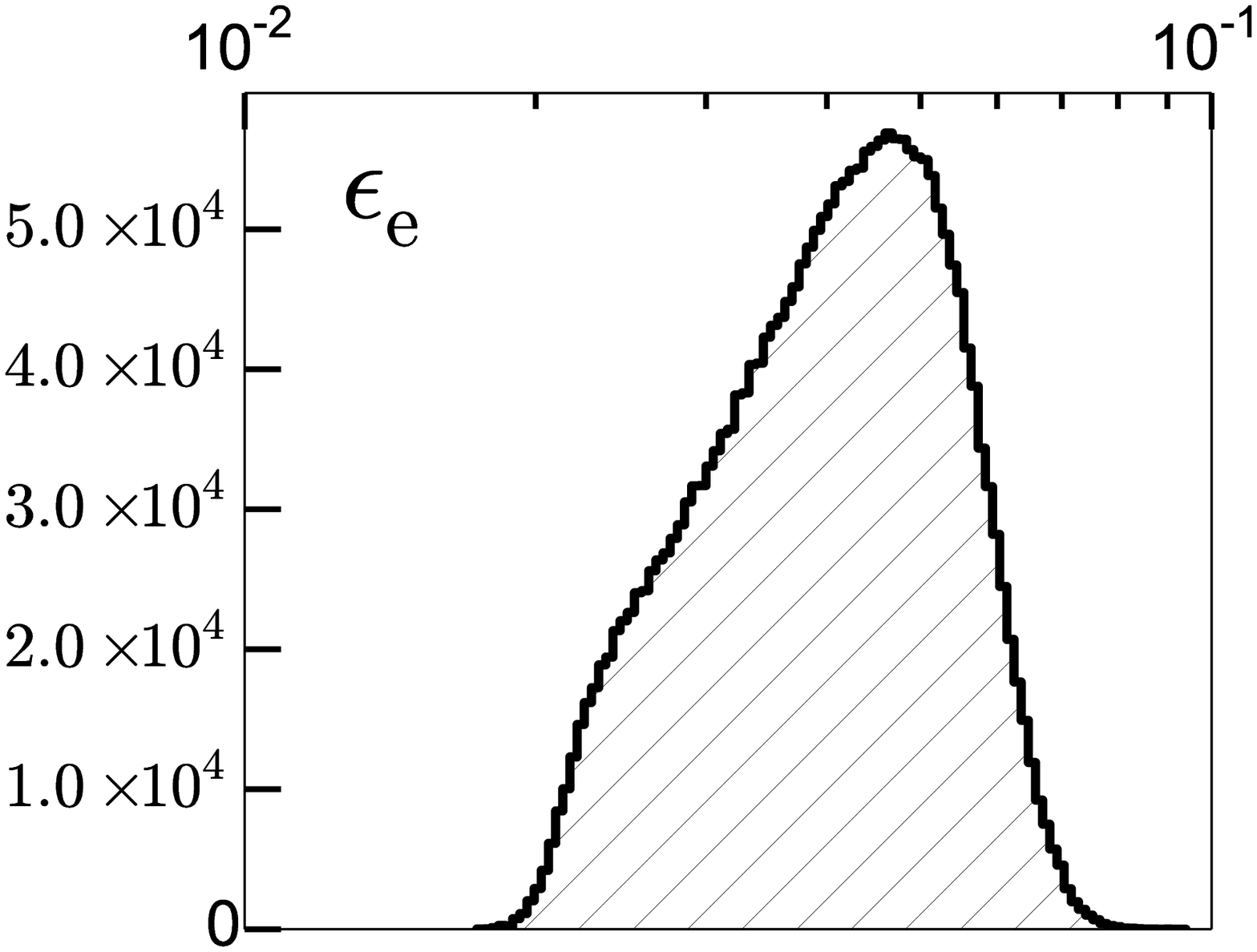} &
 \includegraphics[width=0.30\columnwidth]{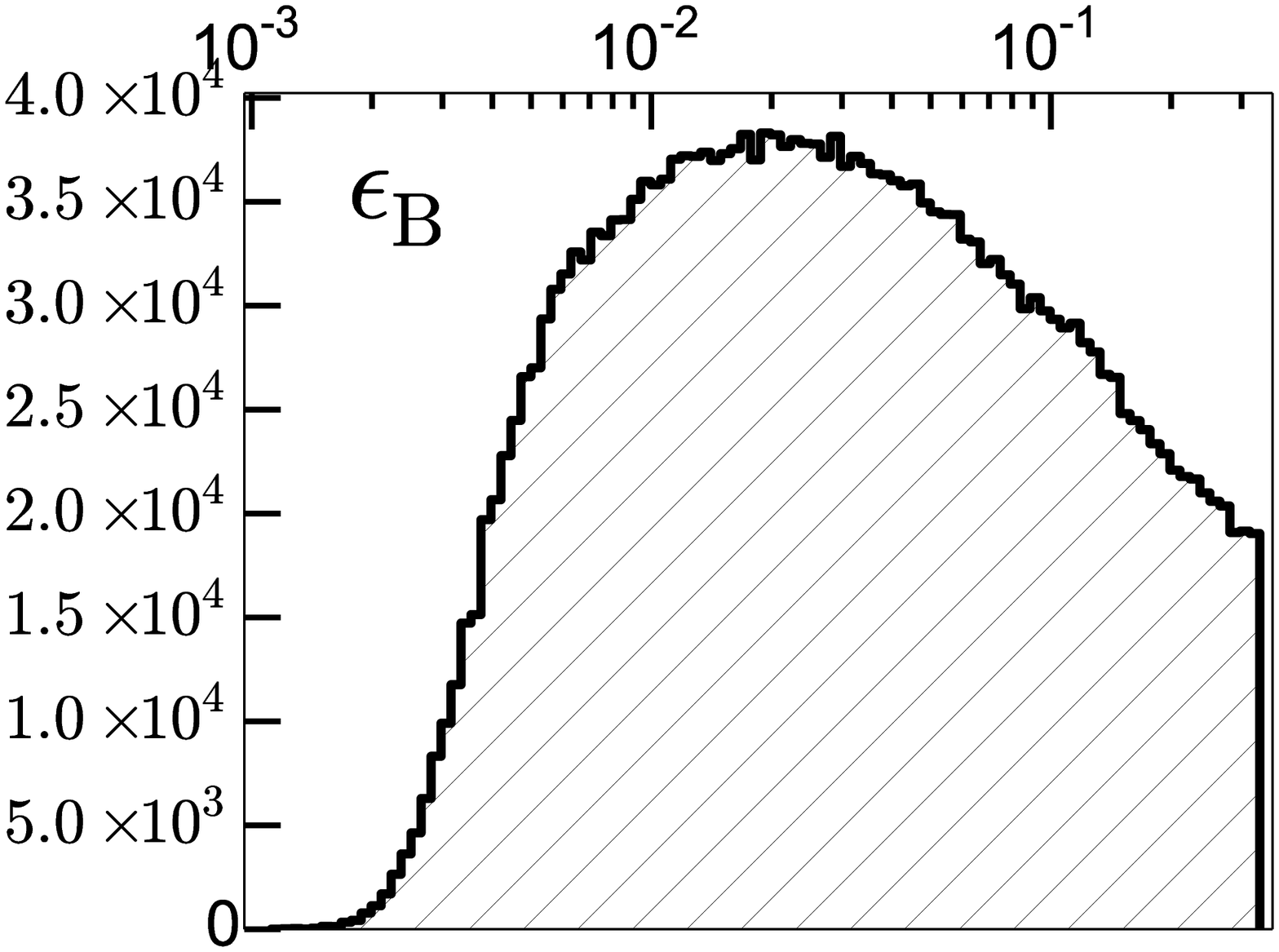} \\
 \includegraphics[width=0.30\columnwidth]{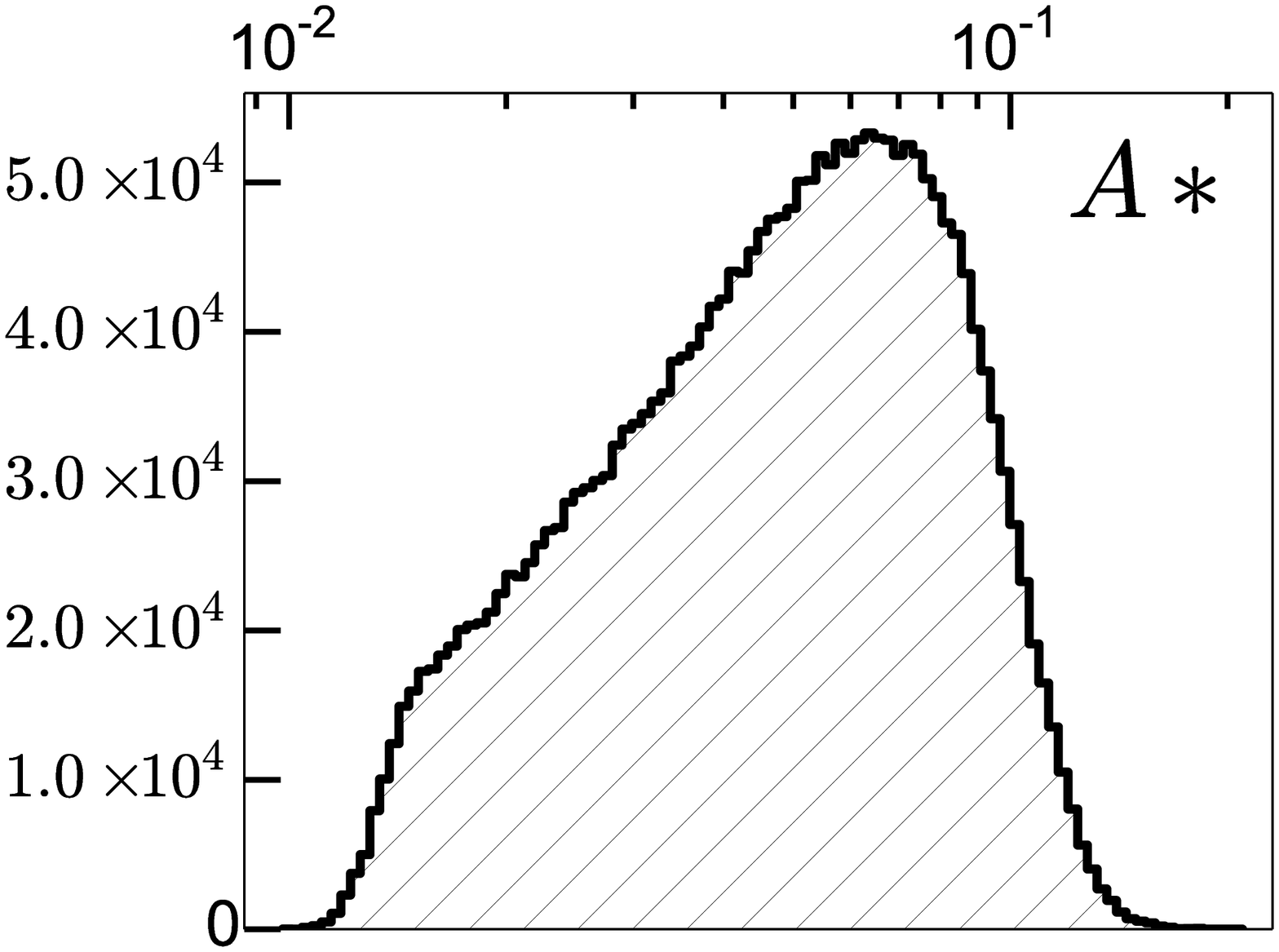} & 
 \includegraphics[width=0.30\columnwidth]{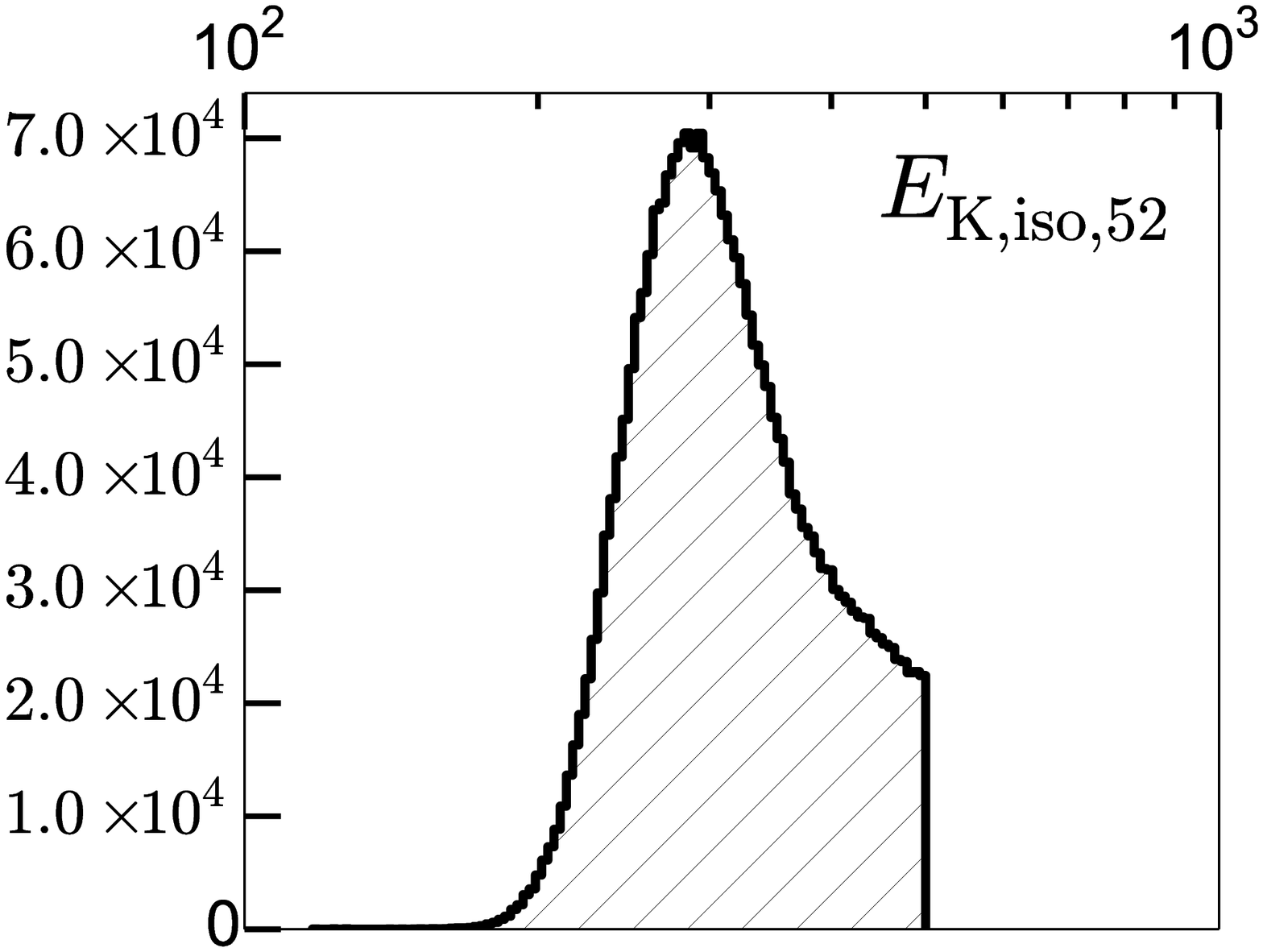} &
 \includegraphics[width=0.30\columnwidth]{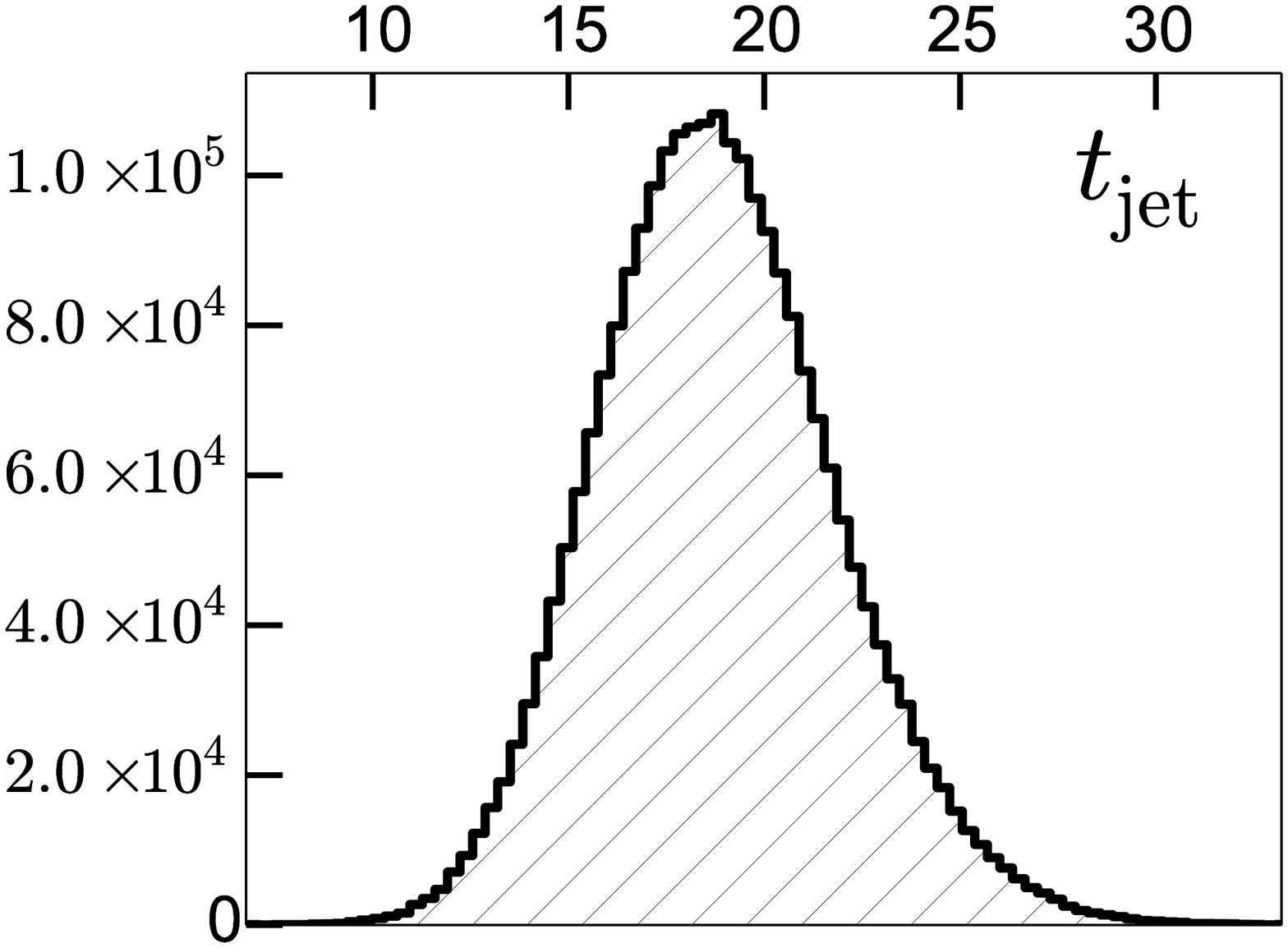} \\ 
 \includegraphics[width=0.30\columnwidth]{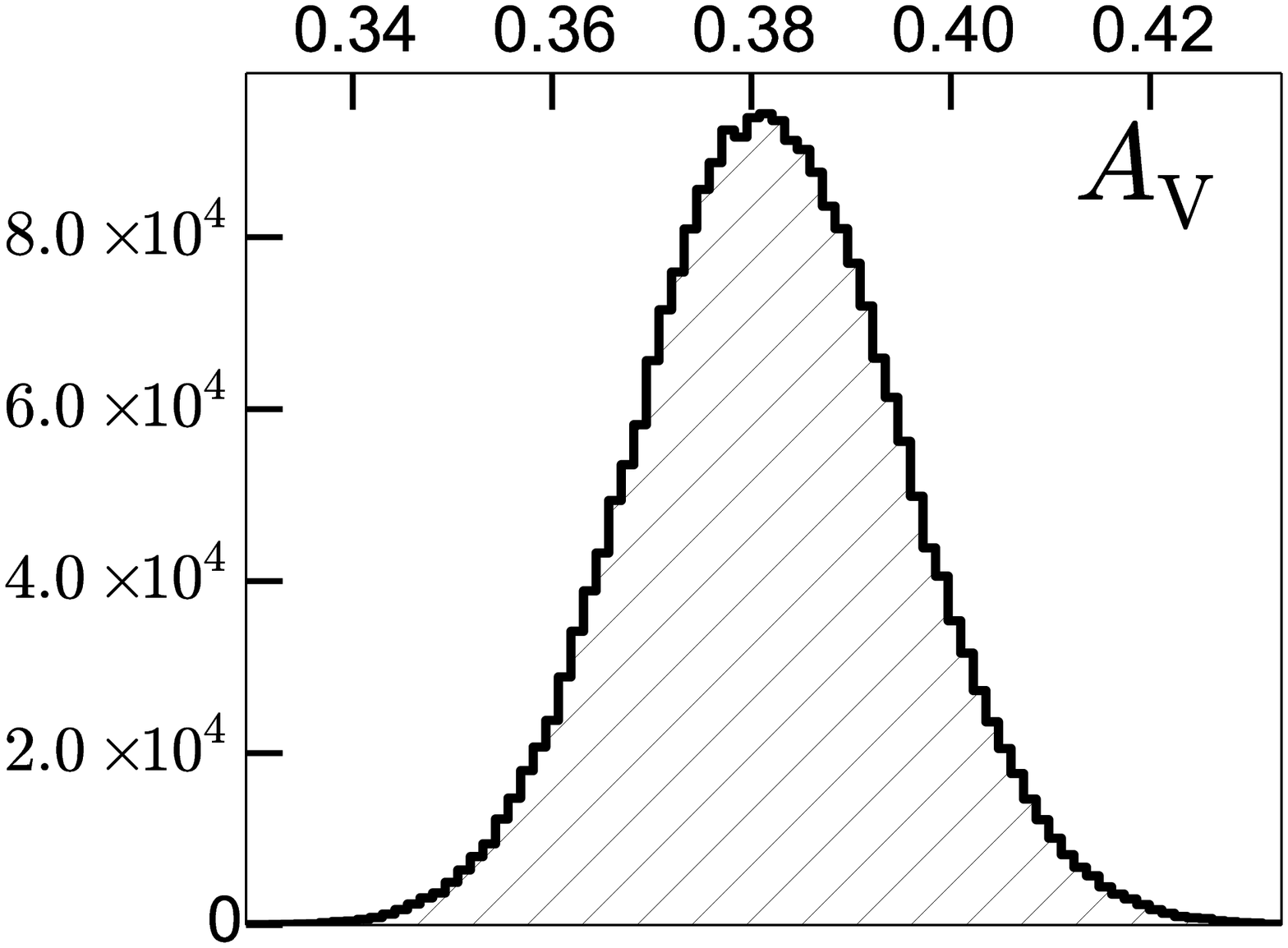} &
 \includegraphics[width=0.30\columnwidth]{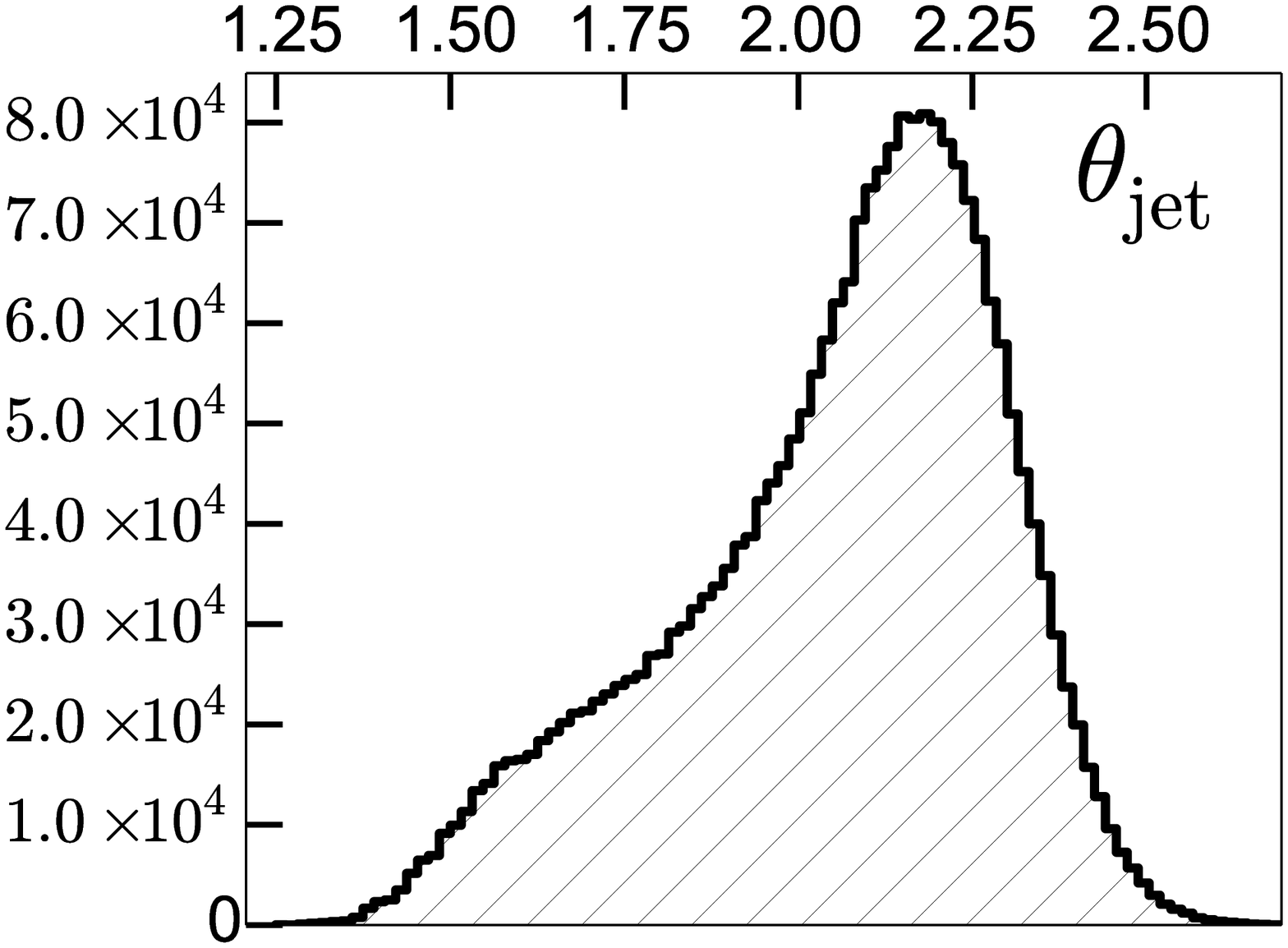}  &
 \includegraphics[width=0.30\columnwidth]{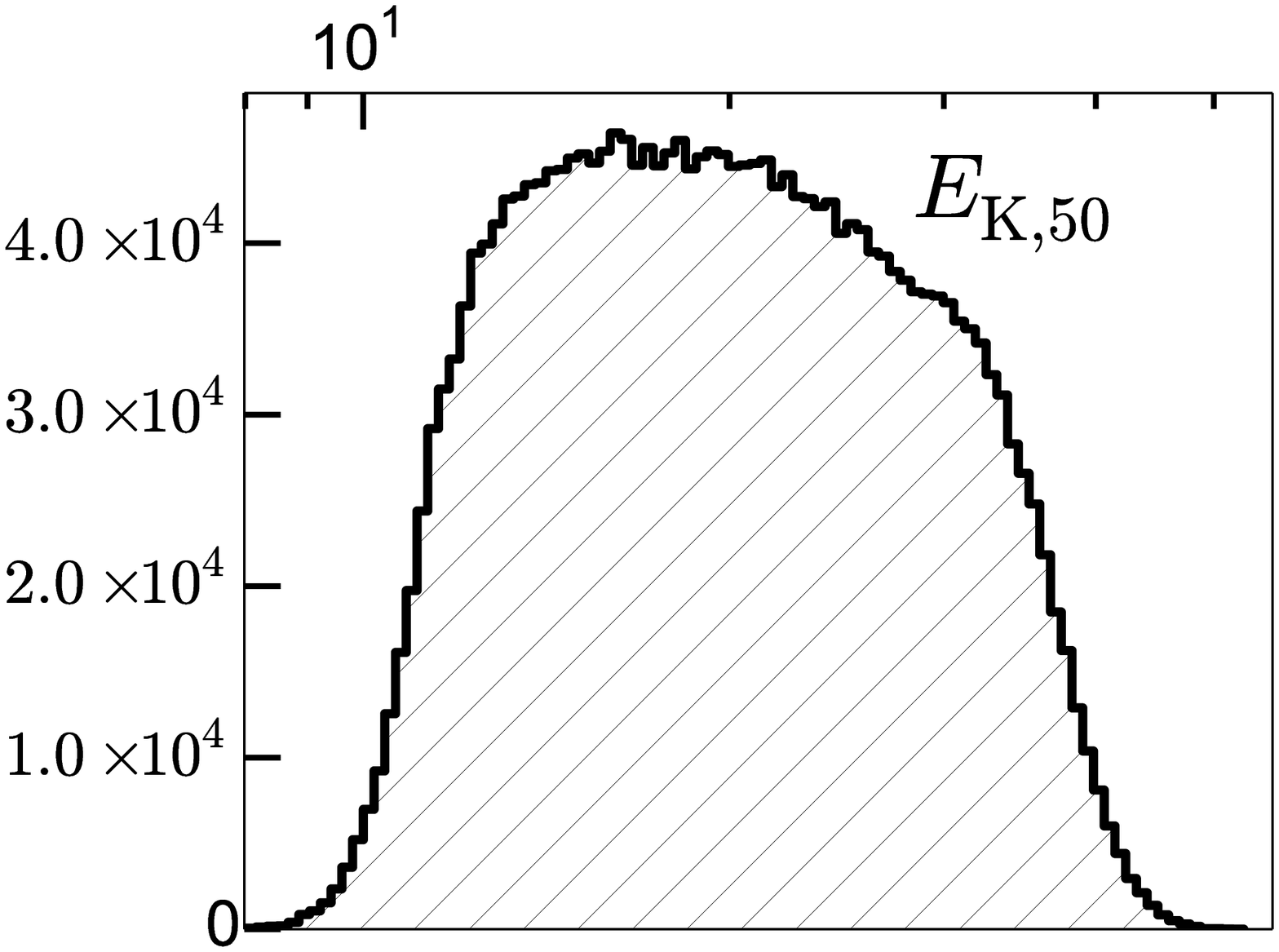} \\  
\end{tabular}
\caption{Posterior probability density functions for the physical parameters for GRB~120326A in 
the wind model from MCMC simulations. We have restricted $E_{\rm K, iso, 52} < 500$, $\epsilon_{\rm 
e} < \nicefrac{1}{3}$, and $\epsilon_{\rm B} < \nicefrac{1}{3}$.
\label{fig:120326A_wind_highp_hists}}
\end{figure}

\begin{figure}
\begin{tabular}{ccc}
\centering
 \includegraphics[width=0.30\columnwidth]{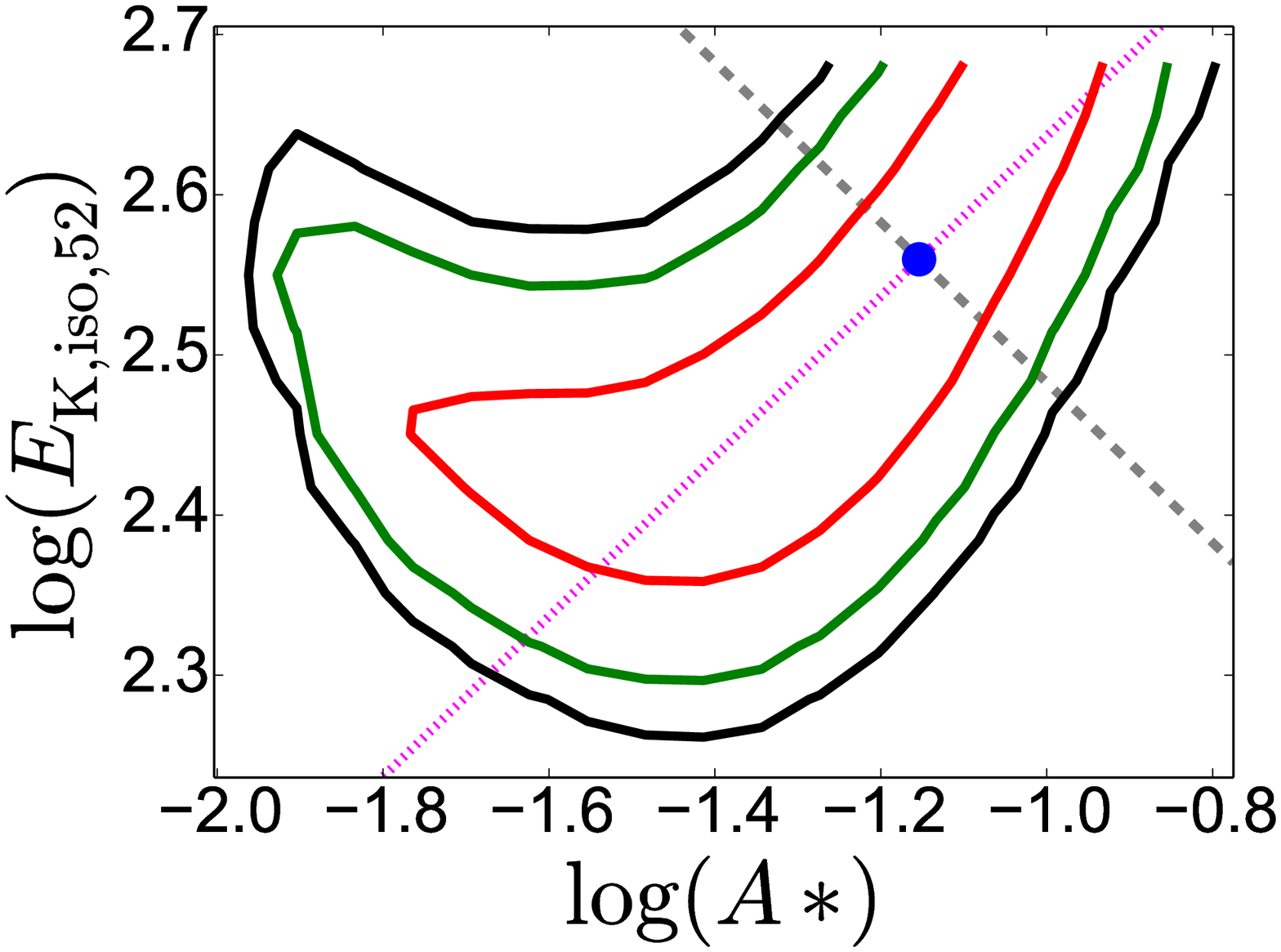} &
 \includegraphics[width=0.30\columnwidth]{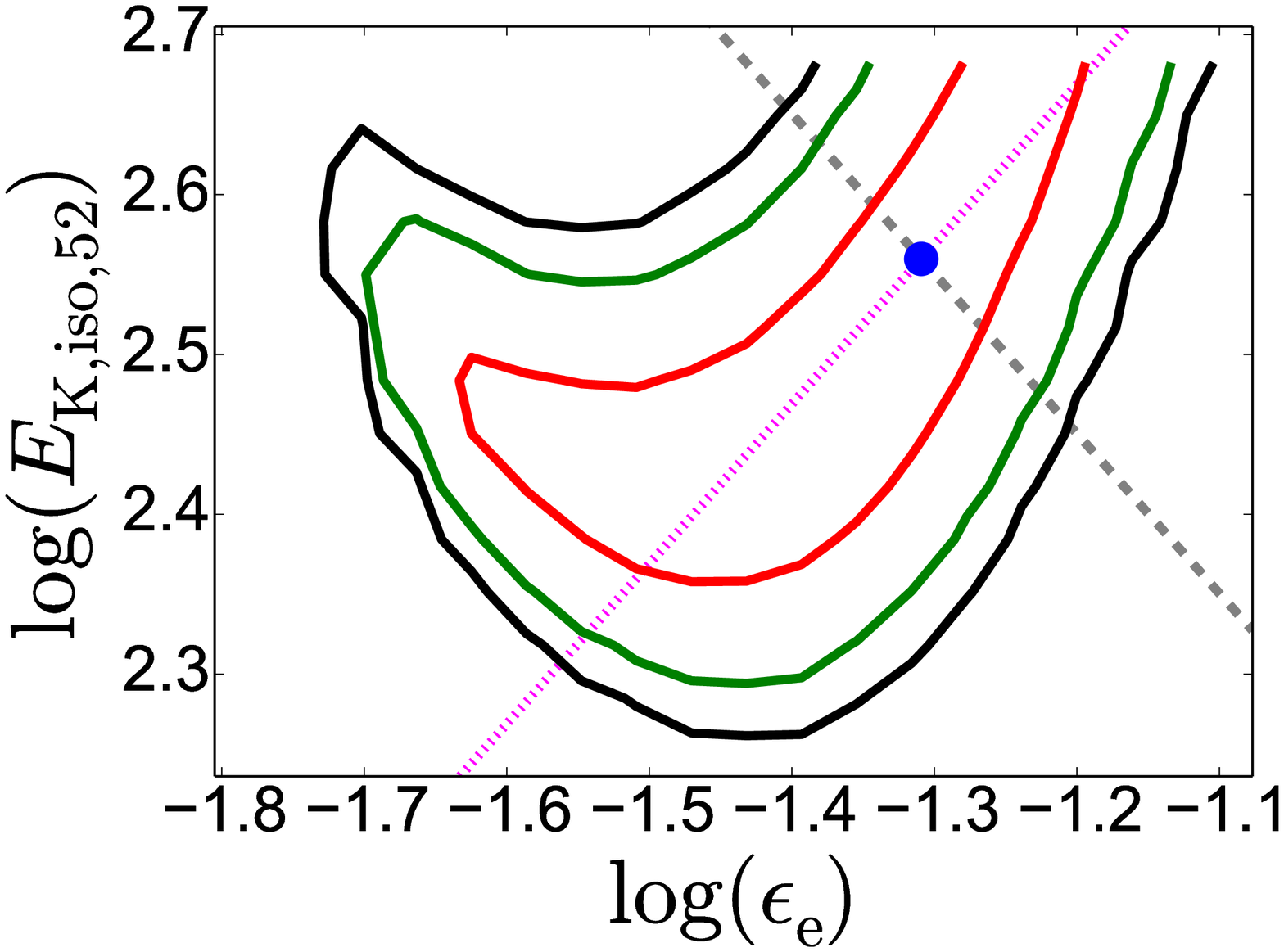} &
 \includegraphics[width=0.30\columnwidth]{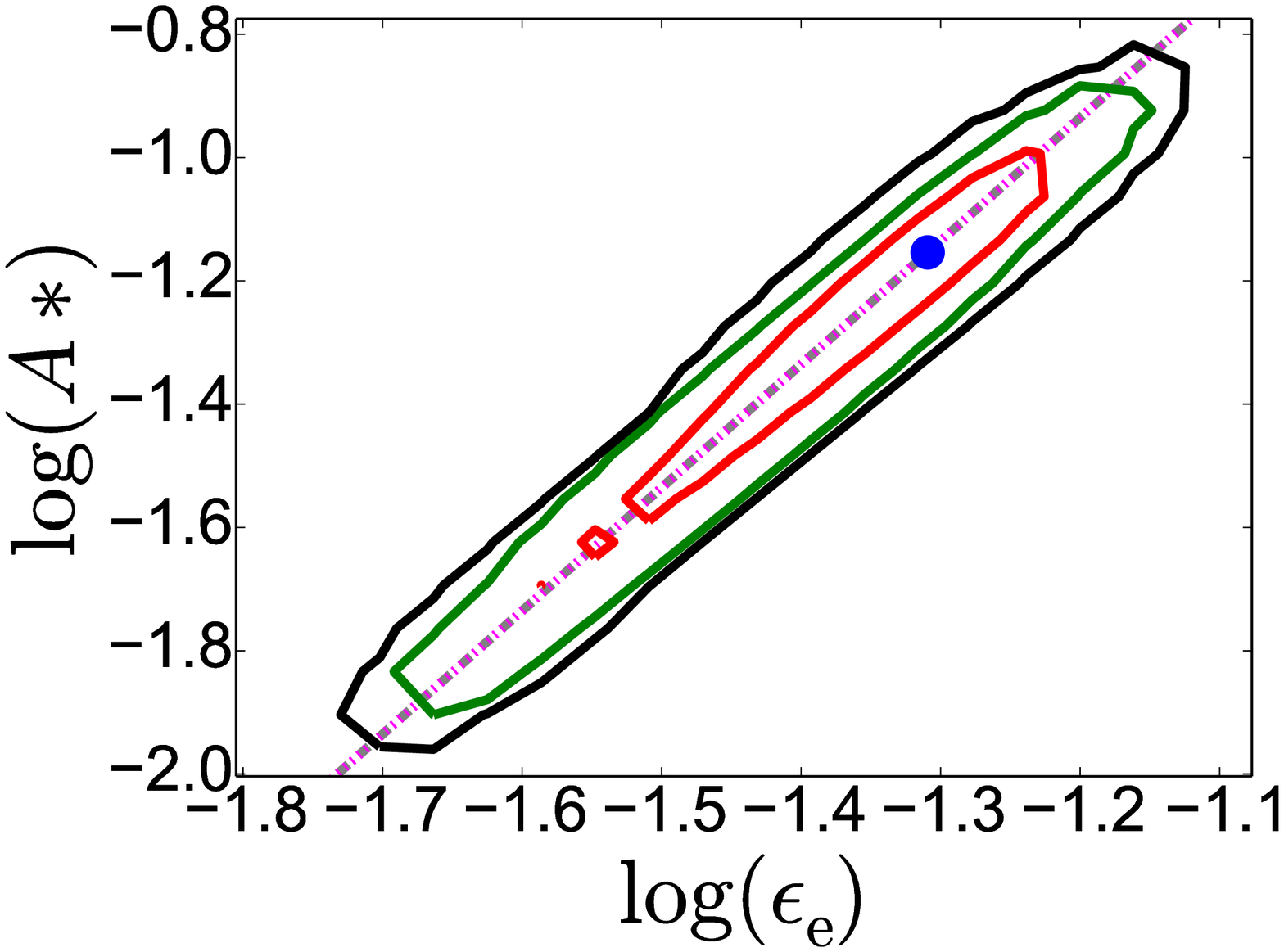} \\
 \includegraphics[width=0.30\columnwidth]{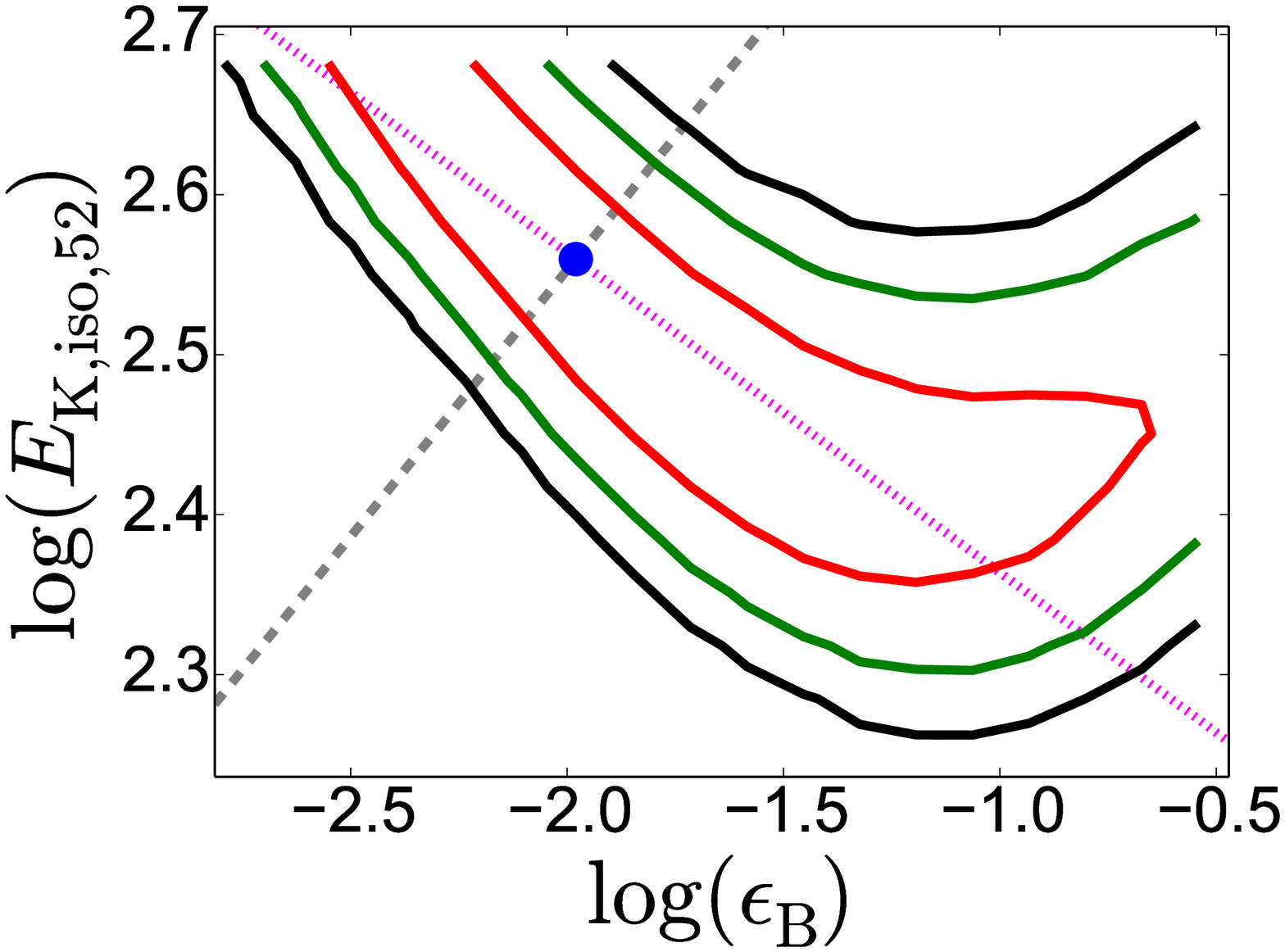} &
 \includegraphics[width=0.30\columnwidth]{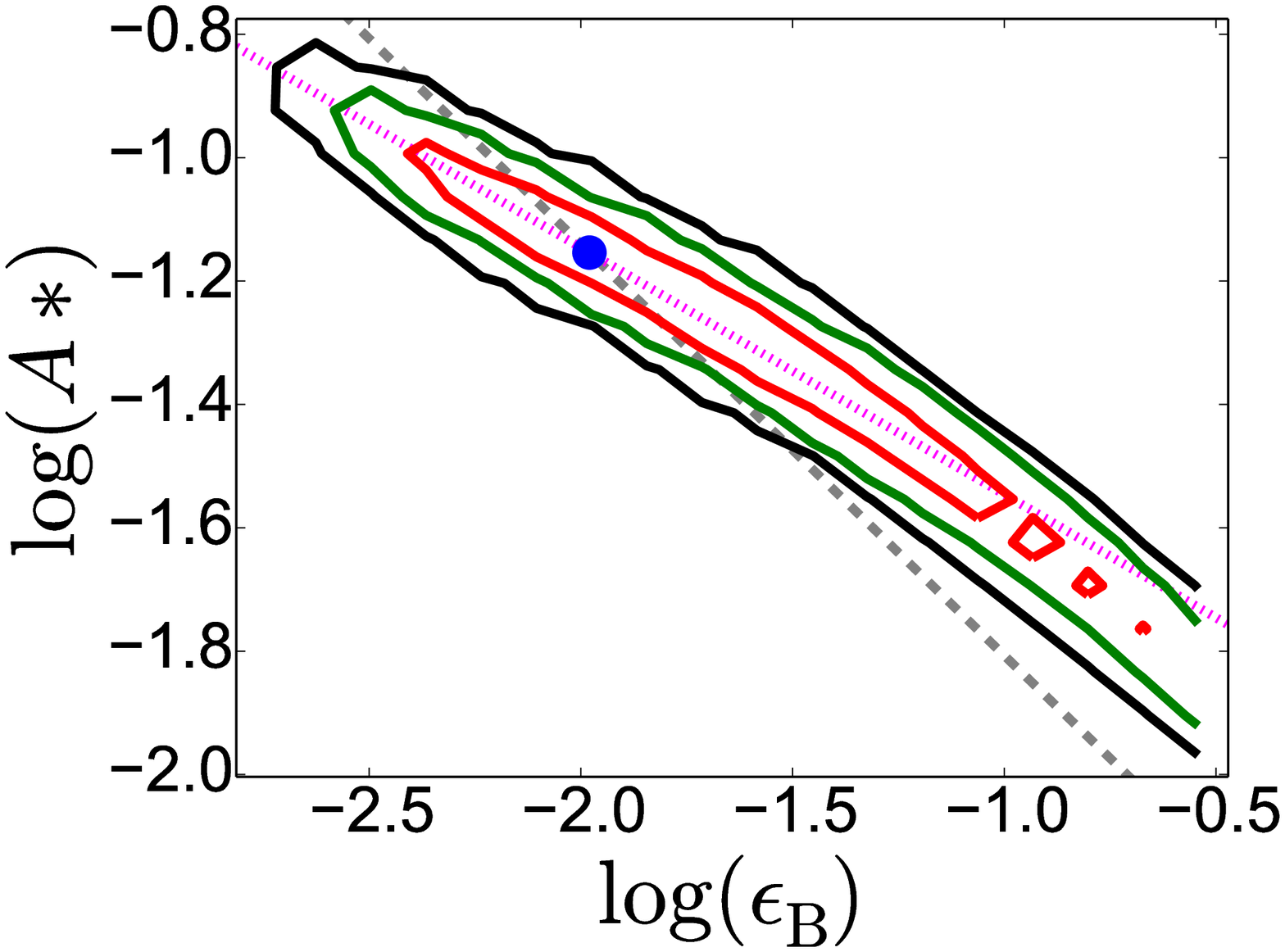} &
 \includegraphics[width=0.30\columnwidth]{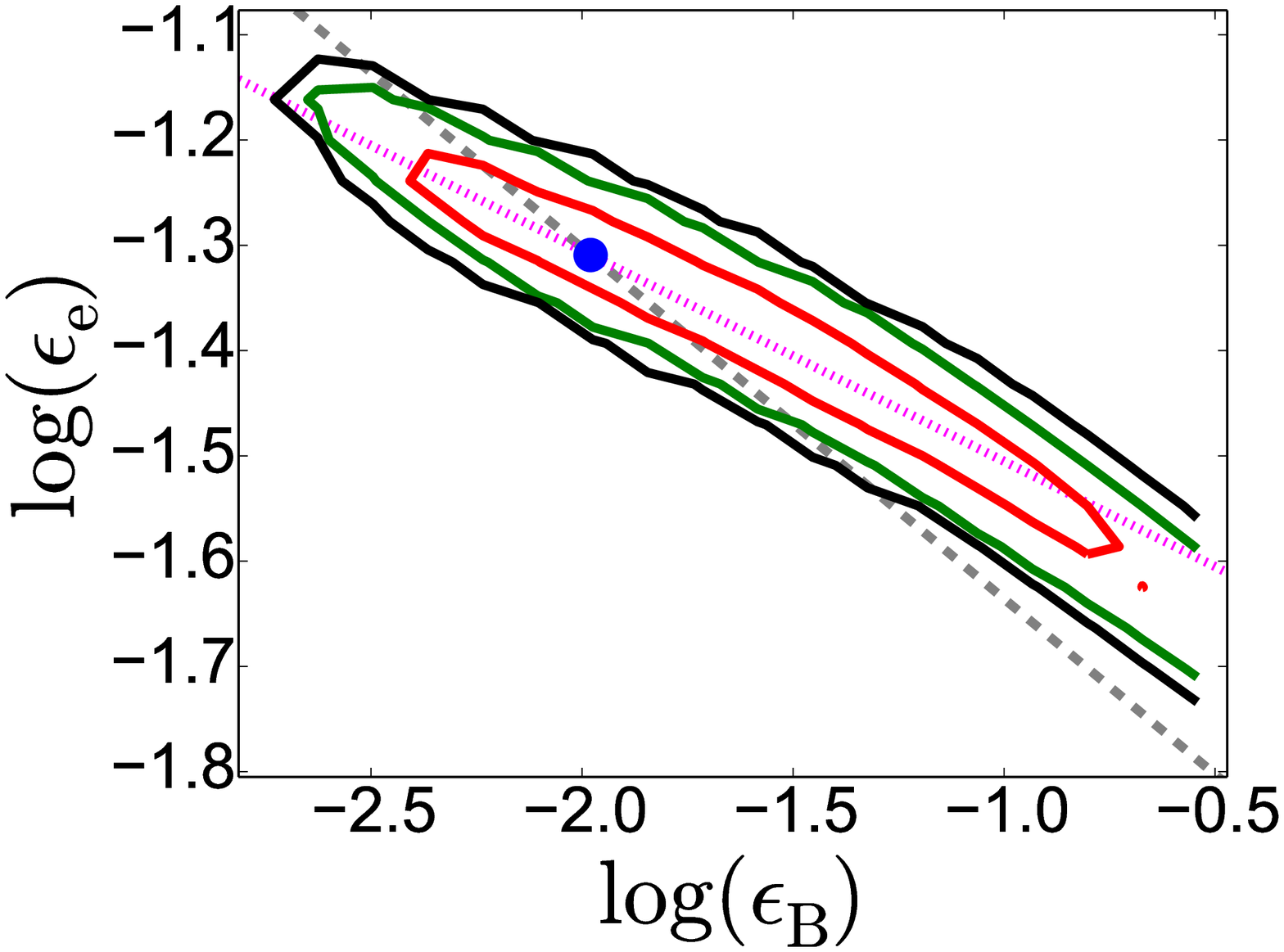} \\
\end{tabular}
\caption{1$\sigma$ (red), 2$\sigma$ (green), and 3$\sigma$ (black) contours for correlations
between the physical parameters, \EK, \Astar, \epse, and \epsb\ for GRB~120326A, in the wind model 
from Monte Carlo simulations. We have restricted $E_{\rm K, iso, 52} < 500$, 
$\epsilon_{\rm e} < \nicefrac{1}{3}$, and $\epsilon_{\rm B} < \nicefrac{1}{3}$. The 
dashed grey lines indicate the expected relations between these parameters when $\nua$ is not fully 
constrained: $\EKiso\propto\Astar^{-1/2}$, $\EKiso\propto \epse^{-1}$, $\Astar\propto 
\epse^{2}$, $\EKiso\propto\epsb^{1/3}$, $\Astar\propto\epsb^{-2/3}$, and $\epse\propto\epsb^{-1/3}$,
normalized to pass through the highest-likelihood point (blue dot), while the dotted magenta lines 
indicate expected relations for changes the value of $\nuc$: $\EKiso\propto\Astar^{1/2}$, 
$\EKiso\propto \epse$, $\Astar\propto \epse^{2}$, $\EKiso\propto\epsb^{-1/5}$, 
$\Astar\propto\epsb^{-2/5}$, and $\epse\propto\epsb^{-1/5}$. $\nua$ falls below the radio band 
before any radio observations took place, and is therefore unconstrained. $\nuc$ lies between the 
optical and X-ray bands and is better constrained; the correlations between the parameters along 
the lines of varying values of $\nuc$ are likely indicative of the strong changes (over two orders 
of magnitude) in the Compton $y$-parameter along these curves. See the on-line version of this 
Figure for additional plots of correlations between these parameters and $p$, $t_{\rm jet}$, 
$\thetajet$, $E_{\rm K}$, $A_{\rm V}$, and $F_{\nu,{\rm host,r^{\prime}}}$. 
\label{fig:120326A_wind_highp_corrplots}}
\end{figure}

From our MCMC simulations, we find $p=2.52\pm0.02$, 
$\EKiso = \left(3.1^{+0.9}_{-0.5}\right)\times10^{54}$\,erg, 
$\Astar=\left(4.8^{+3.3}_{-2.5}\right)\times10^{-2}$, 
$\epse=\left(4.1^{+1.3}_{-1.2}\right)\times10^{-2}$, 
$\epsb=\left(2.8^{+9.4}_{-2.1}\right)\times10^{-2}$, and $\tjet=18.7^{3.1}_{-2.9}$\,d.  Using 
the relation $\thetajet = 0.17\left(2\frac{\tjet*\Astar}{(1+z)E_{\rm K,iso,52}}\right)^{1/4}$ 
for the jet opening angle \citep{cl00}, and the distributions of \EKiso, \dens, and $t_{\rm 
jet}$ from our MCMC simulations (Figure \ref{fig:120326A_wind_highp_hists}), we find $\thetajet 
= 2.1^{+0.2}_{-0.3}$ degrees. Applying the beaming correction, $\Egamma=\Egammaiso 
(1-\cos{\thetajet})$, we find $\Egamma = (2.1\pm0.3)\times10^{49}$\,erg. The beaming-corrected 
kinetic energy is much larger, $\EK = \left(2.0^{+1.0}_{-0.6}\right)\times10^{51}$\,erg. In this 
model, $\nua$ falls below $10^{10}$\,Hz at $1.7\times10^{-2}$\,d and is therefore not probed 
by any of the radio data. Consequently, the model exhibits a degeneracy in its parameters due to to 
the unknown value of $\nua$ (Figure \ref{fig:120326A_wind_highp_corrplots}). This high value of 
$\EK$ also implies a low radiative efficiency, $\eta_{\rm rad} \equiv 
\Egamma/(\EK+\Egamma)\approx1\%$.

We now investigate the effect of energy injection in causing an X-ray/UV/optical re-brightening. 
The X-ray light curve during the re-brightening is located above the cooling frequency. In the 
wind model, the flux density above $\nuc$ is $F_{\nu>\nuc}\propto\EKiso^{(2+p)/4}t^{(2-3p)/4}$ 
\citep{gs02}. For $p=2.5$, this reduces to $F_{\nu>\nuc}\propto\EKiso^{1.13} t^{-1.38}$. 
During energy injection, $E\propto t^{m}$, such that $F_{\nu>\nuc}\propto t^{1.13 m-1.38}$. 
The steep rise ($\alpha_1 = 0.85\pm0.19$; Section \ref{text:basic_considerations:re-brightening}) 
requires $m=2.0\pm0.2$. However $0\le m<1$ is bounded (Section \ref{text:energy_injection}), which 
implies that in this model, energy injection due to a distribution of ejecta energy to lower 
Lorentz factors can not cause the X-ray flux to rise with time. 

Relaxing the requirement $m<1$, our best solution for the multi-wavelength re-brightening for an 
energy injection model requires two periods of energy injection. In the first episode between 
$1.7\times10^{-3}$\,d and $2.5\times10^{-2}$\,d, \EKiso\ increases as $t^{0.5}$ growing by a factor 
of $3.9$ from $1.6\times10^{51}$\,erg to $6.2\times10^{51}$\,erg. In the second episode, 
$\EKiso\propto t^{2.3}$ from $2.5\times10^{-2}$\,d to 0.4\,d, further increasing by a factor of 
over 2000 to its final value of $\EKiso\approx3.6\times10^{54}$\,erg in this period. The resulting 
light curves, which are optimized to match the UV and optical re-brightening, cannot reproduce the 
X-ray light curve prior to the re-brightening (a shallower rise in the X-rays in the ISM model was 
achieved by placing $\numax$ between the optical and X-rays). Due to the extremely large injected 
energy coupled with the fact that the steep rise in the optical violates the bounds on $m$, the wind 
model is a less attractive solution for the multi-wavelength afterglow of \me.

\begin{figure*}
\begin{tabular}{cc}
 \centering
 \includegraphics[width=0.47\textwidth]{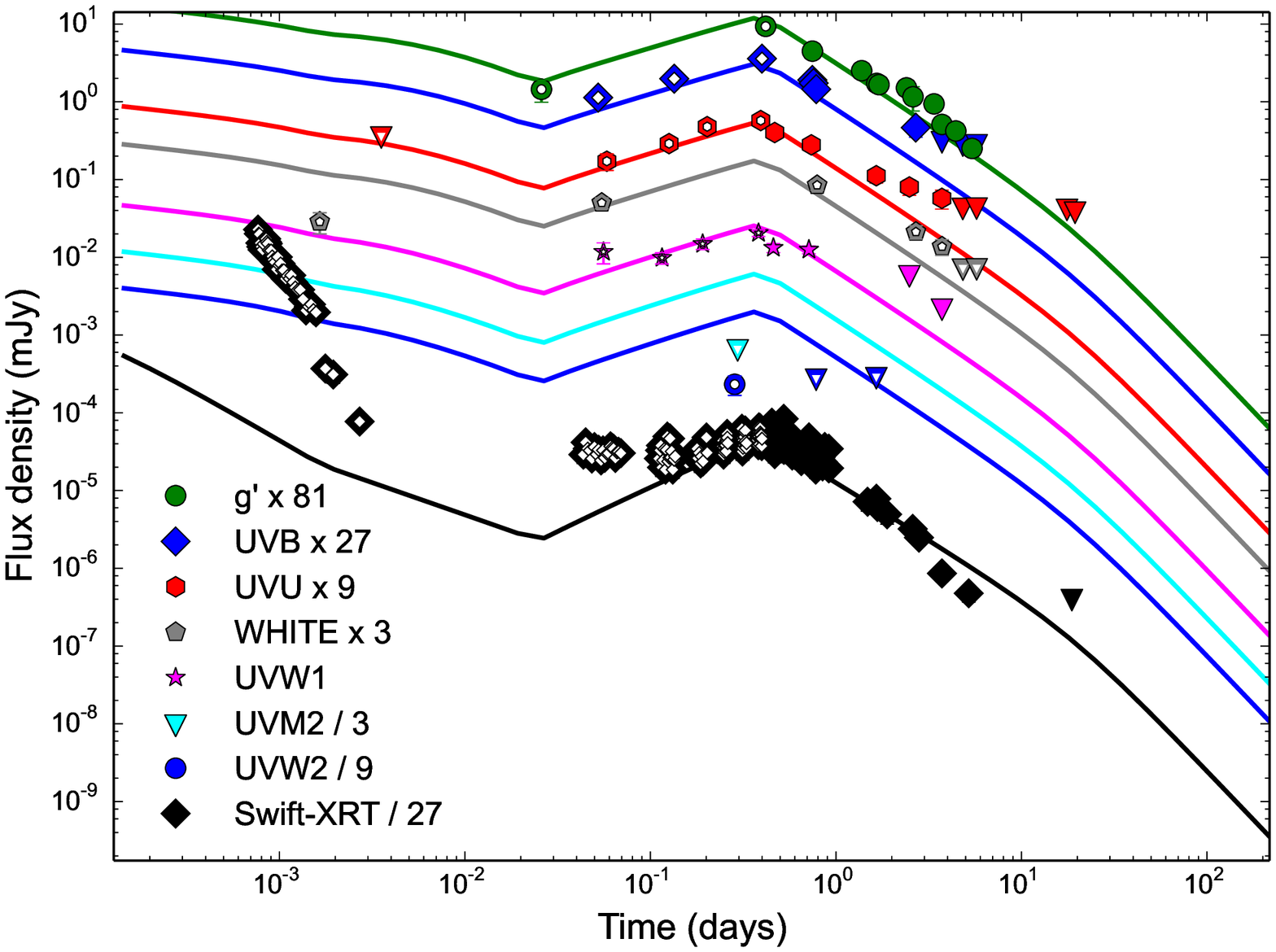} &
 \includegraphics[width=0.47\textwidth]{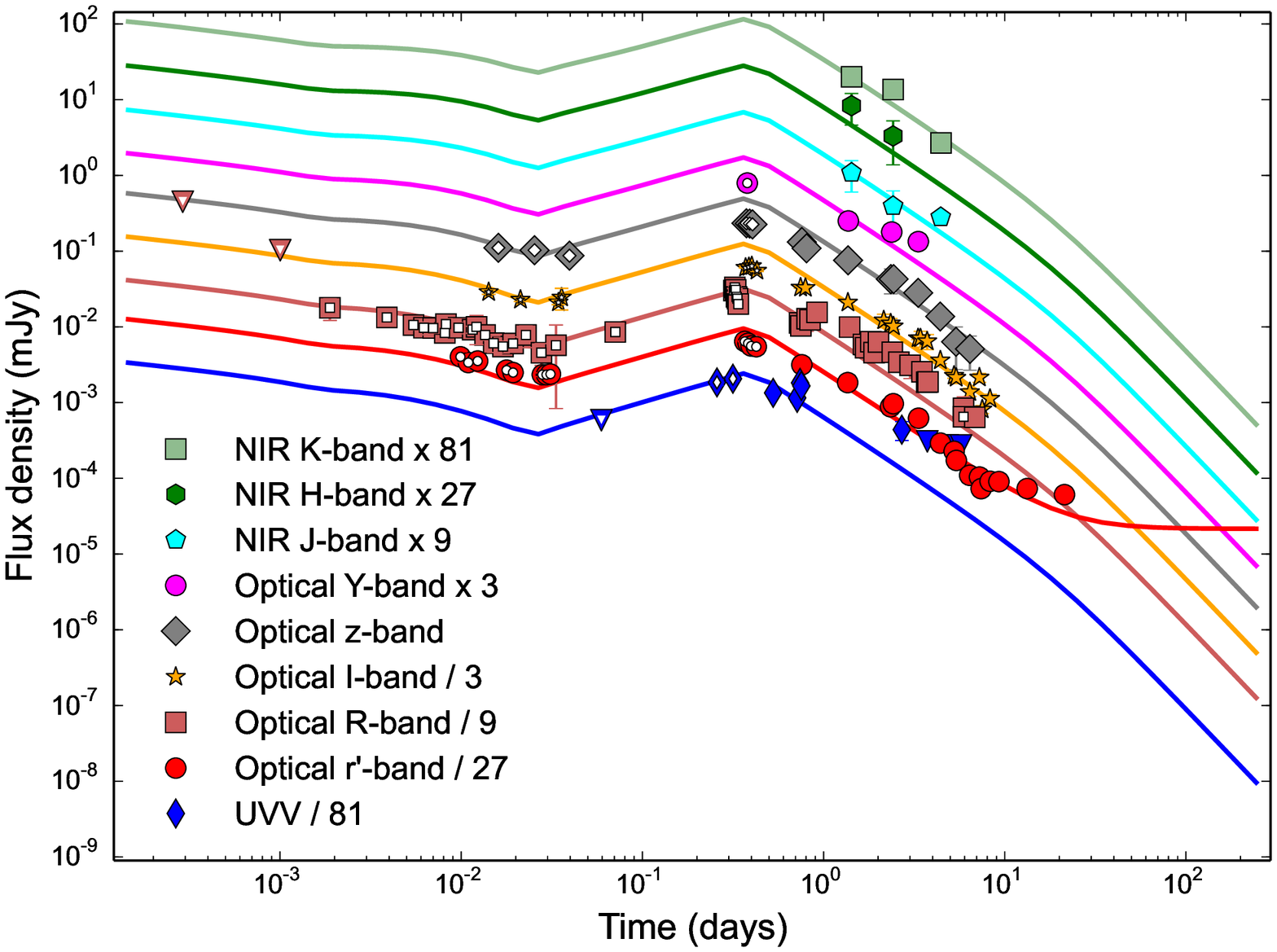} \\
 \includegraphics[width=0.47\textwidth]{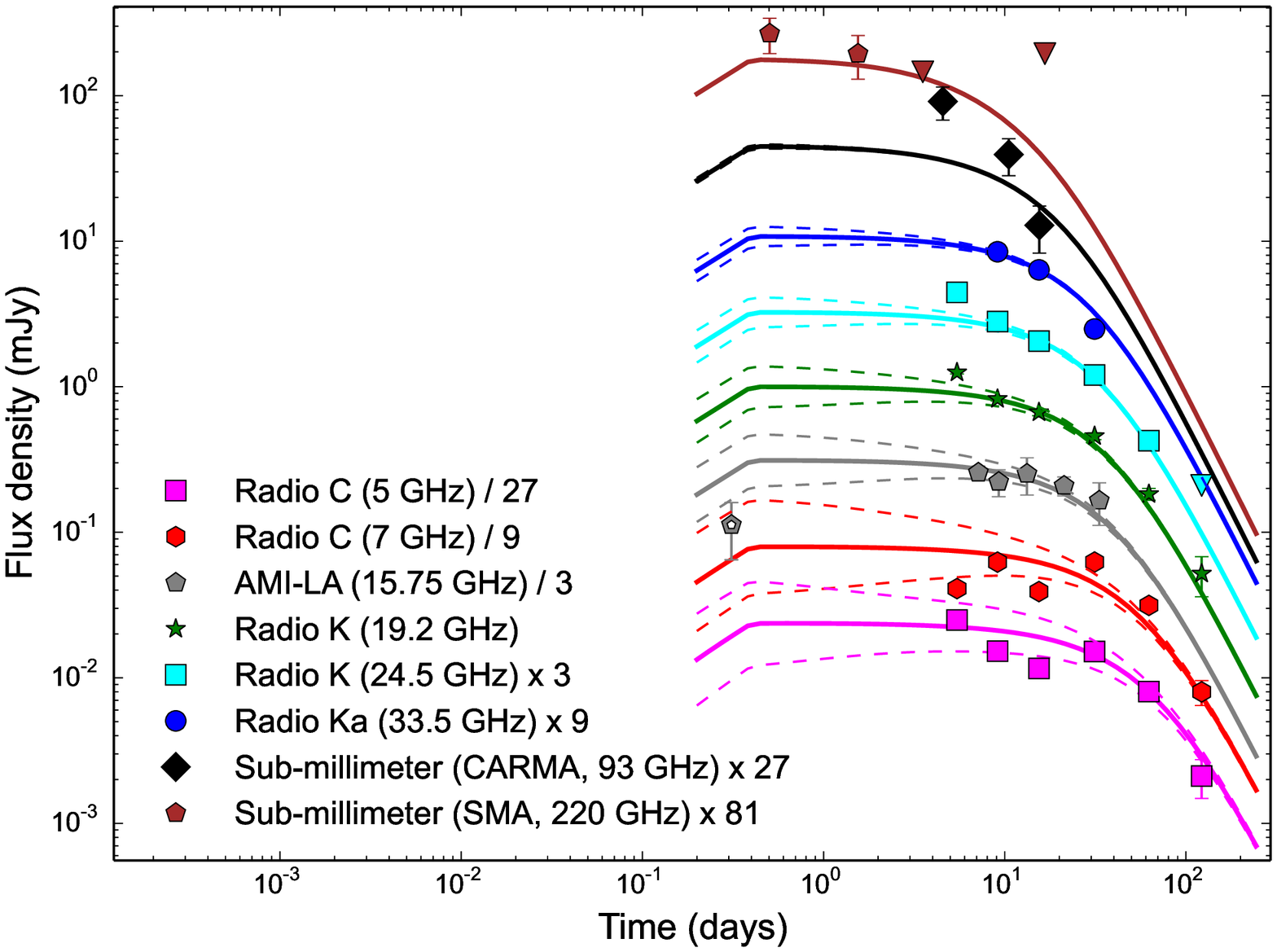} & \\ 
\end{tabular}
\caption{X-ray, UV (top left), optical (top right), and radio (bottom left) light curves of 
GRB~120326A in the wind scenario, with the full afterglow model (solid lines), including energy 
injection before 0.4\,d.
\label{fig:120326A_wind_highp_enj}}
\end{figure*}

\section{A Wind Model for GRB~100418A}
\label{appendix:100418A_wind}
We apply our MCMC analysis described in Section \ref{text:100418A:FS} to explore afterglow models 
with a wind-like circumburst environment for GRB~100418A. The parameters of the highest-likelihood 
model are $p\approx2.1$, $\epse\approx0.33$, $\epsb\approx0.33$, $\Astar\approx0.15$, 
$\EKiso\approx4.0\times10^{51}$\,erg, and $F_{\nu,\rm host, White}\approx2.3\,\mu$Jy, with 
negligible extinction. This model transitions from fast to slow cooling at $1.5$\,d. The spectral 
break frequencies at 1\,d are located at $\nuac\approx5.8$\,GHz, $\nusa\approx32$\,GHz, 
$\nuc\approx6.4\times10^{11}$\,Hz, and $\numax\approx3.6\times10^{12}$\,Hz at 1\,d, with $F_{\rm 
max} \approx50$\,mJy at $\nuc$ at 1\,d and a Compton $y$-parameter of 0.6. 

\begin{figure}
\begin{tabular}{ccc}
 \centering
 \includegraphics[width=0.30\columnwidth]{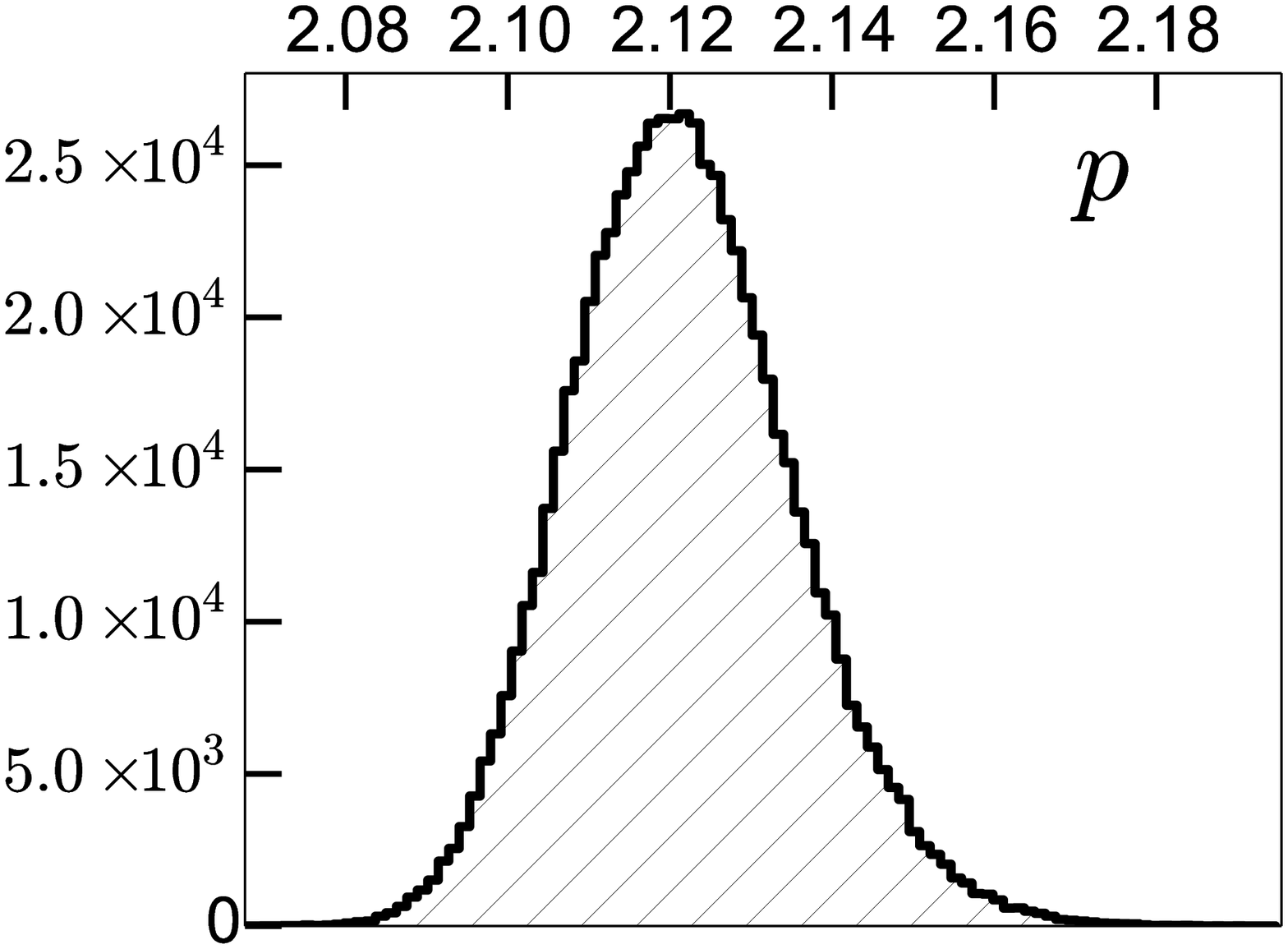} &
 \includegraphics[width=0.30\columnwidth]{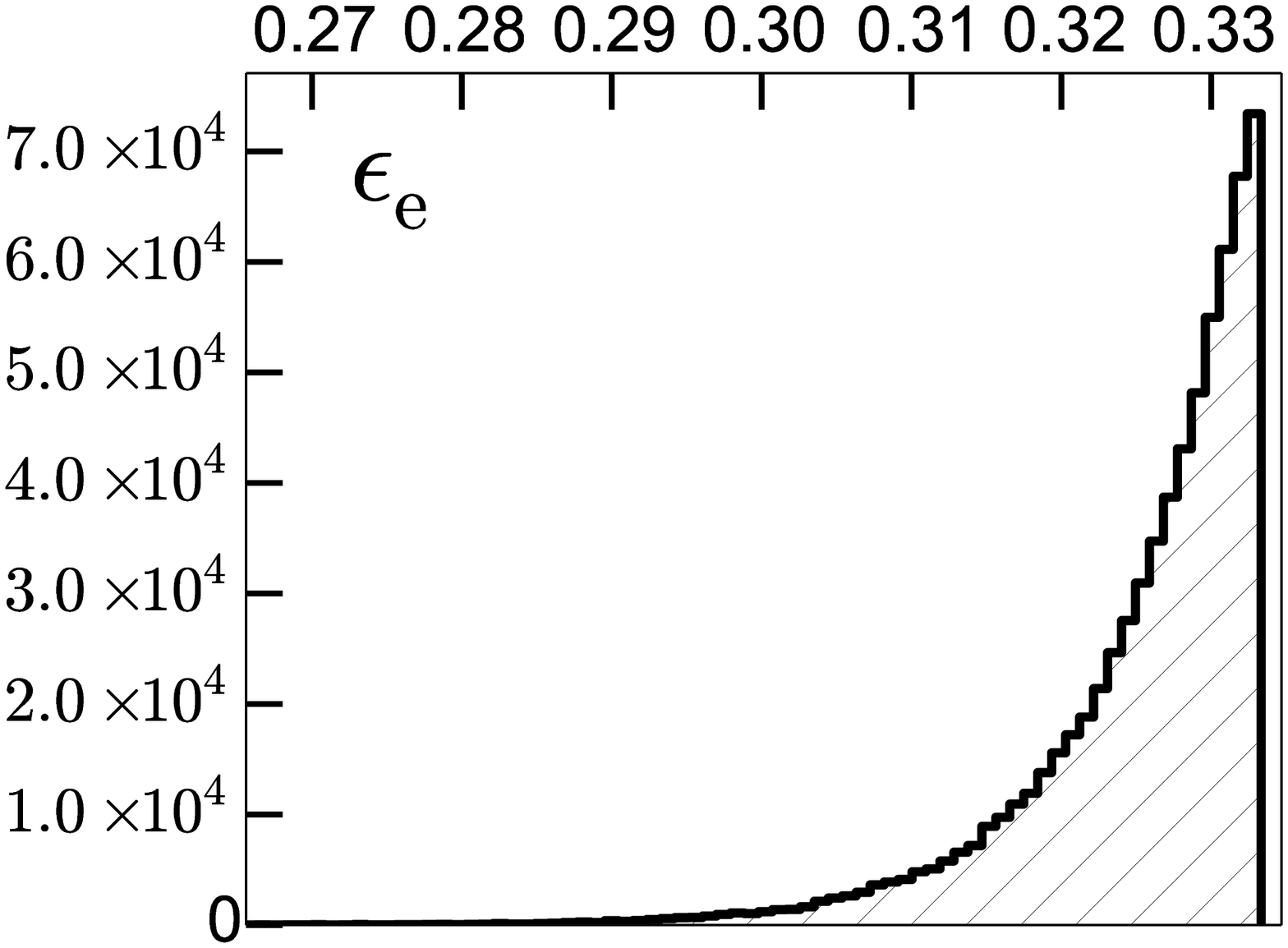} &
 \includegraphics[width=0.30\columnwidth]{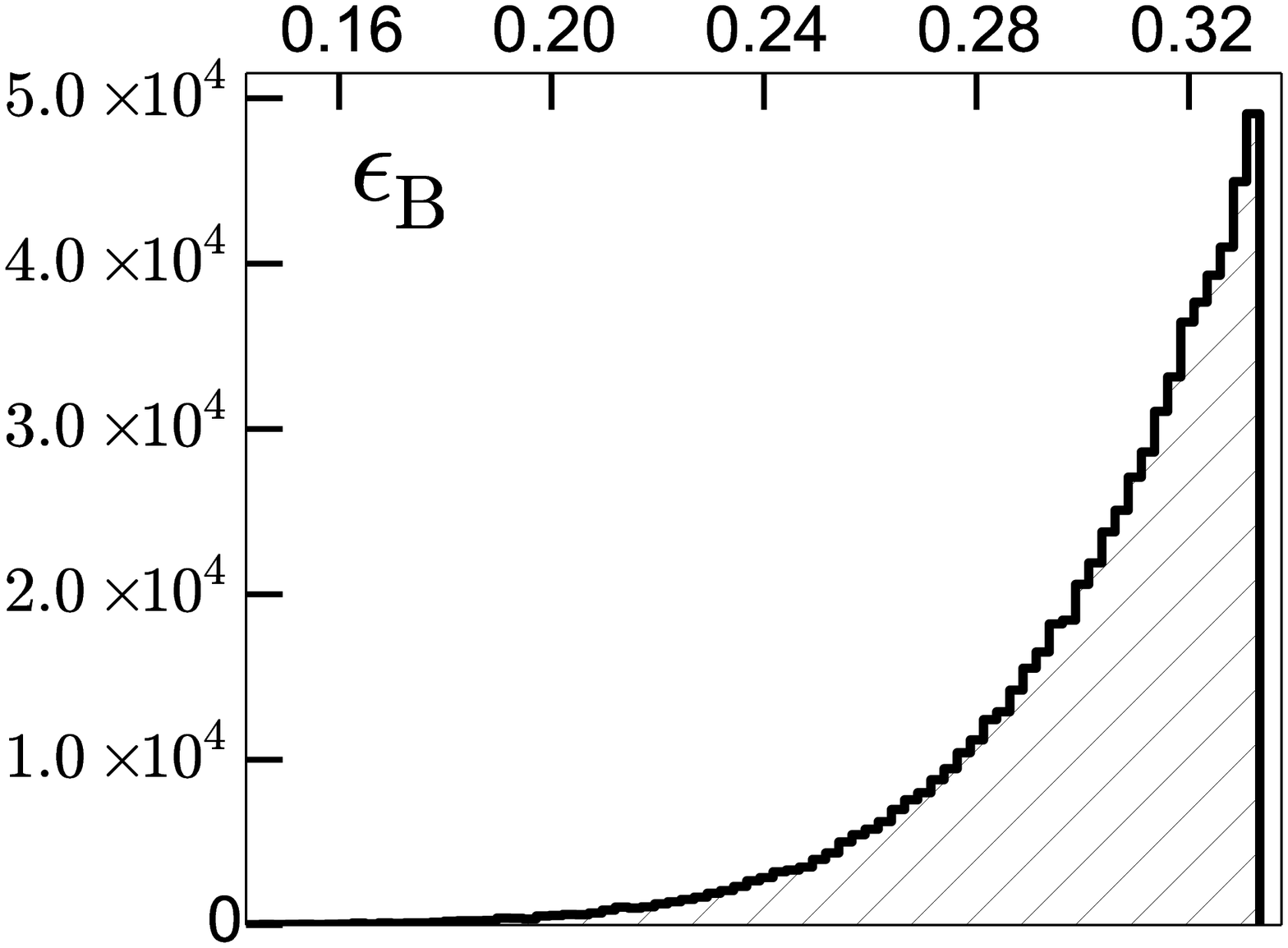} \\
 \includegraphics[width=0.30\columnwidth]{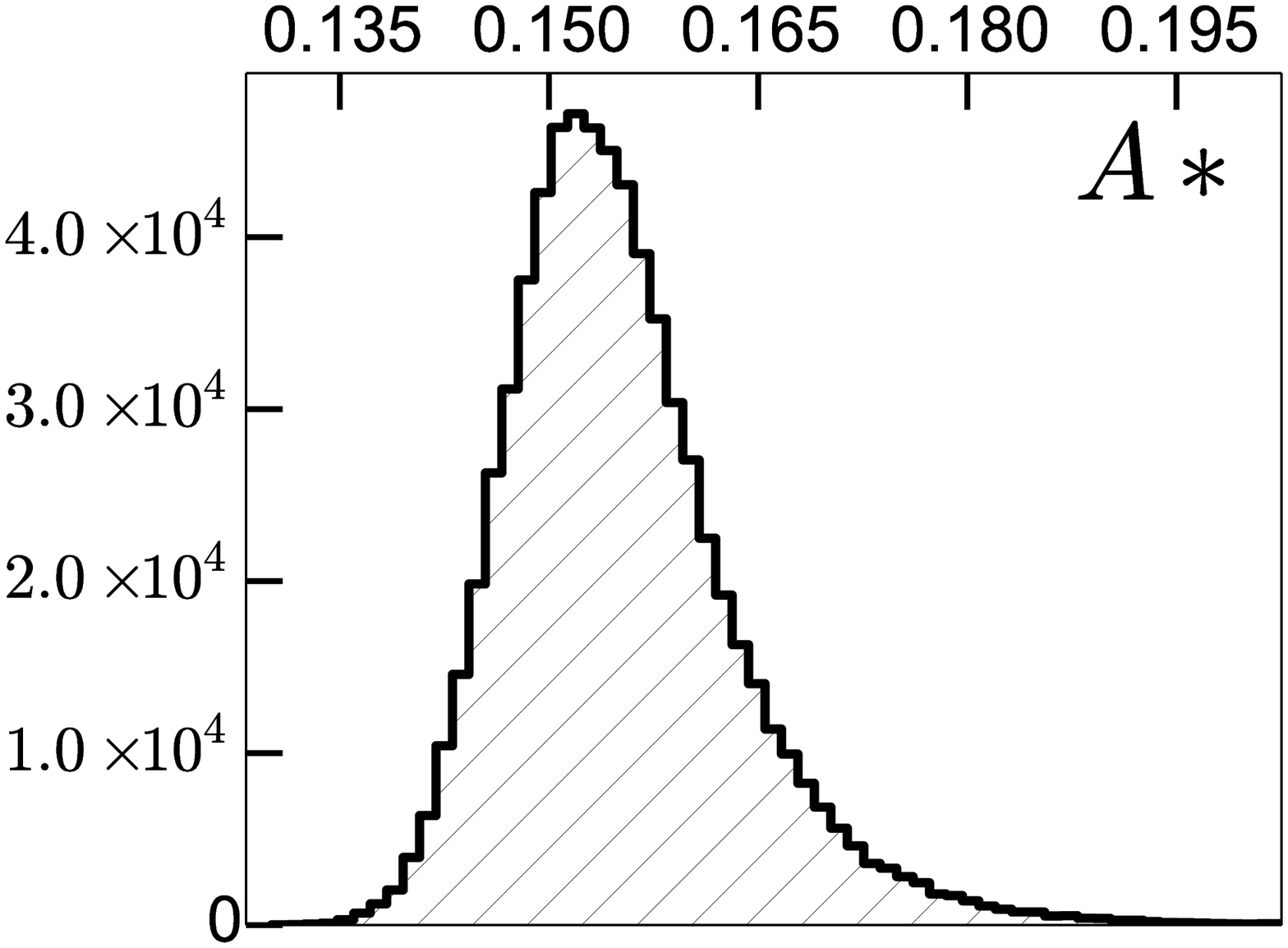} & 
 \includegraphics[width=0.30\columnwidth]{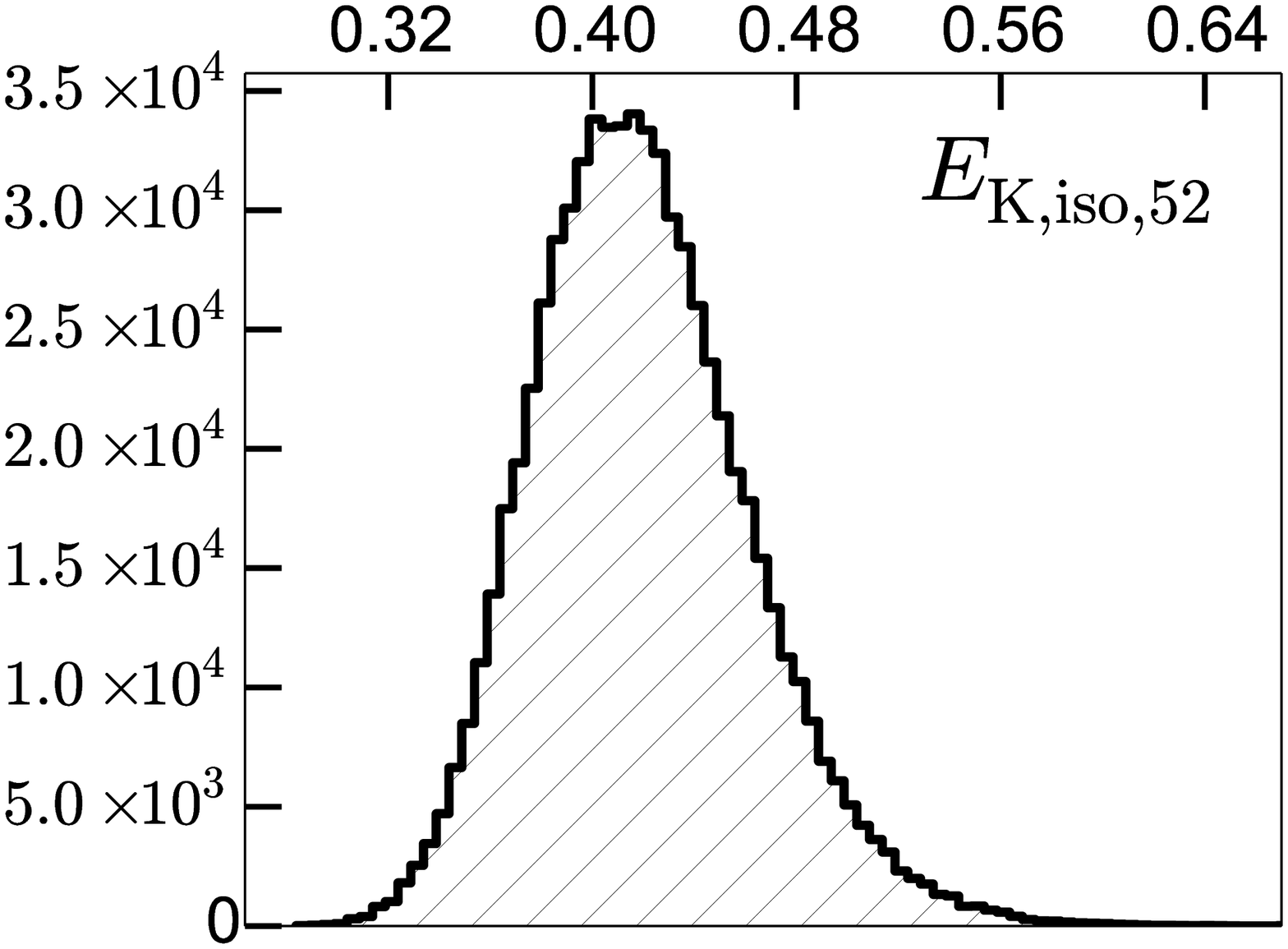} &
 \includegraphics[width=0.30\columnwidth]{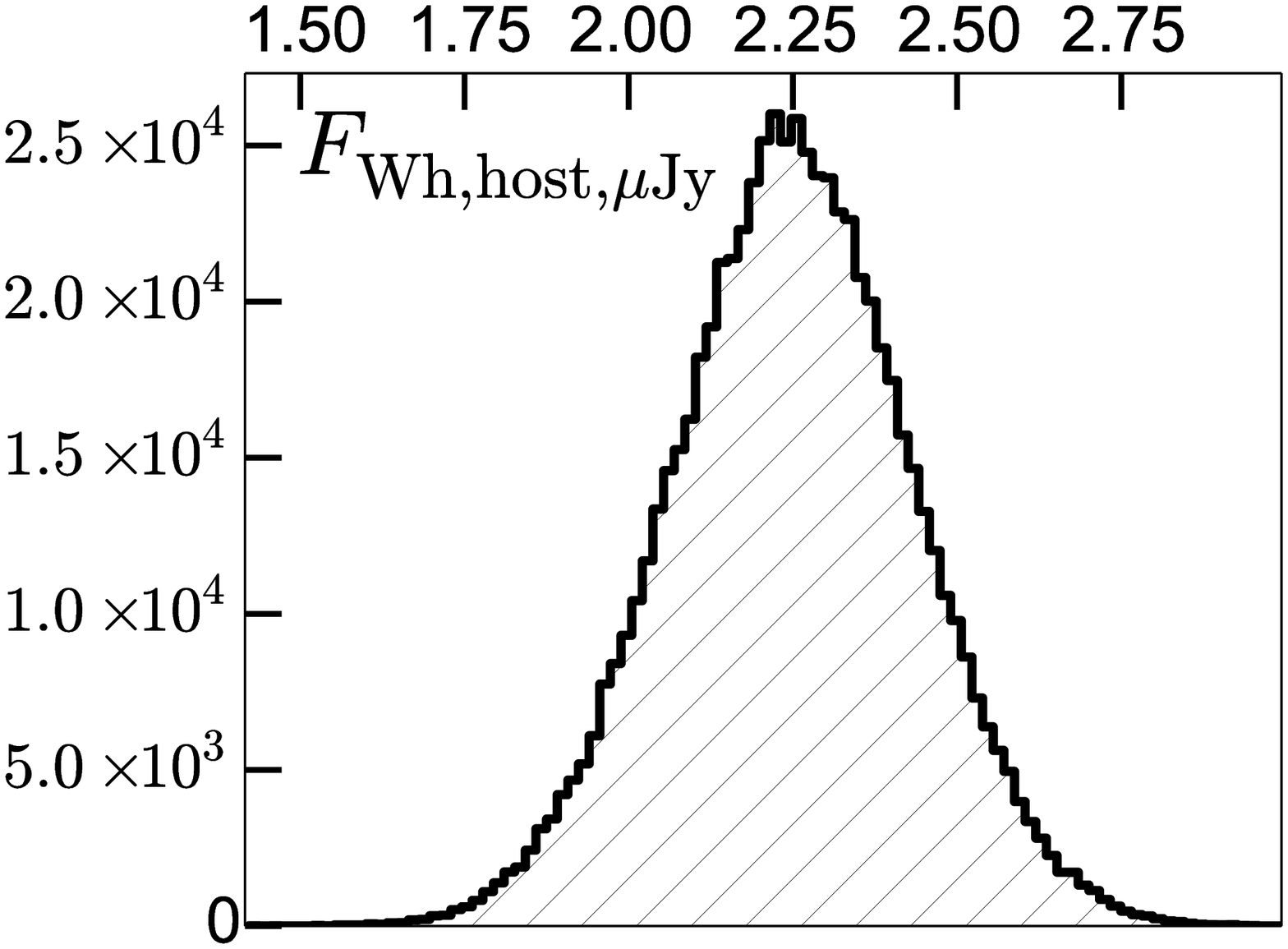} \\  
\end{tabular}
\caption{Posterior probability density functions for the physical parameters for GRB~100418A in 
a wind environment from MCMC simulations. We have restricted $E_{\rm K, iso, 52} 
< 500$, $\epsilon_{\rm e} < \nicefrac{1}{3}$, and $\epsilon_{\rm B} < \nicefrac{1}{3}$. The last 
panel corresponds to the flux density of the host galaxy in the \Swift/\textit{White} band.
\label{fig:100418A_wind_hists}}
\end{figure}

\begin{figure}
\begin{tabular}{ccc}
\centering
 \includegraphics[width=0.30\columnwidth]{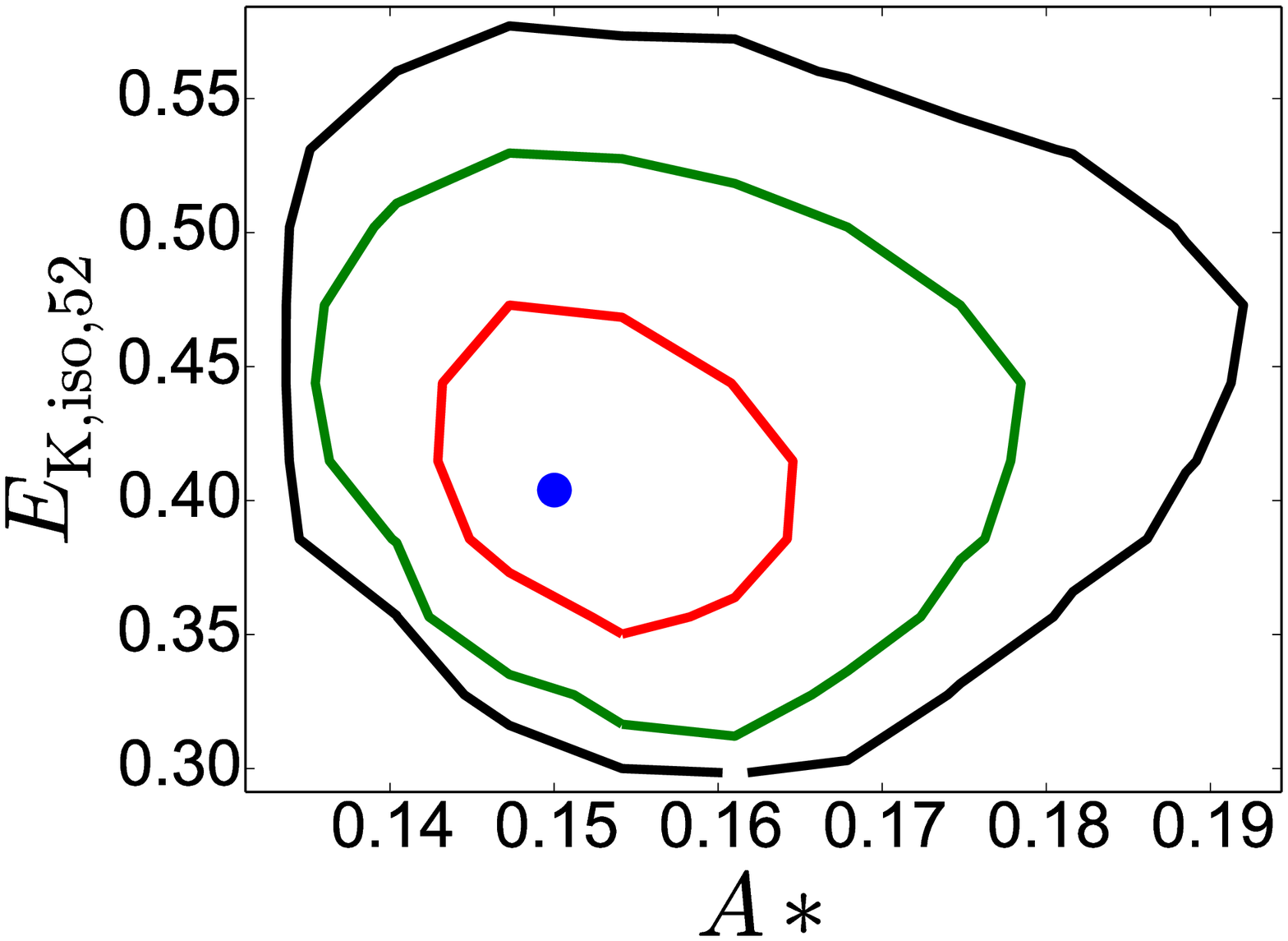} &
 \includegraphics[width=0.30\columnwidth]{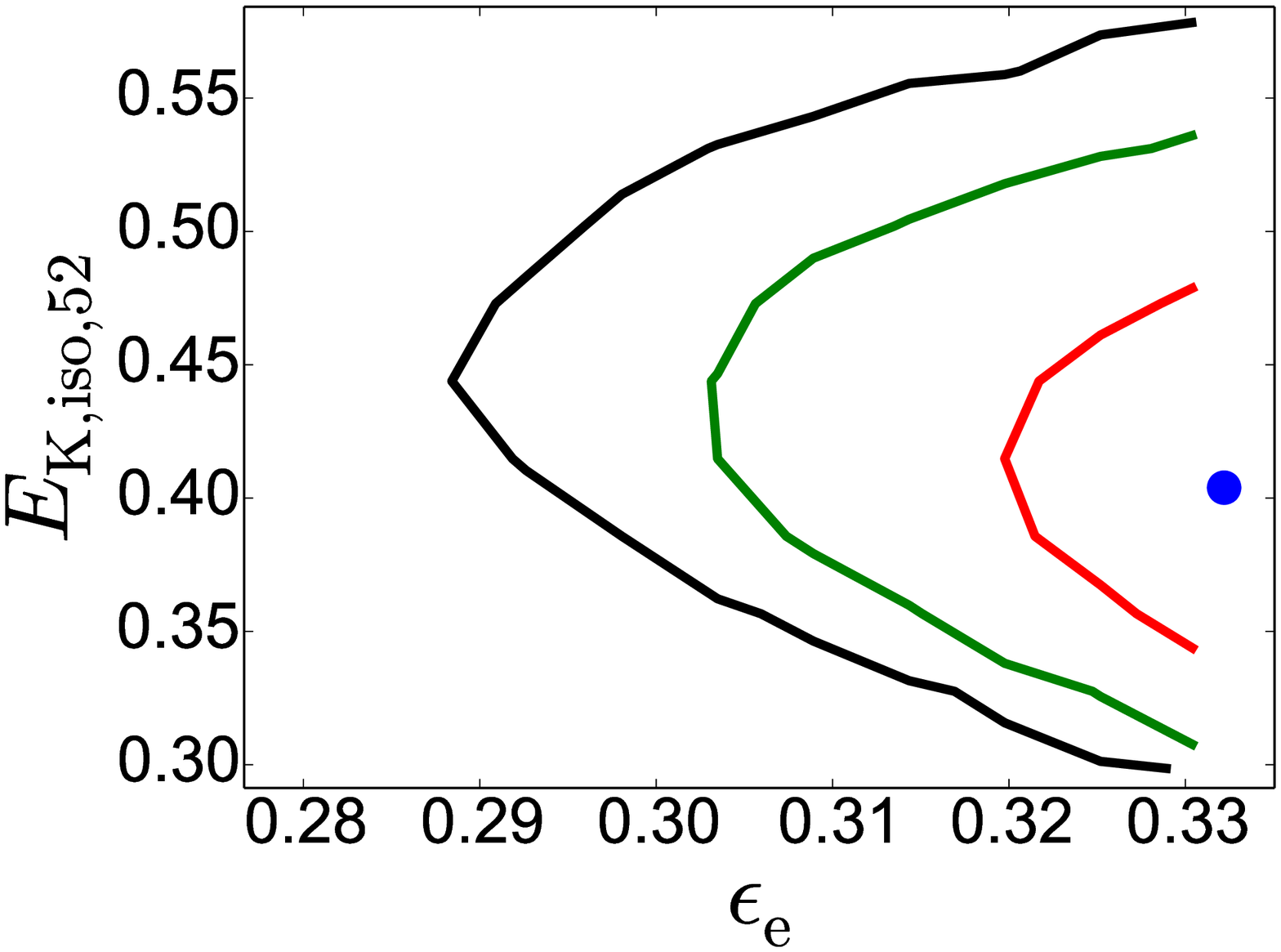} &
 \includegraphics[width=0.30\columnwidth]{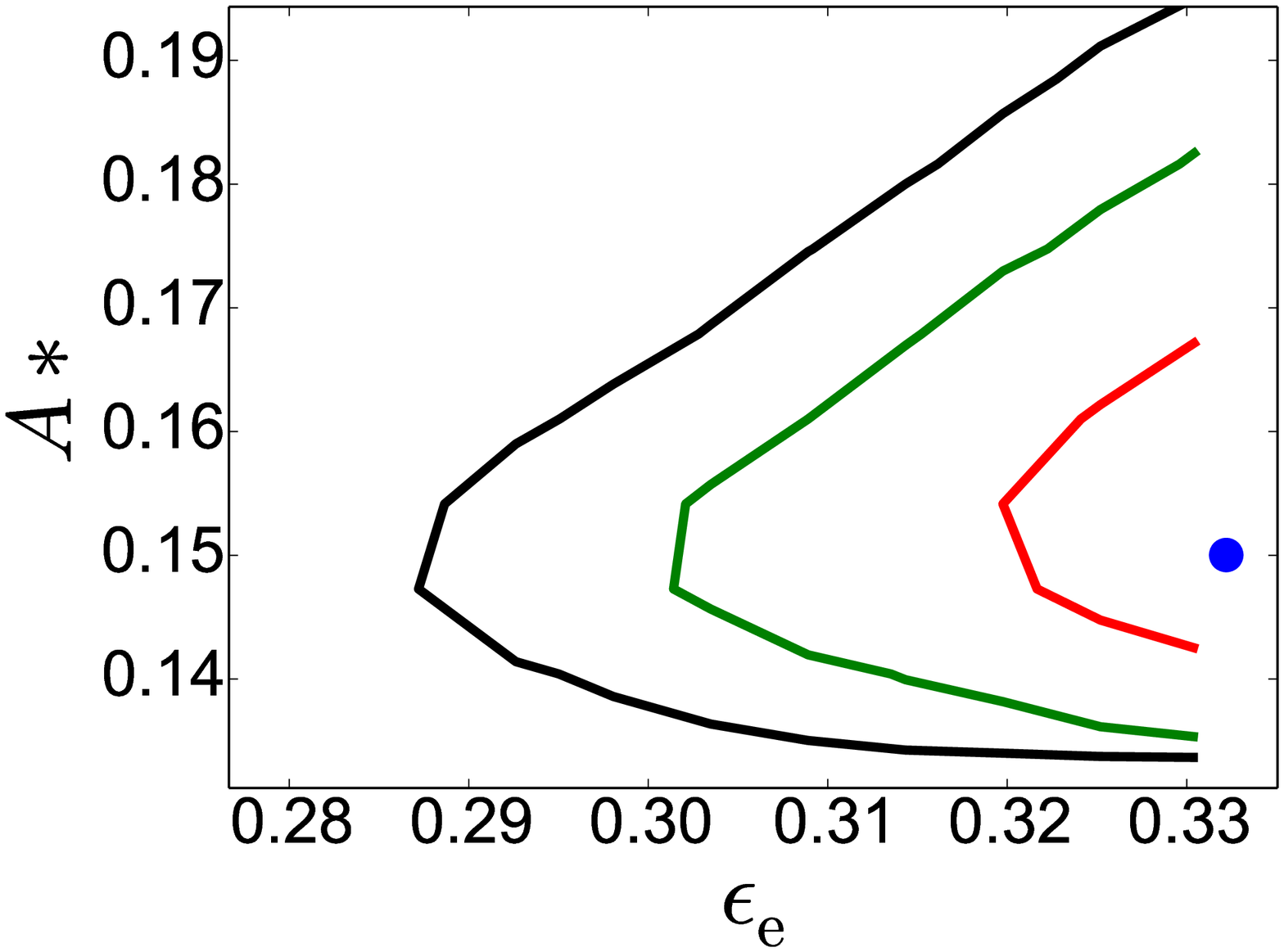} \\
 \includegraphics[width=0.30\columnwidth]{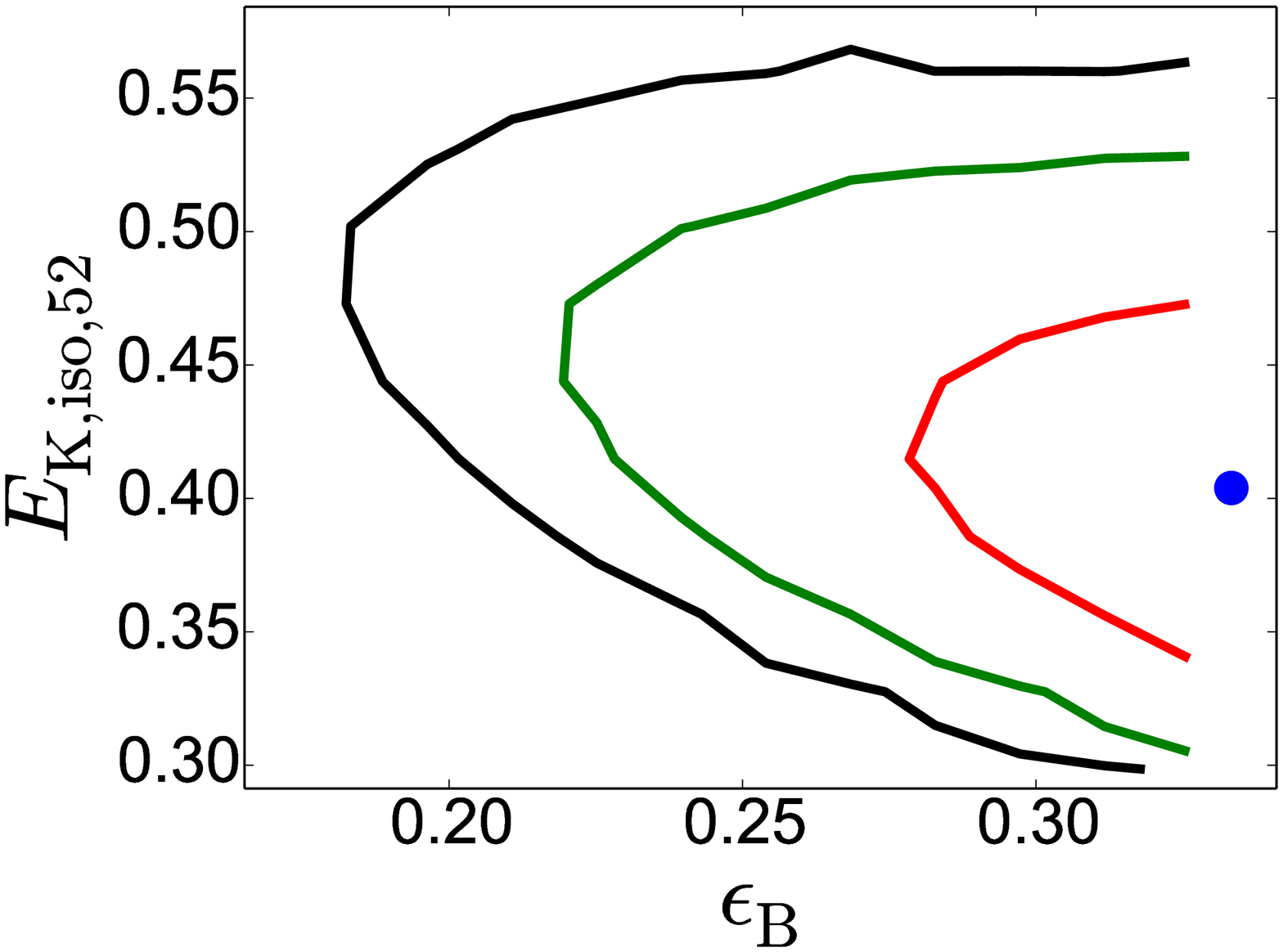} &
 \includegraphics[width=0.30\columnwidth]{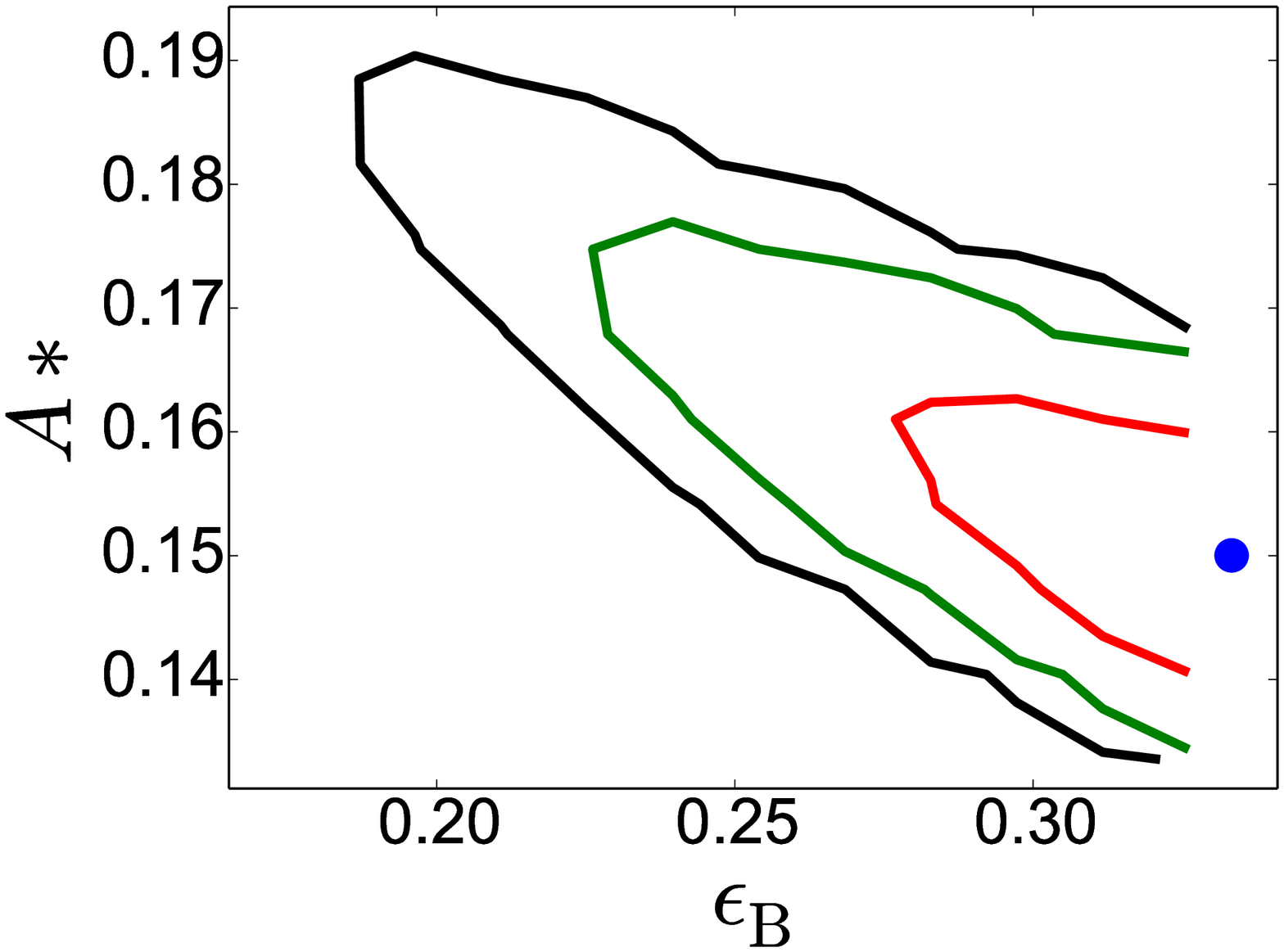} &
 \includegraphics[width=0.30\columnwidth]{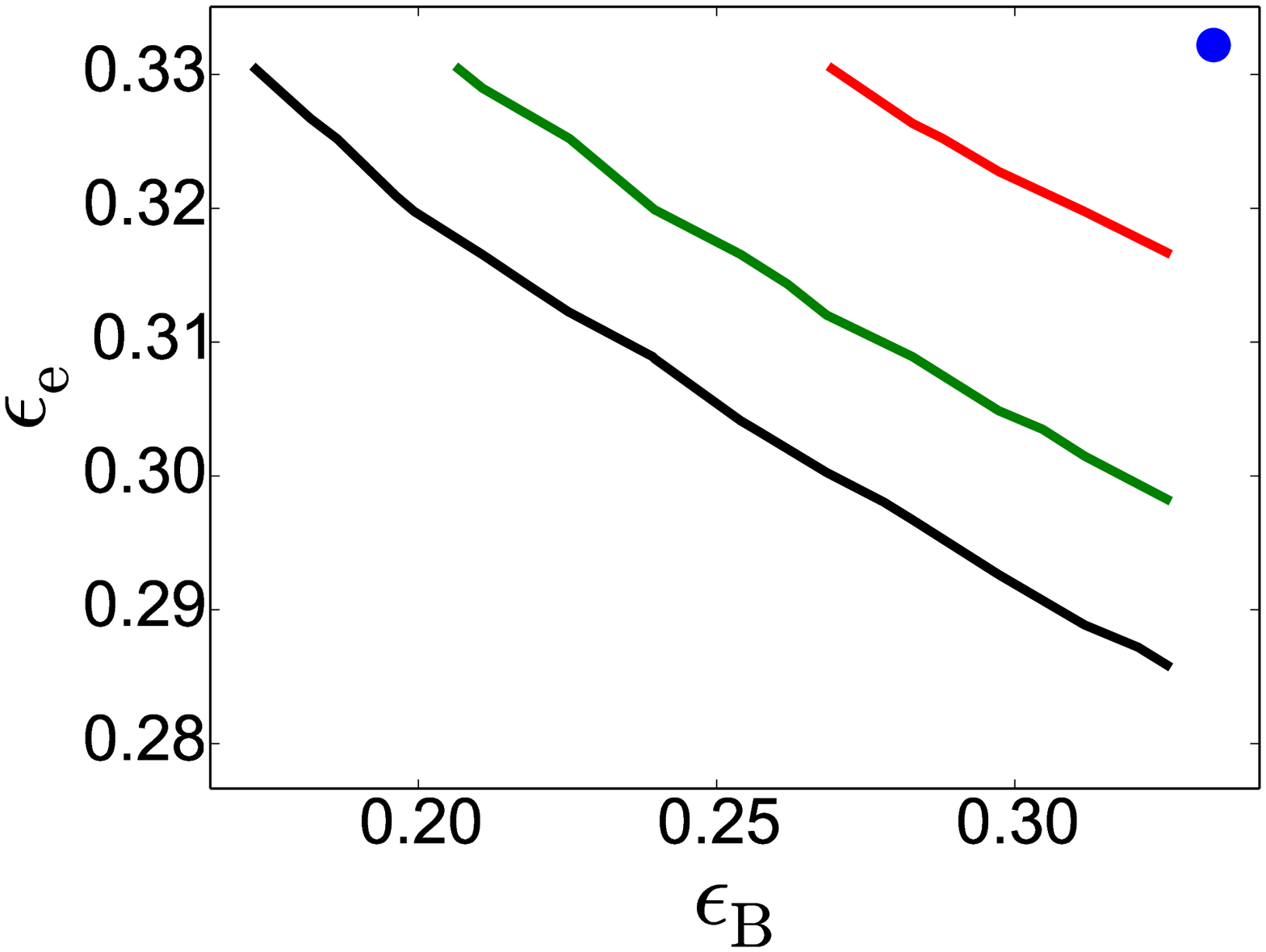} \\
\end{tabular}
\caption{1$\sigma$ (red), 2$\sigma$ (green), and 3$\sigma$ (black) contours for correlations
between the physical parameters, \EKiso, \dens, \epse, and \epsb\ for GRB~100418A, in the wind 
model from Monte Carlo simulations. We have restricted $E_{\rm K, iso, 52} < 500$, $\epsilon_{\rm 
e} < \nicefrac{1}{3}$, and $\epsilon_{\rm B} < \nicefrac{1}{3}$. The highest-likelihood model 
is marked with a blue dot. See the on line version of this Figure for additional plots of 
correlations between these parameters and $p$ and $F_{\nu, \rm host, White}$. 
\label{fig:100418A_wind_corrplots}}
\end{figure}

Like in the ISM model, the optical and X-ray bands are located above both \numax\, and \nuc. This 
model does not require a jet break, and we find $\tjet\gtrsim140$\,d. Thus we cannot constrain 
\thetajet\ in this model, nor correct \Egammaiso\ or \EKiso\ for beaming. The summary statistics 
from our MCMC simulations are $p=2.12\pm0.01$, $\EKiso = (4.2\pm0.4)\times10^{51}$\,erg, 
$\Astar=0.16^{+0.008}_{-0.006}$, $\epse=0.328^{+0.004}_{-0.009}$, and $\epsb=0.31^{+0.03}_{-0.02}$. 
We plot histograms of the posterior density functions for these parameters in Figure 
\ref{fig:100418A_wind_hists} and present contours of the joint posterior density for the physical 
parameters \Astar, \EKiso, \epse, and \epsb\ in Figure \ref{fig:100418A_wind_corrplots}.

\begin{figure*}
\begin{tabular}{cc}
 \centering
 \includegraphics[width=0.47\textwidth]{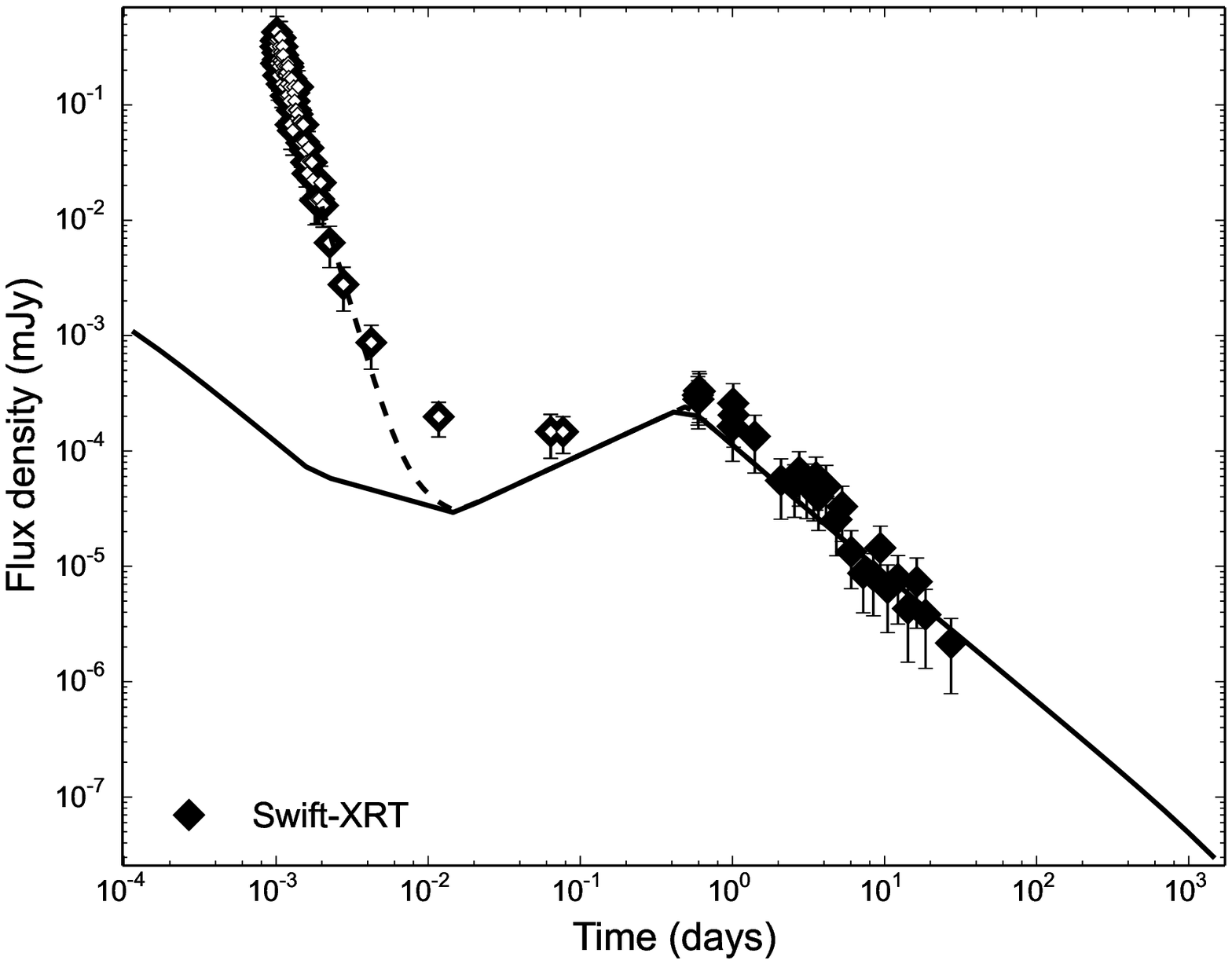} &
 \includegraphics[width=0.47\textwidth]{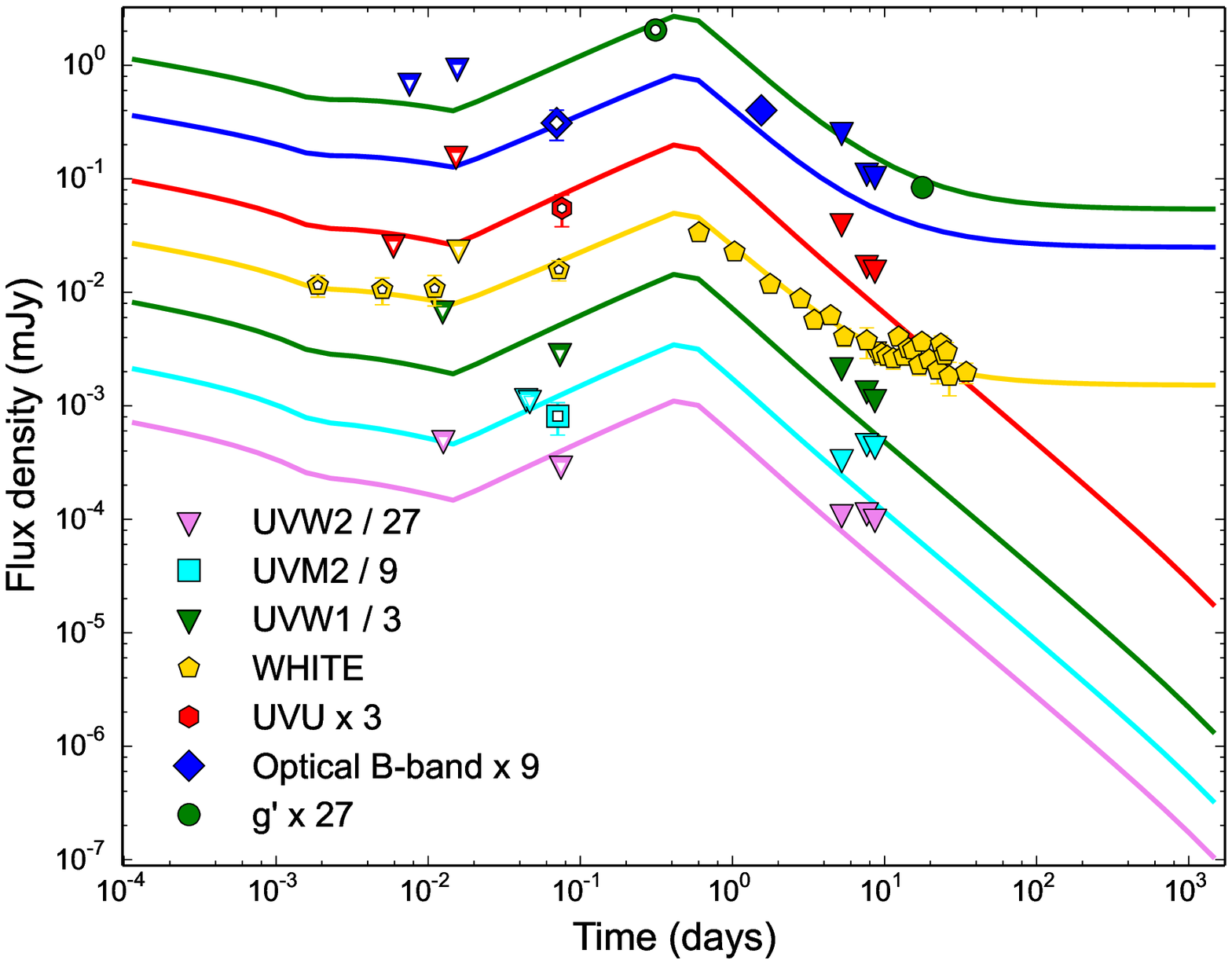} \\
 \includegraphics[width=0.47\textwidth]{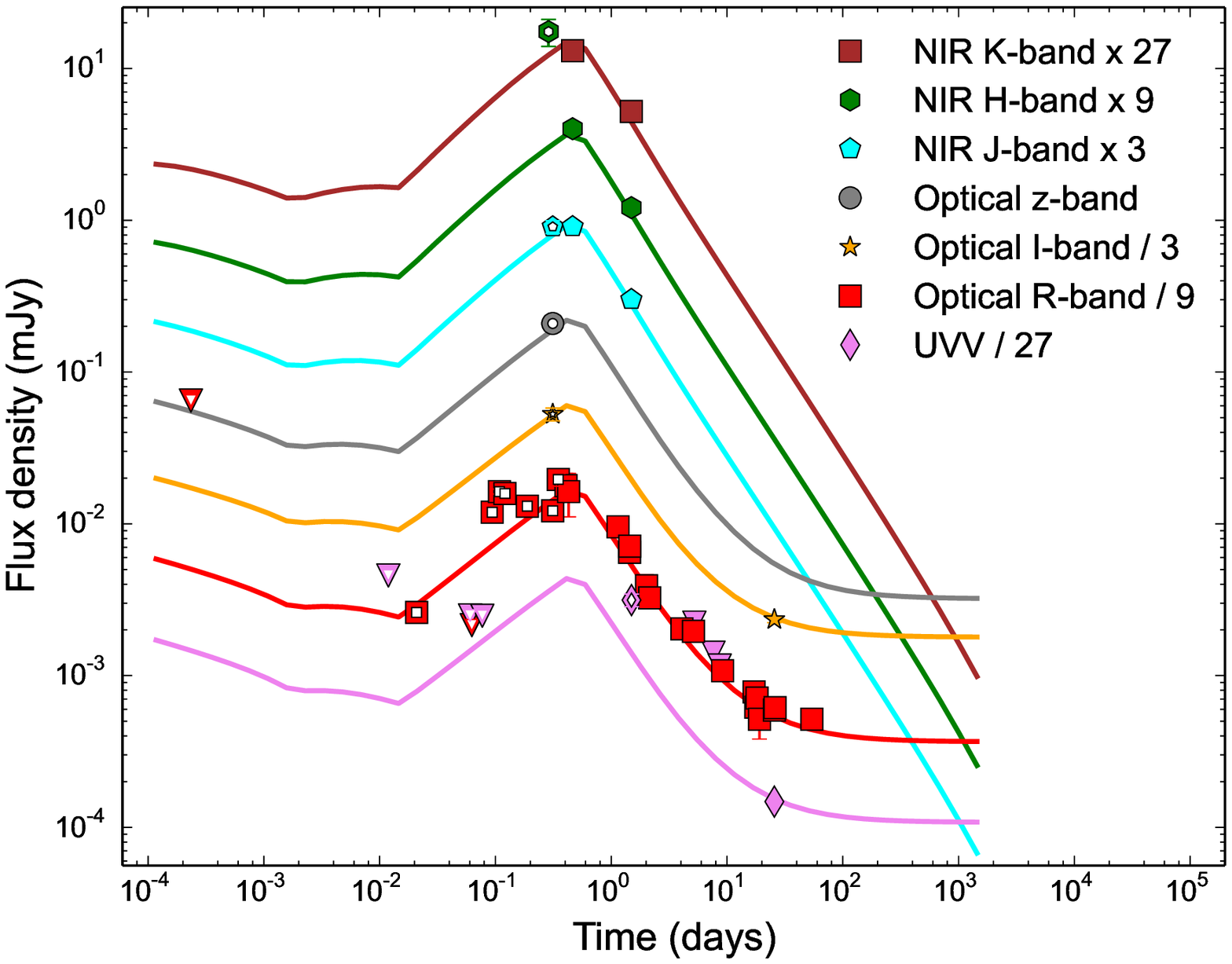} &
 \includegraphics[width=0.47\textwidth]{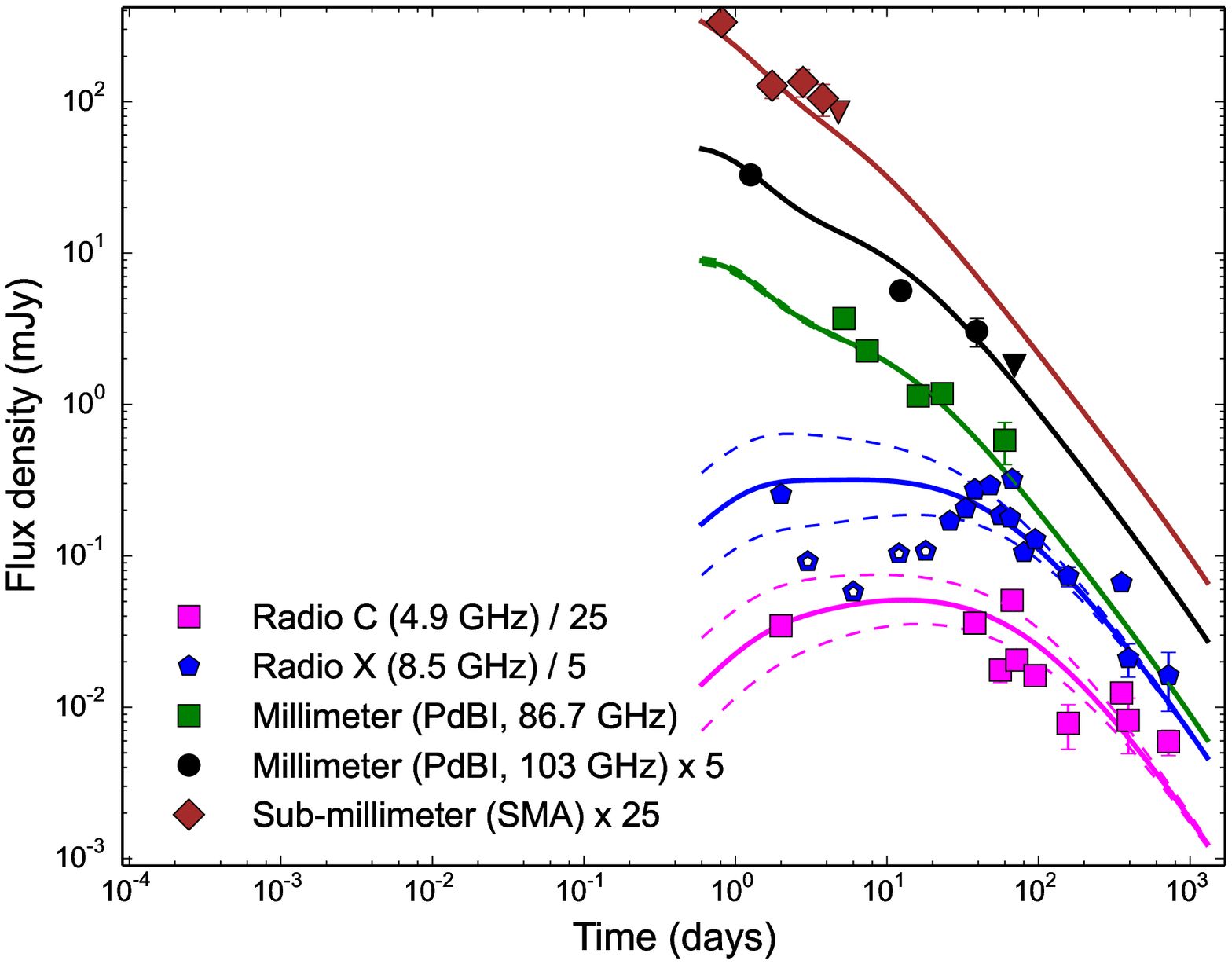} \\ 
\end{tabular}
\caption{X-ray (top left), UV (top right), optical (bottom left), and radio (bottom right) light 
curves of GRB~100418A in the wind scenario, with the full afterglow model (solid lines), including 
energy injection before 0.5\,d.
\label{fig:100418A_wind_enj}}
\end{figure*}

It is challenging to fit both the X-ray and optical light curves before the bump together in 
the energy injection scenario under the wind model. Like for the wind model for \me\ (Appendix 
\ref{appendix:120326A_wind}), we find that the optical light curves before the peak require a 
steeper injection rate than allowed by a distribution of Lorentz factors in the ejecta. In 
particular, our best energy injection model that matches the optical well but slightly 
under-predicts the X-ray data before 0.5\,d (Figure \ref{fig:100418A_wind_enj}), requires $E\propto 
t^{0.7}$ between $1.8\times10^{-3}$\,d and $1.5\times10^{-2}$\,d, steepening to $E\propto t^{1.65}$ 
between $1.5\times10^{-2}$\,d and $0.5$\,d. In this model, the blastwave kinetic energy increases 
by 
a factor of 4.4 between $1.8\times10^{-3}$\,d and $1.5\times10^{-2}$\,d, and another factor of 
$\approx325$ between $1.5\times10^{-2}$\,d and $0.5$\,d, for an overall increase by a factor of 
$1440$. Due to the large injection of energy required to account for the optical re-brightening in 
this model, we do not consider the wind environment to be a likely explanation for the 
multi-wavelength afterglow of GRB~100418A. 

\section{A Wind Model for GRB~100901A}
\label{appendix:100901A_wind}
We apply our MCMC analysis described in Section \ref{text:100901A:FS} to explore afterglow models 
with a wind-like circumburst environment for GRB~100901A. The parameters of the highest-likelihood 
model are $p\approx2.02$, $\epse\approx0.33$, $\epsb\approx0.33$, $\Astar\approx1.7\times10^{-2}$, 
$\EKiso\approx3.8\times10^{53}$, and $\AV \lesssim 0.1$\,mag.
This model transitions from fast to slow cooling at $5.4\times10^{-2}$\,d. The spectral break 
frequencies at 1\,d are located at $\nua\approx3.1\times10^8$\,Hz, 
$\numax\approx1.8\times10^{12}$\,Hz, and $\nuc\approx1.7\times10^{15}$\,Hz, 
with $F_{\rm max} \approx3.8$\,mJy at 1\,d, and a Compton $y$-parameter of 0.6.
Like in the ISM model, \nuc is located between the optical and X-ray bands. Since the light 
curves in a wind environment decline faster than in the ISM case, the jet break in this model is 
later ($\tjet\approx3.5$\,d) compared to the ISM model ($\tjet\approx1$\,d). The jet opening angle 
is $\thetajet\approx1.9$, resulting in a beaming-corrected kinetic energy of 
$2.0\times10^{50}$\,erg.

\begin{figure}
\begin{tabular}{ccc}
 \centering
 \includegraphics[width=0.30\columnwidth]{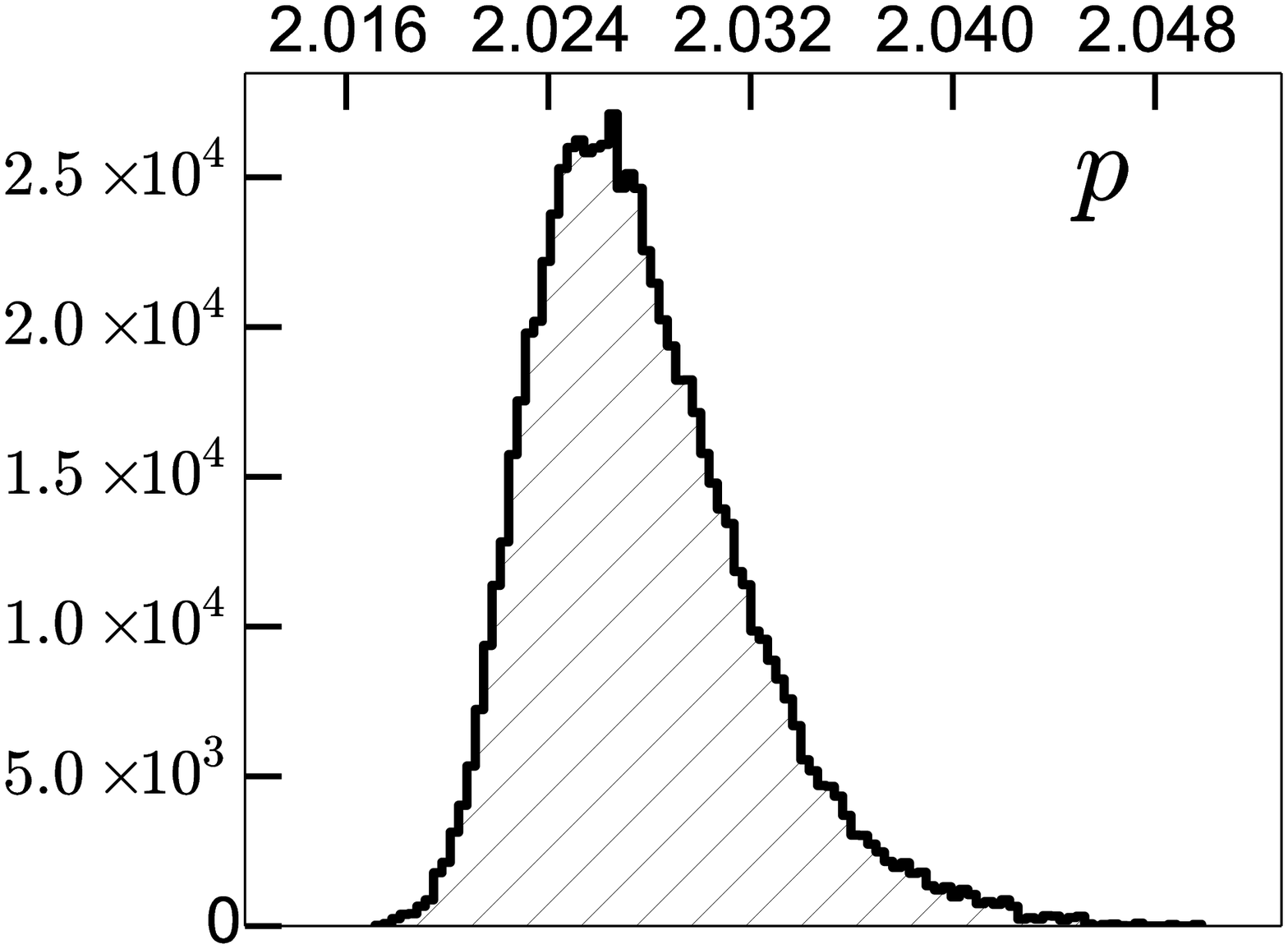} &
 \includegraphics[width=0.30\columnwidth]{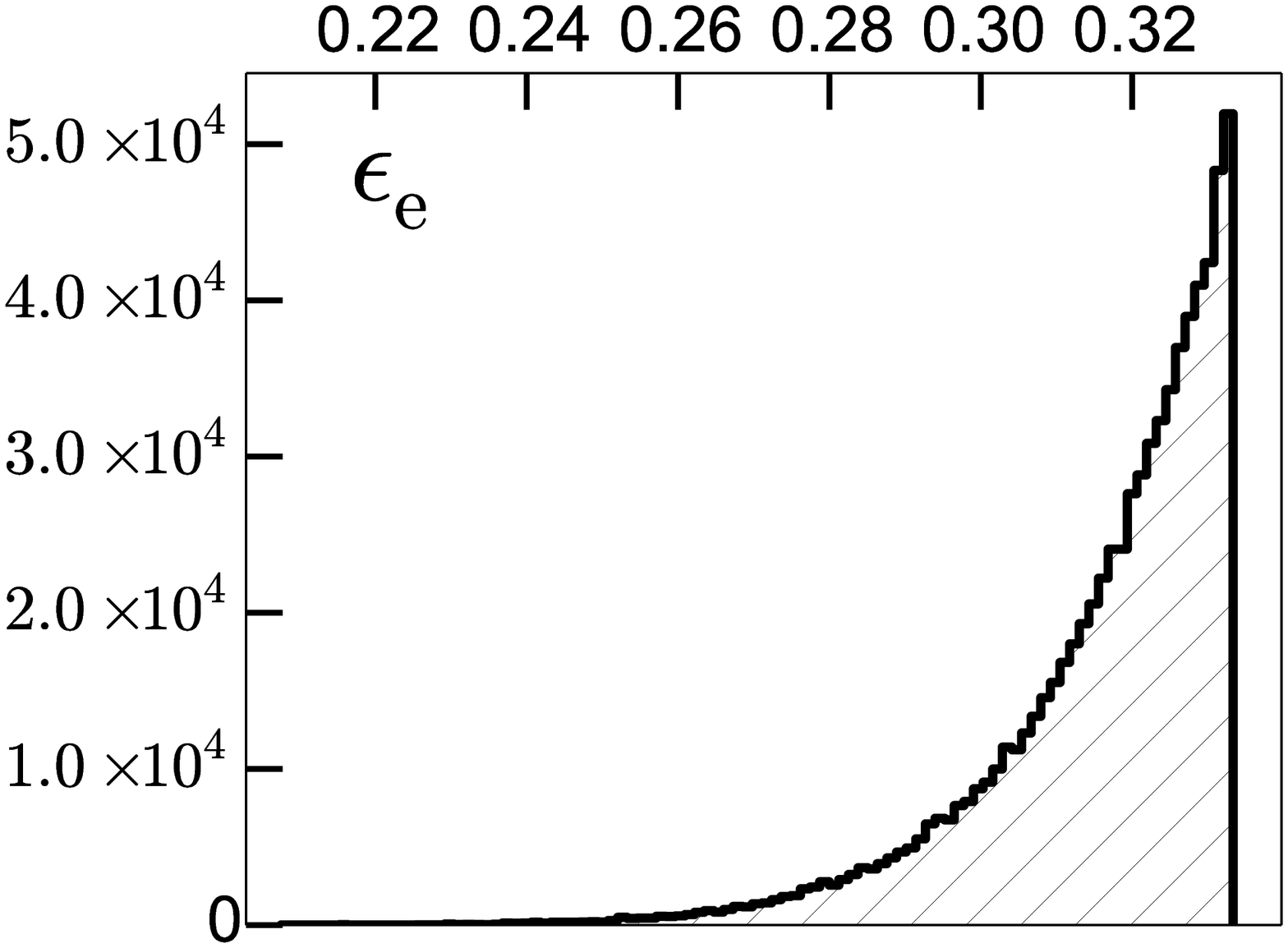} &
 \includegraphics[width=0.30\columnwidth]{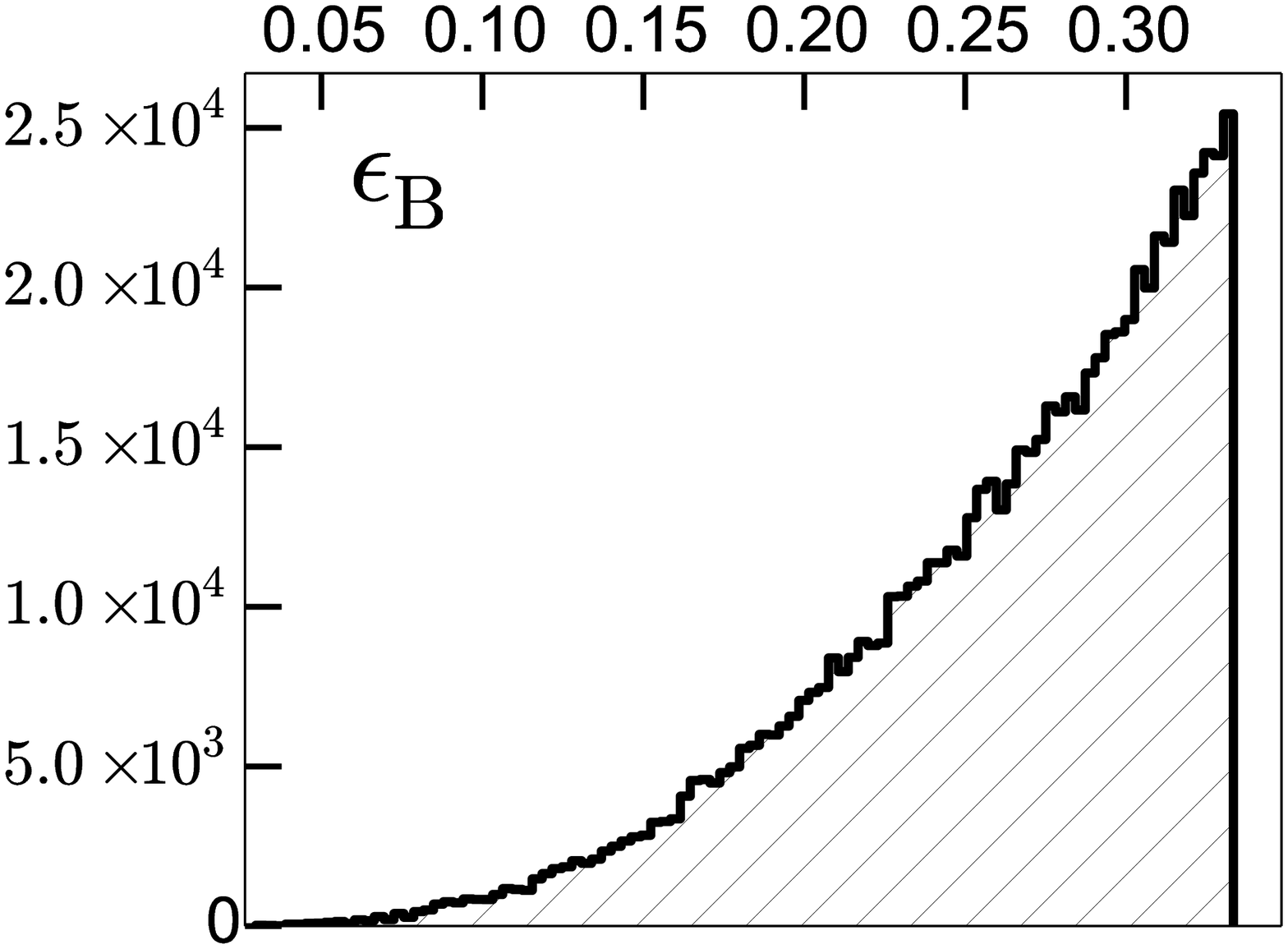} \\
 \includegraphics[width=0.30\columnwidth]{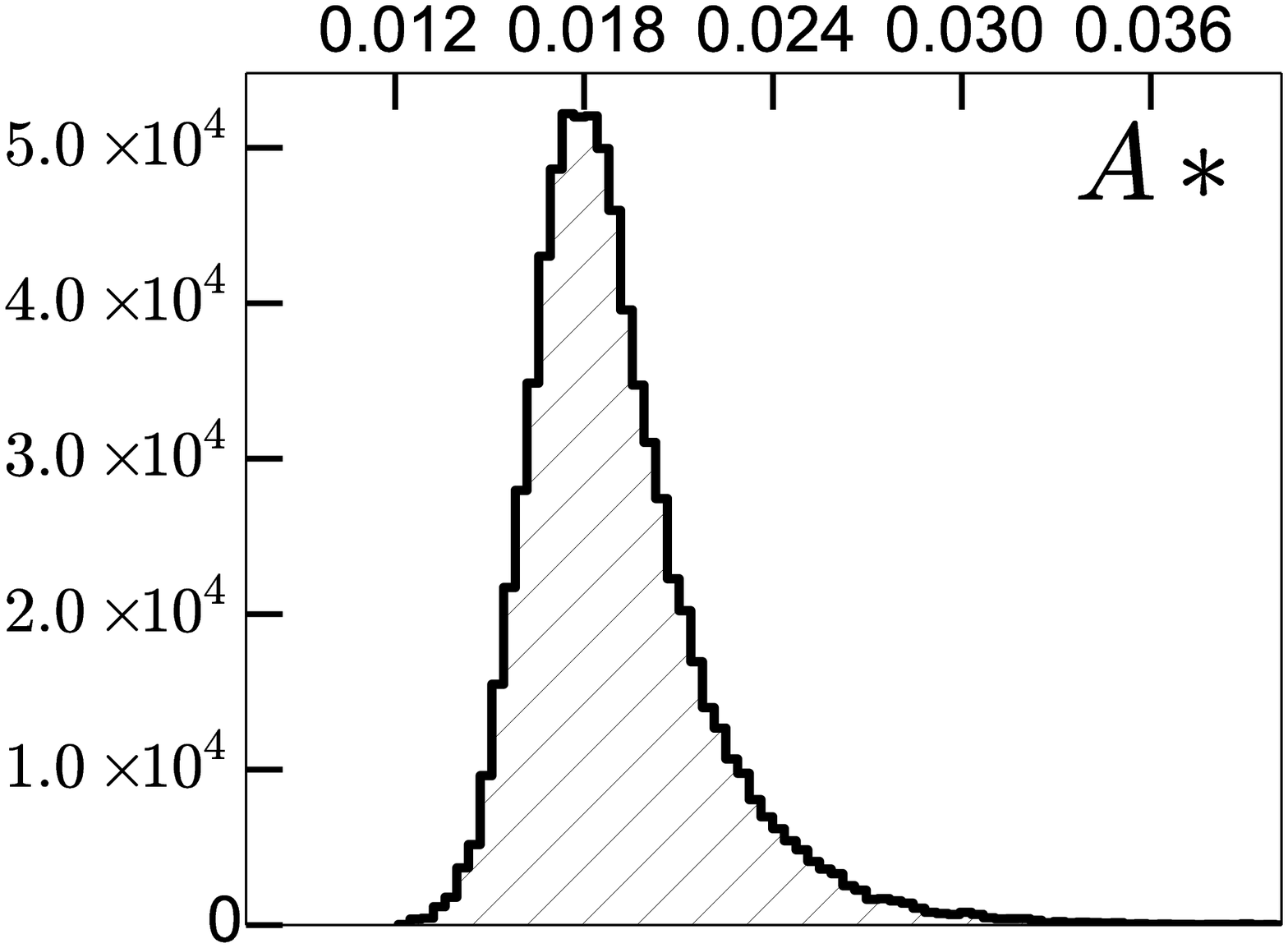} & 
 \includegraphics[width=0.30\columnwidth]{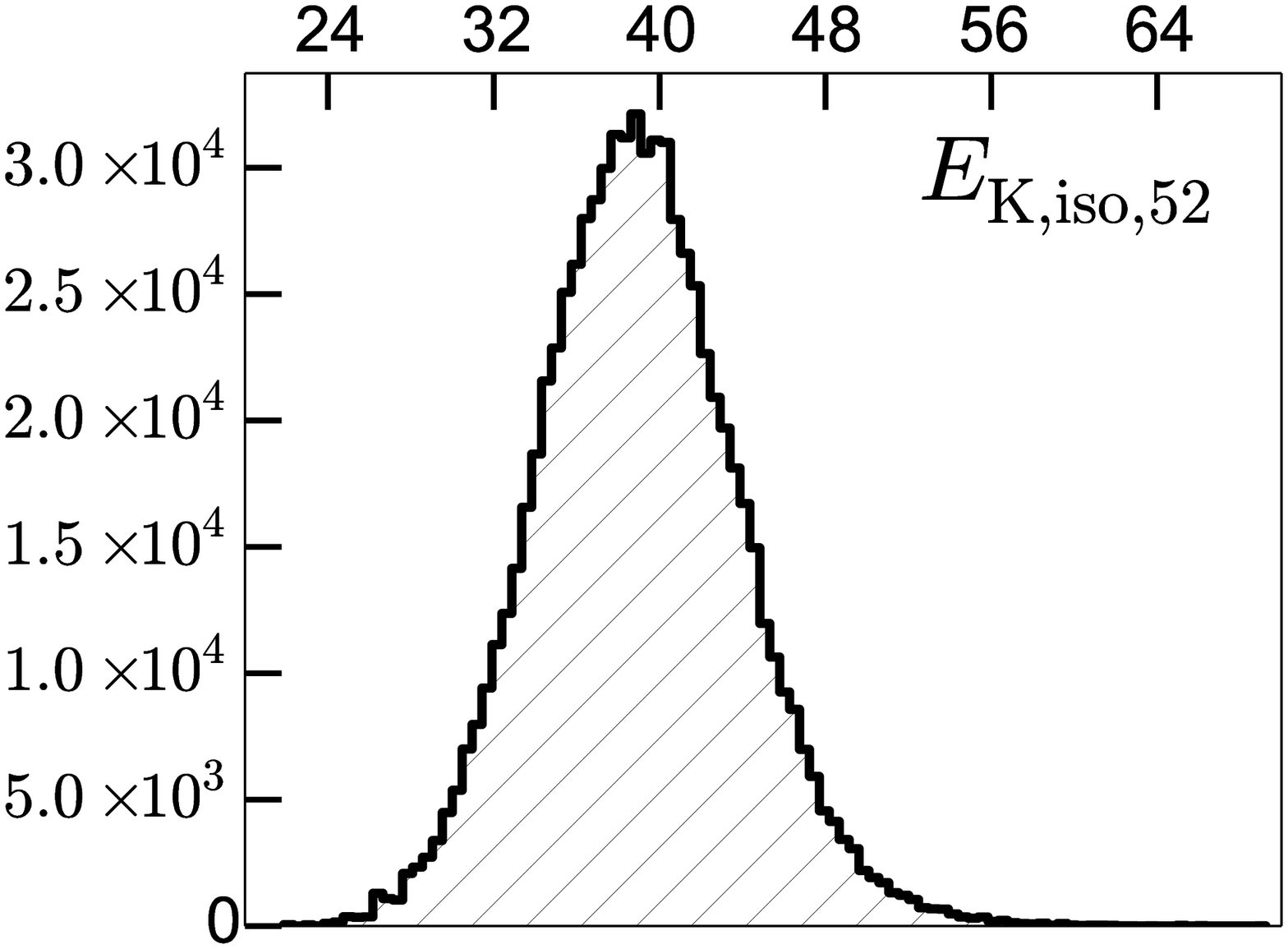} &
 \includegraphics[width=0.30\columnwidth]{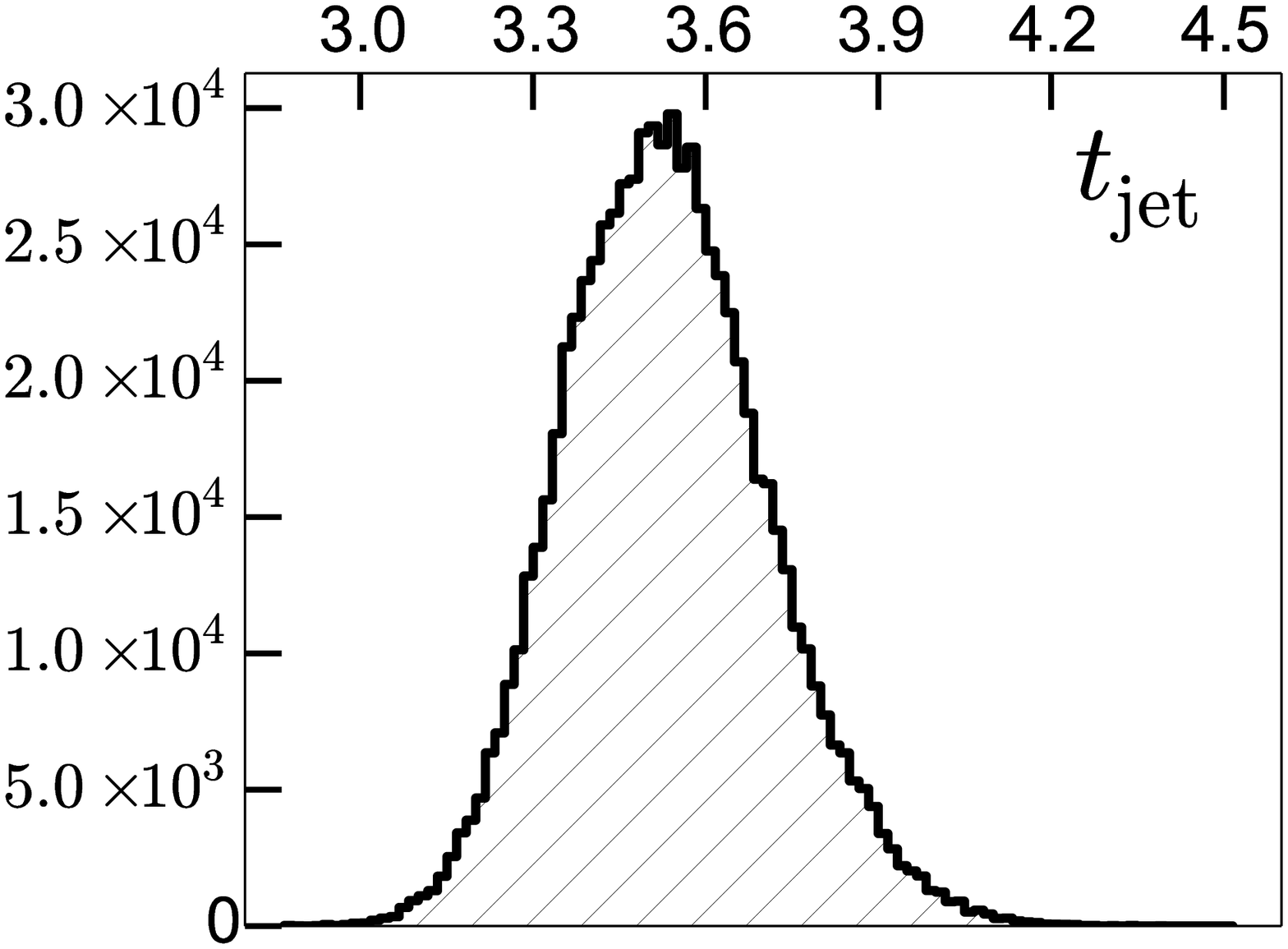} \\
 \includegraphics[width=0.30\columnwidth]{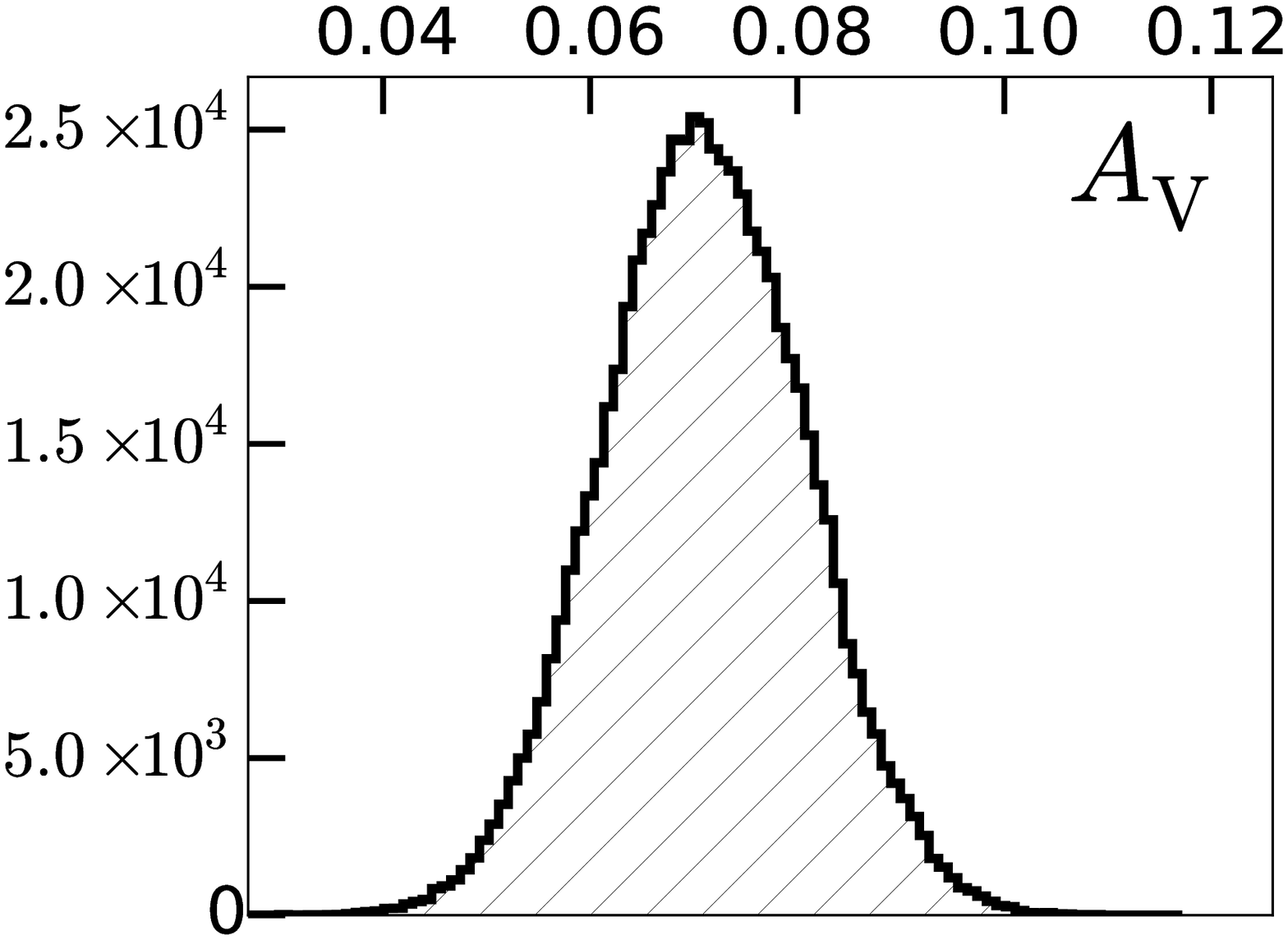} &
 \includegraphics[width=0.30\columnwidth]{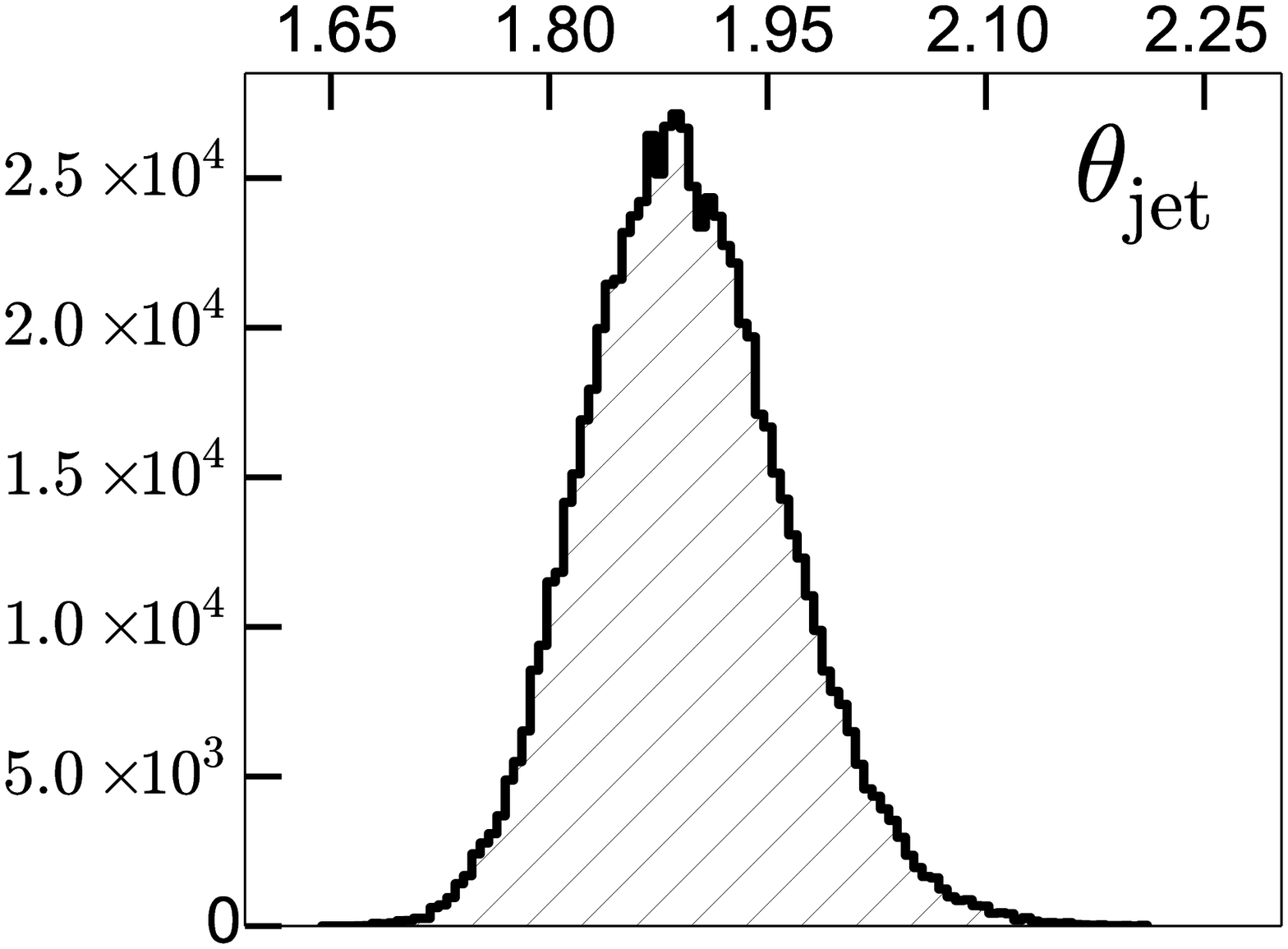} &
 \includegraphics[width=0.30\columnwidth]{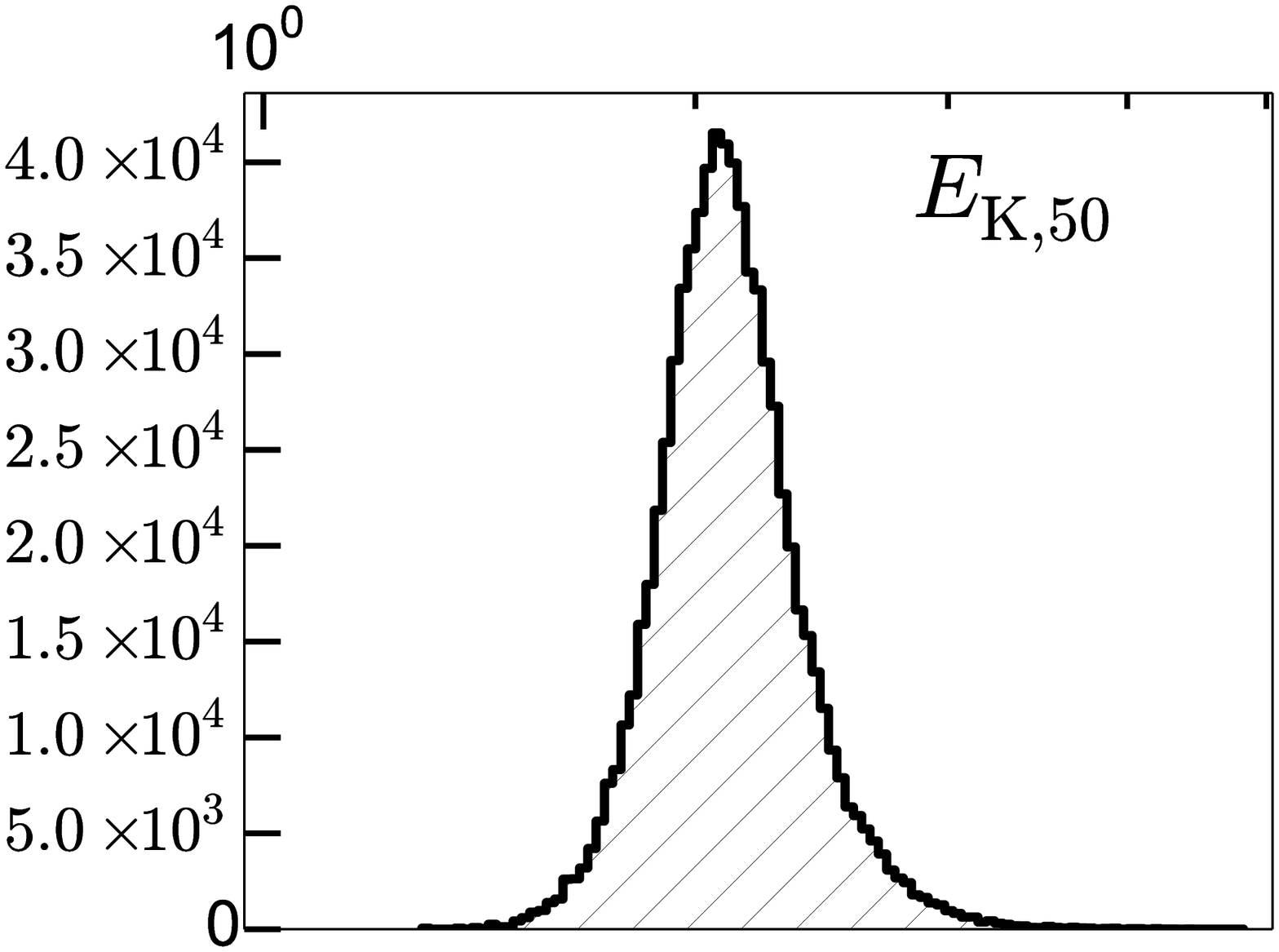} \\
\end{tabular}
\caption{Posterior probability density functions for the physical parameters for GRB~100901A in 
a wind environment from MCMC simulations. We have restricted $E_{\rm K, iso, 52} 
< 500$, $\epsilon_{\rm e} < \nicefrac{1}{3}$, and $\epsilon_{\rm B} < \nicefrac{1}{3}$.
\label{fig:100901A_wind_hists}}
\end{figure}

\begin{figure}
\begin{tabular}{ccc}
 \centering
 \includegraphics[width=0.30\columnwidth]{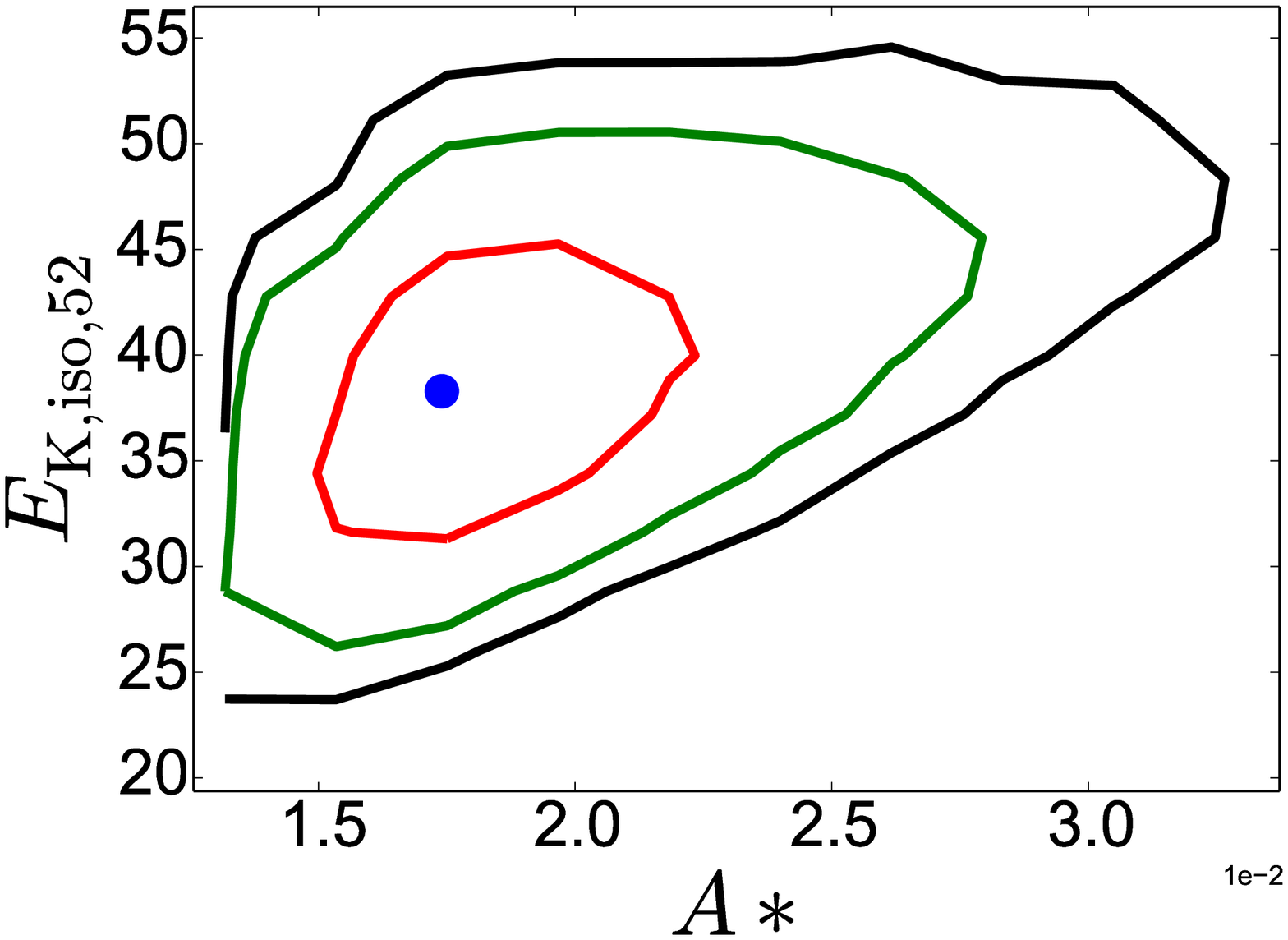} &
 \includegraphics[width=0.30\columnwidth]{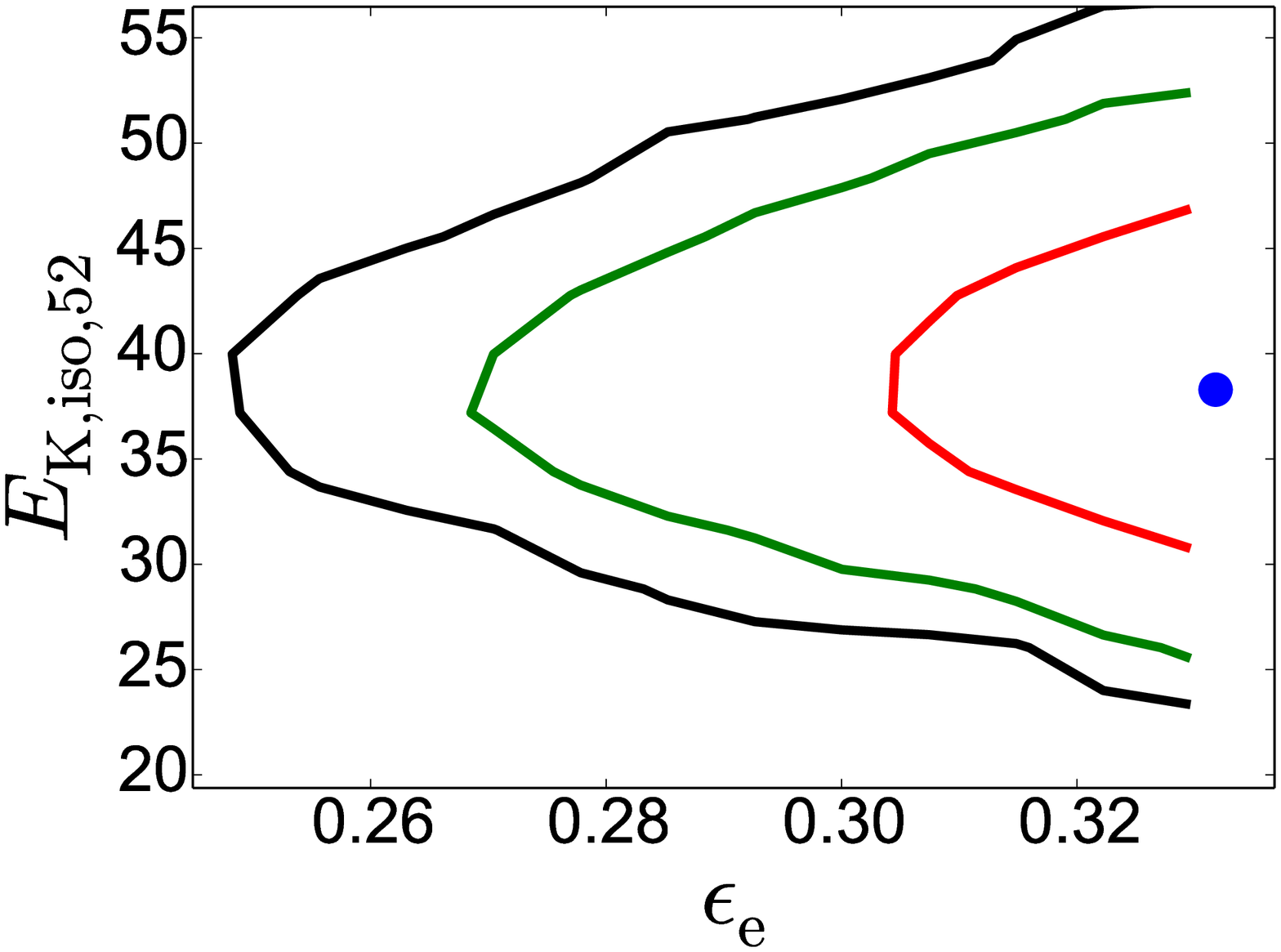} &
 \includegraphics[width=0.30\columnwidth]{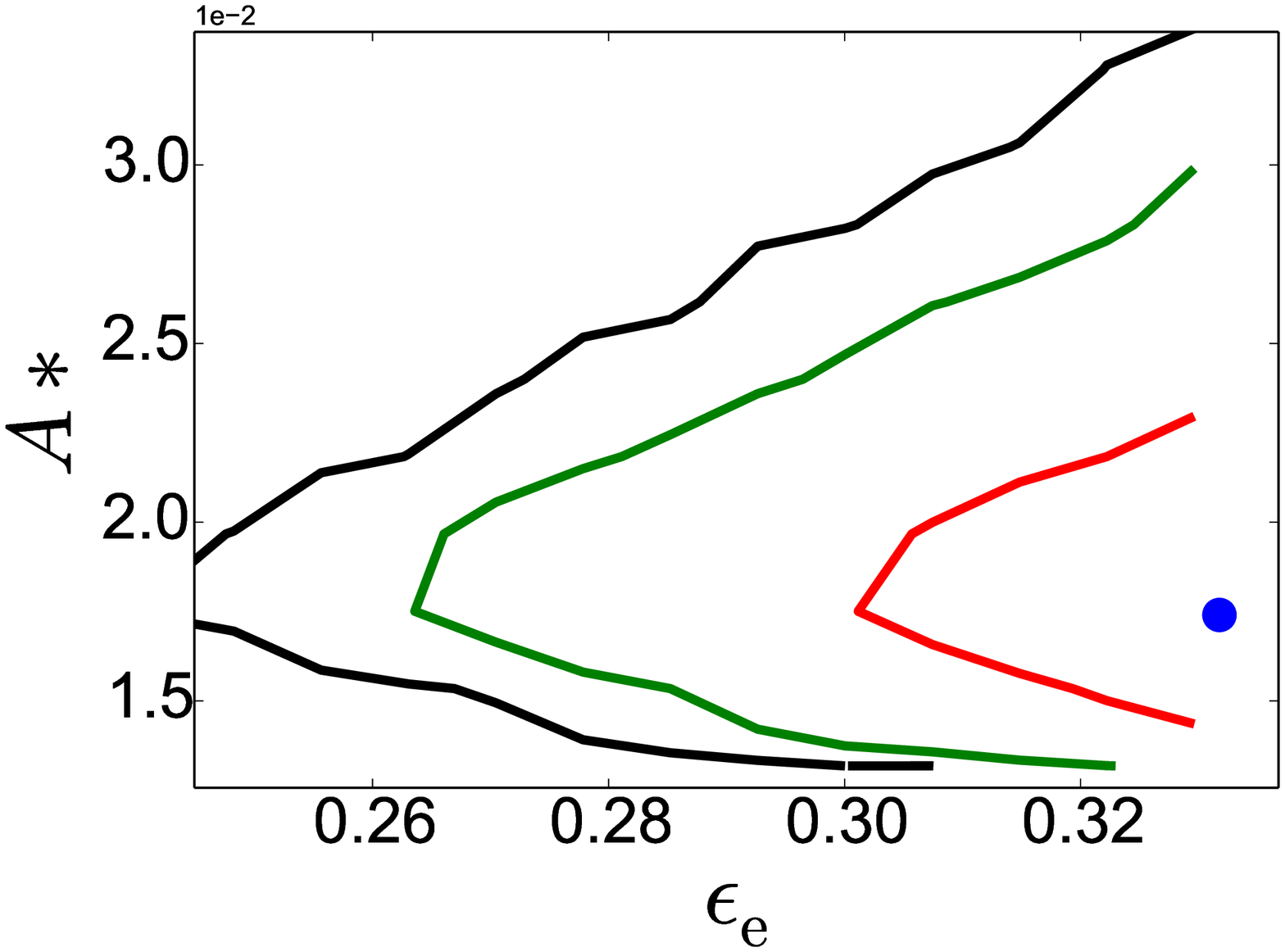} \\
 \includegraphics[width=0.30\columnwidth]{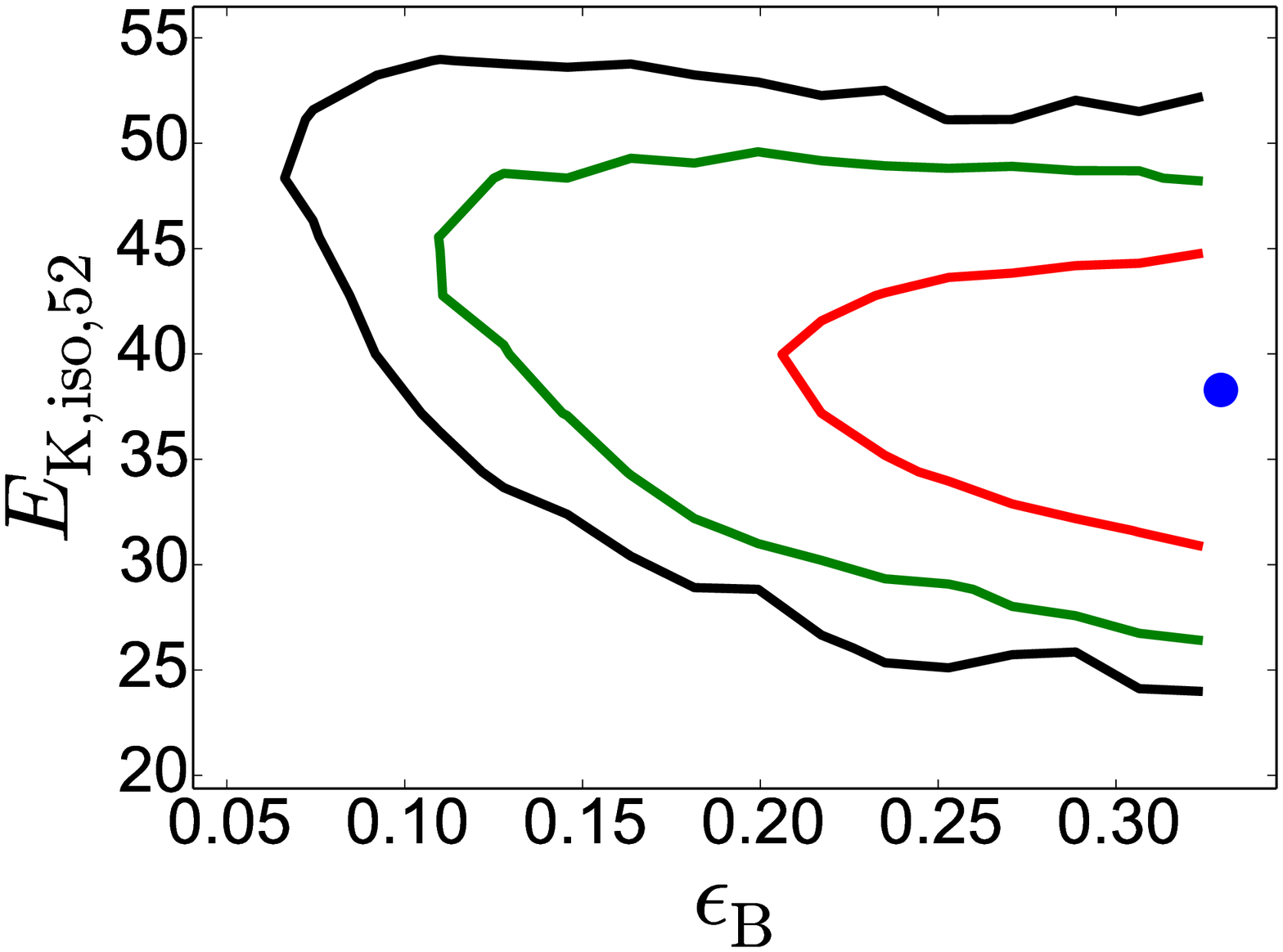} &
 \includegraphics[width=0.30\columnwidth]{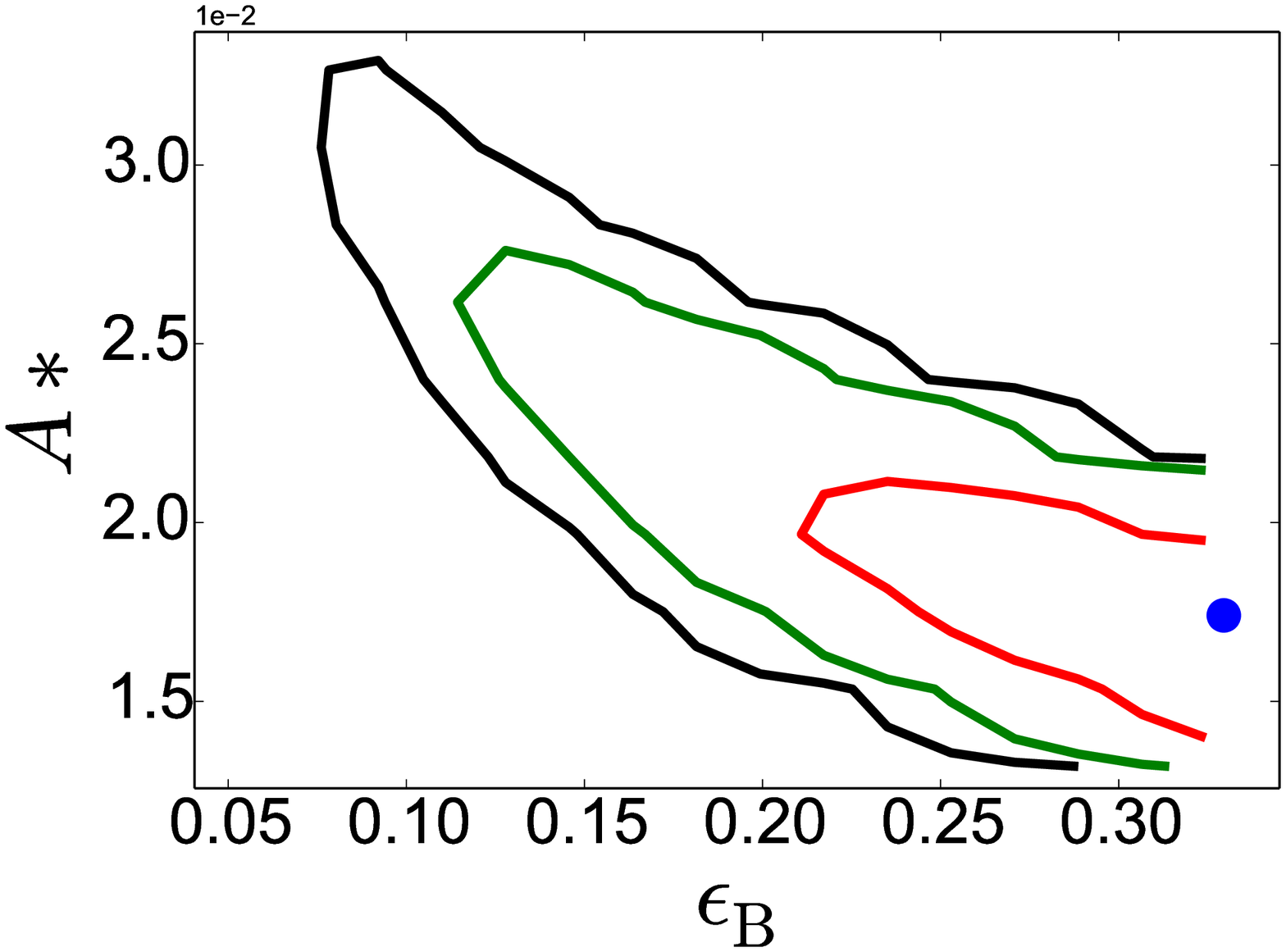} &
 \includegraphics[width=0.30\columnwidth]{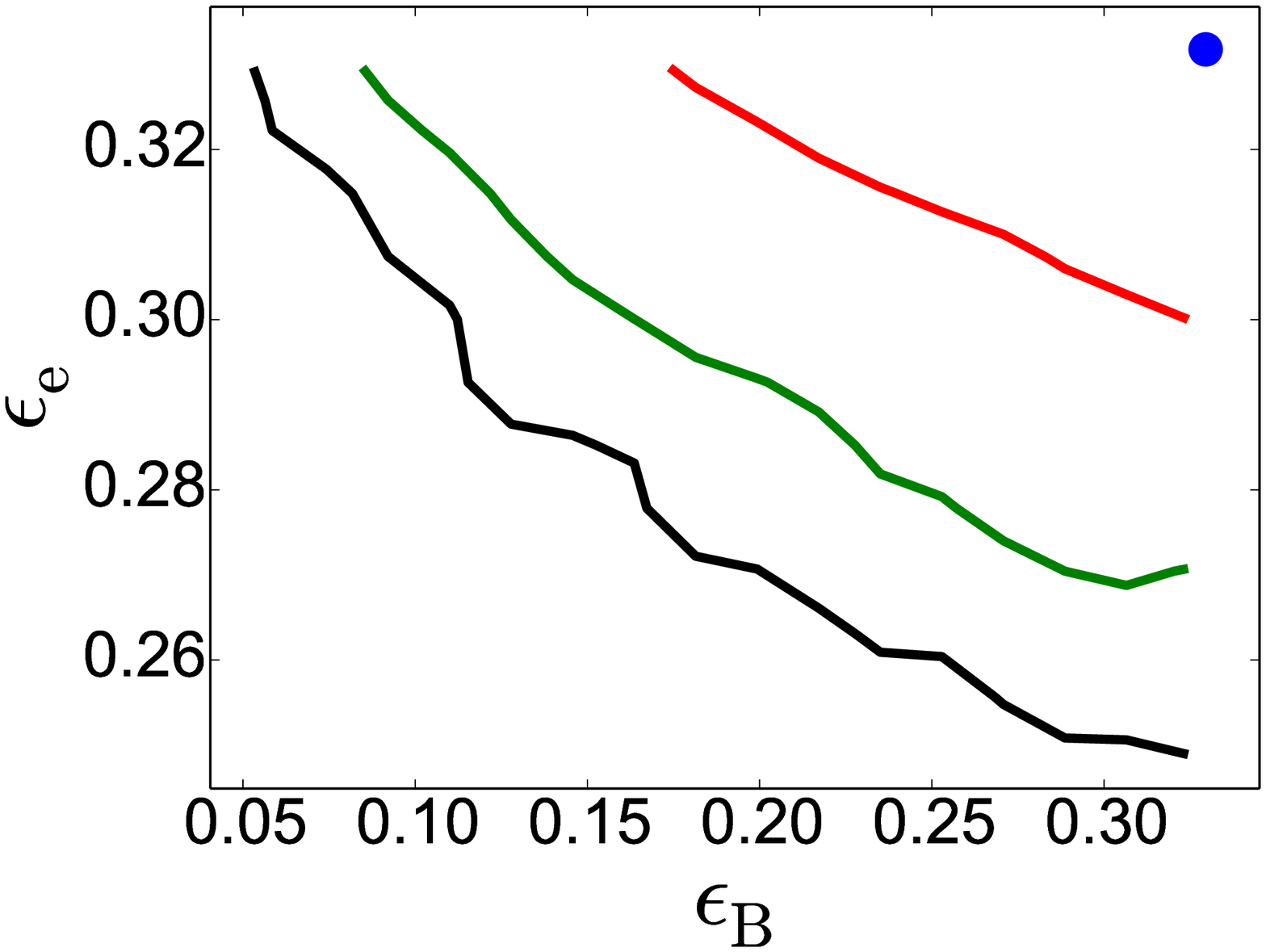} \\
\end{tabular}
\caption{1$\sigma$ (red), 2$\sigma$ (green), and 3$\sigma$ (black) contours for correlations
between the physical parameters, \EKiso, \dens, \epse, and \epsb\ for GRB~100901A, in the wind 
model from Monte Carlo simulations. We have restricted $E_{\rm K, iso, 52} < 500$, $\epsilon_{\rm 
e} < \nicefrac{1}{3}$, and $\epsilon_{\rm B} < \nicefrac{1}{3}$. The highest-likelihood model 
is marked with a blue dot. See the on line version of this Figure for additional plots of 
correlations between these parameters and $p$, $\AV$, $\tjet$, $\thetajet$, and $\EK$.
\label{fig:100901A_wind_corrplots}}
\end{figure}

The summary statistics from our MCMC simulations are $p=2.027^{+0.005}_{-0.003}$, 
$\EKiso = (3.9\pm0.5)\times10^{53}$\,erg, $\Astar=1.9^{+0.3}_{-0.2}\times10^{-2}$, 
$\epse=0.32^{+0.01}_{-0.02}$, $\epsb=0.27^{+0.04}_{-0.07}$, $\AV\lesssim0.1$\,mag, 
$\tjet = 3.5\pm0.2$\,d, $\thetajet = 1.9\pm0.1\degr$, $\EK=(2.1\pm0.2)\times10^{50}\,$erg, 
and $\Egamma = $.
We plot histograms of the posterior density functions for these 
parameters in Figure \ref{fig:100901A_wind_hists} and present contours of the joint posterior 
density for the physical parameters \Astar, \EKiso, \epse, and \epsb\ in Figure 
\ref{fig:100901A_wind_corrplots}.

\begin{figure*}
\begin{tabular}{cc}
 \centering
 \includegraphics[width=0.47\textwidth]{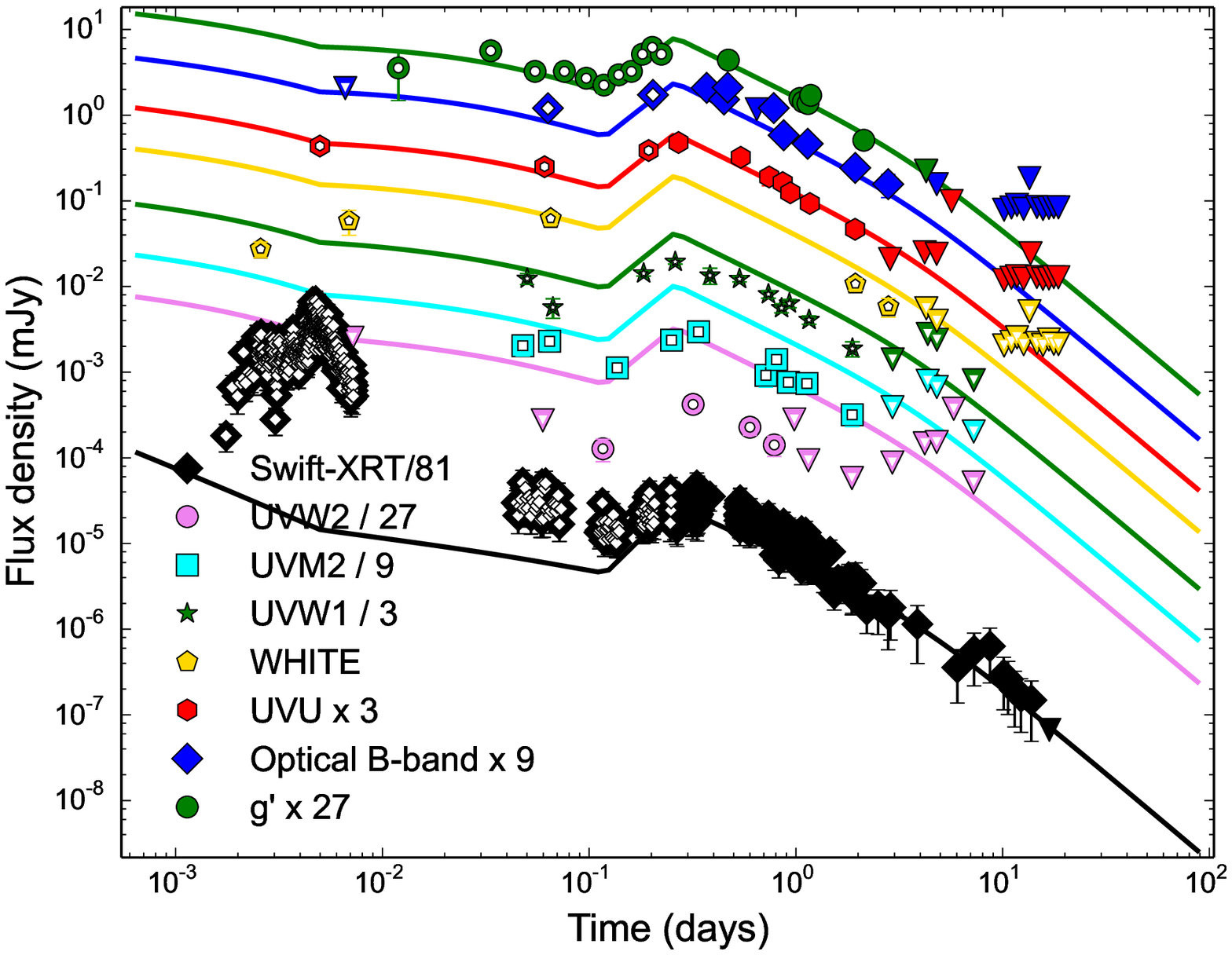} &
 \includegraphics[width=0.47\textwidth]{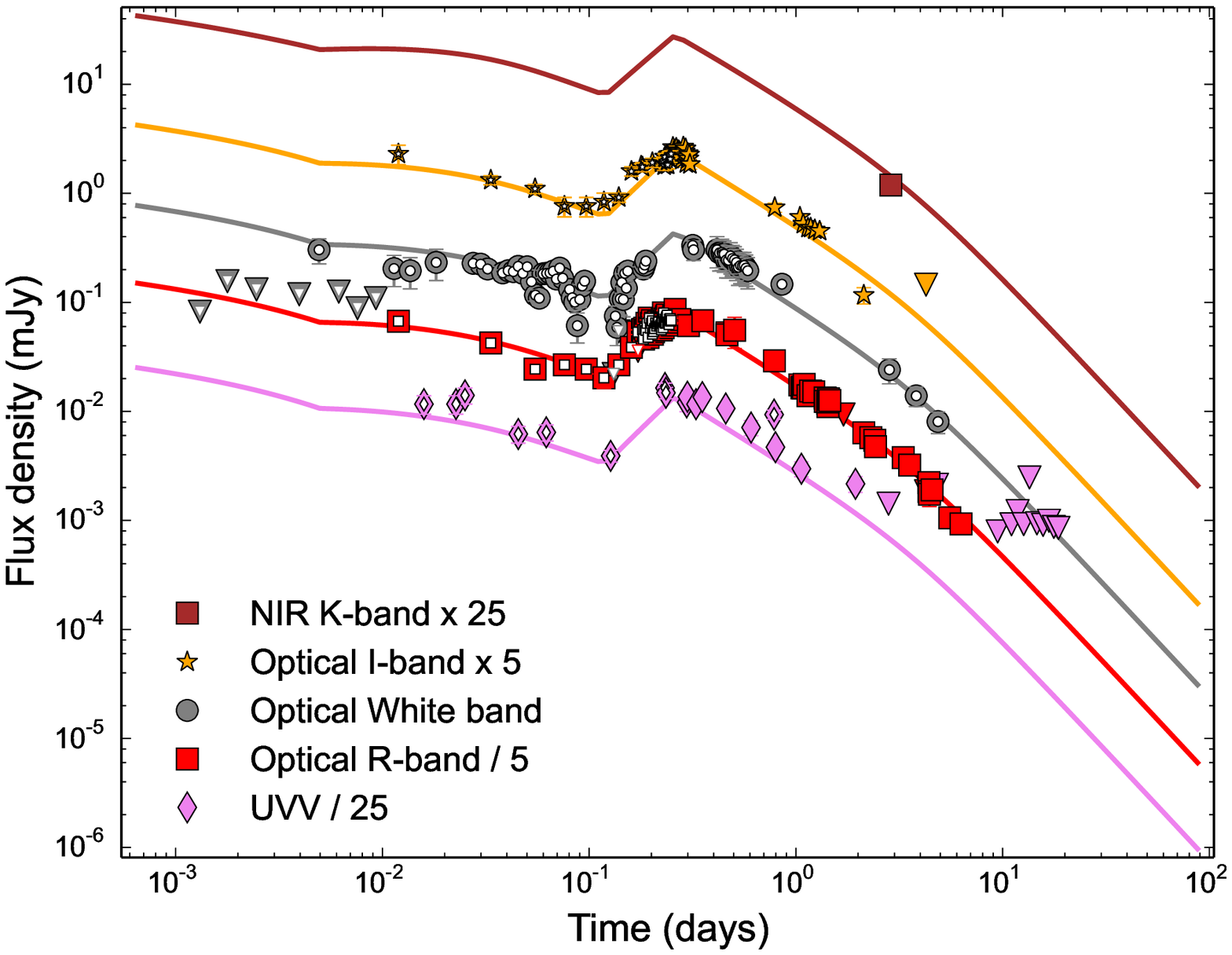} \\
 \includegraphics[width=0.47\textwidth]{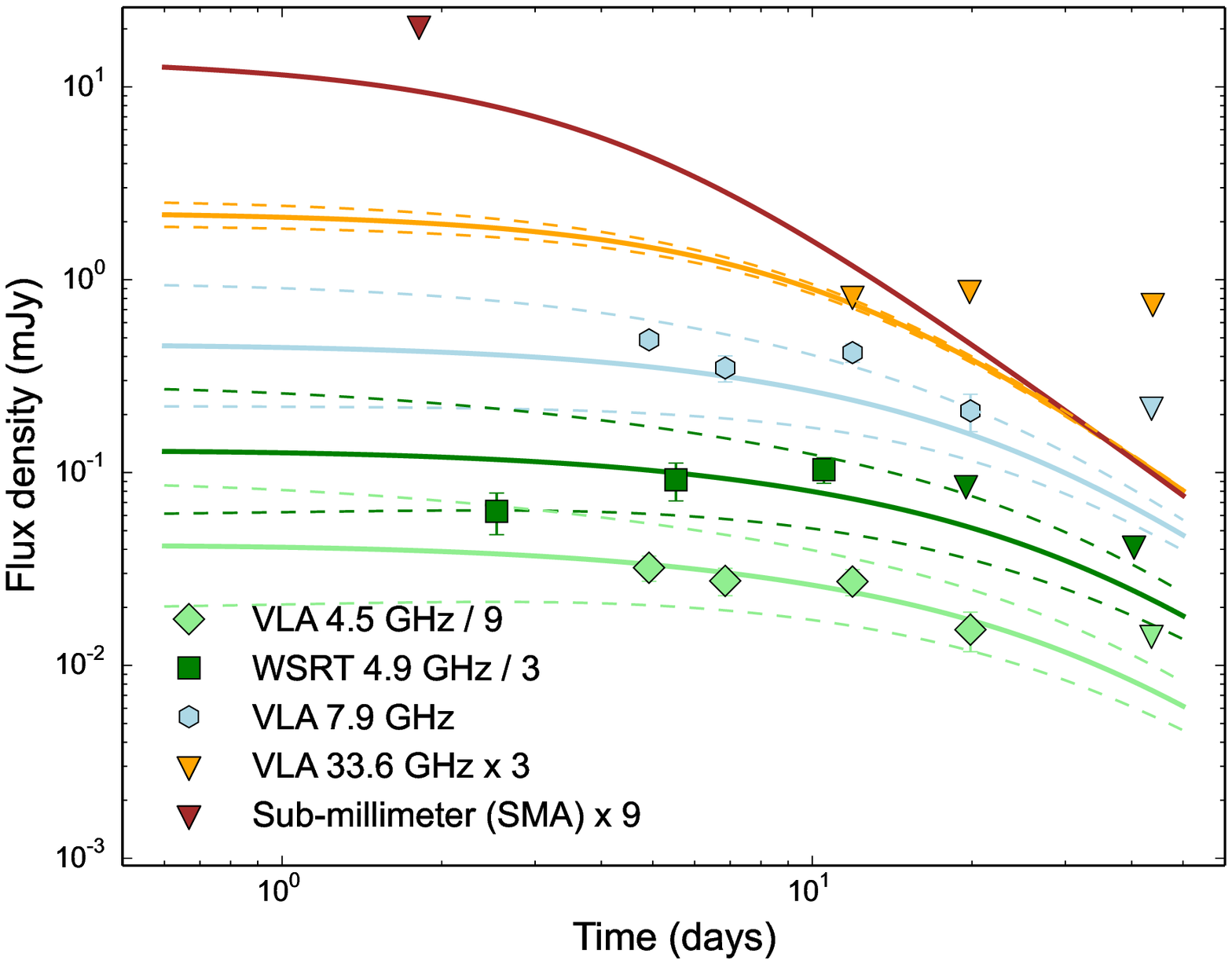} \\ 
\end{tabular}
\caption{X-ray and UV (top left), optical (top right), and radio (bottom left) light 
curves of GRB~100901A in the wind scenario, with the full afterglow model (solid lines), including 
energy injection before 0.26\,d.
\label{fig:100901A_wind_enj}}
\end{figure*}

Like for GRBs~100418A and 120326A, it is challenging to fit both the X-ray and optical light curves 
before the bump together in the energy injection scenario under the wind model. Like for the wind 
model for\me\ (Appendix \ref{appendix:120326A_wind}) and GRB~100418A (Appendix 
\ref{appendix:100418A_wind}), the optical light curves before the peak require a steeper injection 
rate than allowed by a distribution of Lorentz factors in the ejecta. In particular, our best energy 
injection model that matches the optical well but slightly under-predicts the X-ray data before 
0.15\,d (Figure \ref{fig:100901A_wind_enj}), requires $E\propto t^{0.7}$ between $5\times10^{-3}$\,d 
and $0.12$\,d, steepening to $E\propto t^{3.5}$ between $0.12$\,d and $0.26$\,d. In this model, the 
blastwave kinetic energy increases by a factor of 9 between $5\times10^{-3}$\,d and $0.12$\,d, and 
another factor of $15$ between $0.12$\,d and $0.26$\,d, for an overall increase by a factor of 
$\approx140$. Due to the discrepancy in the X-rays, the wind model may be considered a 
marginally viable model for GRB~100901A.

\section{A Wind Model for GRB~120404A}
\label{appendix:120404A_wind}
We apply the methods described in Section \ref{text:120404A:FS} to explore afterglow models with a 
wind-like circumburst environment for GRB~120404A. The parameters of our highest-likelihood model 
are $p\approx2.03$, $\epse \approx 0.33$, $\epsb \approx 0.30$, $A_* \approx 1.9$, $\EKiso \approx 
1.1\times10^{53}$\,erg, $t_{\rm jet} \approx 8.9\times10^{-2}$\,d, and $A_{\rm V} \approx 0.12$. 
This model also remains in spectrum 4 (Figure \ref{fig:120404A_seds}) for the duration of the 
observations, with the ordering $\nuc < \nua < \numax$. The spectral break frequencies are located 
at $\nuac\approx2.4\times10^{10}$\,Hz, $\nusa\approx7.9\times10^{11}$\,Hz, 
$\numax\approx4.0\times10^{13}$\,Hz, and $\nuc\approx1.5\times10^{10}$\,Hz, at 0.1\,d with $F_{\rm 
max} = F_{\nu,{\rm sa}}\approx27$\,mJy. The Compton $y$-parameter is 0.6.

\begin{figure*}
\begin{tabular}{cc}
 \centering
 \includegraphics[width=0.47\textwidth]{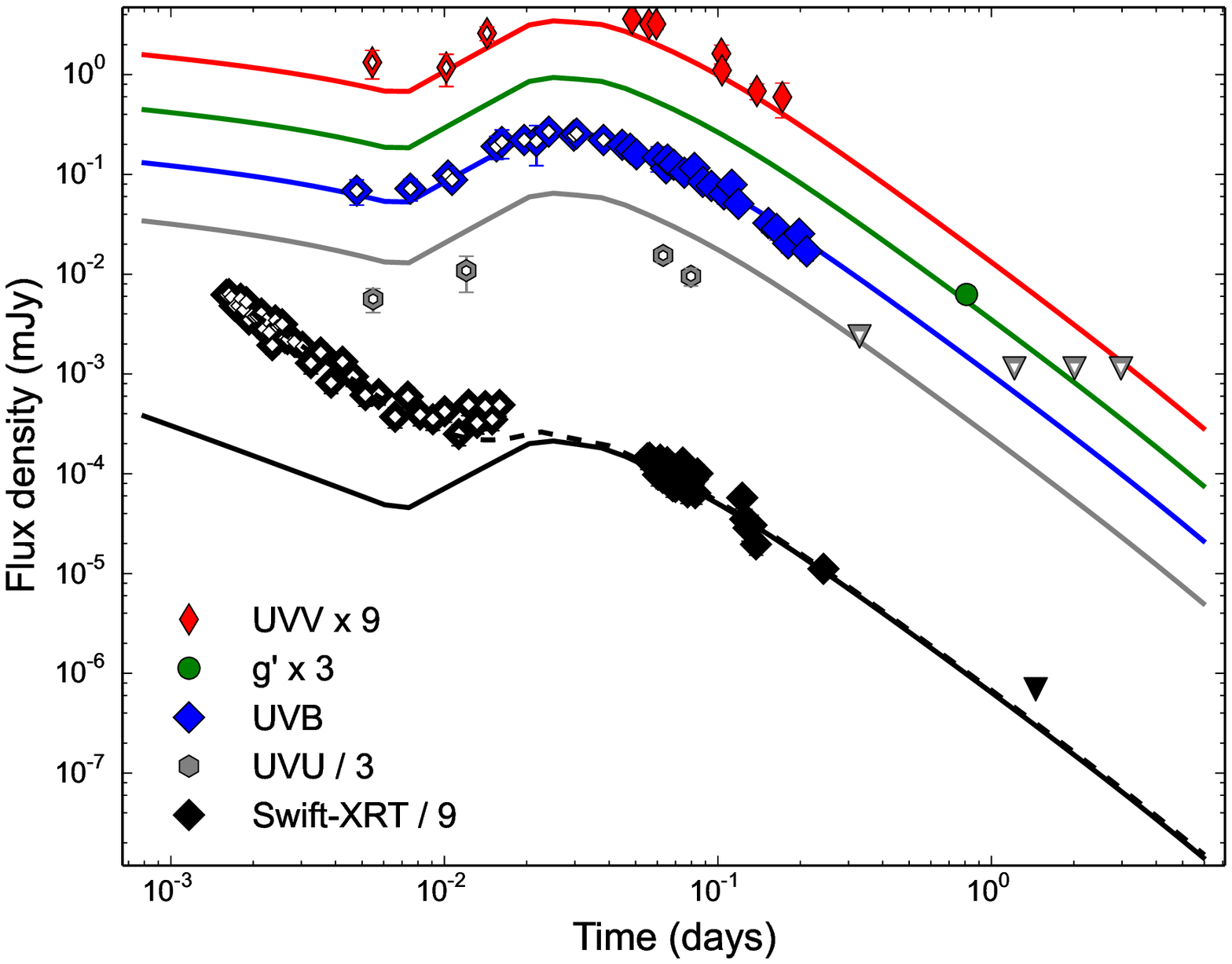} &
 \includegraphics[width=0.47\textwidth]{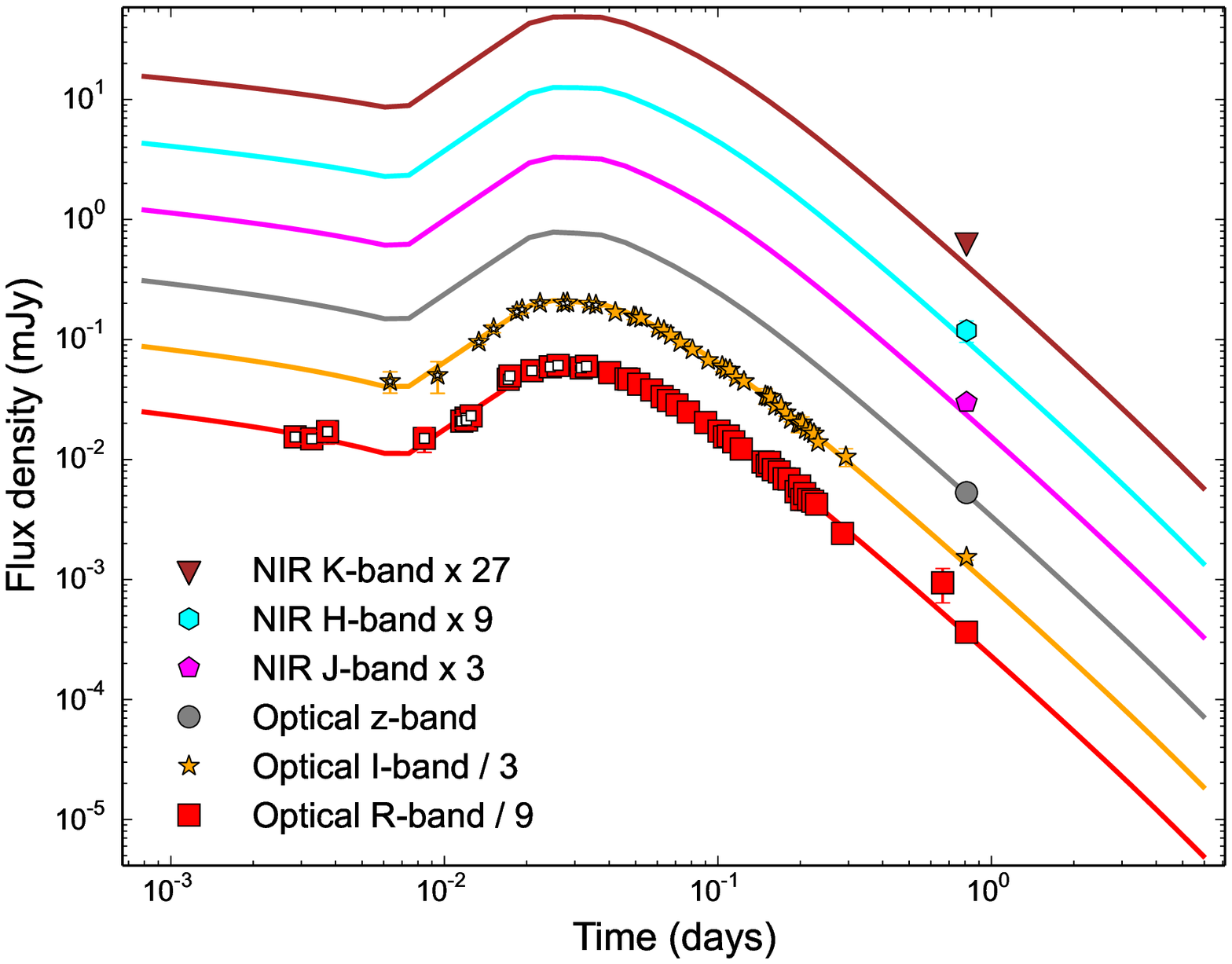} \\
 \includegraphics[width=0.47\textwidth]{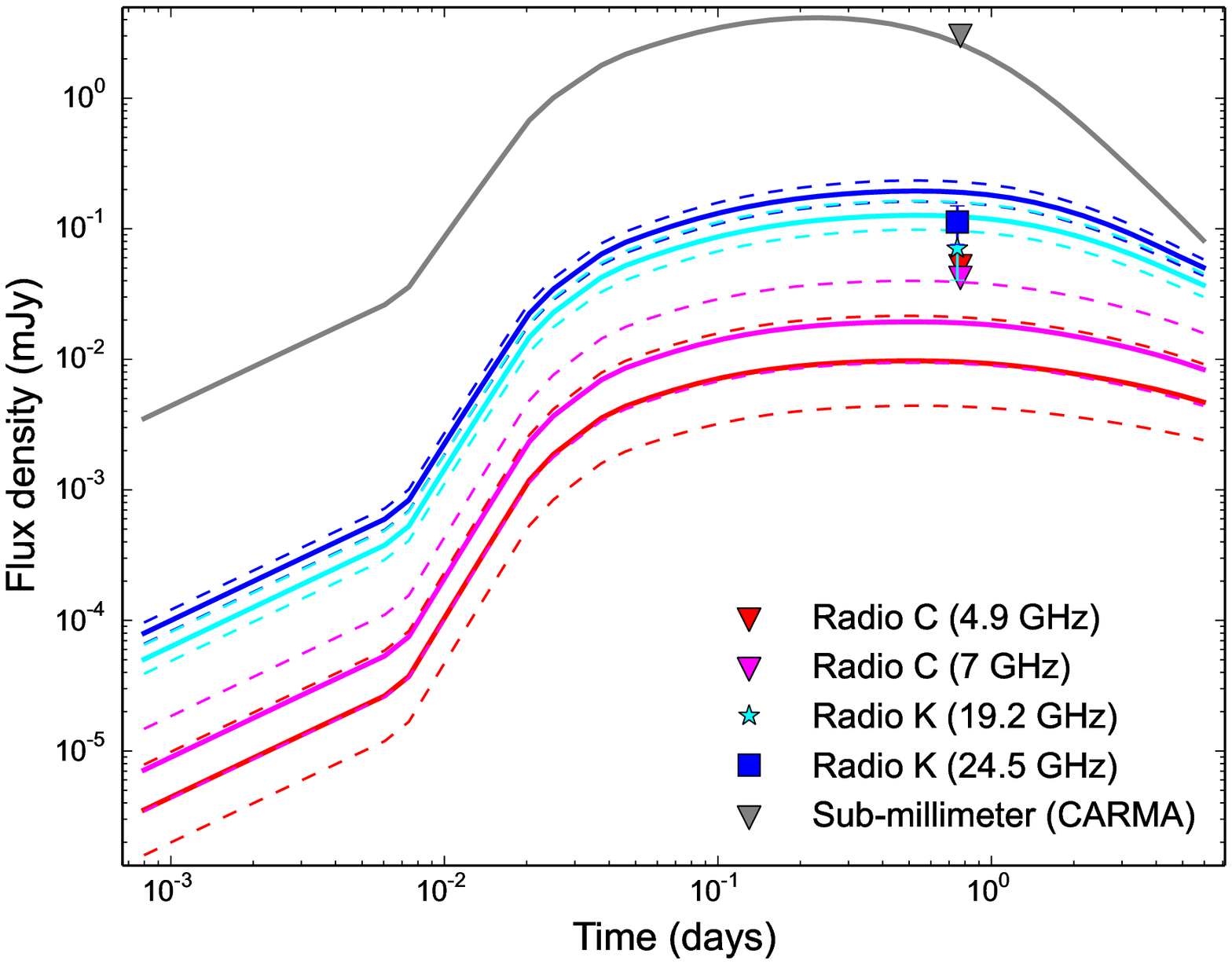} & \\ 
\end{tabular}
\caption{Similar to Figure \ref{fig:120404A_enj}, but for a wind-like circumburst environment. 
The light curves before 0.04\,d are based on the same energy injection model described in 
Section \ref{text:120404A:enj} as applied to the case of the wind medium. See Appendix 
\ref{appendix:120404A_wind} for a discussion.\label{fig:120404A_enj_wind}}
\end{figure*}

\begin{figure}
\begin{tabular}{ccc}
\centering
 \includegraphics[width=0.30\columnwidth]{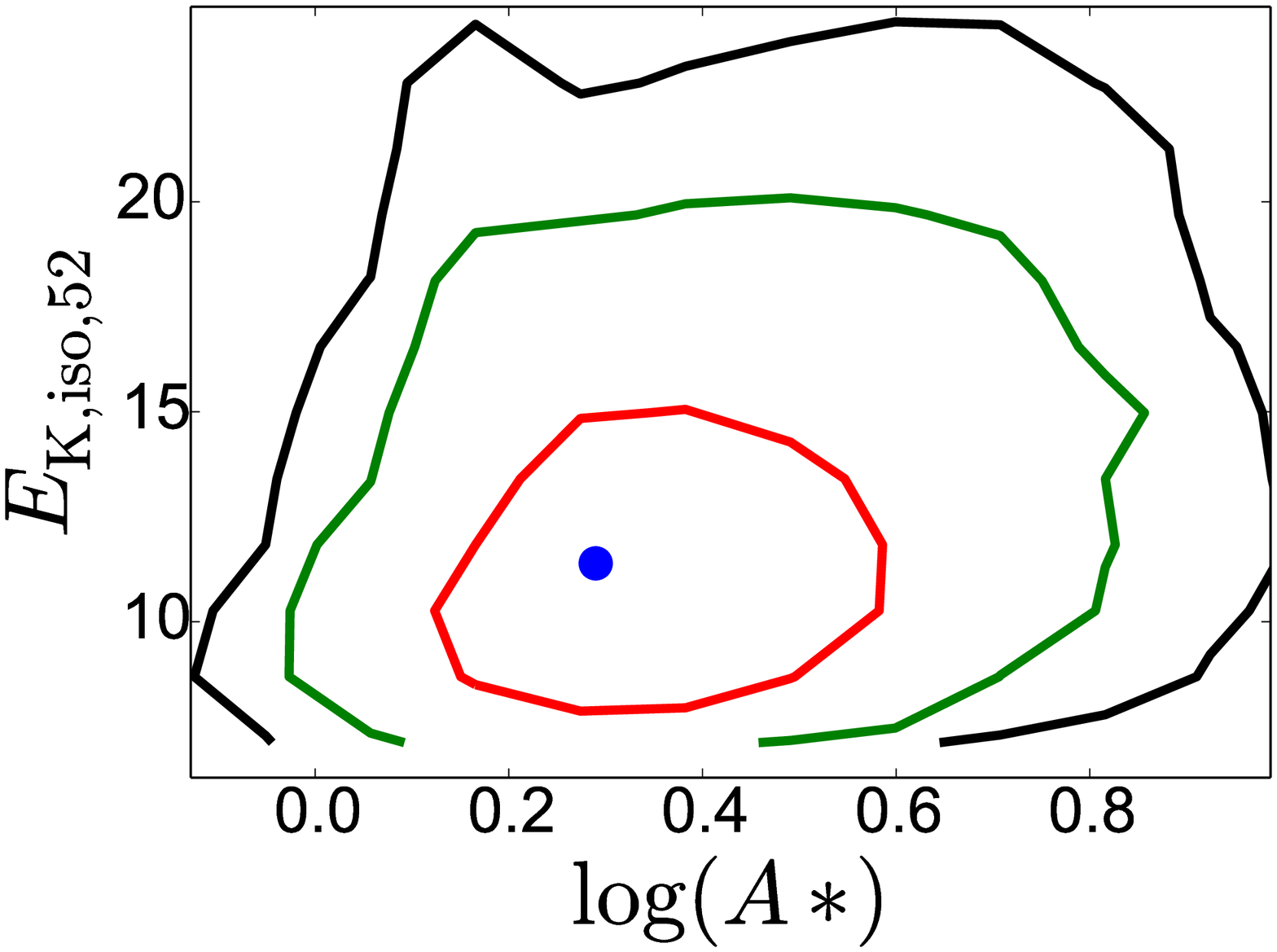} &
 \includegraphics[width=0.30\columnwidth]{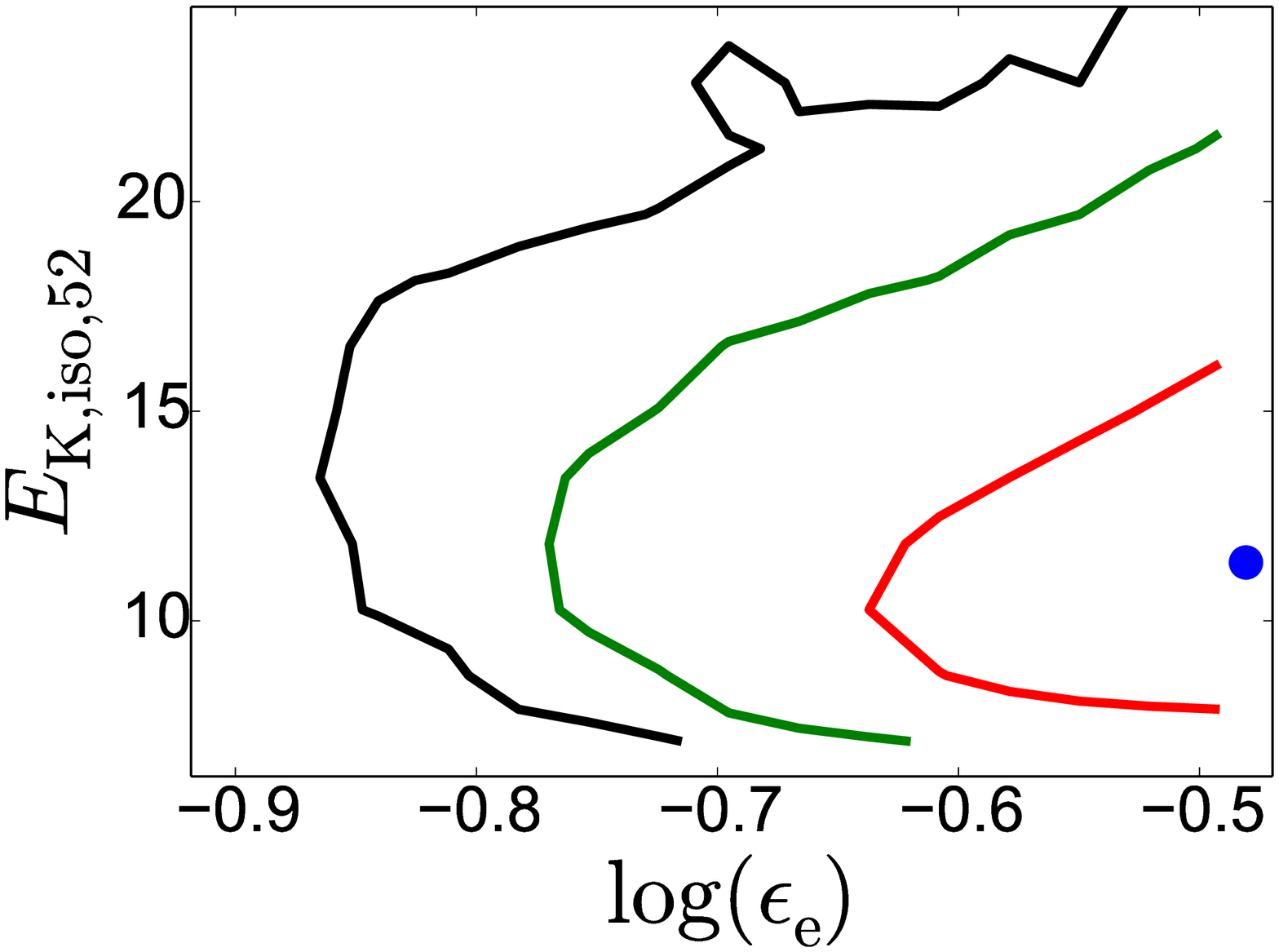} &
 \includegraphics[width=0.30\columnwidth]{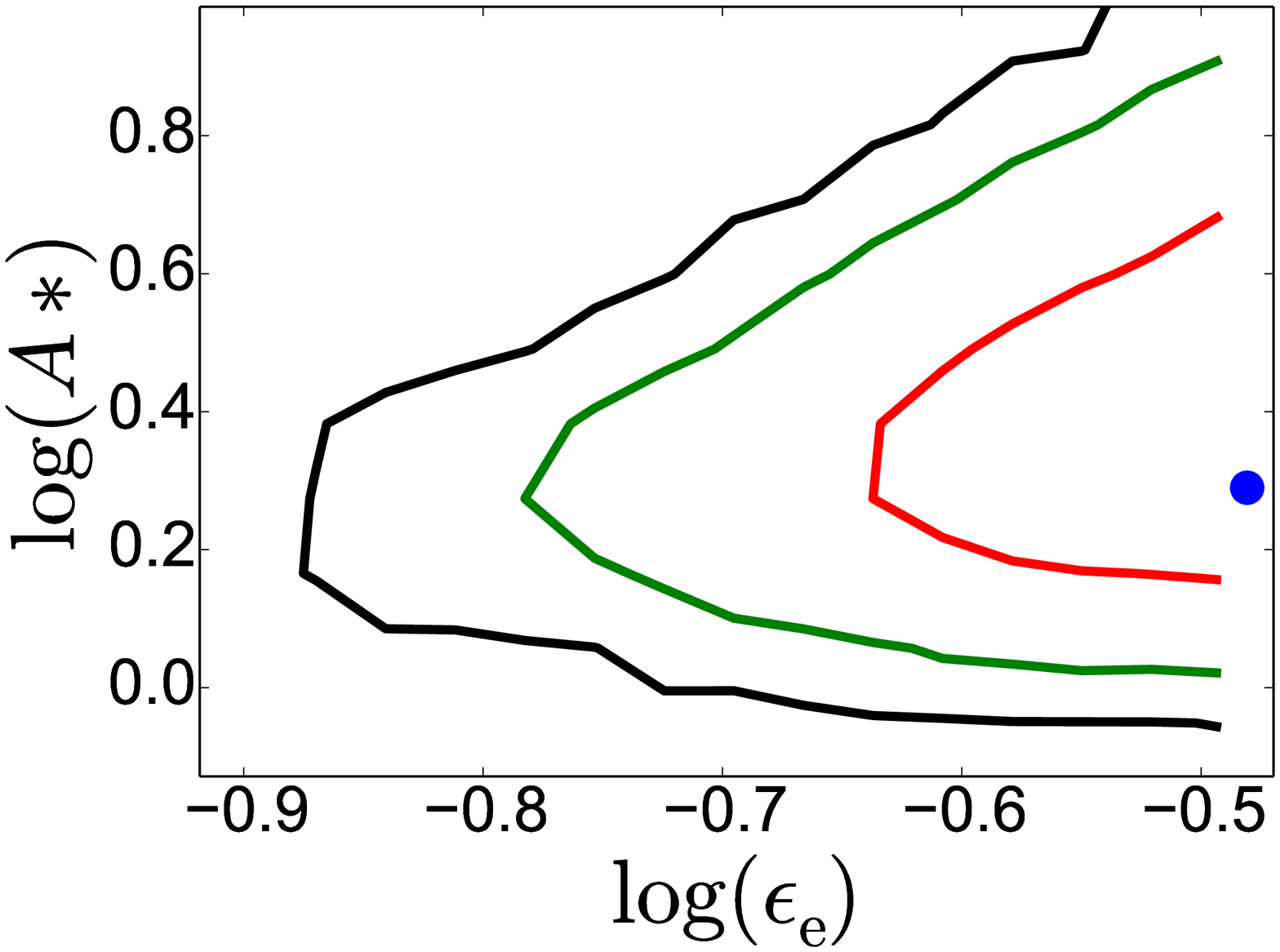} \\
 \includegraphics[width=0.30\columnwidth]{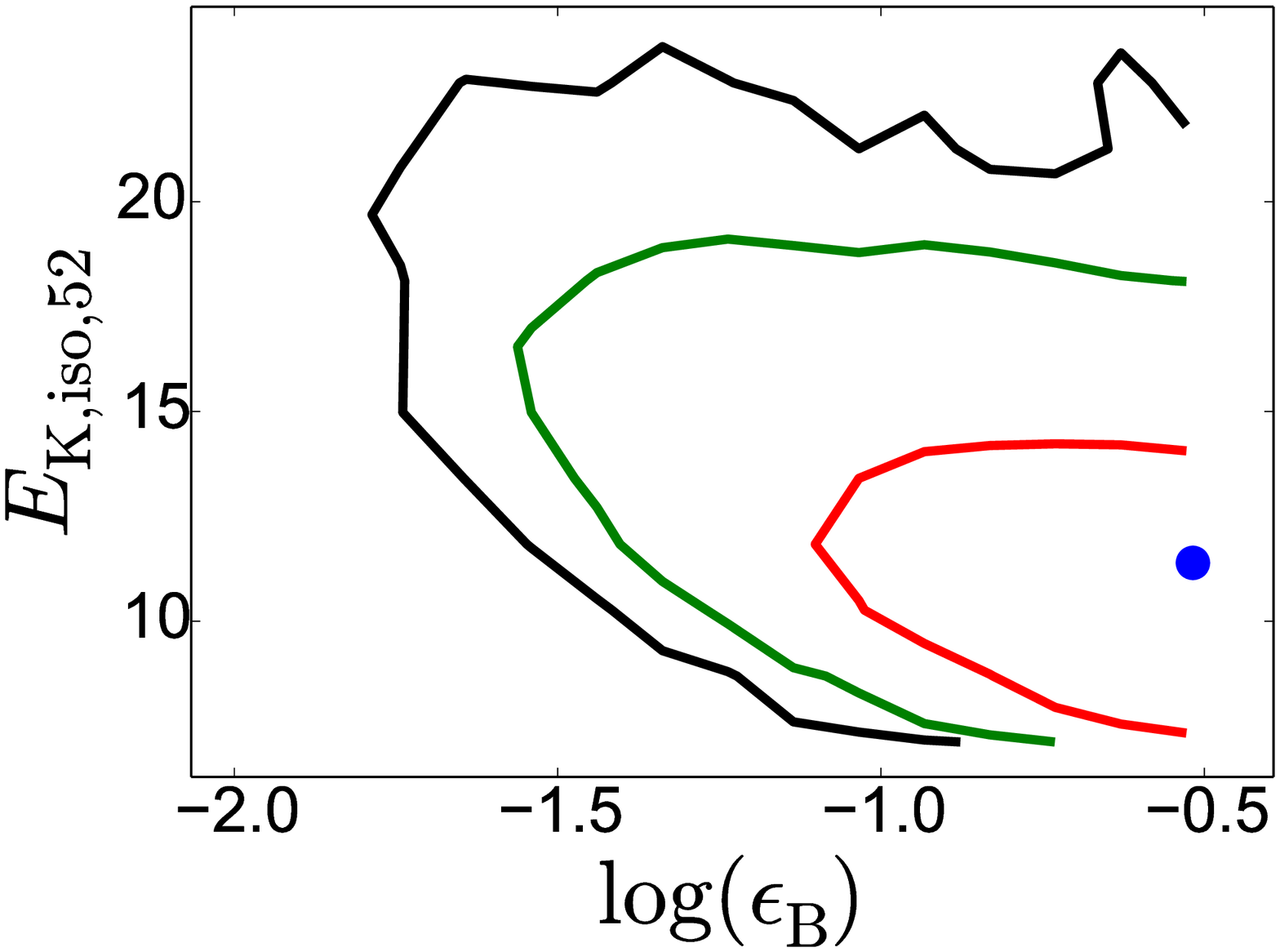} &
 \includegraphics[width=0.30\columnwidth]{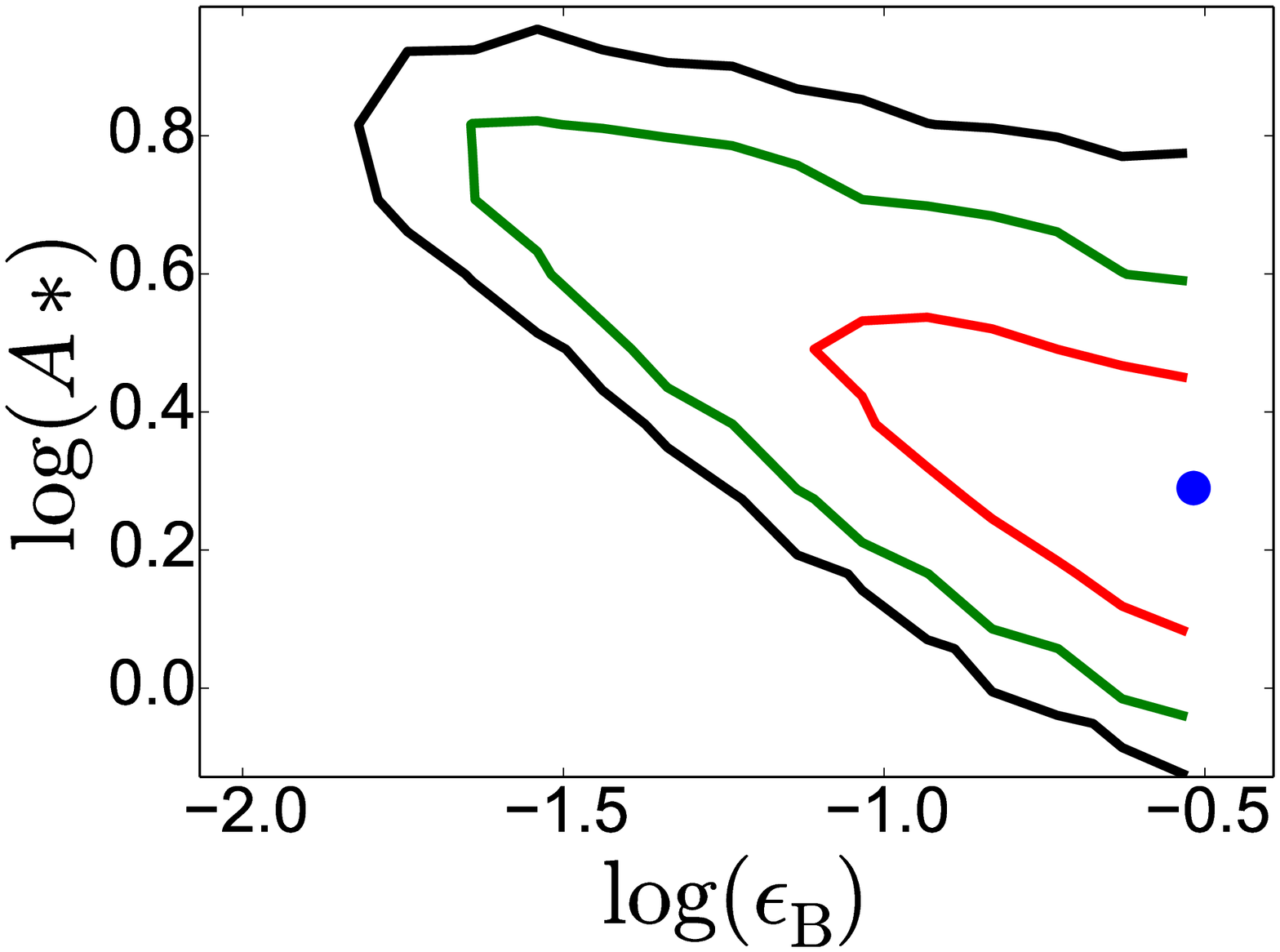} &
 \includegraphics[width=0.30\columnwidth]{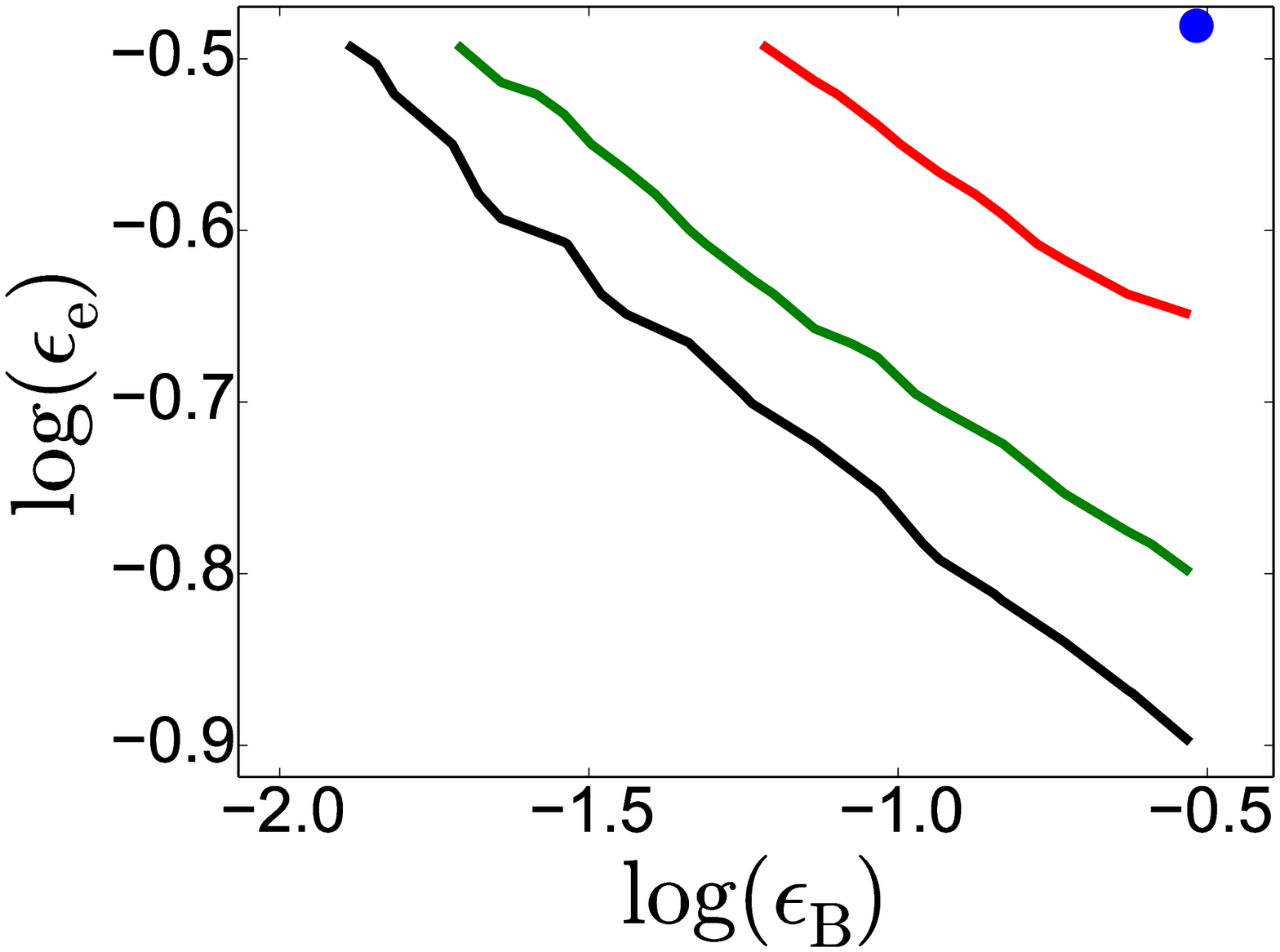} \\
\end{tabular}
\caption{1$\sigma$ (red), 2$\sigma$ (green), and 3$\sigma$ (black) contours for correlations
between the physical parameters, \EKiso, \dens, \epse, and \epsb\ for GRB~120404A, in the wind 
model from Monte Carlo simulations. We have restricted $\epsilon_{\rm e} < \nicefrac{1}{3}$ and 
$\epsilon_{\rm B} < \nicefrac{1}{3}$. The highest-likelihood model is marked with a blue dot. See 
the on line version of this Figure for additional plots of correlations between these parameters and 
$p$, $t_{\rm jet}$, $\thetajet$, $E_{\rm K}$, and $A_{\rm V}$. \label{fig:120404A_wind_corrplots}}
\end{figure}

\begin{figure}
\begin{tabular}{ccc}
 \centering
 \includegraphics[width=0.30\columnwidth]{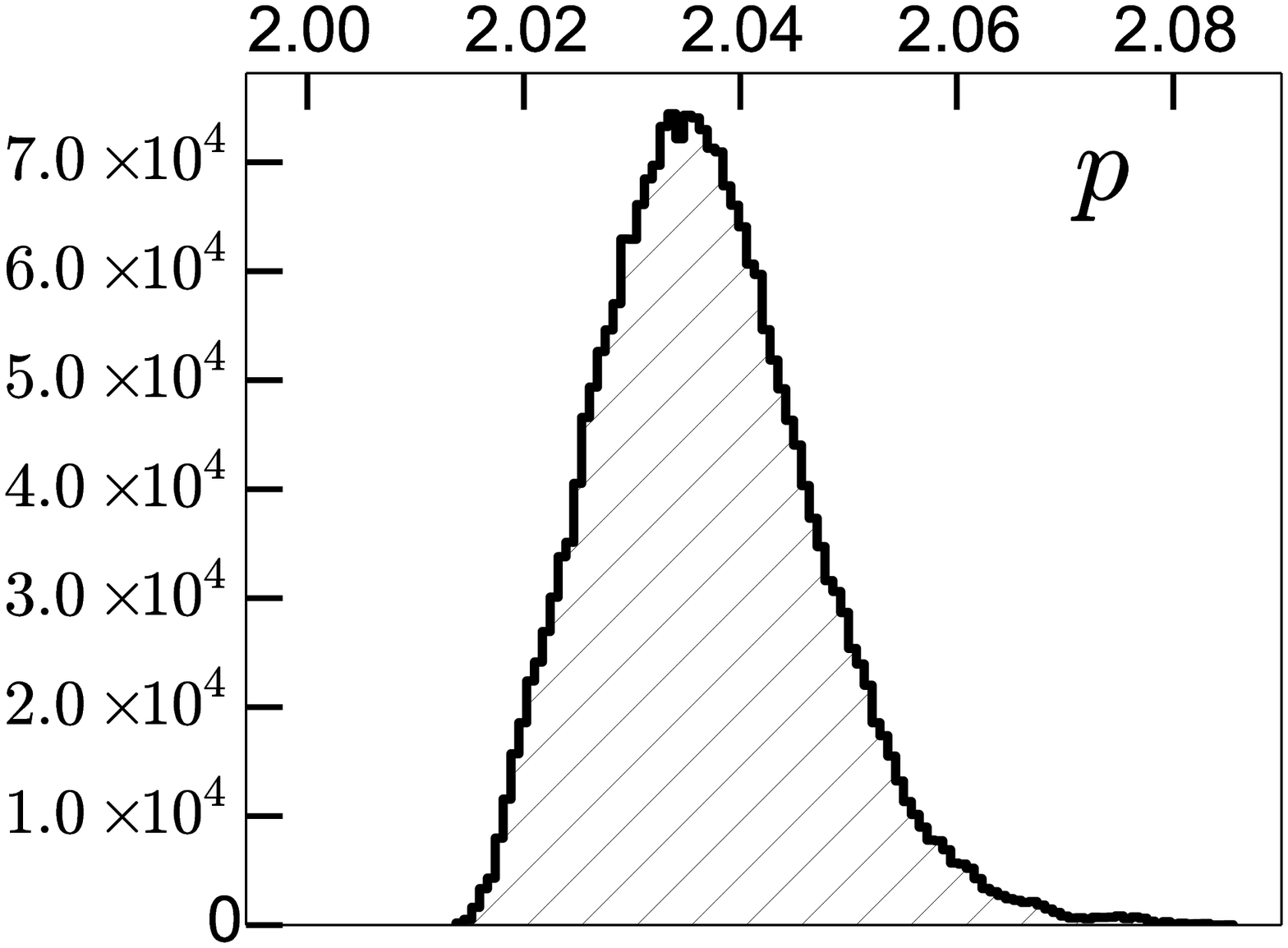} &
 \includegraphics[width=0.30\columnwidth]{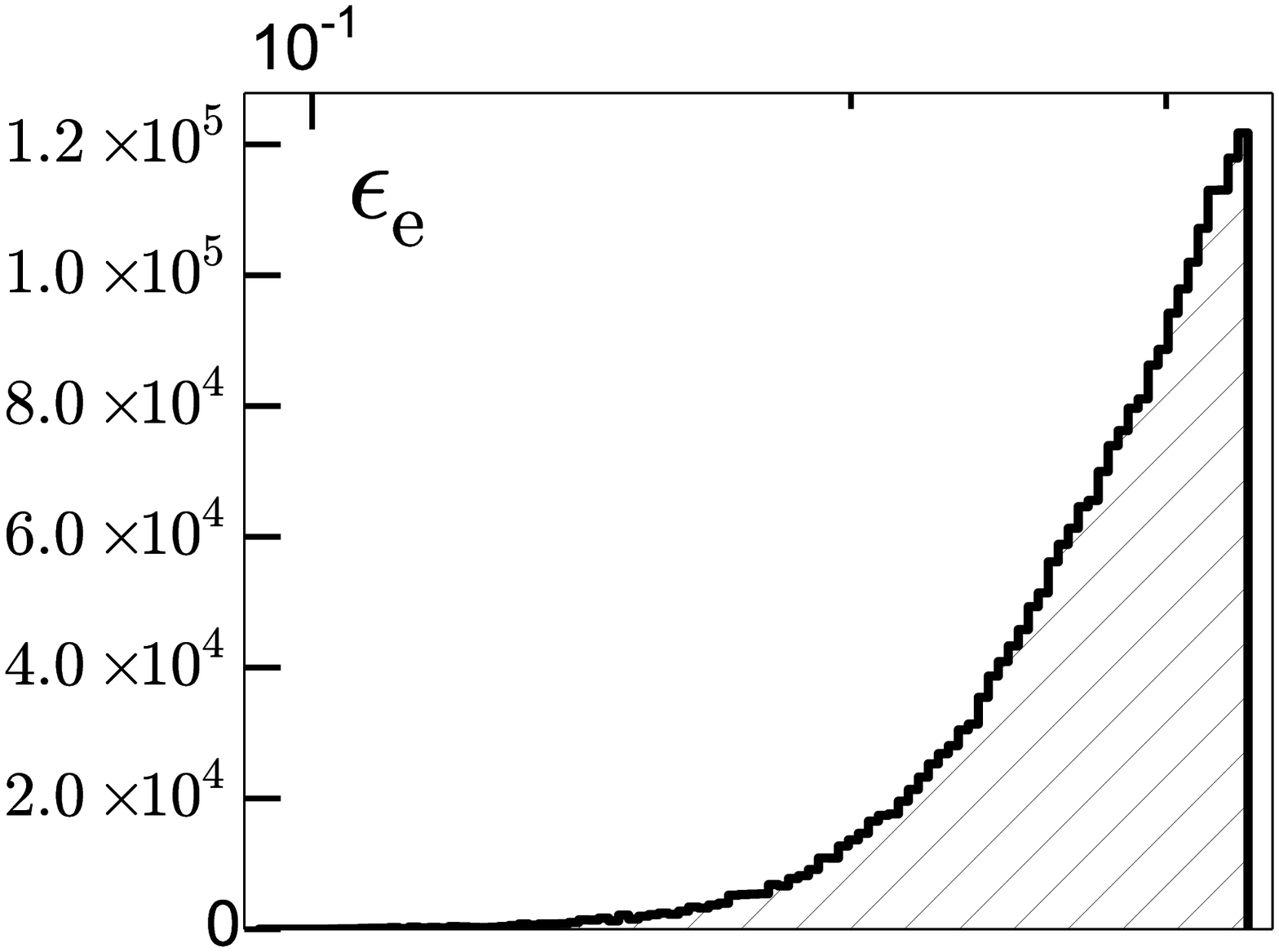} &
 \includegraphics[width=0.30\columnwidth]{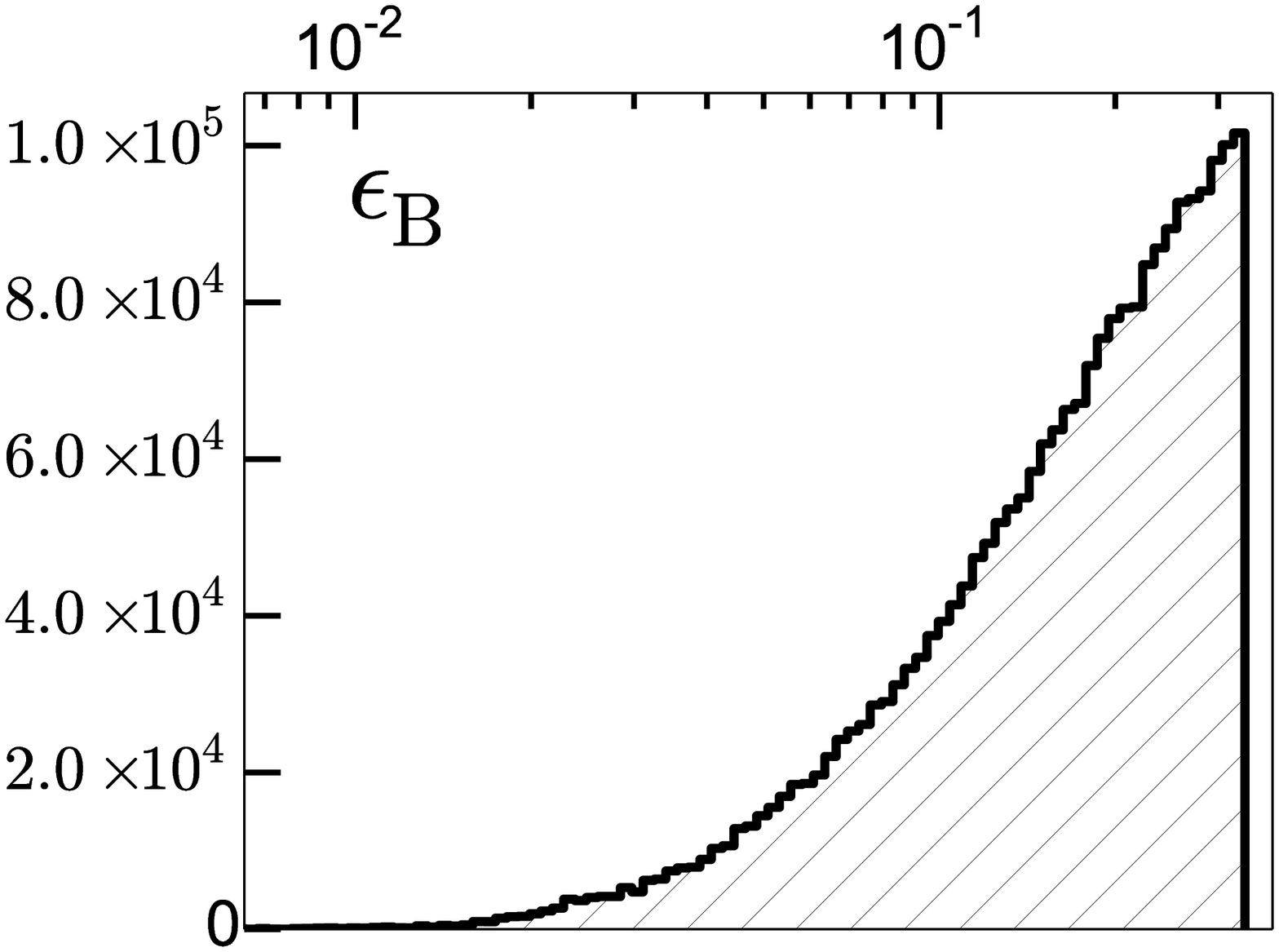} \\
 \includegraphics[width=0.30\columnwidth]{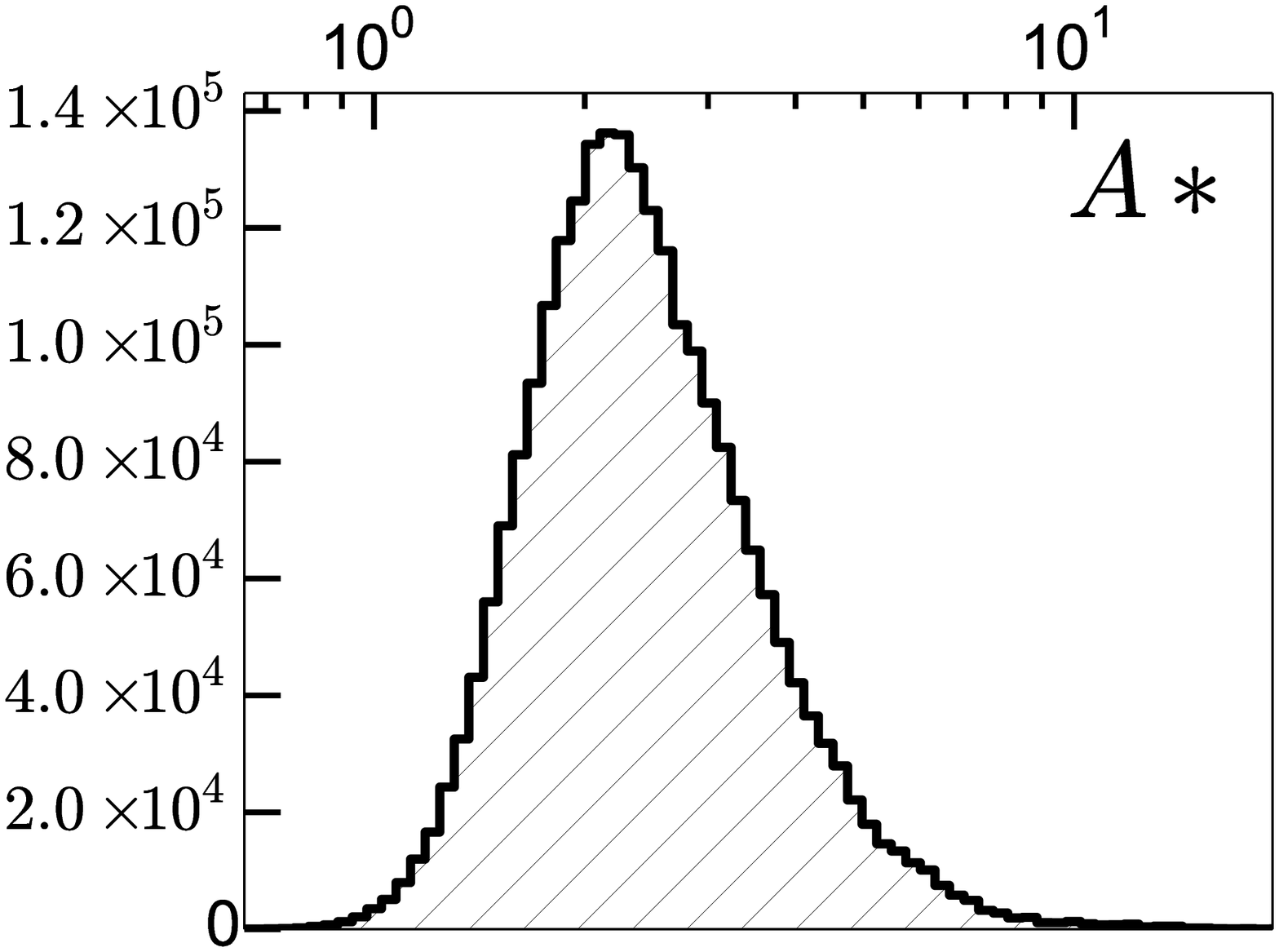} & 
 \includegraphics[width=0.30\columnwidth]{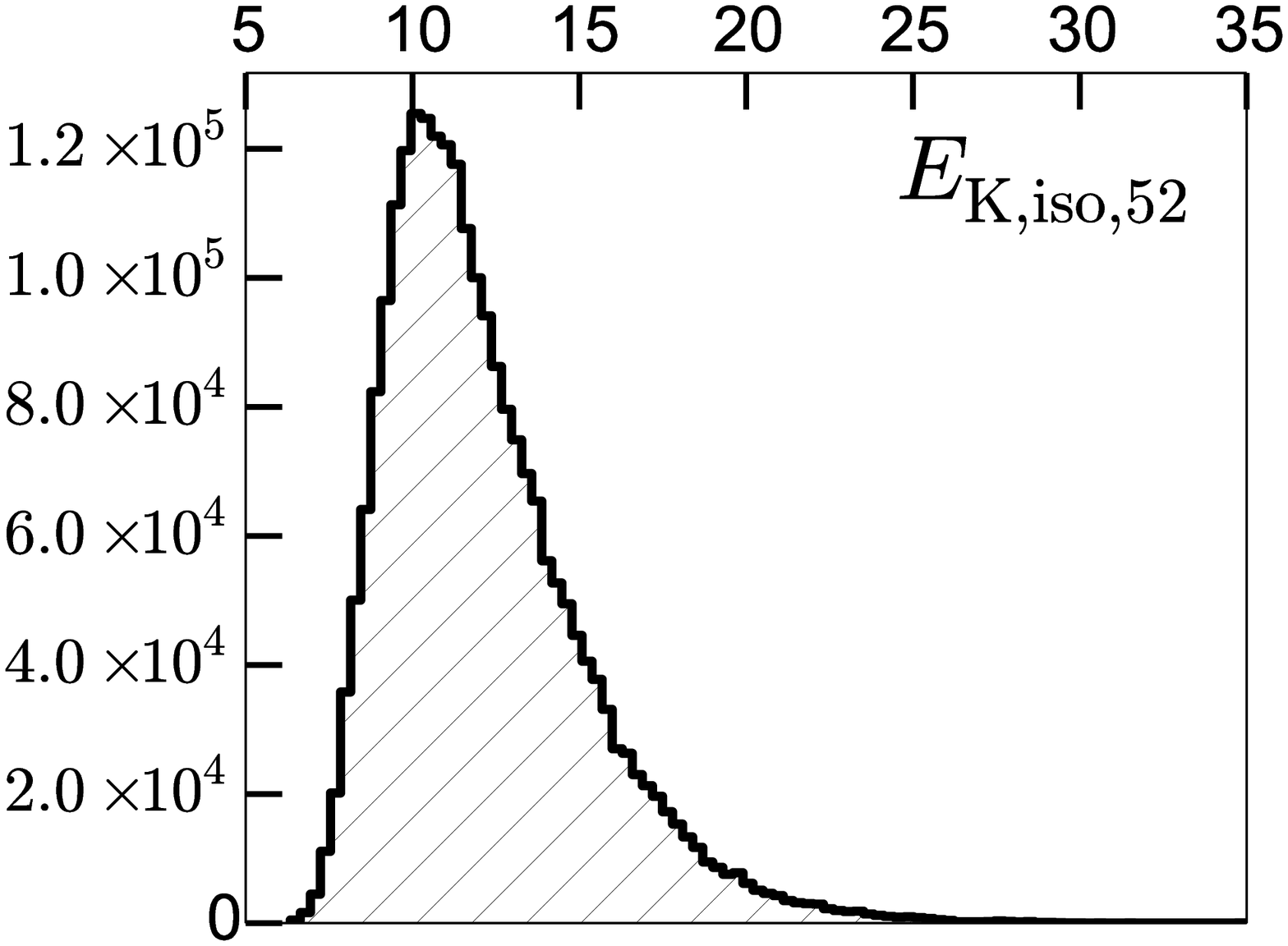} &
 \includegraphics[width=0.30\columnwidth]{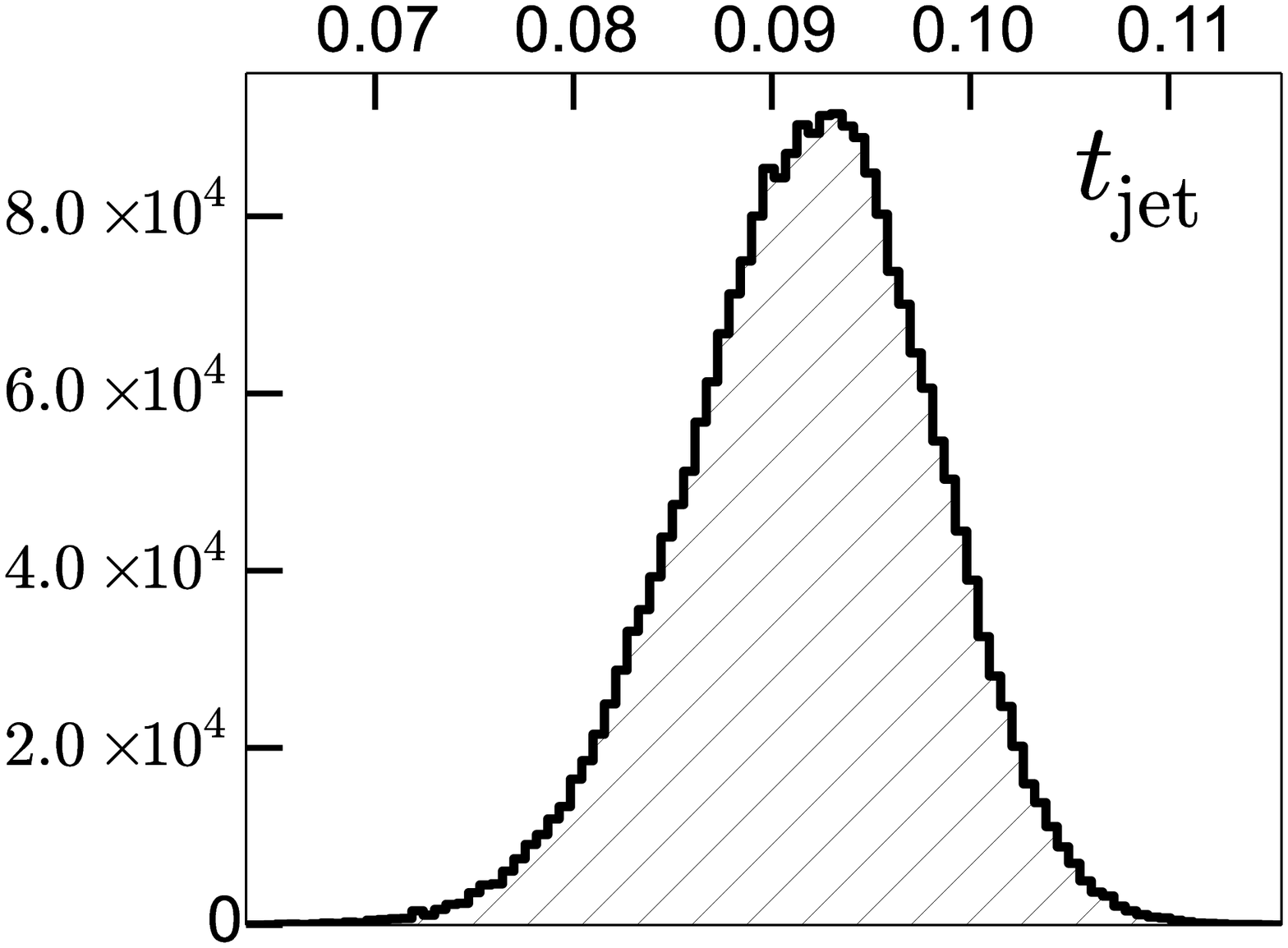} \\ 
 \includegraphics[width=0.30\columnwidth]{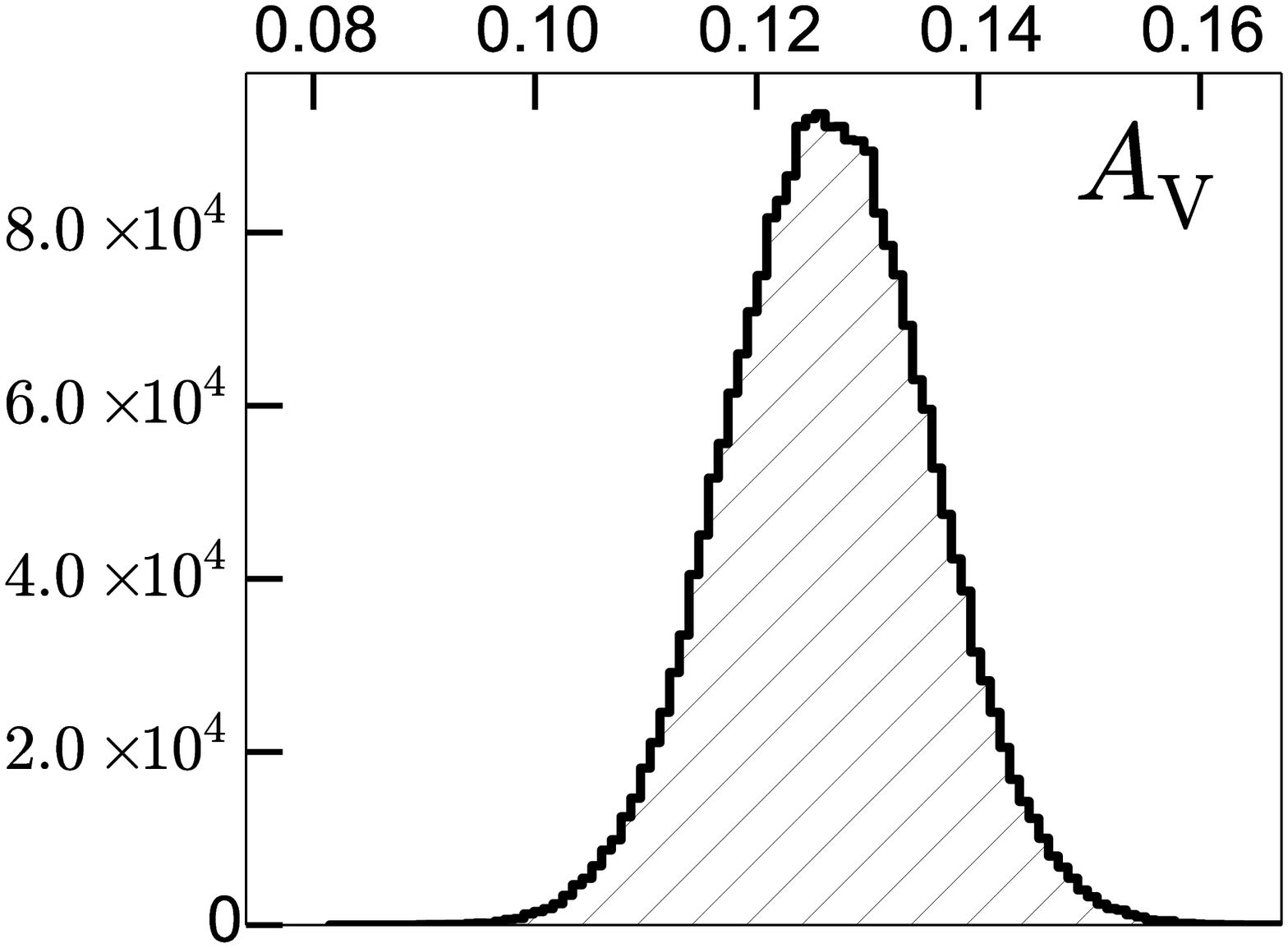} &
 \includegraphics[width=0.30\columnwidth]{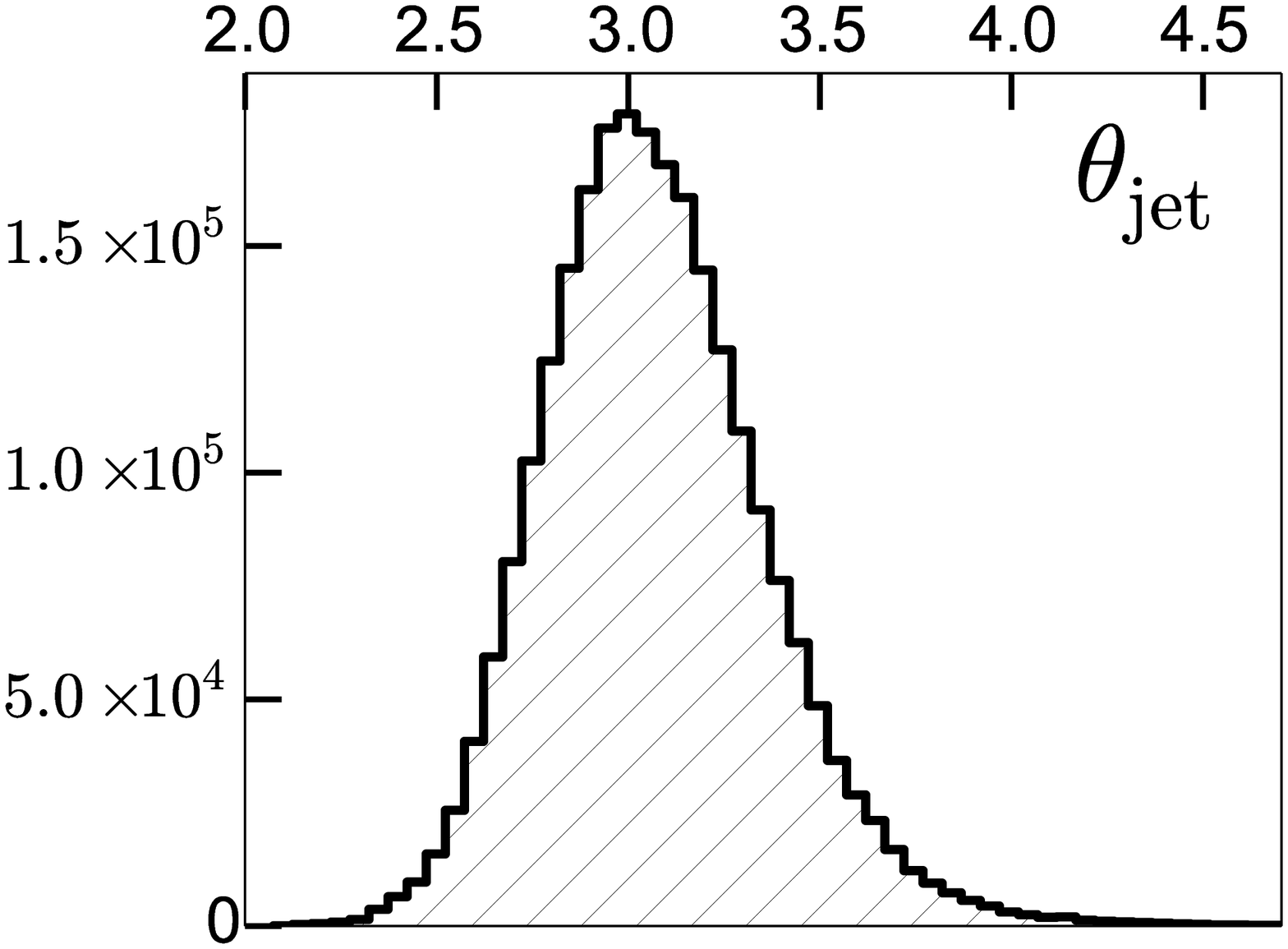}  &
 \includegraphics[width=0.30\columnwidth]{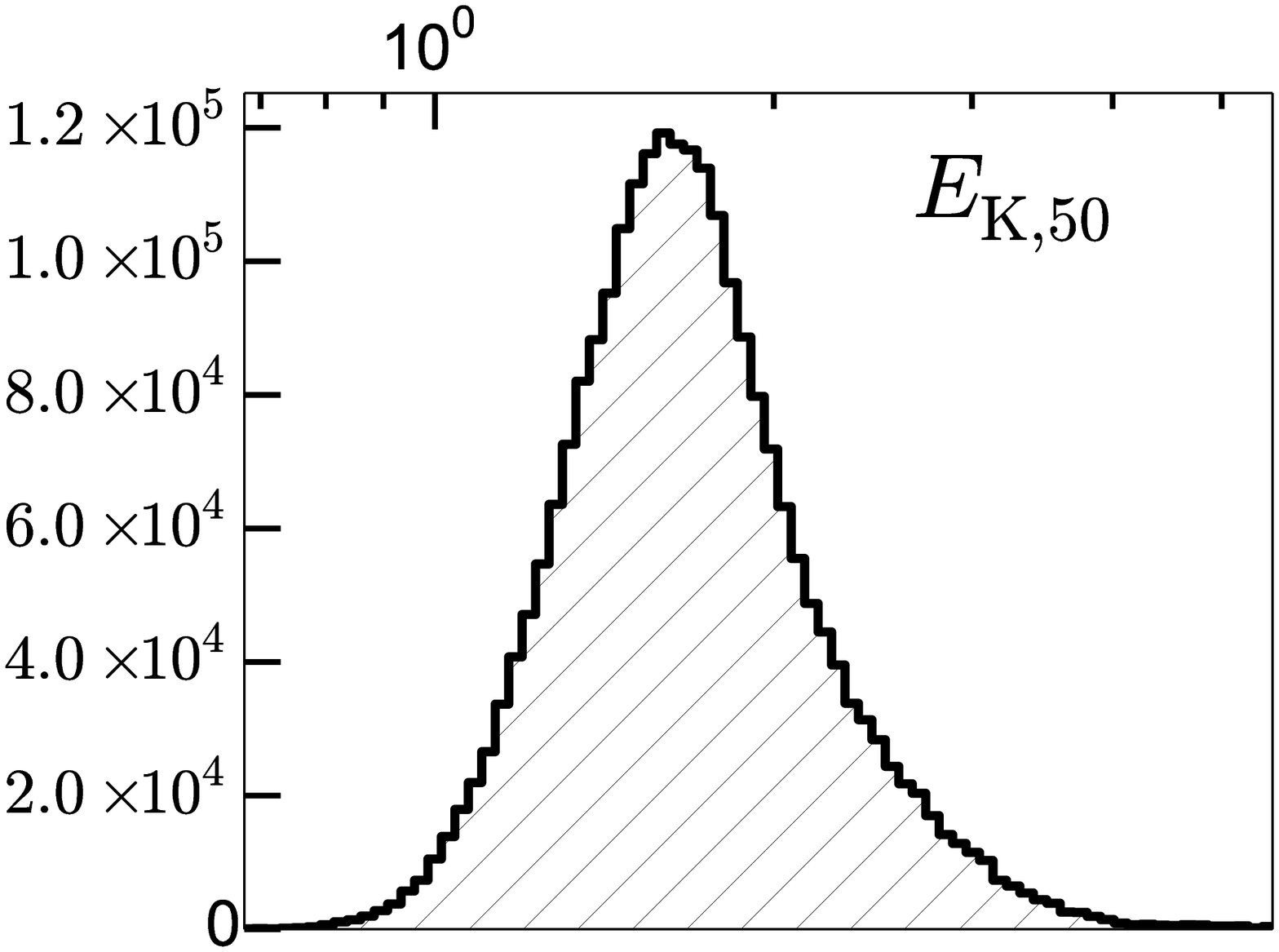} \\  
\end{tabular}
\caption{Posterior probability density functions for the physical parameters for GRB~120404A in 
a wind environment from MCMC simulations. We have restricted $\epsilon_{\rm e} < \nicefrac{1}{3}$ 
and $\epsilon_{\rm B} < \nicefrac{1}{3}$.
\label{fig:120404A_wind_hists}}
\end{figure}

The summary statistics (median and $68\%$ credible intervals) of the posterior density for these 
parameters are $p=2.04\pm0.01$, $\epse=0.28^{+0.03}_{-0.05}$, $\epsb=0.18\pm0.09$, 
$A_*=2.3^{+1.2}_{-0.7}$, $\EKiso=\left(1.2^{+0.3}_{-0.2}\right)\times10^{53}$\,erg, 
$\tjet=(9.2\pm0.6)\times10^{-2}$\,d, and $\AV = 0.13\pm0.01$, corresponding to a jet opening angle 
of $\thetajet=3\degr.0\pm0\degr.3$ and a beaming corrected kinetic energy of $\EK = 
\left(1.6^{0.5}_{-0.3}\right)\times10^{50}$\,erg, the model match the data after the optical peak 
well (Figure \ref{fig:120404A_enj_wind}). We present the correlation contours between the physical 
parameters in Figure \ref{fig:120404A_wind_corrplots} and the marginalized distributions for 
individual parameters in Figure \ref{fig:120404A_wind_hists}. 

The light curves before 0.04\,d in Figure \ref{fig:120404A_enj_wind} are based on the same energy 
injection model as presented in Section \ref{text:120404A:enj}. In this model, the energy increases 
by a factor of $\approx 27$ like in the the ISM case. However, the injection rate is not compatible 
with the maximum rate expected for a wind-like environment, similar to the other wind models 
(Appendices \ref{appendix:120326A_wind}, \ref{appendix:100418A_wind}, and 
\ref{appendix:100901A_wind}). We note that the optical and X-ray frequencies are located above both 
the cooling frequency and \numax, with $\nuc<\numax<\nuopt<\nuX$ in this case, and in this 
regime, the light curves are independent of the circumburst density profile. Thus the measurements 
do not allow us to distinguish between a wind or ISM-like environment in the case of GRB~120404A. 
Clear detections of a wind-like circumburst environment in conjunction with a steep energy 
injection rate in the future will enable us to furthur probe the massive ejecta model of energy 
injection in GRB afterglows, and thereby to further constrain the mechanism responsible for 
plateaus and re-brightening events in GRB afterglows.

\bibliographystyle{apj}
\bibliography{/home/tanmoy/Projects/Edo/Papers/grb_alpha,/home/tanmoy/Projects/Edo/Papers/gcn}

\begin{thebibliography}{}
\expandafter\ifx\csname natexlab\endcsname\relax\def\natexlab#1{#1}\fi

\bibitem[{Amati {et~al.}(2008)Amati, Guidorzi, Frontera, Della~Valle, Finelli,
  Landi, \& Montanari}]{agf+08}
Amati, L., Guidorzi, C., Frontera, F., {et~al.} 2008, \mnras, 391, 577

\bibitem[{{Andreev} {et~al.}(2010{\natexlab{a}}){Andreev}, {Sergeev}, \&
  {Pozanenko}}]{gcn11166}
{Andreev}, M., {Sergeev}, A., \& {Pozanenko}, A. 2010{\natexlab{a}}, GRB
  Coordinates Network, 11166, 1

\bibitem[{{Andreev} {et~al.}(2010{\natexlab{b}}){Andreev}, {Sergeev}, \&
  {Pozanenko}}]{gcn11168}
---. 2010{\natexlab{b}}, GRB Coordinates Network, 11168, 1

\bibitem[{{Andreev} {et~al.}(2010{\natexlab{c}}){Andreev}, {Sergeev}, \&
  {Pozanenko}}]{gcn11191}
---. 2010{\natexlab{c}}, GRB Coordinates Network, 11191, 1

\bibitem[{{Andreev} {et~al.}(2010{\natexlab{d}}){Andreev}, {Sergeev}, \&
  {Pozanenko}}]{gcn11201}
---. 2010{\natexlab{d}}, GRB Coordinates Network, 11201, 1

\bibitem[{{Andreev} {et~al.}(2010{\natexlab{e}}){Andreev}, {Sergeev},
  {Pozanenko}, {Parakhin}, {Velichko}, {Borachok}, \& {Petkov}}]{gcn11200}
{Andreev}, M., {Sergeev}, A., {Pozanenko}, A., {et~al.} 2010{\natexlab{e}}, GRB
  Coordinates Network, 11200, 1

\bibitem[{{Antonelli} {et~al.}(2010){Antonelli}, {Maund}, {Palazzi}, {Goldoni},
  {Vergani}, {Flores}, {de Ugarte Postigo}, {Covino}, {Fynbo}, {Hjorth},
  {Malesani}, {Sollerman}, \& {Thoene}}]{gcn10620}
{Antonelli}, L.~A., {Maund}, J.~R., {Palazzi}, E., {et~al.} 2010, GRB
  Coordinates Network, 10620, 1

\bibitem[{Barthelmy {et~al.}(2005)Barthelmy, Barbier, Cummings, Fenimore,
  Gehrels, Hullinger, Krimm, Markwardt, Palmer, Parsons, Sato, Suzuki,
  Takahashi, Tashiro, \& Tueller}]{bbc+05}
Barthelmy, S.~D., Barbier, L.~M., Cummings, J.~R., {et~al.} 2005, \ssr, 120,
  143

\bibitem[{{Barthelmy} {et~al.}(2012){Barthelmy}, {Sakamoto}, {Markwardt},
  {Baumgartner}, {Cummings}, {Fenimore}, {Gehrels}, {Krimm}, {Palmer}, {Sato},
  {Siegel}, {Stamatikos}, {Tueller}, \& {Ukwatta}}]{gcn13120}
{Barthelmy}, S.~D., {Sakamoto}, T., {Markwardt}, C.~B., {et~al.} 2012, GRB
  Coordinates Network, 13120, 1

\bibitem[{Berger {et~al.}(2000)Berger, Sari, Frail, Kulkarni, Bertoldi, Peck,
  Menten, Shepherd, Moriarty-Schieven, Pooley, Bloom, Diercks, Galama, \&
  Hurley}]{bsf+00}
Berger, E., Sari, R., Frail, D.~A., {et~al.} 2000, \apj, 545, 56

\bibitem[{Berger {et~al.}(2005)Berger, Price, Cenko, Gal-Yam, Soderberg,
  Kasliwal, Leonard, Cameron, Frail, Kulkarni, Murphy, Krzeminski, Piran, Lee,
  Roth, Moon, Fox, Harrison, Persson, Schmidt, Penprase, Rich, Peterson, \&
  Cowie}]{bpc+05}
Berger, E., Price, P.~A., Cenko, S.~B., {et~al.} 2005, \nat, 438, 988

\bibitem[{Bernardini {et~al.}(2011)Bernardini, Margutti, Chincarini, Guidorzi,
  \& Mao}]{bmc+11}
Bernardini, M.~G., Margutti, R., Chincarini, G., Guidorzi, C., \& Mao, J. 2011,
  \aap, 526, A27

\bibitem[{Blandford \& McKee(1977)}]{bm77}
Blandford, R.~D., \& McKee, C.~F. 1977, \mnras, 180, 343

\bibitem[{{Breeveld} \& {Stratta}(2012)}]{gcn13226}
{Breeveld}, A.~A., \& {Stratta}, G. 2012, GRB Coordinates Network, 13226, 1

\bibitem[{Burrows {et~al.}(2005{\natexlab{a}})Burrows, Romano, Falcone,
  Kobayashi, Zhang, Moretti, O'Brien, Goad, Campana, Page, Angelini, Barthelmy,
  Beardmore, Capalbi, Chincarini, Cummings, Cusumano, Fox, Giommi, Hill,
  Kennea, Krimm, Mangano, Marshall, M\'esz\'aros, Morris, Nousek, Osborne,
  Pagani, Perri, Tagliaferri, Wells, Woosley, \& Gehrels}]{brf+05}
Burrows, D.~N., Romano, P., Falcone, A., {et~al.} 2005{\natexlab{a}}, Science,
  309, 1833

\bibitem[{Burrows {et~al.}(2005{\natexlab{b}})Burrows, Hill, Nousek, Kennea,
  Wells, Osborne, Abbey, Beardmore, Mukerjee, Short, Chincarini, Campana,
  Citterio, Moretti, Pagani, Tagliaferri, Giommi, Capalbi, Tamburelli,
  Angelini, Cusumano, Br\"auninger, Burkert, \& Hartner}]{bhn+05}
Burrows, D.~N., Hill, J.~E., Nousek, J.~A., {et~al.} 2005{\natexlab{b}}, \ssr,
  120, 165

\bibitem[{Cenko {et~al.}(2006)Cenko, Fox, Moon, Harrison, Kulkarni, Henning,
  Guzman, Bonati, Smith, Thicksten, Doyle, Petrie, Gal-Yam, Soderberg,
  Anagnostou, \& Laity}]{cfm+06}
Cenko, S.~B., Fox, D.~B., Moon, D.-S., {et~al.} 2006, \pasp, 118, 1396

\bibitem[{Cenko {et~al.}(2010)Cenko, Frail, Harrison, Kulkarni, Nakar, Chandra,
  Butler, Fox, Gal-Yam, Kasliwal, Kelemen, Moon, Ofek, Price, Rau, Soderberg,
  Teplitz, Werner, Bock, Bloom, Starr, Filippenko, Chevalier, Gehrels, Nousek,
  \& Piran}]{cfh+10}
Cenko, S.~B., Frail, D.~A., Harrison, F.~A., {et~al.} 2010, \apj, 711, 641

\bibitem[{Cenko {et~al.}(2011)Cenko, Frail, Harrison, Haislip, Reichart,
  Butler, Cobb, Cucchiara, Berger, Bloom, Chandra, Fox, Perley, Prochaska,
  Filippenko, Glazebrook, Ivarsen, Kasliwal, Kulkarni, LaCluyze, Lopez, Morgan,
  Pettini, \& Rana}]{cfh+11}
---. 2011, \apj, 732, 29

\bibitem[{{Chandra} \& {Frail}(2010{\natexlab{a}})}]{gcn11257}
{Chandra}, P., \& {Frail}, D.~A. 2010{\natexlab{a}}, GRB Coordinates Network,
  11257, 1

\bibitem[{{Chandra} \& {Frail}(2010{\natexlab{b}})}]{gcn10650}
---. 2010{\natexlab{b}}, GRB Coordinates Network, 10650, 1

\bibitem[{Chandra {et~al.}(2008)Chandra, Cenko, Frail, Chevalier, Macquart,
  Kulkarni, Bock, Bertoldi, Kasliwal, Fox, Price, Berger, Soderberg, Harrison,
  Gal-Yam, Ofek, Rau, Schmidt, Cameron, Cowie, Cowie, Roth, Dopita, Peterson,
  \& Penprase}]{ccf+08}
Chandra, P., Cenko, S.~B., Frail, D.~A., {et~al.} 2008, \apj, 683, 924

\bibitem[{Chevalier \& Li(2000)}]{cl00}
Chevalier, R.~A., \& Li, Z.-Y. 2000, \apj, 536, 195

\bibitem[{Chiang \& Dermer(1999)}]{cd99}
Chiang, J., \& Dermer, C.~D. 1999, \apj, 512, 699

\bibitem[{Chincarini {et~al.}(2007)Chincarini, Moretti, Romano, Falcone,
  Morris, Racusin, Campana, Covino, Guidorzi, Tagliaferri, Burrows, Pagani,
  Stroh, Grupe, Capalbi, Cusumano, Gehrels, Giommi, La~Parola, Mangano, Mineo,
  Nousek, O'Brien, Page, Perri, Troja, Willingale, \& Zhang}]{cmr+07}
Chincarini, G., Moretti, A., Romano, P., {et~al.} 2007, \apj, 671, 1903

\bibitem[{Chincarini {et~al.}(2010)Chincarini, Mao, Margutti, Bernardini,
  Guidorzi, Pasotti, Giannios, Della~Valle, Moretti, Romano, D'Avanzo,
  Cusumano, \& Giommi}]{cmm+10}
Chincarini, G., Mao, J., Margutti, R., {et~al.} 2010, \mnras, 406, 2113

\bibitem[{{Chornock} {et~al.}(2010){Chornock}, {Berger}, {Fox}, {Levan},
  {Tanvir}, \& {Wiersema}}]{gcn11164}
{Chornock}, R., {Berger}, E., {Fox}, D., {et~al.} 2010, GRB Coordinates
  Network, 11164, 1

\bibitem[{{Collazzi}(2012)}]{gcn13145}
{Collazzi}, A.~C. 2012, GRB Coordinates Network, 13145, 1

\bibitem[{Covino {et~al.}(2008)Covino, D'Avanzo, Klotz, Perley, Amati, Campana,
  Chincarini, Cucchiara, D'Elia, Guetta, Guidorzi, Kann, K\"upc\"u Yolda\c~s,
  Misra, Olofsson, Tagliaferri, Antonelli, Berger, Bloom, B\"oer, Clemens,
  D'Alessio, Della~Valle, di~Serego~Alighieri, Filippenko, Foley, Fox, Fugazza,
  Fynbo, Gendre, Goldoni, Greiner, Kocevksi, Maiorano, Masetti, Meurs, Modjaz,
  Molinari, Moretti, Palazzi, Pandey, Piranomonte, Poznanski, Primak, Romano,
  Rossi, Roy, Silverman, Stella, Stratta, Testa, Vergani, Vitali, \&
  Zerbi}]{cdk+08}
Covino, S., D'Avanzo, P., Klotz, A., {et~al.} 2008, \mnras, 388, 347

\bibitem[{{Cucchiara} \& {Tanvir}(2012)}]{gcn13217}
{Cucchiara}, A., \& {Tanvir}, N.~R. 2012, GRB Coordinates Network, 13217, 1

\bibitem[{Dai \& Lu(1998)}]{dl98a}
Dai, Z.~G., \& Lu, T. 1998, \aap, 333, L87

\bibitem[{Dall'Osso {et~al.}(2011)Dall'Osso, Stratta, Guetta, Covino,
  De~Cesare, \& Stella}]{dsg+11}
Dall'Osso, S., Stratta, G., Guetta, D., {et~al.} 2011, \aap, 526, A121

\bibitem[{{De Cia} {et~al.}(2010){De Cia}, {Vreeswijk}, \&
  {Jakobsson}}]{gcn11170}
{De Cia}, A., {Vreeswijk}, P.~M., \& {Jakobsson}, P. 2010, GRB Coordinates
  Network, 11170, 1

\bibitem[{de~Ugarte~Postigo {et~al.}(2007)de~Ugarte~Postigo, Fatkhullin,
  J\'ohannesson, Gorosabel, Sokolov, Castro-Tirado, Balega, Spiridonova,
  Jel\'inek, Guziy, P\'erez-Ram\'irez, Hjorth, Laursen, Bersier, Pandey,
  Bremer, Monfardini, Huang, Urata, Ip, Tamagawa, Kinoshita, Mizuno, Arai,
  Yamagishi, Soyano, Usui, Tashiro, Abe, Onda, Aslan, Khamitov, Ozisik,
  Kiziloglu, Bikmaev, Sakhibullin, Burenin, Pavlinsky, Sunyaev, Bhattacharya,
  Kamble, Ishwara~Chandra, \& Trushkin}]{dupfj+07}
de~Ugarte~Postigo, A., Fatkhullin, T.~A., J\'ohannesson, G., {et~al.} 2007,
  \aap, 462, L57

\bibitem[{Dermer {et~al.}(2000{\natexlab{a}})Dermer, B\"ottcher, \&
  Chiang}]{dbc00}
Dermer, C.~D., B\"ottcher, M., \& Chiang, J. 2000{\natexlab{a}}, \apj, 537, 255

\bibitem[{Dermer {et~al.}(2000{\natexlab{b}})Dermer, Chiang, \& Mitman}]{dcm00}
Dermer, C.~D., Chiang, J., \& Mitman, K.~E. 2000{\natexlab{b}}, \apj, 537, 785

\bibitem[{Duffell \& MacFadyen(2014)}]{dm14}
Duffell, P.~C., \& MacFadyen, A.~I. 2014, ArXiv e-prints, arXiv:1407.8250

\bibitem[{Eichler \& Granot(2006)}]{eg06}
Eichler, D., \& Granot, J. 2006, \apjl, 641, L5

\bibitem[{{Elenin} {et~al.}(2010{\natexlab{a}}){Elenin}, {Molotov}, \&
  {Pozanenko}}]{gcn11184}
{Elenin}, L., {Molotov}, I., \& {Pozanenko}, A. 2010{\natexlab{a}}, GRB
  Coordinates Network, 11184, 1

\bibitem[{{Elenin} {et~al.}(2010{\natexlab{b}}){Elenin}, {Molotov}, \&
  {Pozanenko}}]{gcn11234}
---. 2010{\natexlab{b}}, GRB Coordinates Network, 11234, 1

\bibitem[{Evans {et~al.}(2007)Evans, Beardmore, Page, Tyler, Osborne, Goad,
  O'Brien, Vetere, Racusin, Morris, Burrows, Capalbi, Perri, Gehrels, \&
  Romano}]{ebp+07}
Evans, P.~A., Beardmore, A.~P., Page, K.~L., {et~al.} 2007, \aap, 469, 379

\bibitem[{Evans {et~al.}(2009)Evans, Beardmore, Page, Osborne, O'Brien,
  Willingale, Starling, Burrows, Godet, Vetere, Racusin, Goad, Wiersema,
  Angelini, Capalbi, Chincarini, Gehrels, Kennea, Margutti, Morris, Mountford,
  Pagani, Perri, Romano, \& Tanvir}]{ebp+09}
---. 2009, \mnras, 397, 1177

\bibitem[{Falcone {et~al.}(2006)Falcone, Burrows, Lazzati, Campana, Kobayashi,
  Zhang, M\'esz\'aros, Page, Kennea, Romano, Pagani, Angelini, Beardmore,
  Capalbi, Chincarini, Cusumano, Giommi, Goad, Godet, Grupe, Hill, La~Parola,
  Mangano, Moretti, Nousek, O'Brien, Osborne, Perri, Tagliaferri, Wells, \&
  Gehrels}]{fbl+06}
Falcone, A.~D., Burrows, D.~N., Lazzati, D., {et~al.} 2006, \apj, 641, 1010

\bibitem[{Fan \& Piran(2006)}]{fp06}
Fan, Y., \& Piran, T. 2006, \mnras, 369, 197

\bibitem[{Fan \& Wei(2005)}]{fw05}
Fan, Y.~Z., \& Wei, D.~M. 2005, \mnras, 364, L42

\bibitem[{{Filgas} {et~al.}(2010){Filgas}, {Klose}, \& {Greiner}}]{gcn10617}
{Filgas}, R., {Klose}, S., \& {Greiner}, J. 2010, GRB Coordinates Network,
  10617, 1

\bibitem[{Foreman-Mackey {et~al.}(2013)Foreman-Mackey, Hogg, Lang, \&
  Goodman}]{fhlg13}
Foreman-Mackey, D., Hogg, D.~W., Lang, D., \& Goodman, J. 2013, \pasp, 125, 306

\bibitem[{Frail {et~al.}(2000)Frail, Waxman, \& Kulkarni}]{fwk00}
Frail, D.~A., Waxman, E., \& Kulkarni, S.~R. 2000, \apj, 537, 191

\bibitem[{Frail {et~al.}(2003)Frail, Yost, Berger, Harrison, Sari, Kulkarni,
  Taylor, Bloom, Fox, Moriarty-Schieven, \& Price}]{fyb+03}
Frail, D.~A., Yost, S.~A., Berger, E., {et~al.} 2003, \apj, 590, 992

\bibitem[{Friedman \& Bloom(2005)}]{fb05}
Friedman, A.~S., \& Bloom, J.~S. 2005, \apj, 627, 1

\bibitem[{Gehrels {et~al.}(2004)Gehrels, Chincarini, Giommi, Mason, Nousek,
  Wells, White, Barthelmy, Burrows, Cominsky, Hurley, Marshall, M\'esz\'aros,
  Roming, Angelini, Barbier, Belloni, Campana, Caraveo, Chester, Citterio,
  Cline, Cropper, Cummings, Dean, Feigelson, Fenimore, Frail, Fruchter,
  Garmire, Gendreau, Ghisellini, Greiner, Hill, Hunsberger, Krimm, Kulkarni,
  Kumar, Lebrun, Lloyd-Ronning, Markwardt, Mattson, Mushotzky, Norris, Osborne,
  Paczynski, Palmer, Park, Parsons, Paul, Rees, Reynolds, Rhoads, Sasseen,
  Schaefer, Short, Smale, Smith, Stella, Tagliaferri, Takahashi, Tashiro,
  Townsley, Tueller, Turner, Vietri, Voges, Ward, Willingale, Zerbi, \&
  Zhang}]{gcg+04}
Gehrels, N., Chincarini, G., Giommi, P., {et~al.} 2004, \apj, 611, 1005

\bibitem[{Ghirlanda {et~al.}(2007)Ghirlanda, Nava, Ghisellini, \&
  Firmani}]{gngf07}
Ghirlanda, G., Nava, L., Ghisellini, G., \& Firmani, C. 2007, \aap, 466, 127

\bibitem[{{Gorbovskoy} {et~al.}(2010){Gorbovskoy}, {Lipunov}, {Kornilov},
  {Belinski}, {Shatskiy}, {Tyurina}, {Kuvshinov}, {Balanutsa}, {Chazov},
  {Kortunov}, {Kuznetsov}, {Zimnukhov}, {Kornilov}, {Ivanov}, {Chuvalaev},
  {Poleschuk}, {Konstantinov}, {Lenok}, {Gres}, {Yazev}, {Budnev}, {Yurkov},
  {Sergienko}, {Varda}, {Kudelina}, {Tlatov}, {Parhomenko}, {Dormidontov},
  {Sennik}, {Krushinski}, {Zalozhnich}, {Kopytova}, \& {Popov}}]{gcn11178}
{Gorbovskoy}, E., {Lipunov}, V., {Kornilov}, V., {et~al.} 2010, GRB Coordinates
  Network, 11178, 1

\bibitem[{{Gorbovskoy} {et~al.}(2012){Gorbovskoy}, {Lipunov}, {Kornilov},
  {Kuvshinov}, {Belinski}, {Tyurina}, {Shatskiy}, {Balanutsa}, {Zimnukhov},
  {Kuznetsov}, {Chazov}, {Kuznetsov}, {Sankovich}, {Yurkov}, {Sergienko},
  {Varda}, {Sinyakov}, {Tlatov}, {Parhomenko}, {Dormidontov}, {Sennik},
  {Krushinski}, {Zalozhnich}, {Popov}, {Bourdanov}, {Punanova}, {Levato},
  {Saffe}, {Mallamaci}, {Lopez}, \& {Podest}}]{gcn13230}
---. 2012, GRB Coordinates Network, 13230, 1

\bibitem[{Gorbovskoy {et~al.}(2012)Gorbovskoy, Lipunova, Lipunov, Kornilov,
  Belinski, Shatskiy, Tyurina, Kuvshinov, Balanutsa, Chazov, Kuznetsov,
  Zimnukhov, Kornilov, Sankovich, Krylov, Ivanov, Chvalaev, Poleschuk,
  Konstantinov, Gress, Yazev, Budnev, Krushinski, Zalozhnich, Popov, Tlatov,
  Parhomenko, Dormidontov, Senik, Yurkov, Sergienko, Varda, Kudelina,
  Castro-Tirado, Gorosabel, S\'anchez-Ram\'irez, Jelinek, \& Tello}]{gll+12}
Gorbovskoy, E.~S., Lipunova, G.~V., Lipunov, V.~M., {et~al.} 2012, \mnras, 421,
  1874

\bibitem[{Granot {et~al.}(2006)Granot, K\"onigl, \& Piran}]{gkp06}
Granot, J., K\"onigl, A., \& Piran, T. 2006, \mnras, 370, 1946

\bibitem[{Granot \& Kumar(2006)}]{gk06}
Granot, J., \& Kumar, P. 2006, \mnras, 366, L13

\bibitem[{Granot {et~al.}(2001)Granot, Miller, Piran, Suen, \& Hughes}]{gmp+01}
Granot, J., Miller, M., Piran, T., Suen, W., \& Hughes, P. 2001, in Gamma-Ray
  Bursts in the Afterglow Era, ed. E.~Costa, F.~Frontera, \& J.~Hjorth, ESO
  ASTROPHYSICS SYMPOSIA (http://arxiv.org/pdf/astro-ph/0103038.pdf: Springer
  Berlin Heidelberg), 312--314

\bibitem[{Granot {et~al.}(2002)Granot, Panaitescu, Kumar, \& Woosley}]{gpkw02}
Granot, J., Panaitescu, A., Kumar, P., \& Woosley, S.~E. 2002, \apjl, 570, L61

\bibitem[{Granot \& Sari(2002)}]{gs02}
Granot, J., \& Sari, R. 2002, \apj, 568, 820

\bibitem[{Greisen(2003)}]{gre03}
Greisen, E.~W. 2003, Information Handling in Astronomy - Historical Vistas,
  285, 109

\bibitem[{Guidorzi {et~al.}(2015)Guidorzi, Dichiara, Frontera, Margutti,
  Baldeschi, \& Amati}]{gdf+15}
Guidorzi, C., Dichiara, S., Frontera, F., {et~al.} 2015, \apj, 801, 57

\bibitem[{{Guidorzi} {et~al.}(2012){Guidorzi}, {Melandri}, \&
  {Japelj}}]{gcn13209}
{Guidorzi}, C., {Melandri}, A., \& {Japelj}, J. 2012, GRB Coordinates Network,
  13209, 1

\bibitem[{Guidorzi {et~al.}(2007)Guidorzi, Vergani, Sazonov, Covino, Malesani,
  Molkov, Palazzi, Romano, Campana, Chincarini, Fugazza, Moretti, Tagliaferri,
  Llorente, Gorosabel, Antonelli, Capalbi, Cusumano, D'Avanzo, Mangano,
  Masetti, Meurs, Mineo, Molinari, Morris, Nicastro, Page, Perri, Sbarufatti,
  Stratta, Sunyaev, Troja, \& Zerbi}]{gvs+07}
Guidorzi, C., Vergani, S.~D., Sazonov, S., {et~al.} 2007, \aap, 474, 793

\bibitem[{Guidorzi {et~al.}(2013)Guidorzi, Mundell, Harrison, Margutti,
  Sudilovsky, Zauderer, Kobayashi, Cucchiara, Melandri, Pandey, Berger,
  Bersier, D'Elia, Gomboc, Greiner, Japelj, Kopa\v~c, Kumar, Malesani, Mottram,
  O'Brien, Rau, Smith, Steele, Tanvir, \& Virgili}]{gmh+13}
Guidorzi, C., Mundell, C.~G., Harrison, R., {et~al.} 2013, \mnras,
  arXiv:1311.4340

\bibitem[{Harrison {et~al.}(2001)Harrison, Yost, Sari, Berger, Galama,
  Holtzman, Axelrod, Bloom, Chevalier, Costa, Diercks, Djorgovski, Frail,
  Frontera, Hurley, Kulkarni, McCarthy, Piro, Pooley, Price, Reichart, Ricker,
  Shepherd, Schmidt, Walter, \& Wheeler}]{hys+01}
Harrison, F.~A., Yost, S.~A., Sari, R., {et~al.} 2001, \apj, 559, 123

\bibitem[{Hascoet {et~al.}(2015)Hascoet, Beloborodov, Daigne, \&
  Mochkovitch}]{hbdm15}
Hascoet, R., Beloborodov, A.~M., Daigne, F., \& Mochkovitch, R. 2015, ArXiv
  e-prints, arXiv:1503.08333

\bibitem[{Hasco\"et {et~al.}(2012)Hasco\"et, Daigne, \& Mochkovitch}]{hdm12}
Hasco\"et, R., Daigne, F., \& Mochkovitch, R. 2012, \aap, 541, A88

\bibitem[{{Hentunen} {et~al.}(2010){Hentunen}, {Nissinen}, \&
  {Salmi}}]{gcn11173}
{Hentunen}, V.-P., {Nissinen}, M., \& {Salmi}, T. 2010, GRB Coordinates
  Network, 11173, 1

\bibitem[{{Hentunen} {et~al.}(2012){Hentunen}, {Nissinen}, \&
  {Salmi}}]{gcn13119}
---. 2012, GRB Coordinates Network, 13119, 1

\bibitem[{Ho {et~al.}(2004)Ho, Moran, \& Lo}]{hml04}
Ho, P.~T.~P., Moran, J.~M., \& Lo, K.~Y. 2004, \apjl, 616, L1

\bibitem[{Hou {et~al.}(2014)Hou, Geng, Wang, Wu, Huang, Dai, \& Lu}]{hgw+14}
Hou, S.~J., Geng, J.~J., Wang, K., {et~al.} 2014, \apj, 785, 113

\bibitem[{{Im} {et~al.}(2010){Im}, {Choi}, {Jun}, {Kang}, {Urata}, {Choi},
  {Sakamoto}, {Tanvir}, \& {Levan}}]{gcn11208}
{Im}, M., {Choi}, C., {Jun}, H., {et~al.} 2010, GRB Coordinates Network, 11208,
  1

\bibitem[{{Immler} {et~al.}(2010){Immler}, {Barthelmy}, {Baumgartner},
  {Beardmore}, {Campana}, {D'Elia}, {Evans}, {Gelbord}, {Godet}, {Gronwall},
  {Guidorzi}, {Holland}, {Hoversten}, {Littlejohns}, {Marshall}, {O'Brien},
  {Osborne}, {Pagani}, {Page}, {Palmer}, {Pritchard}, {Rowlinson},
  {Sbarufatti}, {Siegel}, {Stamatikos}, \& {Starling}}]{gcn11159}
{Immler}, S., {Barthelmy}, S.~D., {Baumgartner}, W.~H., {et~al.} 2010, GRB
  Coordinates Network, 11159, 1

\bibitem[{{Jang} {et~al.}(2012){Jang}, {Im}, \& {Urata}}]{gcn13139}
{Jang}, M., {Im}, M., \& {Urata}, Y. 2012, GRB Coordinates Network, 13139, 1

\bibitem[{Jester {et~al.}(2005)Jester, Schneider, Richards, Green, Schmidt,
  Hall, Strauss, Vanden~Berk, Stoughton, Gunn, Brinkmann, Kent, Smith, Tucker,
  \& Yanny}]{jsr+05}
Jester, S., Schneider, D.~P., Richards, G.~T., {et~al.} 2005, \aj, 130, 873

\bibitem[{Jin {et~al.}(2007)Jin, Yan, Fan, \& Wei}]{jyfw07}
Jin, Z.~P., Yan, T., Fan, Y.~Z., \& Wei, D.~M. 2007, \apjl, 656, L57

\bibitem[{Kalberla {et~al.}(2005)Kalberla, Burton, Hartmann, Arnal, Bajaja,
  Morras, \& P\"oppel}]{kbh+05}
Kalberla, P.~M.~W., Burton, W.~B., Hartmann, D., {et~al.} 2005, \aap, 440, 775

\bibitem[{{Kann} {et~al.}(2010{\natexlab{a}}){Kann}, {Laux}, \&
  {Stecklum}}]{gcn11236}
{Kann}, D.~A., {Laux}, U., \& {Stecklum}, B. 2010{\natexlab{a}}, GRB
  Coordinates Network, 11236, 1

\bibitem[{{Kann} {et~al.}(2010{\natexlab{b}}){Kann}, {Ludwig}, \&
  {Stecklum}}]{gcn11246}
{Kann}, D.~A., {Ludwig}, F., \& {Stecklum}, B. 2010{\natexlab{b}}, GRB
  Coordinates Network, 11246, 1

\bibitem[{{Klotz} {et~al.}(2012){Klotz}, {Gendre}, {Boer}, \&
  {Atteia}}]{gcn13108}
{Klotz}, A., {Gendre}, B., {Boer}, M., \& {Atteia}, J.~L. 2012, GRB Coordinates
  Network, 13108, 1

\bibitem[{Kobayashi(2000)}]{kob00}
Kobayashi, S. 2000, \apj, 545, 807

\bibitem[{Kobayashi \& Sari(2000)}]{ks00}
Kobayashi, S., \& Sari, R. 2000, \apj, 542, 819

\bibitem[{Kobayashi \& Zhang(2003)}]{kz03a}
Kobayashi, S., \& Zhang, B. 2003, \apj, 597, 455

\bibitem[{Kong {et~al.}(2010)Kong, Wong, Huang, \& Cheng}]{kwhc10}
Kong, S.~W., Wong, A.~Y.~L., Huang, Y.~F., \& Cheng, K.~S. 2010, \mnras, 402,
  409

\bibitem[{{Kopac} {et~al.}(2010){Kopac}, {Dintinjana}, \& {Gomboc}}]{gcn11177}
{Kopac}, D., {Dintinjana}, B., \& {Gomboc}, A. 2010, GRB Coordinates Network,
  11177, 1

\bibitem[{{Kuin} {et~al.}(2012){Kuin}, {Holland}, \& {Siegel}}]{gcn13114}
{Kuin}, N.~P.~M., {Holland}, S., \& {Siegel}, M.~H. 2012, GRB Coordinates
  Network, 13114, 1

\bibitem[{Kumar \& Piran(2000)}]{kp00a}
Kumar, P., \& Piran, T. 2000, \apj, 532, 286

\bibitem[{{Kuroda} {et~al.}(2010{\natexlab{a}}){Kuroda}, {Hanayama}, {Miyaji},
  {Watanabe}, {Yanagisawa}, {Yoshida}, {Ohta}, \& {Kawai}}]{gcn11258}
{Kuroda}, D., {Hanayama}, H., {Miyaji}, T., {et~al.} 2010{\natexlab{a}}, GRB
  Coordinates Network, 11258, 1

\bibitem[{{Kuroda} {et~al.}(2010{\natexlab{b}}){Kuroda}, {Hanayama}, {Miyaji},
  {Watanabe}, {Yanagisawa}, {Yoshida}, {Ohta}, \& {Kawai}}]{gcn11205}
---. 2010{\natexlab{b}}, GRB Coordinates Network, 11205, 1

\bibitem[{{Kuroda} {et~al.}(2010{\natexlab{c}}){Kuroda}, {Yanagisawa},
  {Shimizu}, {Toda}, {Nagayama}, {Yoshida}, {Ohta}, \& {Kawai}}]{gcn11172}
{Kuroda}, D., {Yanagisawa}, K., {Shimizu}, Y., {et~al.} 2010{\natexlab{c}}, GRB
  Coordinates Network, 11172, 1

\bibitem[{{Kuroda} {et~al.}(2010{\natexlab{d}}){Kuroda}, {Yanagisawa},
  {Shimizu}, {Toda}, {Nagayama}, {Yoshida}, {Ohta}, \& {Kawai}}]{gcn11189}
---. 2010{\natexlab{d}}, GRB Coordinates Network, 11189, 1

\bibitem[{{Lacluyze} {et~al.}(2012){Lacluyze}, {Haislip}, {Ivarsen}, {Maturi},
  {Reichart}, {Moore}, {Cromartie}, {Egger}, {Foster}, {Frank}, {Nysewander},
  {Oza}, {Speckhard}, {Trotter}, \& {Crain}}]{gcn13109}
{Lacluyze}, A., {Haislip}, J., {Ivarsen}, K., {et~al.} 2012, GRB Coordinates
  Network, 13109, 1

\bibitem[{Landolt(2009)}]{lan09}
Landolt, A.~U. 2009, \aj, 137, 4186

\bibitem[{Laskar {et~al.}(2013)Laskar, Berger, Zauderer, Margutti, Soderberg,
  Chakraborti, Lunnan, Chornock, Chandra, \& Ray}]{lbz+13}
Laskar, T., Berger, E., Zauderer, B.~A., {et~al.} 2013, \apj, 776, 119

\bibitem[{Laskar {et~al.}(2014)Laskar, Berger, Tanvir, Zauderer, Margutti,
  Levan, Perley, Fong, Wiersema, Menten, \& Hrudkova}]{lbt+14}
Laskar, T., Berger, E., Tanvir, N., {et~al.} 2014, \apj, 781, 1

\bibitem[{Lazzati \& Perna(2007)}]{lp07}
Lazzati, D., \& Perna, R. 2007, \mnras, 375, L46

\bibitem[{Lazzati {et~al.}(2002)Lazzati, Rossi, Covino, Ghisellini, \&
  Malesani}]{lrc+02}
Lazzati, D., Rossi, E., Covino, S., Ghisellini, G., \& Malesani, D. 2002, \aap,
  396, L5

\bibitem[{Li {et~al.}(2012)Li, Liang, Tang, Chen, Xi, L\"u, Gao, Zhang, Zhang,
  Yi, Lu, L\"u, \& Wei}]{llt+12}
Li, L., Liang, E.-W., Tang, Q.-W., {et~al.} 2012, \apj, 758, 27

\bibitem[{Liang {et~al.}(2007)Liang, Zhang, \& Zhang}]{lzz07}
Liang, E.-W., Zhang, B.-B., \& Zhang, B. 2007, \apj, 670, 565

\bibitem[{Liang {et~al.}(2006)Liang, Zhang, O'Brien, Willingale, Angelini,
  Burrows, Campana, Chincarini, Falcone, Gehrels, Goad, Grupe, Kobayashi,
  M\'esz\'aros, Nousek, Osborne, Page, \& Tagliaferri}]{lzo+06}
Liang, E.~W., Zhang, B., O'Brien, P.~T., {et~al.} 2006, \apj, 646, 351

\bibitem[{Lipkin {et~al.}(2004)Lipkin, Ofek, Gal-Yam, Leibowitz, Poznanski,
  Kaspi, Polishook, Kulkarni, Fox, Berger, Mirabal, Halpern, Bureau, Fathi,
  Price, Peterson, Frebel, Schmidt, Orosz, Fitzgerald, Bloom, van Dokkum,
  Bailyn, Buxton, \& Barsony}]{log+04}
Lipkin, Y.~M., Ofek, E.~O., Gal-Yam, A., {et~al.} 2004, \apj, 606, 381

\bibitem[{Liu {et~al.}(2008)Liu, Wu, \& Lu}]{lwl08}
Liu, X.~W., Wu, X.~F., \& Lu, T. 2008, \aap, 487, 503

\bibitem[{{Malesani} \& {Palazzi}(2010)}]{gcn10631}
{Malesani}, D., \& {Palazzi}, E. 2010, GRB Coordinates Network, 10631, 1

\bibitem[{Malesani {et~al.}(2007)Malesani, Covino, D'Avanzo, D'Elia, Fugazza,
  Piranomonte, Ballo, Campana, Stella, Tagliaferri, Antonelli, Chincarini,
  Della~Valle, Goldoni, Guidorzi, Israel, Lazzati, Melandri, Pellizza, Romano,
  Stratta, \& Vergani}]{mcd+07}
Malesani, D., Covino, S., D'Avanzo, P., {et~al.} 2007, \aap, 473, 77

\bibitem[{Mangano {et~al.}(2007)Mangano, Holland, Malesani, Troja, Chincarini,
  Zhang, La~Parola, Brown, Burrows, Campana, Capalbi, Cusumano, Della~Valle,
  Gehrels, Giommi, Grupe, Guidorzi, Mineo, Moretti, Osborne, Pandey, Perri,
  Romano, Roming, \& Tagliaferri}]{mhm+07}
Mangano, V., Holland, S.~T., Malesani, D., {et~al.} 2007, \aap, 470, 105

\bibitem[{Margutti {et~al.}(2011{\natexlab{a}})Margutti, Bernardini,
  Barniol~Duran, Guidorzi, Shen, \& Chincarini}]{mbbd+11}
Margutti, R., Bernardini, G., Barniol~Duran, R., {et~al.} 2011{\natexlab{a}},
  \mnras, 410, 1064

\bibitem[{Margutti {et~al.}(2010{\natexlab{a}})Margutti, Guidorzi, Chincarini,
  Bernardini, Genet, Mao, \& Pasotti}]{mgc+10}
Margutti, R., Guidorzi, C., Chincarini, G., {et~al.} 2010{\natexlab{a}},
  \mnras, 406, 2149

\bibitem[{Margutti {et~al.}(2010{\natexlab{b}})Margutti, Genet, Granot,
  Barniol~Duran, Guidorzi, Chincarini, Mao, Schady, Sakamoto, Miller, Olofsson,
  Bloom, Evans, Fynbo, Malesani, Moretti, Pasotti, Starr, Burrows, Barthelmy,
  Roming, \& Gehrels}]{mgg+10}
Margutti, R., Genet, F., Granot, J., {et~al.} 2010{\natexlab{b}}, \mnras, 402,
  46

\bibitem[{Margutti {et~al.}(2011{\natexlab{b}})Margutti, Chincarini, Granot,
  Guidorzi, Berger, Bernardini, Gehrels, Soderberg, Stamatikos, \&
  Zaninoni}]{mcg+11}
Margutti, R., Chincarini, G., Granot, J., {et~al.} 2011{\natexlab{b}}, \mnras,
  417, 2144

\bibitem[{Margutti {et~al.}(2013)Margutti, Zaninoni, Bernardini, Chincarini,
  Pasotti, Guidorzi, Angelini, Burrows, Capalbi, Evans, Gehrels, Kennea,
  Mangano, Moretti, Nousek, Osborne, Page, Perri, Racusin, Romano, Sbarufatti,
  Stafford, \& Stamatikos}]{mzb+13}
Margutti, R., Zaninoni, E., Bernardini, M.~G., {et~al.} 2013, \mnras, 428, 729

\bibitem[{{Marshall} {et~al.}(2010){Marshall}, {Beardmore}, {Gelbord},
  {Holland}, {Hoversten}, {Kennea}, {Littlejohns}, {Markwardt}, {O'Brien},
  {Pagani}, {Page}, {Palmer}, {Rowlinson}, {Stroh}, {Ukwatta}, \&
  {Vetere}}]{gcn10612}
{Marshall}, F.~E., {Beardmore}, A.~P., {Gelbord}, J.~M., {et~al.} 2010, GRB
  Coordinates Network, 10612, 1

\bibitem[{Marshall {et~al.}(2011)Marshall, Antonelli, Burrows, Covino,
  de~Pasquale, Evans, Fugazza, Holland, Liang, O'Brien, Oates, Osborne, Pagani,
  Sakamoto, Siegel, Wu, \& Zhang}]{mab+11}
Marshall, F.~E., Antonelli, L.~A., Burrows, D.~N., {et~al.} 2011, \apj, 727,
  132

\bibitem[{Melandri {et~al.}(2014)Melandri, Virgili, Guidorzi, Bernardini,
  Kobayashi, Mundell, Gomboc, Dintinjana, Hentunen, Japelj, Kopa\v~c, Kuroda,
  Morgan, Steele, Quadri, Arici, Arnold, Girelli, Hanayama, Kawai, Miku\v~z,
  Nissinen, Salmi, Smith, Strabla, Tonincelli, \& Quadri}]{mvg+14}
Melandri, A., Virgili, F.~J., Guidorzi, C., {et~al.} 2014, \aap, 572, A55

\bibitem[{{Moin} {et~al.}(2010){Moin}, {Tingay}, {Phillips}, {Taylor},
  {Wieringa}, \& {Martin}}]{gcn10832}
{Moin}, A., {Tingay}, S., {Phillips}, C., {et~al.} 2010, GRB Coordinates
  Network, 10832, 1

\bibitem[{Moin {et~al.}(2013)Moin, Chandra, Miller-Jones, Tingay, Taylor,
  Frail, Wang, Reynolds, \& Phillips}]{mcm+13}
Moin, A., Chandra, P., Miller-Jones, J.~C.~A., {et~al.} 2013, \apj, 779, 105

\bibitem[{Monfardini {et~al.}(2006)Monfardini, Kobayashi, Guidorzi, Carter,
  Mundell, Bersier, Gomboc, Melandri, Mottram, Smith, \& Steele}]{mkg+06}
Monfardini, A., Kobayashi, S., Guidorzi, C., {et~al.} 2006, \apj, 648, 1125

\bibitem[{{Moody} {et~al.}(2010){Moody}, {Laney}, {Pearson}, \&
  {Pace}}]{gcn10665}
{Moody}, J.~W., {Laney}, D., {Pearson}, R., \& {Pace}, C. 2010, GRB Coordinates
  Network, 10665, 1

\bibitem[{{Morgan}(2012)}]{gcn13143}
{Morgan}, A.~N. 2012, GRB Coordinates Network, 13143, 1

\bibitem[{Nakar \& Granot(2007)}]{ng07}
Nakar, E., \& Granot, J. 2007, \mnras, 380, 1744

\bibitem[{Nousek {et~al.}(2006)Nousek, Kouveliotou, Grupe, Page, Granot,
  Ramirez-Ruiz, Patel, Burrows, Mangano, Barthelmy, Beardmore, Campana,
  Capalbi, Chincarini, Cusumano, Falcone, Gehrels, Giommi, Goad, Godet,
  Hurkett, Kennea, Moretti, O'Brien, Osborne, Romano, Tagliaferri, \&
  Wells}]{nkg+06}
Nousek, J.~A., Kouveliotou, C., Grupe, D., {et~al.} 2006, \apj, 642, 389

\bibitem[{O'Brien {et~al.}(2006)O'Brien, Willingale, Osborne, Goad, Page,
  Vaughan, Rol, Beardmore, Godet, Hurkett, Wells, Zhang, Kobayashi, Burrows,
  Nousek, Kennea, Falcone, Grupe, Gehrels, Barthelmy, Cannizzo, Cummings, Hill,
  Krimm, Chincarini, Tagliaferri, Campana, Moretti, Giommi, Perri, Mangano, \&
  LaParola}]{owo+06}
O'Brien, P.~T., Willingale, R., Osborne, J., {et~al.} 2006, \apj, 647, 1213

\bibitem[{Pagani {et~al.}(2006)Pagani, Morris, Kobayashi, Sakamoto, Falcone,
  Moretti, Page, Burrows, Grupe, Retter, Racusin, Kennea, Campana, Romano,
  Tagliaferri, Hill, Angelini, Cusumano, Goad, Barthelmy, Chincarini, Wells,
  Giommi, Nousek, \& Gehrels}]{pmk+06}
Pagani, C., Morris, D.~C., Kobayashi, S., {et~al.} 2006, \apj, 645, 1315

\bibitem[{{Page} \& {Immler}(2010)}]{gcn11171}
{Page}, K.~L., \& {Immler}, S. 2010, GRB Coordinates Network, 11171, 1

\bibitem[{Panaitescu \& Kumar(2000)}]{pk00}
Panaitescu, A., \& Kumar, P. 2000, \apj, 543, 66

\bibitem[{Panaitescu \& Kumar(2001)}]{pk01}
---. 2001, \apj, 554, 667

\bibitem[{Panaitescu \& Kumar(2002)}]{pk02}
---. 2002, \apj, 571, 779

\bibitem[{Panaitescu \& M\'esz\'aros(1998)}]{pm98}
Panaitescu, A., \& M\'esz\'aros, P. 1998, \apj, 501, 772

\bibitem[{Panaitescu {et~al.}(2006)Panaitescu, M\'esz\'aros, Gehrels, Burrows,
  \& Nousek}]{pmg+06}
Panaitescu, A., M\'esz\'aros, P., Gehrels, N., Burrows, D., \& Nousek, J. 2006,
  \mnras, 366, 1357

\bibitem[{Panaitescu {et~al.}(2013)Panaitescu, Vestrand, \& Wo\'zniak}]{pvw13}
Panaitescu, A., Vestrand, W.~T., \& Wo\'zniak, P. 2013, \mnras, 433, 759

\bibitem[{{Pandey} \& {Zheng}(2010)}]{gcn11179}
{Pandey}, S.~B., \& {Zheng}, W. 2010, GRB Coordinates Network, 11179, 1

\bibitem[{Pei(1992)}]{pei92}
Pei, Y.~C. 1992, \apj, 395, 130

\bibitem[{{Perley} {et~al.}(2012){Perley}, {Alatalo}, \& {Horesh}}]{gcn13175}
{Perley}, D.~A., {Alatalo}, K., \& {Horesh}, A. 2012, GRB Coordinates Network,
  13175, 1

\bibitem[{Perna {et~al.}(2006)Perna, Armitage, \& Zhang}]{paz06}
Perna, R., Armitage, P.~J., \& Zhang, B. 2006, \apjl, 636, L29

\bibitem[{Phinney(1989)}]{phi89}
Phinney, E.~S. 1989, in IAU Symposium, Vol. 136, The Center of the Galaxy, ed.
  M.~Morris, 543

\bibitem[{Piro {et~al.}(1998)Piro, Amati, Antonelli, Butler, Costa, Cusumano,
  Feroci, Frontera, Heise, in~'t Zand, Molendi, Muller, Nicastro, Orlandini,
  Owens, Parmar, Soffitta, \& Tavani}]{paa+98}
Piro, L., Amati, L., Antonelli, L.~A., {et~al.} 1998, \aap, 331, L41

\bibitem[{{Pritchard} \& {Immler}(2010)}]{gcn11176}
{Pritchard}, T.~A., \& {Immler}, S. 2010, GRB Coordinates Network, 11176, 1

\bibitem[{Proga \& Zhang(2006)}]{pz06}
Proga, D., \& Zhang, B. 2006, \mnras, 370, L61

\bibitem[{Rees \& Meszaros(1998)}]{rm98}
Rees, M.~J., \& Meszaros, P. 1998, \apjl, 496, L1

\bibitem[{Rhoads(1999)}]{rho99}
Rhoads, J.~E. 1999, \apj, 525, 737

\bibitem[{Romano {et~al.}(2006)Romano, Moretti, Banat, Burrows, Campana,
  Chincarini, Covino, Malesani, Tagliaferri, Kobayashi, Zhang, Falcone,
  Angelini, Barthelmy, Beardmore, Capalbi, Cusumano, Giommi, Goad, Godet,
  Grupe, Hill, Kennea, La~Parola, Mangano, M\'esz\'aros, Morris, Nousek,
  O'Brien, Osborne, Parsons, Perri, Pagani, Page, Wells, \& Gehrels}]{rmb+06}
Romano, P., Moretti, A., Banat, P.~L., {et~al.} 2006, \aap, 450, 59

\bibitem[{Roming {et~al.}(2005)Roming, Kennedy, Mason, Nousek, Ahr, Bingham,
  Broos, Carter, Hancock, Huckle, Hunsberger, Kawakami, Killough, Koch,
  McLelland, Smith, Smith, Soto, Boyd, Breeveld, Holland, Ivanushkina, Pryzby,
  Still, \& Stock}]{rkm+05}
Roming, P.~W.~A., Kennedy, T.~E., Mason, K.~O., {et~al.} 2005, \ssr, 120, 95

\bibitem[{{Rujopakarn} \& {Flewelling}(2012)}]{gcn13106}
{Rujopakarn}, W., \& {Flewelling}, H. 2012, GRB Coordinates Network, 13106, 1

\bibitem[{{Rumyantsev} {et~al.}(2010){Rumyantsev}, {Shakhovkoy}, \&
  {Pozanenko}}]{gcn11255}
{Rumyantsev}, V., {Shakhovkoy}, D., \& {Pozanenko}, A. 2010, GRB Coordinates
  Network, 11255, 1

\bibitem[{{Sahu} {et~al.}(2012){Sahu}, {Anupama}, \& {Pandey}}]{gcn13185}
{Sahu}, D.~K., {Anupama}, G.~C., \& {Pandey}, S.~B. 2012, GRB Coordinates
  Network, 13185, 1

\bibitem[{{Sahu} {et~al.}(2010{\natexlab{a}}){Sahu}, {Arora}, {Singh}, \&
  {Kartha}}]{gcn11197}
{Sahu}, D.~K., {Arora}, S., {Singh}, N.~S., \& {Kartha}, S.~S.
  2010{\natexlab{a}}, GRB Coordinates Network, 11197, 1

\bibitem[{{Sahu} {et~al.}(2010{\natexlab{b}}){Sahu}, {Bhatt}, \&
  {Arora}}]{gcn11175}
{Sahu}, D.~K., {Bhatt}, B.~C., \& {Arora}, S. 2010{\natexlab{b}}, GRB
  Coordinates Network, 11175, 1

\bibitem[{Salvaterra {et~al.}(2009)Salvaterra, Della~Valle, Campana,
  Chincarini, Covino, D'Avanzo, Fern\'andez-Soto, Guidorzi, Mannucci, Margutti,
  Th\"one, Antonelli, Barthelmy, de~Pasquale, D'Elia, Fiore, Fugazza, Hunt,
  Maiorano, Marinoni, Marshall, Molinari, Nousek, Pian, Racusin, Stella, Amati,
  Andreuzzi, Cusumano, Fenimore, Ferrero, Giommi, Guetta, Holland, Hurley,
  Israel, Mao, Markwardt, Masetti, Pagani, Palazzi, Palmer, Piranomonte,
  Tagliaferri, \& Testa}]{sdvc+09}
Salvaterra, R., Della~Valle, M., Campana, S., {et~al.} 2009, \nat, 461, 1258

\bibitem[{Sari \& Esin(2001)}]{se01}
Sari, R., \& Esin, A.~A. 2001, \apj, 548, 787

\bibitem[{Sari \& M\'esz\'aros(2000)}]{sm00}
Sari, R., \& M\'esz\'aros, P. 2000, \apjl, 535, L33

\bibitem[{Sari {et~al.}(1996)Sari, Narayan, \& Piran}]{snp96}
Sari, R., Narayan, R., \& Piran, T. 1996, \apj, 473, 204

\bibitem[{Sari {et~al.}(1999)Sari, Piran, \& Halpern}]{sph99}
Sari, R., Piran, T., \& Halpern, J.~P. 1999, \apjl, 519, L17

\bibitem[{Sari {et~al.}(1998)Sari, Piran, \& Narayan}]{spn98}
Sari, R., Piran, T., \& Narayan, R. 1998, \apjl, 497, L17+

\bibitem[{Sault {et~al.}(1995)Sault, Teuben, \& Wright}]{stw95}
Sault, R.~J., Teuben, P.~J., \& Wright, M.~C.~H. 1995, in Astronomical Society
  of the Pacific Conference Series, Vol.~77, Astronomical Data Analysis
  Software and Systems IV, ed. R.~A. Shaw, H.~E. Payne, \& J.~J.~E. Hayes, 433

\bibitem[{Schaefer {et~al.}(2003)Schaefer, Gerardy, H\"oflich, Panaitescu,
  Quimby, Mader, Hill, Kumar, Wheeler, Eracleous, Sigurdsson, M\'esz\'aros,
  Zhang, Wang, Hessman, \& Petrosian}]{sgh+03}
Schaefer, B.~E., Gerardy, C.~L., H\"oflich, P., {et~al.} 2003, \apj, 588, 387

\bibitem[{Shao \& Dai(2007)}]{sd07}
Shao, L., \& Dai, Z.~G. 2007, \apj, 660, 1319

\bibitem[{Shen {et~al.}(2008)Shen, Barniol~Duran, \& Kumar}]{sbdk08}
Shen, R.-F., Barniol~Duran, R., \& Kumar, P. 2008, \mnras, 384, 1129

\bibitem[{{Siegel} \& {Marshall}(2010)}]{gcn10625}
{Siegel}, M.~H., \& {Marshall}, F. 2010, GRB Coordinates Network, 10625, 1

\bibitem[{{Siegel} {et~al.}(2012){Siegel}, {Barthelmy}, {Burrows}, {Gehrels},
  {Grupe}, {Hoversten}, {Marshall}, {Palmer}, {Sakamoto}, {Sbarufatti}, \&
  {Swenson}}]{gcn13105}
{Siegel}, M.~H., {Barthelmy}, S.~D., {Burrows}, D.~N., {et~al.} 2012, GRB
  Coordinates Network, 13105, 1

\bibitem[{{Sposetti} \& {Immler}(2010)}]{gcn11213}
{Sposetti}, S., \& {Immler}, S. 2010, GRB Coordinates Network, 11213, 1

\bibitem[{{Stratta} {et~al.}(2012){Stratta}, {Barthelmy}, {Baumgartner},
  {Beardmore}, {Campana}, {D'Elia}, {Evans}, {Guidorzi}, {Holland},
  {Hoversten}, {Kennea}, {Lien}, {Mangano}, {Marshall}, {Maselli}, {Melandri},
  {O'Brien}, {Pagani}, {Page}, {Romano}, {Starling}, {Stroh}, \&
  {Zhang}}]{gcn13208}
{Stratta}, G., {Barthelmy}, S.~D., {Baumgartner}, W.~H., {et~al.} 2012, GRB
  Coordinates Network, 13208, 1

\bibitem[{{Sudilovsky} {et~al.}(2012){Sudilovsky}, {Nicuesa Guelbenzu}, \&
  {Greiner}}]{gcn13129}
{Sudilovsky}, V., {Nicuesa Guelbenzu}, A., \& {Greiner}, J. 2012, GRB
  Coordinates Network, 13129, 1

\bibitem[{Tagliaferri {et~al.}(2005)Tagliaferri, Goad, Chincarini, Moretti,
  Campana, Burrows, Perri, Barthelmy, Gehrels, Krimm, Sakamoto, Kumar,
  M\'esz\'aros, Kobayashi, Zhang, Angelini, Banat, Beardmore, Capalbi, Covino,
  Cusumano, Giommi, Godet, Hill, Kennea, Mangano, Morris, Nousek, O'Brien,
  Osborne, Pagani, Page, Romano, Stella, \& Wells}]{tgc+05}
Tagliaferri, G., Goad, M., Chincarini, G., {et~al.} 2005, \nat, 436, 985

\bibitem[{{Tello} {et~al.}(2012){Tello}, {Sanchez-Ramirez}, {Gorosabel},
  {Castro-Tirado}, {Rivero}, {Gomez-Velarde}, \& {Klotz}}]{gcn13118}
{Tello}, J.~C., {Sanchez-Ramirez}, R., {Gorosabel}, J., {et~al.} 2012, GRB
  Coordinates Network, 13118, 1

\bibitem[{Toma {et~al.}(2006)Toma, Ioka, Yamazaki, \& Nakamura}]{tiyn06}
Toma, K., Ioka, K., Yamazaki, R., \& Nakamura, T. 2006, \apjl, 640, L139

\bibitem[{Totani(1998)}]{tot98}
Totani, T. 1998, \apjl, 502, L13

\bibitem[{Uhm \& Zhang(2014)}]{uz14}
Uhm, Z.~L., \& Zhang, B. 2014, \apj, 789, 39

\bibitem[{Uhm {et~al.}(2012)Uhm, Zhang, Hasco\"et, Daigne, Mochkovitch, \&
  Park}]{uzh+12}
Uhm, Z.~L., Zhang, B., Hasco\"et, R., {et~al.} 2012, \apj, 761, 147

\bibitem[{{Ukwatta} {et~al.}(2010){Ukwatta}, {Sakamoto}, {Gehrels}, \&
  {Dhuga}}]{gcn11198}
{Ukwatta}, T.~N., {Sakamoto}, T., {Gehrels}, N., \& {Dhuga}, K.~S. 2010, GRB
  Coordinates Network, 11198, 1

\bibitem[{{Ukwatta} {et~al.}(2012){Ukwatta}, {Barthelmy}, {Baumgartner},
  {Cummings}, {Fenimore}, {Gehrels}, {Krimm}, {Markwardt}, {Palmer},
  {Sakamoto}, {Sato}, {Stamatikos}, {Stratta}, \& {Tueller}}]{gcn13220}
{Ukwatta}, T.~N., {Barthelmy}, S.~D., {Baumgartner}, W.~H., {et~al.} 2012, GRB
  Coordinates Network, 13220, 1

\bibitem[{{Updike} {et~al.}(2010{\natexlab{a}}){Updike}, {Hartmann}, \& {de
  Pree}}]{gcn10637}
{Updike}, A.~C., {Hartmann}, D.~H., \& {de Pree}, C. 2010{\natexlab{a}}, GRB
  Coordinates Network, 10637, 1

\bibitem[{{Updike} {et~al.}(2010{\natexlab{b}}){Updike}, {Hartmann}, {Keel}, \&
  {Darnell}}]{gcn11174}
{Updike}, A.~C., {Hartmann}, D.~H., {Keel}, W., \& {Darnell}, E.
  2010{\natexlab{b}}, GRB Coordinates Network, 11174, 1

\bibitem[{{Updike} {et~al.}(2010{\natexlab{c}}){Updike}, {Hartmann}, \&
  {Murphy}}]{gcn10619}
{Updike}, A.~C., {Hartmann}, D.~H., \& {Murphy}, B. 2010{\natexlab{c}}, GRB
  Coordinates Network, 10619, 1

\bibitem[{Urata {et~al.}(2014)Urata, Huang, Takahashi, Im, Yamaoka, Tashiro,
  Kim, Jang, \& Pak}]{uht+14}
Urata, Y., Huang, K., Takahashi, S., {et~al.} 2014, \apj, 789, 146

\bibitem[{{van der Horst} {et~al.}(2010{\natexlab{a}}){van der Horst},
  {Kamble}, {Wijers}, {Rol}, {Kouveliotou}, \& {Wiersema}}]{gcn10647}
{van der Horst}, A.~J., {Kamble}, A.~P., {Wijers}, R.~A.~M.~J., {et~al.}
  2010{\natexlab{a}}, GRB Coordinates Network, 10647, 1

\bibitem[{{van der Horst} {et~al.}(2010{\natexlab{b}}){van der Horst},
  {Wiersema}, {Kamble}, {Wijers}, {Rol}, \& {Kouveliotou}}]{gcn11221}
{van der Horst}, A.~J., {Wiersema}, K., {Kamble}, A.~P., {et~al.}
  2010{\natexlab{b}}, GRB Coordinates Network, 11221, 1

\bibitem[{{Volnova} {et~al.}(2010){Volnova}, {Pozanenko}, {Ibrahimov},
  {Hafizov}, \& {Satovski}}]{gcn11266}
{Volnova}, A., {Pozanenko}, A., {Ibrahimov}, M., {Hafizov}, B., \& {Satovski},
  B. 2010, GRB Coordinates Network, 11266, 1

\bibitem[{{Volnova} {et~al.}(2012){Volnova}, {Sinyakov}, {Varda}, \&
  {Molotov}}]{gcn13198}
{Volnova}, A., {Sinyakov}, E., {Varda}, D., \& {Molotov}, I. 2012, GRB
  Coordinates Network, 13198, 1

\bibitem[{Wang {et~al.}(2006)Wang, Li, \& M\'esz\'aros}]{wlm06}
Wang, X.-Y., Li, Z., \& M\'esz\'aros, P. 2006, \apjl, 641, L89

\bibitem[{Wei \& Lu(1998)}]{wl98}
Wei, D.~M., \& Lu, T. 1998, \apj, 505, 252

\bibitem[{Willingale {et~al.}(2010)Willingale, Genet, Granot, \&
  O'Brien}]{wggo10}
Willingale, R., Genet, F., Granot, J., \& O'Brien, P.~T. 2010, \mnras, 403,
  1296

\bibitem[{Wilson {et~al.}(2003)Wilson, Eikenberry, Henderson, Hayward, Carson,
  Pirger, Barry, Brandl, Houck, Fitzgerald, \& Stolberg}]{weh+03}
Wilson, J.~C., Eikenberry, S.~S., Henderson, C.~P., {et~al.} 2003, in Society
  of Photo-Optical Instrumentation Engineers (SPIE) Conference Series, Vol.
  4841, Society of Photo-Optical Instrumentation Engineers (SPIE) Conference
  Series, ed. M.~Iye \& A.~F.~M. Moorwood, 451--458

\bibitem[{{Xin} {et~al.}(2012{\natexlab{a}}){Xin}, {Wei}, {Qiu}, {Wang},
  {Deng}, {Wu}, \& {Han}}]{gcn13150}
{Xin}, L.~P., {Wei}, J.~Y., {Qiu}, Y.~L., {et~al.} 2012{\natexlab{a}}, GRB
  Coordinates Network, 13150, 1

\bibitem[{{Xin} {et~al.}(2012{\natexlab{b}}){Xin}, {Wei}, {Qiu}, {Wang},
  {Deng}, {Wu}, \& {Han}}]{gcn13221}
---. 2012{\natexlab{b}}, GRB Coordinates Network, 13221, 1

\bibitem[{Xin {et~al.}(2012)Xin, Pozanenko, Kann, Xu, Gorosabel, Leloudas, Wei,
  Andreev, Qin, Ibrahimov, Han, de~Ugarte~Postigo, Qiu, Deng, Volnova,
  Jakobsson, Castro-Tirado, Aceituno, Fynbo, Wang, Sanchez-Ramirez, Kouprianov,
  Zheng, Tello, \& Wu}]{xpk+12}
Xin, L.~P., Pozanenko, A., Kann, D.~A., {et~al.} 2012, \mnras, 422, 2044

\bibitem[{{Yoshida} {et~al.}(2010){Yoshida}, {Sasada}, {Komatsu}, \&
  {Kawabata}}]{gcn11190}
{Yoshida}, M., {Sasada}, M., {Komatsu}, T., \& {Kawabata}, K.~S. 2010, GRB
  Coordinates Network, 11190, 1

\bibitem[{Yost {et~al.}(2003)Yost, Harrison, Sari, \& Frail}]{yhsf03}
Yost, S.~A., Harrison, F.~A., Sari, R., \& Frail, D.~A. 2003, \apj, 597, 459

\bibitem[{Yost {et~al.}(2002)Yost, Frail, Harrison, Sari, Reichart, Bloom,
  Kulkarni, Moriarty-Schieven, Djorgovski, Price, Goodrich, Larkin, Walter,
  Shepherd, Fox, Taylor, Berger, \& Galama}]{yfh+02}
Yost, S.~A., Frail, D.~A., Harrison, F.~A., {et~al.} 2002, \apj, 577, 155

\bibitem[{Yu \& Dai(2007)}]{yd07}
Yu, Y.~W., \& Dai, Z.~G. 2007, \aap, 470, 119

\bibitem[{{Zauderer} {et~al.}(2012){Zauderer}, {Laskar}, \&
  {Berger}}]{gcn13231}
{Zauderer}, A., {Laskar}, T., \& {Berger}, E. 2012, GRB Coordinates Network,
  13231, 1

\bibitem[{Zhang {et~al.}(2006)Zhang, Fan, Dyks, Kobayashi, M\'esz\'aros,
  Burrows, Nousek, \& Gehrels}]{zfd+06}
Zhang, B., Fan, Y.~Z., Dyks, J., {et~al.} 2006, \apj, 642, 354

\bibitem[{Zhang \& M\'esz\'aros(2001)}]{zm01}
Zhang, B., \& M\'esz\'aros, P. 2001, \apjl, 552, L35

\bibitem[{Zhang \& M\'esz\'aros(2002)}]{zm02}
---. 2002, \apj, 566, 712

\bibitem[{Zhang {et~al.}(2007)Zhang, Liang, Page, Grupe, Zhang, Barthelmy,
  Burrows, Campana, Chincarini, Gehrels, Kobayashi, M\'esz\'aros, Moretti,
  Nousek, O'Brien, Osborne, Roming, Sakamoto, Schady, \& Willingale}]{zlp+07}
Zhang, B., Liang, E., Page, K.~L., {et~al.} 2007, \apj, 655, 989

\bibitem[{{Zhao} {et~al.}(2012){Zhao}, {Mao}, {Xu}, \& {Bai}}]{gcn13122}
{Zhao}, X.-H., {Mao}, J., {Xu}, D., \& {Bai}, J.-M. 2012, GRB Coordinates
  Network, 13122, 1

\bibitem[{Zou {et~al.}(2005)Zou, Wu, \& Dai}]{zwd05}
Zou, Y.~C., Wu, X.~F., \& Dai, Z.~G. 2005, \mnras, 363, 93

\end{thebibliography}

\end{document}